\documentclass{cernyrep} 
\pdfoutput=1
\usepackage{texnames}
\usepackage[T1]{fontenc}
\usepackage[bookmarks, colorlinks=true, linktoc=page, pdftex, linkcolor=blue, citecolor=blue, urlcolor=blue]{hyperref}
\usepackage{import}
\begin{document}
\begin{flushright}
CERN-TH-2016-113
\end{flushright}
\title{Physics at a 100 TeV $pp$ collider: Higgs and EW symmetry
  breaking studies}
 
\IfFileExists{../authors.tex}{\author{
Editors: \\
R.~Contino$^{1,2}$,\, 
D.~Curtin$^{3}$,\,
A.~Katz$^{1,4}$,\,
M.~L.~Mangano$^{1}$,\,
G.~Panico$^{5}$,\,
M.~J.~Ramsey-Musolf$^{6,7}$,\, 
G.~Zanderighi$^{1}$\,
\\[0.4cm]
Contributors: \\
C.~Anastasiou$^{8}$,\,
W.~Astill$^{9}$,\,
G.~Bambhaniya$^{21}$,\, 
J.~K.~Behr$^{10,11}$,\,
W.~Bizon$^{9}$,\,
P.~S.~Bhupal~Dev$^{12}$,\,
D.~Bortoletto$^{10}$,\,
D.~Buttazzo$^{22}$\,  
Q.-H.~Cao$^{13,14,15}$,\,
F.~Caola$^{1}$,\,
J.~Chakrabortty$^{16}$,\, 
C.-Y.~Chen$^{17,18,19}$,\, 
S.-L.~Chen$^{15,20}$,\, 
D.~de~Florian$^{23}$,\, 
F.~Dulat$^{8}$,\,
C.~Englert$^{24}$,\,  
J.~A.~Frost$^{10}$,\, 
B.~Fuks$^{25}$,\,  
T.~Gherghetta$^{26}$,\, 
G.~Giudice$^{1}$,\,
J.~Gluza$^{27}$,\,  
N.~Greiner$^{28}$,\,
H.~Gray$^{29}$,\,
N.~P.~Hartland$^{10}$,\,
V.~Hirschi$^{30}$,\,
C.~Issever$^{10}$,\,
T.~Jeli\'nski$^{27}$,\,  
A.~Karlberg$^{9}$,\,
J.~H.~Kim$^{31,32,33}$,\, 
F.~Kling$^{34}$,\,  
A.~Lazopoulos$^{8}$,\,
S.~J.~Lee$^{35,36}$,\, 
Y.~Liu$^{13}$,\,
G.~Luisoni$^{1}$,\,
O.~Mattelaer$^{37}$,\,
J.~Mazzitelli$^{23,38}$,\, 
B.~Mistlberger$^{1}$,\,
P.~Monni$^{9}$,\,
K.~Nikolopoulos$^{39}$,\,  
R.~N~Mohapatra$^{3}$,\,
A.~Papaefstathiou$^{1}$,\,
M.~Perelstein$^{40}$,\,  
F.~Petriello$^{41}$,\,  
T.~Plehn$^{42}$,\,  
P.~Reimitz$^{42}$,\, 
J.~Ren$^{43}$,\, 
J.~Rojo$^{10}$,\,
K.~Sakurai$^{37}$,\, 
T.~Schell$^{42}$,\, 
F.~Sala$^{44}$,\, 
M.~Selvaggi$^{45}$,\, 
H.-S.~Shao$^{1}$,\,
M.~Son$^{31}$,\,
M.~Spannowsky$^{37}$,\,
T.~Srivastava$^{16}$,\,
S.-F.~Su$^{34}$,\,
R.~Szafron$^{46}$,\, 
T.~Tait$^{47}$,\,   
A.~Tesi$^{48}$,\,   
A.~Thamm$^{49}$,\,  
P.~Torrielli$^{50}$,\,
F.~Tramontano$^{51}$,\,
J.~Winter$^{52}$,\,
A.~Wulzer$^{53}$,\,
Q.-S.~Yan$^{54,55,56}$,\,
W.~M.~Yao$^{57}$,\,
Y.-C.~Zhang$^{58}$,\,
X.~Zhao$^{54}$,\,
Z.~Zhao$^{54,59}$,\,
Y.-M.~Zhong$^{60}$
\vspace*{2cm}
}

\institute{
$^{1}$ CERN, TH Department, CH-1211 Geneva, Switzerland. \\
$^{2}$ EPFL, Lausanne, Switzerland. \\
$^{3}$Maryland Center for Fundamental Physics, Department of Physics, University of Maryland, College
Park, MD 20742, USA\\
$^{4}$Universit\'{e} de Gen\`{e}ve, Department of Theoretical Physics and Center for Astroparticle Physics,
24 quai E. Ansermet, CH-1211, Geneva 4, Switzerland\\
$^{5}$ IFAE, Universitat Aut\`onoma de Barcelona, E-08193 Bellaterra, Barcelona\\
$^{6}$ Amherst Center for Fundamental Interactions, Physics Department,University of Massachusetts Amherst, Amherst, MA 01003, USA\\
$^{7}$ Kellogg Radiation Laboratory, California Institute of Technology, Pasadena, CA 91125 USA\\
$^{8}$ Institute for Theoretical Physics, ETH Z\"urich,  8093 Z\"urich, Switzerland \\
$^{9}$ Rudolf Peierls Centre for Theoretical Physics, 1 Keble Road, University of Oxford, UK \\
$^{10}$ Physics Department, 1 Keble Road, University of Oxford, United Kingdom\\
$^{11}$ Deutsches Elektronen-Synchrotron (DESY), Notkestrasse 85, D-22607 Hamburg, Germany\\
$^{12}$ Max-Planck-Institut f\"ur Kernphysik, Saupfercheckweg 1,
D-69117 Heidelberg, Germany\\
$^{13}$ Department of Physics and State Key Laboratory of Nuclear Physics and Technology, Peking University, Beijing 100871, China\\
$^{14}$ Collaborative Innovation Center of Quantum Matter, Beijing 100871, China\\
$^{15}$ Center for High Energy Physics, Peking University, Beijing 100871, China\\
$^{16}$  Department of Physics, Indian Institute of Technology, Kanpur-208016, India\\
$^{17}$ Department of Physics, Brookhaven National Laboratory, Upton, New York 11973, USA\\
$^{18}$ Department of Physics and Astronomy, University of Victoria, Victoria, British Columbia V8P 5C2, Canada\\
$^{19}$ Perimeter Institute for Theoretical Physics, Waterloo, Ontario N2J 2W9, Canada\\
$^{20}$ Key Laboratory of Quark and Lepton Physics (MoE) and Institute of Particle Physics, Central China Normal University, Wuhan 430079, China\\
$^{21}$ Theoretical Physics Division, Physical Research Laboratory, Ahmedabad-380009, India\\
$^{22}$ Physik-Institut, Universit\"at Z\"urich, CH-8057 Z\"urich, Switzerland \\ 
$^{23}$ International Center for Advanced Studies (ICAS), UNSAM, Campus Miguelete, 25 de Mayo y Francia, (1650) Buenos Aires, Argentina\\
$^{24}$ SUPA, School of Physics and Astronomy, University of Glasgow,
Glasgow G12 8QQ, UK \\
$^{25}$ Sorbonne Universit\'es, UPMC Univ.~Paris 06, UMR 7589, LPTHE, F-75005 Paris, France.  CNRS, UMR 7589, LPTHE, F-75005 Paris, France \\
$^{26}$ School of Physics and Astronomy, University of Minnesota, Minneapolis, MN 55455, USA\\
$^{27}$ Institute of Physics, University of Silesia, Uniwersytecka 4, 40-007 Katowice, Poland\\
$^{28}$ Physik Institut, Universit\"at Z\"urich, Winterthurerstrasse 190, CH-8057 Z\"urich, Switzerland \\
$^{29}$ CERN, EP Department, CH-1211 Geneva, Switzerland. \\
$^{30}$ SLAC, National Accelerator Laboratory, 2575 Sand Hill Road, Menlo Park, CA 94025-7090, USA \\
$^{31}$ Department of Physics, Korea Advanced Institute of Science and Technology, 335 Gwahak-ro, Yuseong-gu, Daejeon 305-701, Korea \\
$^{32}$ Center for Theoretical Physics of the Universe, IBS, 34051 Daejeon, Korea\\
$^{33}$ Center for Axion and Precision Physics Research, IBS, 34141 Daejeon, Korea \\
$^{34}$ Department of Physics, University of Arizona, Tucson, AZ 85721, USA\\
$^{35}$ Department of Physics, Korea University, Seoul 136-713, Korea\\
$^{36}$ School of Physics, Korea Institute for Advanced Study, Seoul 130-722, Korea\\
$^{37}$ IPPP, Department of Physics, University of Durham, Science Laboratories, South Road, Durham, DH1 3LE, UK\\
$^{38}$ Departamento de Fisica and IFIBA, FCEyN, Universidad de
Buenos Aires, (1428) Pabellon 1 Ciudad Universitaria, Capital Federal,
Argentina \\
$^{39}$ School of Physics and Astronomy, University of Birmingham,
Birmingham, B15 2TT, United Kingdom \\
$^{40}$Laboratory for Elementary Particle Physics, Cornell University,
Ithaca, NY 14853, USA\\
$^{41}$ 
Northwestern University, Department of Physics and Astronomy, 2145
Sheridan Road, 
Evanston, Illinois 60208-3112, USA\\
$^{42}$ Institut f\"ur Theoretische Physik, Universit\"at Heidelberg, Germany \\
$^{43}$ Department of Physics, University of Toronto, Toronto, Ontario, Canada M5S1A7\\
$^{44}$ 2LPTHE, CNRS, UMR 7589, 4 Place Jussieu, F-75252, Paris, France\\
$^{45}$ Centre for Cosmology, Particle Physics and Phenomenology CP3, Universit\'e Catholique de Louvain, Chemin du Cyclotron, 1348 Louvain la Neuve, Belgium\\
$^{46}$ Department of Physics, University of Alberta, Edmonton, AB
T6G 2E1, Canada\\
$^{47}$ Department of Physics and Astronomy, University of California, Irvine, CA 92697, USA\\
$^{48}$ Enrico Fermi Institute, University of Chicago, Chicago, IL 60637, USA\\
$^{49}$ PRISMA Cluster of Excellence and Mainz Institute for Theoretical Physics, Johannes Gutenberg University, 55099 Mainz, Germany\\
$^{50}$ Dipartimento di Fisica, Universit\`a di Torino, and INFN, Sezione di Torino, \\
Via P.~Giuria~1, I-10125, Turin, Italy \\
$^{51}$ Universit\`a di Napoli ``Federico II'' and INFN, Sezione di
Napoli, 80126 Napoli, Italy \\
$^{52}$Department of Physics and Astronomy, Michigan State University,
East Lansing, MI 48824, USA\\
$^{53}$Dipartimento di Fisica e Astronomia, Universit\`{a} di Padova and
INFN, Sezione di Padova, via Marzolo 8, I-35131 Padova, Italy\\
$^{54}$ School of Physical Sciences, University of Chinese Academy of Sciences, Beijing 100049, P. R. China\\
$^{55}$ Center for High-Energy Physics, Peking University, Beijing 100871, P. R. China\\
$^{56}$ Center for future high energy physics, Chinese Academy of Sciences 100049, P. R. China\\
$^{57}$ Lawrence Berkeley National Lab (LBNL), One Cyclotron Rd, Berkeley, CA94720, USA\\
$^{58}$ Service de Physique Th\'eorique, Universit\'e Libre de Bruxelles, Boulevard du Triomphe, CP225, 1050 Brussels, Belgium\\
$^{59}$ Department of Physics, University of Siegen, 57068 Siegen, Germany\\
$^{60}$ C.N. Yang Institute for Theoretical Physics, Stony Brook University, Stony Brook, New York 11794, USA\\
}
}{}

\maketitle 

\begin{abstract}
This report summarises the physics opportunities for the study of
Higgs bosons and the dynamics of electroweak symmetry breaking at
the 100~TeV $pp$ collider.
\end{abstract}
 
\tableofcontents
 
\newpage 

\newcommand{\met}{E\!\!\!\!/_T} 
\def\iab{ab$^{-1}$}
\def\ifb{fb$^{-1}$}
\def\ipb{pb$^{-1}$}
\def \gsim{\mathrel{\vcenter
     {\hbox{$>$}\nointerlineskip\hbox{$\sim$}}}}
\def \lsim{\mathrel{\vcenter
     {\hbox{$<$}\nointerlineskip\hbox{$\sim$}}}}

%
\section{Foreword}
A 100~TeV $pp$ collider is under consideration, by the high-energy
physics community~\cite{benedikt,cepc_website},
as an important target for the future development
of our field, following the completion of the LHC and High-luminosity
LHC physics programmes. The physics opportunities and
motivations for such an ambitious
project were recently reviewed in~\cite{Arkani-Hamed:2015vfh}. The
general considerations on the strengths and reach of very high energy
hadron colliders have been introduced long ago in the classic pre-SSC
EHLQ review~\cite{Eichten:1984eu}, and a possible framework to establish the
luminosity goals of such accelerator was presented recently
in~\cite{Hinchliffe:2015qma}.

The present document is the result of an extensive study, carried out
as part of the Future Circular Collider (FCC) study towards a Conceptual
Design Report, which includes separate Chapters dedicated to Standard
Model physics~\cite{SMreport}, physics of the Higgs boson and EW symmetry
breaking (this Cjapter), physics beyond the Standard
Model~\cite{Golling:2016gvc}, physics of heavy ion
collisions~\cite{Dainese:2016gch} and physics with the FCC injector
complex~\cite{inj-report}. Studies on the physics programme of an
$e^+e^-$ collider (FCC-ee) and $ep$ collider (FCC-eh)
at the FCC facility are proceeding
in parallel, and preliminary results are documented
in~\cite{Gomez-Ceballos:2013zzn} (for FCC-ee) and
in~\cite{AbelleiraFernandez:2012cc} (for the LHeC precursor of FCC-eh).

\section{Introduction} 
\label{sec:intro}
Despite its impressive success in accounting for a wide range of
experimental observations, the Standard Model (SM) leaves many of our most
important questions unanswered:
\begin{itemize}
\item Why does the universe contain more matter than anti-matter?
\item What is the identity of the dark matter and what are its interactions?
\item Why are the masses of neutrinos so much smaller than those of all other known elementary fermions?
\end{itemize}
The discovery~\cite{Aad:2012tfa,Chatrchyan:2012xdj}
of the Higgs-like scalar~\cite{Englert:1964et,Higgs:1964ia,Higgs:1964pj,Guralnik:1964eu,Higgs:1966ev,Kibble:1967sv} at the LHC highlights
additional theoretical puzzles.  The scalar sector is the ``obscure"
sector of the SM, in the sense that it is the least understood part of
the theory.  The principles dictating its structure are still
unclear. This is to be contrasted with the gauge sector, which
logically follows from an elegant symmetry principle and has all the
features of a fundamental structure. Not surprisingly, many of the
open problems of the SM are connected to the Higgs sector. For
example, the stability of the Higgs mass~\cite{Weisskopf:1939zz} and
of the Electroweak (EW) scale in general against UV-sensitive
radiative corrections motivates additional symmetry structures near
the TeV scale. To address these questions, possible theoretical extensions of
the SM have been proposed. Their experimental manifestations can be
direct, via the 
production of new particles, or indirect, via deviations of the Higgs
properties from their SM predictions.

With its higher energy and the associated increase in parton
luminosity, a 100 TeV pp collider would provide an unprecedented
potential to both detect new particles, and to explore in detail the
Higgs boson properties, uniquely complementing the capabilities at the
LHC and possible future $e^+e^-$ colliders. This Chapter is dedicated
to a first assessment of this potential.

In the first Section we review what is known today about the
production properties of the 125~GeV Higgs boson at 100~TeV. We
present evidence that the increased energy does not introduce
uncertainties larger than those already established from the studies
for the
LHC~\cite{Dittmaier:2011ti,Dittmaier:2012vm,Heinemeyer:2013tqa}.
Furthermore, the large production rates available at 100~TeV open new
opportunities to optimize the balance between statistical
uncertainties, background contamination, and systematic uncertainties
of both theoretical and experimental origin.  The second Section
illustrates these ideas with a few concrete examples of possible
precision attainable at 100~TeV. These are not
intended to provide a robust and definitive assessment of the ultimate
goals; this would be premature, since both the theoretical landscape
(higher-order corrections, resummations, PDFs, and event simulation
tools in general) and the future detectors' performance potential are
far from being known. Rather, these examples suggest possible new
directions, which on paper and in the case of idealized analysis
scenarios offer exciting opportunities to push the precision and the
reach of Higgs physics into a domain that will hardly be attainable by
the LHC (although some of these ideas might well apply to the HL-LHC
as well).

The third Section addresses the determination of the Higgs self-coupling
and the measurement of the Higgs potential. This is important for
several reasons. In the SM, the shape of the Higgs potential is
completely fixed by the mass and vacuum expectation value of the Higgs
field.  Therefore, an independent measurement of the trilinear and
quadrilinear Higgs self-interactions provides important additional
tests of the validity of the SM.  This test is quite
non-trivial. Indeed, as discussed in the final Section, in many
Beyond-the-SM (BSM) scenarios sizable corrections to the Higgs
self-couplings are predicted, which, in some cases, can lead to large
deviations in multi-Higgs production processes but not in other
observables. In these scenarios, an analysis of the non-linear Higgs
couplings can be more sensitive to new-physics effects than other
direct or indirect probes~\cite{Grober:2010yv,Contino:2012xk}.
This Section includes an overview of the production rates for
multiple Higgs production, including those of associated production
and in the vector-boson fusion channel.
This is followed by a detailed up-to-date study of the possible
precision with which the triple Higgs coupling can be measured, and a
first assessment of the potential to extract information on the
quartic coupling.

Determining the structure of the Higgs potential is also important to
understand the features of the EW phase transition, whose properties
can have significant implications for cosmology. For instance, a
strong first order transition could provide a viable scenario to
realize baryogenesis at the EW scale (see for
example~\cite{Trodden:1998ym} and references therein).  In the SM the
EW transition is known to be rather weak (for a Higgs mass $m_h
\sim 70-80$~GeV, only a cross-over is predicted), so that it is not
suitable for a successful baryogenesis. Many BSM scenarios, however,
predict modifications in the Higgs potential that lead to first order
EW transitions, whose strength could allow for a viable baryogenesis.
An additional aspect related to the structure of the Higgs potential
is the issue of the stability of the EW vacuum (see for instance
Ref.~\cite{Degrassi:2012ry}). The final Section of this Chapter will
address these questions in great detail.
This Section will also study the impact of studies in the Higgs sector
on the issue of Dark Matter, on the origin of neutrino masses, and on
naturalness. Extensions of the SM affecting the Higgs sector of the
theory often call for the existence of additional scalar degrees of
freedom, either fundamental or emergent.  Such \emph{Beyond the SM
  (BSM) Higgs sectors} frequently involve new singlet or
electroweak-charged fields, making their discovery at the LHC
challenging. The prospects for their direct observation at 100~TeV
will be presented in the final part of the Section.

The results and observations presented throughout this document, in
addition to put in perspective the crucial role of a 100~TeV $pp$
collider in clarifying the nature of the Higgs boson and electroweak
symmetry breaking, can be used as benchmarks to define detector
performance goals, or to exercise new analysis concepts (focused, for
example, on the challenge of tagging multi-TeV objects such as top and
bottom quarks, or Higgs and gauge bosons).  Equally important, they
will hopefully trigger complete analyses, as well as new ideas and
proposals for interesting observables. Higgs physics at 100~TeV will
not just be a larger-statistics version of the LHC, it will have the
potential of being a totally new ballgame.

\clearpage 
\section{SM Higgs production}
\label{sec:SM}
We discuss in this Section the 125~GeV SM Higgs boson production
properties at 100 TeV, covering total rates and kinematical
distributions. Multiple Higgs production is discussed in
Section~\ref{sec:HH}.

For ease of reference, and for the dominant production channels,
we summarize in Table~\ref{table:HiggsXS} the central values of the
total cross sections that will be described in more detail below. The
increases with respect to the LHC energy are very large, ranging from 
a factor of $\sim 10$ for the $VH$ $(V=W,Z)$ associated production, to
a factor of $\sim 60$ for the $t\bar{t}H$ channel. As will be shown in
this section, much larger increases are expected for kinematic
configurations at large transverse momentum.

With these very large rate increases, it is important to verify that
the relative accuracy of the predictions does not deteriorate.  We
shall therefore present the current estimates of theoretical
systematics, based on the available calculations of QCD and
electroweak perturbative corrections, and on the knowledge of the
proton parton distribution functions (PDFs).  With the long time
between now and the possible operation of the FCC-hh, the results
shown here represent only a crude and conservative
picture of the precision that will
eventually be available. But it is extremely encouraging that, already
today, the typical systematical uncertainties at 100 TeV, whether due
to missing higher-order effects or to PDFs, are comparable to those at 14
TeV. This implies that, in perspective, the FCC-hh has a great
potential to perform precision measurements of the Higgs boson.
A first assessment of this potential will be discussed in the next Section.

In addition to the standard production processes, we document, in the
last part of this Section, the rates of rarer channels of
associated production (e.g. production with multiple gauge bosons).
These processes could allow independent tests of the Higgs boson
properties, and might provide channels with improved signal over
background, with possibly reduced systematic uncertainties. We hope
that the first results shown here will trigger some dedicated
phenomenological analysis. For a recent overview of Higgs physics at
33 and 100~TeV, see also~\cite{Baglio:2015wcg}.

\begin{table}[h!]
\begin{center}
\def\arraystretch{1.5}
\begin{tabular}{ l | c | c | c | c| c }
  &  $gg\to H$  & $VBF$ 
  & $HW^\pm$  & $HZ$ 
  & $t\bar{t}H$
  \\
  & (Sect~\ref{sec:H_ggH}) &  (Sect~\ref{sec:H_vbf})
  & (Sect~\ref{sec:H_VH}) & (Sect~\ref{sec:H_VH})
  &  (Sect~\ref{sec:H_ttH})
\\   \hline
$\sigma$(pb) & 802 & 69 & 15.7 & 11.2 & 32.1
\\
$\sigma$(100 TeV)/$\sigma$(14 TeV) &  16.5 & 16.1  & 10.4 & 11.4 & 52.3
\\
\end{tabular}
\end{center}
\caption{Upper row: cross sections at a $100\,$TeV collider for the production of a SM Higgs
  boson in $gg$ fusion, vector boson fusion, associated production
  with $W$ and $Z$ bosons, and associated production with a $t\bar{t}$
  pair. Lower row: rate
  increase relative to 14~TeV~\cite{HXSWG}. The details of the
  individual processes are described in the relevant subsections.
}
\label{table:HiggsXS}
\end{table}

\subsection{Inclusive $gg\to H$ production}
\label{sec:H_ggH}

In this section we analyse the production of a Standard Model Higgs boson via the gluon fusion production mode at a 100 TeV proton proton collider. 
As at the LHC with 13 TeV this particular production mode represents the dominant channel for the production of Higgs bosons.

We relate perturbative QFT predictions to the cross section at a $\sqrt{S}=$ 100 TeV collider using the general factorisation formula 
\begin{equation}
\label{eq:sigma}
\sigma = \tau \sum_{ij} \int_{\tau}^1  \frac{dz}{z}\int_{\frac{\tau}{z}}^1 \frac{dx} {x}  f_i \left(x \right) f_j\left(\frac{\tau}{z x} \right) 
\frac{\hat{\sigma}_{ij}(z)}{z}
 \,,
\end{equation} 
where  $\hat{\sigma}_{ij}$ are the partonic cross sections for
producing a Higgs boson from a scattering of partons $i$ and $j$, and $f_i$ and
$f_j$ are the corresponding parton densities. We have defined the ratios
\begin{equation}
\tau = \frac{m_H^2}{S} {\rm~~and~~} z = \frac{m_H^2}{s}\,.
\end{equation}
Here, $s$ is the partonic center of mass energy.
In the wake of the LHC program tremendous efforts have been made from the phenomenology community to improve the theoretical predictions for the Higgs boson cross section. In this section we want to briefly review the various ingredients for a state of the art prediction for the FCC and discuss the associated uncertainties.
To this end we split the partonic cross section as follows.
\begin{equation}\label{eq:master}
\hat{\sigma}_{ij} \simeq R_{LO}\,\left(\hat{\sigma}_{ij,EFT} + \hat{\sigma}_{ij,EW}\right) + \delta\hat{\sigma}_{ij,ex;t,b,c}^{LO}+\delta\hat{\sigma}_{ij,ex;t,b,c}^{NLO}\, + \delta_{t}\hat{\sigma}_{ij,EFT}^{NNLO}.
\end{equation}

The relatively low mass of the Higgs boson in comparison to the top threshold allows the use of an effective theory in which we regard a limit of infinite top quark mass and only consider the effects of massless five-flavour QCD on the gluon fusion cross section. This effective theory is described by an effective Lagrangian~\cite{Georgi:1977gs,Wilczek1977,Shifman1978,Inami1983,Spiridonov:1988md}
\begin{equation}\label{eq:L_eff}
\mathcal{L}_{\text{EFT}}=\mathcal{L}_{\textrm{QCD},5}-\frac{1}{3\pi}C \, H\, G_{\mu\nu}^a G_a^{\mu\nu},
\end{equation}
where the Higgs boson is coupled to the Yang-Mills Lagrangian of QCD via a Wilson coefficient~\cite{Chetyrkin:1997un,Chetyrkin:2005ia,Schroder:2005hy} and $\mathcal{L}_{\textrm{QCD},5}$ is the QCD Lagrangian with five massless quark flavours.
The cross section $\hat{\sigma}_{ij,EFT}$ is the partonic cross section for Higgs production computed in this effective theory. It captures the dominant part of the gluon fusion production mode~\cite{Dawson:1990zj,Harlander2002,Ravindran:2003um,Anastasiou2002}. Recently, it was computed through N$^3$LO in perturbation theory~\cite{Anastasiou2015}. 

Effects due to the fact that the top mass is finite need to be included in order to make precision predictions for the inclusive Higgs boson production cross section.
At LO and NLO in QCD the full dependence on the quark masses is known~\cite{Aglietti:2006tp,Spira1995,Harlander2005a,Anastasiou:2009kn,Djouadi:1991tka,Graudenz1993,Bonciani2007,Anastasiou:2006hc}.

First, to improve the behaviour of the effective theory cross section we rescale $\hat{\sigma}_{ij,EFT}$ with the constant ratio
\begin{equation}
R_{\text{LO}} \equiv \frac{\sigma_{ex;t}^{\text{LO}}}{\sigma_{\text{EFT}}^{\text{LO}}}\,,
\label{eq:KLO}
\end{equation}
where $\sigma_{ex;t}^{\text{LO}}$ is the leading order cross section in QCD computed under the assumption that only the top quark has a non-vanishing Yukawa coupling. $\sigma_{\text{EFT}}^{\text{LO}}$ is the leading order effective theory cross section. In order to also include important effects due to the non-vanishing Yukawa coupling of the bottom and charm quark we correct our cross section prediction at LO with $\delta\hat{\sigma}_{ij,ex;t,b,c}^{LO}$ and at NLO with $\delta\hat{\sigma}_{ij,ex;t,b,c}^{NLO}$. These correction factors account for the exactly known mass dependence at LO and NLO beyond the rescaled EFT.
The exact mass dependence at NNLO is presently unknown. However, corrections due to the finite top mass beyond the rescaled EFT have been computed as an expansion in the inverse top mass~\cite{Harlander2010,Pak:2009dg,Harlander:2009mq}. We account for these effects with the term $\delta_{t}\hat{\sigma}_{ij,EFT}^{NNLO}$.

Besides corrections due to QCD it is important to include electroweak corrections to the inclusive production cross section.
The electroweak corrections to the LO cross section at first order in the weak coupling were computed~\cite{Actis2008a,Aglietti2004a} and an approximation to mixed higher order corrections at first order in the weak as well as the strong coupling exists~\cite{Anastasiou2009b}. We account for these corrections with $\hat{\sigma}_{ij,EW}$.

Next, we study the numerical impact of the aforementioned contributions on the Higgs boson cross section at 100 TeV and estimate the respective uncertainties. We implemented the effects mentioned above into a soon to be released version of the code $\mathtt{iHixs}$~\cite{Anastasiou2011,Anastasiou2012} and evaluated the cross section with the setup summarised in table~\ref{tab:setup}.
\begin{table}[!thb]
\center
\begin{tabular}{cc}
\hline
    
    $\sqrt{S}$	&  100TeV                      \\
$m_h$  		&  125GeV                      \\
PDF 			&  {\tt PDF4LHC15\_nnlo\_100}                      \\
$a_s(m_Z)$ 	&  0.118                    \\
$m_t(m_t)$	&  162.7  ($\overline{MS}$)\\
$m_b(m_b)$	&  4.18  ($\overline{MS}$)\\
$m_c(3GeV)$	&  0.986  ($\overline{MS}$)\\
$\mu=\mu_R=\mu_F$	&  62.5  ($=m_h/2$)\\
\hline
\end{tabular}
\caption{Setup}
\label{tab:setup}
\end{table}
Throughout the following analysis we choose parton distribution functions provided by ref.~\cite{Rojo2015}. For a detailed analysis of the various sources of uncertainties at 13 TeV we refer the interested reader to ref.~\cite{Anastasiou:2016cez}.

\subsubsection{Effective Theory}
The Higgs boson cross section is plagued by especially large perturbative QCD corrections. 
The dominant part of these corrections is captured by the effective field theory description of the cross section introduced in eq.~\eqref{eq:L_eff}. 
As a measure for the uncertainty of the partonic cross section due to the truncation of the perturbative series we regard the dependence of the cross section on the unphysical scale $\mu$ of dimensional regularisation. We will choose a central scale $\mu_{\text{central}}=\frac{m_h}{2}$ for the prediction of the central value of our cross section and vary the scale in the interval $\mu\in\left[\frac{m_h}{4},m_h\right]$ to obtain an estimate of the uncertainty due to missing higher orders.

First, we investigate the dependence of $\hat{\sigma}_{ij,EFT}$ computed through different orders in perturbation theory on the hadronic center of mass energy $S$ as plotted in fig.~\ref{fig:es}. 
\begin{figure}[ht!]
\begin{center}
\includegraphics[width=0.8\textwidth]{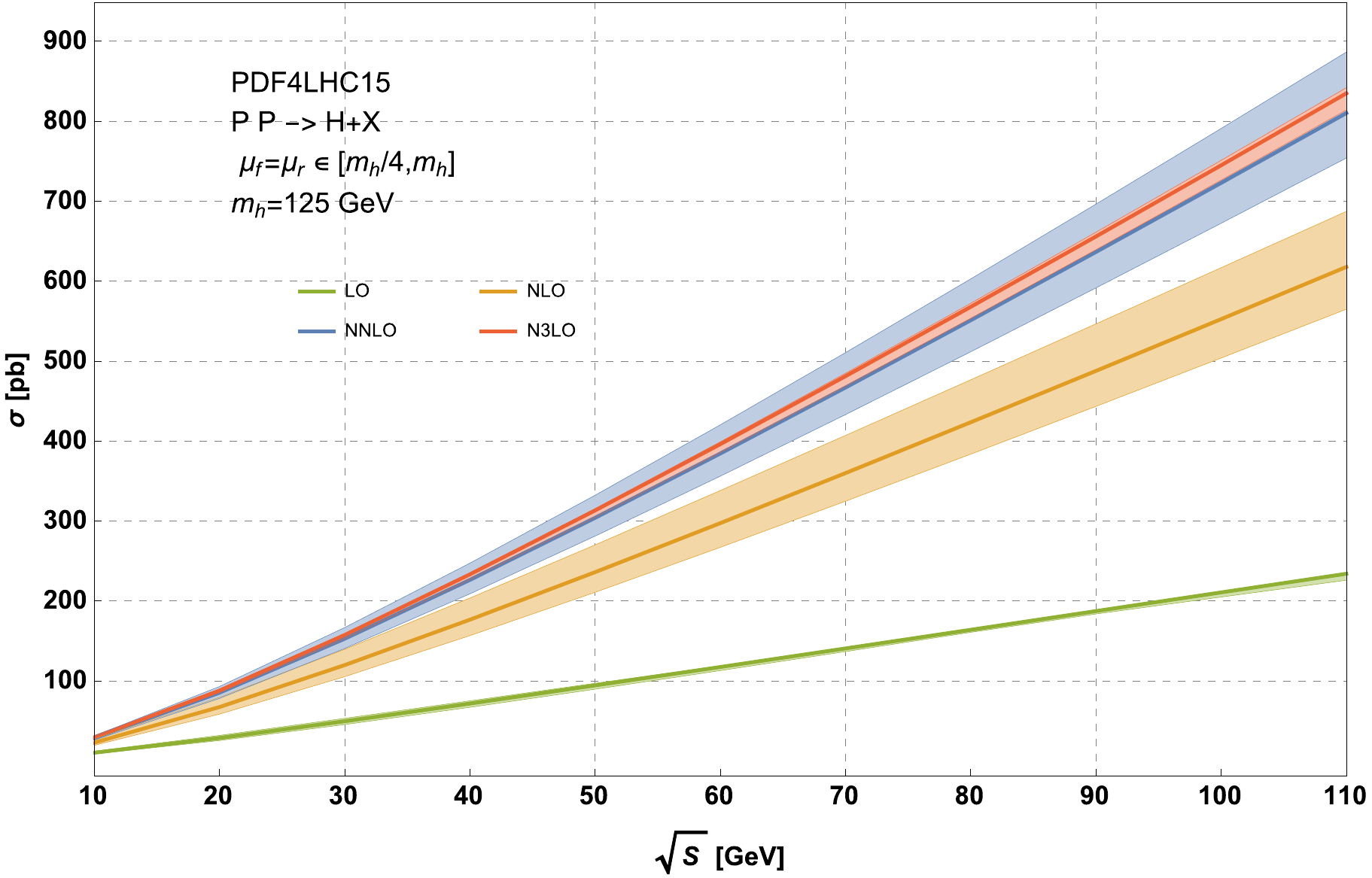}
\end{center}
\caption{\label{fig:es} The effective theory gluon fusion cross section at all perturbative orders through N$^3$LO in the scale interval $\left[\frac{m_h}{4},m_h\right]$ as a function of the collider energy $\sqrt{S}$.}
\end{figure}
One can easily see that an increase of the center of mass energy leads to a more than linear increase of the production cross section.
Furthermore, we observe that higher orders in perturbation theory play an important role for precise predictions for the Higgs boson cross section. The lower orders dramatically underestimate the cross section and particular the scale uncertainty. Only with the recently obtained N${}^{3}$LO corrections~\cite{Anastasiou2015} the perturbative series finally stabilises and the uncertainty estimate due to scale variations is significantly reduced.

In fig.~\ref{fig:kfac} we plot the effective theory $K$-factor for various orders in perturbation theory.
\begin{equation}
K^{(n)}=\frac{\sigma_{\text{EFT}}^{\text{N$^n$LO}}}{\sigma_{\text{EFT}}^{\text{LO}}}.
\end{equation}
Here, $\sigma_{\text{EFT}}^{\text{N$^n$LO}}$ is the hadronic Higgs production cross section based on the effective theory prediction through N$^n$LO. 
One can easily see that QCD corrections become slightly more important as we increase the center of mass energy.
The relative size of the variation of the cross section due to variation of the common scale $\mu$ is roughly independent of the center of mass energy of the proton collider. 
\begin{figure}[ht!]
\begin{center}
\includegraphics[width=0.8\textwidth]{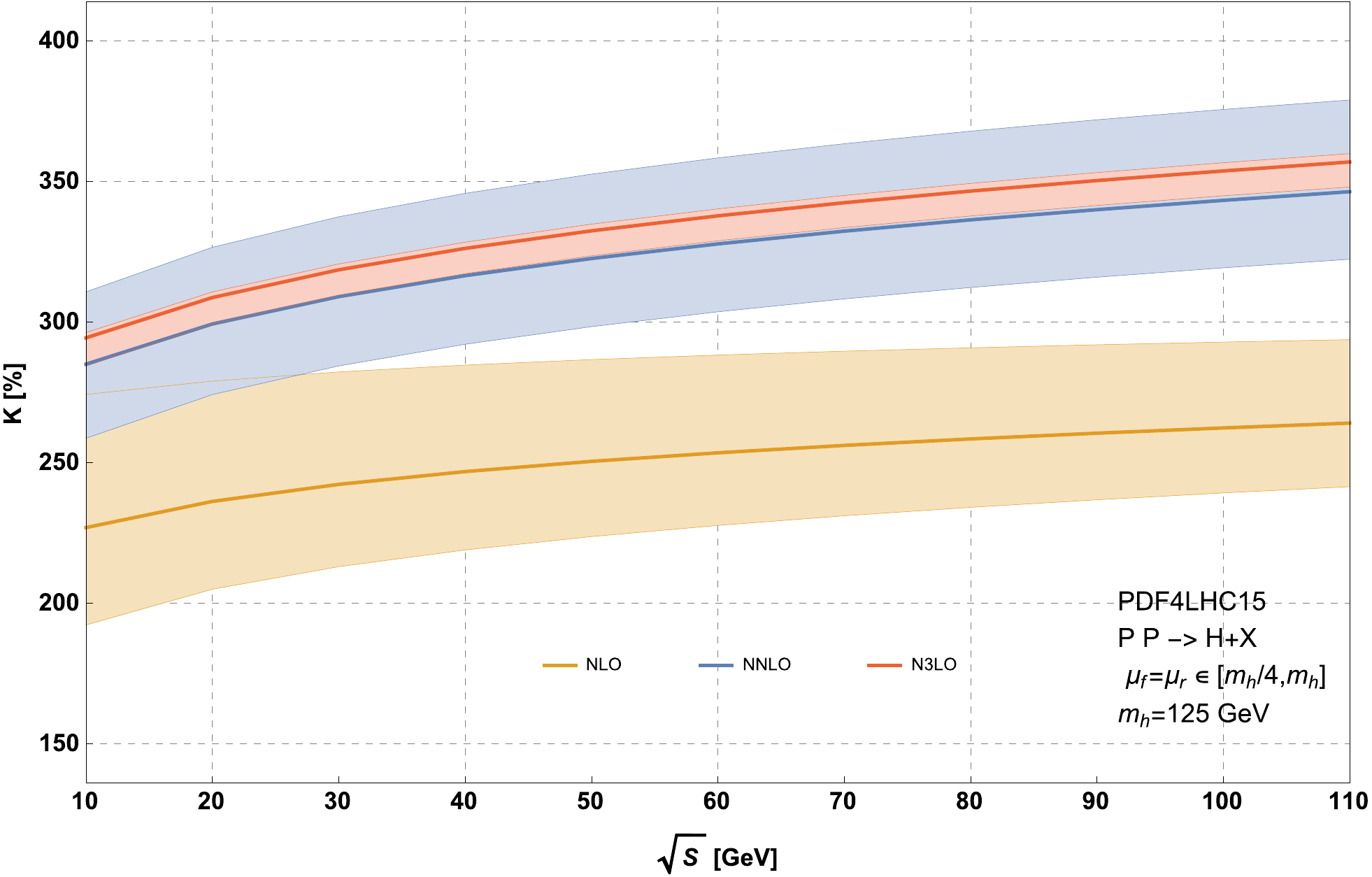}
\end{center}
\caption{\label{fig:kfac} QCD $K$-factor for the effective theory Higgs production cross section as a function of the hadronic center of mass energy.}
\end{figure}

\subsubsection{Quark Mass Effects}
First, let us discuss the quality of the effective theory approach considered above. The cross section obtained with this approach corresponds to the leading term in an expansion of the partonic cross section in $\delta=\frac{s}{4m_t^2}$. In fig~\ref{fig:lumi} we plot the gluon luminosity for Higgs production as a function of $z$.
The area that represents the production of gluons with a partonic center of mass energy larger than $2m_t$ is shaded in red and in green for the complement. 
In $\sim 96 \% $ of all events in which a gluon pair has large enough energy to produce a Higgs boson the expansion parameter $\delta$ is smaller than one and the effective theory can be expected to perform reasonably well. In comparison, at 13 TeV $\delta$ is smaller than one for $\sim 98 \%$ of all gluon pairs that are produced with a center of mass energy larger than the Higgs boson mass.
\begin{figure}[!ht]
\begin{center}
\includegraphics[width=0.8\textwidth]{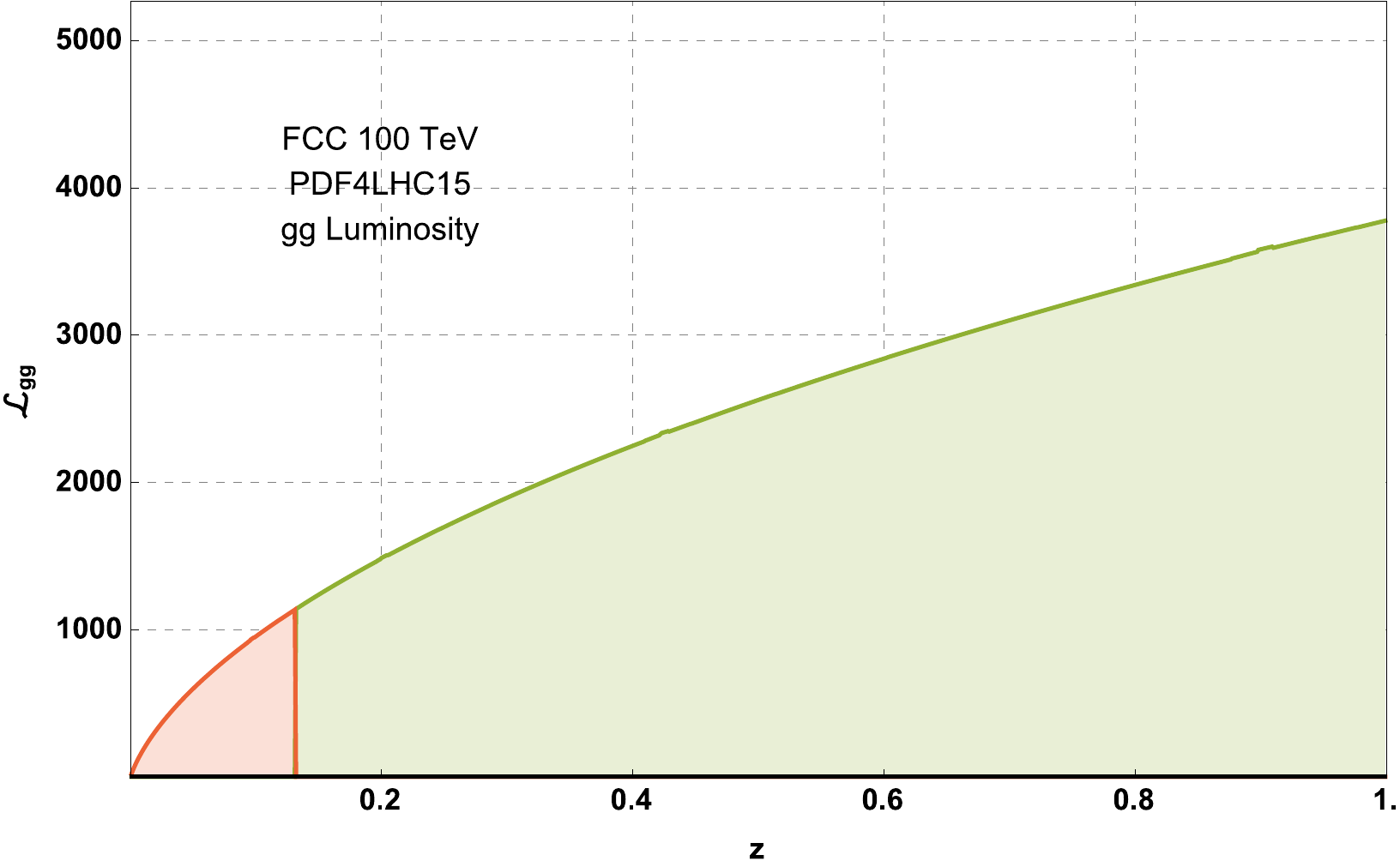}
\end{center}
\caption{\label{fig:lumi} Higgs production gluon luminosity at a 100 TeV proton proton collider. The area shaded in red corresponds to partonic center of mass energy larger than $2 m_t$ and the green area to partonic center of mass energy less than $2m_t$.}
\end{figure}

Next, let us asses the performance of the rescaled effective theory quantitatively.
The rescaling by the ratio $R_{LO}=1.063$ provides a reasonable approximation of the cross section with full top mass dependence. If we consider the exact corrections due to the top quark through NLO we find only a mild correction of $2.8 \%$ on top of the rescaled effective theory NLO cross section. 
At NNLO the exact dependence of the QCD cross section on the top quark mass is unknown and only higher order terms in the expansion in $\delta$ are available. These amount to $1.1\%$ of the total cross section. Following the recommendation of ref.~\cite{Harlander2010} we assign a matching uncertainty of $\delta_{\frac{1}{m_t}}=\pm 1\%$  due to the incomplete NNLO corrections.

Of considerable importance are effects due to the interference of amplitudes coupling light quarks to the Higgs and amplitudes with the usual top quark Yukawa interaction. 
At LO and NLO we find destructive interference of these contributions and we include them as part of $\hat{\sigma}_{ij,ex;t,b,c}^{NLO}$. Currently, no computation of interference effects of light and heavy quark amplitudes at NNLO is available and we asses the uncertainty due to these missing contributions via
\begin{equation}
\delta_{tbc}=\pm \Bigg| \frac{\sigma_{ex;t}^{\text{NLO}}-\sigma_{ex;t,b,c}^{\text{NLO}}}{\sigma_{ex;t}^{\text{NLO}}}\Bigg|  \frac{R_{\text{LO}} \delta\sigma_{\text{EFT}}^{\text{NNLO}}}{ \sigma}=\pm 0.8\%.
\end{equation}
Here, $\sigma_{ex;t}^{\text{NLO}}$ and $\sigma_{ex;t,b,c}^{\text{NLO}}$ are the hadronic cross sections based on NLO partonic cross sections containing mass effects from the top quark only and mass effects from the top, bottom and charm quark respectively. $\delta\sigma_{\text{EFT}}^{\text{NNLO}}$ is the NNLO correction to the cross section resulting from the effective theory partonic cross section.

Parametric uncertainties due to the imprecise knowledge of the quark masses are small and we neglect in all further discussions.

\subsubsection{Electroweak corrections}
Electroweak corrections at $\mathcal{O}(\alpha)$ were computed in
ref.~\cite{Actis2008a}. 
These corrections contain only virtual contributions and are thus
independent of the energy.  
We currently include them as 
\begin{equation}
\hat{\sigma}_{ij,EW}= \kappa_{EW} \times \hat{\sigma}_{ij,EFT},
\end{equation}
where $\kappa_{EW}$ is the rescaling factor arising due to the
electroweak corrections. 
Electroweak corrections beyond $\mathcal{O}(\alpha)$ where
approximated in ref.~\cite{Anastasiou2009b}. We also include those
effects and assign residual uncertainty of $\delta_{\text{EW}}=\pm 1
\%$ on the total cross section due to missing higher order mixed
electroweak and QCD corrections~\cite{Anastasiou:2016cez}. 

Electroweak corrections for Higgs production in gluon fusion in
association with a jet were computed by~\cite{Keung:2009bs}. These
turn out to be negligible for the inclusive cross section. 

\subsubsection{$\alpha_S$ and PDF uncertainties}
The strong coupling constant and the parton distribution functions are
quantities that are extracted from a large set of diverse
measurements. Naturally, there is an uncertainty associated with these
quantities that has to be taken into account when deriving predictions
for the Higgs boson production cross section.  
Here, we follow the prescription outlined by the PDF4LHC working group
in ref.~\cite{Rojo2015} to derive the PDF and $\alpha_S$ uncertainty
for the Higgs production cross section.  
We find
\begin{equation}
\delta_{\text{PDF}}=\pm2.5 \%,\hspace{1cm}\delta_{\alpha_S}=\pm 2.9\%.
\end{equation}
In fig.~\ref{fig:pdferrvar} we plot the PDF and $\alpha_S$ uncertainty for the effective theory cross section as a function of the scale $\mu$ normalised to its central value.
 \begin{figure}[ht!]
\begin{center}
\includegraphics[width=0.8\textwidth]{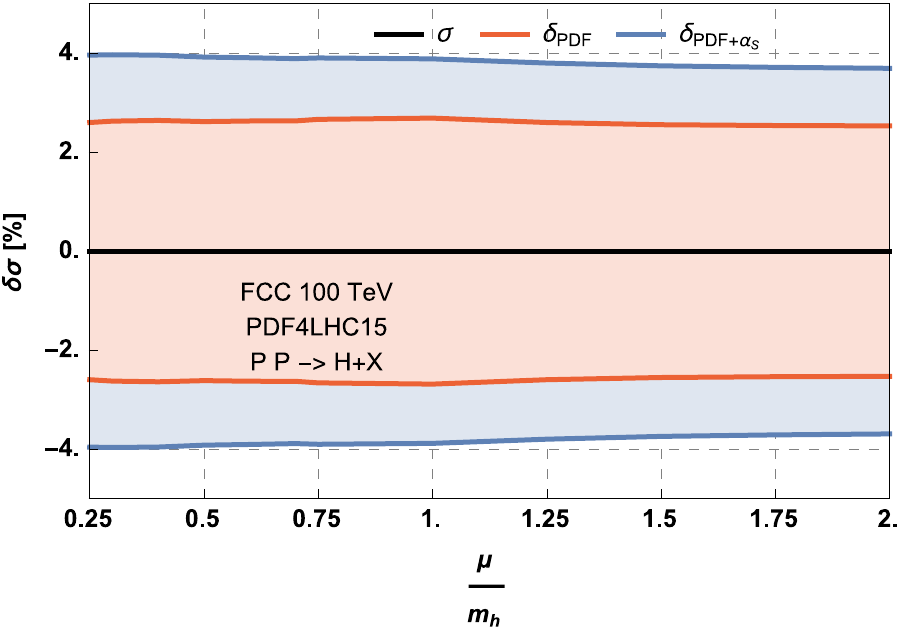}
\end{center}
\caption{\label{fig:pdferrvar} PDF and $\alpha_S$ uncertainty of the effective theory cross section as a function of the pertubative scale $\mu$ normalised to the central value of $\sigma_{\text{EFF}}(\mu)$.}
\end{figure}

We want to remark that the predictions obtained here are subject to the choice of the parton distribution functions. Especially choosing parton distribution functions and strong coupling constant according to ref.~\cite{Alekhin2014,Alekhin2012} results in quantitatively different predictions. This discrepancy is not covered by current uncertainty estimates and should be resolved.

Currently, parton distributions are obtained using cross sections computed up to at most NNLO. 
As we combine these NNLO parton distribution functions with the effective theory cross section computed at N${}^3$LO we have to assign an uncertainty for the miss-match.  
As a measure for this uncertainty $\delta_{\text{PDF-theo}}$ we use 
\begin{equation}
\delta_{\text{PDF-theo}}=\pm \frac{1}{2}\left| \frac{\sigma_{\text{EFT, NNLO}}^{\text{NNLO}}-\sigma_{\text{EFT, NLO}}^{\text{NNLO}}}{\sigma_{\text{EFT, NNLO}}^{\text{NNLO}}}\right| = \pm 2.7\%.
\end{equation}
Here, $\sigma_{\text{EFT, N$^n$LO}}^{\text{NNLO}}$ is the hadronic cross section resulting from the convolution of the effective theory NNLO partonic cross section with N$^n$LO parton distribution functions. For both orders we use PDF sets provided by the PDF4LHC working group~\cite{Rojo2015}.

\subsubsection{Summary}
In this section we have discussed state-of-the-art predictions for the inclusive Higgs boson production cross section via gluon fusion at a $100\,$TeV proton-proton 
collider.  This inclusive cross section will be accessible experimentally at percent level precision and an in-depth theoretical understanding of this observable is 
consequently paramount to a successful Higgs phenomenology program at the FCC. 

Already now we are in the position to derive high-precision predictions for this cross section. The current state-of-the-art prediction with its associated uncertainties is:
\begin{equation}
\label{eq:all}
\sigma= 802 \text{ pb}\, ^{+6.1\%}_{-7.2 \%}(\delta_\text{theo})^{+2.5\%}_{-2.5 \%}(\delta_{\text{PDF}})^{+2.9\%}_{-2.9 \%}(\delta_{ \alpha_s}).
\end{equation}
A more detailed summary of all sources of uncertainties we included can be found in tab.~\ref{tab:uncertainties}. In eq.~\eqref{eq:all} we combined all but the PDF and $\alpha_S$ uncertainty linearly to obtain one theoretical uncertainty $\delta_{\text{theo}}$ for the gluon fusion Higgs production cross section at 100 TeV.
\begin{table}
\center
\begin{tabular}{ccccccc}
\hline
$\delta_{\text{PDF}}$  & $\delta_{\alpha_S}$  & $\delta_{\text{scale}}$ & $\delta_{\text{PDF-theo}}$ &  $\delta_{\text{EW}}$ &  $\delta_{\text{tbc}}$ & $\delta_{\frac{1}{m_t}}$  \\
$\pm$ 2.5\% & $\pm$ 2.9\% & ${}^{+0.8\% }_{-1.9\%}$ & $\pm$ 2.7\% & $\pm$ 1\% & $\pm$ 0.8\% & $\pm$ 1\% \\
\hline
\end{tabular}
\caption{\label{tab:uncertainties} Various sources of uncertainties of the inclusive gluon fusion Higgs production cross section at a 100 TeV proton-proton collider.}
\end{table}
It is interesting to see how the inclusive cross section is comprised of the different contributions discussed above. The breakdown of the cross section is
\begin{eqnarray}
802 \text{pb} 
&=& 223.7 \,\text{pb } \hspace{1cm} (\text{LO},\, R_{LO} \times \text{EFT})\nonumber\\
&+& 363.1  \,\text{pb }\hspace{1cm} (\text{NLO},\, R_{LO} \times \text{EFT})\nonumber\\
&-&    37.2  \,\text{pb} \hspace{1.25cm} ((t,b,c),\,  \text{exact NLO})\nonumber\\
&+&  181.1  \,\text{pb  } \hspace{1cm} (\text{NNLO},\, R_{LO} \times \text{EFT})\nonumber\\
&+& 8.2  \,\text{pb          }  \hspace{1.4cm} (\text{NNLO},\, 1/m_t)\nonumber\\
&+& 39.5  \,\text{pb       } \hspace{1.2cm} (\text{Electro-Weak})\nonumber\\
&+& 23.6  \,\text{pb        } \hspace{1.2cm} (\text{N3LO},\, R_{LO} \times \text{EFT}).
\end{eqnarray}

The experimental and theoretical advances in anticipation of a 100 TeV collider will help to elevate the inclusive Higgs production cross section to an unprecedented level of precision that will enable future collider studies to tackle the precision frontier.
Improvements of the experimental methods and extraction methods as well as refined theoretical predictions will lead to more precise determinations of the strong coupling constant and of the parton distributions. This will serve to greatly reduce the dominant sources of uncertainty that plague the Higgs cross section at the current level of precision.
One of the most important advances for precision in anticipation of a 100 TeV collider will be the extraction of N$^3$LO  parton distribution functions. This will unlock the full benefit of the N$^3$LO calculation of partonic cross section and lead to a significant reduction of the residual uncertainty.
Another milestone for theoretical predictions will be computation of the NNLO partonic cross sections with full dependence on the quark masses. This computation would simultaneously shrink the uncertainties due to $\delta_{tbc}$ as well as $\delta_{\frac{1}{m_t}}$.
Furthermore, an improved understanding of electroweak effects will be highly desirable. In particular a full calculation of the mixed QCD and electroweak corrections to Higgs production will lead to a better control of the residual uncertainties and bring the inclusive Higgs cross section to an even higher level of precision.

\clearpage
\subsection{Higgs plus jet and Higgs $p_T$ spectrum in $gg\to H$}
\label{sec:H_ggHpt}
In this section we study the production of Higgs in gluon fusion in association
with one extra jet and more in general we analyze the transverse momentum spectrum of the Higgs.
Results in this section are obtained using MCFM~\cite{Campbell:2010ff} and~\cite{Caola:2015wna}.

\subsubsection{Jet veto efficiencies}
At 100 TeV, extra jet radiation is enhanced and a significant fraction of Higgs boson events
is produced in association with one or more extra jets. To quantify this statement, in Fig.~\ref{fig:efficiency}
we plot the jet veto efficiency at 100 TeV, defined as the fraction of exactly 0-jet events in
the total Higgs sample
\begin{equation}
\epsilon(p_{\rm{t,veto}}) \equiv 1-\frac{\Sigma_{\rm 1-jet,incl} (p_{\rm t,veto})}{\sigma_{\rm tot}},
\end{equation}
as well as the one-jet inclusive cross section as a function of the jet transverse momentum requirement,
$\Sigma_{\rm 1-jet,incl} (p_{\rm t,veto}) \equiv \int_{p_{\rm t,veto}}^{\infty}
\frac{{\rm d} \sigma}{{\rm d}p_{\rm t,jet}} {\rm d} p_{\rm t,jet}$. 
\begin{figure}[h]
\includegraphics[width=0.45\textwidth]{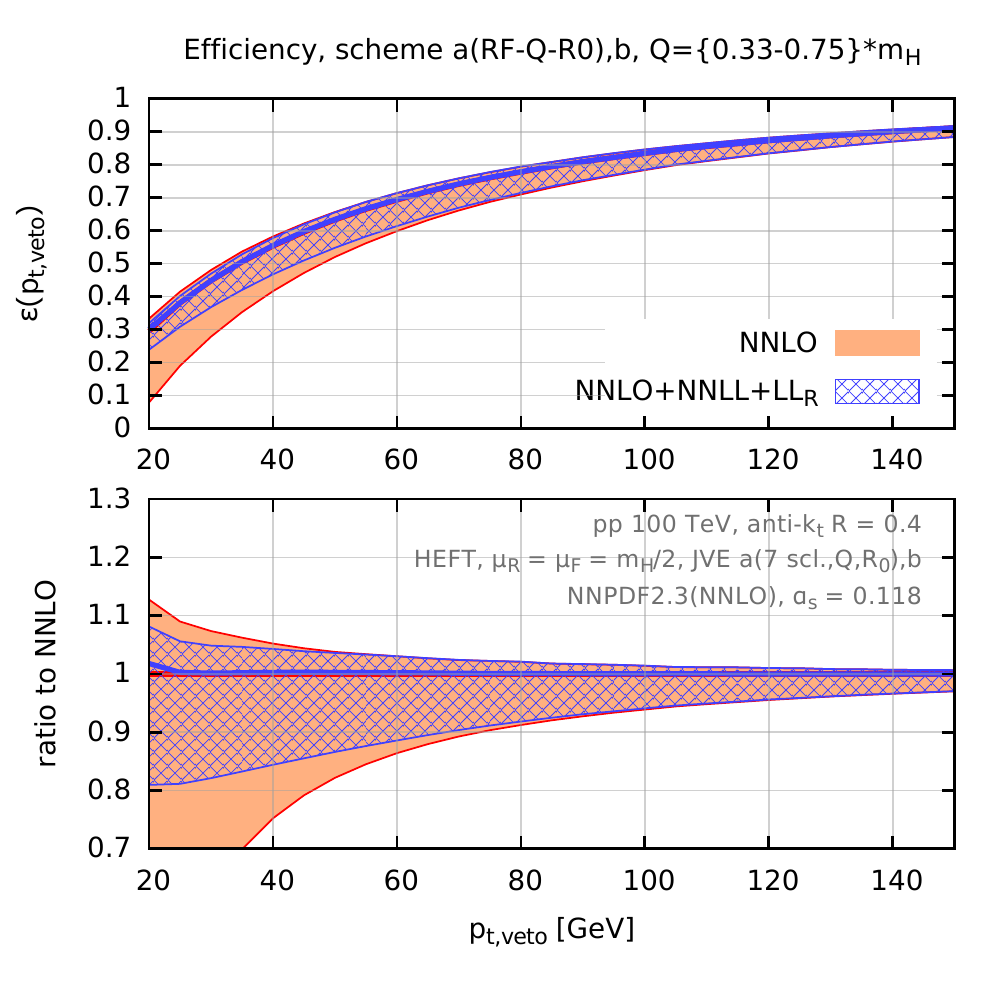}
\includegraphics[width=0.45\textwidth]{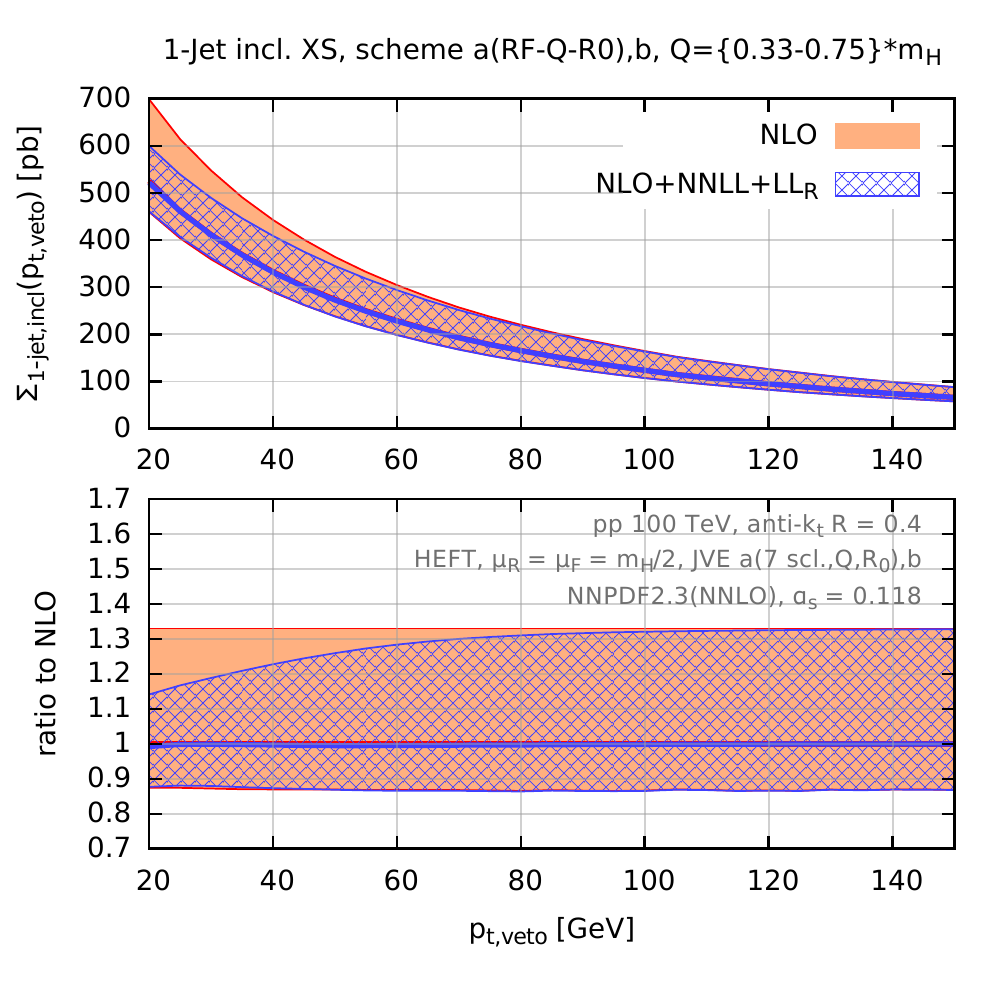}
\caption{Jet veto efficiency (left) and 1-jet inclusive cross section for Higgs production
in gluon fusion at 100 TeV, see text for details.}\label{fig:efficiency}
\end{figure}
Throughout this section, jets are reconstructed with the anti-$k_t$
algorithm with $R=0.4$. No rapidity cut on the jet is applied. The
efficiency and one-jet cross section shown in
Fig. ~\ref{fig:efficiency} are computed both in pure fixed-order
perturbation theory (red/solid) and matched to NNLL
$\ln m_H/p_{\rm t,veto}$ and LL jet-radius $\ln R$ resummation
(blue/hatched). The uncertainties are obtained with the
Jet-Veto-Efficiency method, see~\cite{Banfi:2015pju} for details.
For a jet $p_t$ of $\sim 60~{\rm GeV}$, Fig.~\ref{fig:efficiency}
shows that about $30\%$ of the total Higgs cross section comes from
events with one or more jets. Also, for jet transverse momenta larger
than $\sim 60$~GeV it is also clear from Fig.~\ref{fig:efficiency}
that pure fixed-order perturbation theory provides an excellent
description of the jet efficiencies and cross sections. All-order
resummation effects become sizable at smaller transverse momenta,
where however soft physics effects like underlying event and MPI may
play an important role at the centre-of-mass energies considered in
this report.

\subsubsection{The Higgs $p_T$ spectrum}
We now study the Higgs cross-section as a function of a cut on the transverse momentum of the Higgs
boson $\sigma(p_{T,H}>p_{t,{\rm cut}})$. 
\begin{figure}
\includegraphics[width=0.45\textwidth]{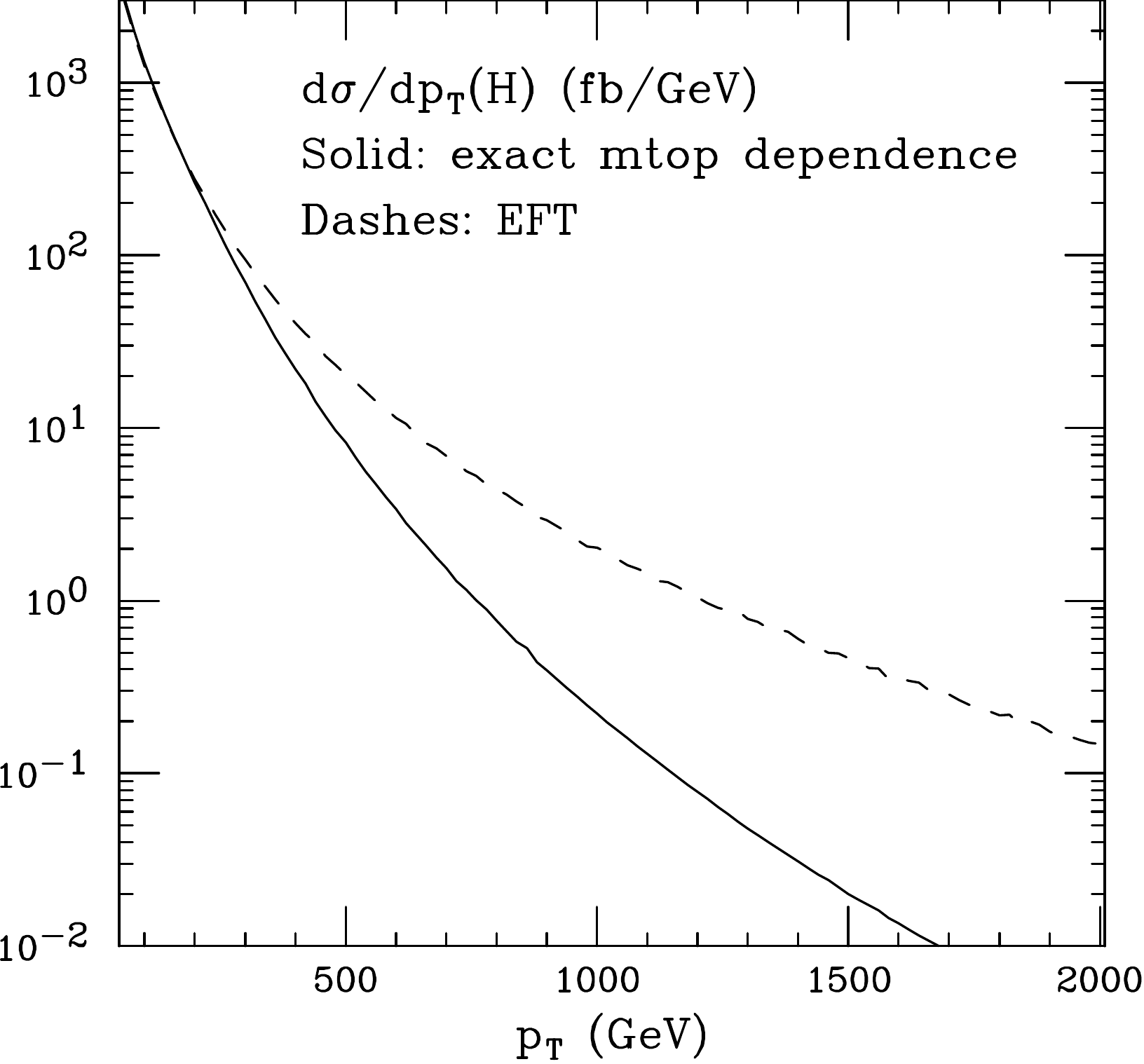}
\hfil
\includegraphics[width=0.45\textwidth]{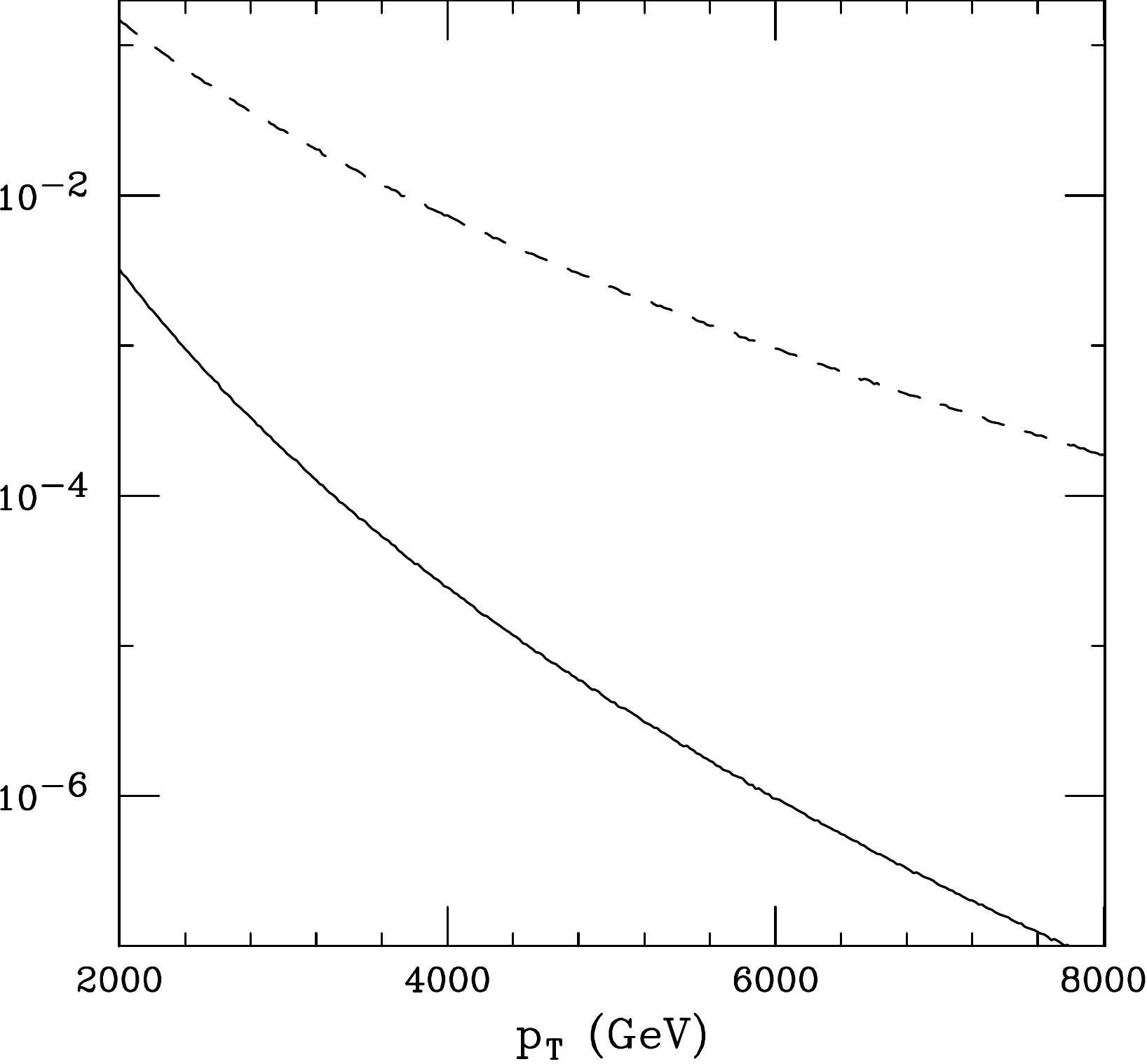}
\\
\includegraphics[width=0.45\textwidth]{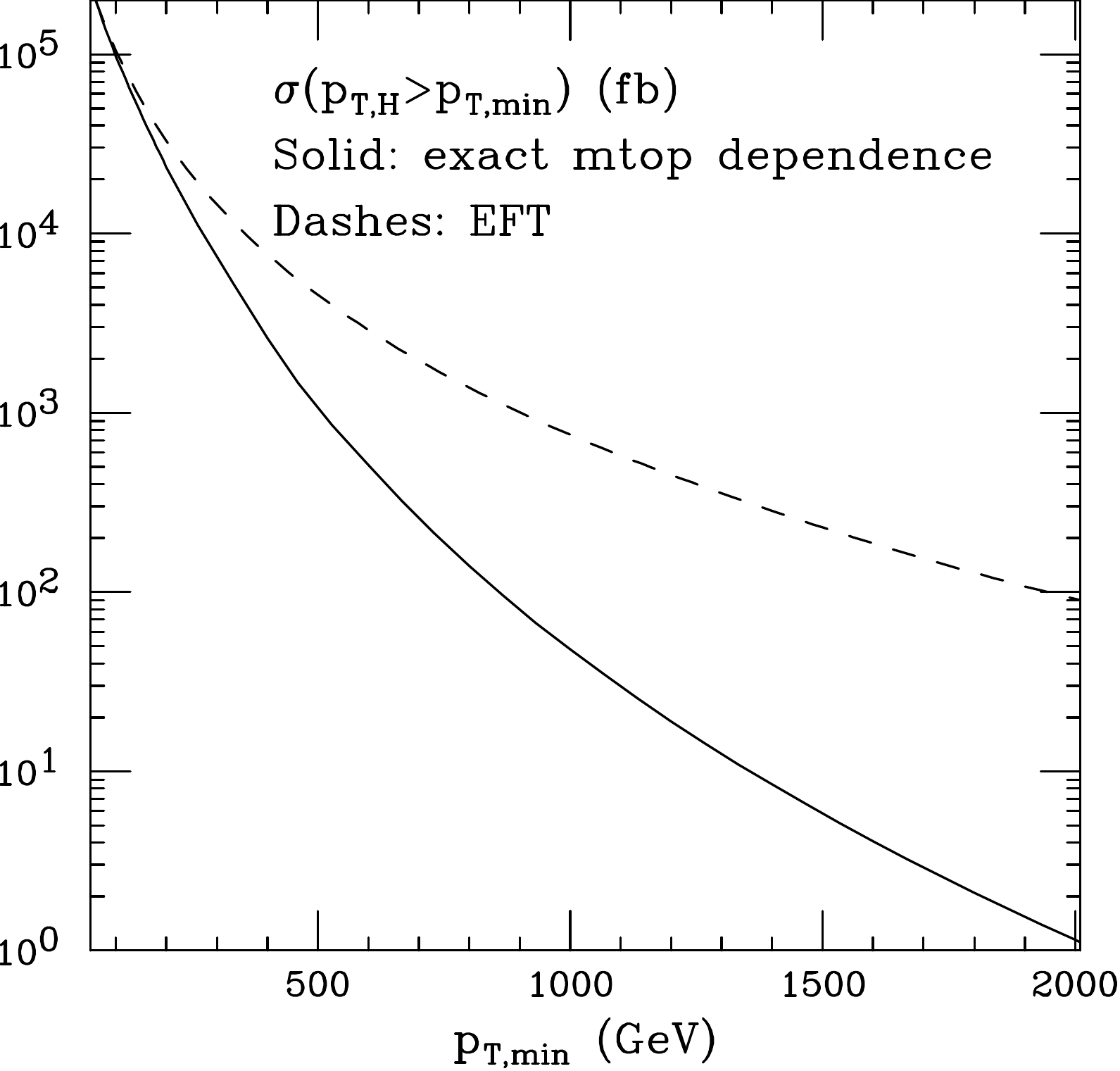}
\hfil
\includegraphics[width=0.45\textwidth]{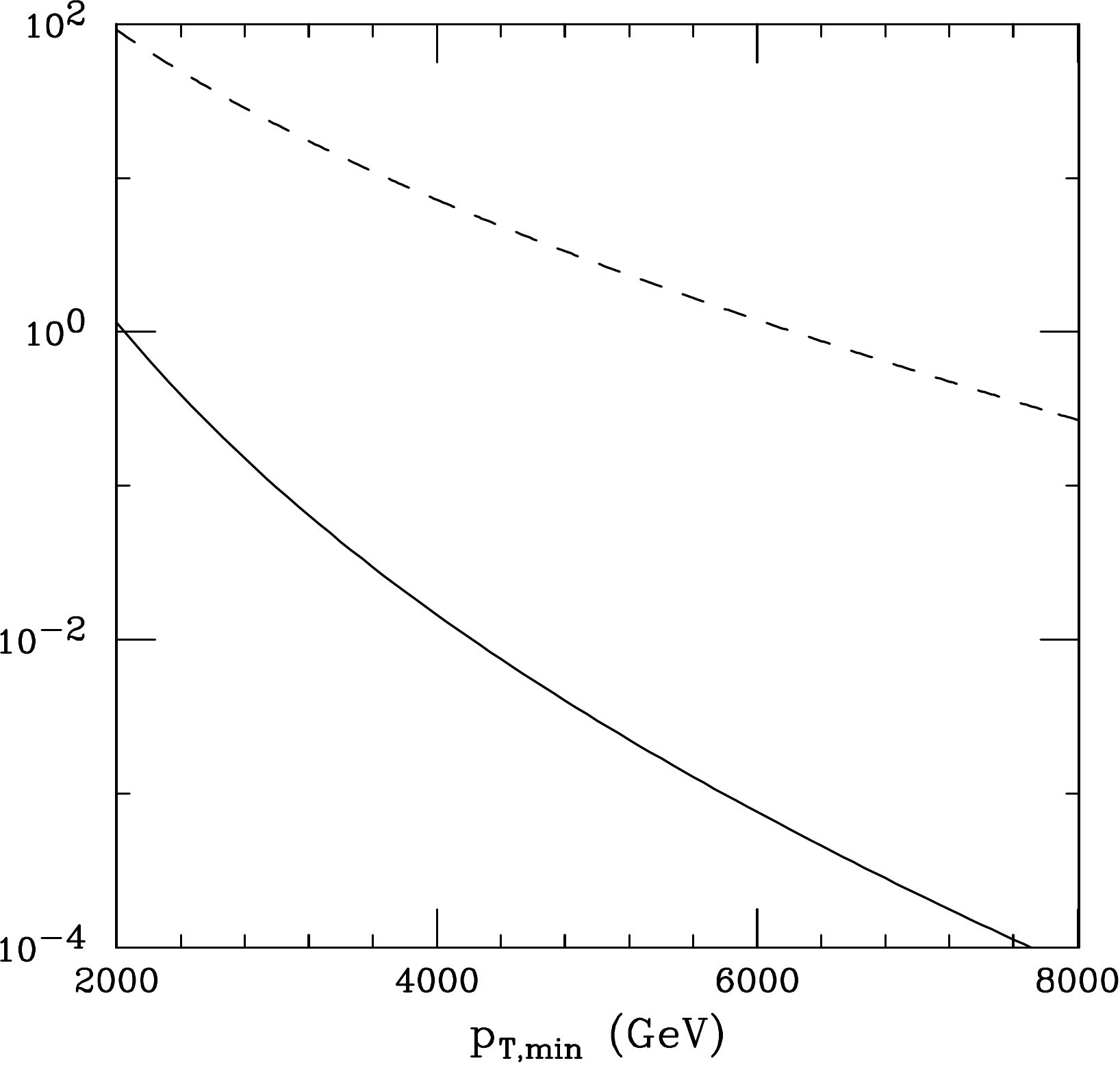}
\caption{Higgs differential (top) and integrated (bottom) $p_t$
  spectrum, comparing the results of the calculation with the exact
  $m_{top}$ dependence, and in the effective field theory (EFT) approximation.}\label{fig:sigma_pt}
\end{figure}
Recently, NNLO predictions for the Higgs transverse momentum spectrum
became
available~\cite{Boughezal:2015dra,Caola:2015wna,Boughezal:2015aha,Chen:2014gva,Chen:2016vqn}.
Unfortunately, all these computations are performed in the Higgs
Effective Theory approximation, where the top quark is integrated out
and the Higgs couples directly to gluons via a point-like effective
interaction. As such, they are only reliable for energy scales well
below the top mass. In the full theory, the Higgs transverse momentum
distribution is only known at leading order. In
Fig.~\ref{fig:sigma_pt}, we compare the LO distributions for the
effective $m_t\to\infty$ and full (resolved top) theory.  This figure
clearly shows the breakdown of the Higgs Effective Theory at high
$p_t$. Finite top quark effects at high $p_t$ are more important than
perturbative QCD corrections, despite the latter being large. To
quantify this statement, we show in the left panel of
Fig.~\ref{fig:scale_Hpt} the LO, NLO and NNLO predictions for the
transverse momentum spectrum in the effective theory. We see that QCD
corrections can lead to $\sim 100\%$ corrections, while
Fig.~\ref{fig:sigma_pt} shows that the full theory deviates from the
effective one by 1-2 order of magnitudes in the high $p_t$
regime. Because of this, in this section we will use the LO prediction
with full quark mass dependence, which as we already said is the best
result available right now. Given its LO nature, these predictions are
affected by a very large scale uncertainty. To choose an optimal scale
for LO predictions, we study the perturbative convergence in the
effective theory. In Fig.~\ref{fig:scale_Hpt} we show the impact of
higher order corrections for the central scale $\mu =
{\sqrt{m_H^2+p_{t,H}^2}}/{2}$. With this choice, the impact of higher
order corrections is somewhat reduced, and we will use this as a
default for all the predictions in this section.
Fig.~\ref{fig:scale_Hpt} suggests that this should be a good
approximation in the whole $p_t$ range considered here up to a factor
of about 2.
\begin{figure}
\includegraphics[width=0.5\textwidth]{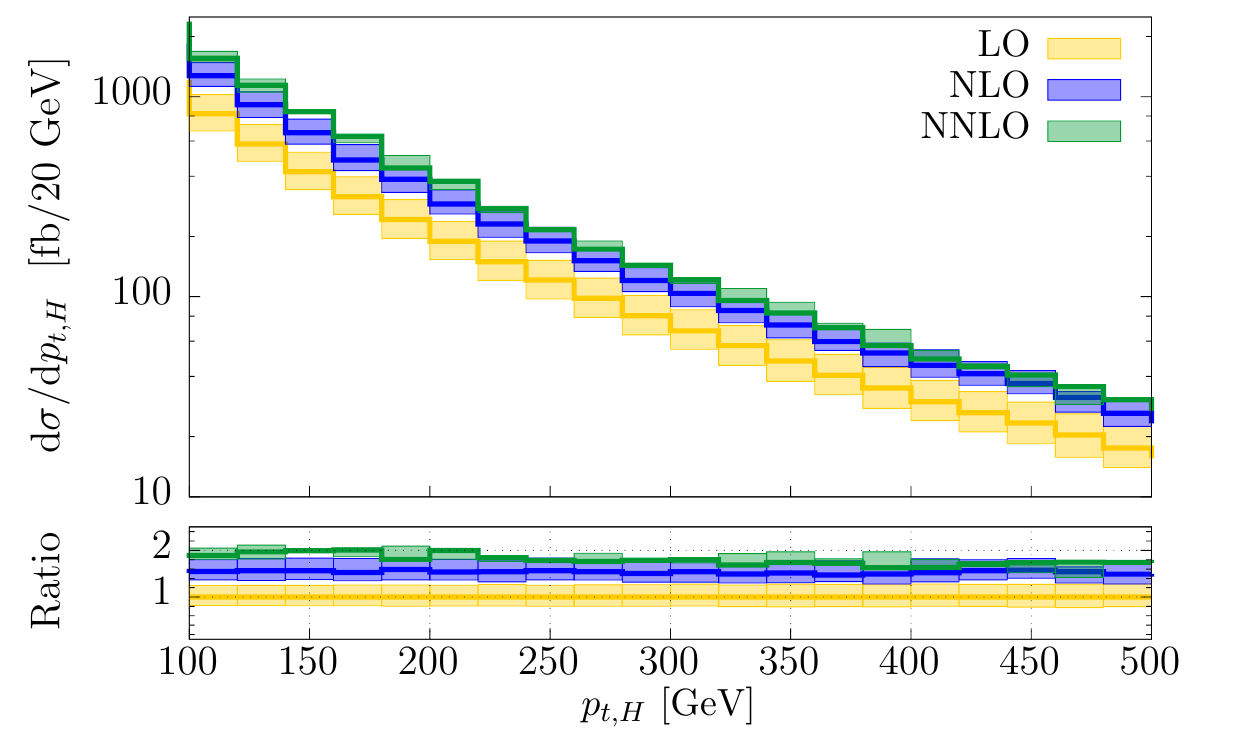}
\includegraphics[width=0.5\textwidth]{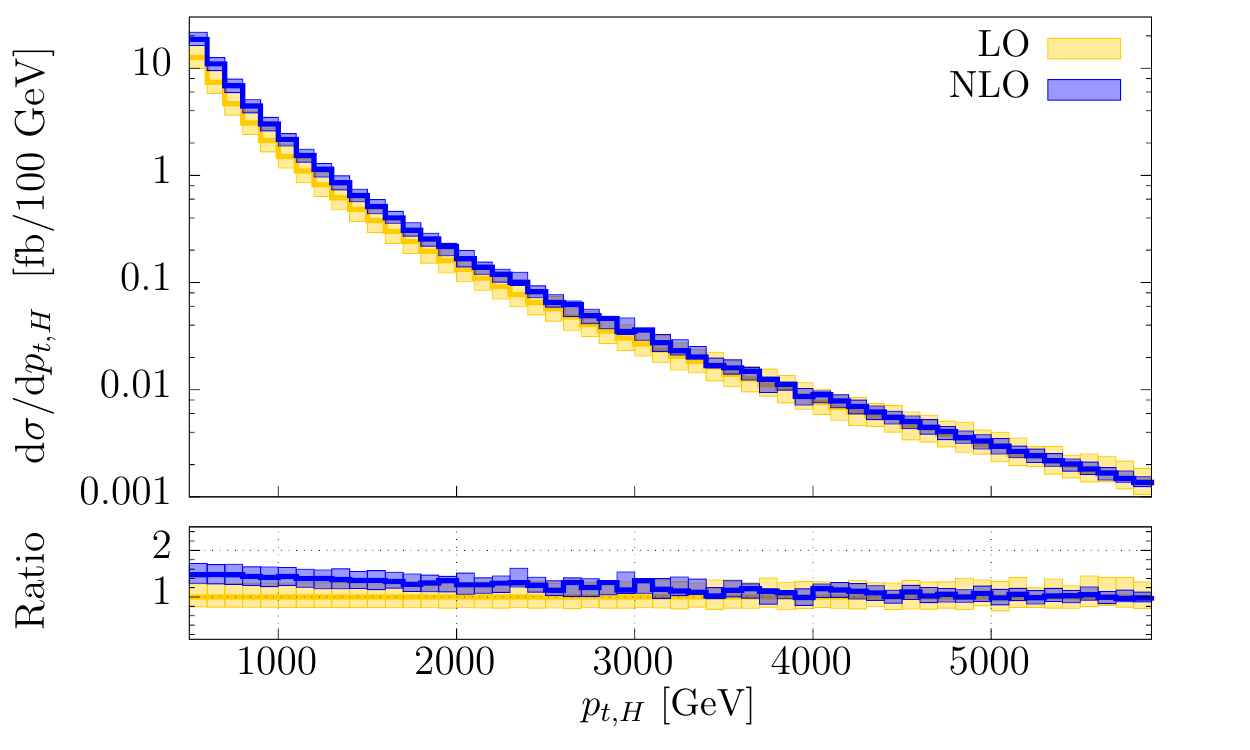}
\caption{Scale variation study. Note that at high $p_t$ there is a
  large uncertainty 
coming from PDFs.}\label{fig:scale_Hpt}
\end{figure}

From the result in Fig.~\ref{fig:sigma_pt} it is clear that even for
very large values of the Higgs transverse momentum the cross section is non
negligible. This, combined with a projected luminosity target in the
ab$^{-1}$ range, will allow for detailed studies of Higgs boson production at
very high transverse momentum in all the major decay channels.  To
quantify this statement, in Tab.~\ref{tab:sigmaBR_pt} we report the
value of the transverse momentum cut $p_{t,cut}$ for which
$\sigma(p_{t,H}>p_{t,cut})$ is larger than $\sim$ 1~fb/1~ab. 
Fig.~\ref{fig:sigma_pt} also indicates that at a 100 TeV collider a
detailed study of the structure of the $ggH$ coupling would be
possible through an analysis of the Higgs transverse momentum shape. Indeed,
it will be possible to investigate the energy dependence of the $ggH$
coupling from scales $\sim m_H$ all the way up to the multi-TeV
regime. This can provide valuable information on possible BSM effects
in the Higgs sector, see e.g.~\cite{Arnesen:2008fb} for a general
discussion and~\cite{Grojean:2013nya} for a more targeted analysis
at a 100 TeV collider.
In this context, it may also be interesting to study the channel
decomposition of the full result. For our scale choice, this
is shown in Fig.~\ref{fig:sigma_pt_ch}. We see a 
cross-over between a $gg$-dominated regime to a $qg$ dominated regime
around $\sim$ 2.5 TeV. 
We conclude a general analysis of differential distributions for 
Higgs production in association with one extra jet by showing
in Fig.~\ref{fig:yHyj} the Higgs and jet rapidity distributions at 100 TeV
compared with the same at 13 TeV. It is clear that a wider rapidity
coverage is desirable at 100 TeV.

\begin{figure}
\centering
\includegraphics[width=0.5\textwidth]{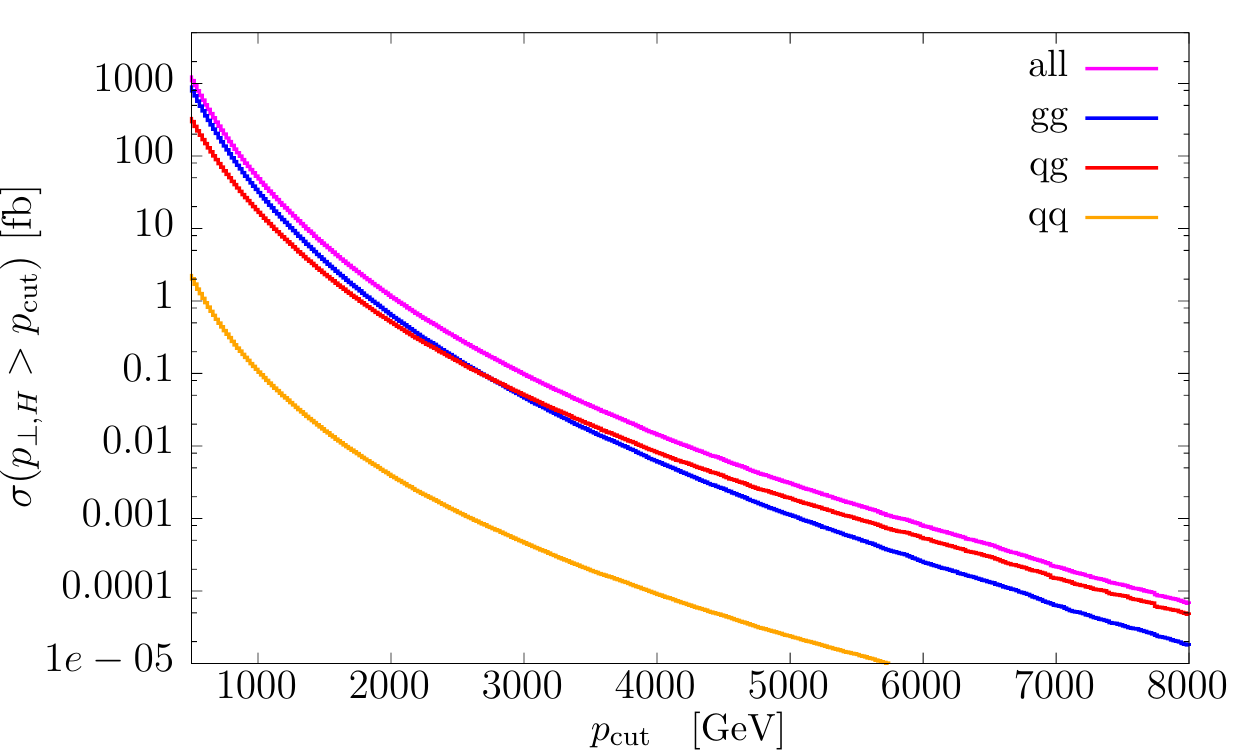}
\label{fig:sigma_pt_ch}\caption{Channel decomposition of the 
Higgs total cross section, as a function of the Higgs transverse
momentum. See text for details.}
\end{figure}

\begin{table}
\centering
\begin{tabular}{|c|c|c|}
\hline
& $\sigma(p_t>p_{t,cut})\times BR = 1~{\rm fb}$ & $\sigma(p_t>p_{t,cut})\times BR = 1~{\rm ab}$ \\
\hline
$H\to b\bar b$ & $p_{t,cut} = 1860~{\rm GeV}$ & $p_{t,cut} = 5380~{\rm GeV}$ \\
$H\to \tau\bar \tau$ & $p_{t,cut} = 1240~{\rm GeV}$ & $p_{t,cut} = 3950~{\rm GeV}$ \\
$H\to \mu^+\mu^-$ & $p_{t,cut} = 340~{\rm GeV}$ & $p_{t,cut} = 1570~{\rm GeV}$ \\
\hline
$H\to c\bar c$ & $p_{t,cut} = 1070~{\rm GeV}$ & $p_{t,cut} = 3520~{\rm GeV}$ \\
$H\to s\bar s$ & $p_{t,cut} = 350~{\rm GeV}$ & $p_{t,cut} = 1600~{\rm GeV}$ \\
\hline
$H\to gg$ & $p_{t,cut} = 1320~{\rm GeV}$ & $p_{t,cut} = 4130~{\rm GeV}$ \\
$H\to \gamma\gamma$ & $p_{t,cut} = 620~{\rm GeV}$ & $p_{t,cut} = 2350~{\rm GeV}$ \\
$H\to Z\gamma$ & $p_{t,cut} = 570~{\rm GeV}$ & $p_{t,cut} = 2200~{\rm GeV}$ \\
$H\to W^+W^-$ & $p_{t,cut} = 1560~{\rm GeV}$ & $p_{t,cut} = 4700~{\rm GeV}$ \\
$H\to ZZ$ & $p_{t,cut} = 1050~{\rm GeV}$ & $p_{t,cut} = 3470~{\rm GeV}$ \\
\hline
\end{tabular}
\caption{Cross-section times branching ratio as a function of $p_{t,cut}$. Each entry
corresponds to the $p_{t,cut}$ value. No $VV\to 4l$ branching ratio included.}\label{tab:sigmaBR_pt}
\end{table}

\begin{figure}
\includegraphics[width=0.49\textwidth]{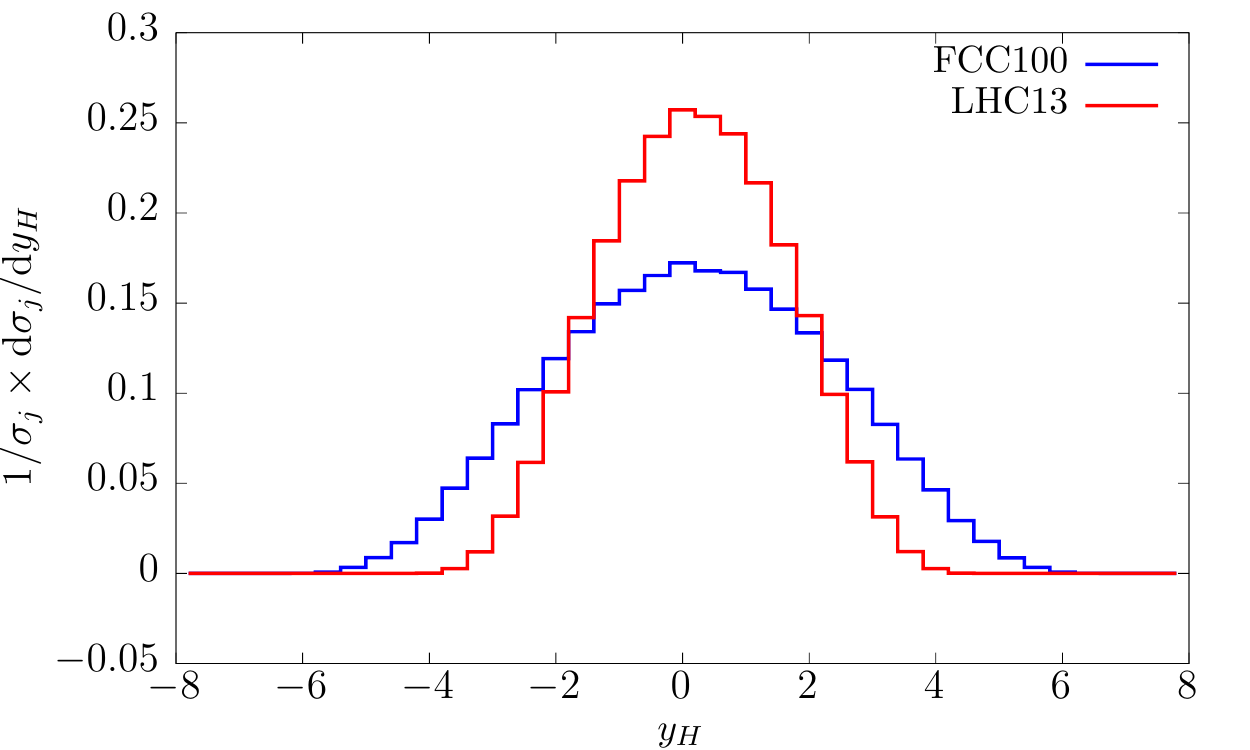}
\includegraphics[width=0.49\textwidth]{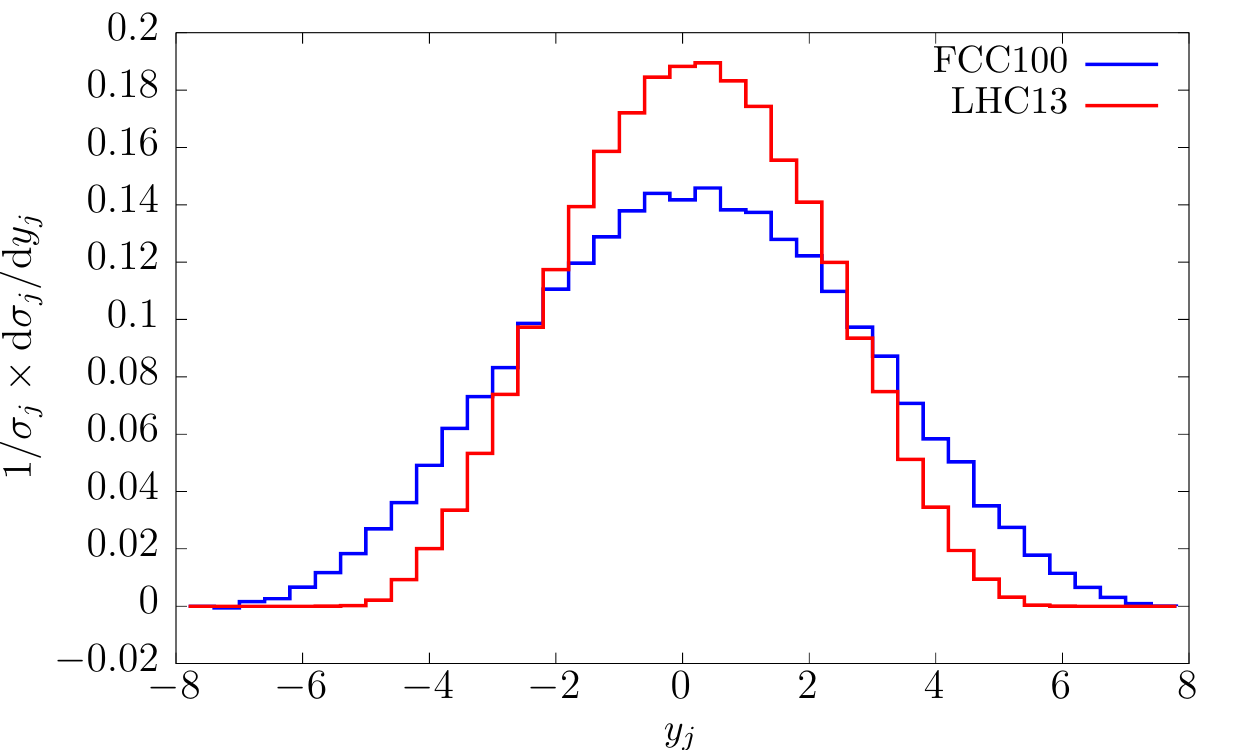}
\caption{Higgs and leading jet rapidity distributions, normalized to the total
cross section.}
\label{fig:yHyj}
\end{figure}

Finally, we consider differential distributions of Higgs decay
products. As a case of study, we consider the $H\to WW$ channel and
study the kinematics distributions of the final state leptons.
We consider two scenarios, one with a mild cut $p_{\perp,H} > 60~$GeV
on the Higgs transverse momentum and one with a much harder cut 
$p_{\perp,H} > 1~$TeV. For reference, the total cross section for 
$pp\to H\to WW \to 2l2\nu$ in the two cases is $\sigma=470/0.1~\mathrm{fb}$ for
the low/high cut. Results are shown in Fig.~\ref{fig:ggHpt-hww}. 
While the di-lepton invariant mass shape is very stable with respect
to the $p_t$ cut, both the di-lepton $p_t$ and azimuthal separation shapes
change significantly. As expected, the $p_{t,ll}$ spectrum shifts towards
higher values of $p_t$. The characteristic peak at small $\phi$ of the
lepton azimuthal separation becomes more and more pronounced as the
Higgs $p_t$ increases.

\begin{figure}
\centering
\includegraphics[width=0.49\textwidth]{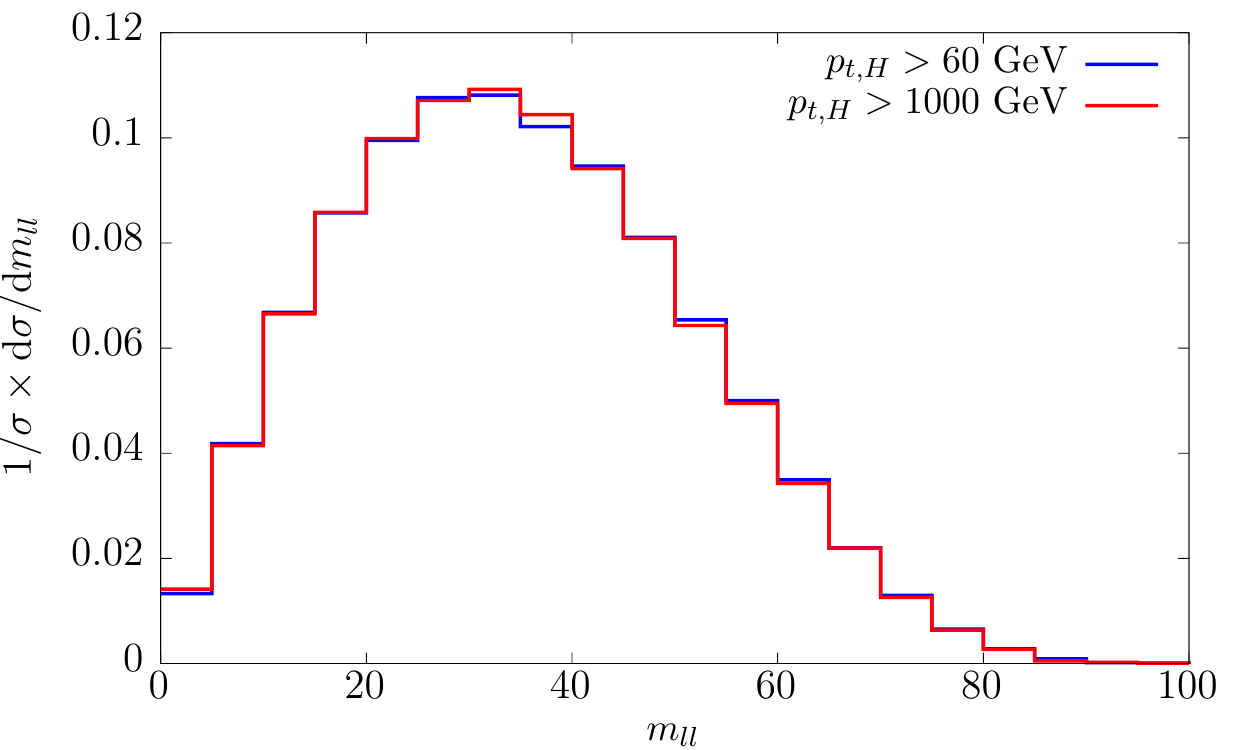}
\includegraphics[width=0.49\textwidth]{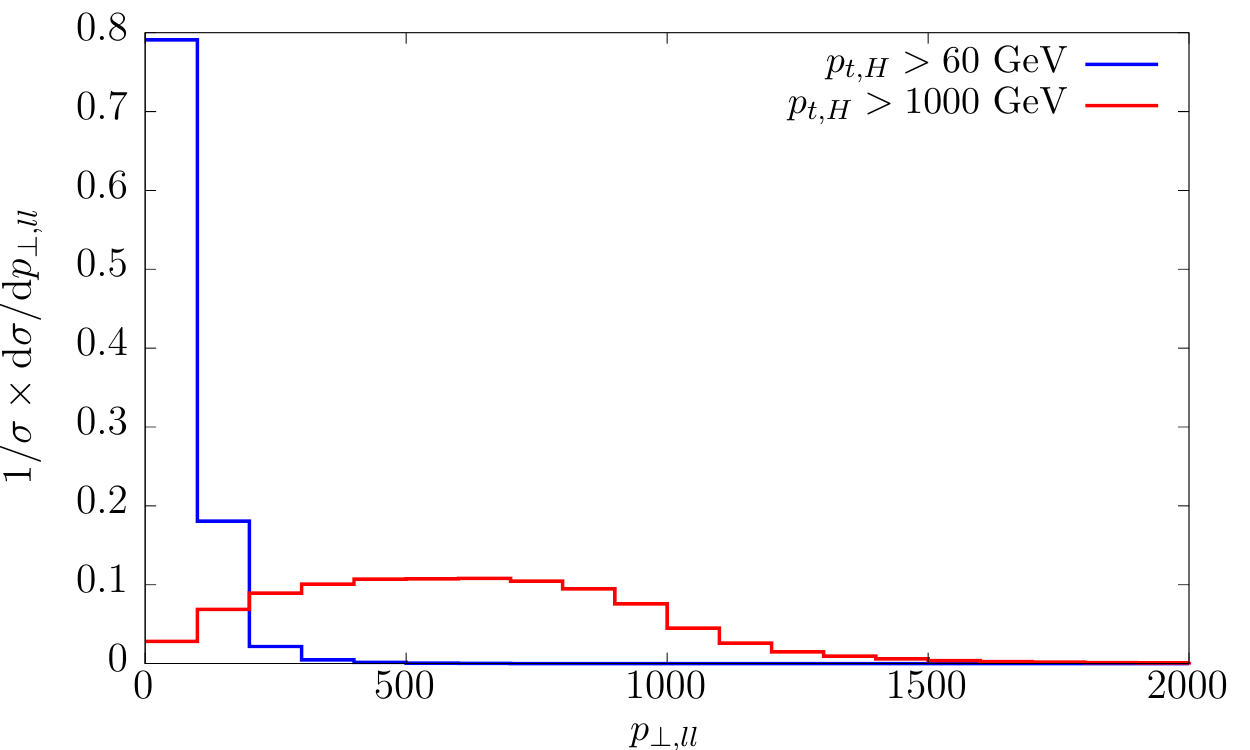}
\includegraphics[width=0.49\textwidth]{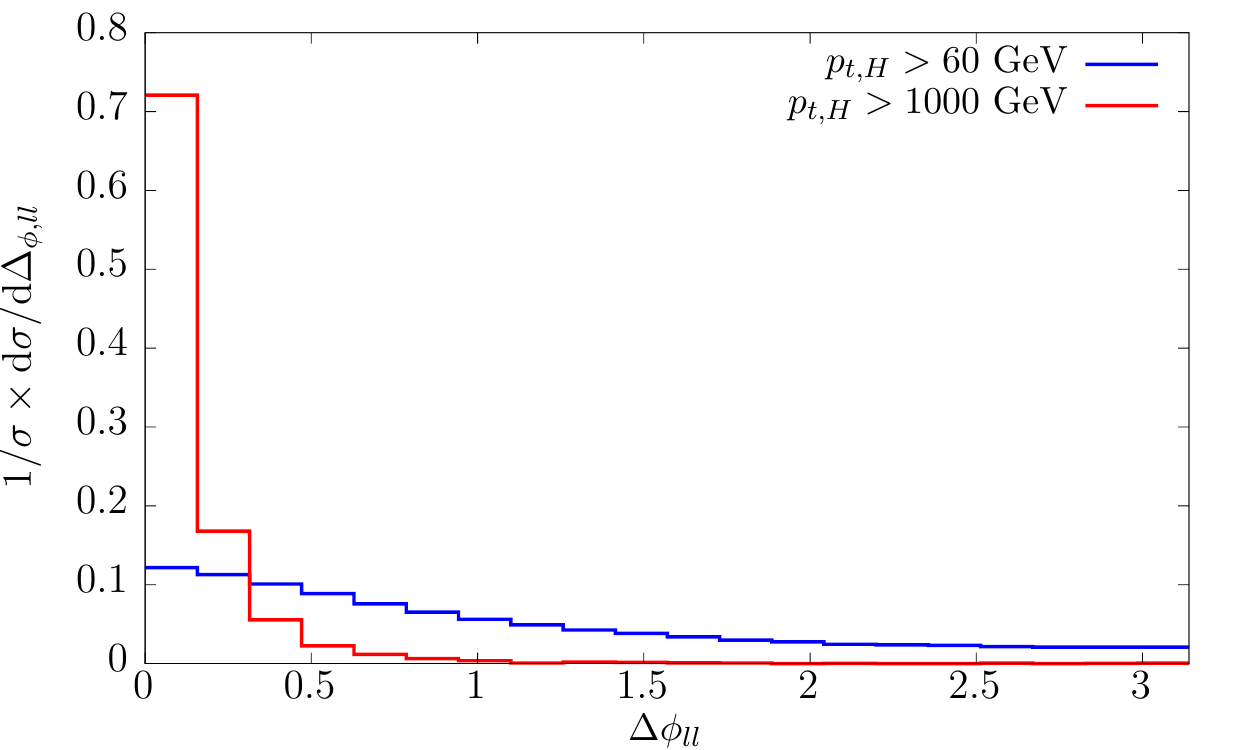}
\caption{Normalized differential distributions for the $pp\to H\to WW\to 2l2\nu$ process.
Results for two values of cuts on the Higgs transverse momentum are shown, see text for details.}
\label{fig:ggHpt-hww}
\end{figure}


\clearpage
\subsection{Higgs plus jets production in $gg\to H$}
\label{sec:H_ggHjets}

In this section we present NLO QCD results for the production of a
Standard Model Higgs boson in association with up to three jets in
gluon-gluon fusion (GGF). If not stated differently, the computations are
done in the approximation of an infinitely heavy top quark using the
same effective field theory Lagrangian presented in Eq.~\eqref{eq:L_eff}.

Gluon--gluon fusion is not only the largest Higgs boson production
channel, but, as already shown in Section~\ref{sec:H_ggH}, it is also
characterized by very large higher-order corrections. Although less
dramatic than in the fully inclusive case of Higgs boson production,
the production in association with jets also suffers from large
corrections due to NLO effects. In this section we will study how this
changes when the center-of-mass energy increases from 14 to 100~TeV.

Gluon fusion is also the largest background for Higgs boson production
through vector boson fusion (VBFH). Despite the very peculiar
experimental signature of the VBFH channel, whose topology allows to
define fiducial cuts which reduce the backgrounds dramatically, the
contamination from GGF remains a very important aspect at LHC
energies. It is therefore interesting to study the impact of typical
VBFH-type selection cuts on the GGF background also at FCC energies.

Another important aspect to keep in mind, is the limited range of
validity of the effective field theory description, in which the top
quark is integrated out. As already shown for the inclusive case in
previous sections, finite top quark and bottom quark mass effects can
become large when the transverse energy is large enough to resolve the
quark loop that couples the Higgs boson to gluons. At the end of this
section, we will investigate the impact of these corrections presenting
LO results in the full theory.

\subsubsection{Computational setup}
\label{sec:ggHjets_setup}

The computation is performed using the setup developed for an
analogous analysis at 8 and 13 TeV~\cite{Greiner:2015jha}, and is
based on the automated tools
\textsc{GoSam}~\cite{Cullen:2011ac,Cullen:2014yla} and
\textsc{Sherpa}~\cite{Gleisberg:2008ta}, linked via the interface
defined in the Binoth Les Houches
Accord~\cite{Binoth:2010xt,Alioli:2013nda}.

The one-loop amplitudes are generated with
\textsc{GoSam}, and are based on an algebraic generation of
$d$-dimensional integrands using a Feynman diagrammatic approach.  The
expressions for the amplitudes are generated employing
\textsc{QGraf}~\cite{Nogueira:1991ex},
\textsc{Form}~\cite{Vermaseren:2000nd,Kuipers:2012rf} and
\textsc{Spinney}~\cite{Cullen:2010jv}. For the reduction of the tensor
integrals at running time, we used
\textsc{Ninja}~\cite{vanDeurzen:2013saa,Peraro:2014cba}, which is an
automated package carrying out the integrand reduction via Laurent
expansion~\cite{Mastrolia:2012bu}, and
\textsc{OneLoop}~\cite{vanHameren:2010cp} for the evaluation of the
scalar integrals. Unstable phase space points are detected
automatically and reevaluated with the tensor integral library
\textsc{Golem95}~\cite{Heinrich:2010ax,Binoth:2008uq,Cullen:2011kv}.
The tree-level matrix elements for the Born and real-emission
contribution, and the subtraction terms in the
Catani-Seymour approach~\cite{Catani:1996vz} have been evaluated within
\textsc{Sherpa} using the matrix element generator
\textsc{Comix}~\cite{Gleisberg:2008fv}.

Using this framework we stored NLO events in the form of \textsc{Root}
Ntuples. Details about the format of the Ntuples generated by
\textsc{Sherpa} can be found in~\cite{Bern:2013zja}.
The predictions presented in the following were computed using Ntuples
at 14 and 100 TeV with generation cuts specified by
\[
  p_{T,\,\mathrm{jet}}\;>\;25\:\mathrm{GeV}\qquad\mbox{and}\qquad
  |\eta_{\mathrm{jet}}|\;<\;10\,,
\]
and for which the Higgs boson mass $m_H$ and the Higgs vacuum
expectation value $v$ are set to $m_H=125\:\mathrm{GeV}$ and
$v=246\:\mathrm{GeV}$, respectively. To improve the efficiency in
performing the VBFH analysis using the selection cuts described below,
a separate set of Ntuples was generated. This set includes an
additional generation cut on the invariant mass of the two leading
transverse momentum jets. To generated large dijet masses from
scratch, we require $m_{j_1j_2}>1600\:\mathrm{GeV}$.

The results reported here are obtained by clustering jets with the
anti-$k_T$ algorithm~\cite{Cacciari:2005hq,Cacciari:2008gp} employing
a cone radius of $R=0.4$. We utilized the implementation as provided
by the \textsc{FastJet} package~\cite{Cacciari:2011ma}, and also
relied on using the CT14nlo PDF set~\cite{Dulat:2015mca} in the
calculations presented here.
In order to assess the impact of varying the transverse momentum
threshold for the jets, we apply four different cuts at
$$p_{T,\,\mathrm{jet}}\;>\;30,\,50,\,100~~\mbox{and}~~300~\mathrm{GeV}\,,$$
and keep the same cut on $\eta_{\mathrm{jet}}$a s in the Ntuples
generation. For the VBFH analysis, we then apply additional cuts as
described further below in Section~\ref{sec:ggHjets_vbfresults}

\begin{table}[t!]
\centering\small
\begin{tabular}{|l||c|c||c|c||c|}
  \hline
  Numbers in pb
  & $\sigma_{\mathrm{LO}}^{14\,\mathrm{TeV}}$
  & $\sigma_{\mathrm{NLO}}^{14\,\mathrm{TeV}}$
  & $\sigma_{\mathrm{LO}}^{100\,\mathrm{TeV}}$
  & $\sigma_{\mathrm{NLO}}^{100\,\mathrm{TeV}}$
  & \scriptsize{NLO Ratio}\phantom{\Big|}\\
  \hline
  \multicolumn{6}{|c|}{$\mathrm{H}\!+\!1$ jet\phantom{\Big|}}\\
  \hline
  $p_{T,\,\mathrm{jet}}>30~\mathrm{GeV}$  \phantom{\Big|} & $9.39^{+38\%}_{-26\%}$ & $15.4^{+15\%}_{-15\%}$ & $ 217^{+21\%}_{-17\%}$ & $ 336^{+10\%}_{-9\%}$ & $21.8$\\
  $p_{T,\,\mathrm{jet}}>50~\mathrm{GeV}$  \phantom{\Big|} & $5.11^{+39\%}_{-26\%}$ & $8.49^{+15\%}_{-15\%}$ & $ 135^{+22\%}_{-18\%}$ & $ 215^{+11\%}_{-10\%}$ & $25.3$\\
  $p_{T,\,\mathrm{jet}}>100~\mathrm{GeV}$ \phantom{\Big|} & $1.66^{+40\%}_{-27\%}$ & $2.73^{+15\%}_{-16\%}$ & $58.2^{+24\%}_{-19\%}$ & $92.1^{+11\%}_{-11\%}$ & $33.7$\\
  $p_{T,\,\mathrm{jet}}>300~\mathrm{GeV}$ \phantom{\Big|} & $0.11^{+43\%}_{-28\%}$ & $0.17^{+15\%}_{-16\%}$ & $8.51^{+28\%}_{-21\%}$ & $13.2^{+11\%}_{-11\%}$ & $77.6$\\
  \hline
  \multicolumn{6}{|c|}{$\mathrm{H}\!+\!2$ jets\phantom{\Big|}}\\
  \hline
  $p_{T,\,\mathrm{jet}}>30~\mathrm{GeV}$  \phantom{\Big|} & $3.60^{+57\%}_{-34\%}$                       & $5.40^{+12\%}_{-18\%}$                       & $ 148^{+40\%}_{-27\%}$ &  $174^{-2\%}_{-8\%}$  & $32.2$\\
  $p_{T,\,\mathrm{jet}}>50~\mathrm{GeV}$  \phantom{\Big|} & $1.25^{+58\%}_{-34\%}$                       & $1.96^{+15\%}_{-19\%}$                       & $65.0^{+41\%}_{-27\%}$ & $83.7^{+3\%}_{-11\%}$ & $42.7$\\
  $p_{T,\,\mathrm{jet}}>100~\mathrm{GeV}$ \phantom{\Big|} & $0.22^{+58\%}_{-34\%}$                       & $0.36^{+17\%}_{-20\%}$                       & $17.7^{+42\%}_{-28\%}$ & $24.6^{+8\%}_{-13\%}$ & $68.3$\\
  $p_{T,\,\mathrm{jet}}>300~\mathrm{GeV}$ \phantom{\Big|} & $6.35\cdot10^{-3}\phantom{i}^{+57\%}_{-34\%}$ & $1.03\cdot10^{-2}\phantom{i}^{+17\%}_{-20\%}$ & $1.41^{+43\%}_{-28\%}$ & $2.07^{+10\%}_{-14\%}$ & $202.9$\\
  \hline
  \multicolumn{6}{|c|}{$\mathrm{H}\!+\!3$ jets\phantom{\Big|}}\\
  \hline
  $p_{T,\,\mathrm{jet}}>30~\mathrm{GeV}$  \phantom{\Big|} & $1.22^{+76\%}_{-40\%}$                       & $1.77^{+9\%}_{-21\%}$                       & $89.0^{+58\%}_{-34\%}$ & $84.3^{-24\%}_{-5\%}$ & $47.6$\\
  $p_{T,\,\mathrm{jet}}>50~\mathrm{GeV}$  \phantom{\Big|} & $0.29^{+75\%}_{-40\%}$                       & $0.46^{+15\%}_{-23\%}$                       & $29.8^{+58\%}_{-34\%}$ & $32.9^{-10\%}_{-10\%}$ & $71.5$\\
  $p_{T,\,\mathrm{jet}}>100~\mathrm{GeV}$ \phantom{\Big|} & $3.07\cdot10^{-2}\phantom{i}^{+74\%}_{-40\%}$ & $4.95\cdot10^{-2}\phantom{i}^{+19\%}_{-23\%}$ & $5.61^{+57\%}_{-34\%}$ & $7.04^{+1\%}_{-14\%}$ & $142.1$\\
  $p_{T,\,\mathrm{jet}}>300~\mathrm{GeV}$ \phantom{\Big|} & $2.97\cdot10^{-4}\phantom{i}^{+71\%}_{-39\%}$ & $4.86\cdot10^{-4}\phantom{i}^{+20\%}_{-23\%}$ & $0.24^{+56\%}_{-34\%}$ & $0.34^{+9\%}_{-16\%}$ & $700.2$\\
  \hline
\end{tabular}
\caption{\label{table:ggHjets_ggf_incl_XS}%
  Total inclusive cross sections for the production of a Higgs
  boson in association with one, two or three jets at LO and NLO in
  QCD. Numbers are reported for center-of-mass energies of 14 and 100
  TeV and four choices of transverse momentum cuts on the jets, namely
  $p_{T,\,\mathrm{jet}}>30,\,50,\,100$ and $300~\mathrm{GeV}$. The
  last column shows the ratios between the NLO cross sections at the
  two center-of-mass energies. The uncertainty estimates are obtained
  from standard scale variations.}
\end{table}

The renormalization and factorization scales were set equal, and are
defined as
\begin{equation}
  \mu_\mathrm{R}\,=\,\mu_\mathrm{F}\,=\,\frac{\hat{H}^\prime_T}{2}\;=\;
  \frac{1}{2}\left(\sqrt{m_H^2+p_{T,H}^2}+\sum_{i}|p_{T,\,j_i}|\right)\,.
\end{equation}
The sum runs over all partons accompanying the Higgs boson in
the event. Theoretical uncertainties are estimated in the standard way
by varying the central scale by factors of $0.5$ and $2$.

\subsubsection{Gluon fusion results}
\label{sec:ggHjets_ggfresults}

We start by summarizing in Table~\ref{table:ggHjets_ggf_incl_XS} the
total inclusive cross sections for the production of a Higgs boson in
gluon-gluon fusion accompanied by one, two or three additional jets.
We show results at LO and NLO in QCD for $pp$ collisions at 14 and
100~TeV. Furthermore, the total cross sections are given for four
different $p_{T,\,\mathrm{jet}}$ cuts on the jets.
In the last column we show the ratio of the NLO result for 100~TeV
over the NLO result for 14~TeV.  This ratio significantly increases
when the $p_{T,\,\mathrm{jet}}$ cut is tightened, and it also strongly
increases as a function of the jet multiplicities. This can be easily
understood by the fact that in a 100~TeV environment, the cuts appear
much less severe than for 14~TeV; their impact on the lower energy is
therefore larger. For the same reasons, this pattern is also found for
the number of jets. With rising center-of-mass energy, it becomes
easier to produce additional jets, which leads to the enhancement of
the inclusive cross section ratio.

\begin{figure}[t!]
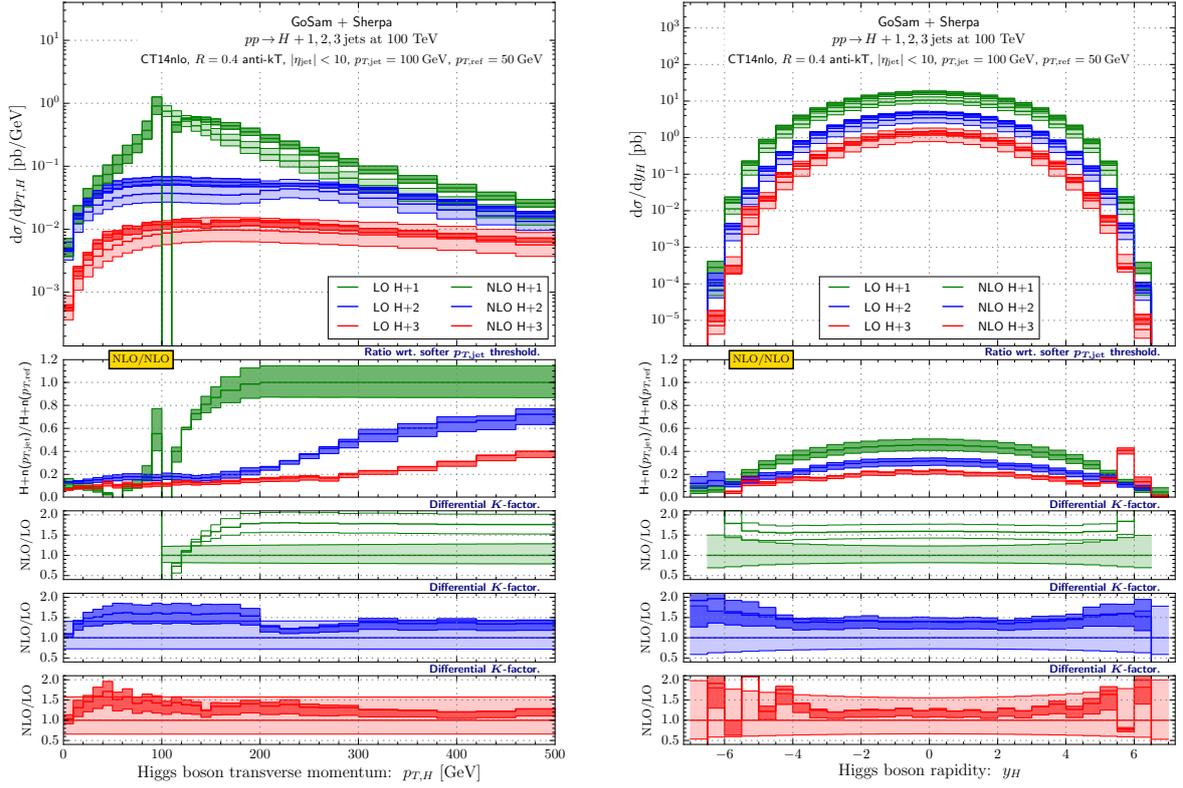

  \centering
  \includegraphics[width=0.49\textwidth]{figs/ggHjets/pTH/%
    Multidist_H+123_pT100vs50eta100mult_higgs_pt_100-E100}
  \hfill
  \includegraphics[width=0.49\textwidth]{figs/ggHjets/yH/%
    Multidist_H+123_pT100vs50eta100mult_higgs_y_40-E100}
  \caption{\label{fig:ggHjets_ggfpTHyH}
    The transverse momentum spectrum $p_{T,H}$ and the rapidity
    distribution $y_H$ of the Higgs boson at 100~TeV for the three
    production modes $\mathrm{H}\!+\!1,2,3$ jets. Results are shown at
    LO and NLO including the effect from standard scale variations and
    imposing a jet threshold of $p_{T,\,\mathrm{jet}}>100~\mathrm{GeV}$.
    The second panel depicts the NLO ratios taken wrt.~reference
    results obtained with $p_{T,\,\mathrm{jet}}>50~\mathrm{GeV}$; the
    other ratio plot panels display the differential $K$-factors for
    the different jet multiplicities.}
\end{figure}

Turning to more exclusive observables,
Figure~\ref{fig:ggHjets_ggfpTHyH} shows (to the left) the transverse
momentum distribution and (to the right) the rapidity distribution of
the Higgs boson at 100 TeV with a transverse momentum requirement on
the jets of $p_{T,\,\mathrm{jet}}>100~\mathrm{GeV}$. The different
colours denote the various jet multiplicities. The brighter bands show
the LO predictions with their respective uncertainties, whereas the
NLO results are displayed by darker bands. As we deal with fixed-order
predictions, we observe for the $p_{T,H}$ distributions -- as
expected -- Sudakov shoulder effects decreasing in their extent at
$p_T\sim100,\,200$ and $300~\mathrm{GeV}$ for the one-jet, two-jet and
three-jet final states, respectively.
The uppermost ratio plot shows the results for
$p_{T,\,\mathrm{jet}}>100~\mathrm{GeV}$ divided by the corresponding
results of the same jet multiplicity, but with a
$p_{T,\,\mathrm{jet}}$ threshold of $50~\mathrm{GeV}$. As expected
this ratio gets smaller for higher jet multiplicities, which means
the more jets are present, the more sensitively the cross section
changes in response to a jet threshold increase. In the one-jet case
we find that the ratio turns one for $p_{T,H}>200~\mathrm{GeV}$.
Below this value, the $50~\mathrm{GeV}$ threshold sample contains event
topologies that are absent for $p_{T,\,\mathrm{jet}}>100~\mathrm{GeV}$.
The ratio will hence be smaller than one. For example, a
configuration consisting of a jet with $p_T=99.9~\mathrm{GeV}$ and a real
emission of size $p_T=99.8~\mathrm{GeV}$ will be present for
$50~\mathrm{GeV}$ thresholds but be missed by the higher
$p_{T,\,\mathrm{jet}}$ sample. Lastly we note that the size of the
$K$-factors decreases for jettier final states. We also observe that
the 100~TeV environment allows for a wide range of Higgs boson
rapidities independent of the jet multiplicity. One easily gains two
absolute units wrt.~the capabilities of the LHC.

\begin{figure}[t!]
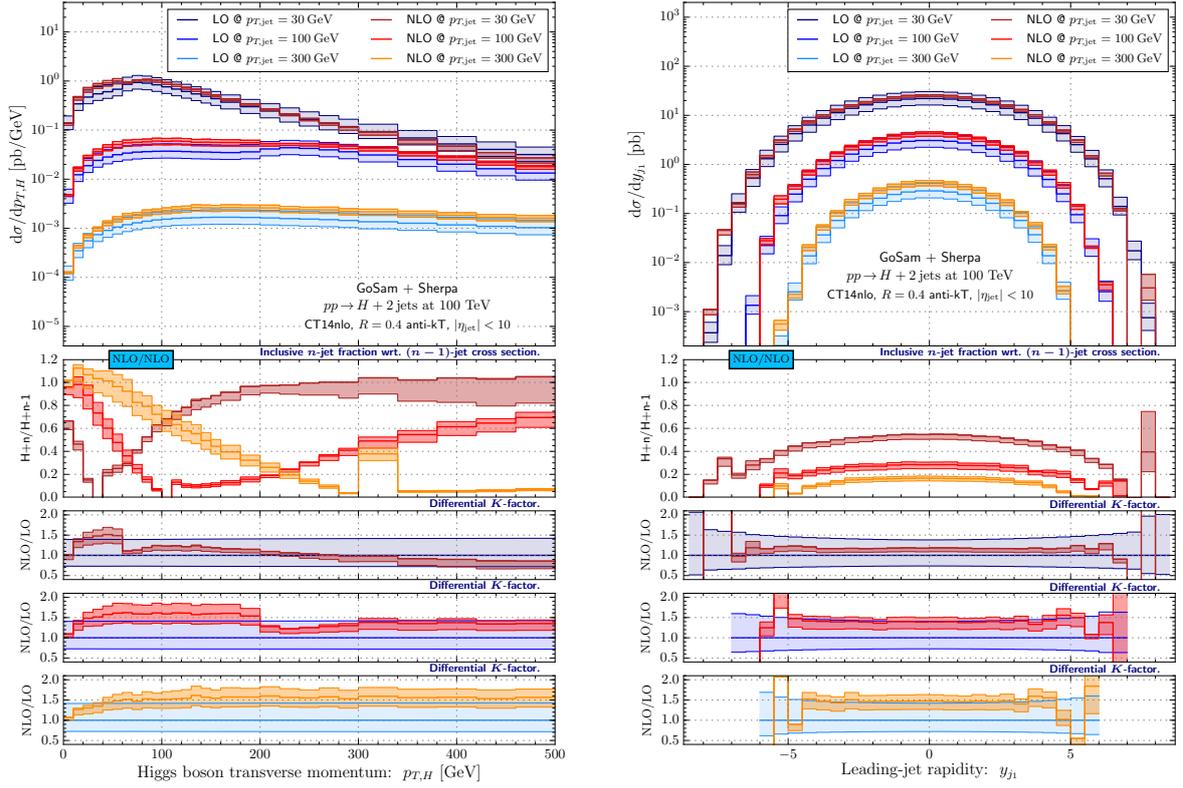

  \centering
  \includegraphics[width=0.49\textwidth]{figs/ggHjets/pTH/%
    Multidist_H+2_pTXeta100mult_higgs_pt_100-E100}
  \hfill
  \includegraphics[width=0.49\textwidth]{figs/ggHjets/yjet1/%
    Multidist_H+2_pTXeta100mult_jet_y_1_40-E100}
  \caption{\label{fig:ggHjets_ggfpTH_jcut}
    The transverse momentum spectrum $p_{T,H}$ of the Higgs boson
    and the rapidity distribution $y_{j_1}$ of the leading jet, both
    of which shown at LO and NLO, and for different transverse
    momentum requirements on the jets in $\mathrm{H}\!+\!2$-jet
    scatterings as produced at a 100~TeV $pp$ collider. The comparison
    to the $\mathrm{H}\!+\!1$-jet case at NLO is visualized in the first
    ratio plot, followed by the canonical NLO versus LO ratio plots
    for the different $p_{T,\,\mathrm{jet}}$ values. All uncertainty
    envelopes originate from standard scale variations by factors of
    two.}
\end{figure}

The left plot of Figure~\ref{fig:ggHjets_ggfpTH_jcut} shows again
transverse momentum distributions of the Higgs boson, however in this
case we only consider the curves for $\mathrm{H}\!+\!2$ jets at 100
TeV. Here, we examine the impact of tightening the transverse momentum
cut on the jets. The typical shoulder present for
$p_{T,\,\mathrm{jet}}>30~\mathrm{GeV}$ progressively disappears for
increasing values of $p_{T,\,\mathrm{jet}}$ such that the corresponding
distribution for $p_{T,\,\mathrm{jet}}>300~\mathrm{GeV}$ becomes almost
flat in the range from $100$ to $500~\mathrm{GeV}$. In the right plot
of Figure~\ref{fig:ggHjets_ggfpTH_jcut}, the analogous comparison for
the rapidity of the leading jet is presented. As expected, the
successively harder jet constraints lead to a more central production
of the jets reducing the rapidity range where the differential cross
section is larger than $1~\mathrm{fb}$ by about six units.
In both plots of Figure~\ref{fig:ggHjets_ggfpTH_jcut}, the first ratio
plot highlights the behaviour of the fraction between the inclusive
results for $\mathrm{H}\!+\!2$~jets and $\mathrm{H}\!+\!1$~jet.
While for the Higgs boson transverse momentum, this fraction
varies considerably and can reach one in phase space regions of
near-zero as well as large $p_{T,H}$ (earmarking the important
two-jet regions), for the leading jet rapidity, the maximum occurs
always at $y_{j_1}=0$ decreasing from about $0.6$ to $0.2$ once the
jet transverse momentum cut is tightened. This shows that the leading
jet tends to be produced more centrally in events of higher jet
multiplicity. An increase of the transverse momentum cut also has
consequences on the size and shape of the NLO corrections. This is
shown in the lower insert plots. In general, for sharper cuts, the
higher-order corrections become larger but flatter over the considered
kinematical range. Similar results are also obtained for the
$\mathrm{H}\!+\!3$-jet process.

\begin{figure}[t!]
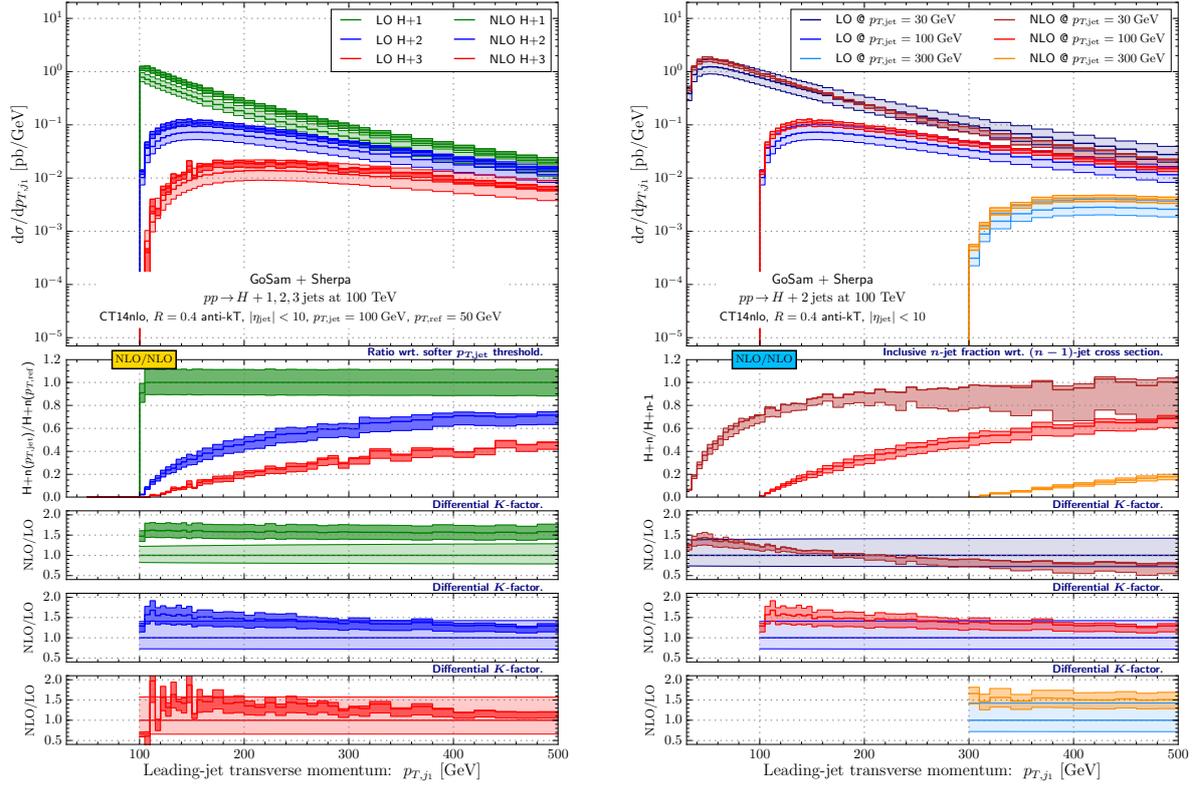

  \centering
  \includegraphics[width=0.49\textwidth]{figs/ggHjets/pTjet1/%
    Multidist_H+123_pT100vs50eta100mult_jet_pT_1_100-E100}
  \hfill
  \includegraphics[width=0.49\textwidth]{figs/ggHjets/pTjet1/%
    Multidist_H+2_pTXeta100mult_jet_pT_1_100-E100}
  \caption{\label{fig:ggHjets_ggfpTj1}
    The transverse momentum distribution $p_{T,\,j_1}$ of the leading
    jet at an FCC energy of 100~TeV for the three production modes of
    $\mathrm{H}\!+\!1,2,3$ jets (left) and varying jet-$p_T$ thresholds
    exemplified for the case of $\mathrm{H}\!+\!2$-jet production
    (right). The layout of the plot to the left (right) is the same as
    used in Figure~\ref{fig:ggHjets_ggfpTHyH}
    (Figure~\ref{fig:ggHjets_ggfpTH_jcut}).}
\end{figure}

Figure~\ref{fig:ggHjets_ggfpTj1} focuses on the leading jet transverse
momentum. The plot on the left hand side compares predictions for
$\mathrm{H}\!+\!1,2,3$ jets with one another at LO and NLO for a jet
threshold of $p_{T,\,\mathrm{jet}}>100~\mathrm{GeV}$. The scheme of
the lower ratio plots is equal to the one of
Figure~\ref{fig:ggHjets_ggfpTHyH}. For
$p_{T,j_1}\approx300~\mathrm{GeV}$, we see that 60\% (30\%) of the
inclusive two-jet (three-jet) events (using the reference jet
threshold) have a second jet at or above a transverse momentum of
$100~\mathrm{GeV}$. The plot on the right hand side instead shows the
effect of the different jet transverse momentum constraints for
$\mathrm{H}\!+\!2$ jets production at 100 TeV, following the colour
convention and the scheme of Figure~\ref{fig:ggHjets_ggfpTH_jcut}.
For lower jet thresholds, the two-jet cross section rises quickly with
increasing lead-jet $p_T$ to the same order of magnitude as the
one-jet cross section. We find that an increase of the jet-$p_T$
constraint helps slow down this behavior sufficiently.

\begin{figure}[t!]
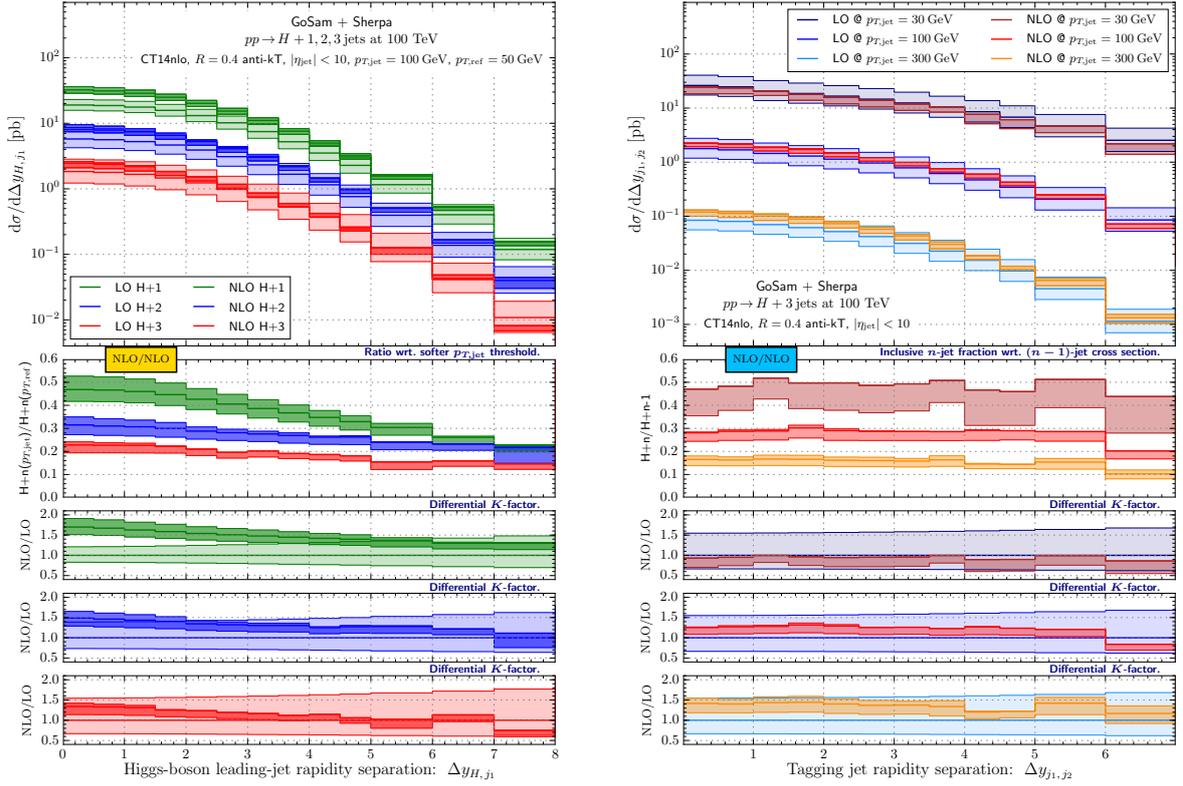

  \centering
  \includegraphics[width=0.49\textwidth]{figs/ggHjets/dy/%
    Multidist_H+123_pT100vs50eta100mult_higgs_jet_dy_1_20-E100}
  \hfill
  \includegraphics[width=0.49\textwidth]{figs/ggHjets/dy/%
    Multidist_H+3_pTXeta100mult_jet_jet_dy_12_20-E100}
  \caption{\label{fig:ggHjets_ggfdy}
    The rapidity separation $\Delta y_{H,\,j_1}$ between the
    Higgs boson and the hardest jet in $\mathrm{H}\!+\!1,2,3$-jet
    production and the rapidity separation $\Delta y_{j_1,\,j_2}$
    between the two hardest jets in $\mathrm{H}\!+\!3$-jet production
    at a 100~TeV proton--proton collider. The spectra for the latter
    are shown for varying jet-$p_T$ thresholds. Again, the layout of
    the left and the right plot corresponds to the layout employed in
    Figure~\ref{fig:ggHjets_ggfpTHyH} and \ref{fig:ggHjets_ggfpTH_jcut},
    respectively.}
\end{figure}

The plots of Figure~\ref{fig:ggHjets_ggfdy} show the rapidity
separation between the Higgs boson and the leading jet (on the left)
and between the two leading jets (on the right). In the former
case, the distributions show the results for the three different final
state multiplicities, whereas in the latter case, the curves refer to
the $\mathrm{H}\!+\!3$-jet process and compare the impact of the
different jet transverse momentum cuts. For both observables, the
large production rates and the huge available phase space allow to
have differential cross sections, which for separations as large as
three units in $\Delta y$, are only a factor of two smaller than the
ones at zero rapidity separation. Independent of the jet multiplicity,
both NLO corrections as well as tighter jet definitions trigger
enhancements in the $\Delta y_{H,\,j_1}$ distribution (left panel) for
configurations where the Higgs boson and the leading jet are close in
rapidity. For the $\Delta y_{j_1,\,j_2}$ variable (right panel), a
rather uniform behaviour is found while changing the jet threshold:
the three-jet over two-jet fraction as well as the $K$-factors remain
rather constant over the entire $\Delta y$ range.

\begin{figure}[t!]
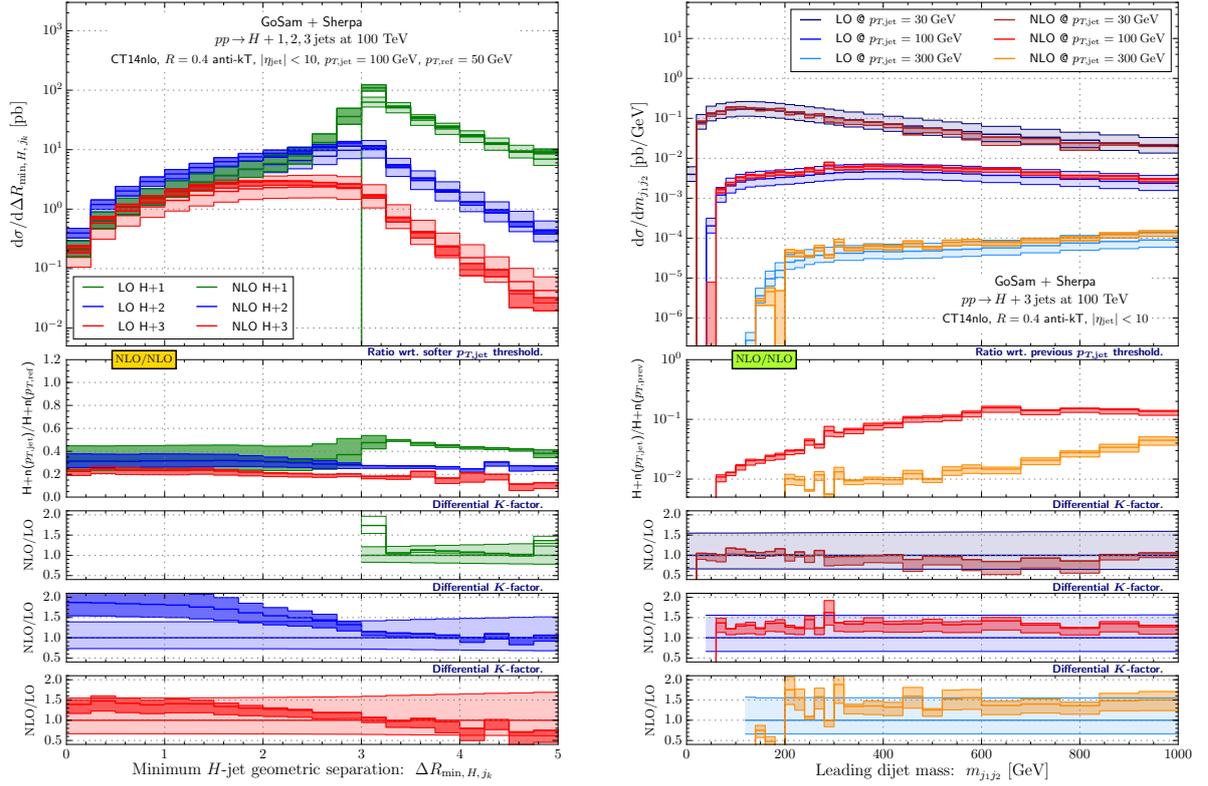

  \centering
  \includegraphics[width=0.49\textwidth]{figs/ggHjets/dRmin/%
    Multidist_H+123_pT100vs50eta100mult_higgs_jet_dR_min-E100}
  \hfill
  \includegraphics[width=0.49\textwidth]{figs/ggHjets/mdijet/%
    Multidist_H+3_pTReta100mult_jet_jet_mass_12_100-E100}
  \caption{\label{fig:ggHjets_ggfdRmin}
    The geometric separation $\Delta R_{\mathrm{min},\,H,\,j_k}$
    between the Higgs boson and the jet closest to it, and the
    invariant mass distribution of the leading dijet system at a
    100~TeV proton--proton collider. For the former, distributions
    are shown for $\mathrm{H}\!+\!1,2,3$-jet production, while for the
    latter, the jet-$p_T$ thresholds are varied to show the
    corresponding distributions obtained from $\mathrm{H}\!+\!3$-jet
    events. The colour coding and plot layout is as previously
    described with the only exception that the upper ratios in the
    left panel are taken between successive $p_{T,\,\mathrm{jet}}$
    results for the same jet multiplicity.}
\end{figure}

Additional two-particle observables are presented in
Figure~\ref{fig:ggHjets_ggfdRmin}. The left plot shows the radial
separation between the Higgs boson and the closest jet for
$\mathrm{H}\!+\!1,2$ and $3$-jet production. As expected, with an
increasing number of jets in the final state, the average radial
separation between the Higgs boson and the closest jet decreases. As a
consequence, for small radius separations, the contributions from all
three jet multiplicities are of similar size, whereas for values
larger than $\pi$, it is the lower multiplicities that dominate for
obvious kinematical reasons. We also take from the first ratio plot
that a higher jet threshold leads to more centrally produces jets such
that there is a small rate increase at low $\Delta R_\mathrm{min}$ for
two- and higher jet multiplicities. Furthermore, the NLO corrections
can be as large as 100\% and in case of $\mathrm{H}\!+\!3$ jets, the
$K$-factor can as well be significantly smaller than one. The right
plot presents predictions for the dijet invariant mass of the two
leading jets in $\mathrm{H}\!+\!3$-jet production for different jet
transverse momentum thresholds. It is interesting to observe that
because of the vast phase space, the distributions fall off very
slowly and for a transverse momentum threshold of
$p_{T,\,\mathrm{jet}}>300~\mathrm{GeV}$, the maximum of the
distribution actually lies above $1~\mathrm{TeV}$. Looking at the
impact of higher-order contributions in the lower three panels, we
observe that these corrections slightly increase for larger jet
thresholds but for all three choices, the $K$-factor remains flat to a
good approximation. In the second panel we instead consider the cross
section ratios at NLO between successive $p_{T,\,\mathrm{jet}}$. They
show that any jet threshold step-up by a factor of three results in a
reduction of at least one order of magnitude.

\subsubsection{Results with vector boson fusion selection criteria}
\label{sec:ggHjets_vbfresults}

In order to quantify the number of GGF background events passing the
VBFH selection criteria, we present results for which we applied the
following VBFH-type cuts on top of the baseline set of constraints
defined in the previous section:
\begin{equation}
\label{eq:ggHjets_vbfcuts}
m_{j_1j_2}\;>\;1600~\mathrm{GeV}\,,\qquad
\left|\Delta y_{j_1,\,j_2}\right|\;>\;6.5\,,\qquad
y_{j_1}\cdot y_{j_2}\;<\;0\,.
\end{equation}
In Table~\ref{table:ggHjets_vbf_incl_XS} the total cross sections for
a center-of-mass energy of 100~TeV are summarized. Differential
distributions will be discussed in a slightly different context in one
of the later sections, see Section~\ref{sec:H_vbf}, together with the
results obtained from the VBF@NNLO computations.

\begin{table}[t!]
\centering\small
\begin{tabular}{|l||c|c|}
  \hline
  Numbers in pb
  & $\sigma_{\mathrm{LO}}^{100\,\mathrm{TeV}}$
  & $\sigma_{\mathrm{NLO}}^{100\,\mathrm{TeV}}$\phantom{\Big|}\\
  \hline
  \multicolumn{3}{|c|}{$\mathrm{H}\!+\!2$ jets\phantom{\Big|}}\\
  \hline
  $p_{T,\,\mathrm{jet}}>30~\mathrm{GeV}$  \phantom{\Big|} & $4.60^{+56\%}_{-33\%}$ & $4.70^{-17\%}_{-7\%}$ \\
  $p_{T,\,\mathrm{jet}}>50~\mathrm{GeV}$  \phantom{\Big|} & $1.71^{+56\%}_{-34\%}$ & $1.98^{-6\%}_{-11\%}$ \\
  $p_{T,\,\mathrm{jet}}>100~\mathrm{GeV}$ \phantom{\Big|} & $0.26^{+57\%}_{-34\%}$ & $0.31^{-3\%}_{-13\%}$ \\
  $p_{T,\,\mathrm{jet}}>300~\mathrm{GeV}$ \phantom{\Big|} & $5.10\cdot10^{-3}\phantom{i}^{+58\%}_{-34\%}$ & $6.20\cdot10^{-3}\phantom{i}^{-1\%}_{-14\%}$ \\
  \hline
\end{tabular}
\caption{\label{table:ggHjets_vbf_incl_XS}%
  Total inclusive cross sections for the production of a Higgs
  boson in association with two jets at LO and NLO in QCD after the
  application of the VBF selection criteria stated in
  Eq.~\ref{eq:ggHjets_vbfcuts}. Numbers are reported for a
  center-of-mass energy of 100 TeV and four choices of transverse
  momentum cuts, namely $p_{T,\,\mathrm{jet}}>30,\,50,\,100$ and
  $300~\mathrm{GeV}$. The uncertainty envelopes are obtained from
  standard scale variations.}
\end{table}

\subsubsection{Finite quark mass effects}
\label{sec:ggHjets_meffresults}

It is well known that the infinitely large top quark mass
approximation has a restricted validity range, and that for energies
large enough to resolve the massive top quark loop, the deviations
start to become sizeable. In order to quantify better the effects due
to finite quark masses, in the following we compare LO predictions in the
effective theory with computations in the full theory.  We consider
here only massive top quarks running in the loop. The effect of
massive bottom quarks for a center-of-mass energy of 100~TeV can be
safely neglected. For the top quark mass, we use $m_t=172.3~\mathrm{GeV}$.
Compared to the results shown in the previous section, we now impose a
more restrictive cut on the pseudo-rapidity of the jets, demanding
$|\eta_\mathrm{jet}|<4.4$; the impact of this cut is however fairly
minimal on the observables that we are considering.

\begin{table}[b!]
\centering\small
\begin{tabular}{|l||c|c|c||c|}
  \hline
  Numbers in pb
  & $\sigma_{\mathrm{LO}}^{100\,\mathrm{TeV}}$
  & $\sigma_{\mathrm{NLO}}^{100\,\mathrm{TeV}}$
  & $\sigma_{\mathrm{LO, full}}^{100\,\mathrm{TeV}}$
  & $\sigma_{\mathrm{LO, full}}^{100\,\mathrm{TeV}}/\sigma_{\mathrm{LO}}^{100\,\mathrm{TeV}}$\phantom{\Big|}\\
  \hline
  \multicolumn{5}{|c|}{$\mathrm{H}\!+\!2$ jets\phantom{\Big|}}\\
  \hline
  $p_{T,\,\mathrm{jet}}>30~\mathrm{GeV}$  \phantom{\Big|} &
  $124^{+39\%}_{-27\%}$ & $156^{+3\%}_{-10\%}$ & $120^{+39\%}_{-26\%}$ & $0.968$ \\
  $p_{T,\,\mathrm{jet}}>50~\mathrm{GeV}$  \phantom{\Big|} &
  $57.3^{+40\%}_{-27\%}$ & $76.5^{+6\%}_{-11\%}$ & $52.2^{+40\%}_{-27\%}$ & $0.911$ \\
  $p_{T,\,\mathrm{jet}}>100~\mathrm{GeV}$ \phantom{\Big|} &
  $16.5^{+41\%}_{-28\%}$ & $23.3^{+9\%}_{-13\%}$ & $13.1^{+41\%}_{-27\%}$ & $0.794$\\
  $p_{T,\,\mathrm{jet}}>300~\mathrm{GeV}$ \phantom{\Big|} &
  $1.40^{+43\%}_{-28\%}$ & $2.05^{+10\%}_{-14\%}$ & $0.62^{+43\%}_{-28\%}$ & $0.443$\\
  \hline
  \multicolumn{5}{|c|}{$\mathrm{H}\!+\!3$ jets\phantom{\Big|}}\\
  \hline
  $p_{T,\,\mathrm{jet}}>30~\mathrm{GeV}$  \phantom{\Big|} &
  $70.4^{+56\%}_{-34\%}$ & $72.6^{-15\%}_{-8\%}$ & $63.0^{+56\%}_{-34\%}$ & $0.895$ \\
  $p_{T,\,\mathrm{jet}}>50~\mathrm{GeV}$  \phantom{\Big|} &
  $25.2^{+56\%}_{-34\%}$ & $29.3^{-5\%}_{-11\%}$ & $20.8^{+56\%}_{-34\%}$ & $0.825$ \\
  $p_{T,\,\mathrm{jet}}>100~\mathrm{GeV}$ \phantom{\Big|} &
  $5.13^{+56\%}_{-34\%}$ & $6.57^{+3\%}_{-14\%}$ & $3.46^{+57\%}_{-34\%}$ & $0.674$\\
  $p_{T,\,\mathrm{jet}}>300~\mathrm{GeV}$ \phantom{\Big|} &
  $0.24^{+56\%}_{-34\%}$ & $0.33^{+9\%}_{-16\%}$ & $0.07^{+60\%}_{-35\%}$ & $0.292$\\
  \hline
\end{tabular}
\caption{\label{table:ggHjets_meff_incl_XS}%
  Total inclusive cross sections for the hadro-production of a
  Higgs boson in association with two as well as three jets at a
  center-of-mass energy of $100$~TeV. The LO results as predicted by
  the full theory are shown in the next-to rightmost column, and are
  compared to the results from the effective theory at LO and NLO.
  Note that all cross section are obtained for the basic gluon fusion
  selection, however imposing a narrower jet rapidity requirement of
  $|\eta_\mathrm{jet}|<4.4$. Again, rates are calculated for up to
  four choices of jet-$p_T$ thresholds, namely
  $p_{T,\,\mathrm{jet}}>30,\,50,\,100$ and $300$~GeV. The last column
  lists the ratios between the LO predictions of the full and
  effective theory.}
\end{table}

\begin{figure}[t!]
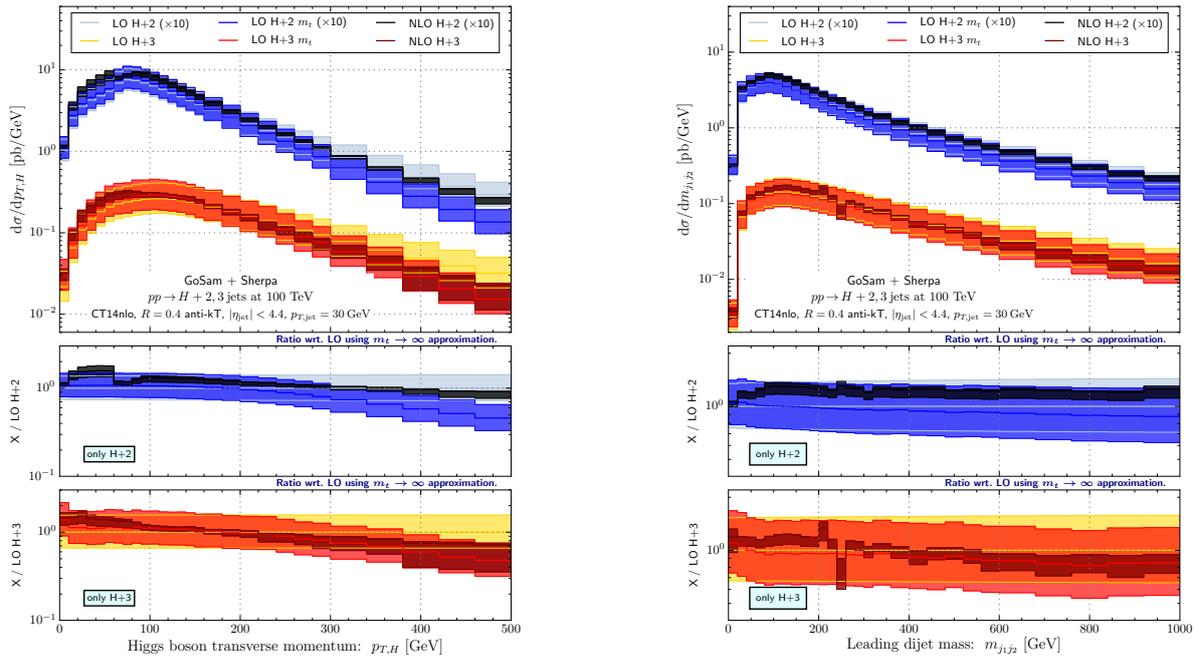

  \centering
  \includegraphics[width=0.45\textwidth]{figs/ggHjets/Meff/%
    Multimeff_H+23_pT30eta100mult_higgs_pt_100-E100}
  \hfill
  \includegraphics[width=0.45\textwidth]{figs/ggHjets/Meff/%
    Multimeff_H+23_pT30eta100mult_jet_jet_mass_12_100-E100}
  \caption{\label{fig:meff_ptH_mj1j2}%
    The impact of finite top quark mass effects in the loop-induced
    emission of a Higgs boson in GGF $\mathrm{H}\!+\!n$-jet production
    at a 100~TeV proton--proton collider. LO results
    based on the full and effective theory as well as NLO results
    using the effective theory are shown (to the left) for the
    transverse momentum spectrum, $p_{T,\,H}$, of the Higgs boson and
    (to the right) for the invariant mass distribution, $m_{j_1j_2}$,
    of the leading dijet system. The $\mathrm{H}\!+\!2$-jet and
    $\mathrm{H}\!+\!3$-jet predictions are grouped in separate ratio
    plots using the respective LO results as their reference.
    Uncertainty bands are derived from standard scale variations.}
\end{figure}

Before turning to the discussion of a handful of differential cross
sections, we compare the predictions for the total inclusive cross
section in the effective theory at LO and NLO, and in the full theory
at LO; Table~\ref{table:ggHjets_meff_incl_XS} lists the results for
various jet-$p_T$ thresholds. We also indicate the reduction of the LO
cross section induced by the incorporation of finite top quark mass
effects. As expected, this reduction becomes more pronounced for
increasing values of $p_{T,\,\mathrm{jet}}$ turning the finite mass
corrections into the dominant effect for
$p_{T,\,\mathrm{jet}}\gtrsim100~\mathrm{GeV}$. This effect becomes
even more dramatic when increasing the multiplicity from
$\mathrm{H}\!+\!2$~jets to $\mathrm{H}\!+\!3$~jets.

Figure~\ref{fig:meff_ptH_mj1j2} shows predictions for the transverse
momentum distribution of the Higgs boson (left), and for the leading
dijet invariant mass (right). We observe that for the transverse
momentum, the finite top mass effects start to become important at
values of $p_{T,H}\approx300~\mathrm{GeV}$. Interestingly, the NLO
corrections for the given scale choice show the same qualitative
behaviour as the LO contribution in the full theory. In particular for
the $\mathrm{H}\!+\!3$-jet process, the full theory and the NLO
effective theory seem to almost give the same prediction, except for
the size of the scale variations.
On the contrary the predictions for the invariant mass distributions
in the full theory only mildly deviate from the ones computed in the
effective theory; again the tails are somewhat softer. For both
$\mathrm{H}\!+\!2$ jets and $\mathrm{H}\!+\!3$ jets, the NLO
predictions remain within the respective scale uncertainties of the LO
results over the whole range shown in the figure. In addition, the
shapes of the NLO effective and LO full theory show a fairly similar
behaviour. In the two-jet case there is however a clear rate
difference due to the considerably greater-than-one $K$-factor.

\begin{figure}[t!]
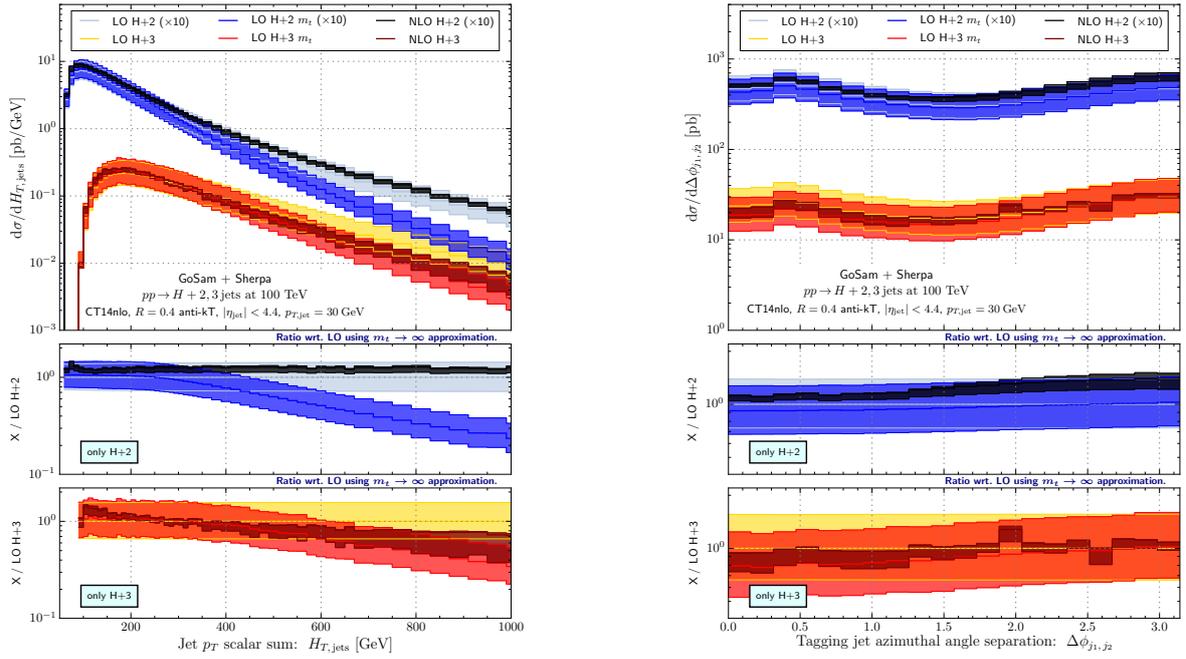

  \centering
  \includegraphics[width=0.45\textwidth]{figs/ggHjets/Meff/%
    Multimeff_H+23_pT30eta100mult_jets_ht_100-E100}
  \hfill
  \includegraphics[width=0.45\textwidth]{figs/ggHjets/Meff/%
    Multimeff_H+23_pT30eta100mult_jet_jet_dphi_12_20-E100}
  \caption{\label{fig:meff_HT_dphij1j2}%
    Finite top quark mass effects in GGF-based $\mathrm{H}\!+\!n$-jet
    production arising from collisions of protons at 100 TeV. The
    jets-only scalar sum of transverse momenta,
    $H_{T,\,\mathrm{jets}}$, and the azimuthal angle distribution,
    $\Delta\phi_{j_1,\,j_2}$, between the leading and subleading jet
    are shown to the left and right, respectively. The differential
    spectra (and associated uncertainties from scale variations) were
    obtained from the full theory at LO, and the effective theory at
    LO (providing the references) and NLO. The upper ratio plot
    contains the $\mathrm{H}\!+\!2$-jet predictions while the lower
    one depicts those for $\mathrm{H}\!+\!3$ jets.}
\end{figure}

The last figure exemplifies how severe deviations can become between
the full and effective theory description for the hardness of
transverse particle/jet production. This is nicely demonstrated by
means of the $H_{T,\,\mathrm{jets}}$ distribution shown to the left of
Figure~\ref{fig:meff_HT_dphij1j2}. The finite top quark mass effects
clearly dominate over the NLO corrections calculated in the effective
theory. The $\mathrm{H}\!+\!2$-jet case in particular demonstrates
the extreme and opposite behaviour of both effects -- a large
$K$-factor on the one side versus an even more effective suppression
of the $H_T$ tail on the other side by about 70-80\%. While the finite
mass effects always suppress the rate for hard jet production, the NLO
corrections can lead to an enhancement as seen for $\mathrm{H}\!+\!2$
jets as well as to a reduction as we observe in the
$\mathrm{H}\!+\!3$-jet case.

As in the previous figure, we contrast two different types of
observables with each other. In the right panel of
Figure~\ref{fig:meff_HT_dphij1j2} we therefore display an angular
correlation, more precisely we show the azimuthal angle separation
between the leading and subleading jet. This kind of observable is
important in precision coupling measurements, and although the
corrections tend to be much smaller, they have to be understood in
detail to satisfy the demand for high precision. Here, both the NLO
corrections and finite mass effects lead to similar shape changes,
suppressing small-angle contributions while the back-to-back
configurations receive an (effective) enhancement (due to the finite
$m_t$ treatment). This time, we notice a slight increase of the $m_t$
effects (from 10\% to 20\%) for the final states of higher jet
multiplicity.

\subsubsection{Conclusions}
\label{sec:ggHjets_conclusions}
In this section we studied the associate production of a Standard
Model Higgs boson in association with up to three jets in gluon--gluon
fusion. We compared LO and NLO QCD predictions for several different
cuts on the transverse momentum of the jets and also produced results
with VBFH-type cuts, which are compared with VBF predictions in
Section~\ref{sec:H_vbf}. Because of the large center-of-mass energy and the
huge available phase space, the production rates become much larger
compared to LHC energies. This can be observed in particular when
comparing the relative size of contributions coming from different
multiplicities. In the last part of this section, we studied the impact
of finite quark mass effects. For typical jet transverse momentum
cuts, which at FCC energies are likely to be of the order of
$100~\mathrm{GeV}$, these corrections are non-negligible and their
impact is in general larger than NLO QCD corrections in the effective
theory. Moreover, all quantities that are related to measuring the
transverse activity of the $\mathrm{H}\!+\!n$-jet processes will
receive significant corrections reducing the cross section in the hard
regions.

\clearpage
 \subsection{Associated $VH$ production}
\label{sec:H_VH}
Associate production of the Higgs and gauge bosons mostly arises from
the Higgs-strahlung processes $q\bar{q} \to V^* \to VH$ ($V=W,Z$). These
provide direct probes of the $VVH$ couplings.

Cross sections for $HV$ associated production in hadron collisions are
studied since long.  For the inclusive cross sections, up to
NNLO QCD corrections are available in the
program vh@nnlo~\cite{Brein:2012ne,Harlander:2013mla,Brein:2011vx}.
Furthermore, programs have been developed for the computation of fully
differential distributions including NNLO QCD
corrections~\cite{Ferrera:2011bk,Ferrera:2013yga,Ferrera:2014lca,Campbell:2016jau},
the EW corrections~\cite{Denner:2014cla,Denner:2011id} and
event generators matching and merging NLO QCD corrections for VH+jets
production to parton
showers~\cite{Luisoni:2013cuh,Frederix:2012ps,Hespel:2015zea}. Finally,
the computation of NNLO QCD corrections matched with parton shower has
been worked out in ref.~\cite{Astill:2016hpa}, reweighting events
samples obtained with the code presented in
ref.~\cite{Luisoni:2013cuh} with the histograms obtained with the
program of ref.~\cite{Ferrera:2014lca} and relying for the level of
accuracy on the theorems presented in~\cite{Hamilton:2013fea}.
The results in Table~\ref{table:HV_fiducial_XS} have been obtained
using vh@nnlo and the
$\rm{NNPDF30\_nnlo\_as\_0118}$~\cite{Ball:2014uwa} pdf set.  The
central renormalisation and factorisation scales have been set both to
the mass of the $HV$ system.  For the estimate of the scale uncertainty
reported in the table, we varied them independently up to a factor of
$3$ with the constraint $\mu_f \cdot \mu_r \le 2$.
\begin{table}[h!]
\centering
\small
\renewcommand\arraystretch{2}
\begin{tabular}{  l | c  c  c | c  c  c  }
 & $\sigma_{tot}$[pb] & $ \delta_{PDF}$[pb] & $ \delta_{scale}$[pb]   & $\sigma_{DY}$[pb] & $\sigma_{ggHV}$[pb] & $\sigma_{top}$[pb] \\
   \hline
HW & $15.710$ & $\pm 0.024$ & $^{+0.010}_{-0.020}$ & $15.403$ & $-$ & $0.307$  \\
HZ & $11.178$ & $\pm 0.022$ & $^{+0.062}_{-0.044}$ & $\phantom{0}8.946$ & $2.069$ & $0.163$  \\
   \hline
\end{tabular}
\caption{Total cross sections $\sigma(VH)$, including up to NNLO QCD corrections, and respective PDF and PDF uncertainties.}
\label{table:HV_fiducial_XS}
\end{table}

Figure~\ref{table:wbbj_fid_XS} shows the cross sections for the $WH$
process, with different selection cuts for the associated jet activity.
\begin{table}[h]
\centering
\small
\begin{tabular}{ | c | c | c | c | c | }
\hline
  \multicolumn{5}{|c|}{$\sigma_{\mathrm{NNLO}}$ [fb] @ 100 TeV\phantom{***}}     \\
\hline
\multicolumn{5}{|c|}{$HW^+(\rightarrow H e^+ \nu_e )$  \phantom{***}}\\
\hline
   no cuts
& no jets with
& at least 1 jet with
& at least 1 jet with
& at least 1 jet with \\
& $p_T> 100$~GeV
& $p_T> 100$~GeV
& $p_T> 500$~GeV
& $p_T> 1$~TeV \\
\hline
$539 $    &  $444$ & $94.7$ & $5.20 $ & $0.817 $ \\

\hline
\multicolumn{5}{|c|}{$HW^-(\rightarrow H e^- \bar{\nu}_e )$ \phantom{***}}\\
\hline
   no cuts
& no jets with
& at least 1 jet with
& at least 1 jet with
& at least 1 jet with \\
& $p_T> 100$~GeV
& $p_T> 100$~GeV
& $p_T> 500$~GeV
& $p_T> 1$~TeV \\
\hline
$425 $    &  $350.6$ & $74.37$ & $3.718$ & $0.541$ \\
\hline
\end{tabular}
\caption{$HW$ fiducial cross sections in fb at NNLO accuracy for the different selection cuts
on the jet activity.}
\label{table:wbbj_fid_XS}
\end{table}

As was the case for the Higgs production processes discussed so far,
collisions at 100 TeV allow to extend the kinematic reach of $VH$
final states to rather extreme configurations, where the $VH$ pair is
produced with huge invariant mass, or at very large $p_T$.  Production
at large invariant mass is of phenomenological interest for several
reasons. For example, it provides the leading source of irreducible background
for the detection of exotic new particles (e.g. new gauge bosons)
decaying to $VH$. Furthermore, prodution
at large invariant mass probes the $VVH$ coupling in the region where
the $Q^2$ of the virtual gauge boson is far off shell. This could
exhibit sensitivity to the presence of higher-dimension effective
operators, potentially beyond what can be tested from the precise
determination of the $H\to VV^*$ decay branching
ratios. Figure~\ref{fig:VH_M} (left panel) shows that, with over
10~\iab\ of integrated luminosity, the SM rate will extend all the way
out to $M(VH)\sim 20$~TeV. In these configurations, with Higgs and
gauge bosons with a transverse momentum of several TeV, it is likely
that one will be able to effectively tag these events through the
$H\to b\bar{b}$ and $V\to$~dijet decay modes, therefore using the
largest available branching ratios!

Another interesting, and complementary, kinematical configuration, is
the prodution of the $VH$ pair at large $p_T$. This will be dominated
by the recoil against a jet, with the $VH$ pair maintaining a small
invariant mass. As shown in the right panel of Figure~\ref{fig:VH_M}, 
there will be rate out to $p_T(VH)$ beyond 7~TeV.
As we shall see in section~\ref{sec:H_prospects_VH}, and in the case
of the abundant $H\to b\bar{b}$ decay modes, these
configurations enjoy a particularly favourable $S/B$ ratio, when
compared to the otherwise dominant QCD background from associated
$Vb\bar{b}$ production. 

\begin{figure}[h]
    \includegraphics[width=0.45\textwidth]{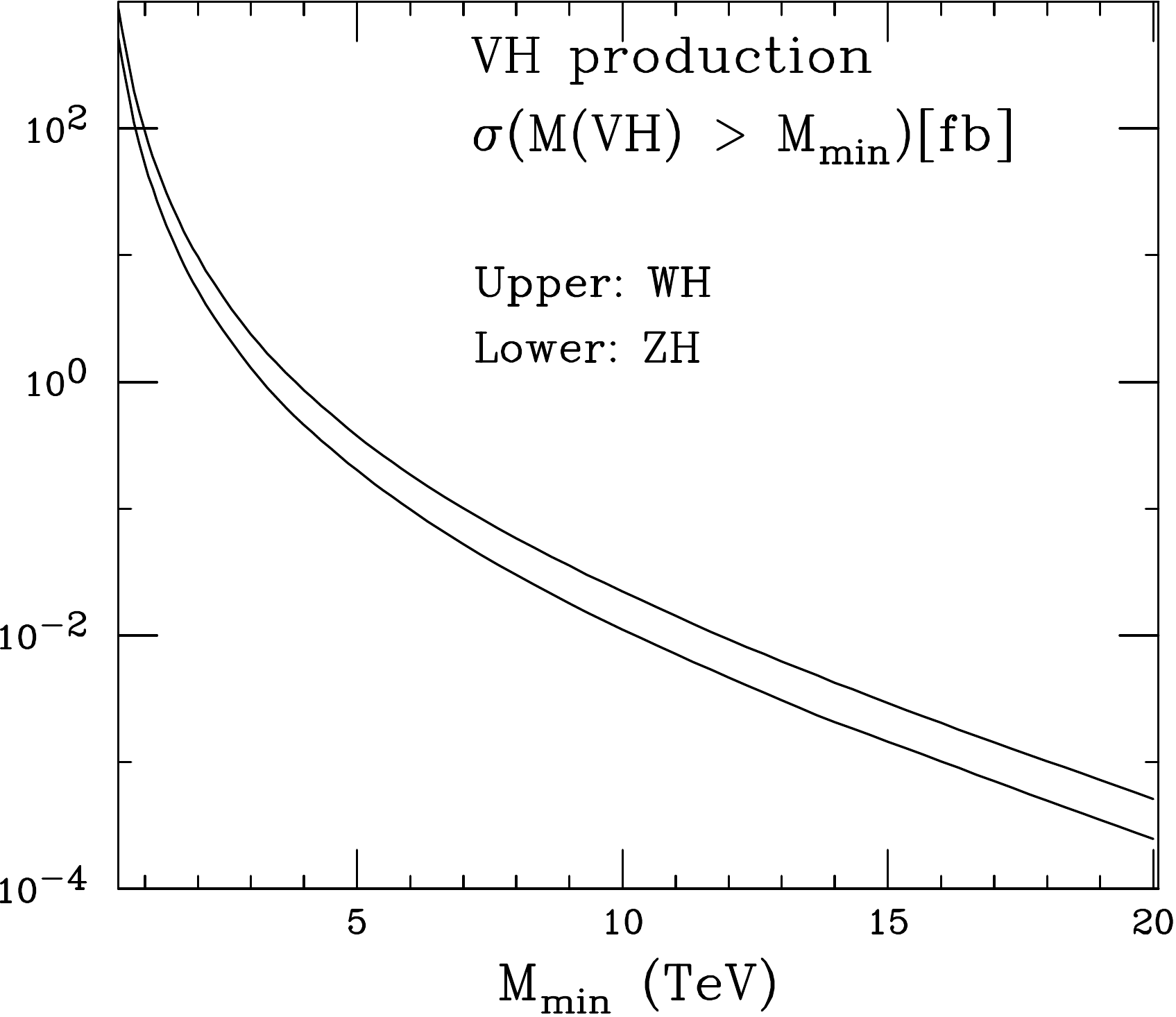}
    \hfill
    \includegraphics[width=0.45\textwidth]{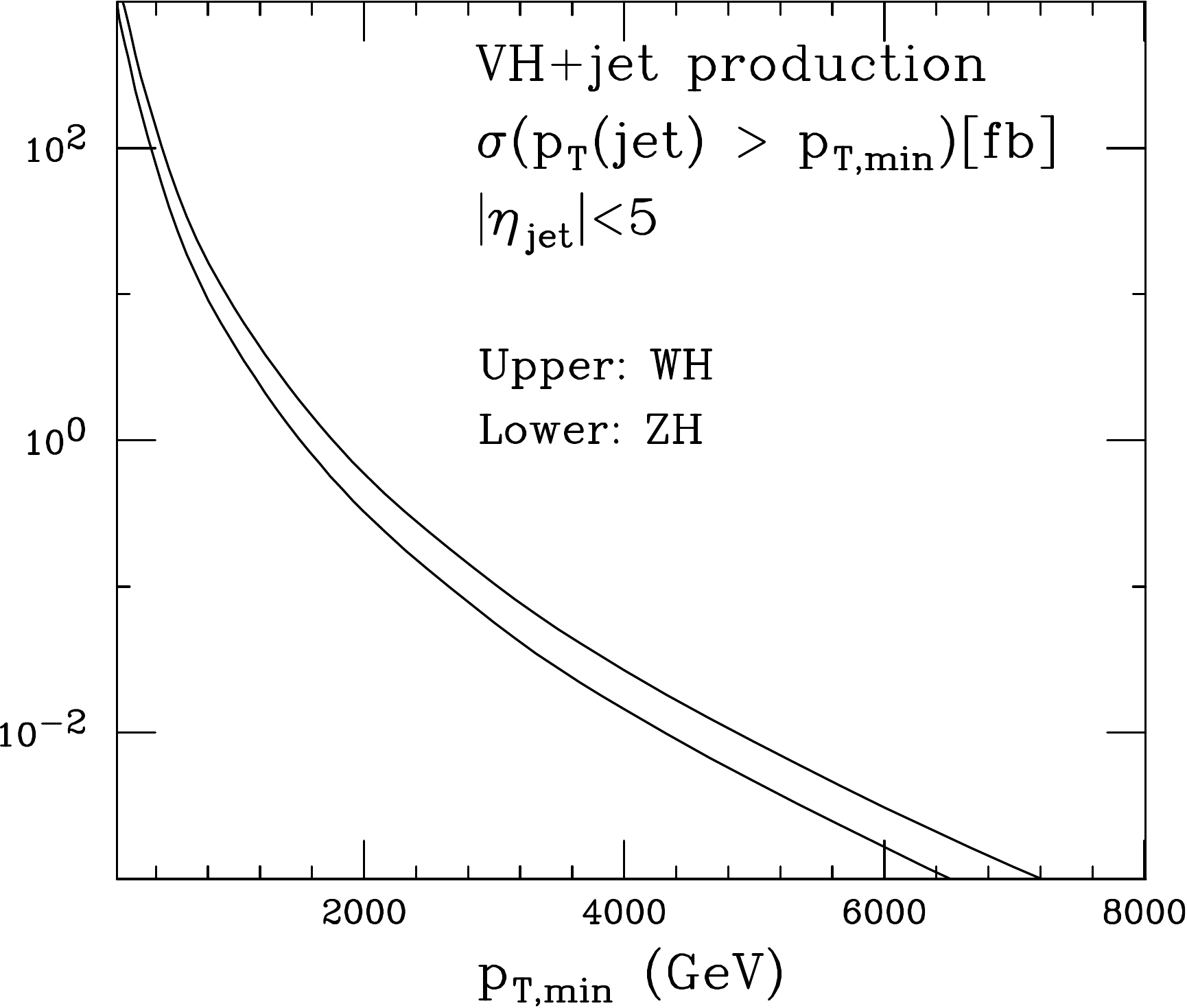}
  \caption{\label{fig:VH_M}
    Left panel: integrated invariant mass distribution of $VH$ pairs,
    at LO. Right panel: integrated transverse momentum spectrum of the
    $VH$ system, recoiling against a jet.}
\end{figure}

Figure~\ref{fig:VH_pt} sows the integrated $p_T$ spectrum of the Higgs
produced in association with a $W^\pm$ boson. The solid line
corresponds to the LO process, in which it is the $W$ that recoils
against the Higgs. The dashed and dotted lines show the contribution
induced by the $q\bar{q} \to WH g$ and
$qg \to WH q$ processes.

\begin{figure}[h]
    \includegraphics[width=0.45\textwidth]{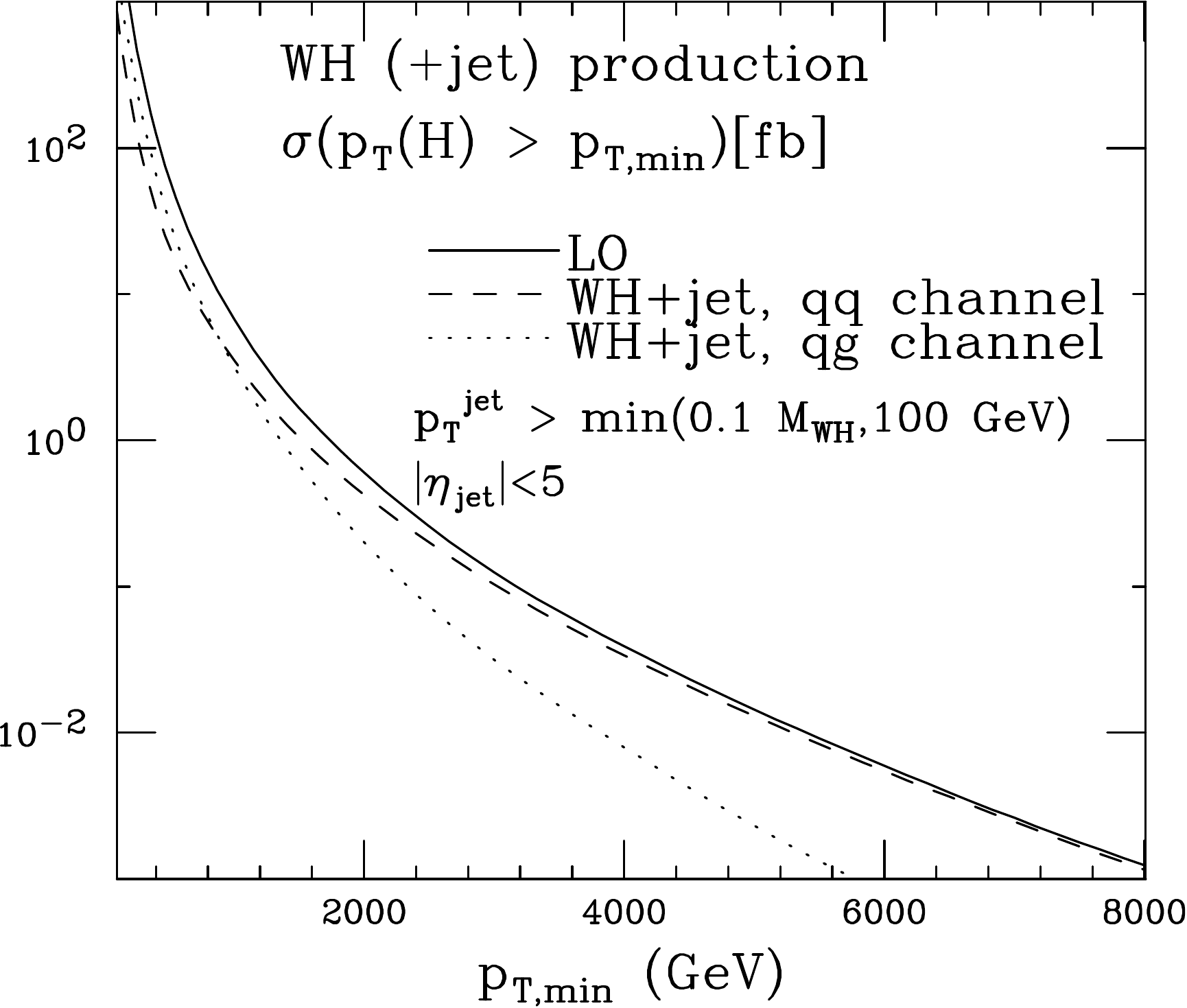}
    \hfill
    \includegraphics[width=0.45\textwidth]{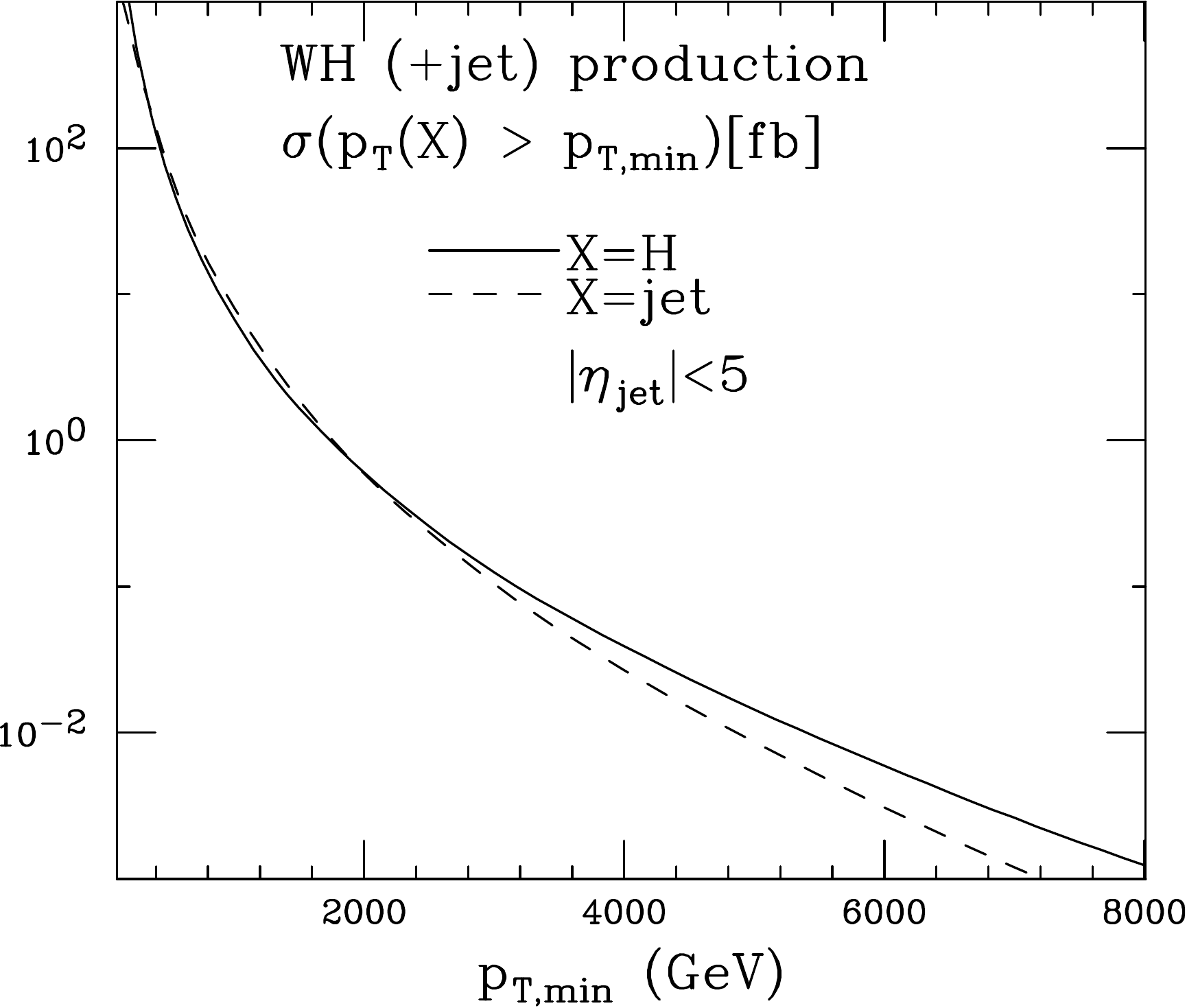}
  \caption{\label{fig:VH_pt}
    Integrated inclusive $p_T$ spectrum of the Higgs boson (left
    panel) and of a jet produced in the $WH+$jet porcess (right panel).}
\end{figure}

To isolate the hard component of these
higher-order corrections, namely to exclude the virtual and
soft/collinear $O(\alpha_s)$ processes that exhibit Born-like kinematics,
we require the radiated parton to have $p_T$ no smaller than 10\% of the
$WH$ mass or of 100~GeV. The results show that Higgs prodution at
large $p_T$ is indeed dominated by Born-like kinematics. This is
confirmed in the left panel of Fig.~\ref{fig:VH_dR}, showing the
back-to-back feature of the $\Delta
R(WH)$ distribution (normalized to 1), for various $p_T(H)$
thresholds. We notice the very different shape of the $\Delta R(WH)$
distribution when events are tagged by the presence of a large $p_T$
jet (see central plot of Fig.~\ref{fig:VH_dR}). 

\begin{figure}[h]
    \includegraphics[width=0.32\textwidth]{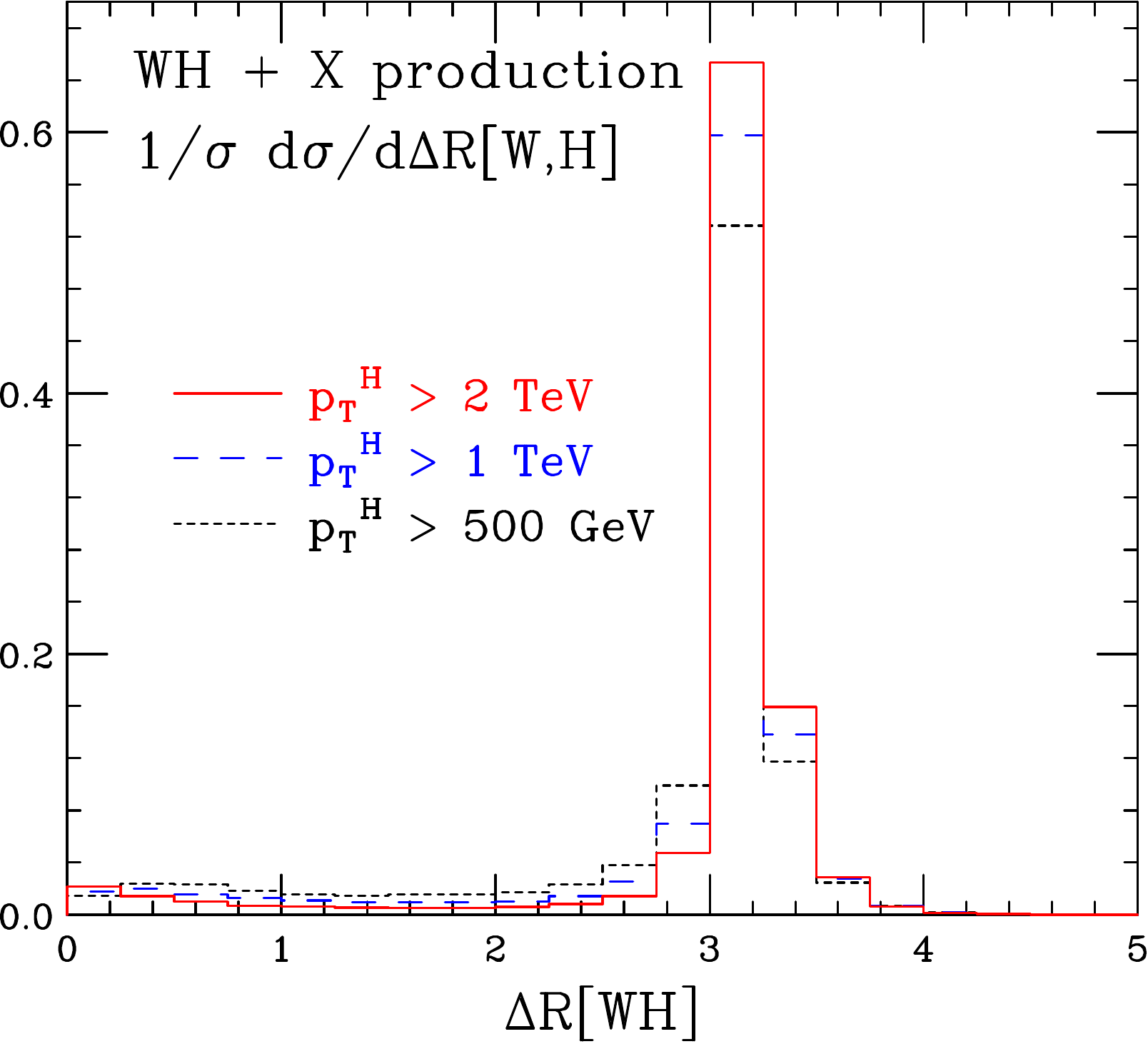}
    \hfill
    \includegraphics[width=0.32\textwidth]{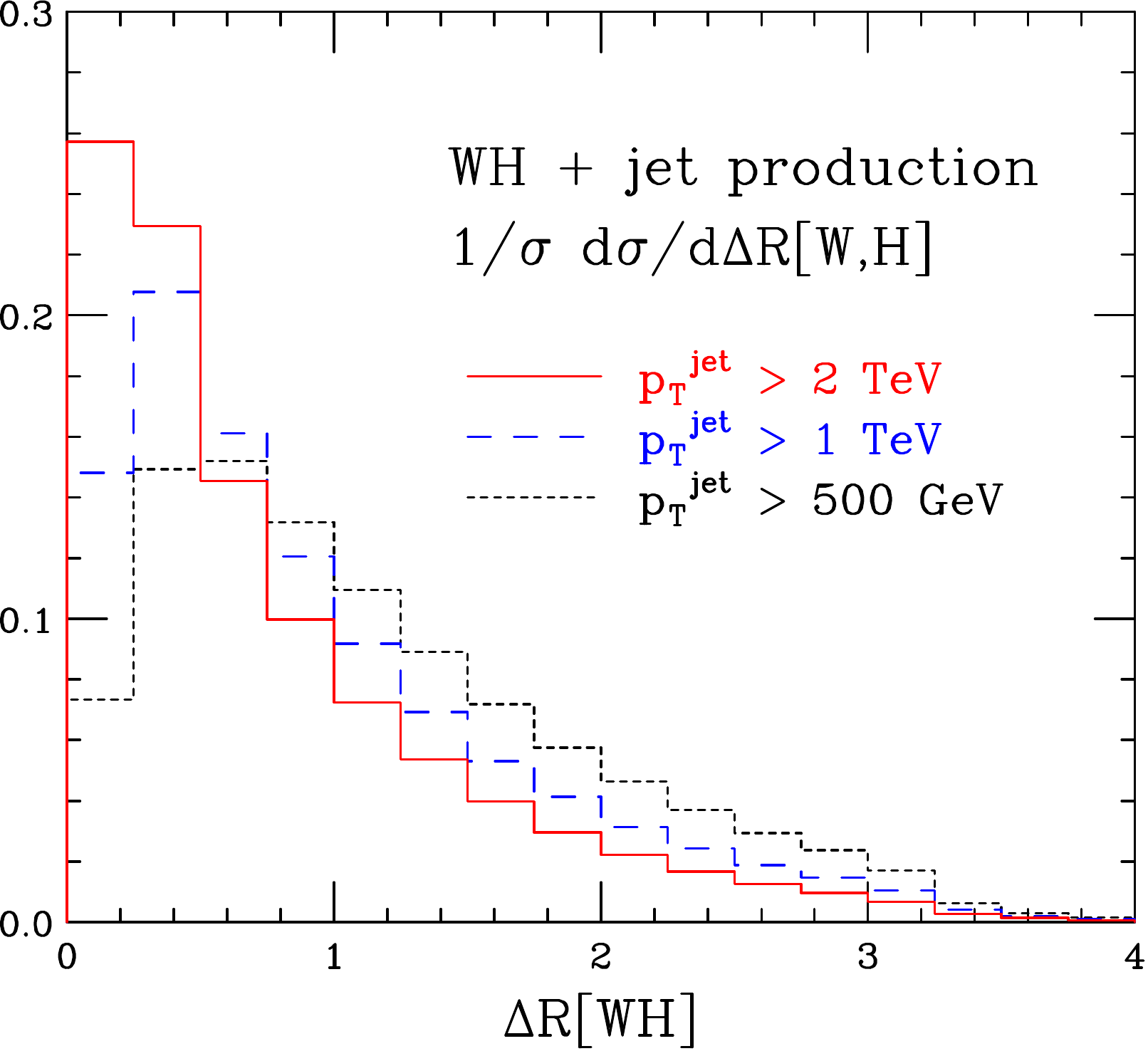}
    \hfill
    \includegraphics[width=0.32\textwidth]{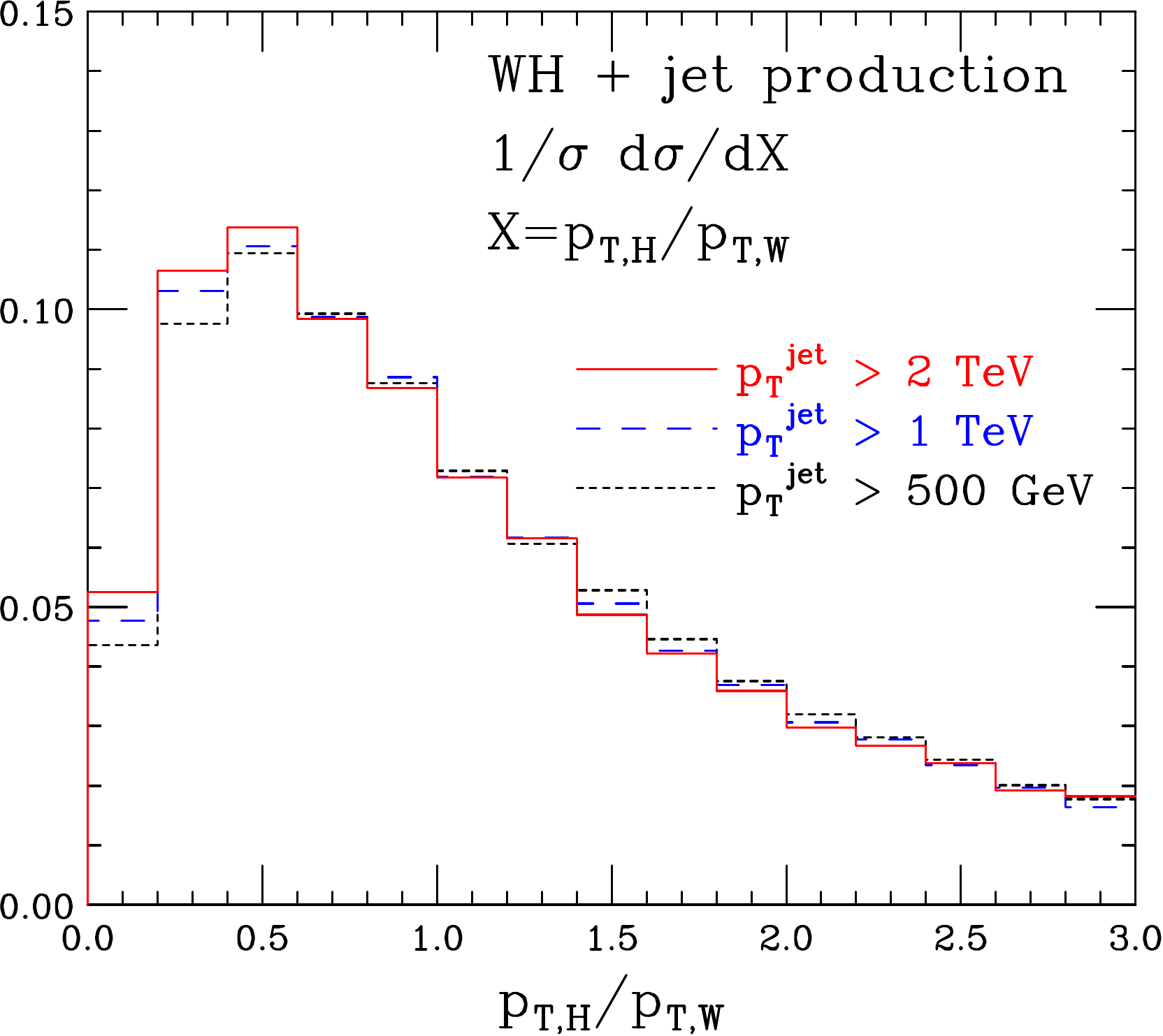}
  \caption{\label{fig:VH_dR}
    $\Delta R$ correlation between the $H$ and $W$ boson, in events
    tagged by the presence of a high-$p_T$ Higgs (left) and of a
    high-$p_T$ jet (central panel). Right panel: transverse momentum of
    the Higgs boson, relative to that of the $W$ boson, in events with
  a high-$p_T$ jet. All distributions are normalized to 1. }
\end{figure}

\clearpage
\subsection{VBF Higgs production}
\label{sec:H_vbf}
In this section we study the production of a Standard Model Higgs
through Vector Boson Fusion (VBFH) at a $100$ TeV proton-proton
collider. As is the case at $13$ TeV, VBFH has the second largest
Higgs production cross section and is interesting on its own for a
multitude of reasons: 1) it is induced already at tree-level; 2) the
transverse momentum of the Higgs is non-zero at lowest order which
makes it suitable for searches for invisible decays; 3) it can be
distiguished from background processes due to a signature of two
forward jets. This last property is very important, as the inclusive
VBF signal is completely drowned in QCD Hjj production. One of the
aims of this section is to study how well typical VBF cuts suppress
this background at a $100$ TeV proton-proton machine.

\subsubsection{Generators}

Fixed order LO predictions for and QCD corrections to VBFH have been
obtained using \textsc{proVBFH}~\cite{Cacciari:2015jma} which is based
on POWHEG's fully differential NLO QCD calculation for Higgs
production in association with three jets via
VBF~\cite{Figy:2007kv,Jager:2014vna}, and an inclusive NNLO QCD
calculation~\cite{Bolzoni:2010xr}. NLO-EW corrections are obtained
with \textsc{HAWK}~\cite{Denner:2014cla,Ciccolini:2007ec}. NLO
interfaced to a Parton Shower (NLO+PS) results have been obtained
using the
\textsc{POWHEG-BOX}~\cite{Nason:2009ai,Nason:2004rx,Frixione:2007vw,Alioli:2010xd}
together with version 6.428 of \textsc{PYTHIA}~\cite{Sjostrand:2006za}
with the Perugia Tune P12~\cite{Skands:2010ak}. QCD Hjj results are
obtained as in section \ref{sec:H_ggHjets}.

\subsubsection{Parameters}
\begin{table}[!thb]
  \center
  \begin{tabular}{cc}
    \hline
    $\sqrt{S}$	&  100 TeV                      \\
    $M_H$  	&  125 GeV                      \\
    PDF 	&  {\tt MMHT2014nnlo68cl} \& {\tt CT14nnlo}                     \\
    $a_s(M_Z)$ 	&  0.118                    \\
    $M_Z$	&  91.1876 GeV  \\
    $M_W$	&  80.385 GeV  \\
    $\Gamma_Z$	&  2.4952 GeV  \\
    $\Gamma_W$	&  2.085 GeV  \\
    $G_F$	&  $1.16637 \times 10^{-5} \mathrm{ GeV}^{-1}$   \\
    $n_f$       &  5 \\
    $\mu^2=\mu_R^2=\mu_F^2$ & $\frac{M_H}{2}\sqrt{\left(\frac{M_H}{2}\right)^2+p_{t,H}^2}$\\
    \hline
  \end{tabular}
  \caption{Setup}
  \label{tab:EWpar}
\end{table}
For the purpose of this study we have used the EW parameters shown in Table \ref{tab:EWpar} together with tree-level electroweak relations to obtain the weak mixing angle, $\theta_W$, and the electromagnetic coupling, $\alpha_{EW}$

\begin{align}
  \sin^2\theta_W = 1 - M_W^2/M_Z^2, \qquad \alpha_{EW} = \sqrt{2}G_F M_W^2 \sin^2\theta_W /\pi.  
\end{align}

We include no off-shell effects for the Higgs Boson but include Breit-Wigner propagators for the $W$ and $Z$ boson. In order to estimate scale uncertainties we vary $\mu$ up and down a factor $2$ while keeping $\mu_R=\mu_F$. We use a diagonal CKM matrix. When reconstructing jets we use the anti-$k_t$ algorithm~\cite{Cacciari:2005hq,Cacciari:2008gp} as implemented in \textsc{FastJet}~\cite{Cacciari:2011ma} with radius parameter $R=0.4$.

For VBFH predictions we have used the \textsc{MMHT2014nnlo68cl}~\cite{Harland-Lang:2014zoa} PDF set and for QCD Hjj predictions we have used the \textsc{CT14nnlo}~\cite{Dulat:2015mca} PDF set as implemented in \textsc{LHAPDF}~\cite{Buckley:2014ana}. In order to include photon induced contributions to the NLO-EW corrections we have employed the \textsc{NNPDF2.3QED}~\cite{Ball:2012cx} PDF set, which includes a photon PDF. It is worth noticing that the relative EW correction factor only very mildly depends on the PDF set, so that the induced error arising from using different PDFs can be safely assumed to be contained in the other theoretical uncertainties. 

\subsubsection{Inclusive VBF production}
Due to the massive vector bosons exchanged in VBFH production the cross section is finite even when both jets become fully unresolved in fixed-order calculations. In Table \ref{tab:vbf_XStot} we present the fully inclusive LO cross section and both NNLO-QCD and NLO-EW corrections. The NNLO-QCD corrections are calculated in the DIS-like approximation (Structure Function Approximation)~\cite{Han:1992hr} where there is no cross-talk between the upper and lower quark line in the VBF diagram. This approximation is exact at LO and NLO but formally excludes a number of diagrams at NNLO. These contributions have been shown to be tiny under typical VBF cuts and can therefore safely be neglected~\cite{Bolzoni:2011cu,Ciccolini:2007ec,Andersen:2007mp}. The NLO-EW corrections include both t- and u-channel contributions but exclude the s-channel contributions to be consistent with the Structure Function Approximation. The s-channel contribution should hence be treated as a background to the VBF signal. 

\begin{table}[!htb]
  \caption{Total VBF cross section including QCD and EW corrections
    and their uncertainties for a $100$ TeV proton-proton collider. $\sigma^{\mathrm{VBF}}$ is obtained using eq. \eqref{EWQCD} where $\sigma_{\mathrm{NNLO QCD}}^{\mathrm{DIS}}$ is the total VBFH cross section computed to NNLO accuracy in QCD, $\delta_{\mathrm{EW}}$ is the relative EW induced corrections and $\sigma_{\gamma}$ is the cross section induced by incoming photons. For comparison, the LO order cross section, $\sigma_{\mathrm{LO}}$, is also shown.}
  \label{tab:vbf_XStot}
  \begin{center}%
    \begin{small}%
      \tabcolsep5pt
      \begin{tabular}{cccccc}%
        \hline
        $\sigma^{\mathrm{VBF}}$[pb] & $\Delta_{\mathrm{scale}}$[\%] &$\sigma_{\mathrm{LO}}$[pb]  &$\sigma_{\mathrm{NNLO QCD}}^{\mathrm{DIS}}$[pb] & $\delta_{\mathrm{EW}}$[\%] & $\sigma_{\gamma}$[pb]
        \\
        \hline
        $69.0$ &$^{+0.85}_{-0.46}$ & $80.6$ & $73.5$ & $-7.3$ & $0.81$ 
        \\
        \hline
      \end{tabular}%
    \end{small}%
  \end{center}%
\end{table}

In order to compute the VBF cross section we we combine the NNLO-QCD and NLO-EW corrections in the following way

\begin{align}
  \sigma^{\mathrm{VBF}} & = \sigma_{\mathrm{NNLO QCD}}^{\mathrm{DIS}} (1 + \delta_{\mathrm{EW}}) + \sigma_{\gamma},
  \label{EWQCD}
\end{align}

where $\sigma_{\mathrm{NNLO QCD}}^{\mathrm{DIS}}$ is the NNLO-QCD prediction in the DIS-like approximation, $\delta_{\mathrm{EW}}$ is the relative EW correction factor and $\sigma_{\gamma}$ is the photon induced contribution. The combined corrections to the LO cross section is about $14\%$ with QCD and EW corrections contributing an almost equal amount. The scale uncertainty $\Delta_{\mathrm{scale}}$ is due to varying $\mu$ by a factor 2 up and down in the QCD calculation alone keeping $\mu_F=\mu_R$. For comparison the total QCD and EW corrections at 14 TeV amount to about $7\%$ and the QCD induced scale variations to about $0.4\%$. 

\subsubsection{VBF cuts}
In order to separate the VBFH signal from the main background of QCD $Hjj$ production we will extend typical VBF cuts used at the LHC to a 100 TeV proton-proton collider. These cuts take advantage of the fact that VBFH production, and VBF production in general, has a very clear signature of two forward jets clearly separated in rapidity. Examining the topology of a typical VBFH production diagram it becomes very clear that this is the case because the two leading jets are essential remnants of the two colliding protons. Since the $p_t$ of the jets will be governed by the mass scale of the weak vector bosons and the energy by the PDFs the jets will typically be very energetic and in opposite rapidity hemispheres.

\begin{figure}[!htb]
  \begin{minipage}{0.49\textwidth}
    \centering
    \includegraphics[width=1\textwidth,page=1]{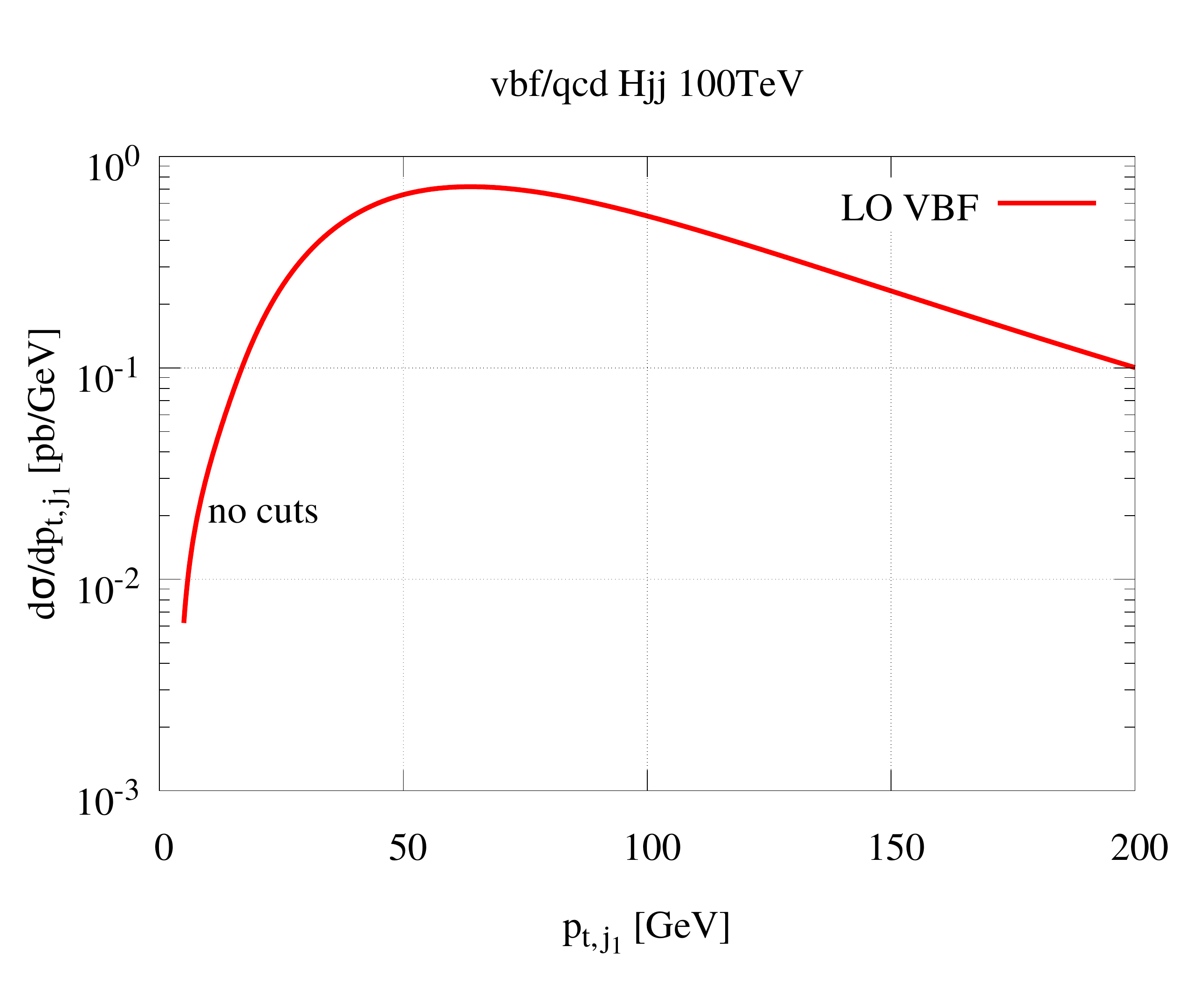}
  \end{minipage}
  \begin{minipage}{0.49\textwidth}
    \centering
    \includegraphics[width=1\textwidth,page=2]{figs/VBF/vbf_100TeV.pdf}
  \end{minipage}
  \caption{\label{fig:VBFptj}Left panel: The $p_t$ of the hardest jet in VBFH production at $100$ TeV. We require at least two jets in the event but apply no other cuts; right panel: The $p_t$ of the second hardest jet in VBFH and QCD Hjj production at $100$ TeV.}
\end{figure}

As is clear from figure \ref{fig:VBFptj} the hardest jet in VBFH
production peaks at around $60$ GeV. As discussed above this value is set by the mass of the weak vector bosons and hence the $p_t$ spectra of the two hardest jets are very similar to what one finds at the LHC. From this point of view and in order to maximise the VBFH cross section one should keep jets with $p_{t,cut} > 30$ GeV. Here we present results for $p_{t,cut} = \{30,50,100\}$ GeV to study the impact of the jet cut on both the VBFH signal and QCD Hjj background. We only impose the cut on the two hardest jets in the event.

To establish VBF cuts at $100$ TeV we first study the variables which are typically used at the LHC. These are the dijet invariant mass, $M_{jj}$, the rapidity separation between the two leading jets, $\Delta y_{jj}$, the separation between the two leading jets in the rapidity-azimuthal angle plane, $\Delta R_{jj}$ and the azimuthal angle between the two leading jet $\phi_{jj}$. In Figure \ref{fig:VBFcuts1} we show $M_{jj}$ and $\Delta y_{jj}$ after applying a cut on the two leading jets of $p_t > 30$ GeV and requiring that the two leading jets are in opposite detector hemispheres. This last cut removes around $60 \%$ of the background while retaining about $80 \%$ of the signal. 

\begin{figure}[!htb]
  \begin{minipage}{0.49\textwidth}
    \centering
    \includegraphics[width=1\textwidth,page=15]{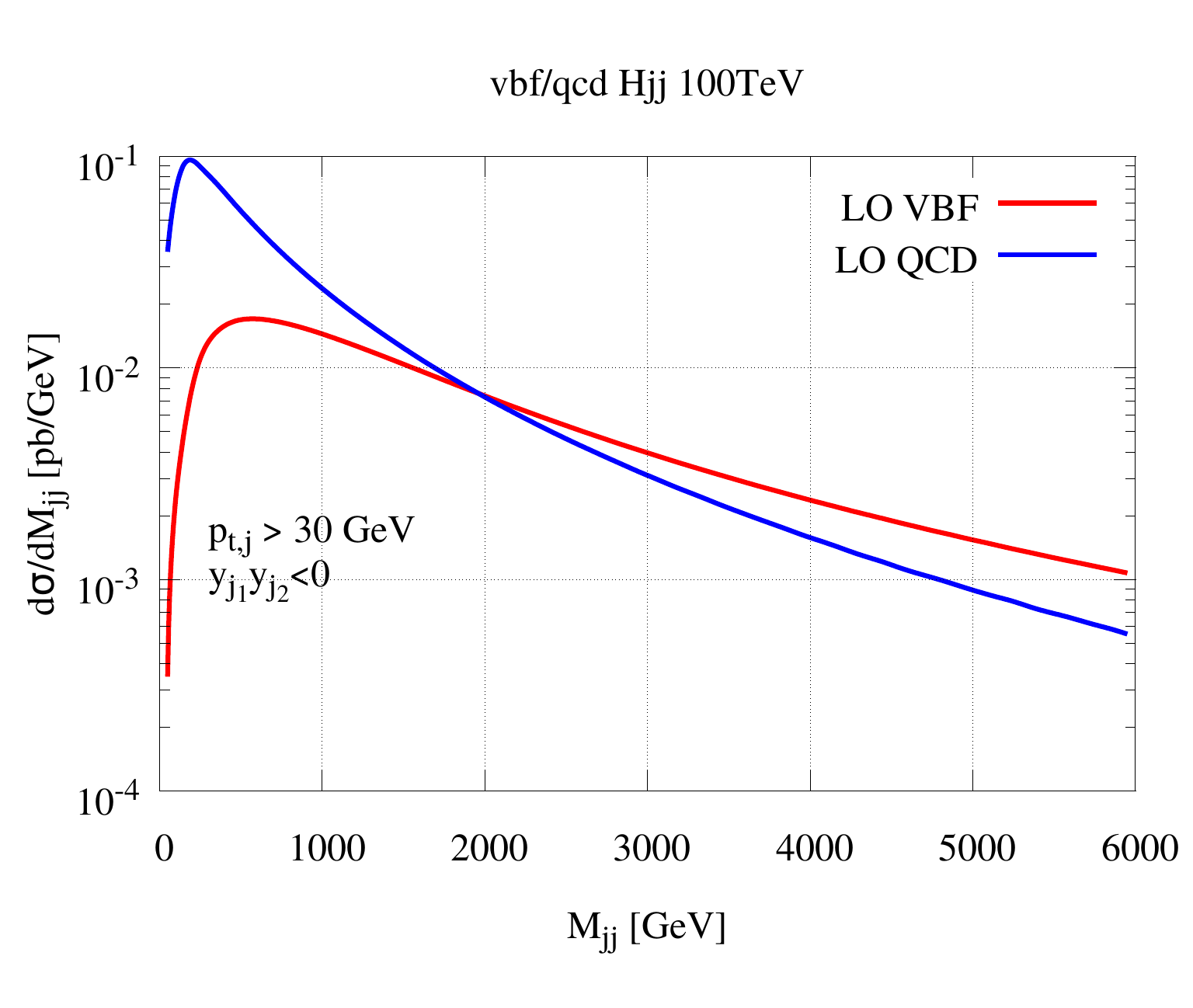}
  \end{minipage}
  \begin{minipage}{0.49\textwidth}
    \centering
    \includegraphics[width=1\textwidth,page=16]{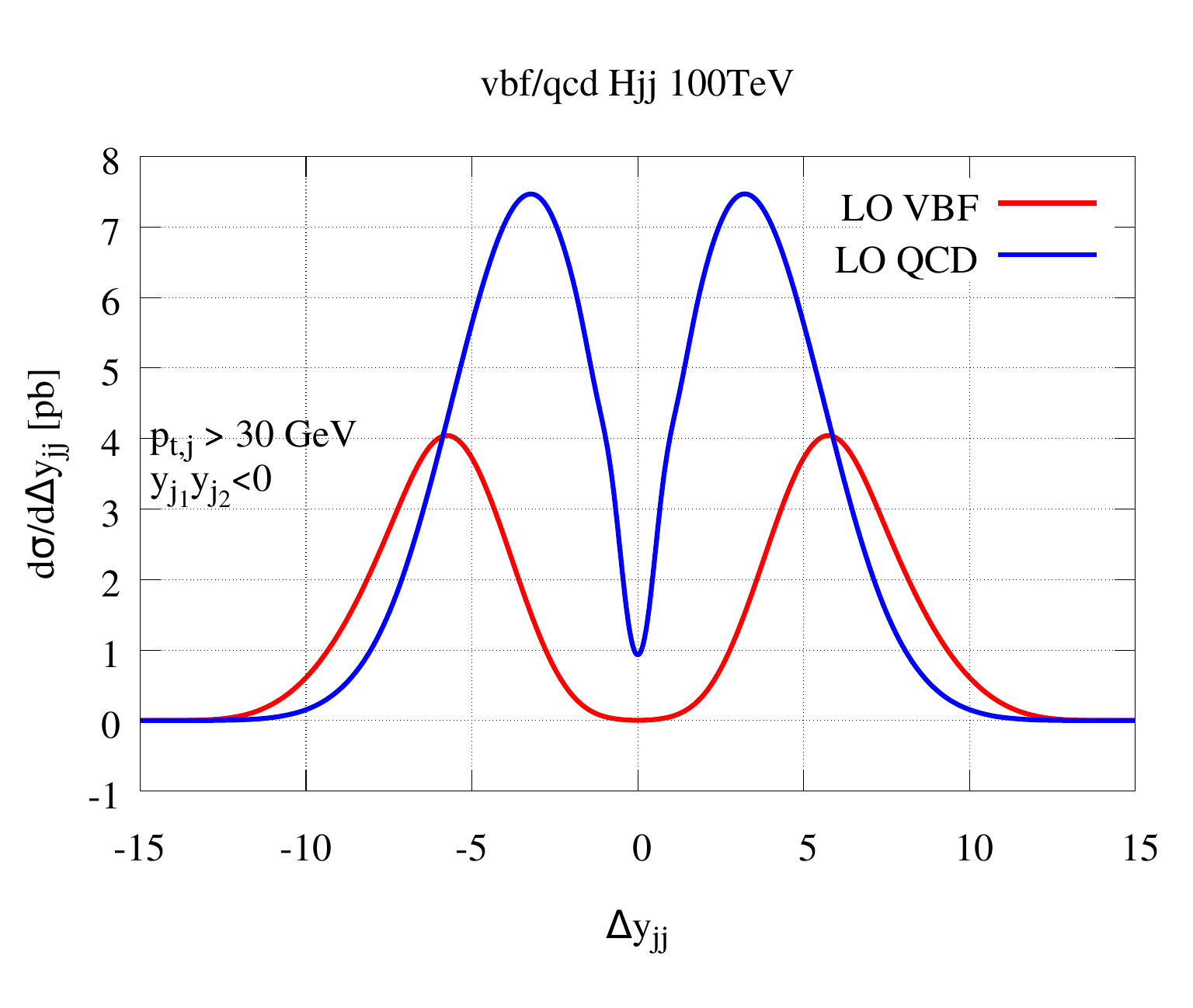}
  \end{minipage}
  \caption{\label{fig:VBFcuts1}Left panel: The invariant dijet mass $M_{jj}$ of the two hardest jets in VBFH and QCD Hjj production at $100$ TeV. right panel: The rapidity separation of the two hardest jets $\Delta y_{jj}$ in VBFH and QCD Hjj production at $100$ TeV.}
\end{figure}

In order to suppress the QCD background a cut of $\Delta y_{jj} > 6.5$ is imposed. This cut also significantly reduces the QCD $M_{jj}$ peak and shifts the VBF peak to about $2400$ GeV. In order to further suppress the QCD background we impose $M_{jj} > 1600$ GeV. After these cuts have been applied, and requiring $p_{t,j} > 30$ GeV, the VBF signal to QCD background ratio is roughly 3 with a total NNLO-QCD VBF cross section of about $12$ pb. From Figure \ref{fig:VBFcuts2} it is clear that one could also impose a cut on $\phi_{jj}$ to improve the suppression whereas a cut on $\Delta R_{jj}$ would not help to achieve that. We hence state the VBF cuts that we will be using throughout this section are

\begin{align}
  M_{j_1 j_2} > 1600 \mathrm{ GeV}, \qquad \Delta y_{j_1 j_2} > 6.5, \qquad y_{j_1}y_{j_2} < 0.
  \label{eq:VBFcuts}
\end{align}

where $j_1$ is the hardest jet in the event and $j_2$ is the second hardest jet. At a 13 TeV machine the VBFH cross section is $\mathcal{O}(1 \mathrm{ pb})$ under typical VBF cuts. 

\begin{figure}[h!]
  \begin{minipage}{0.49\textwidth}
    \centering
    \includegraphics[width=1\textwidth,page=33]{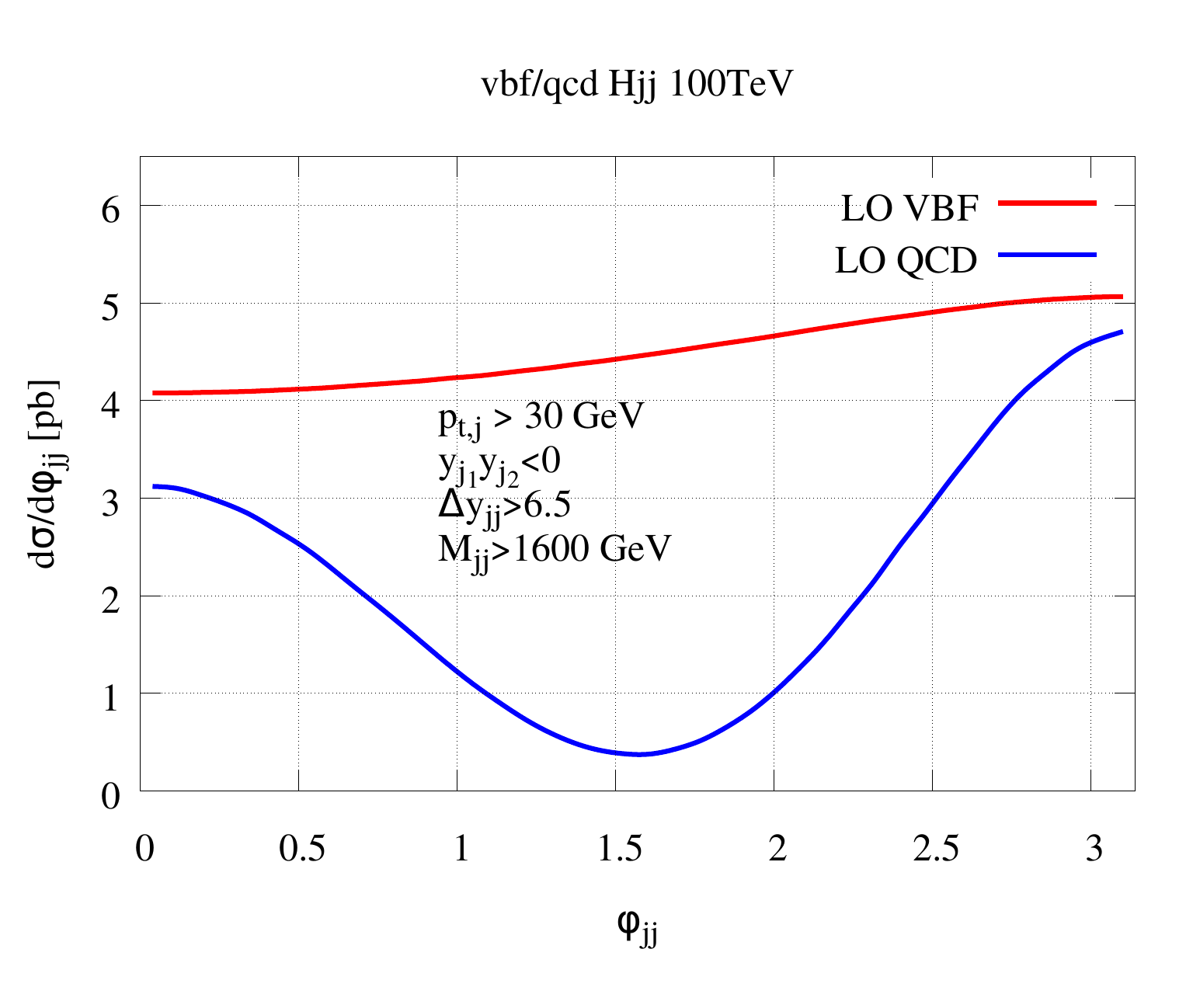}
  \end{minipage}
  \begin{minipage}{0.49\textwidth}
    \centering
    \includegraphics[width=1\textwidth,page=34]{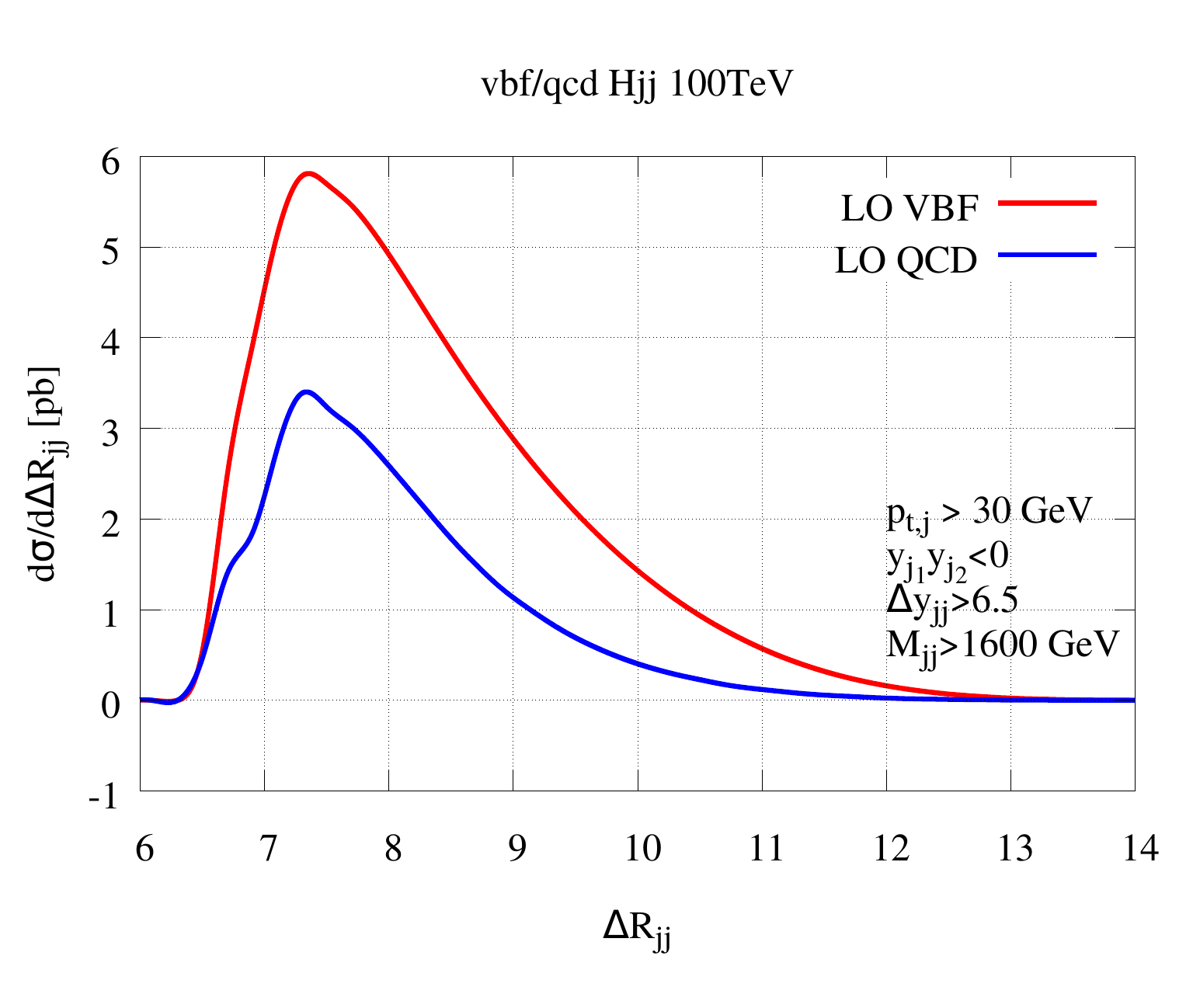}
  \end{minipage}
  \caption{\label{fig:VBFcuts2}Left panel: The azimuthal angle $\phi_{jj}$ between the two hardest jets in VBFH and QCD Hjj production at $100$ TeV. right panel: The rapidity-azimuthal angle separation of the two hardest jets $\Delta R_{jj}$ in VBFH and QCD Hjj production at $100$ TeV.}
\end{figure}

In table \ref{tab:fidVBF} we show the fiducial cross section obtained after applying the VBF cuts of eq. \eqref{eq:VBFcuts} to VBFH and QCD Hjj production. The cross sections are reported at the three different jet $p_t$ cut values $\{30,50,100\}$ GeV. All numbers are computed at LO. It is clear from the table that requiring a somewhat higher jet $p_t$ cut than $30$ GeV leads to a lower $S/\sqrt{B}$ ratio. In going from $30$ GeV to $50$ GeV this reduction is however small. 

\begin{table}[!htb]
  \caption{Fiducial VBFH and QCD Hjj cross sections for a $100$ TeV proton-proton collider at LO under the VBF cuts of \eqref{eq:VBFcuts}. The numbers are obtained using the setup of Table \ref{tab:EWpar} using the {\tt CT14nnlo} PDF. $S/\sqrt{B}$ is defined as the ratio between the VBFH signal and the square root of the QCD background at an integrated luminosity of $20\mbox{ ab}^{-1}$.}
  \label{tab:fidVBF}
  \begin{center}%
    \begin{small}%
      \tabcolsep5pt
      \begin{tabular}{l|ccc}%
        \hline
 & $\sigma(p_{t,j} > 30 \mbox{ GeV})$ [pb] & $\sigma(p_{t,j} > 50 \mbox{ GeV})$ [pb]& $\sigma(p_{t,j} > 100 \mbox{ GeV})$ [pb] 
        \\
        \hline
        VBFH & $14.1$ & $7.51$ & $1.08$ 
        \\
        QCD Hjj & $5.04$ & $1.97$ & $0.331$ 
        \\
        \hline
        $S/\sqrt{B}@(20 \mbox{ ab}^{-1})$ & $28100$ & $24200$ & $8500$ 
        \\
        \hline
      \end{tabular}%
    \end{small}%
  \end{center}%
\end{table}

In table \ref{tab:totVBF} we show for comparison the cross sections obtained after only applying the three jet $p_t$ cuts. As expected the VBFH signal is drowned in the QCD background. It is worth noticing that the $S/\sqrt{B}$ ratio is still very large when one assumes an integrated luminosity of $20\mbox{ ab}^{-1}$ and that it declines as the jet cut is increased. 

\begin{table}[!htb]
  \caption{Total VBFH and QCD Hjj cross sections for a $100$ TeV proton-proton collider at LO with a cut on the two hardest jets. The numbers are obtained using the setup of Table \ref{tab:EWpar} using the {\tt CT14nnlo} PDF. $S/\sqrt{B}$ is defined as the ratio between the VBFH signal and the square root of the QCD background at an integrated luminosity of $20\mbox{ ab}^{-1}$.}
  \label{tab:totVBF}
  \begin{center}%
    \begin{small}%
      \tabcolsep5pt
      \begin{tabular}{l|ccc}%
        \hline
 & $\sigma(p_{t,j} > 30 \mbox{ GeV})$ [pb] & $\sigma(p_{t,j} > 50 \mbox{ GeV})$ [pb]& $\sigma(p_{t,j} > 100 \mbox{ GeV})$ [pb] 
        \\
        \hline
        VBFH & $51.3$ & $28.5$ & $5.25$ 
        \\
        QCD Hjj & $166$ & $78.6$ & $23.9$ 
        \\
        \hline
        $S/\sqrt{B}@(20 \mbox{ ab}^{-1})$ & $17900$ & $14300$ & $4900$ 
        \\
        \hline
      \end{tabular}%
    \end{small}%
  \end{center}%
\end{table}

\subsubsection{Perturbative corrections}
The results shown in the previous section were all computed at LO. Here we briefly investigate the impact of NNLO-QCD, NLO-EW and parton shower corrections to the VBF cross section computed with $p_{t,j} > 30$ GeV and under the VBF cuts of eq. \eqref{eq:VBFcuts} at a 100 TeV collider. We also compare to the NLO-QCD predictions for QCD Hjj production.

In table \ref{tab:vbf_XStot_2} we show the best prediction for $\sigma^{\mathrm{VBF}}$ as obtained by eq. \eqref{EWQCD} and compare it to the same cross section obtained by showering \textsc{POWHEG} events with \textsc{PYTHIA6} but including no effects beyond the parton shower itself. The NLO-EW and NNLO-QCD corrections are found to be of roughly the same order, and amount to a total negative correction of $\sim 23\%$. As was the case for the inclusive cross section the corrections are a factor two larger than at $14$ TeV. Even though the perturbative corrections to QCD Hjj production are negative, the effect of including higher order corrections is that the $S/\sqrt{B}$ ratio at an integrated luminosity of $20\mbox{ ab}^{-1}$ is decreased from $28100$ to $24300$. 

\begin{table}[!htb]
  \caption{Fiducial VBF cross section including QCD and EW corrections
    and their uncertainties for a $100$ TeV proton-proton collider. For comparison the QCD induced Hjj cross section is also shown. At fixed-order QCD corrections are included at NNLO and EW corrections at NLO. }
  \label{tab:vbf_XStot_2}
  \begin{center}%
    \begin{small}%
      \tabcolsep5pt
      \begin{tabular}{l|ccccc}%
        \hline
 Process & $\sigma^{\mathrm{fid}}$[pb] & $\Delta_{\mathrm{scale}}$[\%] &$\sigma_{\mathrm{QCD}}$[pb] & $\delta_{\mathrm{EW}}$[\%] & $\sigma_{\gamma}$[pb]
        \\
        \hline
        VBFH (NNLO-QCD/NLO-EW) & $10.8$ &$\pm 1.0$ & $12.1$ & $-12.6$ & $0.22$ 
        \\
        VBFH (NLO+PS) & $11.9$ &$^{+0.56}_{-0.41}$ & $11.9$ & - & - 
        \\
        QCD Hjj (NLO)& $4.70$ &$^{+0}_{-17}$ & $4.70$ & - & - 
        \\
        \hline
      \end{tabular}%
    \end{small}%
  \end{center}%
\end{table}
In figs. \ref{fig:VBFQCD1}-\ref{fig:VBFQCD4} we show comparisons
between VBFH and QCD Hjj production computed at NNLO and NLO in QCD
respecitvely. We have applied the VBF cuts of
eq. \eqref{eq:VBFcuts}. Also shown is the k-factor for VBFH production
going from LO to NLO and NLO to NNLO. Note that the QCD Hjj
predictions have been obtained with the effective theory setup
described in Section \ref{sec:H_ggH} and hence the $p_t$ spectra
should not be trusted beyond $2 M_t$. Furthermore, in the left plots
of figs. \ref{fig:VBFQCD1} and \ref{fig:VBFQCD2} we notice a large
scale dependence of the QCD Hjj predictions for higher values of the
transverse momentum of the leading jet and of the Higgs boson. This is
probably due to a not optimal choice of scale, which for the downward
variation suffers from large cancellations among the different NLO
contributions (Born, virtual, integrated subtraction terms and real
radiation minus subtraction terms). We observe that increasing the
minimum transverse momentum cut improves the behaviour, whereas
chosing a fixed scale instead of a dynamical one makes it even
worse. The exact origin of this large scale dependence needs further
investigation.

As can be seen from the plots the VBF cuts have suppressed the
background QCD Hjj production in all corners of phasespace. One could
still imagine further optimising these cuts, for example by requiring
$\phi_{jj}$ in the vicinty of $\frac{\pi}{2}$ or a slightly larger
invariant dijet mass. We note in particular that requiring that the
Higgs Boson has a transverse momentum greater than $40$ GeV seems to
favour the VBFH signal. Since a cut on the transverse momentum of the
decay products of the Higgs would no matter what have to be imposed,
this improves the efficiency of the VBF cuts in realistic experimental
setups.

\begin{figure}[!htb]
  \begin{minipage}{0.49\textwidth}
    \centering
    \includegraphics[width=1\textwidth,page=1]{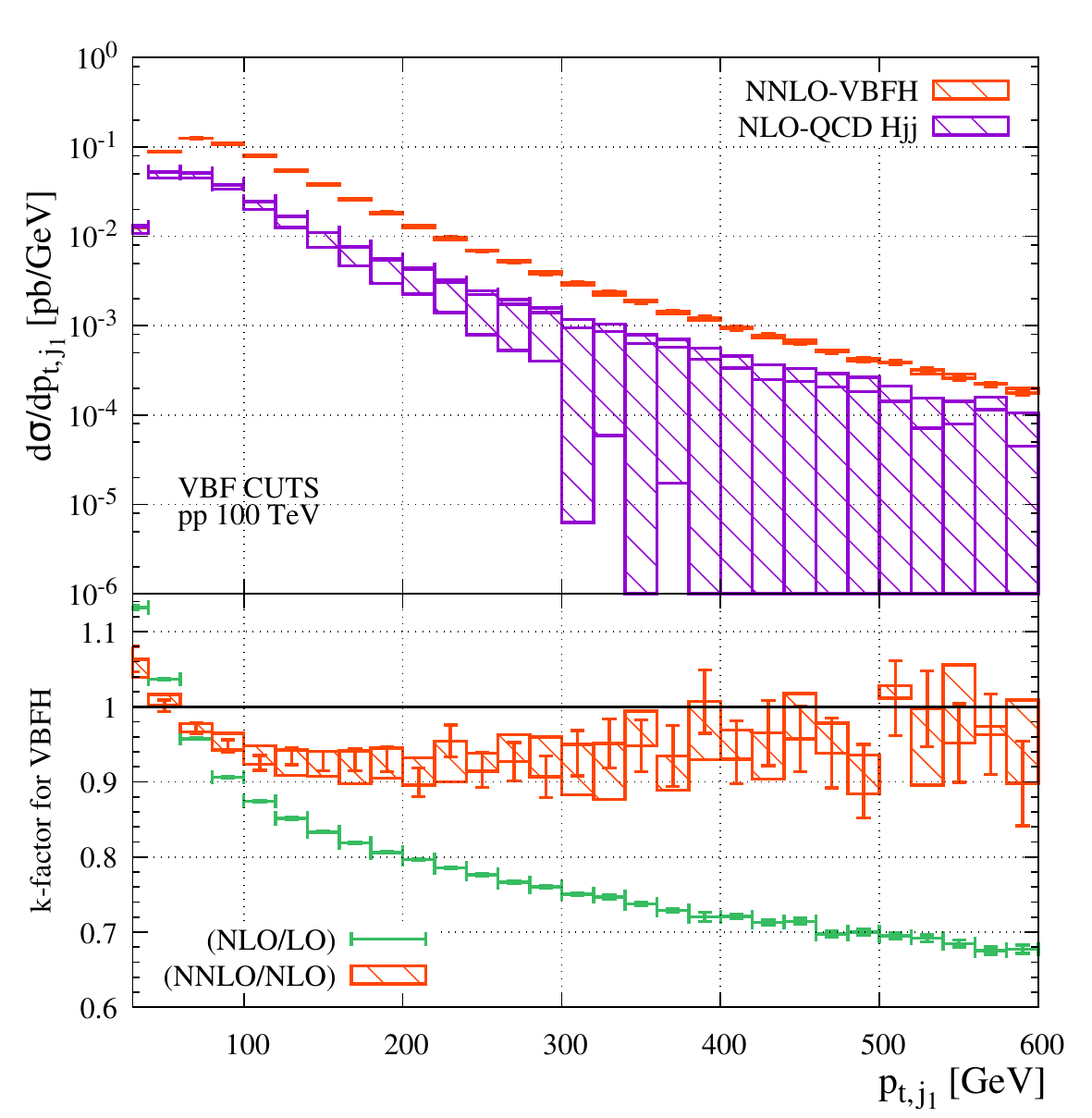}
  \end{minipage}
  \begin{minipage}{0.49\textwidth}
    \centering
    \includegraphics[width=1\textwidth,page=2]{figs/VBF/diff-plots-100TeV-new.pdf}
  \end{minipage}
  \caption{\label{fig:VBFQCD1} Comparison between NNLO predictions for VBFH production at NLO predicitions for QCD Hjj productionunder the VBF cuts of eq. \eqref{eq:VBFcuts}. The bands represent scale uncertainties obtained by varying $\mu_F=\mu_R$ by a factor two up and down. For the VBFH production the statistical uncertainty is represented by the vertical line. No statistical uncertainties are shown for the QCD Hjj result. The lower panel shows the k-factor for VBFH production going to LO to NLO and NLO to NNLO. Left panel: Transverse momentum of the leading jet. Right Panel: Transverse momentum of the subleading jet.}
\end{figure}

\begin{figure}[!htb]
  \begin{minipage}{0.49\textwidth}
    \centering
    \includegraphics[width=1\textwidth,page=3]{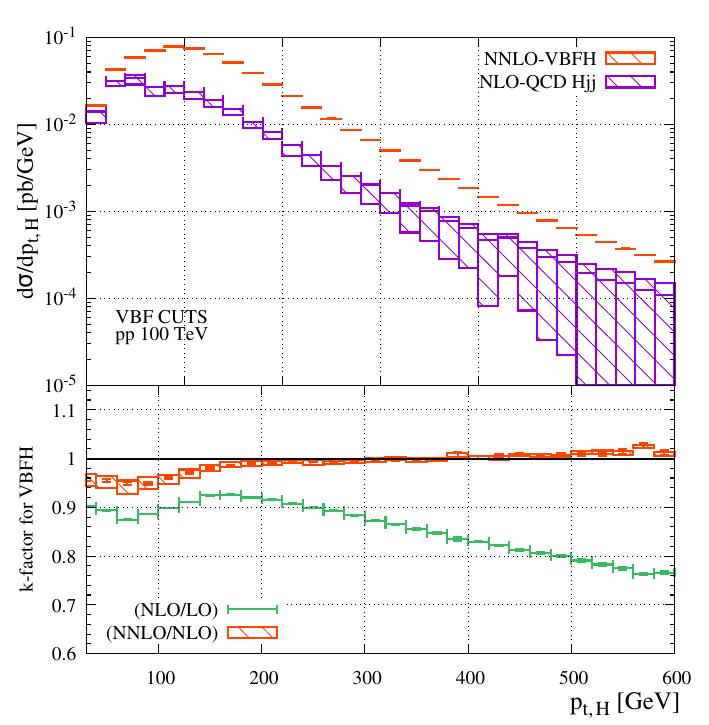}
  \end{minipage}
  \begin{minipage}{0.49\textwidth}
    \centering
    \includegraphics[width=1\textwidth,page=4]{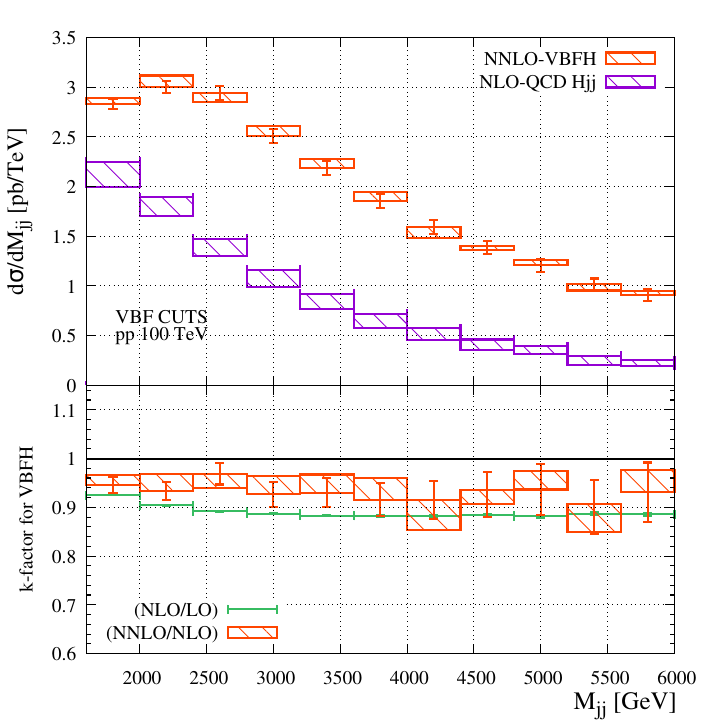}
  \end{minipage}
  \caption{\label{fig:VBFQCD2} Comparison between NNLO predictions for VBFH production at NLO predicitions for QCD Hjj production under the VBF cuts of eq. \eqref{eq:VBFcuts}. The bands represent scale uncertainties obtained by varying $\mu_F=\mu_R$ by a factor two up and down. For the VBFH production the statistical uncertainty is represented by the vertical line. No statistical uncertainties are shown for the QCD Hjj result. The lower panel shows the k-factor for VBFH production going to LO to NLO and NLO to NNLO. Left panel: Transverse momentum of the Higgs Boson. Right Panel: Invariant mass of the dijet pair.}
\end{figure}

\begin{figure}[!htb]
  \begin{minipage}{0.49\textwidth}
    \centering
    \includegraphics[width=1\textwidth,page=5]{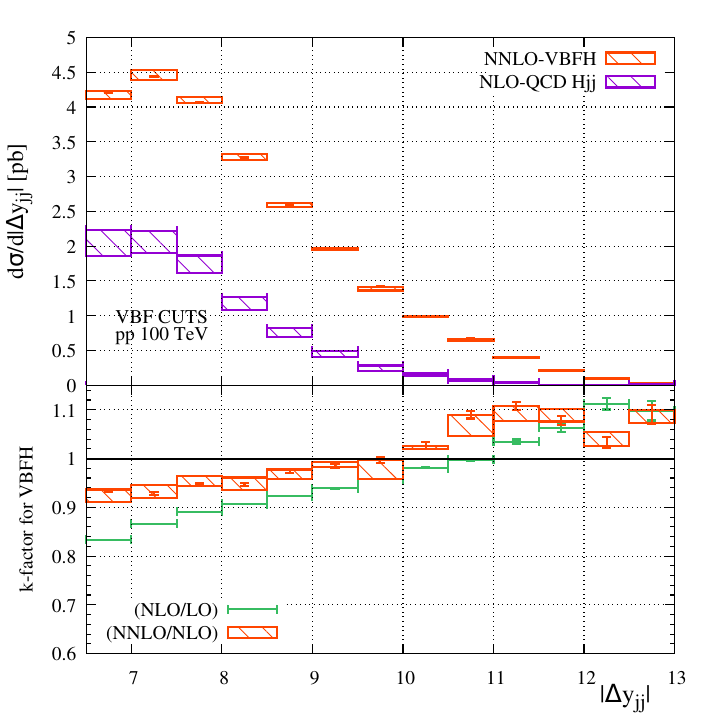}
  \end{minipage}
  \begin{minipage}{0.49\textwidth}
    \centering
    \includegraphics[width=1\textwidth,page=6]{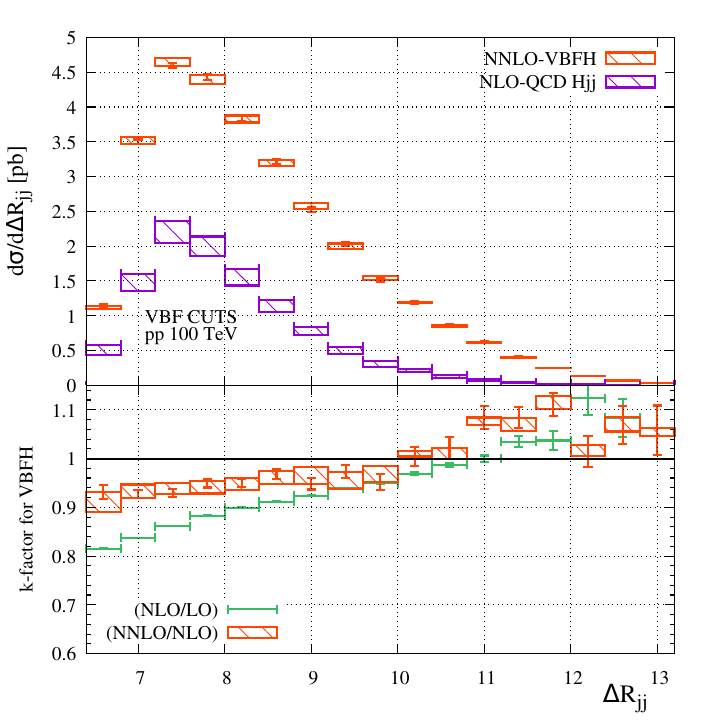}
  \end{minipage}
  \caption{\label{fig:VBFQCD3} Comparison between NNLO predictions for VBFH production at NLO predicitions for QCD Hjj production under the VBF cuts of eq. \eqref{eq:VBFcuts}. The bands represent scale uncertainties obtained by varying $\mu_F=\mu_R$ by a factor two up and down. For the VBFH production the statistical uncertainty is represented by the vertical line. No statistical uncertainties are shown for the QCD Hjj result. The lower panel shows the k-factor for VBFH production going to LO to NLO and NLO to NNLO. Left panel: Absolute value of the rapiridy separation between the two leading jets. Right Panel: Distance between the two leading jets in the rapidity-azimuthal plane.}
\end{figure}

\begin{figure}[!htb]
  \begin{center}
  \begin{minipage}{0.49\textwidth}
    \centering
    \includegraphics[width=1\textwidth,page=7]{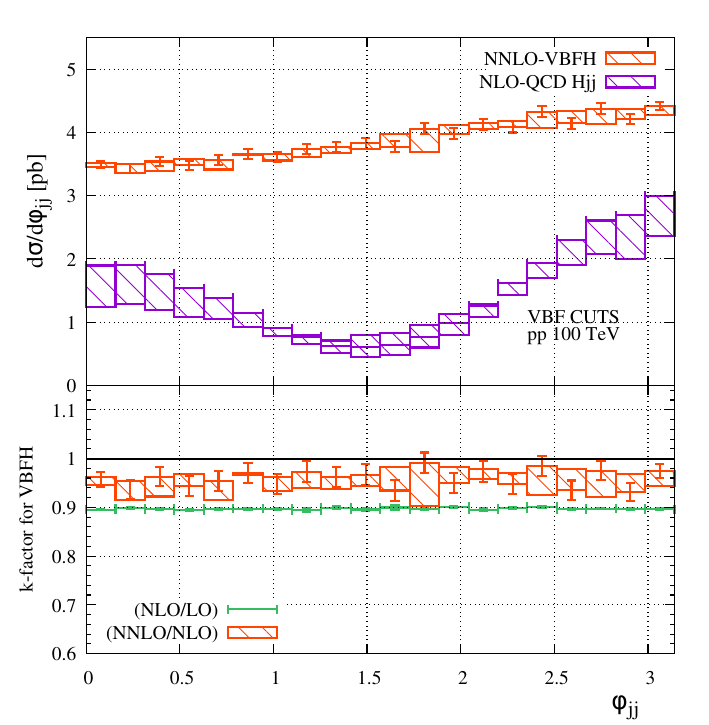}
  \end{minipage}
  \end{center}
  \caption{\label{fig:VBFQCD4} Comparison between NNLO predictions for VBFH production at NLO predicitions for QCD Hjj production under the VBF cuts of eq. \eqref{eq:VBFcuts}. The bands represent scale uncertainties obtained by varying $\mu_F=\mu_R$ by a factor two up and down. For the VBFH production the statistical uncertainty is represented by the vertical line. No statistical uncertainties are shown for the QCD Hjj result. The lower panel shows the k-factor for VBFH production going to LO to NLO and NLO to NNLO. Shown here is the azimuthal angle between the two leading jets.}
\end{figure}

\subsubsection{Differential distributions}
In addition to the distributions already presented, we here show a number of distributions to indicate the kinematical reach of the VBFH channel at 100 TeV. Assuming an integrated luminosity of $20$ ab$^{-1}$ we study how many events will be produced with a Higgs whose transverse momentum exceeds $p_{{t,\mathrm{min}}}$. In figs. \ref{fig:VBFptHaccum} and \ref{fig:VBFptHaccumVBF} we show this distribution for various cut configurations. This variable is particularly interesting in the context of anomalous couplings in the weak sector. It can be seen that even under VBF cuts and requiring hard jets, a number of Higgs bosons with transverse momentum of the order $6$ TeV will be produced in this scenario.

In fig. \ref{fig:VBFptHcorr} we show the same distribution but fully inclusively and at various perturbative orders. Also shown is the k-factor going from LO to NLO and from NLO to NNLO. The perturbative corrections to this variable are modest as it is not sensitive to real radiation at the inclusive level. After applying VBF cuts and jet cuts the low $p_{{t,H}}$-spectrum receives moderate corrections whereas the corrections at larger values of $p_{{t,H}}$ can become very large as indicated in fig. \ref{fig:VBFQCD2}.

In fig. \ref{fig:VBFMjjaccum} we show how many events will be produced with a dijet invariant mass exceeding $M_{\mathrm{min}}$ at various cut configurations. Because the two hardest jets in the VBFH event are typically the proton remnants the invariant dijet mass can become very large. As can be seen from the figure, even after applying VBF cuts and requiring very hard jets hundreds of events with an invariant dijet mass larger than $60$ TeV is expected. This is of interest when probing for BSM physics at the very highest scales. It is also worth noticing that the tail of the distribution is almost unaffected by the VBF cuts, as the VBF cuts are optimised to favour high invariant dijet events. 

\begin{figure}[thb]
  \begin{minipage}{0.49\textwidth}
    \centering
    \includegraphics[width=1\textwidth,page=5]{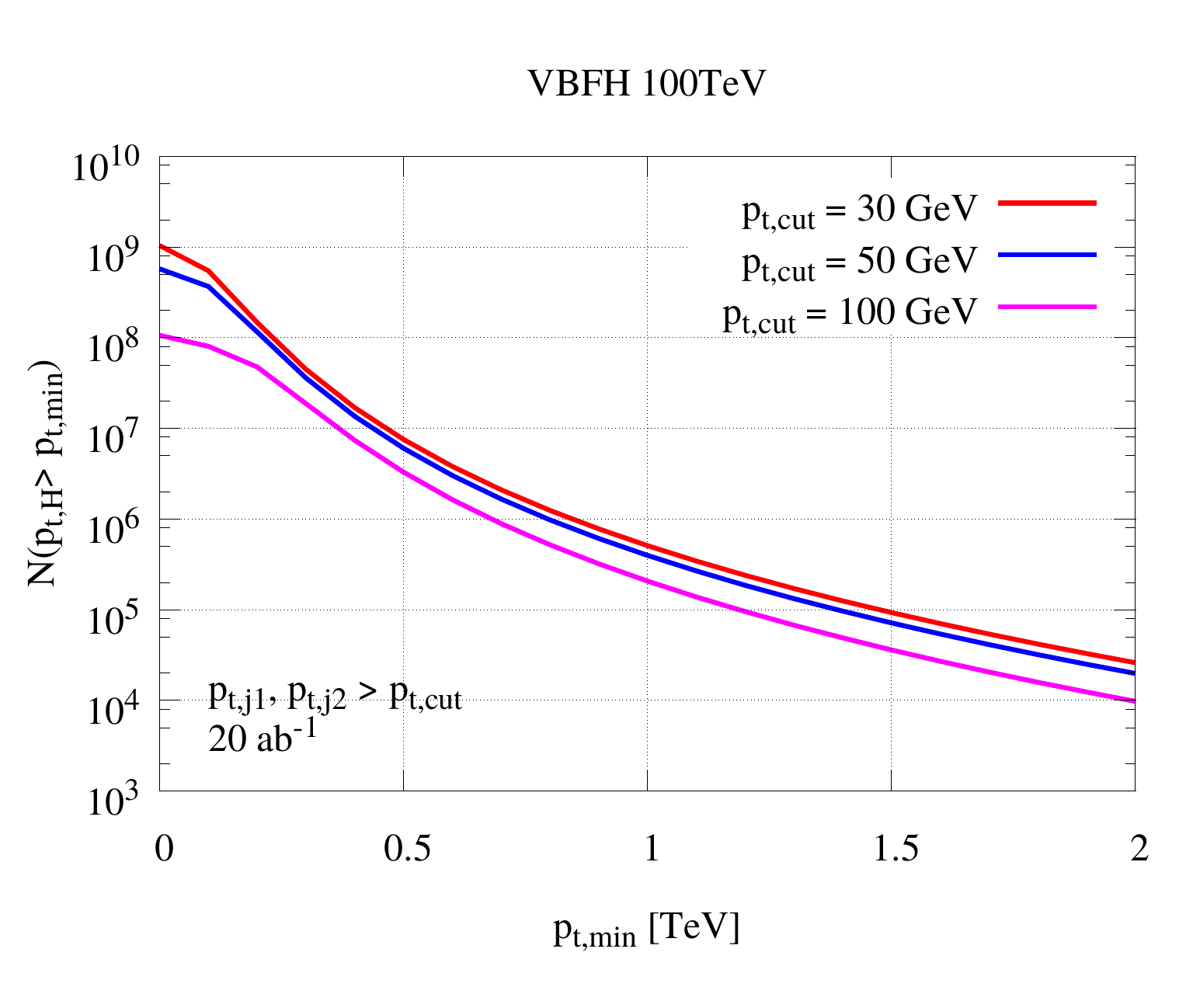}
  \end{minipage}
  \begin{minipage}{0.49\textwidth}
    \centering
    \includegraphics[width=1\textwidth,page=6]{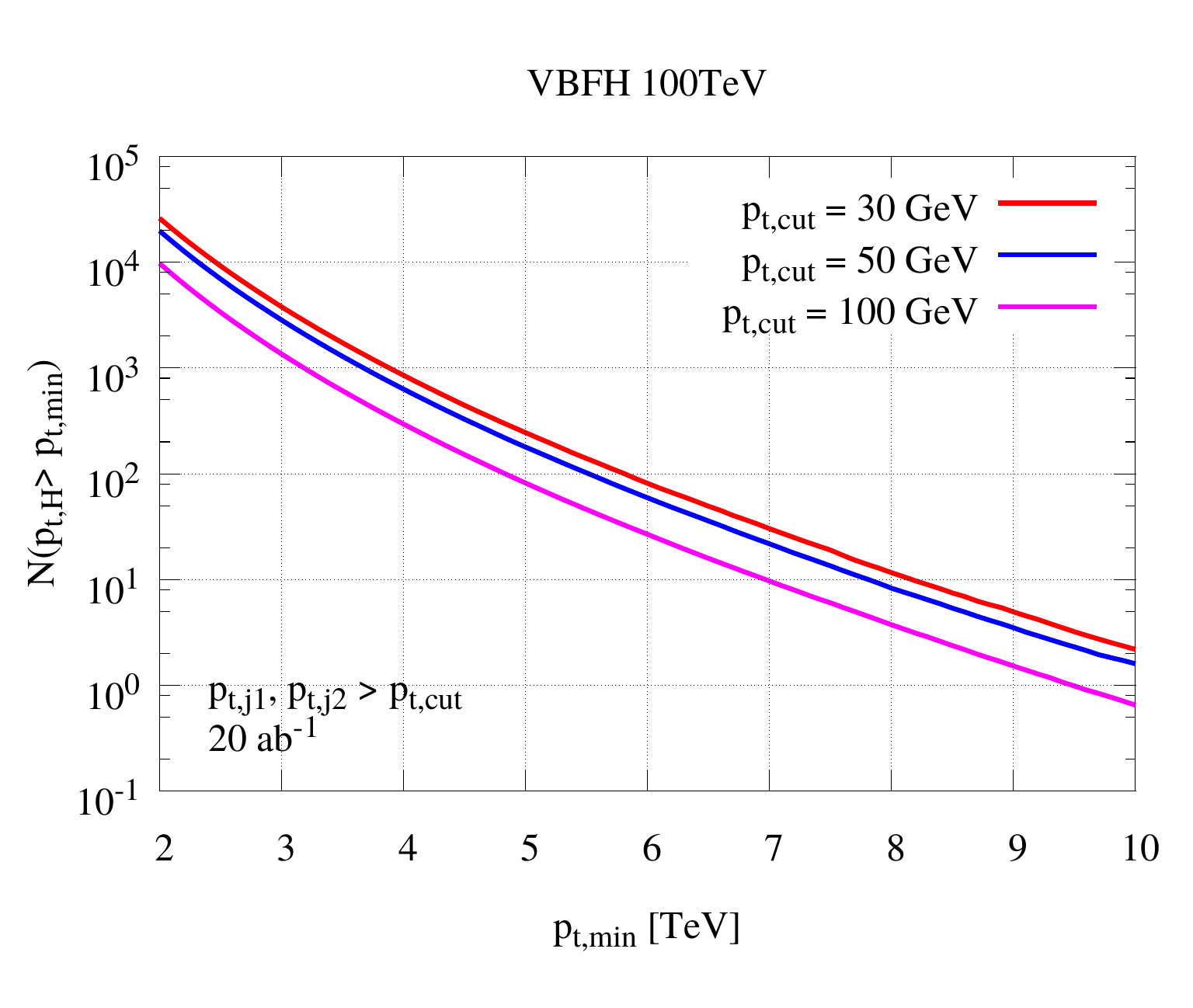}
  \end{minipage}
  \caption{\label{fig:VBFptHaccum} The total number of VBFH events produced with $p_{{t,H}} > p_{{t,\mathrm{min}}}$ at a 100 TeV collider with an integrated luminosity of $20\mbox{ ab}^{-1}$ under three different jet $p_{t}$ cuts. Left panel: $p_{t,H}$ in the range 0-2 TeV. Right panel: $p_{t,H}$ in the range 2-10 TeV.}
\end{figure}

\begin{figure}[thb]
  \begin{minipage}{0.49\textwidth}
    \centering
    \includegraphics[width=1\textwidth,page=7]{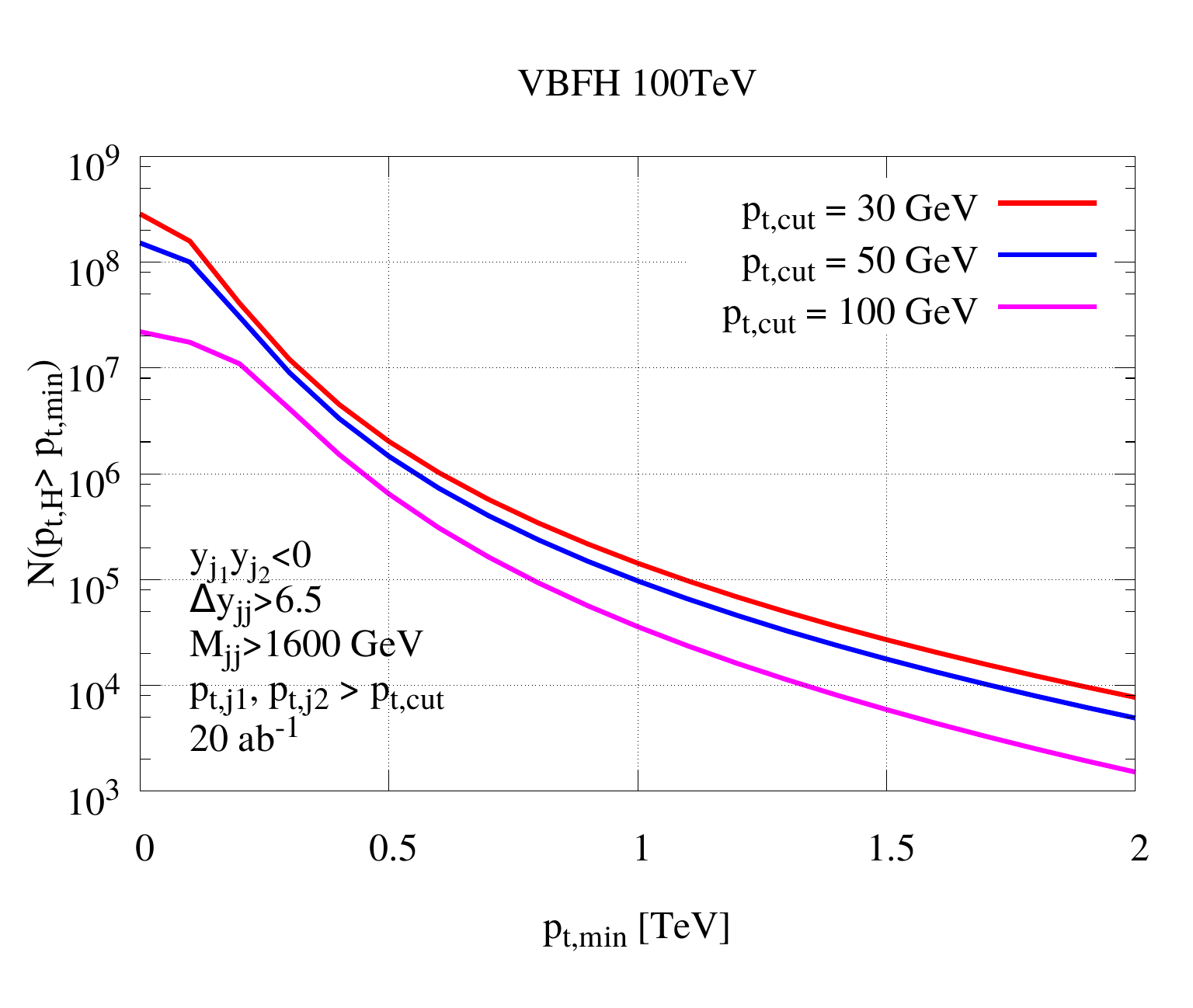}
  \end{minipage}
  \begin{minipage}{0.49\textwidth}
    \centering
    \includegraphics[width=1\textwidth,page=8]{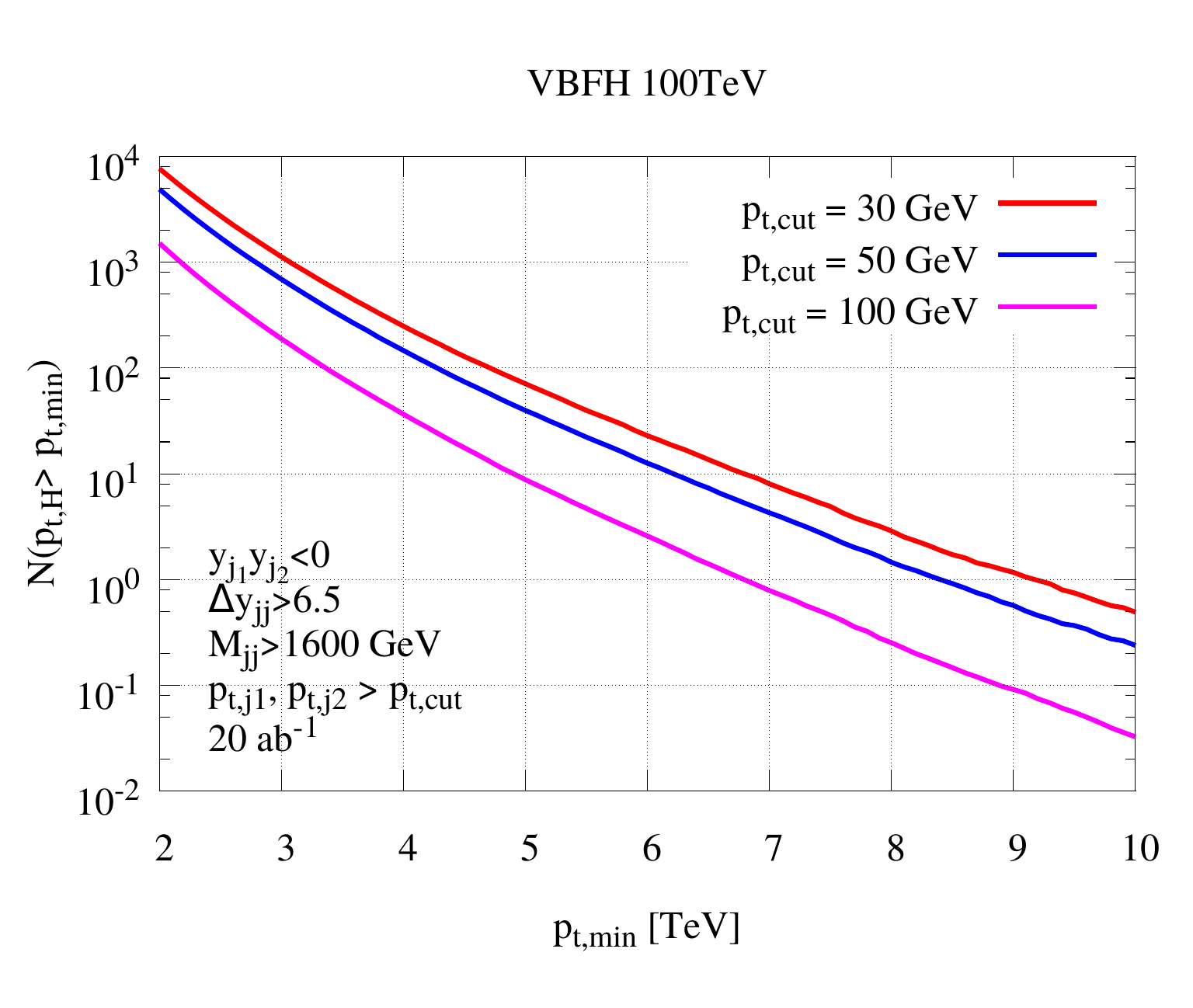}
  \end{minipage}
  \caption{\label{fig:VBFptHaccumVBF} The total number of VBFH events produced with $p_{{t,H}} > p_{{t,\mathrm{min}}}$ at a 100 TeV collider with an integrated luminosity of $20\mbox{ ab}^{-1}$under three different jet $p_{t}$ cuts and with the VBF cuts of eq. \eqref{eq:VBFcuts} applied. Left panel: $p_{t,H}$ in the range 0-2 TeV. Right panel: $p_{t,H}$ in the range 2-10 TeV.}
\end{figure}

\begin{figure}[thb]
  \begin{minipage}{0.49\textwidth}
    \centering
    \includegraphics[width=1\textwidth,page=9]{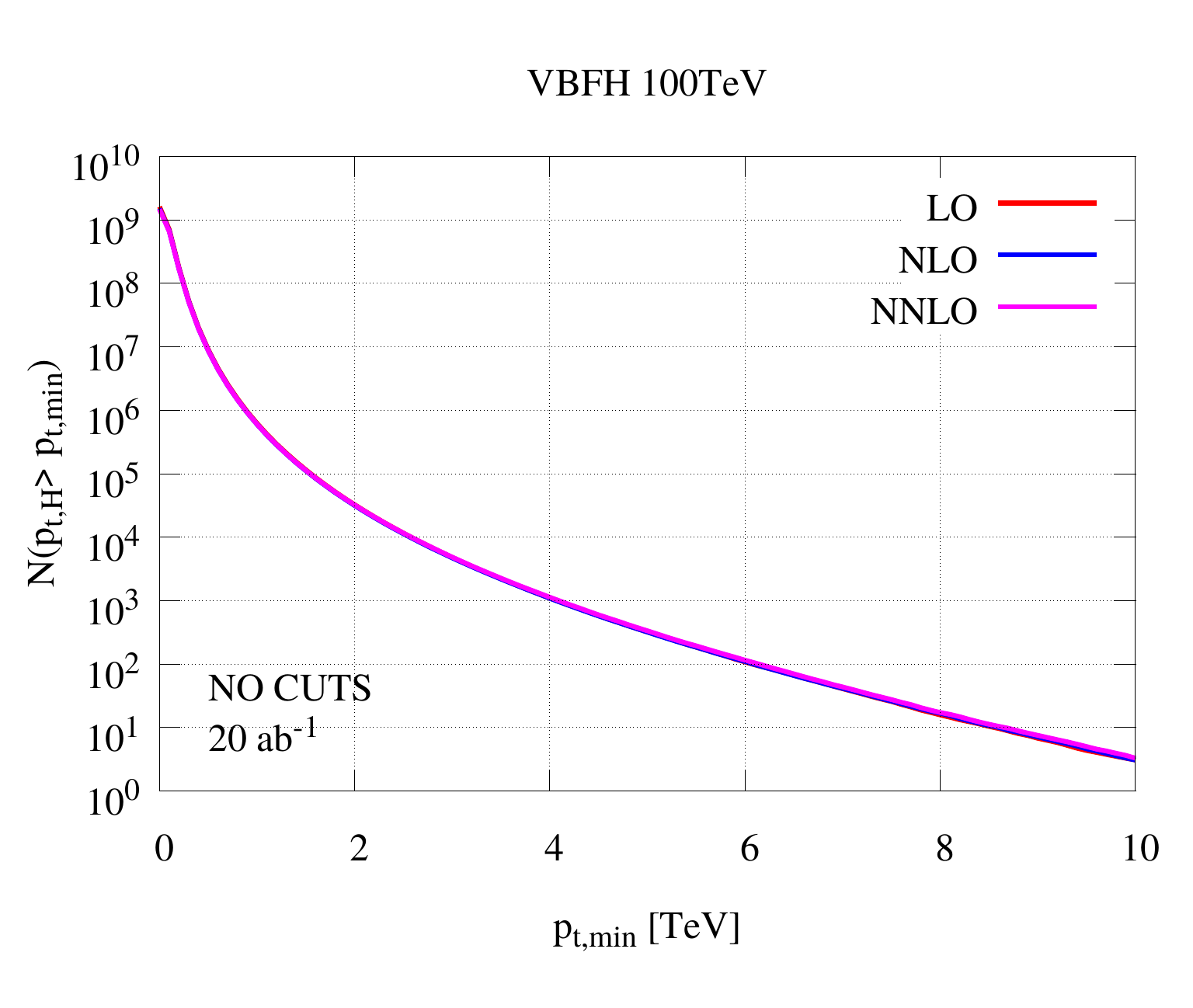}
  \end{minipage}
  \begin{minipage}{0.49\textwidth}
    \centering
    \includegraphics[width=1\textwidth,page=10]{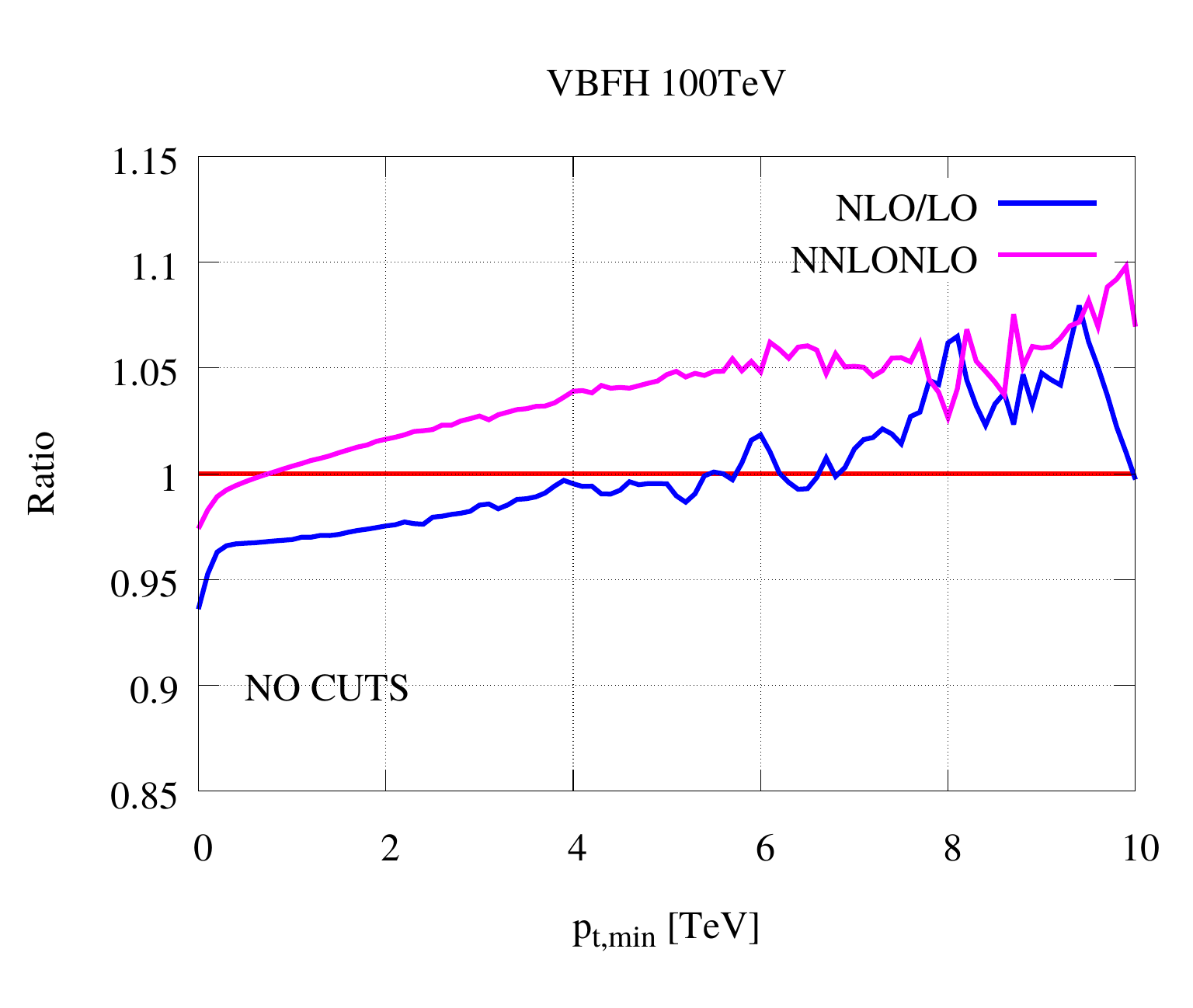}
  \end{minipage}
  \caption{\label{fig:VBFptHcorr} The total number of VBFH events produced with $p_{{t,H}} > p_{{t,\mathrm{min}}}$ at a 100 TeV collider with an integrated luminosity of $20\mbox{ ab}^{-1}$ with no cuts applied. Left panel: Spectrum computed at LO, NLO and NNLO in QCD. Due to the small corrections the difference between the three curves is hard to see by eye. Right panel: The k-factor going from LO to NLO and NLO to NNLO.}
\end{figure}

\begin{figure}[thb]
  \begin{minipage}{0.49\textwidth}
    \centering
    \includegraphics[width=1\textwidth,page=4]{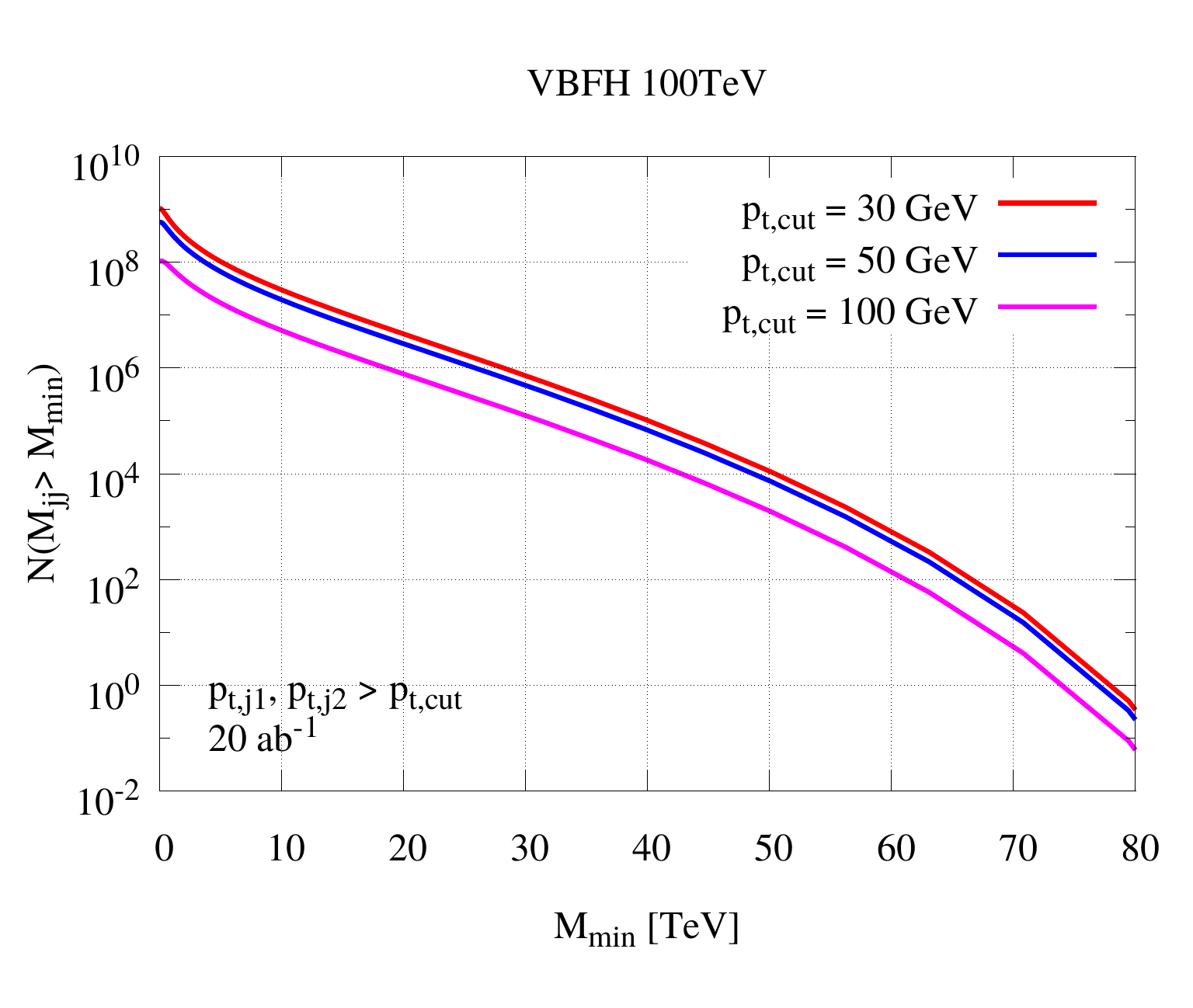}
  \end{minipage}
  \begin{minipage}{0.49\textwidth}
    \centering
    \includegraphics[width=1\textwidth,page=3]{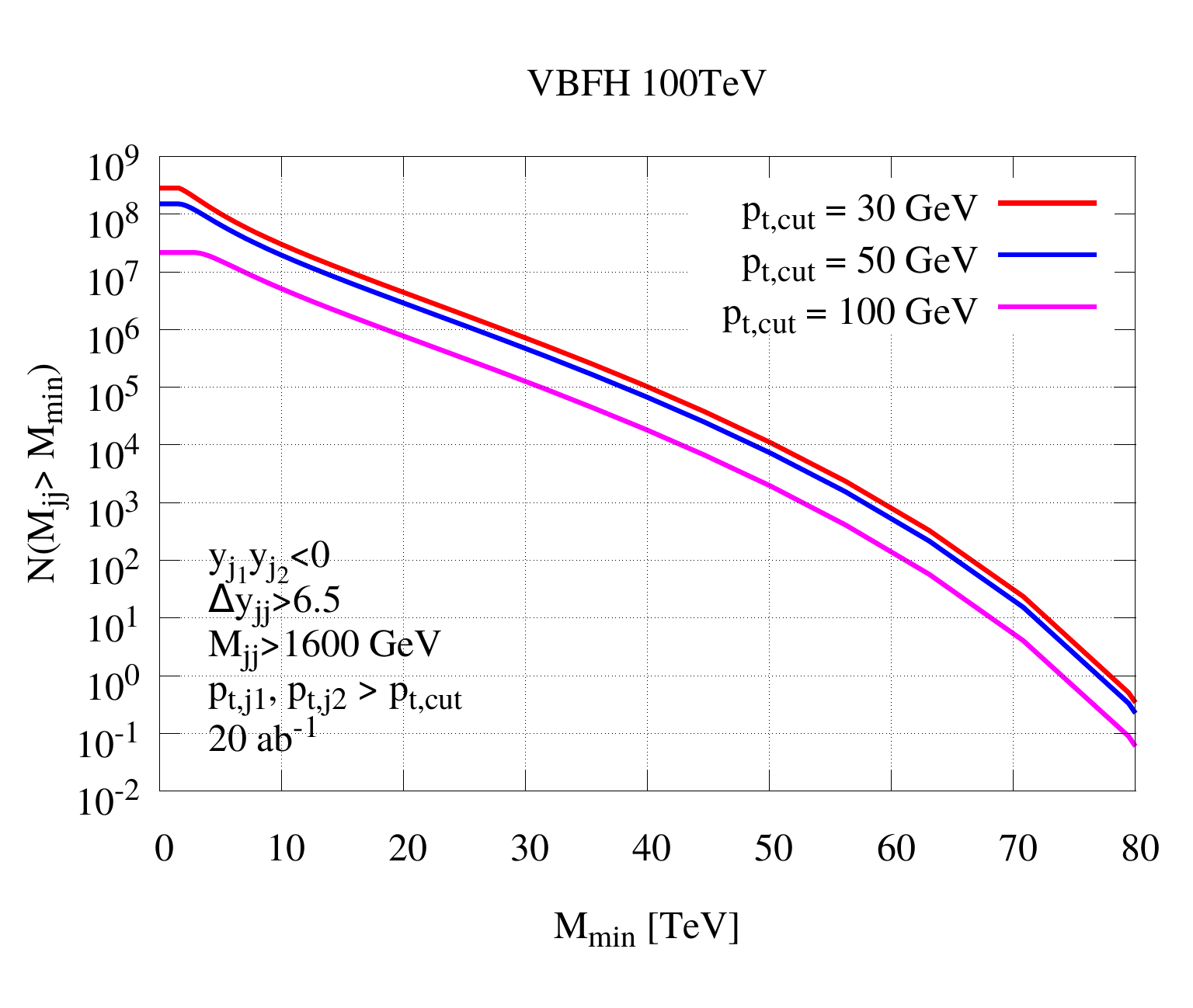}
  \end{minipage}
  \caption{\label{fig:VBFMjjaccum} The total number of VBFH events produced with $M_{j_1 j_2} > M_{\mathrm{min}}$ at a 100 TeV collider with an integrated luminosity of $20 \mbox{ ab}^{-1}$. Left panel: Three different jet $p_{t}$ cuts applied but no VBF cuts applied. Right panel: VBF cuts of eq. \eqref{eq:VBFcuts} and three different jet $p_{t}$ cuts applied.}
\end{figure}

\subsubsection{Detector implications}
The requirement that the two hardest jets are in opposite detector
hemispheres and are separated by at least $6.5$ units of rapidity,
means that a symmetric detector in the style of \textsc{ATLAS} or
\textsc{CMS} must have a rapidity reach well above $3.25$. In fact,
looking at fig. \ref{fig:VBFmaxyj1yj2}, which shows the fraction of
events which satisfy $\mathrm{max}(|y_{j_1}|,|y_{j_2}|) >
y_{\mathrm{min}}$ for various cut configurations, it becomes clear
that a detector with a rapidity reach of $4.5$ would at best only
retain $40\%$ of the VBFH events after VBF cuts are applied.  Since a
jet with $p_{t}=30$~GeV can be produced at a rapidity of
$\sim 8$ whereas a jet with $p_{t}=100$~GeV can only be
produced with rapidities up to $\sim 6.8$, the required rapidity reach
of the detector will also depend on how well soft jets can measured and
controlled at $100\,$TeV. In all cases a rapidity reach above 6 seems to
be desirable.

\begin{figure}[thb]
  \begin{minipage}{0.49\textwidth}
    \centering
    \includegraphics[width=1\textwidth,page=1]{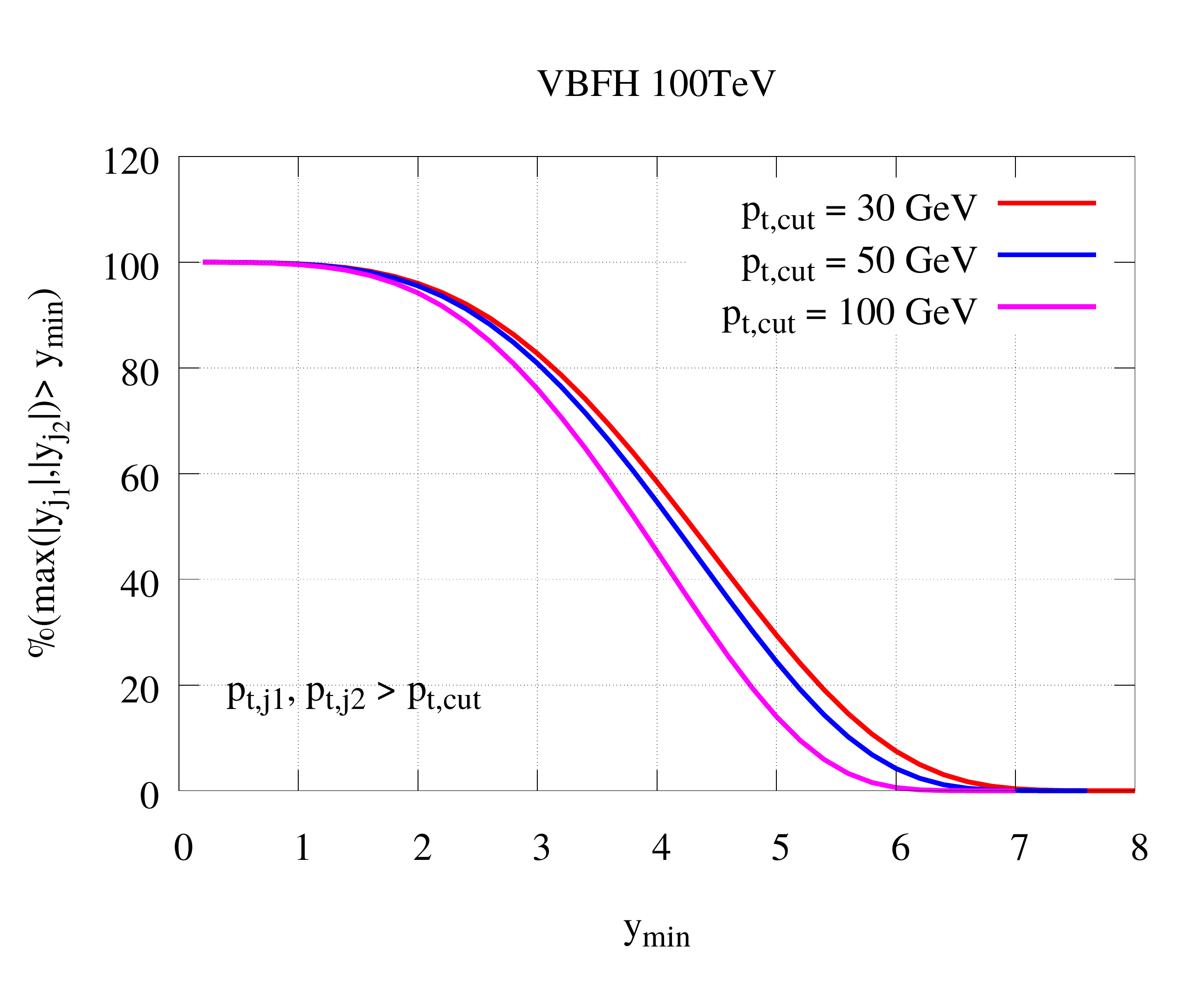}
  \end{minipage}
  \begin{minipage}{0.49\textwidth}
    \centering
    \includegraphics[width=1\textwidth,page=2]{figs/VBF/vbf_100TeV_accumulated.pdf}
  \end{minipage}
  \caption{\label{fig:VBFmaxyj1yj2} The total fraction of events where $\mathrm{max}(|y_{j_1}|,|y_{j_2}|) > y_{\mathrm{min}}$ at a 100 TeV collider.  Left panel: Three different jet $p_{t}$ cuts applied but no VBF cuts applied. Right panel: VBF cuts of eq. \eqref{eq:VBFcuts} and three different jet $p_{t}$ cuts applied.}
\end{figure}

\clearpage
\def\tth{t\bar t H}
\def\ttv{t\bar t V}
\def\ttz{t\bar t Z}
\def\ytop{y_{top}}
\def \gsim{\mathrel{\vcenter
     {\hbox{$>$}\nointerlineskip\hbox{$\sim$}}}}
\def \lsim{\mathrel{\vcenter
     {\hbox{$<$}\nointerlineskip\hbox{$\sim$}}}}
\def\as{\alpha_s}
\def\az{\alpha_0}
\def\mev{\mathrm{MeV}}
\def\gev{\mathrm{GeV}}
\def\tev{\mathrm{TeV}}
\def\fb{\mathrm{fb}}
\def\ab{\mathrm{ab}}
\def\pb{\mathrm{pb}}
\def\sss{\scriptscriptstyle}

\subsection{Associated $\tth$ production}
\label{sec:H_ttH}
The $\tth$ process provides the most direct probe of the interaction
of the Higgs boson with the top quark. Theoretical calculations have
been completed including NLO QCD~\cite{Beenakker:2002nc,Dawson:2003zu}
and EW~\cite{Frixione:2014qaa} corrections. NLO corrections have recently
been extended to the case of unstable top quarks, in the dilepton
final state~\cite{Denner:2015yca}.

In this section, we collect results for the total production cross
sections and for some key kinematical distributions. In particular, we
 update and extend parts of the study presented in
Ref.~\cite{Plehn:2015cta}, where it was shown that tight correlations
between scale and PDF uncertainties lead to very precise predictions for
the ratio of $\tth$ and $\ttz$ production. We focus in this section on
the discussion of rate and theoretical systematics, and in
Section~\ref{sec:H_ytop} we review the prospects for measurements of $y_{top}$.

All results shown here were obtained using the {\sc MadGraph5\_aMC@NLO}
code~\cite{Alwall:2014hca}, 
which includes both NLO QCD and EW corrections, in the case of stable
top quarks. Additional details, and the results for 13~TeV, can be
found in Ref.~\cite{Plehn:2015cta}.
The default parameter set used here is given by:
\begin{center}
 \renewcommand{\arraystretch}{1.0}
\begin{tabular}{cl|cl}\hline
Parameter & value & Parameter & value
\\\hline
$G_{\mu}$ & \texttt{1.1987498350461625} $\cdot$ \texttt{10${}^{-5}$} & $n_{lf}$ & \texttt{5}
\\
$m_{t}$ & \texttt{173.3}    & $v y_{t}$ & \texttt{173.3}
\\
$m_{W}$ & \texttt{80.419}    & $m_{Z}$ & \texttt{91.188}
\\
$m_{H}$ & \texttt{125.0}    & $\alpha^{-1}$ & \texttt{128.930}
\\\hline
\end{tabular}
\end{center}
$\mu_R=\mu_F=\mu_0=\sum_{f\in {\rm final~states}}{m_{T,f}}/2$ is the
default for the central choice of renormalization and factorization
scales, where $m_{T,f}$ is the transverse mass of the final particle
$f$. This scale choice interpolates between the dynamical scales that
were shown in Refs.~\cite{Beenakker:2002nc,Dawson:2003zu} to minimize
the $p_T$ dependence of the NLO/LO ratios for the top and Higgs
spectra.  The scale variation systematics is obtained covering the
standard range $0.5\mu_0 \le \mu_{R,F} \le 2\mu_0$, with $\mu_R$ and
$\mu_F$ varying independently.

Table~\ref{tab:ttH_scale} shows the total cross section results, at
the LO in the EW effects.  The first row shows the results of the
MSTW2008 NLO~\cite{Martin:2009iq} sets, which will be used as a
default for the other results of this section.
The second row uses the more
recent PDF4LHC15~\cite{Butterworth:2015oua} recommendation, which
combines the systematics from
the following NLO PDF sets: NNPDF3.0~\cite{Ball:2014uwa},
MMHT2014~\cite{Harland-Lang:2014zoa} and
CT14~\cite{Dulat:2015mca}. The difference between MSTW2008 and
PDF4LHC15 is at the level of 3\%, which is compatible with the quoted
uncertainty on the $\tth$ cross section. 

Here, and in following tables and figures, we include the results for
the $\ttz$ process as well, and for the $\tth/\ttz$ ratios.  As
discussed in detail in~\cite{Plehn:2015cta}, there are strong
correlations among the sources of systematic uncertainty for these two
processes, leading to very robust predictions for their ratios. In
particular, all the results shown here relative to $\tth/\ttz$ ratios
will enforce the full correlation of the systematics induced by the
PDF variations and by parameters such as $m_{top,H}$, and will assume
likewise a complete correlation between the scale variations.

\begin{table}[b]
 \begin{center}
 \renewcommand{\arraystretch}{1.5}
 \begin{tabular}{c|ccc}
\hline
\rule{0pt}{3.6ex} & $\sigma(t\bar{t}H)$[pb] & $\sigma(t\bar{t}Z)$[pb] &
   $\frac{\displaystyle \sigma(t\bar{t}H)}{\displaystyle \sigma(t\bar{t}Z)}$ \\ 
\hline
  MSTW2008 & 	$33.9^{+7.1\%+2.2\%}_{-8.3\%-2.2\%}$	&
$57.9^{+8.9\%+2.2\%}_{-9.5\%-2.4\%}$	&
$0.585^{+1.3\%+0.3\%}_{-2.0\%-0.2\%}$ \\
 PDF4LHC15 & $32.8^{+6.9\%+1.6\%}_{-8.1\%-1.6\%}$
  & $56.0^{+8.8\%+1.5\%}_{-9.3\%-1.5\%}$
  & $0.586^{+1.3\%+0.12\%}_{-2.0\%-0.12\%}$\\
\hline
\end{tabular}
\end{center}
\caption{Total cross sections $\sigma(\tth)$ and
  $\sigma(\ttz)$ and the ratios
  $\sigma(\tth)/\sigma(\ttz)$ with NLO QCD. The results include the
  renormalization/factorization scale and PDF$+\as$
  uncertainties.}
\label{tab:ttH_scale}
\end{table}
Table~\ref{tab:ttH_scale} shows that the scale uncertainty of the
individual processes, in the range of $\pm 9\%$, is reduced in the
ratio to $\pm 1.5\%$. The PDF uncertainty on
the ratio is at the permille level, and the
comparison of the old MSTW2008 result with the most recent PDF4LHC15
one confirms the reliability of this estimate.

The effect of the NLO EW corrections in two different schemes is shown in
Table~\ref{tab:ttH_EW}.  The shift of the individual $\tth$ and $\ttz$
is at the level of few percent, and depends on the EW scheme and on
the process. The ratio shifts with respect to the LO EW result by less
than 2\%, and the EW scheme dependence is at the permille level. This
suggests that the residual uncertainty of the cross-section ratio 
due to higher-order EW corrections should be significantly
below the percent level.

\setlength{\tabcolsep}{3pt}
\begin{table}
 \renewcommand{\arraystretch}{1.3}
 \begin{tabular}{c|ccc|ccc}
\hline
 & \multicolumn{3}{c|}{$\alpha(m_Z)$ scheme} &
   \multicolumn{3}{c}{$G_\mu$ scheme} \\ 
\rule{0pt}{3.6ex}  & $\sigma(t\bar{t}H)$[pb] & $\sigma(t\bar{t}Z)$[pb] &
   $\frac{\displaystyle \sigma(t\bar{t}H)}{\displaystyle \sigma(t\bar{t}Z)}$ & 
   $\sigma(t\bar{t}H)$[pb] & $\sigma(t\bar{t}Z)$[pb] &
   $\frac{\displaystyle \sigma(t\bar{t}H)}{\displaystyle \sigma(t\bar{t}Z)}$ \\ 
\hline
 NLO QCD 
   & $33.9$ & $57.9$ & $0.585$ 
   & $32.9$ & $56.3$ & $0.585$\\
$\mathcal{O}(\alpha_S^2\alpha^2)$ Weak 
   & $-0.73$   & $-2.15$    & 
   & $0.027$   & $-0.90$    & \\ 
$\mathcal{O}(\alpha_S^2\alpha^2)$ EW 
   & $-0.65$ & $-2.0$ & ~ 
   & $0.14$ & $-0.77$ & ~  \\ 
NLO QCD+Weak 
   & $33.1$ & $55.8$ & $0.594$
   & $32.9$ & $55.4$ & $0.594$\\
NLO QCD+EW 
   & $33.2$ & $55.9$ & $0.594$
   & $33.1$ & $55.6$ & $0.595$\\ \hline
\end{tabular}
\caption{Effect of the EW NLO corrections, in the
  $\alpha(m_Z)$ and $G_{\mu}$ schemes.}
\label{tab:ttH_EW}
\end{table}
\setlength{\tabcolsep}{6pt}

\begin{table}
 \begin{center}
 \renewcommand{\arraystretch}{1.5}
 \begin{tabular}{c|ccc}
\hline
\rule{0pt}{3.6ex} & $\sigma(t\bar{t}H)$[pb] & $\sigma(t\bar{t}Z)$[pb] & $\frac{\displaystyle \sigma(t\bar{t}H)}{\displaystyle \sigma(t\bar{t}Z)}$ \\
\hline
   default & $33.9^{+7.1\%}_{-8.3\%}$	&
   $57.9^{+8.9\%}_{-9.5\%}$	&
   $0.585^{+1.3\%}_{-2.0\%}$ \\ 
   $\mu_0=m_t+m_{H,Z}/2$ & $39.0^{+9.8\%}_{-9.6\%}$ &
   $67.2^{+11\%}_{-11\%}$ &
   $0.580^{+1.2\%}_{-1.8\%}$ \\ 
   $m_t=\ytop v=174.1~\gev$ &
   $33.9$	&
   $57.2$	&
   $0.592$\\ 
   $m_t=\ytop v=172.5~\gev$ &
   $33.7$ &
   $58.6$ &
   $0.576 $\\ 
   $m_H=126.0~\gev$ &
   $33.2$
   &  $57.9$	&
   $0.575$\\ \hline
\end{tabular}
\end{center}
\caption{Results with NLO QCD corrections at 100~TeV by varying some
   parameter values. In the first two rows we include
   the renormalization/factorization scale uncertainties.}
\label{tab:ttH_parsys}
\end{table}

We explore further variations in our default parameter set in
Table~\ref{tab:ttH_parsys}. There, we remove the PDF uncertainties,
which are practically unaffected by these parameter changes.  Choosing
the fixed value $\mu_0=m_t+m_{H,Z}/2$ for the central choice of the
renormalization and factorization scales, modifies the ratio
$\sigma(t\bar{t}H)/\sigma(t\bar{t}Z)$ by $1\%-1.5\%$, consistent with
the range established using the dynamical scale.

For $m_t$, we consider a variation in the range of
$m_t=173.3\pm0.8$~GeV. We notice that $\sigma(\tth)$ is practically
constant. This is due to the anti-correlation between the increase
(decrease) in rate due to pure phase-space, and the decrease
(increase) in the strength of $\ytop$, when the top mass is lower
(higher).  The $\ttz$ process is vice versa directly sensitive to
$m_t$ at the level of $\pm 1.5\%$ over the $\pm 0.8$~GeV range, and
this sensitivity is reflected in the variation of the cross-section
ratio. We notice, however, that if we kept the value of $\ytop$ fixed
when we change $m_t$, the dynamical effect on the rate would be
totally correlated, and the ratio would remain constant to within a
few permille, as shown in Table~\ref{tab:tot_yt}. This shows that the
ratio is only sensitive to the strength of $\ytop$, and only minimally
to the precise value of $m_t$.

Finally, we observe a $\sim 2\%$ shift in $\sigma(\tth)$ (and
therefore in the ratios) when $m_H$ is changed by 1~GeV, which is a
gross underestimate of the precision with which the Higgs mass
is~\cite{Aad:2015zhl} and will soon be known.

In summary, we quote, as the best estimate for the $\tth$ cross
section at 100 TeV (with $m_H=125$~GeV and $m_{top}=173.3$~GeV), the following:
\begin{equation}
\sigma(\tth)[pb] = 32.1 {+6.9\% \atop
-8.1\%}_{scale} \pm{1.6\%}_{PDF4LHC15} \pm{0.3\%}_{mtop}
\end{equation}
This includes the full EW corrections, and accounts for a $\pm
0.8$~GeV uncertainty in the top mass.

\begin{table}
 \begin{center}
 \renewcommand{\arraystretch}{1.3}
 \begin{tabular}{c|ccc}
\hline
\rule{0pt}{3.6ex} & $\sigma(t\bar{t}H)$[pb] & $\sigma(t\bar{t}Z)$[pb] & $\frac{\displaystyle \sigma(t\bar{t}H)}{\displaystyle \sigma(t\bar{t}Z)}$ \\
\hline
  $m_t=174.1~\gev$ & $23.88$	&  $37.99$ & $0.629$\\ 
  $m_t=172.5~\gev$ & $24.21$	&  $38.73$ & $0.625$\\ \hline
\end{tabular}
\end{center}
\caption{LO results for different top masses, keeping
  the top Yukawa coupling fixed at $\ytop v =173.3$ GeV.}
\label{tab:tot_yt}
\end{table}

\subsubsection{Kinematical distributions}
As shown in the study of other production processes, one of
the key features of the 100~TeV collider is the existence of large
production rates even with kinematical configurations at extremely
large energy. For the $\tth$ process, this is well illustrated by
Fig.~\ref{fig:ttH_pt}, which gives the cross sections for production
of the Higgs (left) and top quark (right) above a given $p_T$
threshold. With the expected FCC-hh luminosities, the production will
extend well beyond $p_T\sim 5$~TeV. We note that the spectra are very
stable against scale and PDF systematics, the former staying within a
10\% window.  In Fig.~\ref{fig:ttH_dR} we also plot the integrated
cross section for producing $t$, $\bar{t}$ and $H$ all above a given
$p_T$ threshold, in configurations in which these three objects are
pair-wise separated by $\Delta R>1$ and $2$. 

\begin{figure}
\hspace{-0.05\textwidth}
\includegraphics[width=0.55\textwidth]{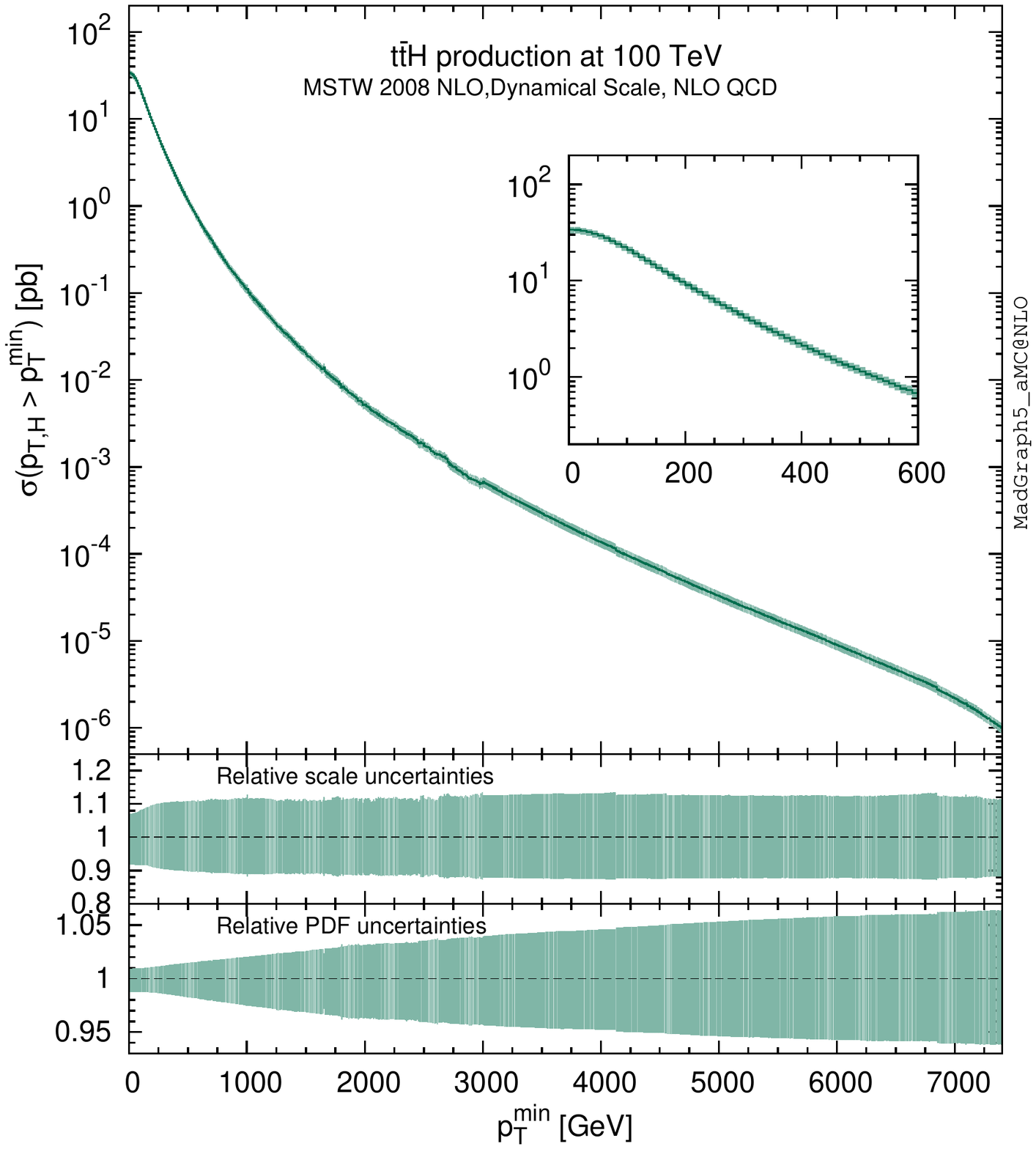}
\hspace*{-0.11\textwidth}
\includegraphics[width=0.55\textwidth]{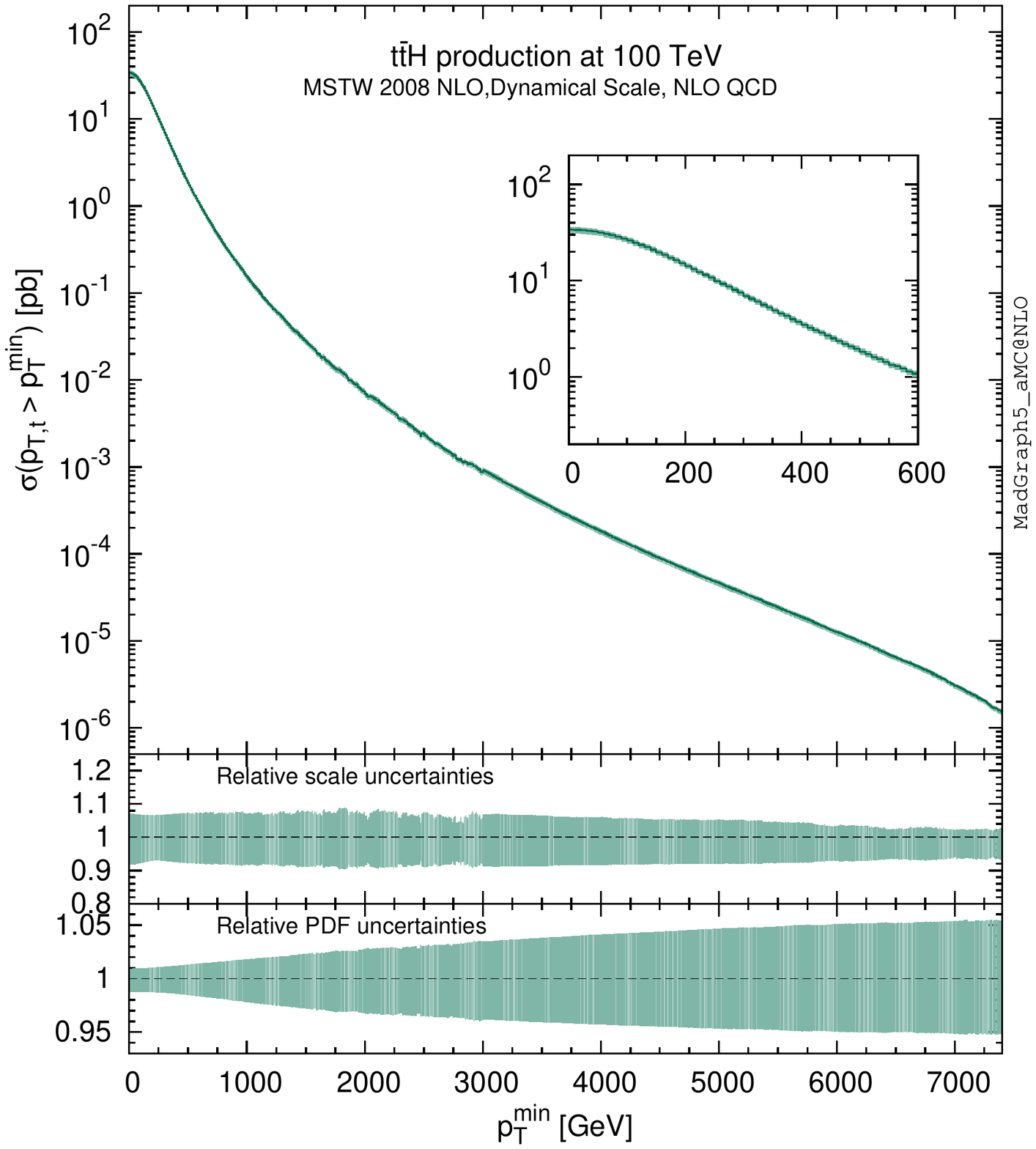}
\vspace{-3.5cm}
\caption{Integrated transverse momentum distributions for the Higgs
  boson (left) and (anti-)top quark (right), in the $t\bar{t}H$
  process at a 100~TeV collider.}
\label{fig:ttH_pt}
\end{figure}

\begin{figure}
\begin{center}
\includegraphics[width=0.55\textwidth]{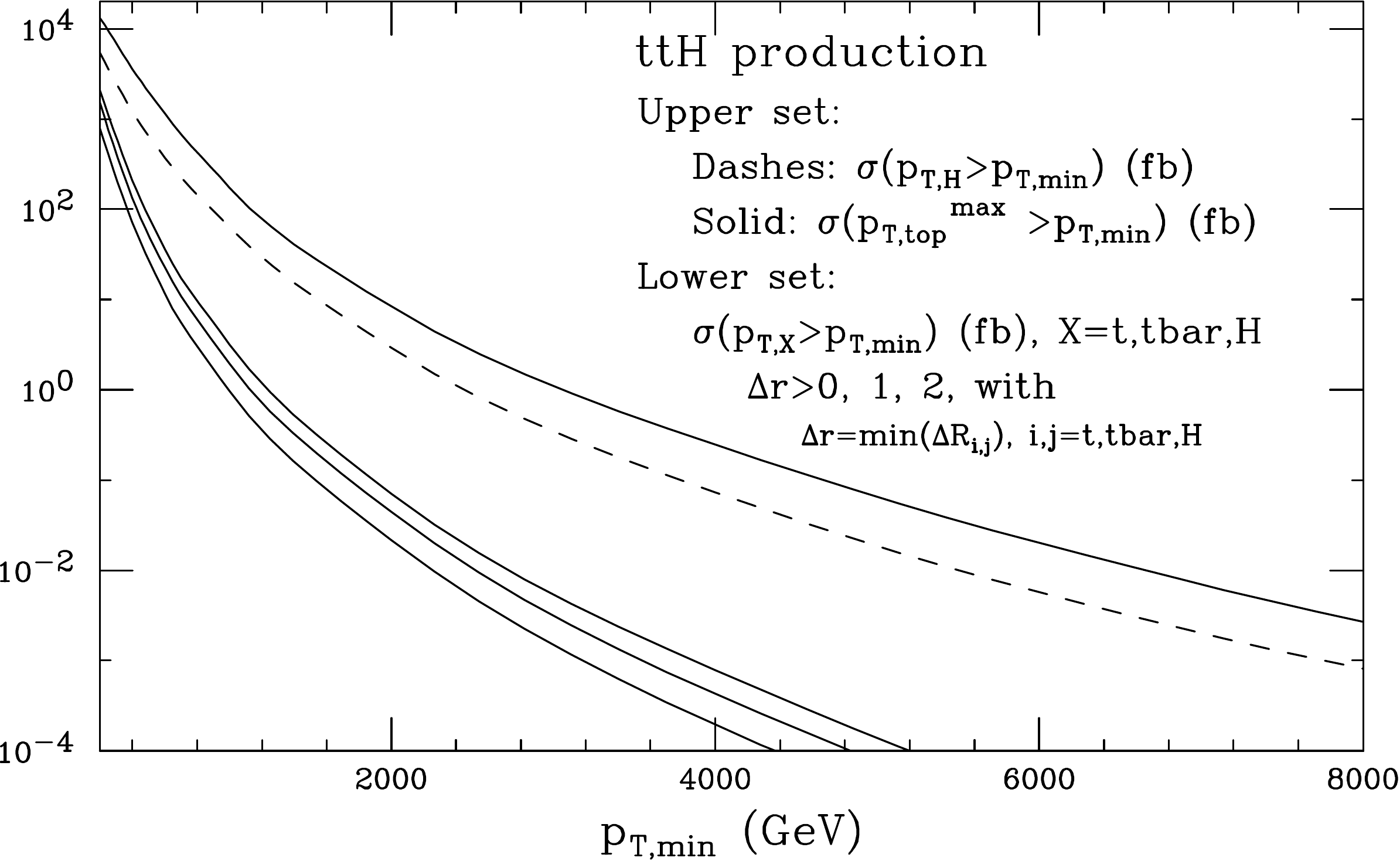}
\caption{Integrated distributions for the $p_T$ of the Higgs
  boson (upper solid line), the maximum $p_T$ of top and antitop
  quarks (upper dashed line), and for the minimum $p_T$ of $t$,
  $\bar{t}$ nd $H$, with different cuts on the $\Delta R$ separation
  among different objects.}
\label{fig:ttH_dR}
\end{center}
\end{figure}

Any experimental analysis, and in particular the boosted approach that
will be used in Section~\ref{sec:H_ytop},
will restrict the phase-space available to the
final states. To preserve the precision in the theoretical prediction
of the ratio of total $\tth$ and $\ttz$ cross sections, it is crucial
to ensure that the reduction in systematics uncertainties carries over
to the description of final states after kinematical cuts have been
applied. The following results will give some concrete examples,
focused on the discussion of scale and PDF systematics.
\begin{figure}
\centering
\hspace{-0.05\textwidth}
\includegraphics[width=0.55\textwidth]{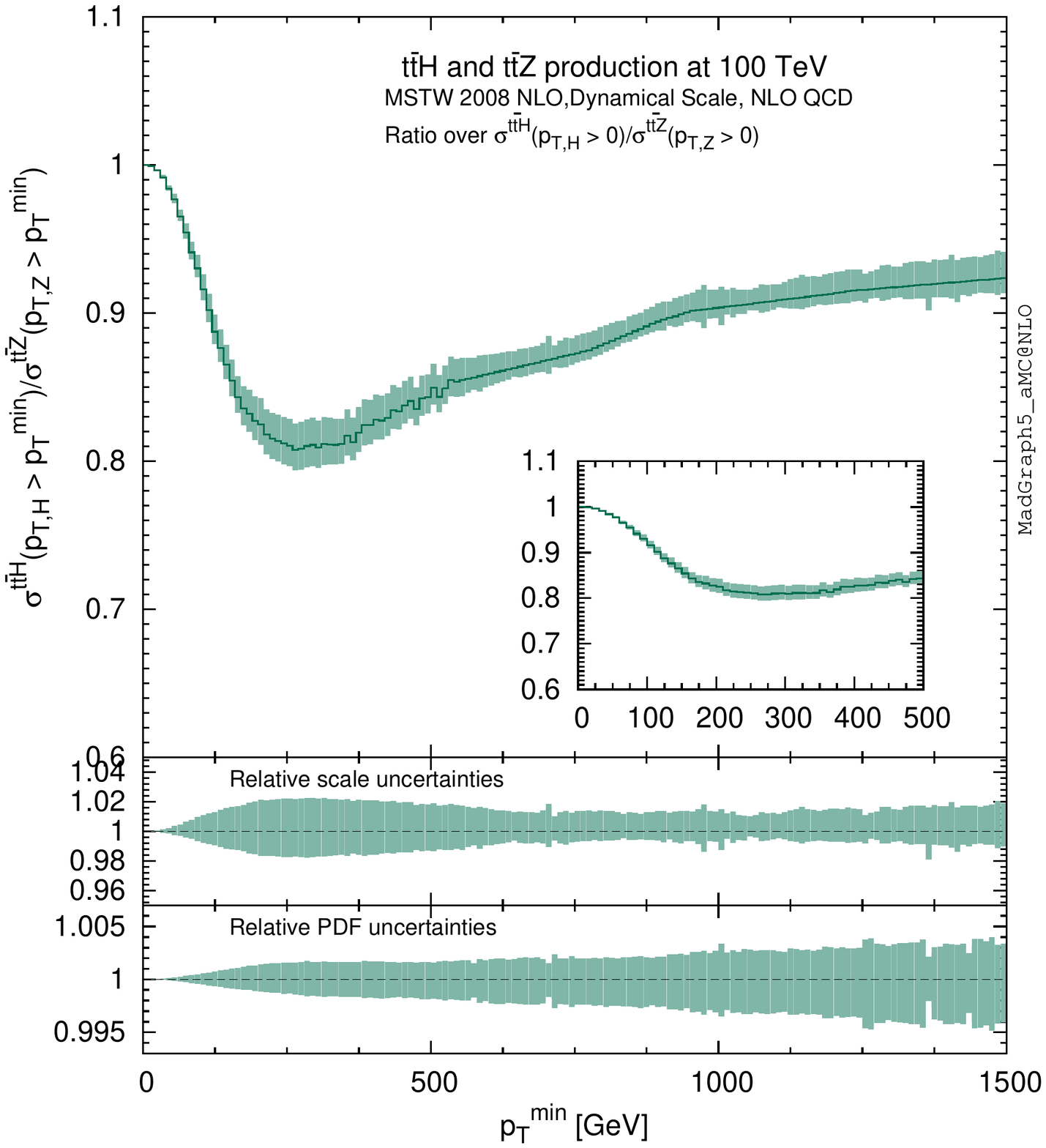}
\hspace*{-0.11\textwidth}
\includegraphics[width=0.55\textwidth]{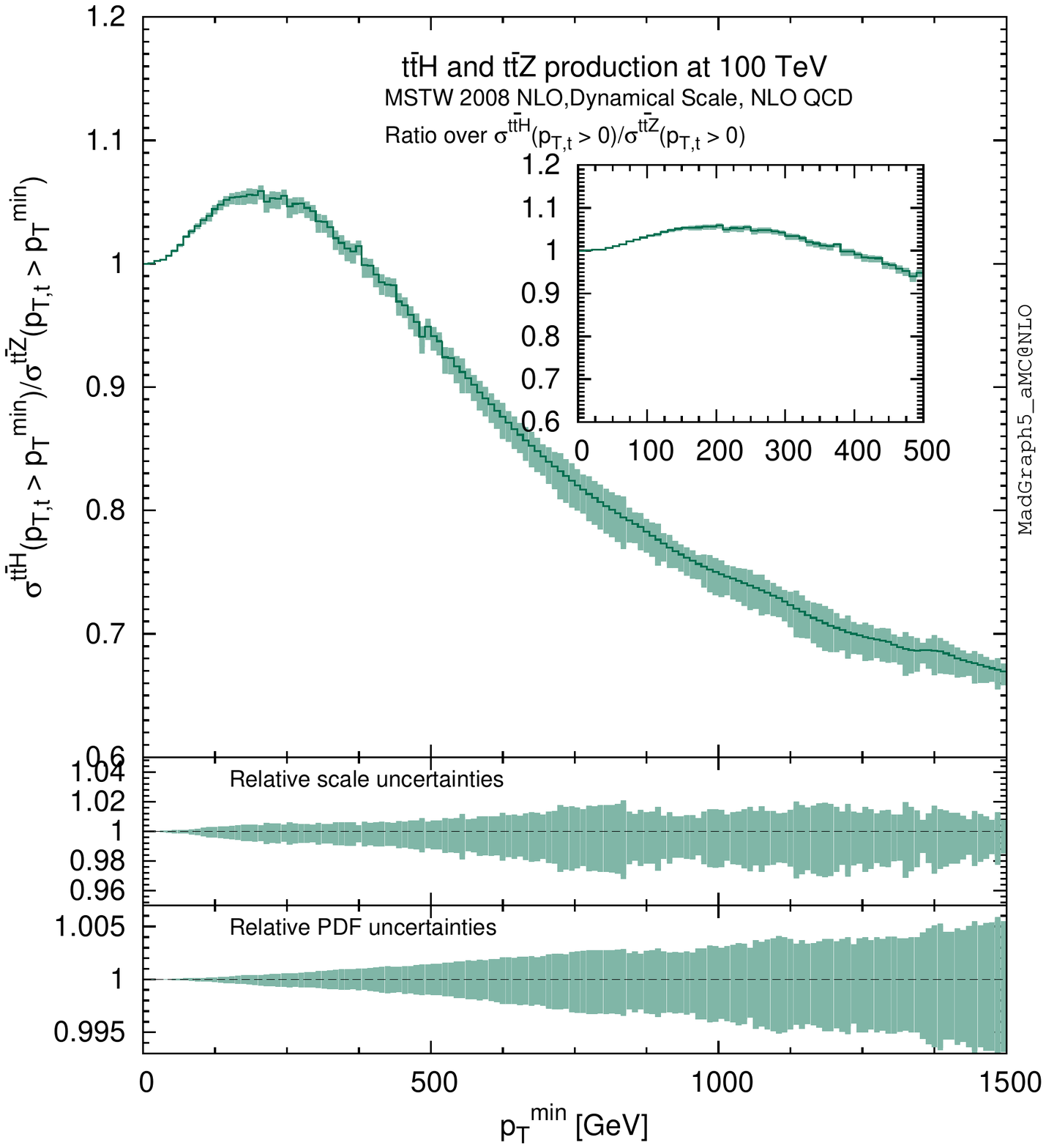}
\vspace{-3.5cm}
\caption{Scale and PDF systematics of ratios of integrated $p_T$
  spectra for different observables, at 100~TeV. From left to right:
  $p_T$ of the boson, $p_T$ of the top quark. }
\label{fig:ptsyst}
\end{figure}

\begin{figure}
\centering
\hspace{-0.05\textwidth}
\includegraphics[width=0.55\textwidth]{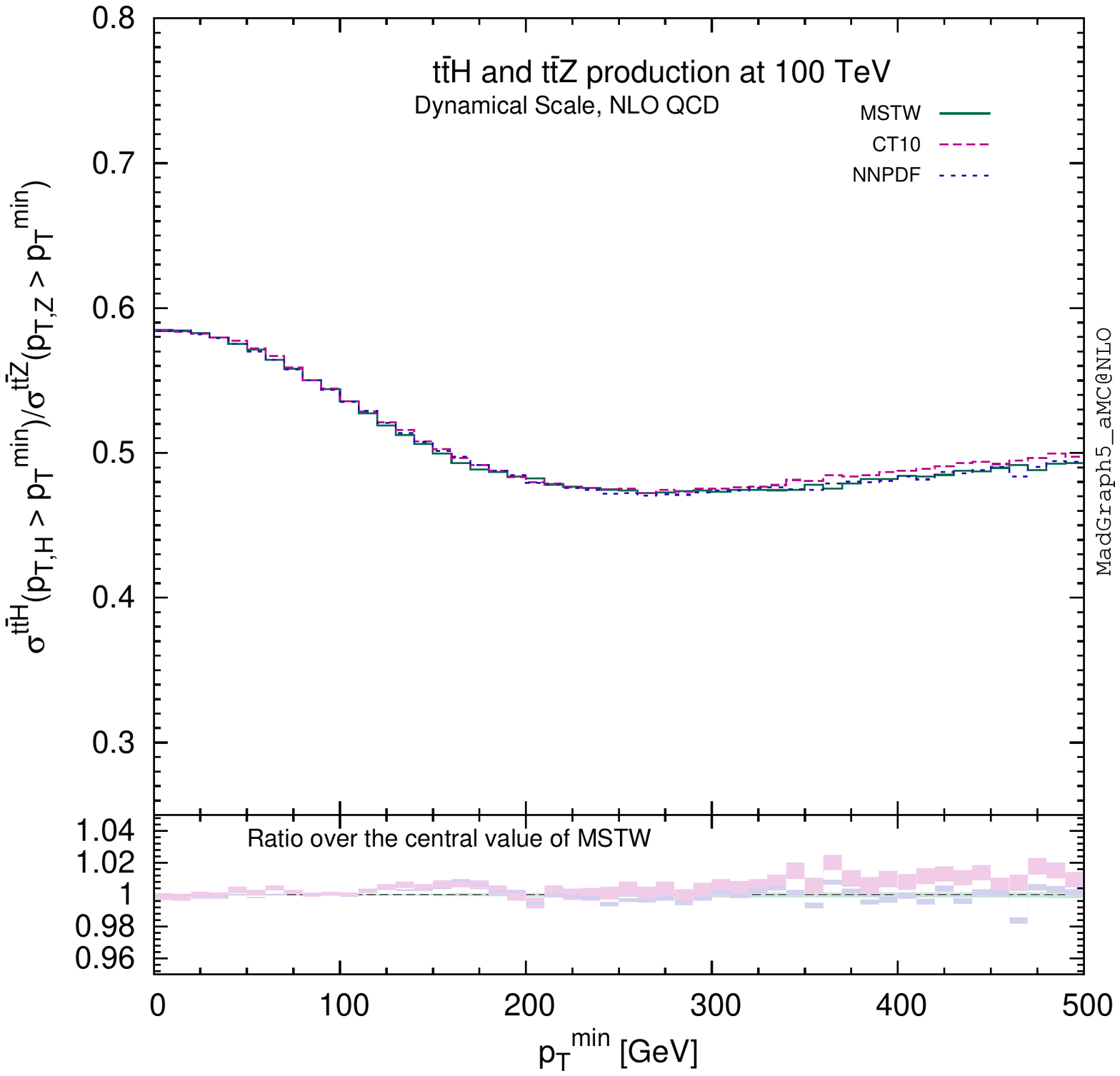}
\hspace*{-0.11\textwidth}
\includegraphics[width=0.55\textwidth]{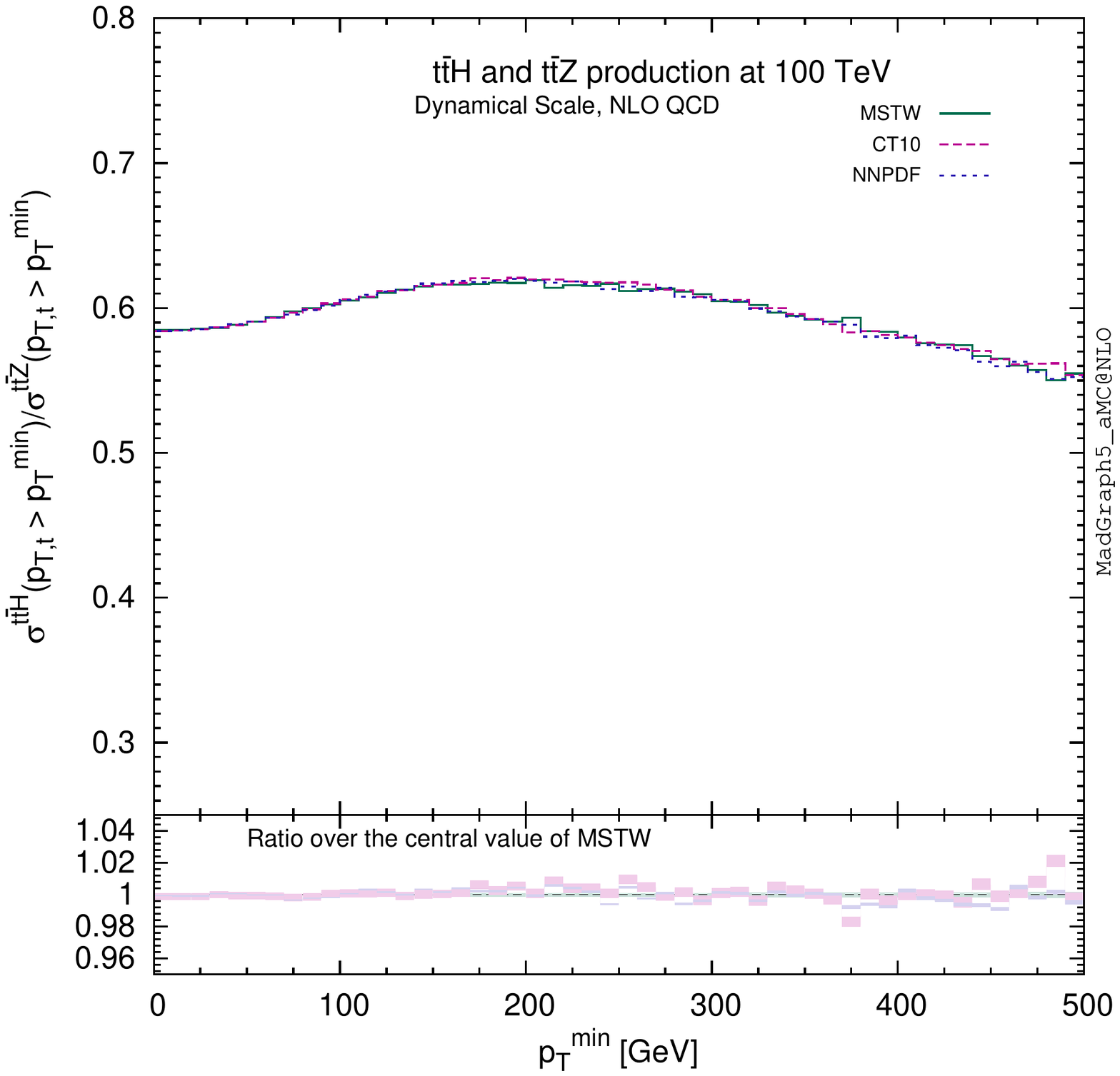}
\vspace{-4.5cm}
\caption{Three PDF sets systematics of ratios of integrated $p_T$
  spectra for different observables, at 100~TeV. From left to right:
  $p_T$ of the boson, $p_T$ of the top quark. }
\label{fig:ptsyst2}
\end{figure}

\begin{figure}
\centering
\hspace{-0.05\textwidth}
\includegraphics[width=0.55\textwidth]{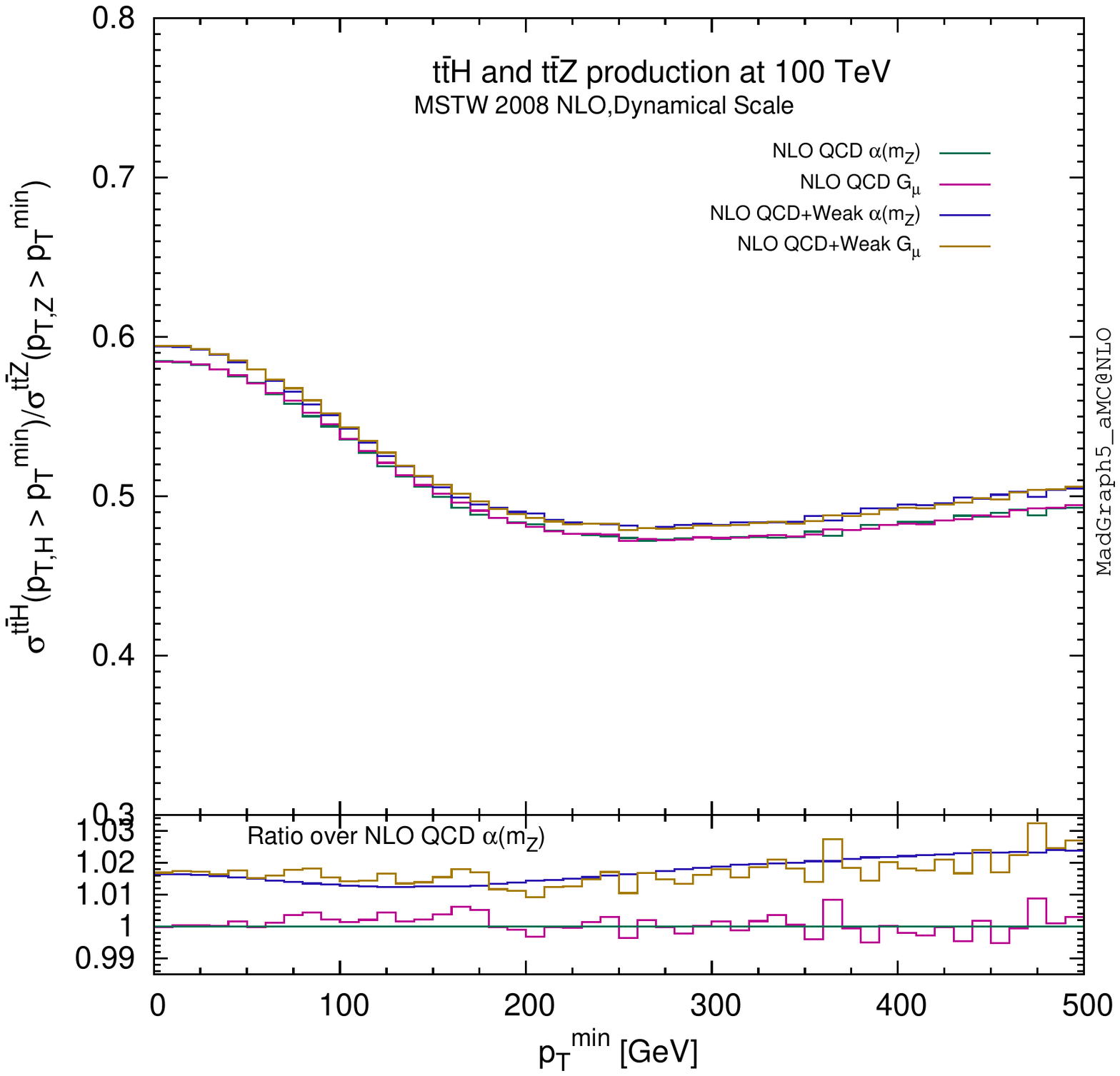}
\hspace*{-0.11\textwidth}
\includegraphics[width=0.55\textwidth]{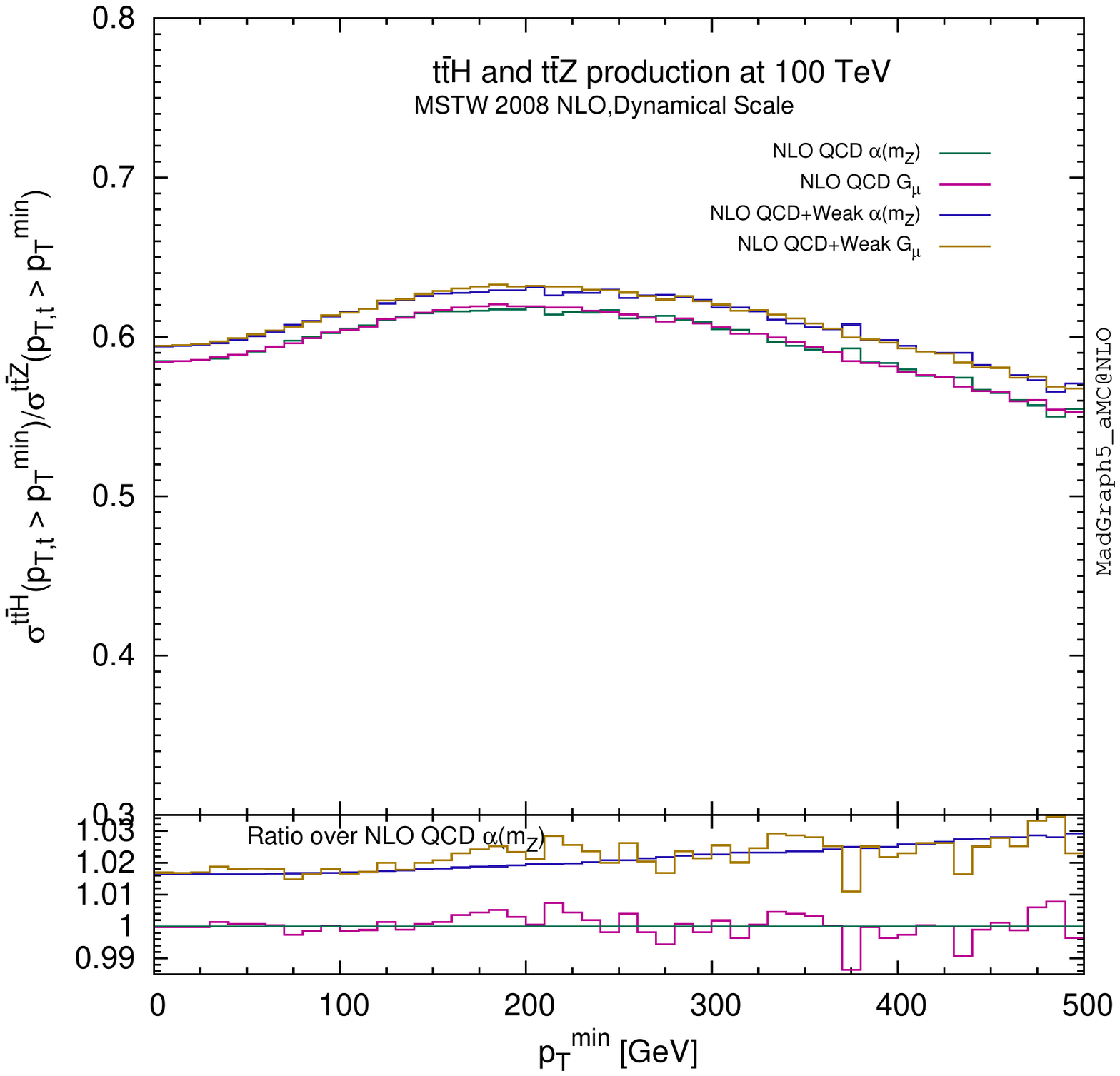}
\vspace{-4.5cm}
\caption{EW scheme dependence and weak corrections of ratios of integrated $p_T$
  spectra for different observables, at 100~TeV. From left to right:
  $p_T$ of the boson, $p_T$ of the top quark. }
\label{fig:ptsyst3}
\end{figure}

\begin{figure}
\centering
\hspace{-0.05\textwidth}
\includegraphics[width=0.55\textwidth]{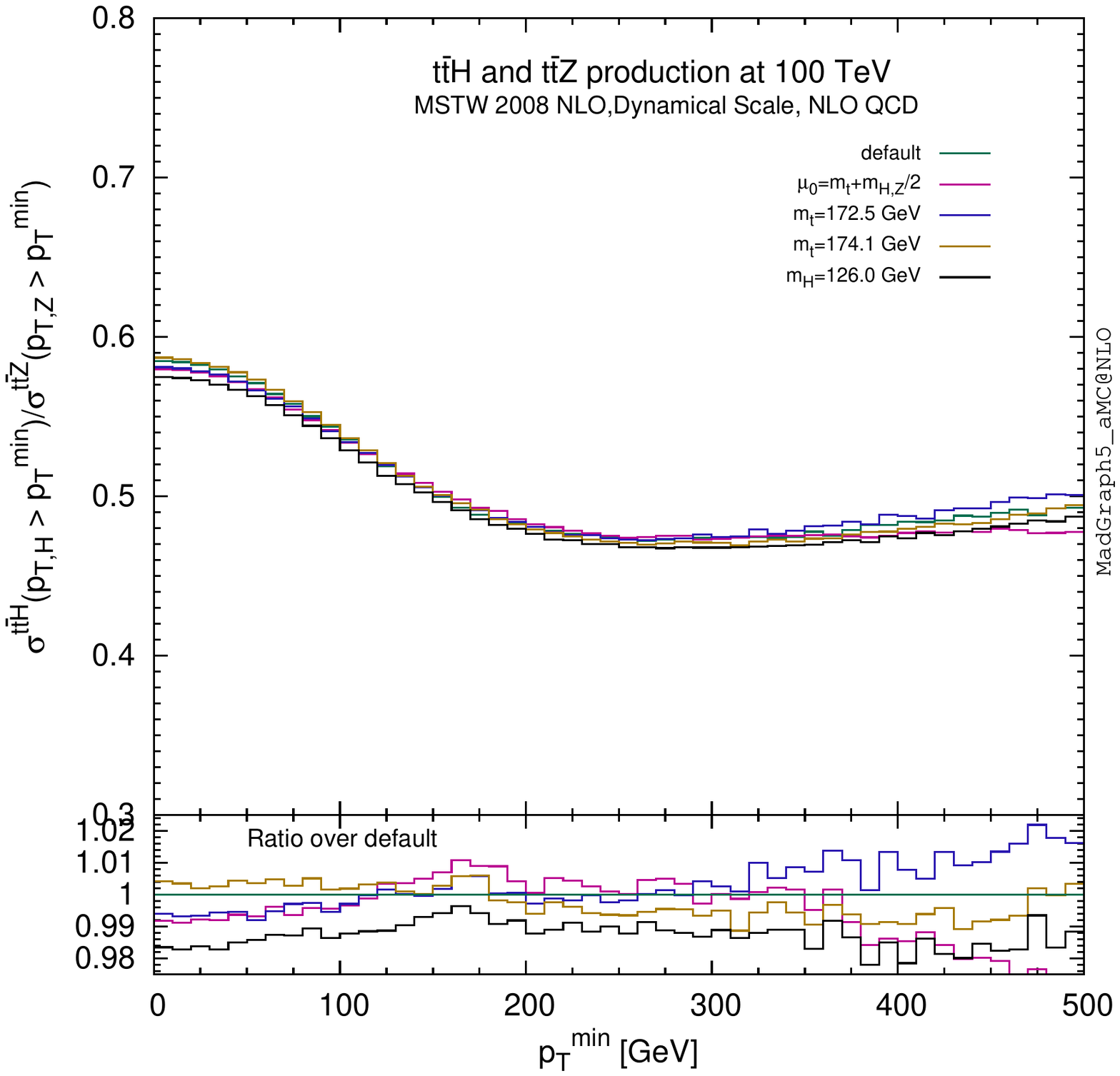}
\hspace*{-0.11\textwidth}
\includegraphics[width=0.55\textwidth]{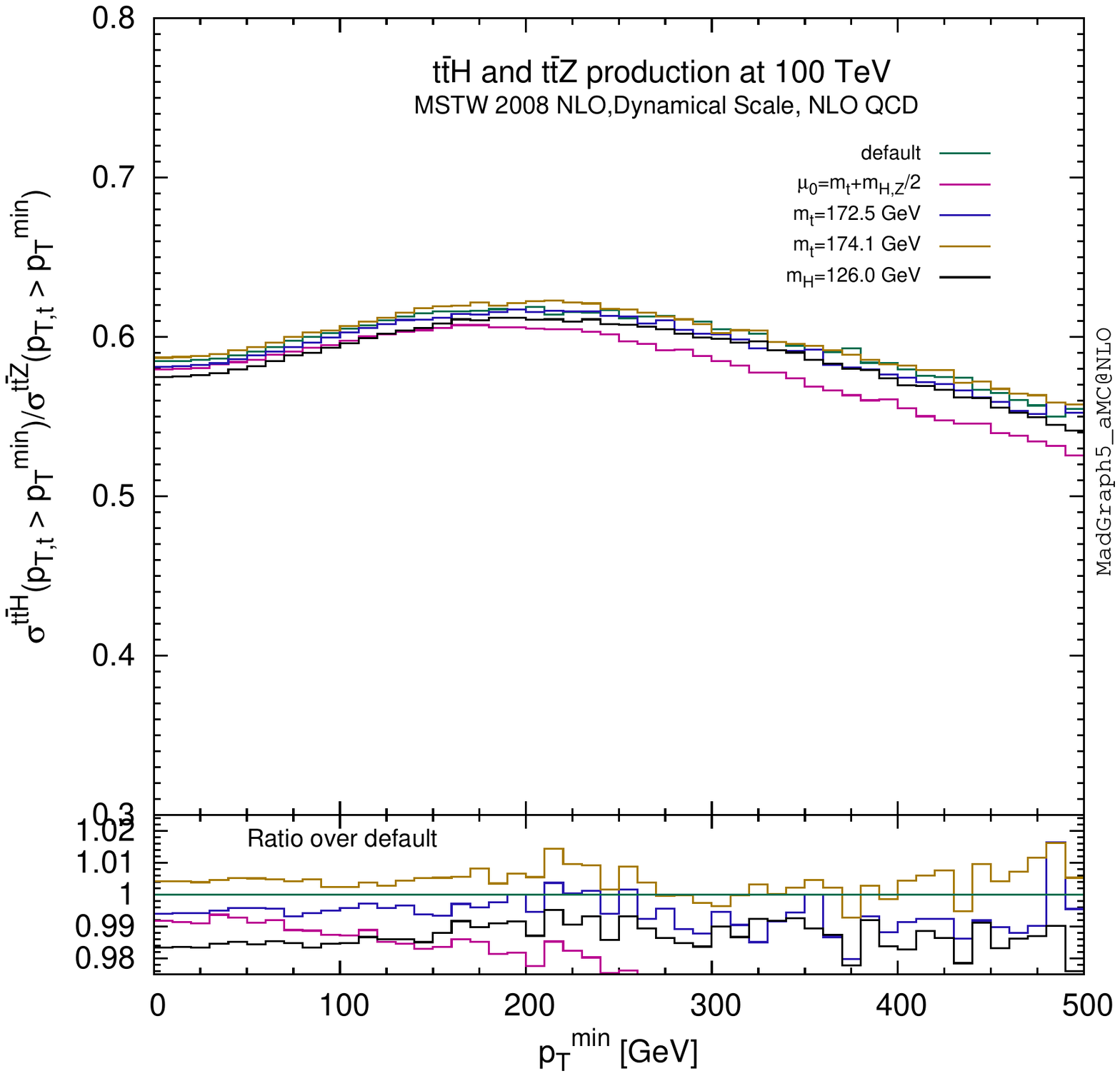}
\vspace{-4.5cm}
\caption{Other systematics of ratios of integrated $p_T$
  spectra for different observables, at 100~TeV. From left to right:
  $p_T$ of the boson, $p_T$ of the top quark. }
\label{fig:ptsyst4}
\end{figure}

We show in Fig.~\ref{fig:ptsyst} the ratio of the integrated $p_T$
spectra of various final-state objects $X$:
$\sigma[\tth](p_{T,X}>p_{T,{\rm min}})/ \sigma[\ttz](p_{T,X}>p_{T,{\rm
    min}})$.  On the left, $X=H(Z)$ for the $\tth$ $(\ttz)$
process. In the middle, $X=t$ and on the right $X$ is the $t\bar{t}$
system.  We normalize the ratios to 1 at $p_{T,{\rm min}}=0$, so that
the resulting uncertainties correspond to the systematics in the
extrapolation of the ratio of differential distributions to the ratio
of the total rates.  The three upper panels show that the ratios are
not a constant, and can change buy up to 20\% up to $p_T = 500$~GeV.
The relative uncertainties, separately for the scale and PDF variation
(MSTW2008 NLO set), are shown in the lower plots.  The scale
uncertainties reach a value of $\pm 2\%$ for the boson $p_T$ spectra,
$\pm 1\%$ for the top, and $\pm 3\%$ for the $p_T$ of the $t\bar{t}$
pair.  The PDF uncertainties remain well below the percent level
throughout.

These results imply that the relative shapes of the $p_T$ spectra  
can be controlled with a precision that remains consistent with
the overall goal of a percent-level extraction of the relative
rates. There is no doubt that future NNLO calculations of both
processes will improve this even further.
Very precise measurements of the shape of the $Z$ boson spectra in
$\ttz$ events using e.g. the very clean leptonic $Z$ decay will also
help confirming the accuracy of the predicted $p_T$ spectra
and reduce a possible left-over uncertainty.

\clearpage
\subsection{Rare production modes}
\label{sec:H_rareprod}

The first section of table \ref{tab:Higgsrare}~\cite{Torrielli:2014rqa}, obtained with the \texttt{MadGraph5\_aMC@NLO} code~\cite{Alwall:2014hca}, reports the rate for associated production of a SM Higgs boson with a single top. The cross section is in excess of 5 picobarns at 100 TeV, and displays a considerable increase with collider energy.

This remarkable growth, together with the sensitivity of this process to the sign of the top Yukawa coupling $y_t$ \cite{Farina:2012xp}, makes this reaction a golden channel for a precise measurement of the latter. It has been shown \cite{Chang:2014rfa} that already at the 14-TeV LHC it is possible to put loose bounds on the sign of $y_t$, mainly with a semileptonically decaying top quark, and in the $H\to b\bar b$ and $H\to\gamma\gamma$ decay channels. At 100 TeV the situation will improve considerably: the NLO cross section for the main irreducible background to $tH(\to\gamma\gamma)j$ production, namely $t\gamma\gamma j$ QCD production, has a growth $\rho$ comparable to that of the signal, hence the significance of the signal, in comparison with the LHC, is expected to scale at least with the square root of the number of events. Moreover, the sensitivity of the signal to $y_t$ is only slightly reduced at 100 TeV with respect to 8 TeV, as shown explicitly in the left panel of figure \ref{fig:Hassoc}.

The second part of table \ref{tab:Higgsrare} and the right panel of figure \ref{fig:Hassoc}~\cite{Torrielli:2014rqa} detail the cross section for a Higgs in association with a pair of gauge bosons (see also \cite{Baglio:2015eon} for a recent analysis). Rates for these channels are smaller than for single top, of the order of a few tens of femtobarns at 100 TeV, but still accessible. Theoretical systematics are typically below 10\%, and the rate growth with energy is mild, compatibly with the fact that these processes are $q\bar q$-driven.

These rare channels are interesting as they can add some power to constrain possible anomalous Higgs couplings to vector-boson (and fermion) pairs, which in turn has implications on the analysis of perturbative unitarity at high energy and strong links with the study of anomalous triple-vector-boson vertices \cite{Corbett:2013pja,Liu:2013vfu}. In particular the $pp\to HW^+W^-$ process, the one with the largest cross section in this category, has been shown \cite{Gabrielli:2013era} to be promising in this respect already at the high-luminosity LHC, and will considerably benefit from the rate increase of a factor of roughly forty at 100 TeV.

\begin{table}[h!]
\begin{center}
\begin{small}
\begin{tabular}{rl | l | l | c}
\hline\hline
 &Process& $\sigma_{\tiny{\mbox{NLO}}}$(8 TeV) [fb]& $\sigma_{\tiny{\mbox{NLO}}}$(100 TeV) [fb]\vspace{1mm}&$\rho$\\\hline\vspace{1mm}
$pp~~\to$ & $Htj$ & $2.07\cdot 10^1~{}^{+2\%}_{-1\%}~{}^{+2\%}_{-2\%}$ & $5.21\cdot 10^3~{}^{+3\%}_{-5\%}~{}^{+1\%}_{-1\%}$ & 252 \vspace{1mm}\\\hline\hline\vspace{1mm}
$pp~~\to$ & $HW^+W^-~(\mbox{4FS})$ & $4.62\cdot 10^0~{}^{+3\%}_{-2\%}~{}^{+2\%}_{-2\%}$ & $1.68\cdot 10^2~{}^{+5\%}_{-6\%}~{}^{+2\%}_{-1\%}$ & 36 \vspace{1mm}\\\hline\vspace{1mm}
$pp~~\to$ & $HZW^\pm$ & $2.17\cdot 10^0~{}^{+4\%}_{-4\%}~{}^{+2\%}_{-2\%}$ & $9.94\cdot 10^1~{}^{+6\%}_{-7\%}~{}^{+2\%}_{-1\%}$ & 46 \vspace{1mm}\\\hline\vspace{1mm}
$pp~~\to$ & $HW^\pm\gamma$ & $2.36\cdot 10^0~{}^{+3\%}_{-3\%}~{}^{+2\%}_{-2\%}$ & $7.75\cdot 10^1~{}^{+7\%}_{-8\%}~{}^{+2\%}_{-1\%}$ & 33 \vspace{1mm}\\\hline\vspace{1mm}
$pp~~\to$ & $HZ\gamma$ & $1.54\cdot 10^0~{}^{+3\%}_{-2\%}~{}^{+2\%}_{-2\%}$ & $4.29\cdot 10^1~{}^{+5\%}_{-7\%}~{}^{+2\%}_{-2\%}$ & 28 \vspace{1mm}\\\hline\vspace{1mm}
$pp~~\to$ & $HZZ$ & $1.10\cdot 10^0~{}^{+2\%}_{-2\%}~{}^{+2\%}_{-2\%}$ & $4.20\cdot 10^1~{}^{+4\%}_{-6\%}~{}^{+2\%}_{-1\%}$ & 38 \vspace{1mm}\\\hline\hline\vspace{1mm}
\end{tabular}
\end{small}
\end{center}
\caption{\label{tab:Higgsrare} Production of a Higgs boson at 8 and 100 TeV. The rightmost column reports the ratio $\rho$ of the 100-TeV to the 8-TeV cross sections \cite{Torrielli:2014rqa}. Theoretical uncertainties are due to scale and PDF variations, respectively. Processes $pp\to Htj$ does not feature any jet cuts.}
\end{table}

\begin{figure}[h!]
\begin{minipage}{0.49\textwidth}
\centering
\includegraphics[width=1\textwidth]{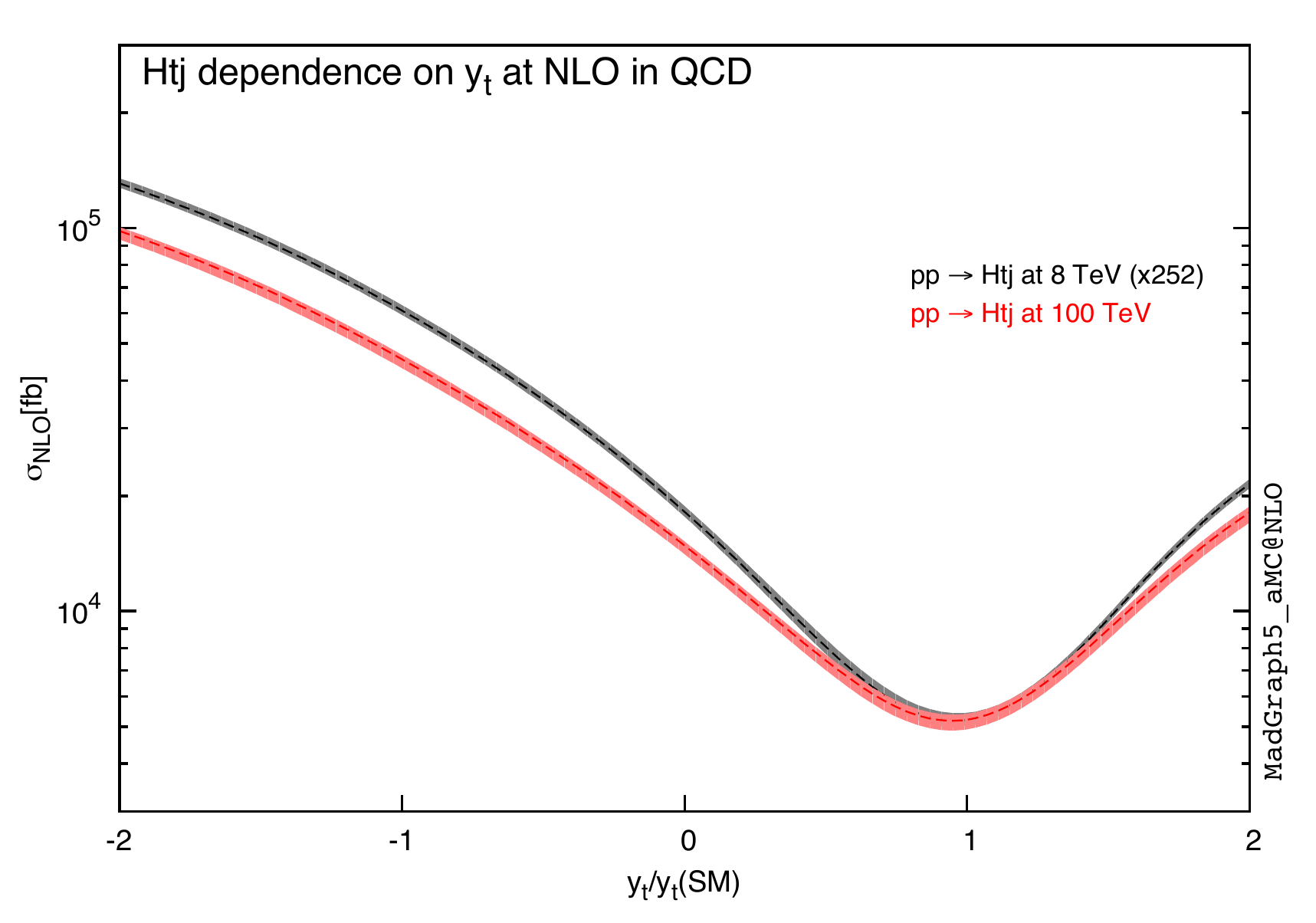}
\end{minipage}
\begin{minipage}{0.49\textwidth}
\centering
\includegraphics[width=1\textwidth]{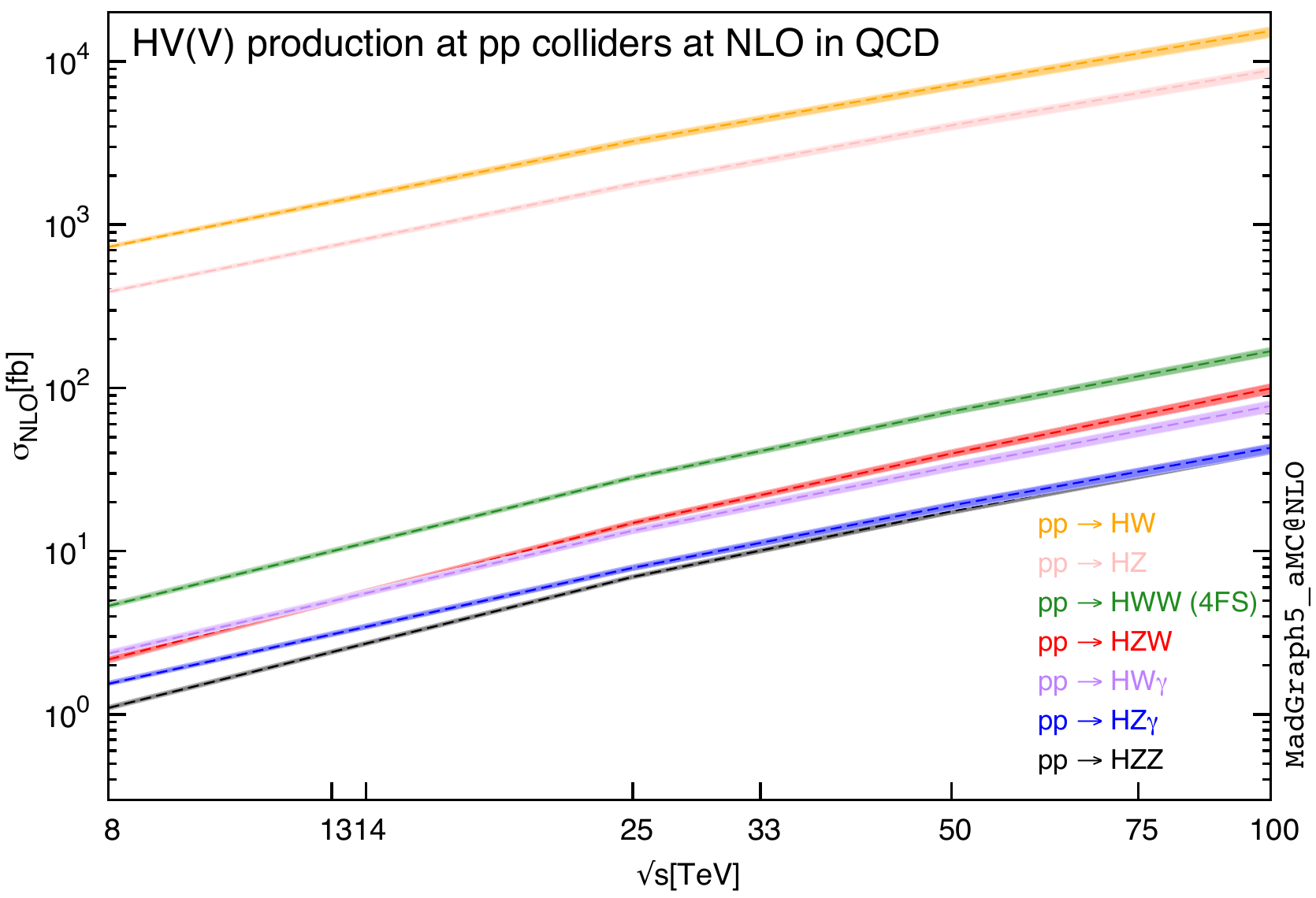}
\end{minipage}
\caption{\label{fig:Hassoc}Left panel: sensitivity of $pp\to Htj$ to $y_t$ at 8 and 100 TeV; right panel: cross section for associated production of a Higgs and up to two electroweak vector bosons \cite{Torrielli:2014rqa}.}
\end{figure}

\clearpage
\section{Prospects for measurements of SM Higgs properties}
\label{sec:H_prospects}
Table~\ref{tab:H_prospects_rates} shows the number of Higgs bosons
produced at 100~TeV with an integrated luminosity of 20~\iab. For
reference, we compare these rates to what was available at the end of
the LHC run~1, and what will be available at the end of the full HL-LHC
programme, namely 3~\iab\ at 14~TeV.

\begin{table}[h]
 \begin{center}
\begin{tabular}{l|c|c|c} 
  \hline
   & $N_{100}$ & $N_{100}/N_8$ & $N_{100}/N_{14} $ \\ \hline 
  $gg\to H$  & $16\times 10^9$ & $4\times 10^4$ & $110$
  \\
  VBF  & $1.6\times 10^9$ & $5\times 10^4$ & $120$
  \\
  $WH$  & $3.2\times 10^8$ & $2\times 10^4$ & $65$
  \\
  $ZH$  & $2.2\times 10^8$ & $3\times 10^4$ & $85$
  \\
  $t\bar{t}H$  & $7.6\times 10^8$ & $3\times 10^5$ & $420$
\\ \hline
\end{tabular}
\end{center}
\caption{Indicative total event rates at 100~TeV ($N_{100}$), and
  statistical increase with 
  respect to the statistics of the LHC run 1 ($N_{8}$) and the HL-LHC
  ($N_{14}$), for various prodution channels. We define here
  $N_{100}=\sigma_{100~\mathrm{TeV}} \times 20$~\iab, 
  $N_{8}=\sigma_{8~\mathrm{TeV}} \times 20$~\ifb,
  $N_{14}=\sigma_{14~\mathrm{TeV}} \times 3$~\iab.    } 
\label{tab:H_prospects_rates}
\end{table}

Naive scaling leads to a potential for improvements in the statistical
precision in the range of few hundreds w.r.t to run~1, and of order
10-20 w.r.t. HL-LHC. As is well known, the HL-LHC itself will already
be systematics dominated for several measurements. But with such a
huge increase in rate and, as we shall see, in kinematic reach, one
can envisage new approaches to both precision measurements and to the
exploration of new phenomena in the production dynamics. Furthermore,
these rates will push the search for rare or forbidden Higgs decays
well beyond the LHC reach.

\begin{figure}
  \begin{center}
    \includegraphics[width=0.50\textwidth]{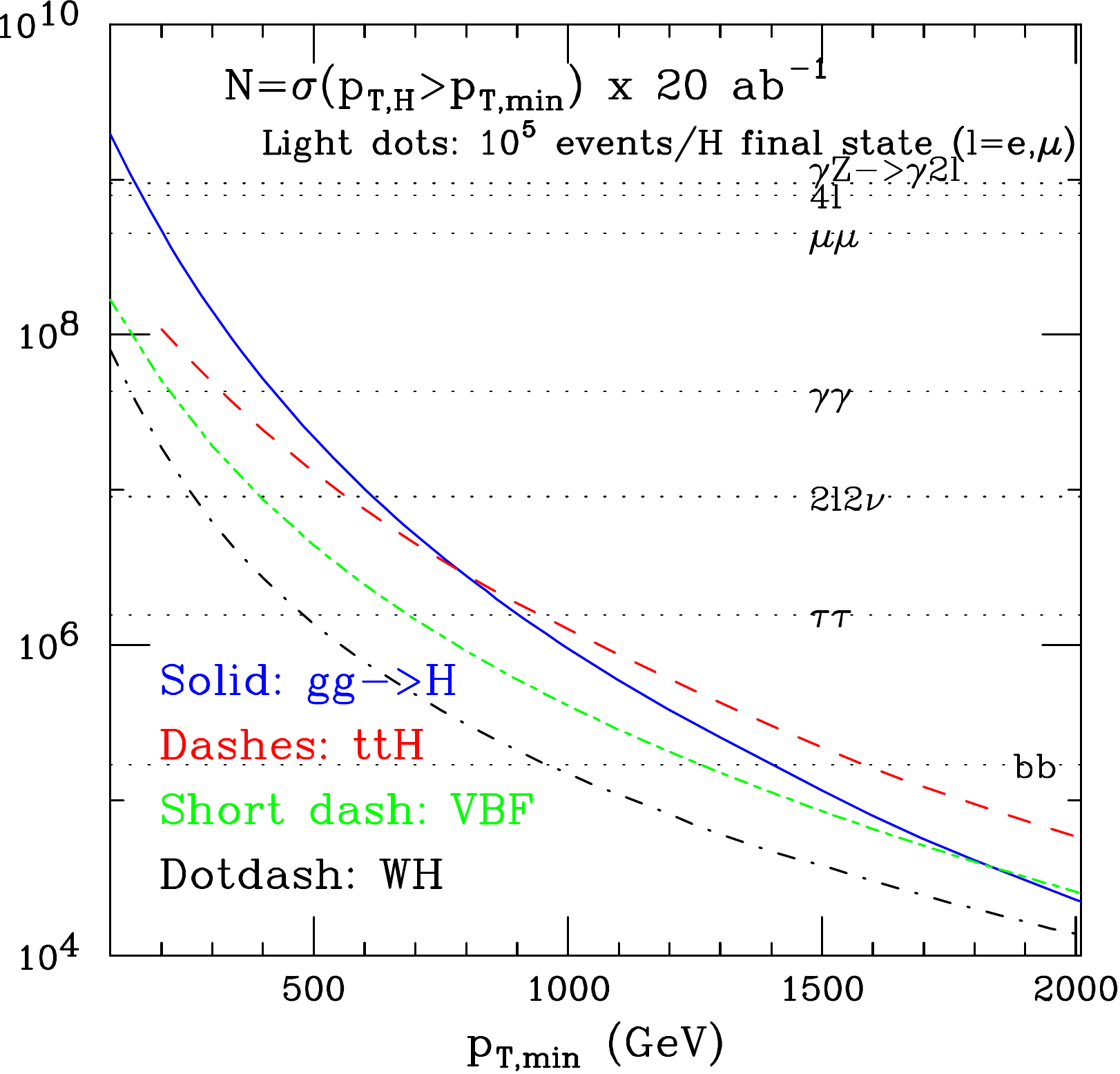}
    \hspace*{-0.02\textwidth}
    \includegraphics[width=0.49\textwidth]{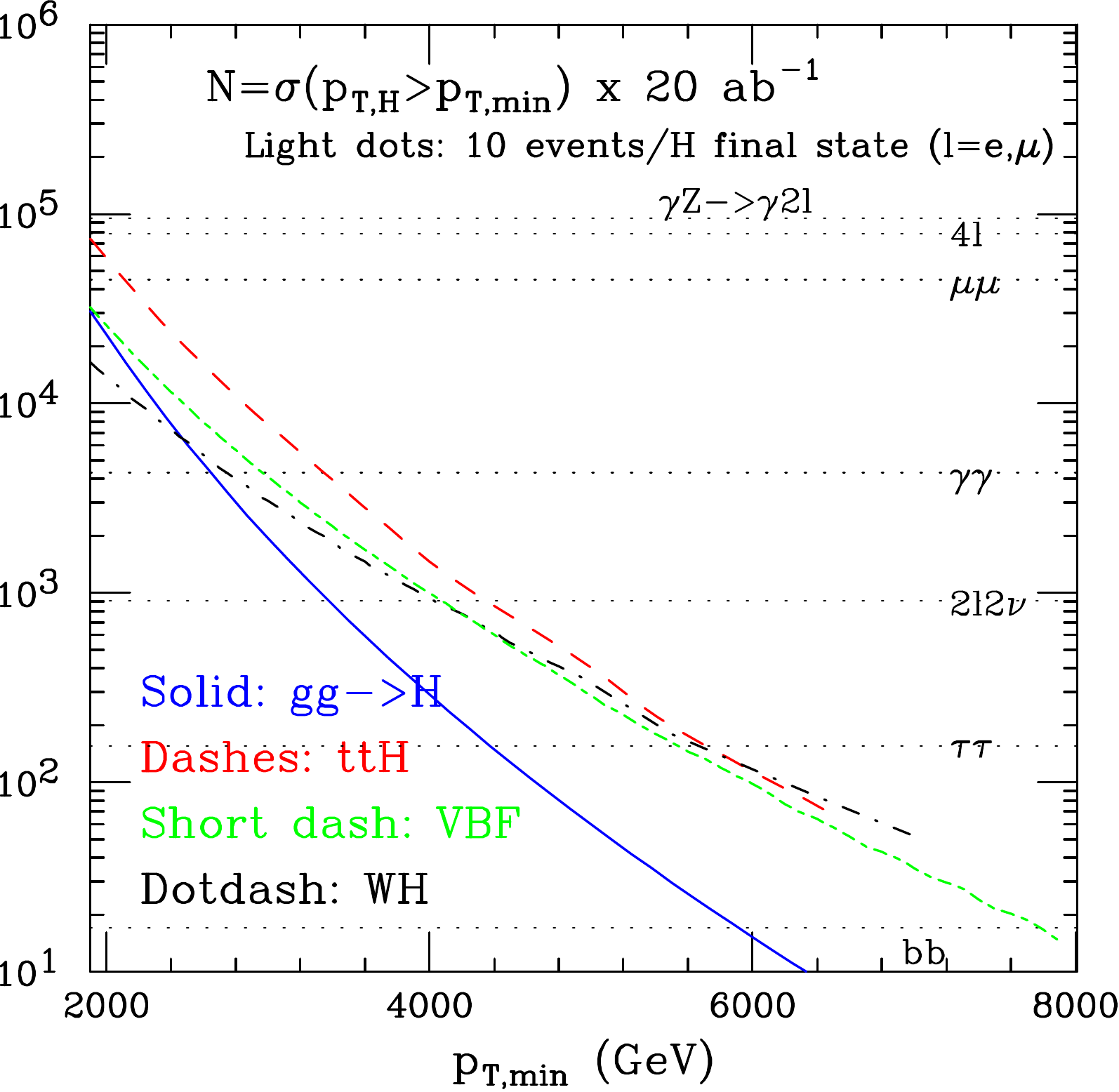}
  \end{center}
  \caption{Integrated Higgs transverse momentum rates, for various
  production channels, with 20~\iab. The light-dotted horizontal lines
  in the left (right) panel
  correspond to the production of $10^5$ (10) events with a Higgs
  decay to the indicated final states.}
\label{fig:allH_pt}
\end{figure}

The most remarkable feature of Higgs production at 100~TeV is not just
the rate increase w.r.t. the LHC, but the extreme kinematical range
over which the Higgs bosons are distributed. Figure~\ref{fig:allH_pt}
shows the integrated $p_H$ spectra for the dominant production
processes, and prompts several important remarks.

To start with, we highlight the remarkable statistics. Horizontal
light-dotted lines in the figures show the $p_T(H)$ values
corresponding to samples of $10^5$ (left) and $10$ (right) events, for
various final states. The former statistics are possibly suitable for
percent-level measurements, the latter indicate the most extreme $p_T$
values at which measurements of Higgs production dynamics are in
principle possible. This could have relevance, for example, in the
context of searches for new physics, where Higgses could either be
part of a signal, or a background.

Secondly, we note that the hierarchy of rates among the different
processes, shown for example in Table~\ref{tab:H_prospects_rates}, is
only valid for the bulk of the production. As $p_T(H)$ grows above
$\sim 500$~GeV, $t\bar{t}H$ emerges as the most abundant source of
large-$p_T$ Higgses. Moving to yet larger $p_T$, even VBF and
eventually associated $VH$ production come to be more important than
$gg\to H$. The key reason for this is the form-factor-like suppression
of the $ggH$ vertex at large virtuality, when the finite-$m_{top}$
effects are properly accounted for.

This observation has important implications for the measurements. For
example, while dedicated cuts are needed to extract the VBF
Higgs-production signal from the inclusive $gg\to H+X$ Higgs sample,
at large $p_T$ the dominant source of irreducible background is top
production. The separation of $t\bar{t}H$ from VBF when $p_T(H)>1$~TeV
can rely on kinematic and event-shape discriminators, which are likely
more powerful and efficient than the usual VBF cuts. This may also
have important implications on the detector, since optimal acceptance
to VBF cuts requires instrumentation in the very difficult forward
$\eta$ region.

Large Higgs $p_T$ values, furthermore, make it possible to consider
using the otherwise disfavoured $H\to b\bar{b}$ decay mode, thanks to
the higher and higher discrimination power of jet-structure
techniques. The ability to use this high-BR decay, extends
considerably the accessible $p_T(H)$ range. Lower-BR final states,
such as $H\to \gamma\gamma$, $ZZ^*$, $Z\gamma$  or $\mu^+\mu^-$, remain
nevertheless usable for precision measurements (i.e. event rates in
excess of $10^4$), over a broad range of $p_T$.

In this Section we shall elaborate in some more detail on these
ideas. One could organize the discussion according to final state
(e.g. addressing the issue of how to best measure a given BR from a
global fit of several production channels), or according to production
channel (e.g. to compare different decays in the same channel, in
order to remove possible production systematics from the precise
determination of BR ratios). We shall adopt a mixed approach and, as
emphasized above, we shall not analyze in quantitative terms all
sources of theoretical and experimental uncertainties. Several of the
studies shown here were done including only the leading relevant order
of pertubation theory. We include the dominant sources of backgrounds,
and make crude, and typically optimistic, assumptions about the
relevant detector performance issues. The key purpose is to show what
is in principle possible, and postpone more rigorous studies to future
work.

\subsection{Higgs acceptance}
We present here some reference results to document the detector
acceptance for Higgs decay final states, as a function of the
pseudorapidity coverage and of the minimum $p_T$ thresholds. These
results can orient the choices in the optimal detector layout.

Figure~\ref{fig:H_2bodyacc} shows the detector acceptance, for
different $p_T$ thresholds, for 2-body Higgs decays (e.g. $H\to
b\bar{b}$, $H\to\gamma\gamma$, $H\to \mu^+\mu^-$).  Each box
corresponds to Higgs bosons produced in $gg$ fusion, at various fixed
values of the Higgs transverse momentum ($p_T(H)=0$, 50, 100, 200, 500
and 1000~GeV). For each $p_T(H)$ value we consider a minimum $p_T$ cut
($p_{T,min}$) for the two decay products (0, 20, 30 and 40~GeV), and
show the acceptance as a function of the largest $|\eta|$
($\eta_{max}$). The acceptance is defined with respect to the total
sample of events produced at the given value of $p_T(H)$.

The largest sensitivity to $p_{T,min}$ is present for values of $p_T(H)$
around $50-100\,$GeV, since the boost in this range will suppress the
acceptace for the decay particle produced in the backward direction
with respect to the Higgs direction. For the largest values of
$p_T(H)$, the acceptance is much less sensitive to $p_{T,min}$, and is well
optimized in the central region $|\eta|<2.5$. 

\begin{figure}[h]
  \begin{center}
    \includegraphics[height=0.45\textwidth]{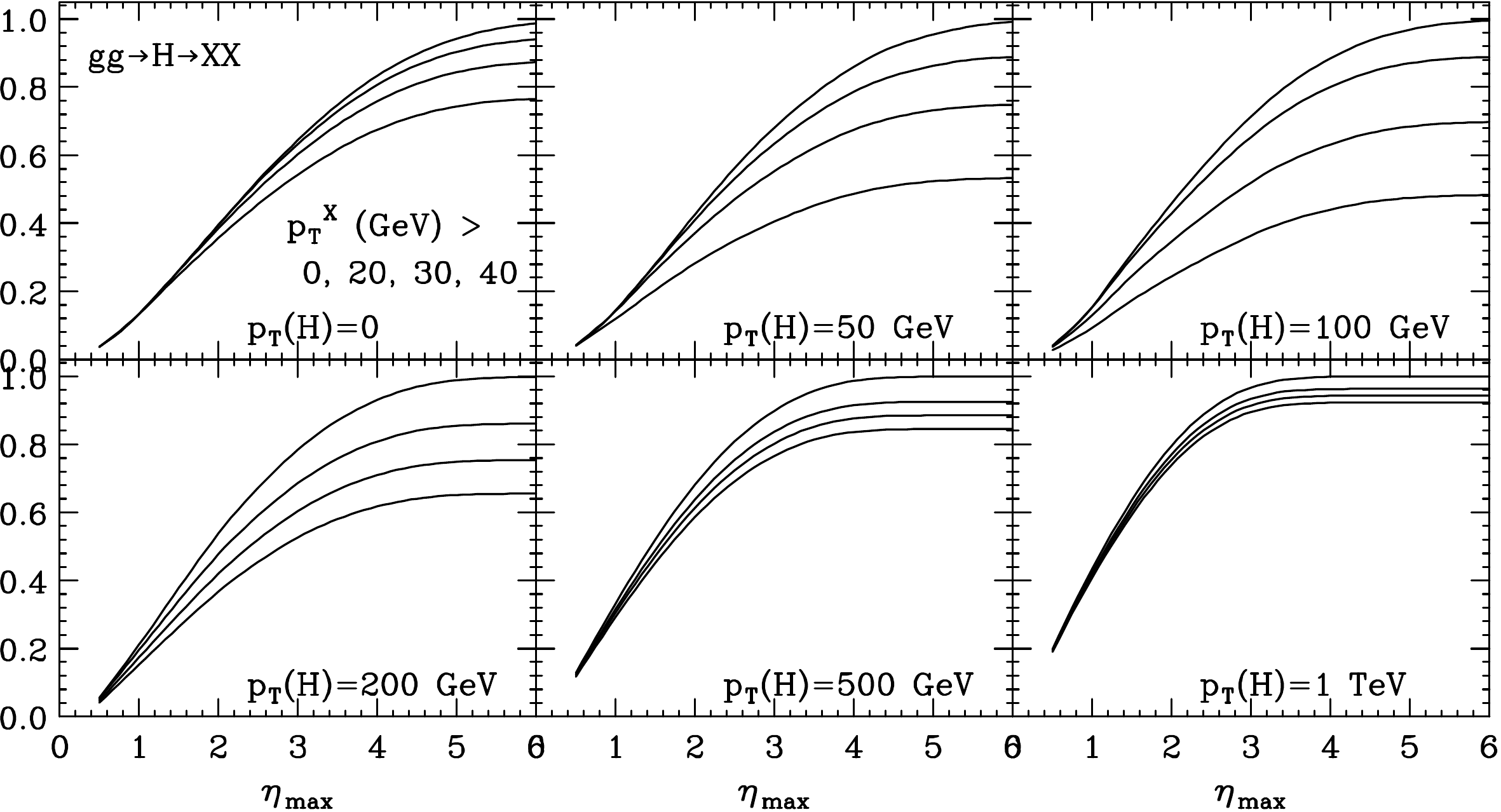}
  \end{center}
  \caption{Detector acceptance, as a function of the maximal
    pseudorapidity coverage $\eta_{max}$, for the 2-body decay of
    Higgs bosons produced in $gg$ fusion at various $p_T$ values (this
    applies, e.g., to $H\to b\bar{b}$, $\gamma\gamma$ or $\mu^+\mu^-$). The
    different lines refer to different thresholds in the minimum $p_T$
  of the decay particles. }
\label{fig:H_2bodyacc}
\end{figure}
\begin{figure}
  \begin{center}
    \includegraphics[height=0.45\textwidth]{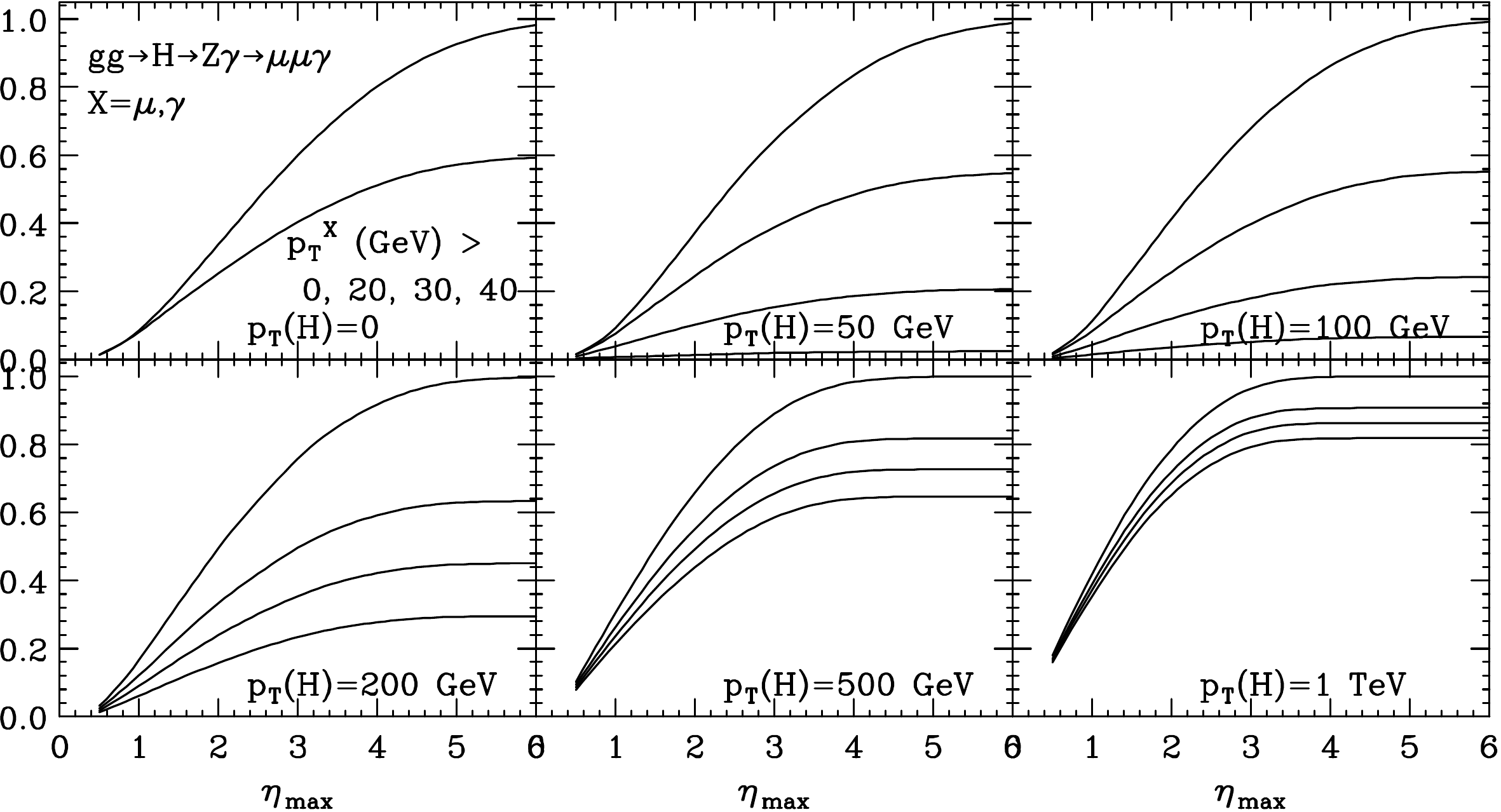}
  \end{center}
  \caption{Same as Fig.~\ref{fig:H_2bodyacc}, for the decay
    $H\to Z\gamma\to\mu^+\mu^-\gamma$. The $p_T$ and $\eta_{max}$ cuts
  apply to both muons and to the photon.}
\label{fig:H_3bodyacc}
\end{figure}

Figure~\ref{fig:H_3bodyacc} shows similar results, for the 3-body
decay $H\to Z\gamma \to \mu^+\mu^- \gamma$. The $p_{T,min}$ and $\eta$
cuts here are applied to all decay products. At $p_T(H)=0$ there is no
acceptance for $p_{T,min}\ge 30$~GeV, since the photon energy in the
$H$ rest frame is of order 30~GeV (up to a negligible effect due to
the finite $Z$ width). For these decays, the overall loss in acceptance due to
the $p_T$ threshold is always significant, as shown by the
large-$\eta$ limit of the distributions. 

The strong $p_{T,min}$ dependence is emphasized even more in the
4-body decays, such as $H\to WW^*\to 2\ell2\nu$ and $H\to ZZ^*\to
4\ell $, whose acceptance plots are shown in Figs.~\ref{fig:H_wwacc}
and~\ref{fig:H_zzacc}. For 4-lepton decays, we consider also the
acceptance of asymmetric cuts, such as those used at the LHC. They
appear as absolutely necessary, at least for $p_T(H)$ values below
$\sim 500$~GeV, since the decay kinematics enhances the spectral
asymmetry, and a uniform cut for all leptons at 20~GeV would lead to
an acceptance at the percent level.

For example, ATLAS~\cite{Aad:2014tca} requires the three leading
leptons to have $p_T$ larger than 10, 15 and 20~GeV, and the fourth
lepton to exceed 6 (if muon) or 7 (if
electron)~GeV. CMS~\cite{Khachatryan:2015yvw} requires the two leading
leptons to have $p_T$ larger than 10 and 20~GeV, and the others to
exceed 5 (if muon) or 7 (if electron)~GeV. We consider here similar
cuts, namely the thresholds $(5,10,15,20)$ or $(10,10,15,20)$. We note
that, for $p_T(H)$ below few hundred GeV, the difference between 5 and
10~GeV for the softest lepton is almost a factor of 2 in
acceptance. We also notice that the acceptance of the fully symmetric
cut $(10,10,10,10)$ is almost identical to that of
$(10,10,15,20)$. This is a result of the decay kinematics. We
stress that for these processes the low-$p_T$ acceptance is far more
important than rapidity coverage, and must be preserved.

In case of $WW^* \to 2\ell 2\nu$ decays, the fiducial regions selected
by ATLAS~\cite{ATLAS:2014aga} and CMS~\cite{CMS:2015obs} require the
thresholds of 10~GeV for the softer lepton, and 20 (CMS) or 22 (ATLAS)
for the leading one. For 100~TeV, we show here the options $(10,10)$,
$(20,20)$ and $(10,20)$. 

\begin{figure}
  \begin{center}
    \includegraphics[height=0.45\textwidth]{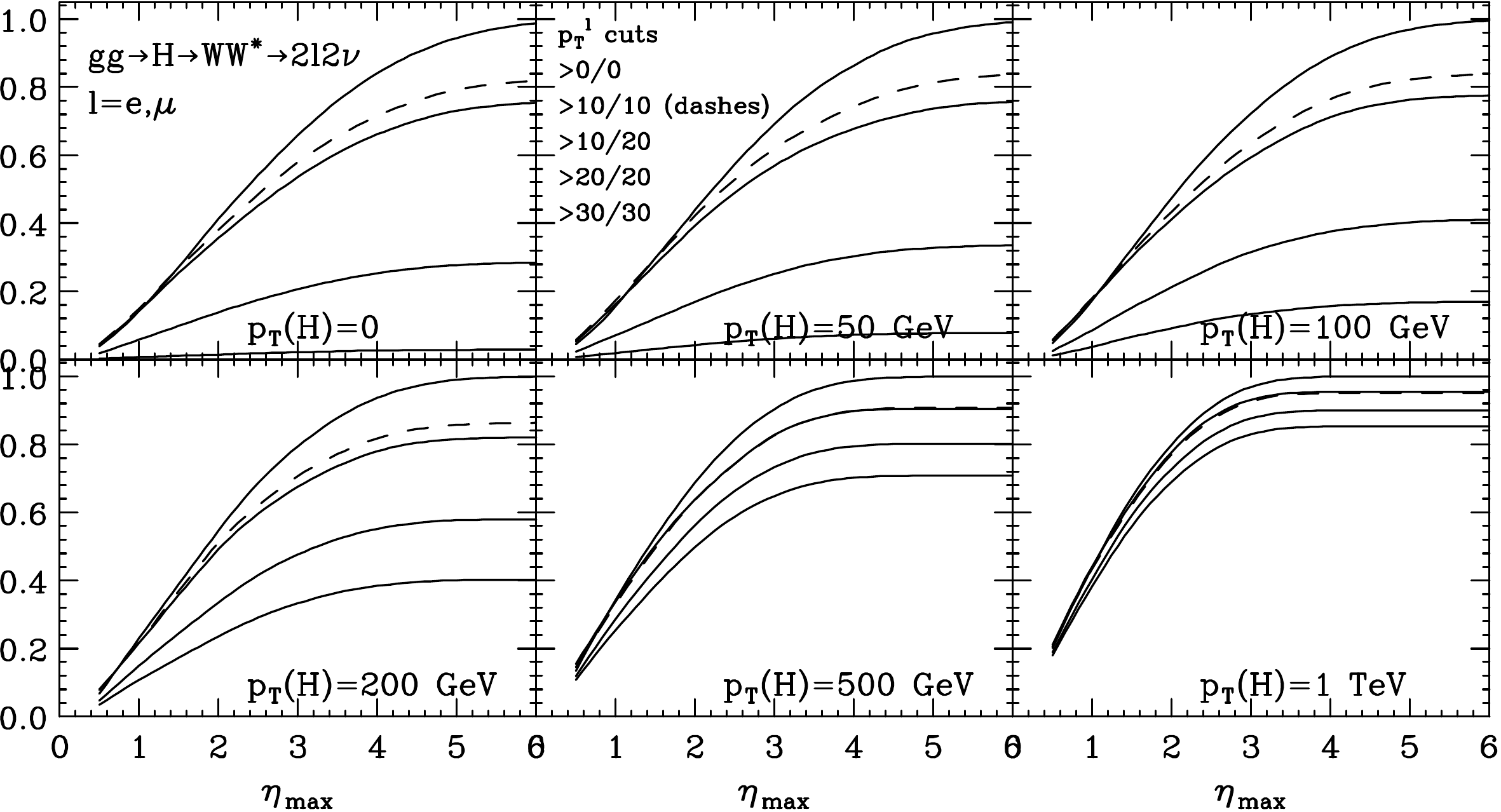}
  \end{center}
  \caption{Same as Fig.~\ref{fig:H_2bodyacc}, for the decay
    $H\to WW^* \to\ell^+\ell^- 2\nu$. The $p_T^\ell$ cuts shown in the
    second inset apply to the
  softest and the hardest of the two charged leptons.}
\label{fig:H_wwacc}
\end{figure}
\begin{figure}
  \begin{center}
    \includegraphics[height=0.45\textwidth]{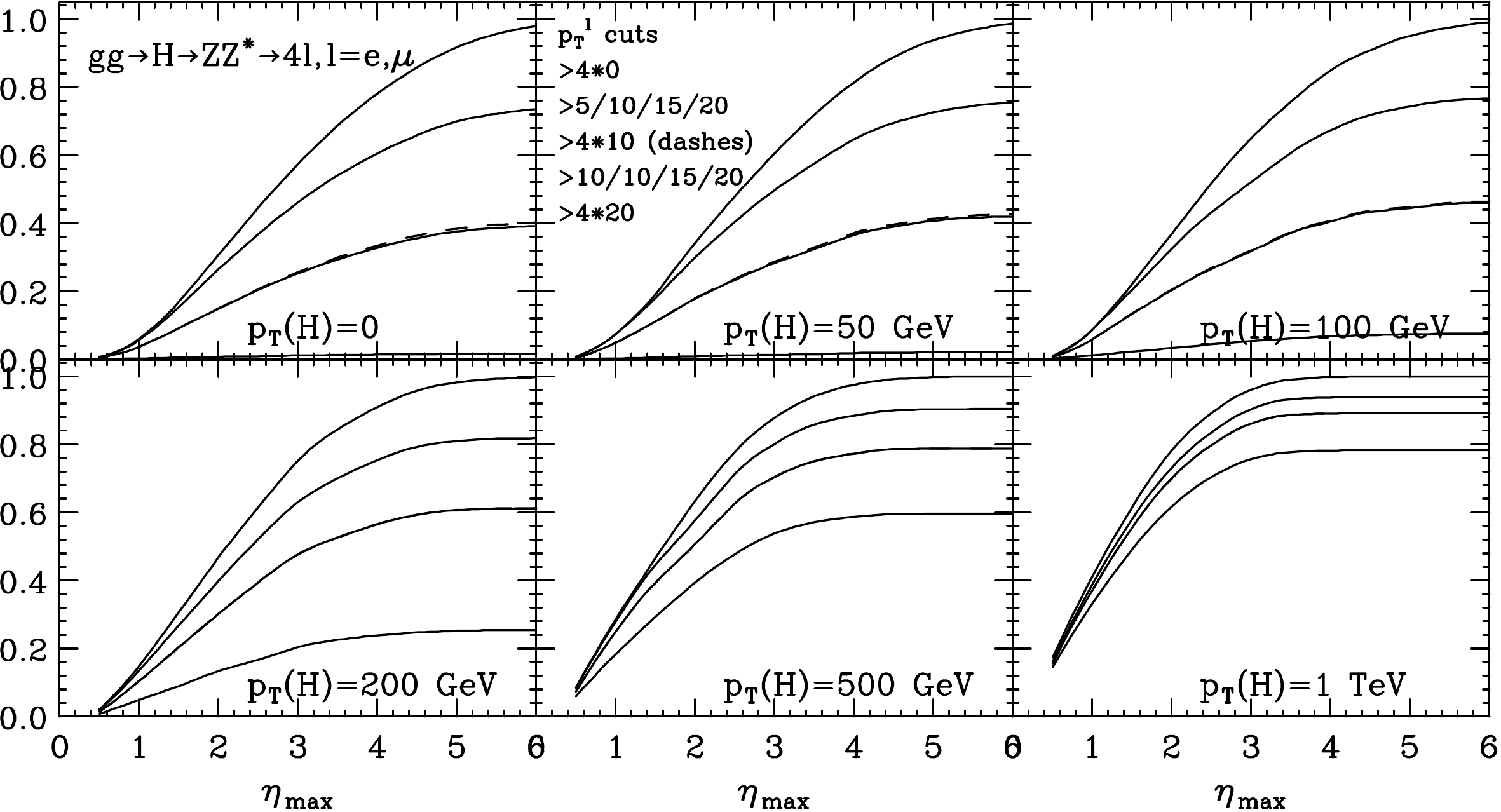}
  \end{center}
  \caption{Same as Fig.~\ref{fig:H_2bodyacc}, for the decay
    $H\to ZZ*\to4$~charged leptons. The $p_T^\ell$ cuts shown in the
    second inset apply to the
  softest through the hardest of the four charged leptons.}
\label{fig:H_zzacc}
\end{figure}

The results for some reference $p_T$ and $|\eta|$ thresholds are
collected in Table~\ref{tab:H_acc}.
\begin{table}
\begin{center}
\begin{tabular}{l | r | rrr || l | r | rrr } 
\hline
$H$ decay & $p_T(H)$ & $|\eta|<2.5$ & $<4$ & $<5$ &
$H$ decay & $p_T(H)$ & $|\eta|<2.5$ & $<4$ & $<5$ 
\\ \hline
2-body
 & 0 & 0.5 & 0.76 & 0.84 & 
$Z\gamma \to \ell\ell\gamma$
 & 0 & 0.33 & 0.51 & 0.57 
\\
$p_T^{min}=30$
 & 50 & 0.48 & 0.68 & 0.74 & 
$p_T^{min}=20$
 & 50 & 0.32 & 0.49 & 0.53
\\
 & 500 & 0.72 & 0.87 & 0.88 & 
 & 500 & 0.66 & 0.81 & 0.82
\\
\hline
$WW^*\to 2\ell2\nu$
 & 0 & 0.17 & 0.25 & 0.28 & 
$ZZ^*\to 4\ell$
 & 0 & 0.20 & 0.33 & 0.38  
\\
$p_T^{min}=20$
 & 50 & 0.21 & 0.30 & 0.33 & 
$p_T^{min}=10$
 & 50 & 0.23 & 0.36 & 0.40
\\
 & 500 & 0.66 & 0.79 & 0.80 & 
 & 500 & 0.63 & 0.77 & 0.79  
\\
\hline 
\end{tabular}
\end{center}
  \caption{Acceptances for various Higgs decay modes, in $gg\to H+X$
    production, as function of Higgs $p_T$. All final state products
    (except the nuetrinos in the $WW^*$ mode) are required to have
    $p_T>p_{T,min}$. }
\label{tab:H_acc}
\end{table}

\subsection{Small-BR $H$ final states at intermediate $p_T$}
\label{sec:H_prospects_H}
We consider here $H$ decays with BRs in the range of
$10^{-3}-10^{-4}$, such as $H\to\gamma\gamma$, $H\to\gamma Z$,
$H\to4\ell$ and $\mu^+\mu^-$. At a fully inclusive level, these events
are produced by the millions; a thorough analysis of the potential for
precise measurements from these large samples requires a detailed
understanding of the experimental environment, starting from the
consideration of the impact of hundreds, if not thousands, of pileup
events. This is work for future studies.

We discuss here instead the possible interest to study these final
states in decays of Higgs bosons produced with $p_T$ values of a few
100~GeV, where rates are still large, but $S/B$ ratios are typically
better than for the inclusive samples, and the experimental
environment is possibly easier (e.g. production is more central than
for the fully inclusive Higgs sample, and the higher $p_T$'s can
improve the reconstruction of the primary vertex and the resolution of
multiple pileup events). A possible target of such studies is a very
precise (percent level of better) measurement of the relative decay
BRs: the production ratio between different final states will in fact
remove several of the dominant systematics intrinsic in the absolute
rate measurements, such as the integrated luminosity or the theoretical
production rate uncertainty.

\subsubsection{$H\to \gamma \gamma$}
\label{sss.htogammagamma}
Figure~\ref{fig:H_dipho} (left plot) shows the $p_T$ spectrum of diphotons from
$H$ decays ($BR=2.3\times 10^{-3}$),
and from the dominant irreducible background, namely QCD
$\gamma\gamma$ production (for a discussion of $pp\to\gamma\gamma$,
see the Volume ``Standard Model physics at 100 TeV''  of this report).
The QCD contribution is constrained by an invariant mass cut,
$|m(\gamma\gamma)-125~\mathrm{GeV}\vert < 4$~GeV. This is rather
conservative even by today standards, where current analyses point at
resolutions in this channel of about 1-2~GeV. But the energy
resolution will degrade at the values of photon
energies considered in the regime of large $p_T(H)$, so we take 4~GeV
as an indicative benchmark. The size of the background, for
a reasonable range of resolution, scales linearly.

The background considered in this plot includes all sources
($q\bar{q}$, $qg$ and $gg$ initial states), and is subject to an
isolation constraint, which plays however a negligible role, since the
diphoton pair at large $p_T$ at this order of perturbation theory
mostly recoils against the partons. We note that the $S/B$ ratio is of
$O(1)$ in this region, much larger than for the low $p_T(H)$ sample,
where it drops well below 1/10.  The statistical precision, in
presence of this background, remains below 1\% up to $p_T(H) \sim
600$~GeV.

\begin{figure}
  \begin{center}
    \includegraphics[height=0.45\textwidth]{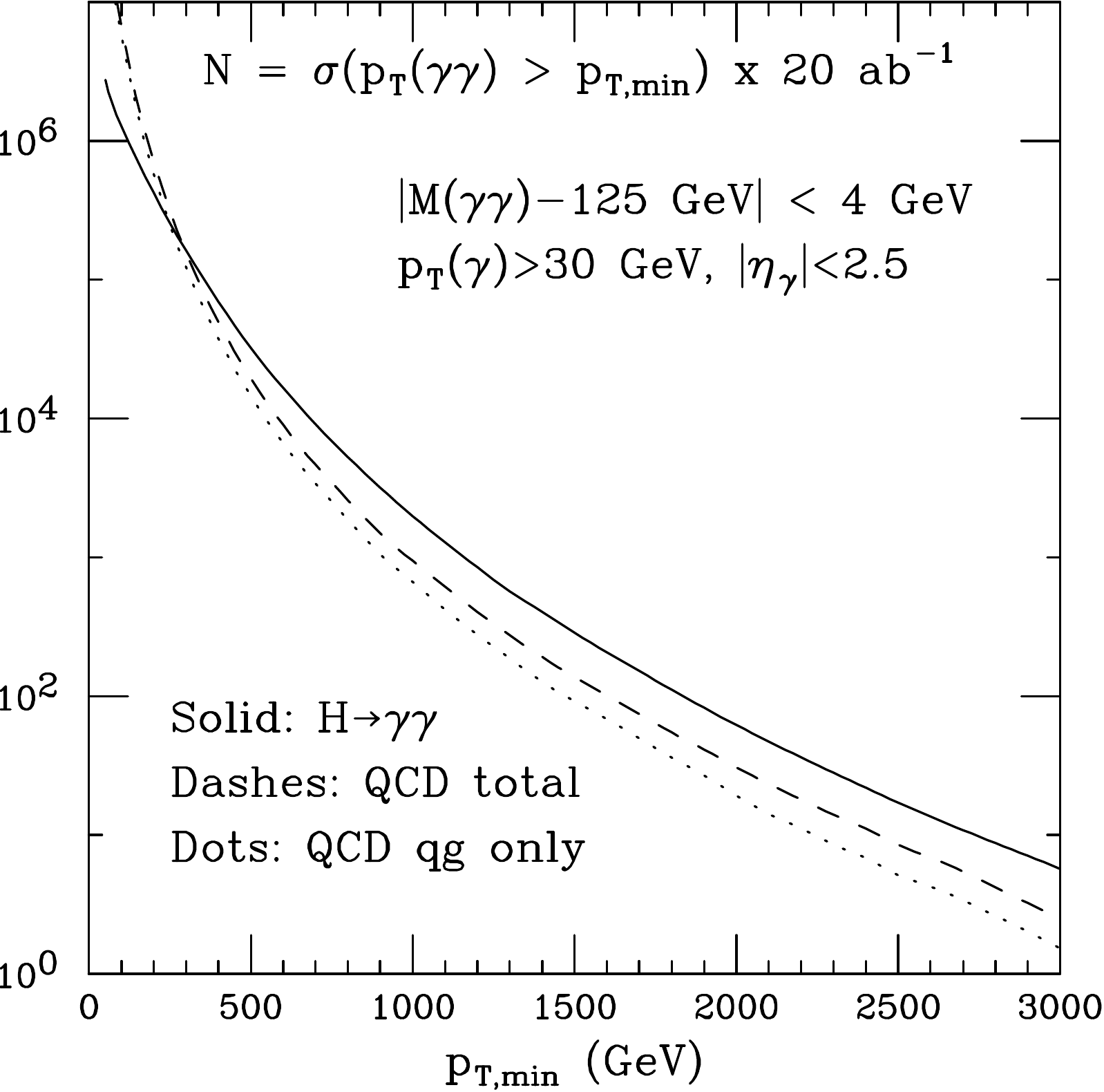}
    \hfill 
    \includegraphics[height=0.45\textwidth]{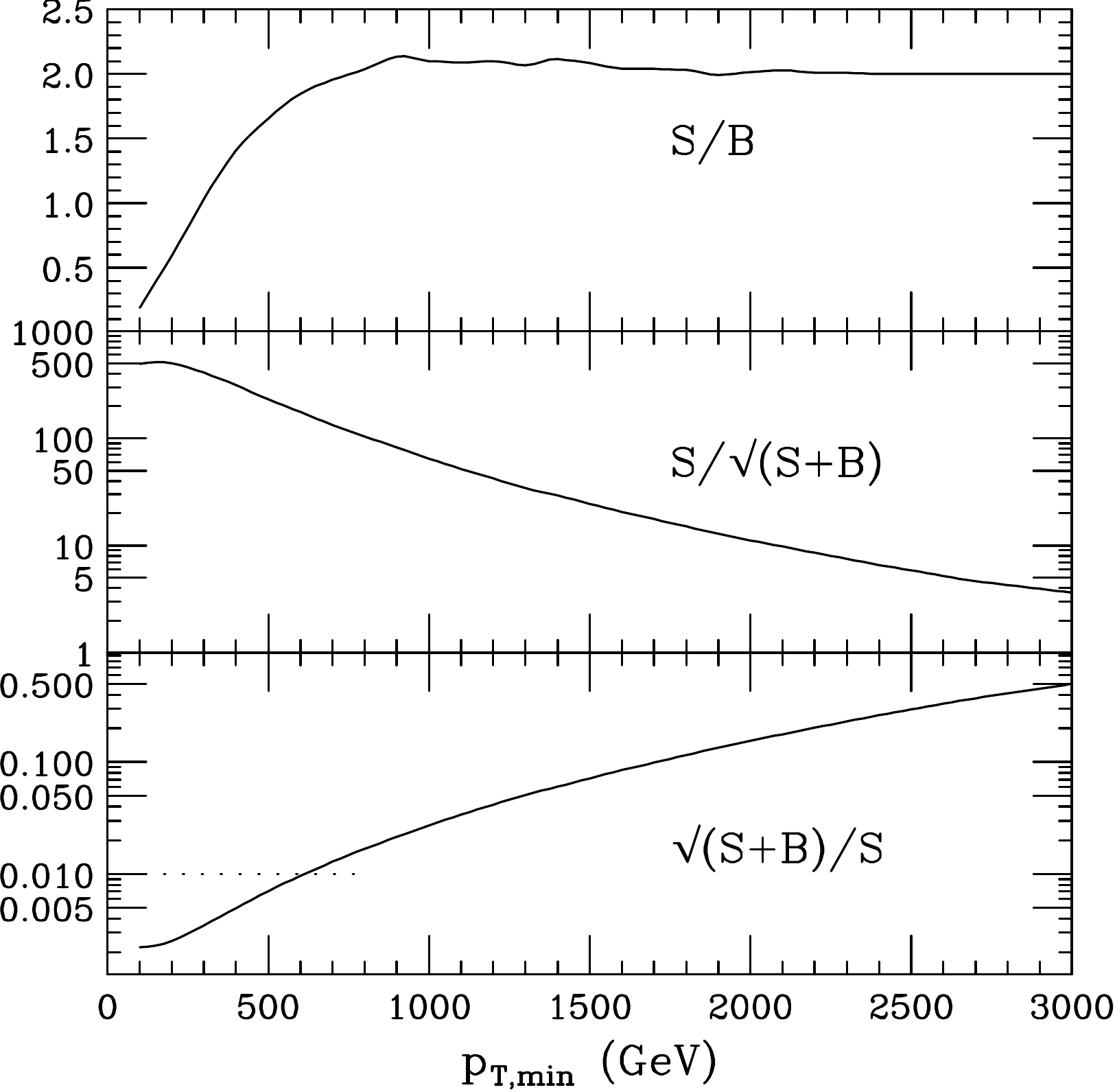}
  \end{center}
  \caption{Left: Integrated transverse momentum rates (20~\iab) for a
    photon pair with mass close to the Higgs mass: signal and QCD
    background. Right: $S/B$, significance of the signal, and potential 
    statistical accuracy of the sample.}
\label{fig:H_dipho}
\end{figure}

\subsubsection{$H\to \mu^+\mu^-$}
Figure~\ref{fig:H_mumu} (left plot) shows the $p_T$ spectrum of
dimuons from $H$ decays ($BR=2.2\times 10^{-4}$), and from the leading
irreducible background, namely Drell-Yan (DY) $\mu^+\mu^-$ production,
dominated by the tail of the $Z^*/\gamma$ distribution (see e.g. the
ATLAS~\cite{Aad:2014xva} and CMS~\cite{Khachatryan:2014aep} analyses).

The DY contribution is constrained by an invariant mass cut,
$|m(\gamma\gamma)-125~\mathrm{GeV}\vert < 1$~GeV. This is better than
the resolution of today's LHC experiments: the signal full width at
half maximum estimated by CMS for events with one central muon, for
example, varies in the range 4-5~GeV~\cite{Khachatryan:2014aep}), but
1~GeV is consistent with the improvement in the muon $p_T$ resolution
by a factor of $O(5)$, projected for the 100~TeV detectors.

The DY background includes $q\bar{q}$ and $qg$ initial
states. Contrary to the $\gamma\gamma$ decay, the $S/B$ for dimuons
deteriorates at larger $p_T(H)$, but still allows for a precision in
the rate measurement better than 2\% for $p_T(H)$ up to $\sim
200$~GeV. This could allow for a 1\% determination of the muon
Yukawa coupling, $y_\mu$,  relative to the $H\gamma\gamma$ coupling. 

\begin{figure}
  \begin{center}
    \includegraphics[height=0.42\textwidth]{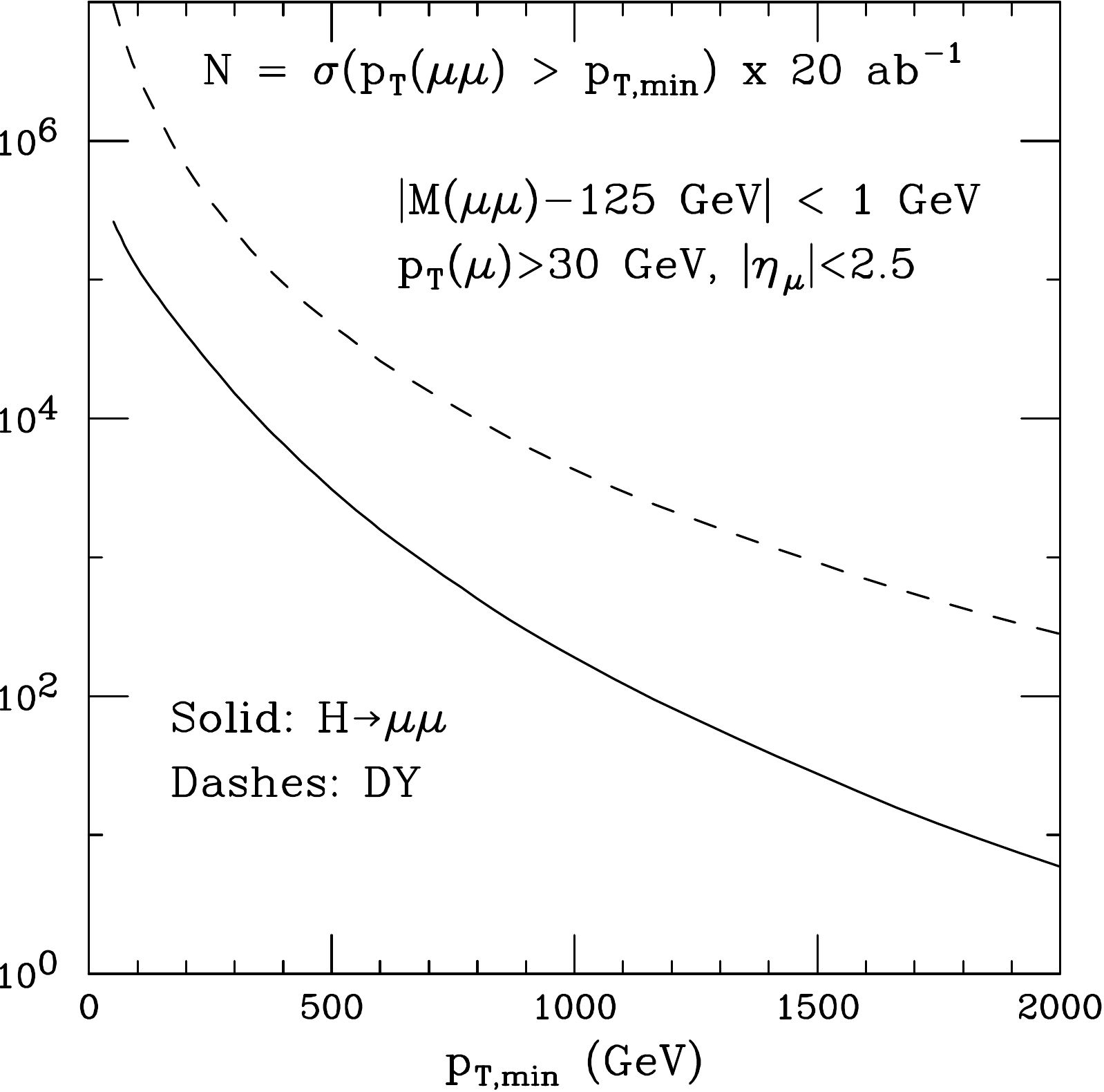}
    \hfill 
    \includegraphics[height=0.42\textwidth]{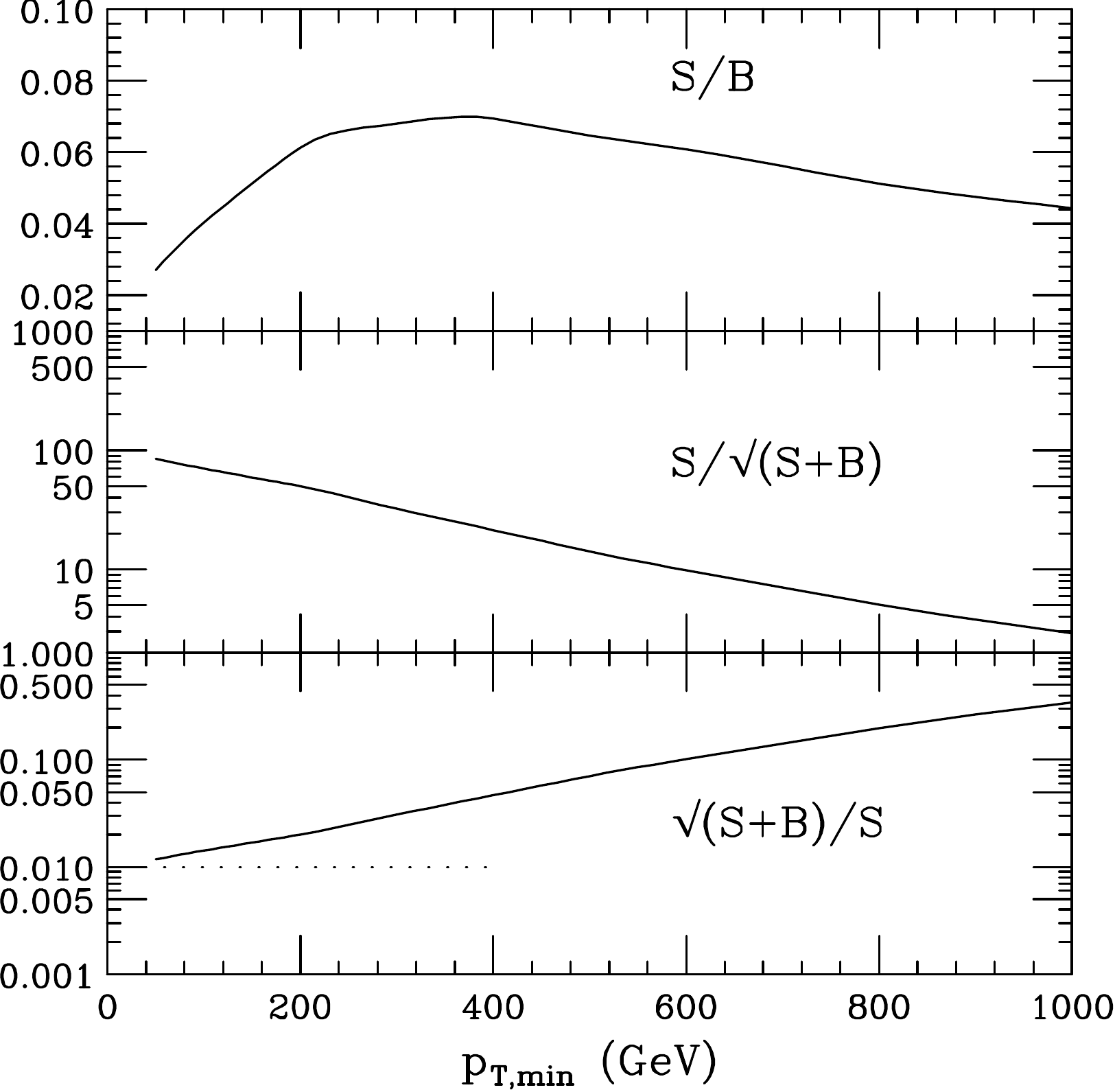}
  \end{center}
  \caption{Left: Integrated transverse momentum rates (20~\iab) for a
    muon pair with mass close to the Higgs mass: signal and DY
    background. Right: $S/B$, significance of the signal, and potential 
    statistical accuracy of the sample.}
\label{fig:H_mumu}
\end{figure}

\subsubsection{$H\to ZZ^*$ and $H\to Z\gamma$ }
We consider here
$H\to ZZ^*$ and
$H\to Z\gamma$, with leptonic decays
of the $Z$ boson to electron or muon pairs
($BR=1.3\times 10^{-4}$ and $BR=1.1\times 10^{-4}$, respectively).
The rates for signals and leading irreducible backgrounds are given in
Figs.~\ref{fig:H_4l} and~\ref{fig:H_Zgam}.

We considered for these plots the following acceptance cuts:
\begin{itemize}
  \item $H\to ZZ^* \to 4 \ell$: $p_T(\ell)>10$~GeV, $\vert \eta(\ell)\vert<2.5$
  \item $H\to Z\gamma \to 2 \ell\gamma$: $p_T(\ell,\gamma)>20$~GeV,
    $\vert \eta(\ell,\gamma)\vert<2.5$
\end{itemize}
We notice that, as shown in the acceptance plots of
Fig.~\ref{fig:H_zzacc}, at large $p_T(H)$
the cut $p_T(\ell)>10$~GeV for all 4 leptons has
an acceptance almost identical to that of the asymmetric cut
10/10/15/20~GeV. With reference to that figure, we also
point out that increasing the $\eta$ range and
reducing the threshold for the $p_T$ of the softest lepton, would each
increase the signal rate by a factor of 2.
  
We assume here once again 4~GeV as mass resolution
for both the 3- and 4-body final states.
For the 4-lepton final state, the $S/B$ ratio was already larger than
1 in the 8~TeV run of the LHC; due to greater increase in the gluon PDF
relative to the quark one, the QCD background at 100 TeV becomes
negligible. A 1\% determination of the rate is statistically possible
for $p_T(H)\lsim 300$ GeV. Likewise, the $S/B$ ratio for
$Z\gamma$ improves significantly as $p_T(H)$ is increased, and becomes
larger than 0.5 above $\sim 300\,$GeV.
In this region the statistical precision is
better than 2\%, allowing for a percent-level measurement of the
$HZ\gamma$ coupling relative to $H\gamma\gamma$.  

\begin{figure}
\hspace{-0.05\textwidth}
  \begin{center}
    \includegraphics[width=0.42\textwidth]{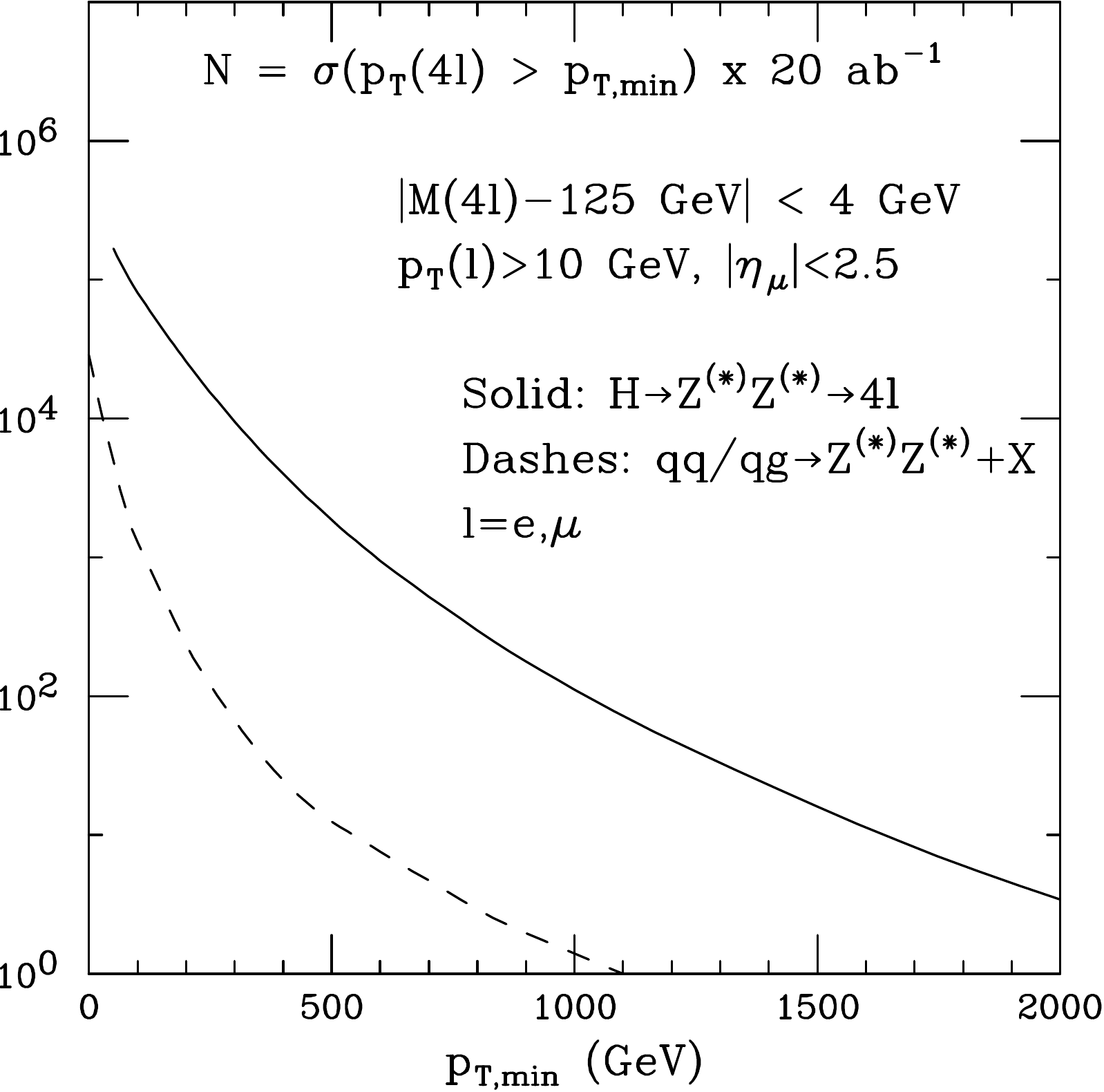}
    \hfill 
    \includegraphics[width=0.47\textwidth]{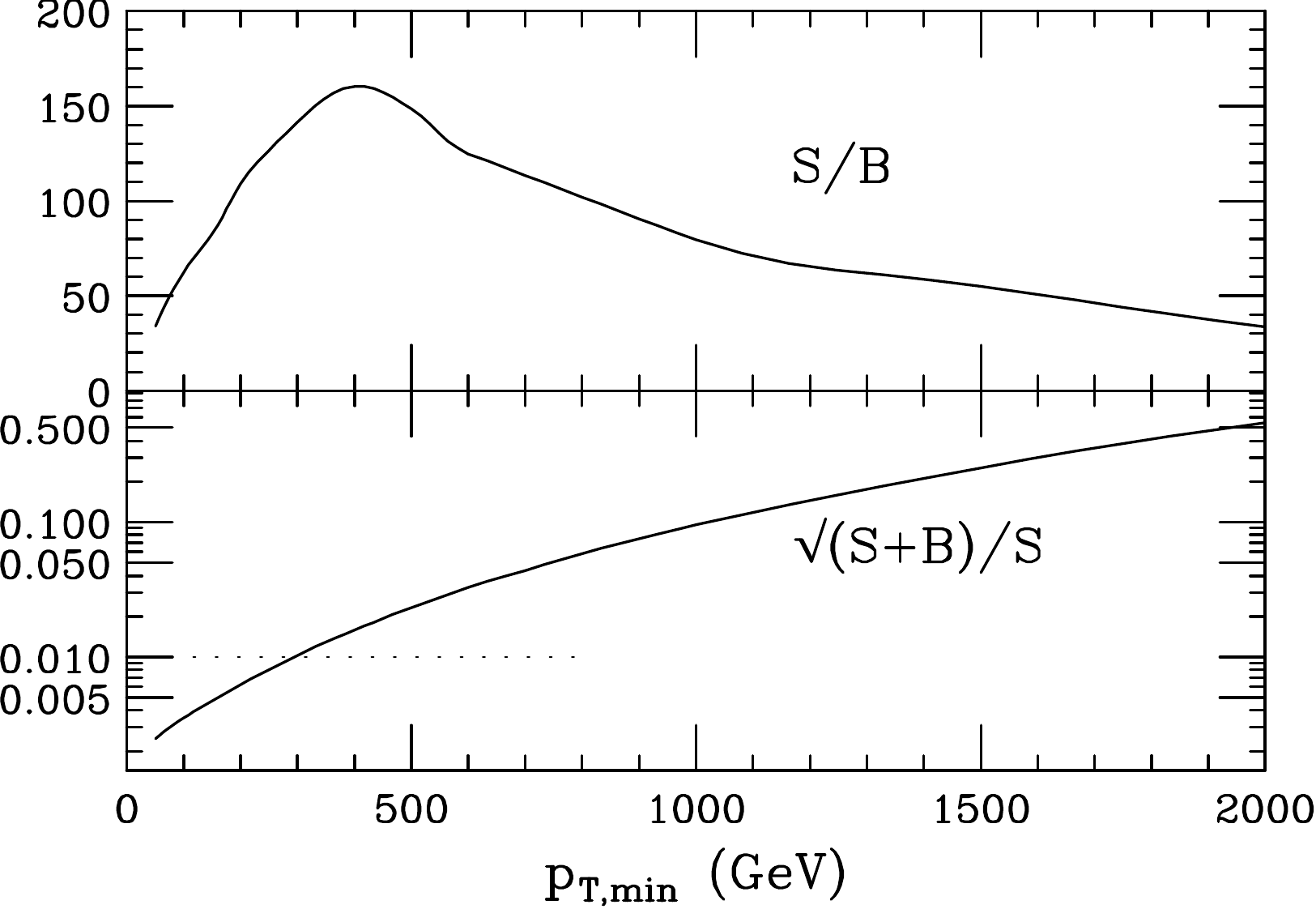}
  \end{center}
  \caption{Integrated transverse momentum rates (20~\iab) for a
    four-lepton final state ($\ell=e,\mu$),
    with mass close to the Higgs mass: signal and QCD
    background.}
\label{fig:H_4l}
\end{figure}

\begin{figure}
  \begin{center}
    \includegraphics[height=0.42\textwidth]{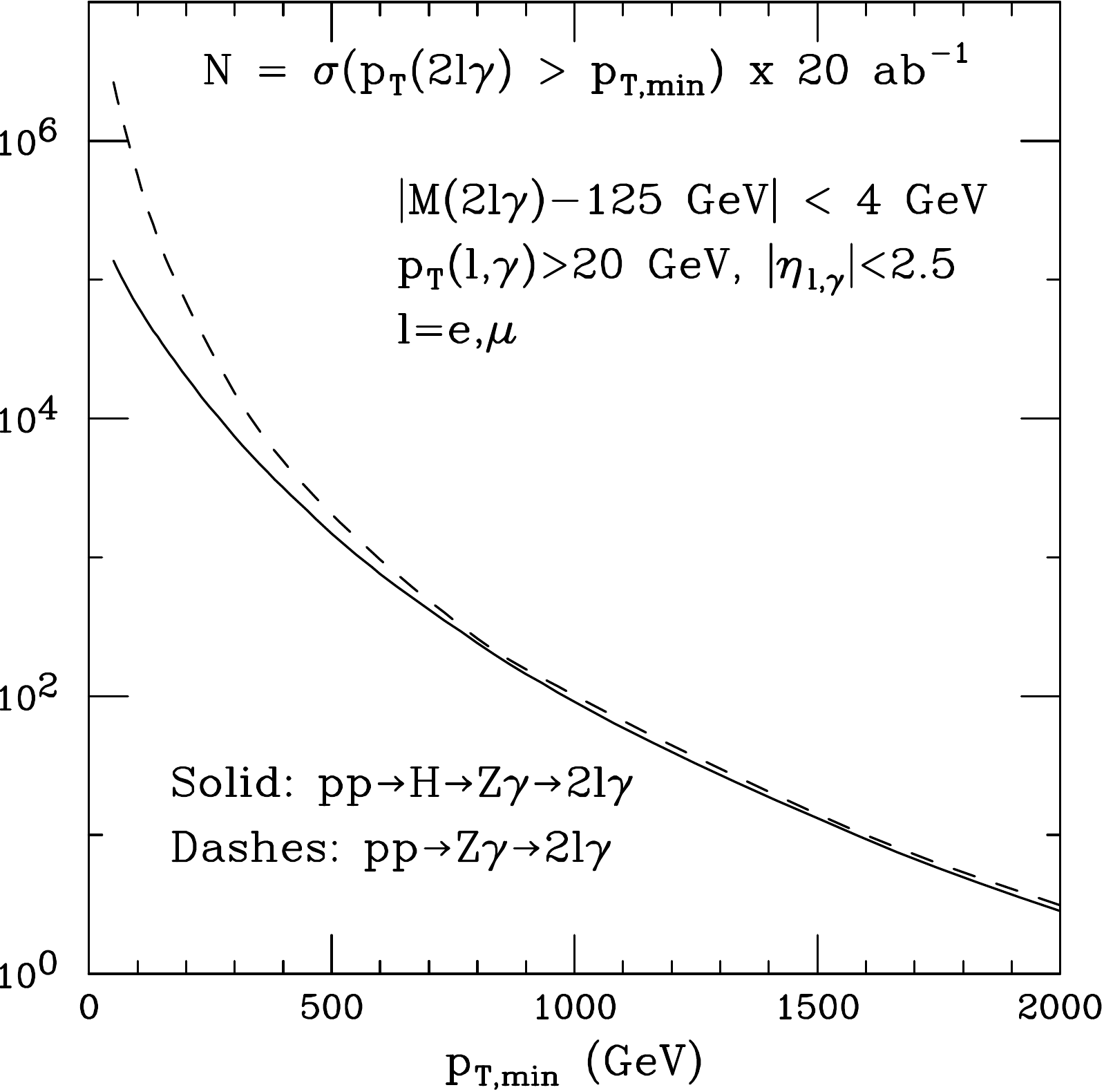}
    \hfill 
    \includegraphics[height=0.42\textwidth]{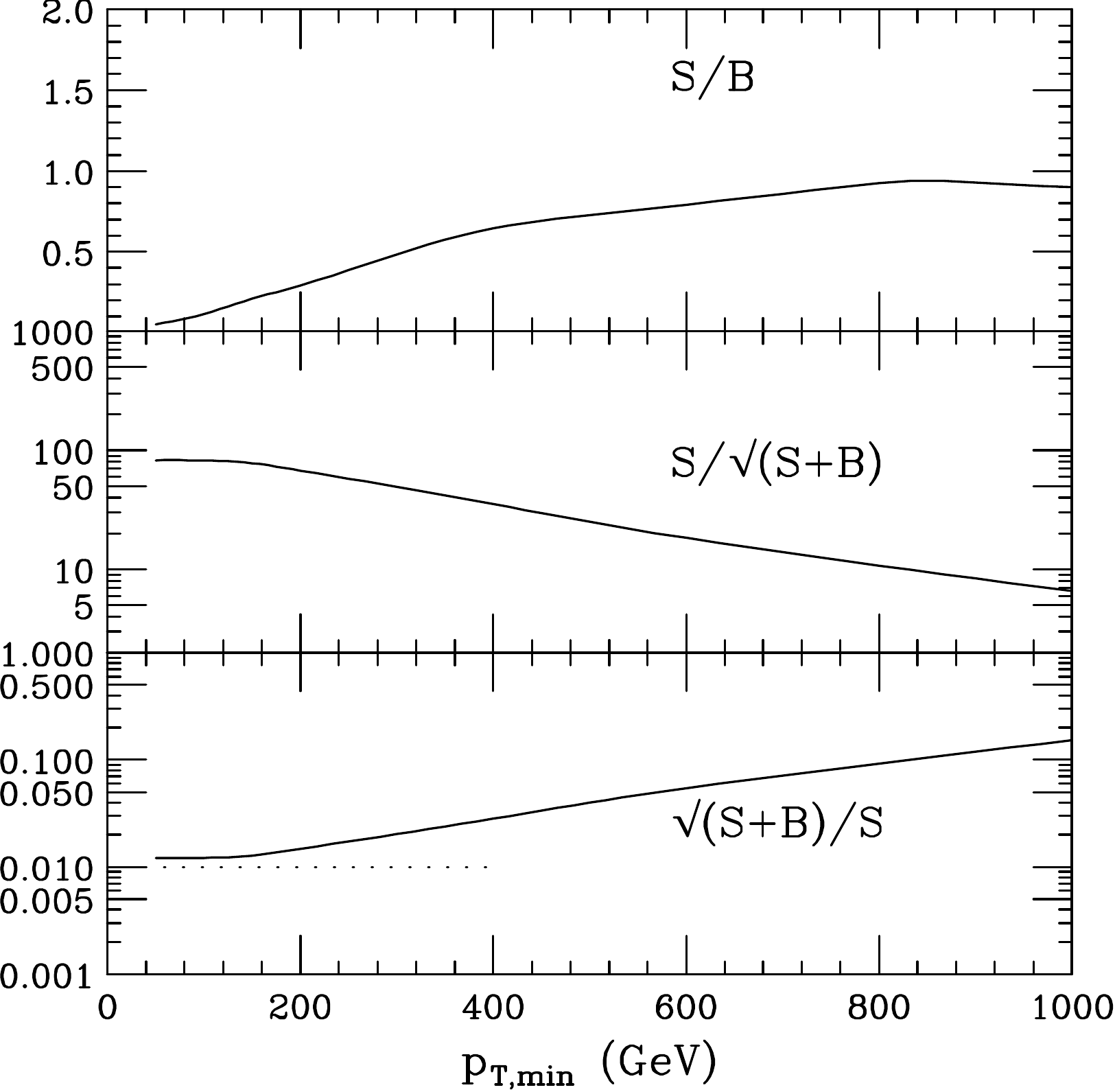}
  \end{center}
  \caption{Left: Integrated transverse momentum rates (20~\iab) for a
    dimuon+photon pair system with mass close to the Higgs mass: signal and QCD
    background. Right: $S/B$, significance of the signal, and potential 
    statistical accuracy of the sample.}
\label{fig:H_Zgam}
\end{figure}

\subsection{Associated $VH$ production}
\label{sec:H_prospects_VH}
We consider here some examples of possible measurements of $WH$
production, in the $H\to b\bar{b}$ final state. As in the previous
discussion, we do not attempt to optimize the detection of the fully
inclusive sample, but examine the opportunities offered by production
in kinematical configuration that are unconventional at the LHC, and
where the 100 TeV collider could offer prospects for interesting new
measurements.

We start by the case of $WH$ production at large invariant mass. As
shown before, this is dominated by the Born-level topologies, with the
$W$ and $H$ recoiling against each other. The largest backgrounds to
the $H\to b\bar{b}$ decay are the QCD associated production of
$Wb\bar{b}$, and the large-mass tail of the $Z$ boson in $WZ^*$, with
$Z^*\to b\bar{b}$. For these kinematics, top quark production is not
an important background. The integrated mass spectra of signal and
backgrounds are shown in the left panel of
Fig.~\ref{fig:prospects_VH_M}. We model the background with a
parton-level calculation, require the $b\bar{b}$ pair to have an
invariant mass in the range of 100--150 GeV, and both $W$ and
$b\bar{b}$ system are in the region $\vert \eta \vert < 2.5$.  The
rates include the branching ratio for the decays $W\to \ell\nu$
($\ell=e,\mu$).

Of course the invariant mass on the $b\bar{b}$ pair provides only a
very crude picture of the potential to suppress the QCD $Wb\bar{b}$
background. The application of the standard $H\to b\bar{b}$ tagging
techniques~\cite{Butterworth:2008iy}, developed for boosts in the
range of few hundred GeV, may require important adaptations and
optimization in the multi-TeV regime, where the whole Higgs-jet is
contained with a cone of radius smaller than $R=0.1$. In the
accompanying SM Volume of this Report~\cite{SMreport}, the tagging of
multi-TeV gauge bosons from the decay of resonances with masses in the
5-40 TeV range is discussed. Gauge boson hadronic decays can be tagged
with efficiencies in the range of 80\%, with suppression factors of
order 20-100 for normal QCD jets of comparable $p_T$. This performance
is comparable to the effectiveness of the naive $m_{bb}$ cut we
applied: the dotted line in Fig.~\ref{fig:prospects_VH_M} shows in
fact the background level obtained by requesting the $b\bar{b}$ pair
to be contained within a jet of radius $R=1$, without any mass
cut. The reduction due to the mass cut is a factor of order 10-20.
The very large $S/B$ shown in Fig.~\ref{fig:prospects_VH_M}
shows that there is plenty of room to cope with the challenge of
identifying these hyper-boosted $H\to b\bar{b}$ jets and rejecting
their backgrounds.

\begin{figure}[h]
    \includegraphics[height=0.45\textwidth]{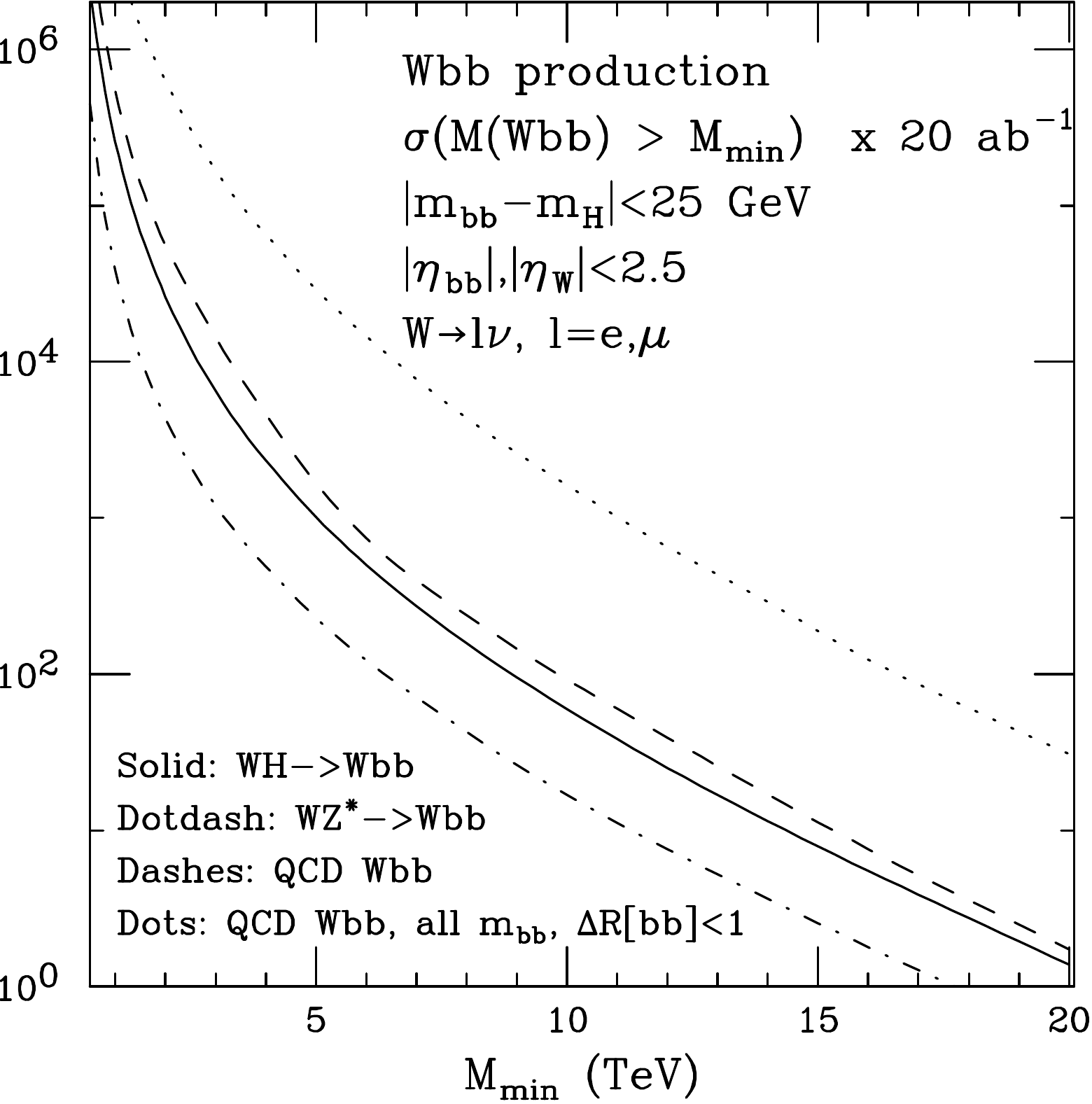}
    \hfill
    \includegraphics[height=0.45\textwidth]{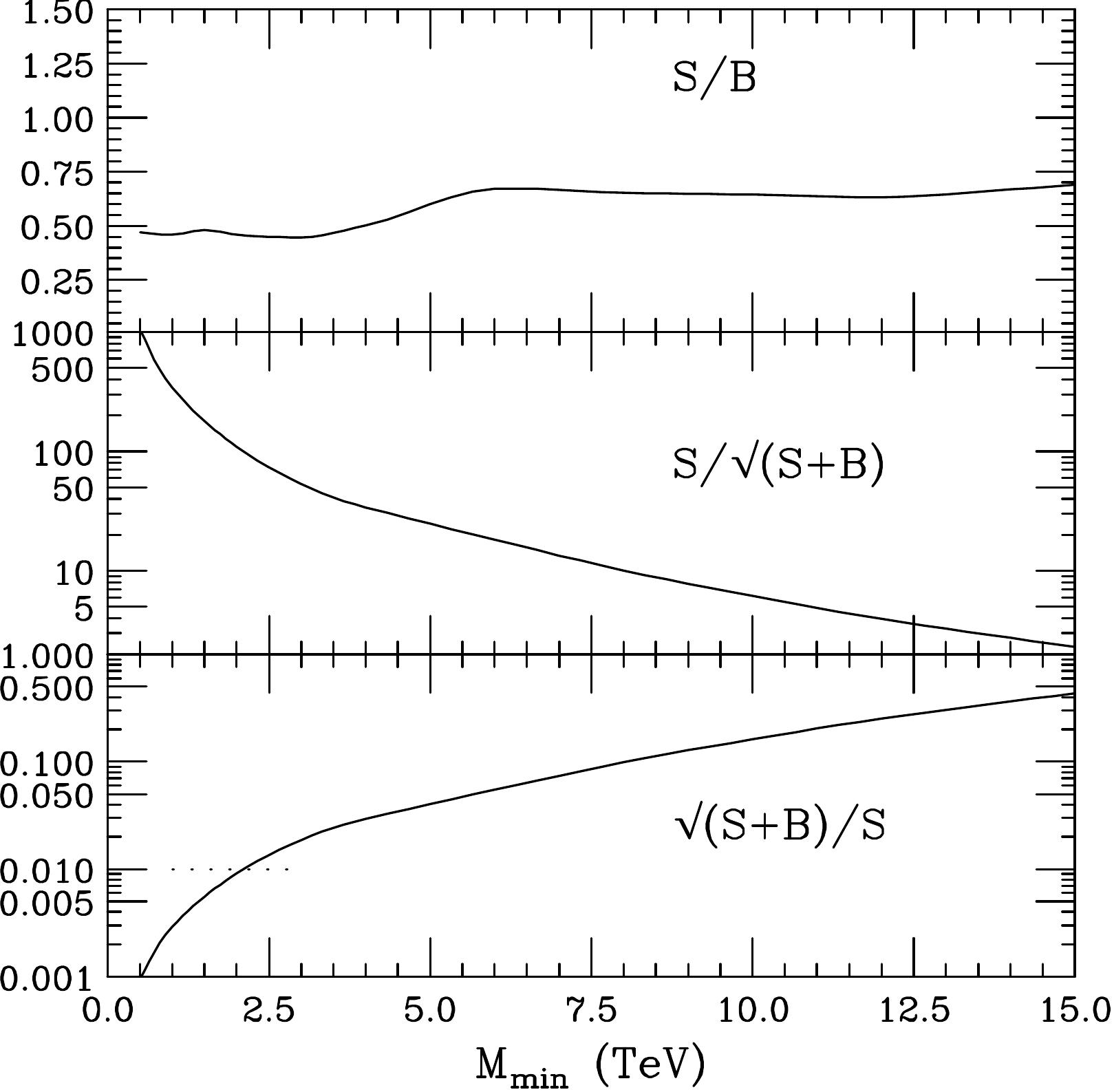}
  \caption{\label{fig:prospects_VH_M}
    Left: Integrated invariant mass rate (20~\iab) for a
    $Wb\bar{b}$ pair, with $W\to \ell\nu$ ($\ell=e,\mu$) and
    $\vert m_{bb} -m_H\vert < 25$~GeV: signal and QCD
    background. Right: $S/B$, significance of the signal, and potential 
    statistical accuracy of the sample.}
\end{figure}

As a further example of possible applications, we consider the other
kinematical configurations of interest, namely $WH$ production in
presence of a high-$p_T$ jet. For the signal, the dominant process if
$q\bar{q} \to g W^* \to g WH$, where the Higgs is simply radiated off
the high-$p_T$ $W$ that recoils against the jet. As shown in
Section~\ref{sec:H_VH}, this leads to a strong correlation between the
$W$ and $H$ direction, resulting in a $\Delta R(WH)$ distribution
peaked at small values. For the background, on the other hand, the
dominant production dynamics is given by the process $qg \to q(g\to
b\bar{b})$, with the $W$ radiated from the initial or final state
quarks. In this case, there is no strong correlation between the
$b\bar{b}$ and the $W$: if anything, they much prefer to be produced
back to back.
\begin{figure}[h]
    \includegraphics[height=0.45\textwidth]{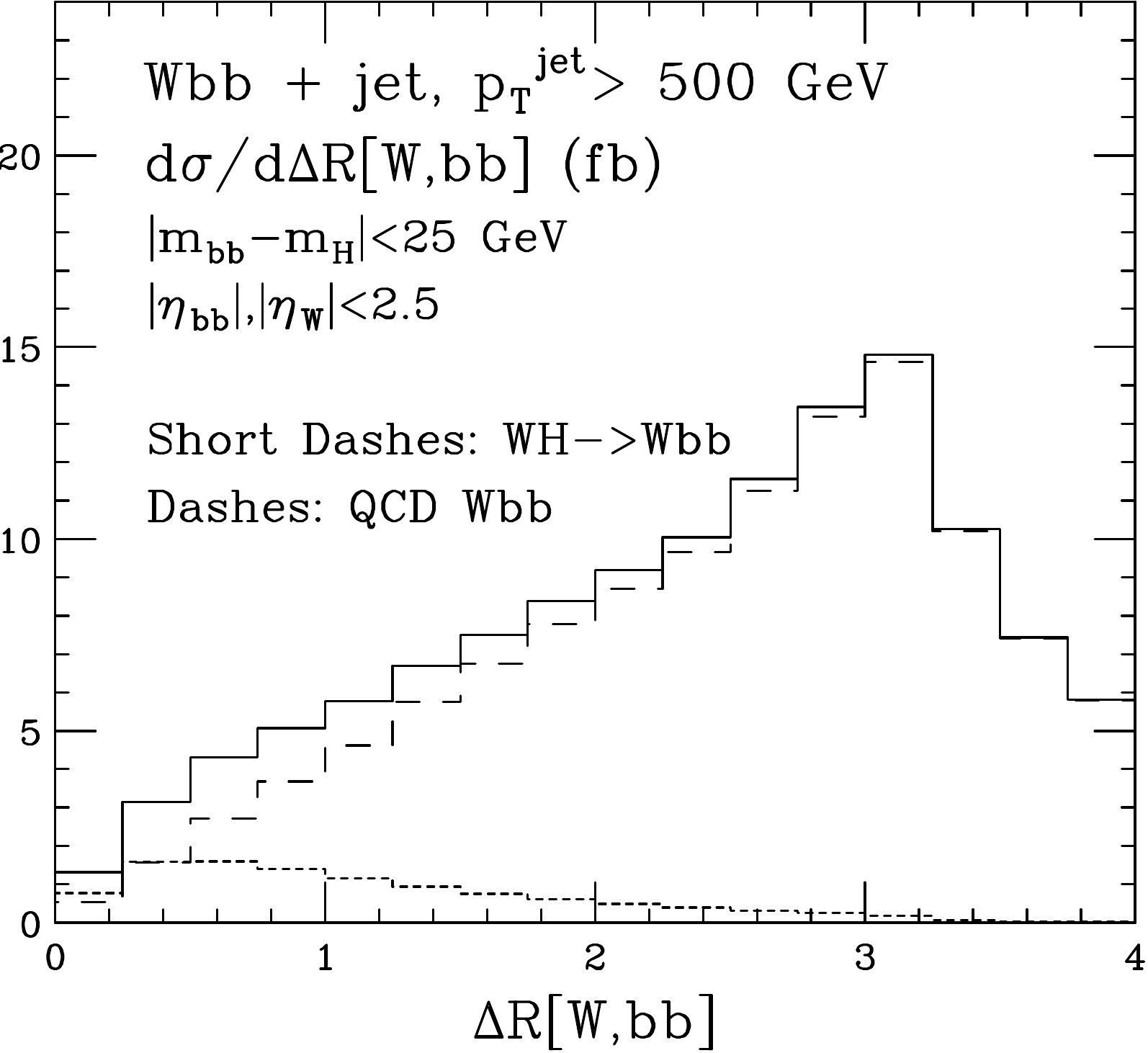}
    \hfill
    \includegraphics[height=0.45\textwidth]{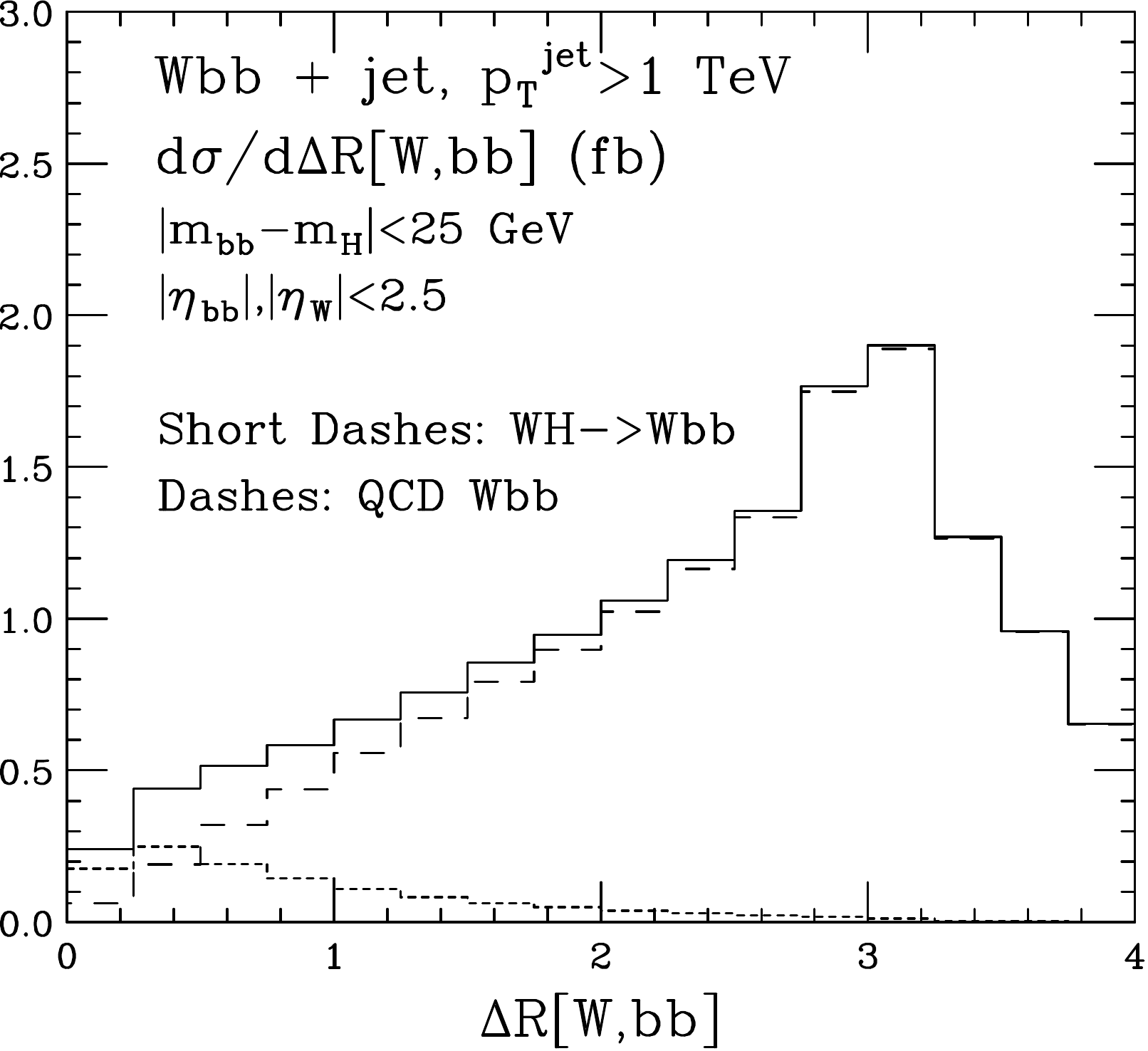}
  \caption{\label{fig:prospects_VH_dR}
    Angular correlation between the $W$ boson and the $b\bar{b}$ pair
    for signal (short dashes) and QCD background (dashes) in final
    states with a jet of $p_T>500$~GeV (left) and 1~TeV (right). The solid curve denotes the sum of signal and QCD background. No
    $W$ branching ratio is included.}
\end{figure}

This is shown very clearly in Fig.~\ref{fig:prospects_VH_dR}, which
shows the $\Delta R(WH)$ distribution for the background (dashed
histograms) and for the Higgs signal (short-dashed histograms), for
different thresholds on the jet $p_T$. A simple cut on $\Delta
R(WH)<1$ leads to an order of magnitude reduction of the background,
while maintaining the largest fraction of the signal. This is shown in
the second panel of Fig.~\ref{fig:prospects_VH_ptj}, where the two
dashed (continuous, red) lines give the background (signal) before and
after the $\Delta R$ cut. The cut brings the $S/B$ ratio to the level
of 1, with sufficient statistics to exceed the percent level precision
in the signal extraction. Further background rejection can likely be
obtained by cutting on the $W$ boson transverse momentum, which is
harder for the signal.

It is clear that more work is needed for a reliable assessment of the
potential for interesting and precise measurements using the associated $VH$
production channels. The application and extension of
$H\to b\bar{b}$ tagging and background rejection techniques
will certainly also lead to valuable input to the detector design
process, both in the calorimeter and tracker areas.  There is also
room for the use of final states other than $b\bar{b}$. We trust that
these topics will be picked up for the studies towards the FCC-hh  Conceptual
Design Report. 

\begin{figure}[h]
\begin{center}
    \includegraphics[height=0.43\textwidth]{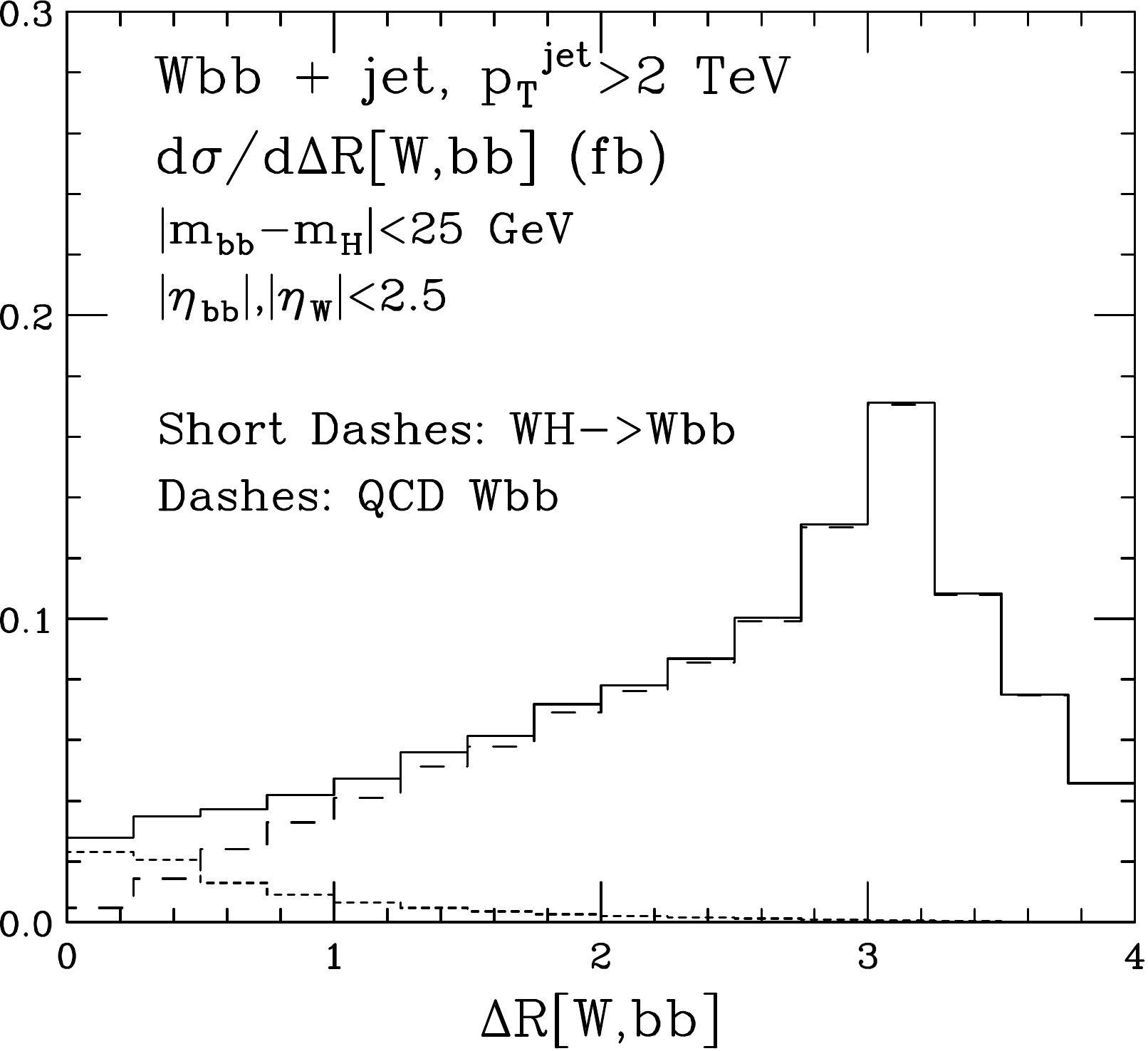}
    \hfill
    \includegraphics[height=0.43\textwidth]{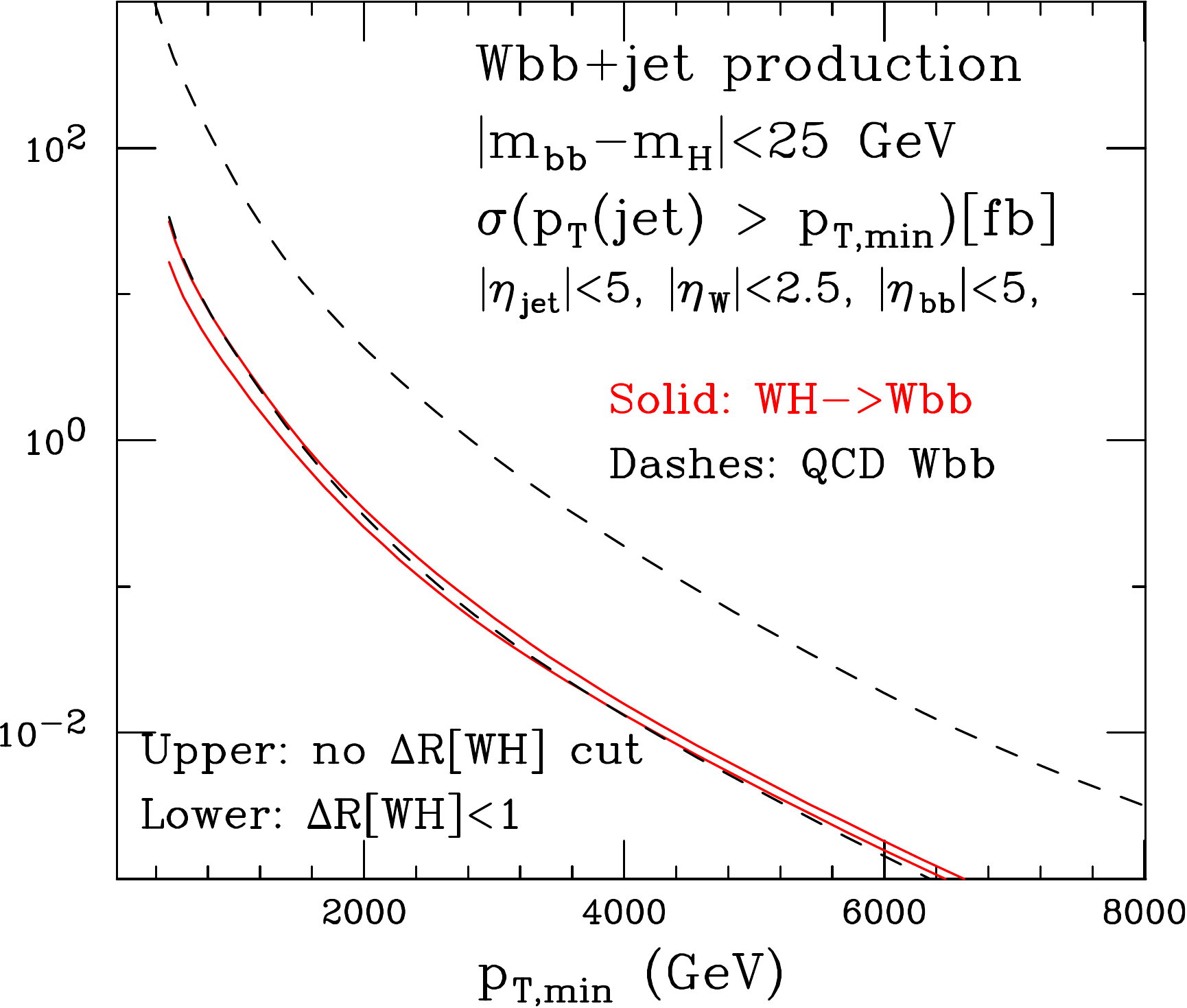}
  \caption{\label{fig:prospects_VH_ptj}
Left panel: same as the previous Figure, with $p_T>2$~TeV. Right:
Signal and background rates, as a function of the jet $p_T$ threshold,
before and after a $\Delta R(WH)<1$ cut.  No
    $W$ branching ratio included.}
\end{center}
\end{figure}

\clearpage
\subsection{Measurement of top Yukawa coupling from the
  $t\bar{t}H/t\bar{t}Z$ ratio}
\label{sec:H_ytop}
The $\tth$ production process can be studied for
a variety of Higgs decay channels.  We collect in
Table~\ref{tab:tthdecays} the event rates for potentially interesting
Higgs decays combined with $\tth$ production, for an integrated
luminosity of 20~\iab\ at 100~TeV. These numbers include the
branching ratio for the mixed lepton-hadron $t\bar{t} \to \ell
\nu_\ell +$~jets decay ($\ell=e,\mu$), in addition to the relevant
Higgs branching ratios.

\begin{table}[b!]
 \begin{center}
\begin{tabular}{c|c|c|c} 
\hline
$H\to 4\ell$ & $H\to \gamma \gamma$ & $ H\to 2\ell 2\nu$ & $H\to
  b\bar{b}$ \\ \hline
$2.6 \cdot 10^4$ & $4.6 \cdot 10^5$ & $2.0 \cdot 10^6$ & $1.2 \cdot 10^8$ \\ \hline
\end{tabular}
\end{center}
\caption{$\tth$ event rates for various Higgs decay modes, with 20~\iab\ at
  100~TeV, assuming $t\bar{t} \to \ell \nu+$jets. Here and for Higgs
  decays, $\ell$ can be either an electron or a muon.} 
\label{tab:tthdecays}
\end{table}
Since analysis cuts and efficiencies will further reduce these rates,
the otherwise very clean $H\to 4\ell$ will hardly meet the target of
the 1\% precision.  In the case of $H\to \gamma\gamma$, basic parton
level cuts such as:
\begin{eqnarray}
p_{T, \gamma,b,j}>25~\gev \; &,& \qquad \vert \eta_{\gamma,b,j} \vert <2.5 
\; , \qquad \Delta R_{jj,bb,bj} > 0.4
\nonumber \\
p_{T,\ell}>20~\gev \; &,& \qquad \vert \eta_\ell \vert <2.5
\end{eqnarray}
leave around $5 \cdot 10^4$ events with 20~\iab, while the
$t\bar{t}\gamma\gamma$ background, subject to a $\vert
m_{\gamma\gamma}-125 \vert < 5~\gev$ cut, is almost a factor of 10
smaller. The $H\to 2\ell 2\nu$ final state has also a potentially
interesting rate, which will deserve a dedicated study.

The large rate for $H\to b\bar{b}$ decays allows to consider boosted
topologies, placing tight cuts on the emerging jets, and drastically
reducing the various sources of backgrounds.  Figure~\ref{fig:ttH_pt}
shows, for example, that requesting $p_{T,H}>500$~GeV gives a rate of
${\cal O}(1)$~pb, or 10M events with 10~\iab. This improved statistics
also allows us to rely on a well-measured and similarly peaked
$t\bar{t}Z \to t\bar{t}\, b\bar{b}$ signal to reduce systematic and
theoretical uncertainties, as anticipated in Section~\ref{sec:H_ttH},
and discussed in detail in Ref.~\cite{Plehn:2015cta}. We summarize
here these findings, and update the results of that work to a broader
range of Higgs $p_T$. We refer to Ref.~\cite{Plehn:2015cta} for the
details.

The analysis models the first \textsc{HEPTopTagger}
application to $t\bar{t}H$ production with $H\to
b\bar{b}$~\cite{Plehn:2009rk},
and builds on the recent improvements in the
\textsc{HEPTopTagger2}~\cite{Kasieczka:2015jma} and in the BDRS Higgs
tagger~\cite{Butterworth:2008iy}, which reduce background sculpting and
increase the signal statistics.


We consider the final states:
\begin{equation}
pp \to t\bar{t}H \to (b j j)\, (\bar{b} \ell \bar{\nu})\, (b\bar{b}),
(b \ell \nu)\, (\bar{b} j j)\, (b\bar{b}) \; . 
\end{equation}
and the leading backgrounds:
\begin{itemize}
\item[] $pp \to t \bar t \, b \bar b$, the main irreducible QCD background
\item[] $pp \to t \bar t Z$, including the $Z$-peak in the $m_{bb}$
  distribution 
\item[] $pp \to t \bar t +$jets with fake-bottoms tags 
\end{itemize}
The analysis requires:
\begin{enumerate}
\item an isolated lepton with $|y_\ell|<2.5$ and
$p_{T,\ell}>15$~GeV.
\item a tagged top ($R=1.8$, $p_{T,j}>200$~GeV, $|y_{j}^{(t)}|<4$)
  without any $b$-tag requirement
\item a tagged Higgs jet with two $b$-tags inside ($R=1.2$,
$p_{T,j}>200$~GeV, $|y_j^{(H)}|<2.5$)
\item a $b$-tagged jet ($R=0.6$, $p_{T,j}>30$~GeV, $|y_b| < 2.5$)
  outside the top and Higgs fat jets, corresponding to
  the top decaying semileptonically.
\end{enumerate}
The $m_{bb}$ distribution provides the sidebands to control the
$t\bar{t}b\bar{b}$ and $t\bar{t}$+jets backgrounds, and a second mass
peak from the $t\bar{t}Z$ mass peak.  All Monte Carlo event samples
are generated at leading order, using MadGraph5~\cite{Alwall:2011uj} with
NNPDF2.3 parton densities~\cite{Ball:2013hta}, showering and hadronization via
Pythia8~\cite{Sjostrand:2014zea} and the fast detector simulation with
Delphes3~\cite{deFavereau:2013fsa,Anderson:2013kxz}.  The jet clustering and the
analysis are done with {FastJet3}~\cite{Cacciari:2011ma}, a modified BDRS
Higgs tagger~\cite{Butterworth:2008iy,Plehn:2009rk} and the
\textsc{HEPTopTagger2}~\cite{Kasieczka:2015jma}.  All $b$-tags require a
parton-level $b$-quark within $\Delta R < 0.3$ and assume a
$b$-tagging efficiency of $50\%$ and a mis-tagging probability of
$1\%$.

Figure~\ref{fig:plots} shows the reconstructed $m_{bb}$ spectrum for
the signal and the backgrounds, varying the $p_T$ threshold of the top
and Higgs tagged fat jets in steps of 100~GeV from 200 up to 500~GeV.
\begin{figure}[h]
\includegraphics[width=0.4\textwidth]{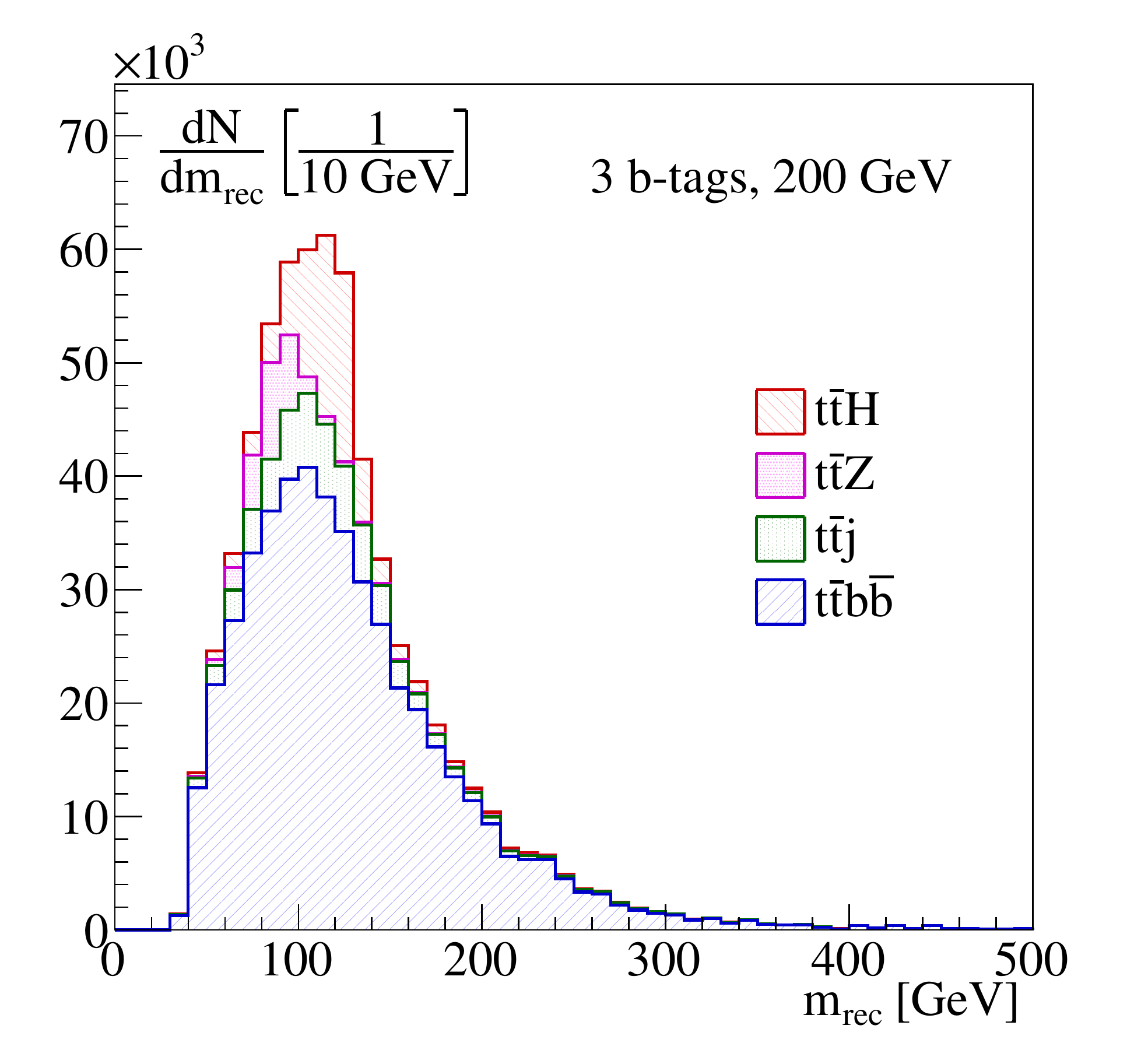} 
\hfill
\includegraphics[width=0.4\textwidth]{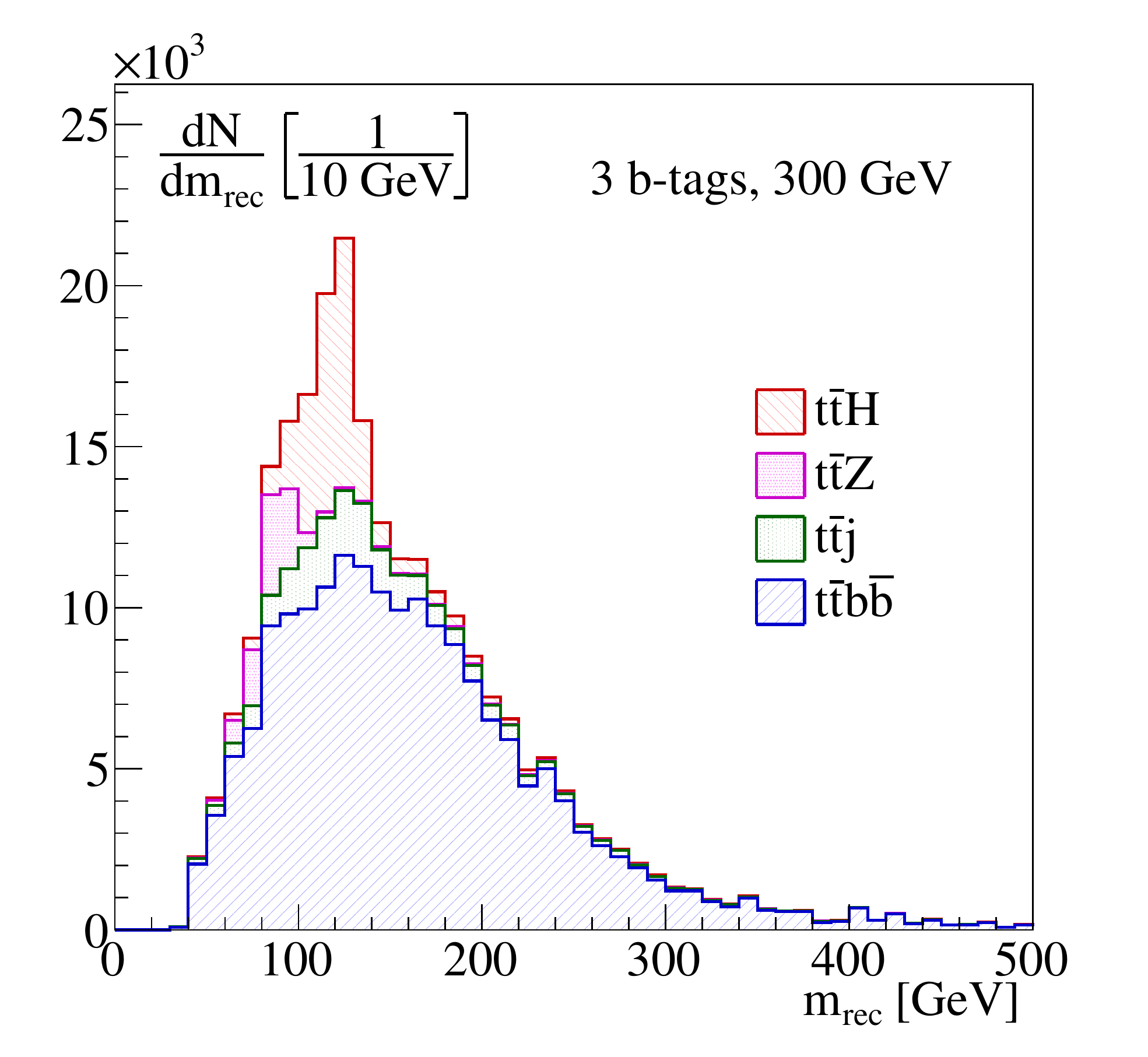}
 \includegraphics[width=0.4\textwidth]{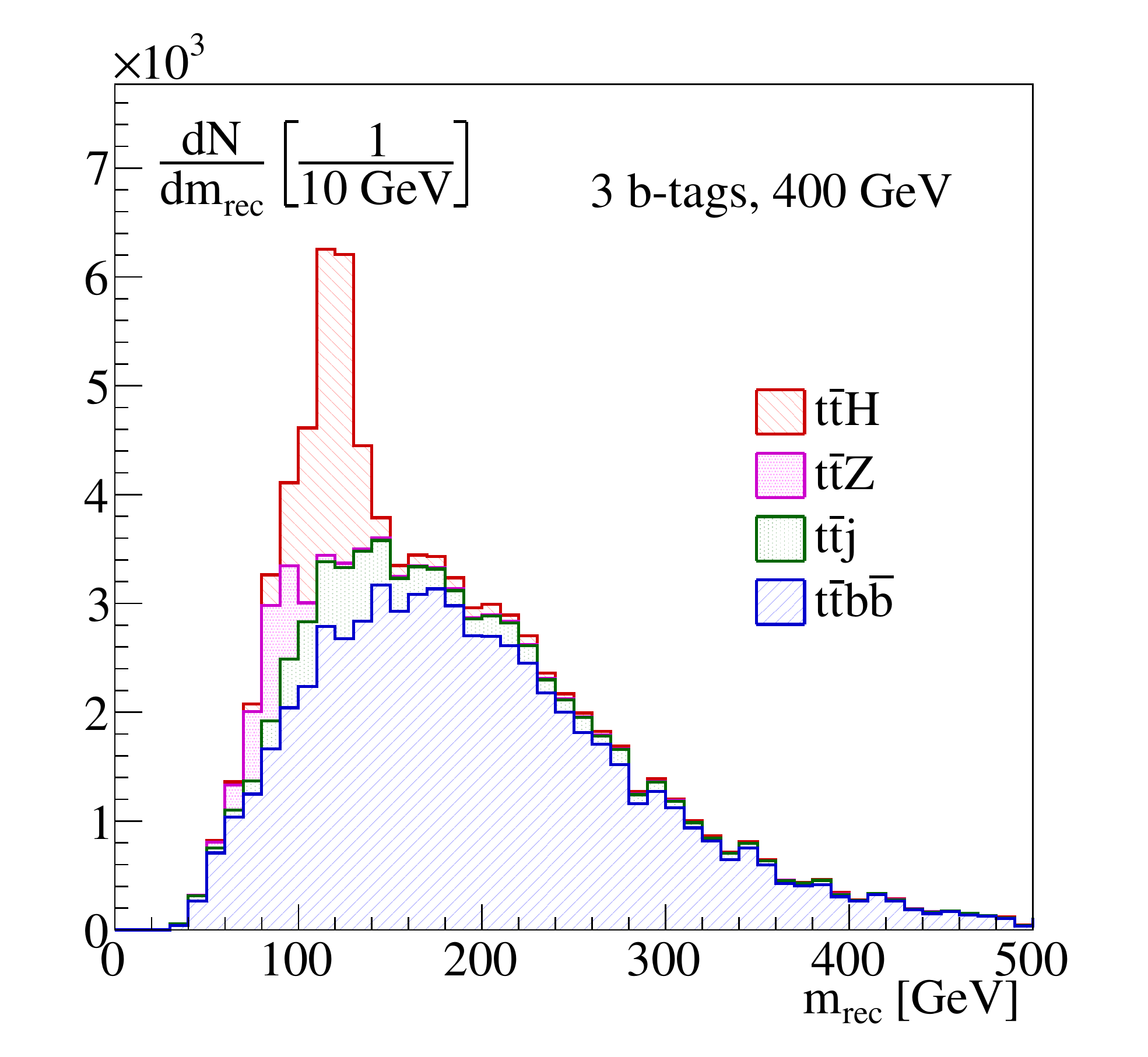}
 \hfill
 \includegraphics[width=0.4\textwidth]{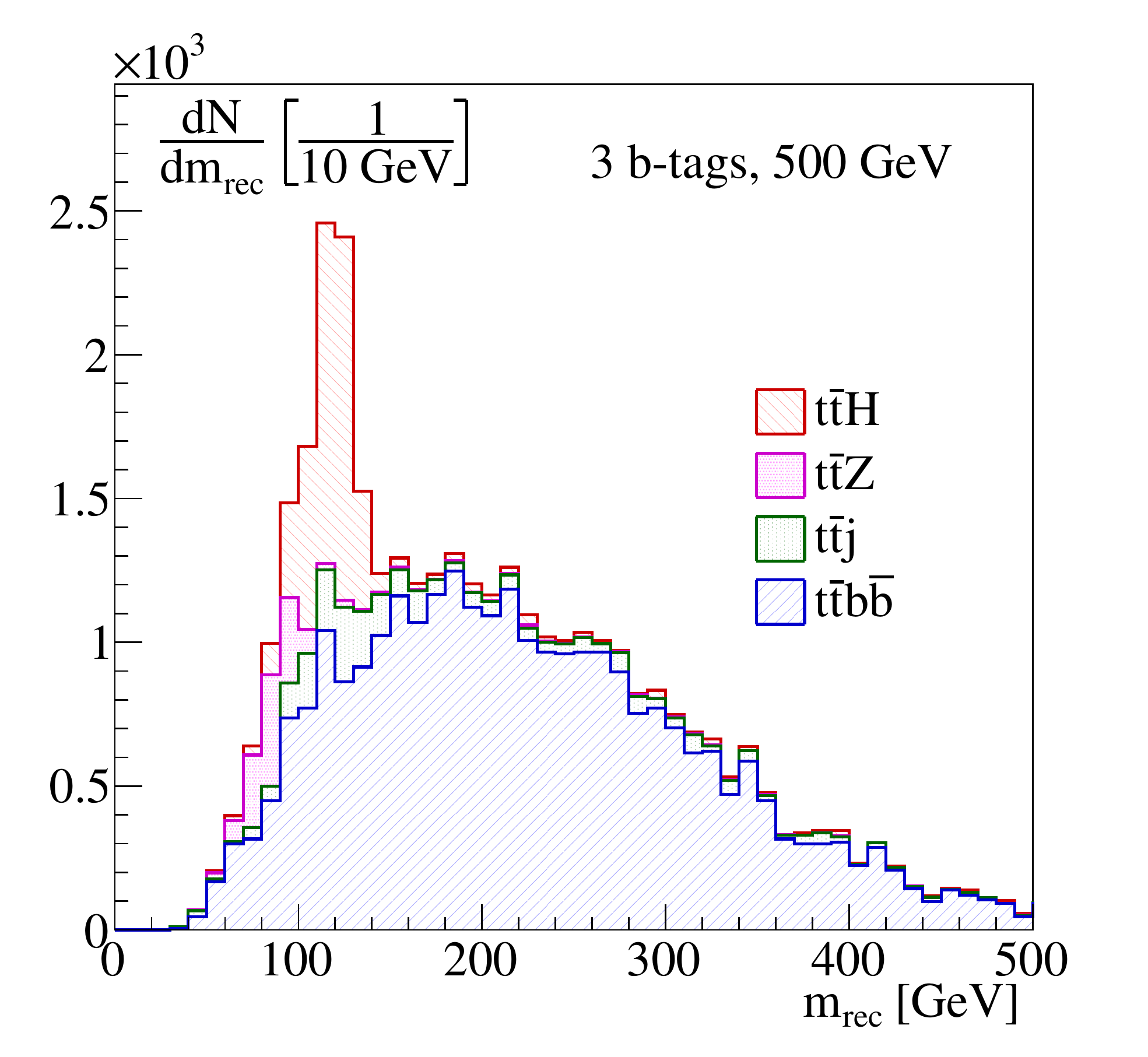}
	\caption{Recontructed $m_{bb}$ for a $p_{T}$ threshold from
          $200$~GeV to 500~GeV (the roughness of the distributions is
          a consequence of limited MC statistics).}
	\label{fig:plots}
\end{figure}

For the 200~GeV cut,
and the signal region $m_{bb} \in [104,136]$~GeV, we arrive at a
signal-to-background ratio around $S/B \approx 1/3$ and a Gaussian
significance $S/\sqrt{B}=120$, assuming an integrated luminosity of
$\mathcal{L} = 20~$\iab. The error on the number of nominally
$N_S = 44700$ signal events is given by two terms. First, we assume
that we can determine $N_S$ from the total number of events $N_S+N_B$
using a perfect determination of $N_B$ from the side bands. Second,
the side band $m_{bb} \in [160,296]$~GeV with altogether
$N_\mathrm{side} = 135000$ events and a relative uncertainty of
$1/\sqrt{N_\mathrm{side}}$ introduces a statistical uncertainty
$\Delta N_B$, altogether leading to
\begin{eqnarray}
\Delta N_S 
&=& 
\left[ \left( \sqrt{N_S + N_B} \right)^2 
      +\left( \Delta N_B \right)^2
\right]^{1/2} \nonumber \\
&=& 
\left[ \left( \sqrt{N_S + N_B} \right)^2 
      +\left( \frac{N_B}{\sqrt{N_\mathrm{side}}} \right)^2
\right]^{1/2} = 0.013 \, N_S \; .
\end{eqnarray}
For the Yukawa coupling this translates into a relative error of
around 1\%. The first term alone would give
$\Delta N_S = 0.010 \, N_S$. \bigskip

The analysis for larger $p_T$ cuts leads to the numbers in the
following table:
\begin{center}
\begin{tabular}{r | rrrr | rrr } 
\hline
$p_{T,\text{min}}$[GeV] & $N_S$ & $N_B$ & $N_S+N_B$ & $N_{\textrm{Sideband}}$ &
                                                                     $\Delta N_S/N_S$
  & $N_S/N_B$ & $N_S/\sqrt{N_B}$
  \\ \hline
\rule{0pt}{3ex} 
250 & 29400 & 74700 & 104000 & 155000 & 0.013 & 0.39 & 107\\
300 & 18800 & 39000 & 57900 & 116000 & 0.014 & 0.48 & 95\\
350 & 13300 & 27500 & 40800 & 79800 & 0.017 & 0.48 & 80\\
400 & 8970 & 16700 & 25600 & 50300 & 0.020 & 0.54 & 69\\
450 & 5950 & 9810 & 15800 & 35100 & 0.023 & 0.61 & 60\\
500 & 3830 & 5730 & 9560 & 24400 & 0.027 & 0.67 & 51\\
\hline 
\end{tabular}
\end{center}
For the signal region we count $N_S$ in the region
with $N_S/N_B>1/5$, for the sideband region we require $N_S/N_B<1/10$.
The corresponding $m_{bb}$ distribution is binned in steps of 10~GeV.
$N_B$ is the sum of all $t\bar{t}b\bar{b}$, $t\bar{t}+$ jets and
$t\bar{t}Z$ events combined. We notice that the precision on the
number of extracted signal events, $\Delta N_S/N_S$, remains at the level of 
1-2\% over a broad range transverse momenta, providing an important
validation of the robustness of the analysis.

More details, and the results of the combined Crystal Ball fit of the
$Z$ and $H$ signals, are given in Ref.~\cite{Plehn:2015cta}.  The
continuum side band and the second peak offer two ways to control the
backgrounds as well as the translation of the $t\bar{t} \, b\bar{b}$
rate into a measurement of the Yukawa coupling. We therefore find that
$y_{top}$ could be measured to around $1\%$ with a 100~TeV collider
and an integrated luminosity of 20~\iab. This is an order of magnitude
improvement over the expected LHC reach, with significantly improved
control over the critical uncertainties.

There exist additional, complementary opportunities offered by the
$\tth$ study.
For example, the $H\to \gamma\gamma$ decay could allow a direct
measurement of the ratio of branching ratios $B(H\to
\gamma\gamma)/B(H\to b\bar{b})$. It would serve as a complementary,
although indirect, probe of the $\tth$ coupling. Furthermore, $H\to
2\ell 2\nu$ could also be interesting, since there is enough rate to
explore the regime $p_{T,H} \gg m_H$, which, especially for the
$e^\pm \mu^\mp \nu\bar{\nu}$ final state, could be particularly
clean.

\subsection{Combined determination of $y_{t}$ and $\Gamma(H)$ from $ttH$
  vs $t\bar{t}t\bar{t}$ production}
\label{sec:H_Hwidth}

Precise information of Higgs boson, e.g. its mass, width, spin,
parity, and couplings, should shed light on new physics beyond
the Standard Model. In this section we discuss the measurements of
two important properties of the Higgs boson, the total width
($\Gamma_H$) and its coupling to top-quark ($y_{Ht\bar{t}}$), through
the $t\bar{t}H$ and $t\bar{t}t\bar{t}$ productions at a 100 TeV $pp$
collider. The top Yukawa-coupling can be measured in the $t\bar{t}H$
production. An ultimate precision of about ~1\% is expected at a
100~TeV $pp$ collider in the channel of $pp\to t\bar{t}H\to t\bar{t}
b\bar{b}$ with an integrated luminosity ($\mathcal{L}$) of $20~{\rm
ab}^{-1}$, assuming the $H\to b\bar b$ branching ratio is the same as in the SM.
However, this assumption may not be valid in NP models; for example, $\Gamma_H$ might differ from the SM value ($\Gamma_H^{\rm SM}$) in the case that the Higgs boson decays into a
pair of invisible particles. It is important to find a new
experimental input to relax the assumption. Four top-quark
($t\bar{t}t\bar{t}$) production provides a powerful tool to probe the
top-quark Yukawa coupling, and in addition, combining the $t\bar{t}H$
and $t\bar{t}t\bar{t}$ productions also determines $\Gamma_H$
precisely~\cite{Cao:2016wib}.

Under the narrow width approximation, the production cross section of $pp\to t\bar{t}H \to t\bar{t}b\bar{b}$ is 
\begin{equation}
\begin{split}
\sigma(pp\to t\bar{t}H \to t\bar{t}b\bar{b}) & = \sigma^{\rm SM}(pp\to t\bar{t}H\to t\bar{t} b\bar{b})\times \kappa_t^2\kappa_b^2\frac{\Gamma_{H}^{\rm SM}}{\Gamma_{H}} \\
& \equiv \sigma^{\rm SM}(pp\to t\bar{t}H\to t\bar{t} b\bar{b})\times\mu^{b\bar{b}}_{t\bar{t}H},
\end{split}
\end{equation}
where $\kappa_t\equiv y_{Htt}/y_{Htt}^{\rm SM}$ and $\kappa_b\equiv y_{Hbb}/y_{Hbb}^{\rm SM}$ are the Higgs coupling scaling factors. 
The signal strength $\mu^{b\bar{b}}_{t\bar{t}H}$, defined as
\begin{equation}
\mu^{bb}_{t\bar{t}H}= \frac{\kappa_t^2 \kappa_b^2 }{R_\Gamma}
\qquad {\rm with}\qquad R_\Gamma\equiv \frac{\Gamma_H}{\Gamma_H^{\rm SM}},
\label{eq:tth}
\end{equation}
is expected to be measured with $1\%$ precision,
$\overline{\mu}_{t\bar{t}H}^{b\bar{b}} = 1.00\pm 0.01$~\cite{Plehn:2015cta}.
Since the $\kappa_t $, $\kappa_b$ and $\Gamma_H$ parameters are independent in $\mu^{b\bar{b}}_{t\bar{t}H}$, one cannot determine them from the $t\bar{t}H$ production alone. Bounds on the $\kappa_t$, $\kappa_b$ and $R_\Gamma$ can be derived from a global analysis of various Higgs production channels. The bottom Yukawa coupling would be measured precisely at electron-positron colliders. Once $\kappa_b$ is known, a correlation between $\kappa_t$ and $R_\Gamma$ is obtained as following
\begin{equation}
\frac{\kappa_t^2}{R_\Gamma} = \overline{\mu}_{t\bar{t}H}.
\label{eq:r2}
\end{equation} 
If the top-quark Yukawa coupling could be directly measured in a single channel, then one can probe $R_\Gamma$ from Eq.~\ref{eq:r2}. 

\begin{figure}[b]
\centering
\includegraphics[scale=0.38]{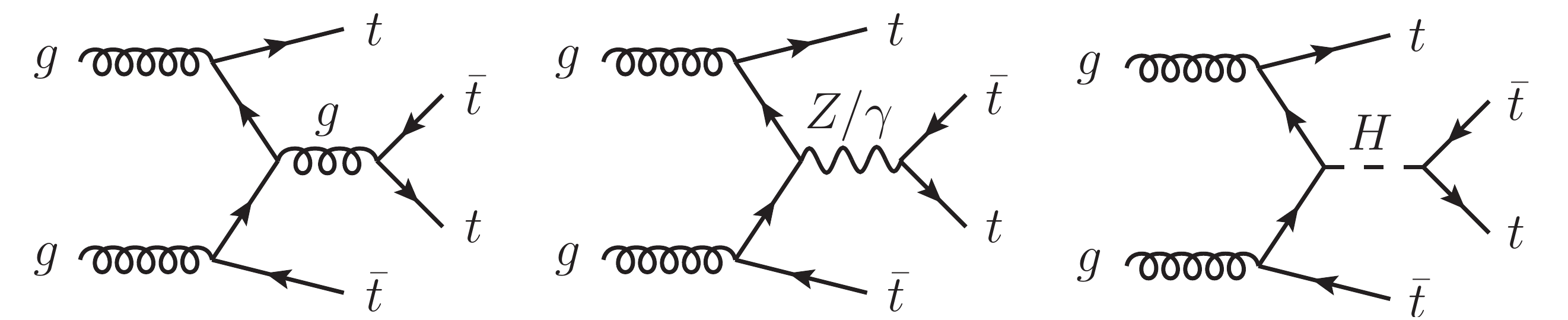}
\caption{Representative Feynman diagrams of the $t\bar{t}t\bar{t}$ production through the QCD interaction and the Higgs boson mediation. }
\label{fig:feyn}
\end{figure}

In the SM the $t\bar{t}t\bar{t}$ production occurs either through a gluon mediator~\cite{Barger:1991vn} or by an off-shell Higgs mediator; see Fig.~\ref{fig:feyn} for the representative Feynman diagrams. Interferences between the QCD diagrams ($t\bar{t}t\bar{t}_g$) and the Higgs diagrams ($t\bar{t}t\bar{t}_H$) are absent at the tree level. We thus name the cross section of the QCD induced channel as $\sigma(t\bar{t}t\bar{t})_g$ and the cross section of the Higgs induced channel as $\sigma(t\bar{t}t\bar{t})_H$. There are two advantages of the Higgs-induced $t\bar{t}t\bar{t}$ production: i) no dependence on the Higgs boson width; ii) the cross section proportional to the top quark Yukawa coupling to the fourth power, i.e.  
\begin{equation}
\sigma(t\bar{t}t\bar{t})_H \propto  \kappa_t^4 \sigma^{\rm SM}(t\bar{t}t\bar{t})_H,
\end{equation}
where $\sigma^{\rm SM}(t\bar{t}t\bar{t})_H$ denotes the SM production cross section. The not-so-small interferences among the three kinds of Feynman diagrams are also accounted. 
Since the QCD and electroweak gauge interactions of top quarks have been well established, we consider only the top Yukawa coupling might differ from the SM value throughout this section. As a result, the cross section of $t\bar{t}t\bar{t}$ production is 
\begin{equation}
\sigma(t\bar{t}t\bar{t}) = \sigma^{\rm SM}(t\bar{t}t\bar{t})_{g+Z/\gamma} + \kappa_t^2 \sigma^{\rm SM}(t\bar{t}t\bar{t})_{\rm int} + \kappa_t^4 \sigma^{\rm SM}(t\bar{t}t\bar{t})_H,
\end{equation}
where
\begin{eqnarray}
\sigma^{\rm SM}(t\bar{t}t\bar{t})_{g+Z/\gamma}&~\propto~& \left|\mathcal{M}_{g} + \mathcal{M}_{Z/\gamma}\right|^2, \nonumber\\
\sigma^{\rm SM}(t\bar{t}t\bar{t})_{H}&\propto& \left|\mathcal{M}_{H}\right|^2, \nonumber\\
\sigma^{\rm SM}(t\bar{t}t\bar{t})_{\rm int} &\propto& \mathcal{M}_{g+Z/\gamma}\mathcal{M}^\dagger_{H}+\mathcal{M}^\dagger_{g+Z/\gamma}\mathcal{M}_{H}.
\end{eqnarray}
We use MadEvent~\cite{Alwall:2007st} to calculate the leading order cross section of  $t\bar{t}t\bar{t}$ production in the SM. The numerical results are summarized as follows:
\begin{align}
&                & \text{14~TeV}~~& &\text{100~TeV}~~& \nonumber\\
&\sigma^{\rm SM}(t\bar{t}t\bar{t})_{g+Z/\gamma}:  & 12.390~{\rm fb}, & &3276~{\rm fb},& \nonumber\\
&\sigma^{\rm SM}(t\bar{t}t\bar{t})_H:  & 1.477~{\rm fb},& &271.3~{\rm fb},& \nonumber\\
&\sigma^{\rm SM}(t\bar{t}t\bar{t})_{\rm int}: & -2.060~{\rm fb},&  &-356.9~{\rm fb}.&
\end{align}
The numerical results shown above are checked with CalcHEP~\cite{Belyaev:2012qa}.
The NLO QCD corrections to the $t\bar{t}t\bar{t}_g$ background is calculated in Ref.~\cite{Maltoni:2015ena}, which is about $4934~{\rm fb}$ with 25\% uncertainty. Unfortunately, as the QCD corrections to the interference and electroweak contributions is not available yet, a tree-level simulation of the signal process is used to estimate the accuracy of Higgs width measurement.  

A special signature of the four top-quark events is the same-sign charged leptons (SSL) from the two same-sign top quarks.  The ATLAS and CMS collaborations have extensively studied the same sign lepton pair signal at the LHC~\cite{ATLAS:2013tma,Chatrchyan:2013fea}. The other two top quarks are demanded to decay hadronically in order to maximize the production rate. Therefore, the topology of the signal event consists of two same-sign charged leptons, four $b$-quarks, four light-flavor quarks, and two invisible neutrinos. In practice it is challenging to identify four $b$-jets. Instead, we demand at least 5 jets are tagged and three of them are identified as $b$-jets. The two invisible neutrinos appear as a missing transverse momentum ($\not{\!\!{\rm E}}_{T}$) in the detector. Thus,  the collider signature of interests to us is two same-sign leptons, at least five jets and three of them tagged as $b$-jets, and a large $\not{\!\!{\rm E}}_{T}$. 

The SM backgrounds for same-sign leptons can be divided into three categories: i) prompt same-sign lepton pair from SM rare process, including 
di-boson and $W^\pm W^\pm jj$; ii) fake lepton, which comes from heavy quark jet, namely $b$-decays, and the dominant one is the $t\bar{t}+X$ events \cite{ATLAS:2012sna}; iii) charge misidentification.
As pointed out by the CMS collaboration~\cite{Chatrchyan:2013fea}, the background from charge mis-identification is generally much smaller and stays below the few-percent level. We thus ignore this type of backgrounds in our simulation and focus on those non-prompt backgrounds $t\bar{t}+X$ and rare SM processes contributions. For four top quark production process another feature worthy being specified is that multiple $b$-jets decay from top quark appear in the final state. Same-sign lepton plus multiple $b$-jets has a significant discrimination with the backgrounds.
From above analysis, it is clear that the major backgrounds are $t\bar{t}+X$ and $W^\pm W^\pm jj$. 
Another SM processes can contribute the same-sign lepton are di-boson, while it can be highly suppressed by the request of multiple jets in the final state. Therefore we focus on the $t\bar{t}+X$, $W^\pm W^\pm jj$ and $t\bar{t}t\bar{t}(g)$ backgrounds below. 
The cross section of the $t\bar{t}$ production is calculated with the next-to-leading-order(NLO) QCD correction using MCFM package~\cite{Campbell:2010ff}. The NLO QCD corrections to the $t\bar{t}Z$ and $t\bar{t}W$ background are taken into account by multiplying the leading order cross sections with a constant $K$-factor; for example, $K_{F} = 1.17$ for the $t\bar{t}Z$ and $K_{F} = 2.20$ for the $t\bar{t}W$ production~\cite{SMreport}. 


Both the signal and background events are generated at the parton level using MadEvent~\cite{Alwall:2007st} at the 100 TeV proton-proton collider. We use Pythia~\cite{Sjostrand:2014zea} to generate parton showering and hadronization effects. The Delphes package~\cite{deFavereau:2013fsa} is used to simulate detector smearing effects in accord to a fairly standard Gaussian-type detector resolution given by $\delta E/E= \mathcal{A}/\sqrt{E/{\rm GeV}}\oplus \mathcal{B}$,
where $\mathcal{A}$ is a sampling term and $\mathcal{B}$ is a constant term.  For leptons we take $\mathcal{A}=5\%$ and $\mathcal{B}=0.55\%$, and for jets we take $\mathcal{A}=100\%$ 
and $\mathcal{B}=5\%$.
We require the charged lepton has a transverse momentum $p^{\ell}_T$ greater than $20$ GeV, rapidity $\left|\eta_{\ell}\right|\leq 2.5~$ and its overlap with jets  $\Delta R_{j\ell} = \sqrt{(\Delta \eta)^2 + (\Delta \phi)^2} \geq 0.4$. The $\not{\!\!{\rm E}}_{T}$ is then defined to balance the total transverse momentum of visible objects.

Figure~\ref{nbj} displays the numbers of reconstructed jets (a) and $b$-tagged jets (b) in the signal and background processes. 
\begin{figure}[tb]
\centering
\hspace*{-0.25cm}
\includegraphics[scale=0.4]{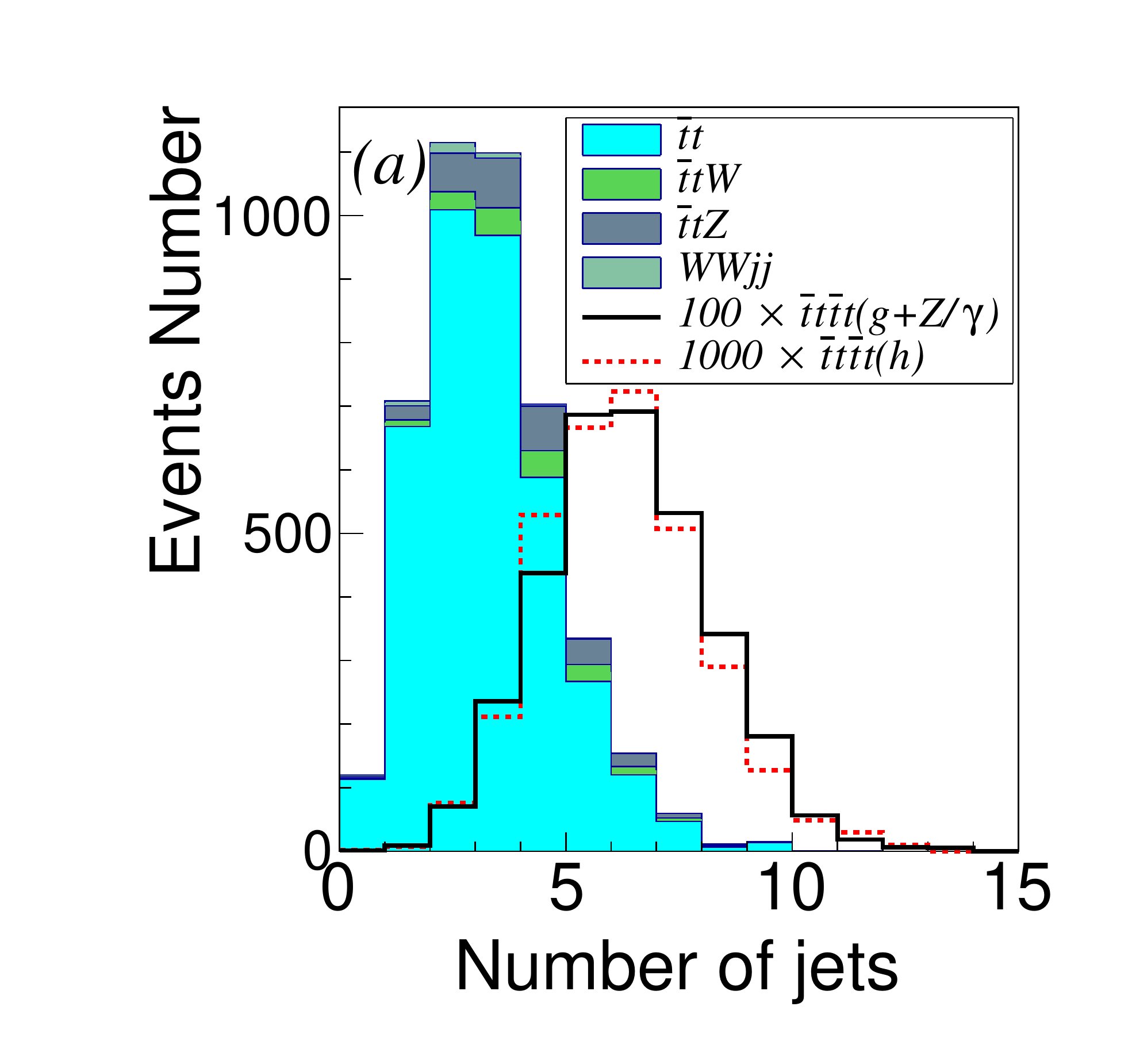}
\hspace*{-0.5cm}
\includegraphics[scale=0.4]{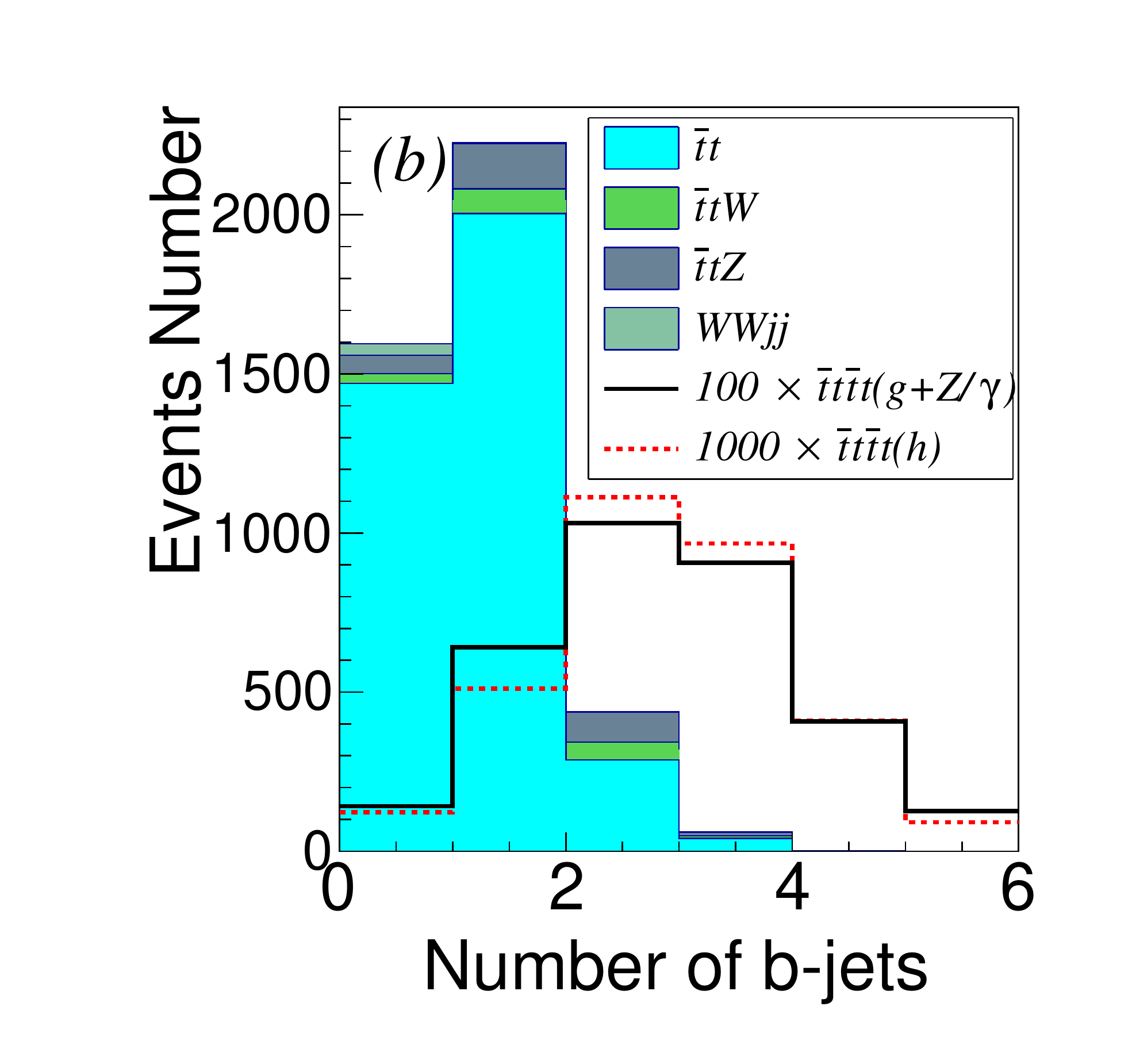}
\caption{The numbers of the reconstructed jets (a) and $b$-tagged jets (b) in the signal and background events at the 100~TeV collider  with an integrated luminosity of $1~{\rm fb}^{-1}$. To better character the signal distribution the cross section has been rescaled to 1000 times. No cuts except for same-sign lepton pair have been applied.}
\label{nbj}
\end{figure}
It is clear that the signal event exhibits often five or more jets. Demanding at least three identified $b$-jets would efficiently reject those SM backgrounds.  In the simulation we impose a set of kinematics cuts as follows:
\begin{align}
& p_T^{j,\ell}\geq 20~{\rm GeV}, && |\eta^{j,\ell}|<2.5, &&\not{\!\!{\rm E}}_{T}\ge 150~{\rm GeV},\nonumber\\
&N_{\ell^\pm}=2, &&N_{\rm jets}\geq 6, &&N_{b-{\rm jets}} \geq 3, \nonumber\\
&m_T \geq 100~{\rm GeV}, &&H_T \geq 800~{\rm GeV}.&& \label{eq:cuts}
\end{align}
Here $m_T$ denotes the transverse mass of the leading charged lepton ($\ell_1$) and the $\not{\!\!{\rm E}}_{T}$, defined as
\begin{equation}
m_T = \sqrt{2p_T^{\ell_1}\not{\!\!{\rm E}}_{T} (1 - \cos \Delta \phi)}, 
\end{equation}
where $\Delta\phi$ is the azimuthal angle between the $\ell_1$ lepton and the $\not{\!\!{\rm E}}_{T}$. The $m_T$ cut is to remove those backgrounds involving  leptonically decayed $W$ bosons. The $H_T$ is the  the scalar sum of the transverse momenta of all the visible particles and the missing energy $\not{\!\!{\rm E}}_{T}$.

Table~\ref{tbl:100tev} shows the numbers of the signal and the background events after a series of kinematics cuts at the 100~TeV proton-proton collider with an integrated luminosity of $1~{\rm ab}^{-1}$. 
\begin{table}[tb]
\centering
\caption{Number of the signal and background events at the $100\,$TeV $pp$ collider with an integrated luminosity of $1~{\rm ab}^{-1}$. 
The kinematics cuts listed in each row are applied sequentially.}
\label{tbl:100tev}
\begin{tabular}{c|c|c|c|c|c|c} \hline
&Basic &SSL&Jets&$\not{\!\!{\rm E}}_{T}$ &$m_T$&$H_T$ \\ \hline
 $\bar{t}t\bar{t}t_H$ &271300&3227.1& 1010.6&412.4&242.8& 222.5 \\ \hline
 $\bar{t}t\bar{t}t_{g+Z/\gamma}$&3276000  &32366.9&11056.5&4193.3 &2620.8&2407.9 \\ \hline
 $\bar{t}t\bar{t}t_{\rm int}$&-356900&-4040.1&-1275.9&-467.5&-273.0&-253.4 \\ \hline
 $\bar{t}t$&$3.22\times 10^{10}$ & 3802170 & 33411 & 0 &0 &0 \\ \hline
 $\bar{t}tW^+$& 2596250& 91917.4&1222.2& 509.2& 356.5& 356.5 \\ \hline
 $\bar{t}tW^-$& 1810460 & 81234.8&1585.5 & 629.4& 399.4 &  387.3 \\ \hline
 $\bar{t}tZ$ &4311270& 306908& 3995.6 & 1109.9&665.9& 621.5 \\ \hline
 $W^\pm W^\pm j j$ &275500& 39097.4& 2.188 &0&0&0 \\ \hline
\end{tabular}
\end{table}
The $t\bar{t}t\bar{t}$ production through the QCD interaction and the production through the Higgs boson mediator share similar kinematics, therefore, both the QCD and Higgs mediated productions exhibit similar efficiencies for each cut shown in Table~\ref{tbl:100tev}; see the second and third columns. It might be possible to distinguish the two contributions using the so-called color pull technique~\cite{Gallicchio:2010sw}. The major backgrounds in the SM are from the $t\bar{t}W^\pm$ and $t\bar{t}Z$ productions. 

After applying the cuts given in Eq.~\ref{eq:cuts}, the $t\bar{t}t\bar{t}$ production from the QCD and electroweak gauge interactions dominates over the SM backgrounds; see Table.~\ref{tbl:100tev}. The $t\bar{t}t\bar{t}$ production in the SM can be measured at a 5$\sigma$ confidence level with an integrated luminosity of 8.95 fb$^{-1}$. We thus expect the $t\bar{t}t\bar{t}$ production to be discovered soon after the operation of the 100 TeV machine.  The great potential enables us to discuss the precision of measuring the top Yukawa coupling in the $t\bar{t}t\bar{t}$ production. We estimate the signal statistical fluctuation as
\begin{align}
\Delta N_S = \sqrt{N_S + N_B},
\end{align}
assuming that the events number satisfies the Gaussian distribution.
The signal uncertainty is $\Delta N_S = 0.0095 N_S$ for $\mathcal{L}=10~{\rm ab}^{-1}$, $\Delta N_S = 0.0067 N_S$ for $\mathcal{L}=20~{\rm ab}^{-1}$, and $\Delta N_S = 0.0055 N_S$ for $\mathcal{L}=30~{\rm ab}^{-1}$, respectively. 
We interpret the uncertainty of the signal event as the uncertainty of the top Yukawa coupling, i.e. 
\begin{align}
\Delta N_S = \delta \kappa_t  \Big[2  \sigma^{\rm SM}(t\bar{t}t\bar{t})_{\rm int} + 4 \sigma^{\rm SM}(t\bar{t}t\bar{t})_H \Big] \times \mathcal{L} +\mathcal{O}(\delta \kappa_t^2),
\end{align}
where $\delta \kappa_t\equiv \kappa_t - 1$ and the SM cross sections refer to the values after all the cuts shown in the last column in Table~\ref{tbl:100tev}.
It yields a precision of $\kappa_t$ measurement as follows: $0.927 \leq \kappa_t  \leq 1.051$ for $\mathcal{L}=10~{\rm ab}^{-1}$, $0.952 \leq \kappa_t  \leq 1.038$ for $\mathcal{L}=20~{\rm ab}^{-1}$, and $0.962 \leq \kappa_t  \leq 1.031$ for $\mathcal{L}=30~{\rm ab}^{-1}$, respectively.

Figure~\ref{fig:limits} displays the correlation between $R_\Gamma$ and $\kappa_t$ imposed by the projected $\overline{\mu}^{b\bar{b}}_{t\bar{t}H}$ measurement~\cite{Plehn:2015cta}; see the red band.  
\begin{figure}[tb]
\centering
\includegraphics[scale=0.5]{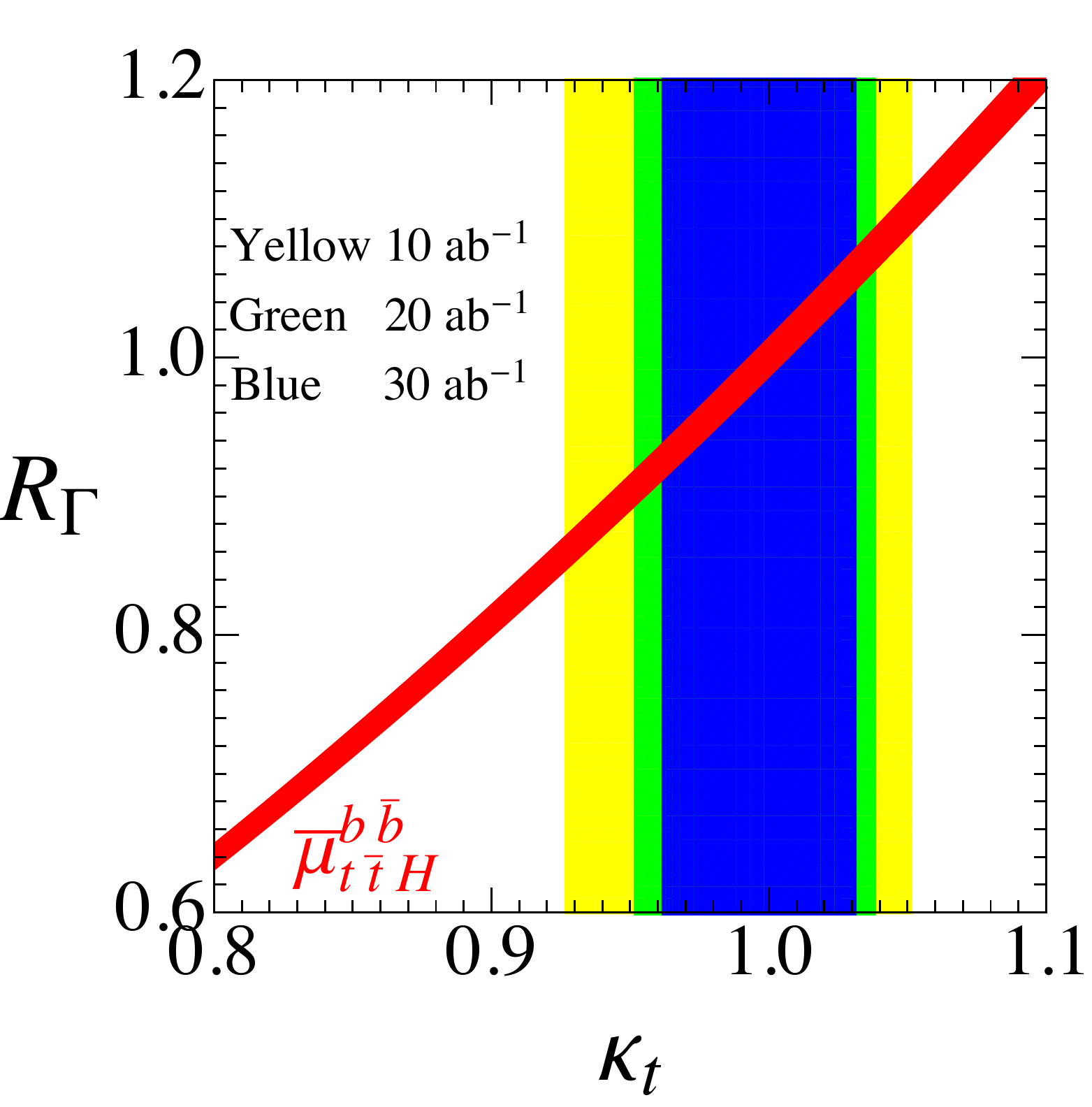}
\caption{Relative uncertainty on the signal strength $\mu_{t\bar{t}H}$ projected in the plane $\kappa_t$ and $R_\Gamma$ at a 100 TeV hadron collider with $20~{\rm ab}^{-1}$ for the Higgs decay modes $H\to b\bar{b}$ (red band). The yellow (green, blue) vertical band denotes the limit $0.927 \leq \kappa_t \leq 1.051$ ( $0.952 \leq \kappa_t \leq 1.038$, $0.962 \leq \kappa_t \leq 1.031$ ) corresponding to the $1\sigma$ signal uncertainty with the integrated luminosity of 10 ab$^{-1}$ (20 ab$^{-1}$, 30 ab$^{-1}$).}  
\label{fig:limits}
\end{figure}
The expectations of the $\kappa_t$ measurement in the $t\bar{t}t\bar{t}$ production are also plotted where the yellow (green, blue) contour region denotes the uncertainty of $\kappa_t$ with $\mathcal{L}=10~{\rm ab}^{-1}$ ($20~{\rm ab}^{-1}$, $30~{\rm ab}^{-1}$), respectively. Combining both the $t\bar{t}H$ and $t\bar{t}t\bar{t}$ productions imposes a tight bound on the Higgs boson width; for example, $0.85~\Gamma_{\rm H}^{\rm SM} \leq \Gamma_{\rm H} \leq 1.12~\Gamma_{\rm H}^{\rm SM}$ for $\mathcal{L}=10~{\rm ab}^{-1}$, $0.89~\Gamma_{\rm H}^{\rm SM} \leq \Gamma_{\rm H} \leq 1.09~\Gamma_{\rm H}^{\rm SM}$ for for $\mathcal{L}=20~{\rm ab}^{-1}$, and $0.91~\Gamma_{\rm H}^{\rm SM} \leq \Gamma_{\rm H} \leq 1.08~\Gamma_{\rm H}^{\rm SM}$ for $\mathcal{L}=30~{\rm ab}^{-1}$, respectively.

\subsection{Rare SM Exclusive Higgs decays}
\label{sec:H_raredec}
The measurement of the rare exclusive decays $H\to V\gamma$, where $V$ denotes a vector meson, would allow a unique probe of the Higgs coupling to light quarks. While the absolute value of the bottom-quark Yukawa coupling can be accessed by measuring $b$-tagged jets in the associated production of the Higgs boson with a $W$ or $Z$ boson, this method becomes progressively more difficult for the lighter-quark couplings.  Advanced charm-tagging techniques may allow some access to the charm-quark Yukawa coupling~\cite{Perez:2015lra}, but no other way of directly measuring even lighter-quark couplings is currently known. The small branching ratios for these exclusive decays renders them inaccessible at future $e^+ e^-$ colliders. The program of measuring these decay modes is therefore unique to hadron-collider facilities.  The large Higgs boson production rate at a proposed 100 TeV collider makes this facility an ideal place to measure these otherwise inaccessible quantities.  

The possibility of measuring rare exclusive Higgs decays was first pointed out in~\cite{Bodwin:2013gca,Kagan:2014ila}, and the theoretical framework for their prediction was further developed in~\cite{Bodwin:2014bpa,Grossmann:2015lea,Koenig:2015pha}. Our discussion follows closely the techniques introduced in these works, and we only summarize the salient features here. We begin our discussion of the theoretical predictions for these modes by introducing the effective Yukawa Lagrangian
\begin{equation}
   {\cal L} = - \sum_q\,\kappa_q\,\frac{m_q}{v}\,H\,\bar{q}_L q_R 
    - \sum_{q\ne q'}\,\frac{y_{qq'}}{\sqrt2}\,H\,\bar{q}_L q'_R + h.c. \,,
\end{equation}
where in the SM $\kappa_q=1$ while the flavor-changing Yukawa couplings $y_{qq'}$ vanish. The effective Lagrangian leads to two categories of exclusive Higgs decays: flavor-conserving decays involving the $\kappa_q$ couplings, where $V=\rho,\omega,\phi,J/\psi,\Upsilon(nS)$, and flavor-violating decays involving the $y_{qq'}$ couplings, where $V=B^{*0}_s,B^{*0}_d,K^{*0},D^{*0}$. In view of the very strong indirect bounds on flavor off-diagonal Higgs couplings to light quarks~\cite{Harnik:2012pb}, the flavor-violating decays $H\to V\gamma$ are bound to be very strongly suppressed. We will therefore restrict our discussion here to flavor-conserving processes.

The exclusive decays $H\to V\gamma$ are mediated by two distinct mechanisms, which interfere destructively.
\begin{itemize}
\item 
In the {\em indirect process}, the Higgs boson decays (primarily through loops involving heavy top quarks or weak gauge bosons) to a real photon $\gamma$ and a virtual $\gamma^*$ or $Z^*$ boson, which then converts into the vector meson $V$. This contribution only occurs for the flavor-conserving decay modes. The effect of the off-shellness of the photon and the contribution involving the $H\gamma Z^*$ coupling are suppressed by $m_V^2/M_H^2$ and hence are very small~\cite{Koenig:2015pha}.
\item 
In the {\em direct process}, the Higgs boson decays into a pair of a quark and an antiquark, one of which radiates off a photon. This mechanism introduces the dependence of the decay amplitude on the $\kappa_q$ parameters. The formation of the vector meson out of the quark-antiquark pair involves some non-trivial hadronic dynamics.
\end{itemize}
The relevant lowest-order Feynman diagrams contributing to the direct and indirect processes are shown in Figure~\ref{figure:diagrams}.

\begin{figure}
\centering
\includegraphics[width=0.28\textwidth]{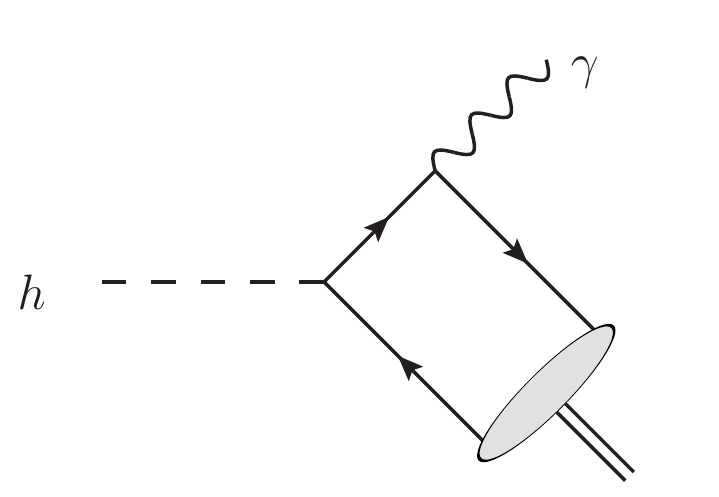}
\includegraphics[width=0.28\textwidth]{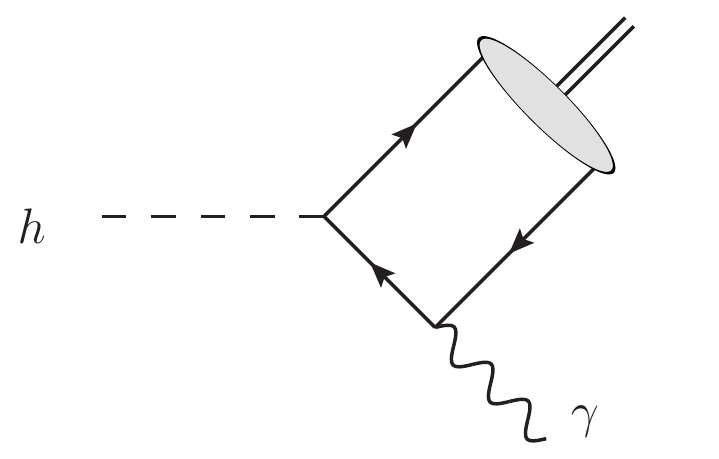}
\includegraphics[width=0.25\textwidth]{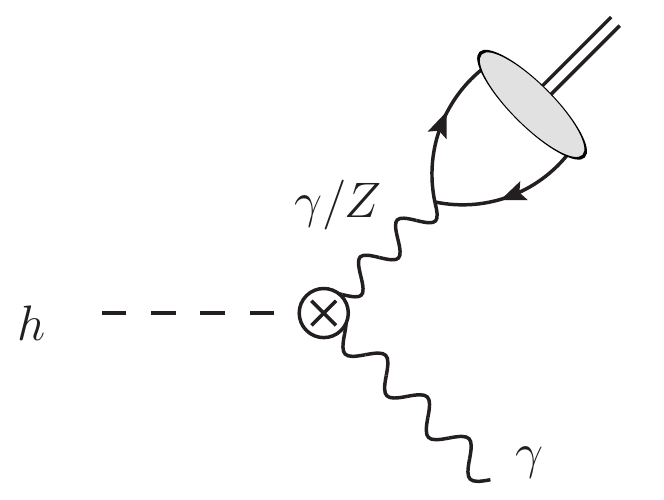}
\caption{\label{figure:diagrams}
Direct (left and center) and indirect (right) contributions to the $H\to V\gamma$ decay amplitude. The blob represents the non-perturbative meson wave function. The crossed circle in the third diagram denotes the off-shell $H\to\gamma\gamma^*$ and $H\to\gamma Z^*$ amplitudes, which in the SM arise first at one-loop order.}
\end{figure}

We begin by outlining the calculation of the indirect amplitude. The virtual photon of $Z$ boson couples to
the vector meson through the matrix element of a local current, which
can be parameterized in terms of a single hadronic parameter: the
vector-meson decay constant $f_V$. This quantity can be obtained
directly from experimental data. In particular, the leptonic decay
rate of the vector meson can be written as 
\begin{equation}
   \Gamma(V\to l^+l^-) = \frac{4\pi Q_V^2 f_V^2}{3 m_V} \alpha^2(m_V) \,,
\end{equation}
where $Q_V$ is the relevant combination of quark electric charges. The
effective couplings $H\gamma\gamma^{*}$ and $H\gamma Z^*$ vertices,
which appear in the indirect amplitude, can be calculated with high
accuracy in the SM. The by far dominant contributions involve loop
diagrams containing heavy top quarks or $W$ bosons. The two-loop
electroweak and QCD corrections to this amplitude are known, and when
combined shift the leading one-loop expression by less than 1\% for
the measured value of the Higgs boson
mass~\cite{Degrassi:2005mc}. However, physics beyond the SM could
affect these couplings in a non-trivial way, either through
modifications of the $Ht\bar t$ and $HW^+W^-$ couplings or by means of
loops containing new heavy particles. The measurement of the
light-quark couplings to the Higgs should therefore be considered
together with the extraction of the effective $H\gamma\gamma$
coupling. As pointed out in~\cite{Koenig:2015pha}, by taking the ratio
of the $H\to V\gamma$ and $H\to\gamma\gamma$ branching fractions one
can remove this sensitivity to unknown new contributions to the
$H\gamma\gamma$ coupling. 

We now consider the theoretical prediction for the direct
amplitude. This quantity cannot be directly related to data, unlike
the indirect amplitude. Two theoretical approaches have been used to
calculate this contribution. The hierarchy $M_H\gg m_V$ implies that
the vector meson is emitted at very high energy $E_V\gg m_V$ in the
Higgs-boson rest frame. The partons making up the vector meson can
thus be described by energetic particles moving collinear to the
direction of $V$. This kinematic hierarchy allows the QCD
factorization approach~\cite{Lepage:1980fj,Efremov:1979qk} to be
utilized. Up to corrections of order $(\Lambda_{\rm QCD}/M_H)^2$ for
light mesons, and of order $(m_V/M_H)^2$ for heavy vector mesons, this
method can be used to express the direct contribution to the $H\to
V\gamma$ decay amplitude as a perturbatively calculable
hard-scattering coefficient convoluted with the leading-twist
light-cone distribution amplitude (LCDA) of the vector meson. This
approach was pursued in~\cite{Koenig:2015pha}, where the full
next-to-leading order (NLO) QCD corrections were calculated and large
logarithms of the form $[\alpha_s\ln(M_H/m_V)]^n$ were resummed at
NLO, and in~\cite{Kagan:2014ila}, where an initial LO analysis was
performed. The dominant theoretical uncertainties remaining after this
calculation are parametric uncertainties associated with the
non-perturbative LCDAs of the vector mesons. Thanks to the high value
$\mu\sim M_H$ of the factorization scale, however, the LCDAs are close
to the asymptotic form $\phi_V(x,\mu)=6x(1-x)$ attained for
$\mu\to\infty$, and hence the sensitivity to yet not well known
hadronic parameters turns out to be mild. For the heavy vector mesons
$V=J/\psi,\Upsilon(nS)$, the quark and antiquark which form the meson
are slow-moving in the $V$ rest frame. This allows the
non-relativistic QCD framework (NRQCD)~\cite{Bodwin:1994jh} to be
employed to facilitate the calculation of the direct amplitude. This
approach was pursued in~\cite{Bodwin:2014bpa}, where the NLO
corrections in the velocity $v$ of the quarks in the $V$ rest frame,
the next-to-leading order corrections in $\alpha_s$, and the
leading-logarithmic resummation of collinear logarithms were
incorporated into the theoretical predictions. The dominant
theoretical uncertainties affecting the results for $H\to
J/\psi\,\gamma$ and $H\to\Upsilon(nS)\,\gamma$ after the inclusion of
these corrections are the uncalculated ${\cal O}(v^4)$ and ${\cal
  O}(\alpha_s v^2)$ terms in the NRQCD expansion. 

\begin{table}
\begin{center}
\begin{tabular}{|c|ccc|}
\hline 
Mode & \multicolumn{3}{c|}{Branching Fraction [$10^{-6}$]} \\
Method &  ~~NRQCD~\cite{Bodwin:2014bpa}~~ & ~~LCDA LO~\cite{Kagan:2014ila}~~
 & ~~LCDA NLO~\cite{Koenig:2015pha}~~ \\
\hline 
$\mbox{Br}(H\to\rho^0\gamma)$ & -- & $19.0\pm 1.5$ & $16.8\pm 0.8$ \\
$\mbox{Br}(H\to\omega\gamma)$ & -- & $1.60\pm 0.17$ & $1.48\pm 0.08$ \\
$\mbox{Br}(H\to\phi\gamma)$ & -- & $3.00\pm 0.13$ & $2.31\pm 0.11$ \\
$\mbox{Br}(H\to J/\psi\,\gamma)$ & $2.79\,_{-0.15}^{+0.16}$ & -- & $2.95\pm 0.17$   \\
$\mbox{Br}(H\to\Upsilon(1S)\,\gamma)$ & $(0.61\,_{-0.61}^{+1.74})\cdot 10^{-3}$ & -- 
 &  $(4.61\,_{-\,1.23}^{+\,1.76})\cdot 10^{-3}$ \\
$\mbox{Br}(H\to\Upsilon(2S)\,\gamma)$ & $(2.02\,_{-1.28}^{+1.86})\cdot 10^{-3}$  & -- 
 & $(2.34\,_{-\,1.00}^{+\,0.76})\cdot 10^{-3}$ \\
$\mbox{Br}(H\to\Upsilon(3S)\,\gamma)$ & $(2.44\,_{-1.30}^{+1.75})\cdot
10^{-3}$ & --  
 & $(2.13\,_{-\,1.13}^{+\,0.76})\cdot 10^{-3}$ \\
\hline 
\end{tabular}
\parbox{15.5cm}
{\caption{\label{table:SM_BRs} 
Theoretical predictions for the $H\to V\gamma$ branching ratios in the
SM, obtained using different theoretical approaches.}} 
\end{center}
\end{table} 

Table~\ref{table:SM_BRs} collects theoretical predictions for the
various $H\to V\gamma$ branching fractions in the SM. The inclusion of
NLO QCD corrections and resummation help to reduce the theoretical
uncertainties. There is in general good agreement between the results
obtained by different groups. The $H\to\phi\gamma$ branching ratio
obtained in~\cite{Koenig:2015pha} is lower than that found
in~\cite{Kagan:2014ila} because of an update of the $\phi$-meson decay
constant performed in the former work. Also, in~\cite{Koenig:2015pha}
the effects of $\rho$--$\omega$--$\phi$ mixing are taken into
account. One observes that the $H\to V\gamma$ branching fractions are
typically of order few times $10^{-6}$, which makes them very
challenging to observe. The most striking feature of the results shown
in the table concerns the $H\to\Upsilon(nS)\,\gamma$ modes, whose
branching fractions are very strongly suppressed. This suppression
results from an accidental and almost perfect cancellation between the
direct and indirect amplitudes, as first pointed out
in~\cite{Bodwin:2013gca}. In the case of $H\to\Upsilon(1S)\,\gamma$
the cancellation is so perfect that the small imaginary part of the
direct contribution induced by one-loop QCD corrections gives the
leading contribution to the decay amplitude. The fact that this
imaginary part was neglected in~\cite{Bodwin:2014bpa} explains why a
too small branching fraction for this mode was obtained there. 

\subsubsection{Experimental prospects}
The considered rare exclusive Higgs boson decays to a quarkonium and a
photon, are -- currently -- the only available means to probe the
quark Yukawa coupling in the first and second generation. The only
exception being, as pointed out earlier, the possibility to implement
advanced charm-tagging techniques, specifically to probe the
charm-quark Yukawa coupling. As a result, these Higgs boson decays are particulary
interesting from an experimental perspective, both as signatures unique 
to the hadron collider programme and as experimental
topologies. Furthermore, similar rare and exclusive decays of the
$W^\pm$ and $Z$ bosons have also attracted
interest~\cite{Mangano:2014xta,Grossmann:2015lea,Huang:2014cxa},
offering a physics 
programme in precision quantum chromodynamics (QCD), electroweak
physics, and physics beyond the SM.

Using 20.3~fb$^{-1}$ of 8\,TeV proton-proton collision data, the ATLAS
Collaboration has performed a search for Higgs and $Z$ boson decays to
$J/\psi\,\gamma$ and $\Upsilon(nS)\,\gamma$
($n=1,2,3$)~\cite{Aad:2015sda}. No significant excess has been
observed and 95\% confidence level upper limits were placed on the
respective branching ratios.  In the $J/\psi\,\gamma$ final state the
limits are $1.5\times10^{-3}$ and $2.6\times10^{-6}$ for the Higgs and
$Z$ boson decays, respectively, while in the
$\Upsilon(1S,2S,3S)\,\gamma$ final states the limits are
$(1.3,1.9,1.3)\times10^{-3}$ and $(3.4,6.5,5.4)\times10^{-6}$,
respectively. The CMS Collaboration has placed a 95\% C.L. upper limit
of $1.5\times10^{-3}$ on the $h\to J/\psi\,\gamma$ branching
ratio~\cite{Khachatryan:2015lga}. In all cases, the SM production rate
for the observed Higgs boson is assumed. Currently, no other direct
experimental constraint on these decays is available.

The scope of these early experimental investigations is two-fold: On
one hand to provide the first direct experimental constraints on these
quantities, and on the other hand to map the experimental challenges
involved in such searches. Looking to the future, the ATLAS
Collaboration estimated the expected sensitivity for Higgs and $Z$
boson decays to a $J/\psi$ and a photon, assuming up to 3000~fb$^{-1}$
of data collected with the ATLAS detector at the centre-of-mass energy
of 14\,TeV, during the operation of the High Luminosity LHC (HL-LHC). The
expected sensitivity for the $h\to J/\psi\,\gamma$ branching ratio,
assuming 300 and 3000~fb$^{-1}$ at 14 TeV, is $153\times 10^{-6}$ and
$44\times 10^{-6}$, respectively~\cite{ATL-PHYS-PUB-2015-043}. The
corresponding sensitivities for the $Z\to J/\psi\,\gamma$ branching
ratios are $7\times 10^{-7}$ and $4.4\times 10^{-7}$,
respectively~\cite{ATL-PHYS-PUB-2015-043}.  In this analysis, the same
overall detector performance as in LHC Run 1 is assumed, while an
analysis optimisation has been performed and a multivariate
discriminant using the same kinematic information as the published
analysis~\cite{Aad:2015sda} has been introduced. The main limiting
factor in reaching SM sensitivity was identified to be the number of
expected signal events, where only about 3 events were expected
following the complete event selection for the complete
HL-LHC. Moreover, as the search sensitivity approaches the SM
expectation for the $h\to J/\psi\,\gamma$ branching ratio, the
contribution from $h\to \mu\mu\gamma$ decays, with a non-resonant
dimuon pair, needs to be included. These can be separated efficiently
from the $h\to J/\psi\,\gamma$ signal, using dimuon mass information.

Moving to the lighter quarks, the Higgs boson coupling to the
strange-quark can be probed through the $h\to\phi\gamma$ decay. The
subsequent $\phi\to K^+K^-$ decay features a large branching ratio of
about 49\% and gives access to a simple final state of a hard photon
recoiling against two collimated high transverse momentum tracks, as
can be seen in Fig.~\ref{fig:KKpt}. With the SM branching ratio
prediction presented in Table~\ref{table:SM_BRs}, about 6.5 events
are expected to be produced with 100 fb$^{-1}$ at 14\,TeV.  For the
first generation quarks, the $h\to\omega\gamma$ and $h\to\rho\gamma$
are being considered, followed by the $\omega\to\pi^+\pi^-\pi^0$ and
$\rho\to\pi^+\pi^-$ decays, both with large branching ratios of about
89\% and 100\%, respectively.  The corresponding expected number of
events, assuming the SM branching ratios for these decays, are about
7.6 and 96, respectively.  The experimental acceptance for these
decays, assuming reasonable geometrical acceptance and transverse
momentum requirements, is expected to range between 40 and
70\%~\cite{phigamma}. It is noted that the search for $\omega\gamma$
and $\rho\gamma$ final states is further complicated due to the large
natural width of the $\rho$ meson and the $\omega$-$\rho$
interference.

\begin{figure}{h}
 \centering
 \includegraphics[width=0.48\textwidth]{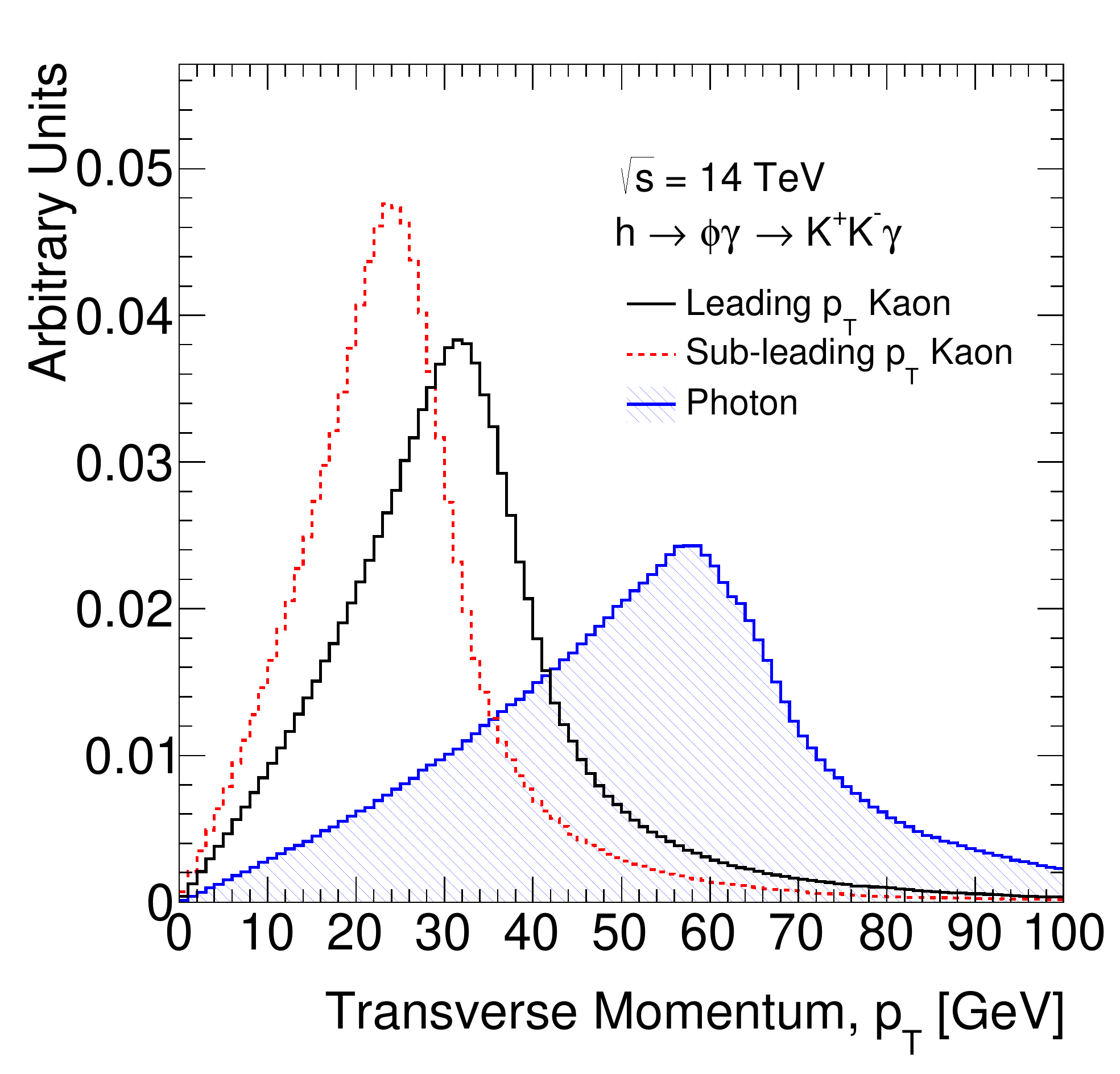}
\caption{Transverse momentum distribution of decay products in $h\to\phi\gamma\to K^+K^-\gamma$ decays~\cite{phigamma}.\label{fig:KKpt}}
\end{figure}

These rare decays to a vector meson and a photon feature very
interesting and experimentally challenging boosted topologies. The
signature is distinct, but the QCD backgrounds require careful
consideration. A primary challenge arises from the trigger
availability to collect the required datasets. In the considered
cases, the decay signature is a photon of large transverse momentum
that is isolated from hadronic activity, recoiling against a narrow
hadronic jet. It is important to consider such signatures, early on in
designing the trigger system. Fast track finding and reconstruction in
the inner detector, available at an early stage in the trigger could
help suppress backgrounds.

At the FCC-hh environment, the large production cross-section for the
signal and the large expected integrated luminosity alleviates the
main issue confronted by the studies at the LHC and the HL-LHC,
namely the small expected yields. With 20 ab$^{-1}$ at the
centre-of-mass energy of 100 TeV, a factor 100 increase in
the produced Higgs boson, with respect to HL-LHC, is expected. This substantial
increase in the signal yield, will also allow for more effective event
categorisation to further suppress the backgrounds. Production based
signatures, like the vector-boson-fusion or production in association
with a leptonically decaying $W$ or $Z$ boson will be
exploited. Furthermore, enhancement in the sensitivity can be expected
by exploiting the boosted regime, where the Higgs boson is produced
with substantial transverse momentum. Early studies on this have been
performed at the LHC~\cite{Aad:2015sda}, and careful evaluation of the
potential at 100 TeV is needed.

\clearpage

\clearpage
\section{Multi-Higgs production}
\label{sec:HH}

\label{sec:HH_intro}

In the previous sections we focused on processes involving the production of a single Higgs boson,
which allow one to test with high accuracy the linear Higgs interactions, most noticeably
those involving gauge bosons and third-generation SM fermions.
These processes, however, cannot be used to directly probe
interactions containing two or more Higgs fields,
whose determination is of primary importance for analyzing the Higgs potential.
Non-linear Higgs vertices can be accessed by looking at channels in which multiple Higgs bosons are produced
either alone or in association with additional objects.
In this section we will consider these channels with the aim of understanding the precision
with which the Higgs potential could be determined at a future $100$~TeV hadron collider.

\subsection{Parametrizing the Higgs interactions}

As we already mentioned, the main aim of the analyses that we will present in this section is to estimate the precision
with which the Higgs potential can be probed through the exploitation of multi-Higgs production processes.
It is thus useful to parametrize the relevant Higgs self-interactions in a general form.
In the language of an effective field theory, we can write the Higgs self-interaction Lagrangian as
\begin{equation}
{\cal L} = -\frac{1}{2} m^2_h h^2 - \lambda_3 \frac{m_h^2}{2v} h^3
- \lambda_4 \frac{m_h^2}{8 v^2}  h^4\,, \label{eq:higgs_pot}
\end{equation}
where $v = 246$~GeV denotes the Higgs vacuum expectation value. The SM Lagrangian is obtained by setting
$\lambda_3 = \lambda_4 = 1$; in this case the terms in Eq.~(\ref{eq:higgs_pot}) provide the whole
Higgs potential. On the contrary, in BSM scenarios, higher-order operators are in general also present,
as for instance contact interactions involving higher-powers of the Higgs field or additional derivatives.

The use of the parametrization in Eq.~(\ref{eq:higgs_pot}) can be fully justified in an effective-field-theory framework
in which an expansion in powers of the momenta is valid. Namely, we assume that each additional derivative in the
effective Lagrangian is accompanied by a factor $1/m_*$, where $m_*$ is a mass scale that broadly characterize a possible
new-physics dynamics. In this way the contribution of higher-derivative terms to low-energy observables
is suppressed by additional powers
of $E^2/m_*^2$, guaranteeing that the effective theory is valid for energy scales $E \ll m_*$. For most of the processes we are
going to consider the kinematic distributions are peaked mostly at threshold. Hence an analysis focusing on the
total cross section can be interpreted
in the effective-field-theory context provided that the new physics is at the TeV scale or beyond
($m_* \gtrsim 1$~TeV).\footnote{Possible issues with the effective description can instead arise in analyses focused on the
high-energy tails of the invariant mass distributions.}
It is important to stress that the parametrization in Eq.~(\ref{eq:higgs_pot}) does not rely on any expansion in powers
of the Higgs field. Operators involving more than four powers of $h$ are in fact irrelevant for the processes we are
considering and can be safely neglected.

In the case in which, in addition to the derivative expansion, we can also rely on an expansion in powers of the Higgs
field, the most relevant new-physics effects can be described in terms of dimension-$6$ operators~\cite{Buchmuller:1985jz,Giudice:2007fh,Grzadkowski:2010es}.
If the Higgs is part of an $SU(2)_L$ doublet $H$, only two effective operator contribute to the modification of the Higgs self-interactions,
namely\footnote{We neglect a third operator ${\cal O}_T = (H^\dagger \overleftrightarrow D_{\mu} H) (H^\dagger \overleftrightarrow D^{\mu} H)$,
since it breaks the custodial symmetry and is constrained by the EW precision measurements to have a very small coefficient.}
\begin{equation}
\Delta {\cal L}_6 \supset \frac{\overline c_H}{2 v^2} \partial_\mu (H^\dagger H) \partial^\mu (H^\dagger H)
- \frac{\overline c_6}{v^2} \frac{m_h^2}{2 v^2} (H^\dagger H)^3\,,
\end{equation}
where $H$ denotes the Higgs doublet. These operators induce corrections to the trilinear and quadrilinear
Higgs interactions, whose size is given by
\begin{equation}
\lambda_3 = 1 - \frac{3}{2} \overline c_H + \overline c_6\,, \qquad \lambda_4 = 1 - \frac{25}{3} \overline c_H + 6\, \overline c_6\,.
\end{equation}

It is important to stress that the operator ${\cal O}_H$ modifies several observables that can be also tested in
single-Higgs processes. For instance, it induces an overall rescaling of the linear couplings of the Higgs field
to the SM gauge bosons and to the fermions. The present LHC data already constrain these corrections
not to exceed the $\sim 10\%$ level. Moreover future lepton colliders could test these effects with very high accuracy,
reaching a precision of the order of a few percent~\cite{Dawson:2013bba}.
The operator ${\cal O}_6$, on the other hand, modifies only the Higgs self-interactions, and it can thus be tested directly only
in collider processes involving multiple Higgs production. Notice that, if the relevant new-physics effects are entirely due
to ${\cal O}_6$, the deviations in the trilinear and quadrilinear Higgs couplings are correlated.

Analogously to  the parametrization of the Higgs potential,  possible deviations
in the Higgs couplings to the gauge bosons can be parametrized by
\begin{equation}
{\mathcal L} = \left(m_W^2 W_\mu W^\mu + \frac{m_Z^2}{2} Z_\mu Z^\mu\right)
\left(1 + 2 c_V \frac{h}{v} + c_{2V} \frac{h^2}{v^2}\right)\,.
\end{equation}
These interactions are relevant for interpreting double Higgs production in the VBF channel. The SM Lagrangian
is recovered for $c_V = c_{2V} = 1$.

\subsubsection{Estimate of the size of new-physics corrections}

As we already mentioned, the measurement of multi-Higgs production processes can provide a significant test of the
validity of the SM. It is however also important to assess its impact in context of BSM scenarios.
In this case, multi-Higgs processes can be used to discover new-physics effects or, in
the case of a good agreement with the SM prediction, can be translated into exclusions on the parameter space of BSM
models. Obviously the impact on the various BSM scenarios crucially depends on the size of the expected deviations in
the Higgs potential and in the other Higgs couplings and on the possibility of disentangling these effects from
other possible corrections due to the presence of new resonances. In the following we will provide some estimates of these effects
in a few motivated BSM contexts.~\footnote{For additional details see Ref.~\cite{Azatov:2015oxa}.} 

As a preliminary observation,  notice that in BSM scenarios multiple Higgs production can be modified
in different ways. A first obvious effect comes from non-standard Higgs interactions, that is 
modified couplings already present in the SM or 
new (non-renormalizable) interactions.~\footnote{For example, new multi-Higgs interactions
are typically present in theories where the Higgs boson is a composite state of new strongly-coupled dynamics.
In these scenarios the non-linear Higgs dynamics implies the presence of new non-renormalizable Higgs
interactions (see for instance Ref.~\cite{Panico:2015jxa}).}
An additional effect can arise from the presence of new resonances, which can 
contribute through tree-level or loop diagrams.
In particular, if the new-physics is light, multi-Higgs production can receive resonant contributions from the on-shell production
of one or more resonances which afterwards decay into multi-Higgs final states.
Obviously, in the latter case a different search strategy must be employed to study the direct production of new states,
either through multi-Higgs production channels or in related processes. Searches for resonant double-Higgs production
have already been performed at the LHC~\cite{Khachatryan:2015yea,CMS:2015zug,ATLAS:2014rxa}.

If the new physics is relatively heavy, it is still useful to perform a non-resonant search for multi-Higgs production using 
a parametrization in terms of effective Higgs couplings.
Indeed, since multi-Higgs production cross sections are typically peaked not far from the kinematic
threshold, a new-physics scale $m_* \gtrsim 1$~TeV is high enough to ensure that resonant production
gives a subleading contribution to the total rates.  Therefore, the impact of the new physics on the total cross section can be
reliably described in terms of effective operators.


Let us now discuss the expected size of the corrections to the Higgs vertices. For definiteness we will concentrate on the
effects relevant for Higgs pair production (triple Higgs production can be analyzed along the same lines). The set of Higgs interactions
relevant for the various production channels is~\cite{Contino:2013kra,Panico:2015cst}
\arraycolsep = 2pt
\begin{eqnarray}
{\mathcal L} & \supset & \left(m_W^2 W_\mu W^\mu + \frac{m_Z^2}{2} Z_\mu Z^\mu\right)
\left(1 + 2 c_V \frac{h}{v} + c_{2V} \frac{h^2}{v^2}\right)
- \lambda_3 \frac{m_h^2}{2v} h^3\nonumber\\
&& -\; m_t \overline t t \left( 1 + c_t \frac{h}{v} + c_{2t} \frac{h^2}{2 v^2}\right)
+ \frac{g_s^2}{4 \pi^2}\left(c_g \frac{h}{v} + c_{2g} \frac{h^2}{2v^2}\right) G_{\mu\nu}^a G^{a\, \mu\nu}\,,\label{eq:nonlin_lagr}
\end{eqnarray}
where the SM corresponds to $c_V = c_{2V} = 1$, $c_t = 1$, $\lambda_3 = 1$ and $c_{2t} = c_g = c_{2g} = 0$.
Notice that, in addition to the dependence on the Higgs trilinear interaction, Higgs pair production is influenced by
several other vertex modifications. In particular, the gluon fusion process is also sensitive to corrections to the Yukawa coupling of the top quark
(and, in a much milder way, of the bottom quark), as well as to the presence of new contact interactions with gluons, which
could arise from loop contributions of new heavy states. In the case of Vector Boson Fusion (VBF), 
modified Higgs couplings to the gauge fields are also relevant.

In generic new-physics scenarios, corrections to all these couplings are present and can have comparable size.
Let us start by estimating these corrections in theories where the Higgs arises as a Nambu-Goldstone boson (NGB) from some new
strongly-coupled dynamics. In this case the SILH power counting implies~\cite{Giudice:2007fh,Panico:2015jxa} 
\begin{equation}\label{eq:ch_corr}
\delta \lambda_3, \delta c_V, \delta c_{2V}, \delta c_t, c_{2t} \sim \frac{v^2}{f^2}\,,
\qquad
\delta c_g, c_{2g} \sim \frac{v^2}{f^2}\frac{\lambda^2}{g_*^2}\,,
\end{equation}
where $g_*$ denotes the typical coupling strength of the strong dynamics, while $f$ is defined as $f = m_*/g_*$.
The corrections to couplings that are forbidden by the Goldstone symmetry are suppressed if the latter is broken
by a small amount. In particular, contact interactions with gluons are generated proportional to (the square of) some weak spurion coupling $\lambda$,
while corrections to the top Yukawa coupling and to the trilinear Higgs coupling are suppressed respectively by the factors
$y_t$ and $m_h^2/2v^2$ (notice that these factors have been already included in the definition of Eq.~(\ref{eq:nonlin_lagr})).

It is apparent from Eq.~(\ref{eq:ch_corr}) that in theories respecting the  SILH power counting the corrections to the various Higgs couplings are
all of comparable order. Higgs pair production is thus affected by all these effects simultaneously. In order to disentangle them
and extract the Higgs self-interactions one thus needs to use additional measurements (as for instance single-Higgs
production channels) and to adopt a more refined analysis strategy which makes use of  kinematic distributions~\cite{Azatov:2015oxa}.

There are however other new-physics scenarios where the corrections to the Higgs self-couplings can be enhanced and become larger than those
to the other couplings.
One scenario of this kind is obtained by assuming that the Higgs
is a generic composite state (not a NGB as assumed before) from a new strongly-coupled dynamics.
In this case the corrections to the Higgs self-interactions are enhanced by a factor $2 v^2 g_*^2/m_h^2$ compared to the
SILH case. One thus expects $\delta \lambda_3 \sim g_*^2 v^4/f^2 m_h^2$, which can be sizable even if $v^2/f^2 \ll 1$
(in which case the corrections to the linear Higgs couplings are small). The price to pay for this enhancement, however,
is an additional tuning that is required to keep the higgs mass small, since one would naturally expect $m_h^2 \sim m_*^2$.

Another scenario which leads to large corrections mainly to the Higgs self-couplings is obtained by considering
a new strong dynamics coupled to the SM through a Higgs portal~\cite{Azatov:2015oxa}: ${\cal L}_{int} = \lambda H^\dagger H O$, where
$O$ is a composite operator and $\lambda$ is the coupling strength. In this case one finds
\begin{equation}
\delta c_V \sim \delta c_{2V} \sim \delta c_t \sim c_{2t} \sim \frac{\lambda^2}{g_*^4} \frac{v^2}{f^2}\,,
\qquad
\delta \lambda_3 \sim \frac{2 v^2 \lambda}{m_h^2} \frac{\lambda^2}{g_*^4} \frac{v^2}{f^2}\,.
\end{equation}
The corrections to the Higgs trilinear self-coupling can be dominant if $\lambda > m_h^2/(2 v^2) \simeq 0.13$.
In this scenario it is thus possible to obtain $\delta \lambda_3 \sim 1$, while keeping the corrections to the other Higgs couplings
at the few percent level.

For other possible new physics giving rise to a modified Higgs
potential see also Section~\ref{sec:BSM} of this report.

\subsubsection{Production cross sections and summary of results}

To conclude this introduction, we present an overview of the various multi-Higgs production channels and we quickly
summarize the results of the analyses that will be presented in details in the following subsections.

Table~\ref{table:Higgspair}, extracted from the results of Ref.~\cite{HXSWG}, reports the rates for
SM Higgs pair and triple production, including channels of associated production with jets, gauge bosons and top quarks.
\begin{table}[tb]
\centering
\def\arraystretch{1.5}
\begin{tabular}{ @{\;} c | c | c | c @{\;}}
process & $\sigma(14~\textrm{TeV})\ \ (\textrm{fb})$ & $\sigma(100~\textrm{TeV})\ \ (\textrm{fb})$ & accuracy\\
\hline
\hline
$HH$ (ggf) & $45.05^{+4.4\%}_{-6.0\%} \pm 3.0\% \pm 10\%$ & $1749^{+5.1\%}_{-6.6\%} \pm 2.7\% \pm 10\%$ & \! NNLL matched to NNLO\!\\
\hline
$HHjj$ (VBF)\! & $1.94^{+2.3\%}_{-2.6\%} \pm 2.3\%$ & $80.3^{+0.5\%}_{-0.4\%} \pm 1.7\%$ & NLO\\
\hline
$HHZ$ & $0.415^{+3.5\%}_{-2.7\%} \pm 1.8\%$ & $8.23^{+5.9\%}_{-4.6\%} \pm 1.7\%$ & NNLO\\
$HHW^+$ & $0.269^{+0.33\%}_{-0.39\%} \pm 2.1\%$ & $4.70^{+0.90\%}_{-0.96\%} \pm 1.8\%$ & NNLO\\
$HHW^-$ & $0.198^{+1.2\%}_{-1.3\%} \pm 2.7\%$ & $3.30^{+3.5\%}_{-4.3\%} \pm 1.9\%$ & NNLO\\
\hline
$HHt\bar{t}$ & $0.949^{+1.7\%}_{-4.5\%} \pm 3.1\%$ & $82.1^{+7.9\%}_{-7.4\%} \pm 1.6\%$ & NLO\\
$HHtj$ & $0.0364^{+4.2\%}_{-1.8\%} \pm 4.7\%$ & $4.44^{+2.2\%}_{-2.6\%} \pm 2.4\%$ & NLO\\
\hline
$HHH$ & $0.0892^{+14.8\%}_{-13.6\%} \pm 3.2\%$ & $4.82^{+12.3\%}_{-11.9\%} \pm 1.8\%$ & NLO
\end{tabular}
\caption{Cross sections for production of two or three SM Higgs
bosons, including associated production channels, at a $14$~TeV and $100$~TeV hadron collider~\cite{HXSWG}. 
The cross sections are computed by choosing $\mu = M_{hh}/2$ ($\mu =
M_{hhh}/2$ in the case of triple production). 
The error intervals correspond to scale variation and PDF + $\alpha_s$
uncertainty. In $HH$ production in the gluon-fusion 
channel a conservative $10\%$ uncertainty is included to take into
account the effects of the infinite top-mass approximation 
(see Section~\ref{sec:status_ggHH}).}
\label{table:Higgspair}
\end{table}
The dependence of the production rates on the center-of-mass (COM) energy of the collider is shown in Fig.~\ref{fig:xsec_HH_COM}.
\begin{figure}[tb]
\centering
\includegraphics[width=0.55\textwidth]{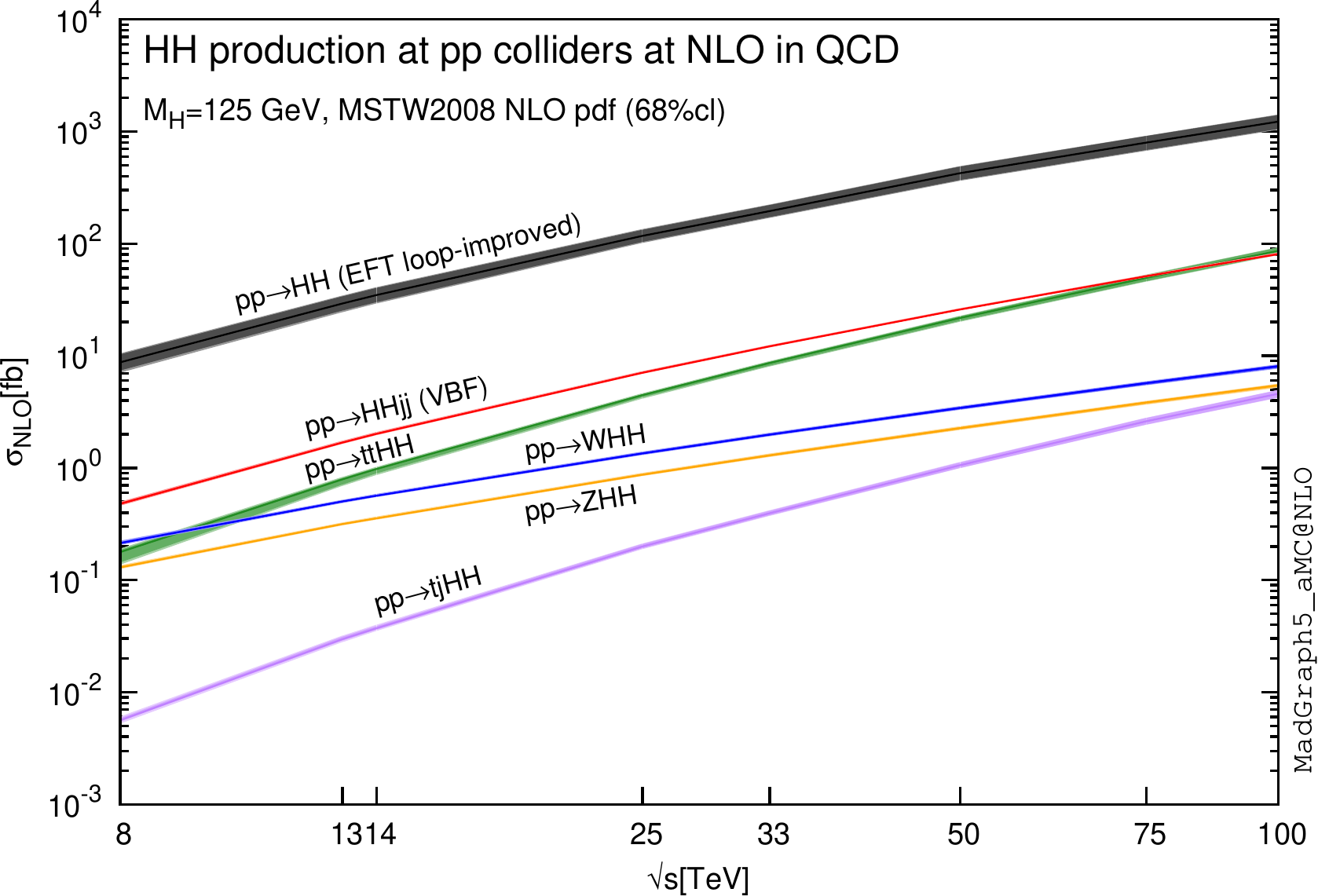}
\caption{Cross sections as a function of the collider COM energy. From ref.~\cite{Frederix:2014hta}.}
\label{fig:xsec_HH_COM}
\end{figure}
As for single-Higgs production, the dominant channel for Higgs pair production is gluon fusion, with a rate of $1750$~fb,
which constitutes more than $90\%$ of the total production rate. With respect to the $14$~TeV LHC, the gluon-fusion rate
is enhanced by a factor $\sim 40$. The second more significant channel is pair production in association with a top pair,
whose cross section is $82$~fb, closely followed by VBF with a rate of $80$~fb. Notice that the relative importance of these two
channels is reversed with respect to the $14$~TeV LHC case, where VBF was about twice larger than $HH\bar t t$.
The remaining pair production modes, in association with a gauge boson or with $tj$, play a secondary role, since their cross section
is at most $\sim 8$~fb. Finally, triple Higgs production has a cross section around $5$~fb.

As we already mentioned, the main aim of the analyses reported in this section is to determine the precision with which the
SM production rates and the Higgs self-couplings can be measured. It is thus important to analyze the dependence of the
cross section on the Higgs self-couplings. The production rates for the Higgs pair production channels are shown in Fig.~\ref{fig:xsec_lambda3}
as a function of the trilinear Higgs coupling $\lambda_3$. 
\begin{figure}[!t]
\centering
\includegraphics[width=0.55\textwidth]{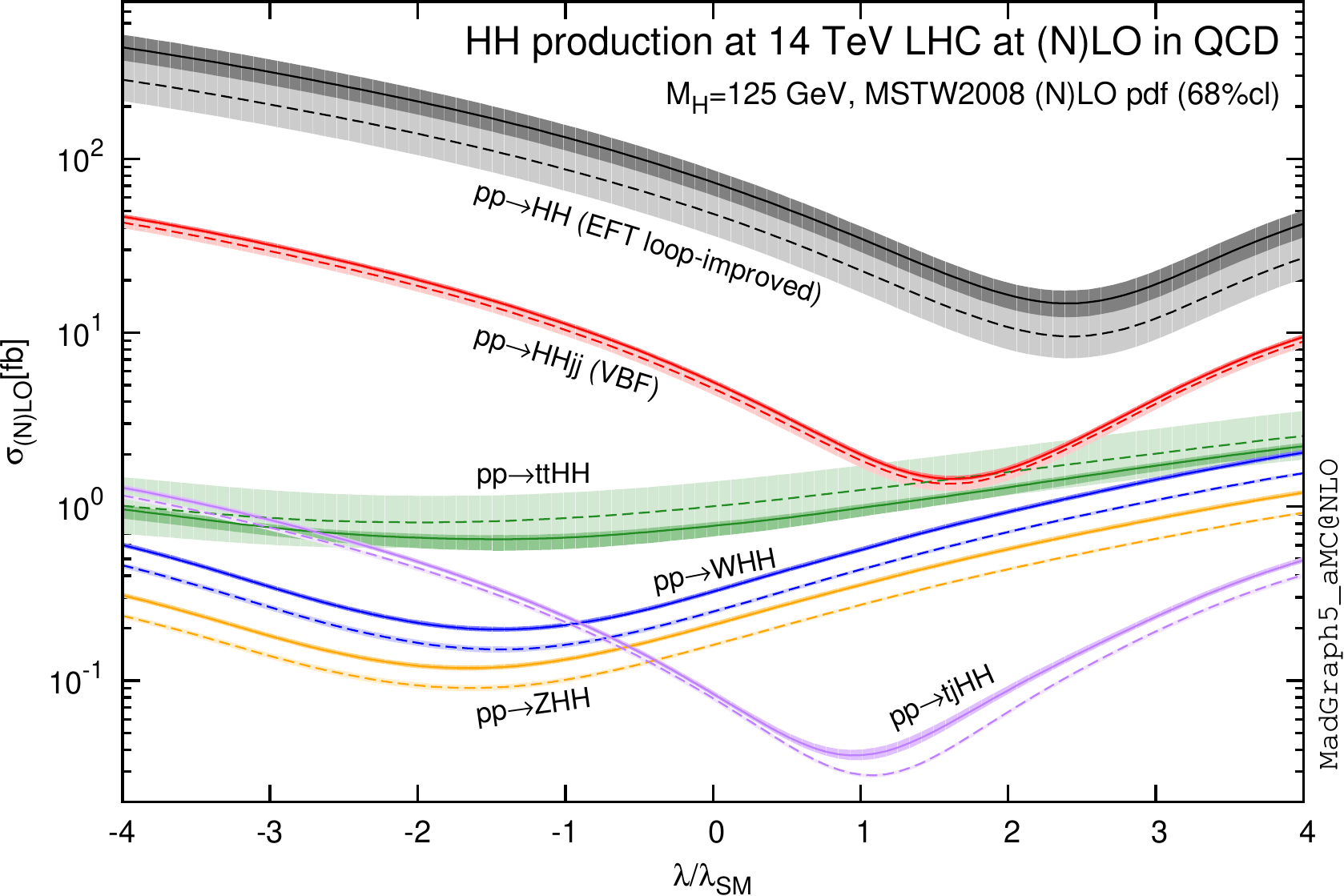}
\caption{Dependence of total cross sections on the Higgs trilinear coupling at $14\ \mathrm{TeV}$. From ref.~\cite{Frederix:2014hta}.}
\label{fig:xsec_lambda3}
\end{figure}
Although the plot shows the rates for the $14$~TeV LHC, it is approximately valid also at $100$~TeV. 
One can see that for $\lambda_3 \sim 1$, {\it i.e.}~for values close to the SM one,
a significant reduction in the cross section is present in the gluon-fusion and VBF channels and, even more, in the $HHtj$ channel.
This feature decreases the signal significance for the SM case. However, it allows one to more easily differentiate scenarios with
a modified trilinear coupling (especially if $\lambda_3 < 1$), since in these cases a large increase in the cross section is present.

In the following we will present a few analyses focused on the most important multi-Higgs production channels.
Here we summarize the main results. In particular, the expected precisions on the extraction of the SM signal cross section
and the Higgs self-couplings are listed in Table~\ref{table:HHprecision}.
\begin{table}
\centering
\def\arraystretch{1.5}
\begin{tabular}{l|c|c}
process & precision on $\sigma_{SM}$ & $68\%$ CL interval  on Higgs self-couplings\\
\hline
\hline
$HH \rightarrow b\overline b \gamma\gamma$ & $3\%$ & $\lambda_3 \in [0.97, 1.03]$\\
\hline
$HH \rightarrow b\overline b b \overline b$ & $5\%$ & $\lambda_3 \in [0.9, 1.5]$\\
\hline
$HH \rightarrow b\overline b 4 \ell$ & $O(25\%)$ & $\lambda_3 \in [0.6, 1.4]$ \\
$HH \rightarrow b\overline b \ell^+ \ell^-$ & $O(15\%)$ & $\lambda_3 \in [0.8,  1.2]$\\
$HH \rightarrow b\overline b \ell^+ \ell^- \gamma$ & $-$ & $-$\\
\hline
$HHH \rightarrow b\bar b b\bar b \gamma\gamma$ & $O(100\%)$ & $\lambda_4 \in [-4, +16]$
\end{tabular}
\caption{Expected precision (at $68\%$ CL) on the SM cross section and $68\%$ CL interval on the Higgs trilinear and quartic self-couplings (in SM units).
All the numbers are obtained for an integrated luminosity of $30$~ab$^{-1}$ and do not take into account possible systematic errors.}
\label{table:HHprecision}
\end{table}

Due to the sizable cross section, the gluon-fusion mode lends itself to the exploitation of several final states. As at the $14$~TeV LHC,
the $b\bar b \gamma\gamma$ final state remains the ``golden'' channel, since it retains a significant signal rate and allows one to
efficiently keep the backgrounds  under control. From this channel a statistical precision of the order of $1-2\%$ is expected on the
SM signal cross section, while the Higgs trilinear coupling could be determined with a precision of order $3-4\%$.
These numbers have to be compared with the precision expected at a possible future high-energy lepton collider,
at which the Higgs trilinear coupling is expected to be measurable with a precision $\sim 16\%$ for a COM energy $\sim 1$~TeV
and $2$~ab$^{-1}$ integrated luminosity~\cite{Tian:2013,Kurata:2013,Fujii:2015jha}.
A better precision, of around $12\%$, is only achievable with a $3$~TeV collider and
$2$~ab$^{-1}$ integrated luminosity~\cite{linssen,Abramowicz:2013tzc}.
Other final states, namely $b\bar b b\bar b$ and final states containing leptons, can also lead to a measurement of the SM signal,
although in these cases the expected significance is lower than in the $b\bar b \gamma\gamma$ channel.


Finally, the Higgs quartic self-coupling can be probed through the triple Higgs production channel. In this case the most promising
final state seems to be $b\bar b b\bar b \gamma\gamma$, whose cross section is however small. This channel could
allow an order-one determination of the SM production rate and could constrain the quartic coupling in the range
$\lambda_4 \in [ -4,  +16]$.

\subsection{Double Higgs production from gluon fusion}
\label{sec:HH_gf}

We start the presentation of the analyses of the various Higgs pair production channels by considering the gluon-fusion
process, which, as we saw, provides the dominant contribution to the total rate. At $100$~TeV, the gluon fusion cross section
computed at NNLL (matched to NNLO) accuracy is $1750$~fb~\cite{HXSWG}. At present, this result is affected by a significant uncertainty
(of the order of $10\%$) due to the fact that the NLO and NNLO contributions are only known in the infinite top mass limit.
A discussion of the current status of the computations and of the sources of uncertainties will be provided in
Subsection~\ref{sec:status_ggHH}.

In the SM the gluon fusion process receives contributions from two types of diagrams (see Fig.~\ref{fig:diagr_ggHH}).
\begin{figure}[t!]
\centering
\includegraphics[height=.12\textwidth]{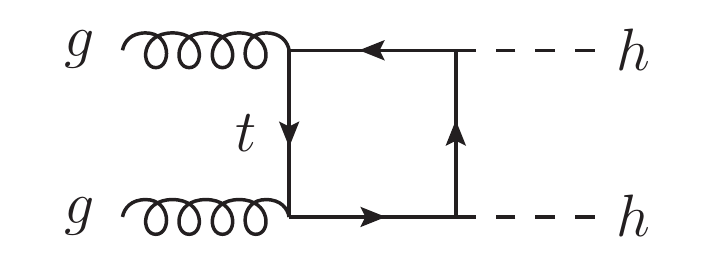}
\hspace{2em}
\includegraphics[height=.12\textwidth]{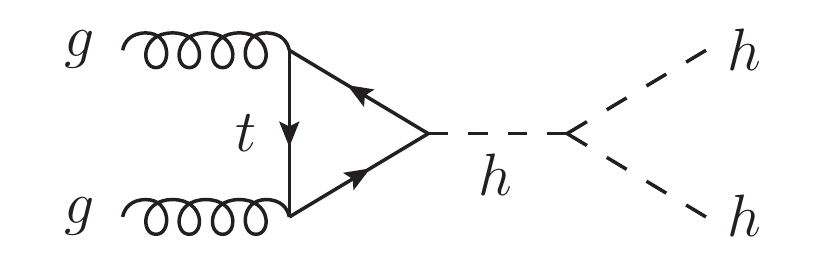}
\caption{Diagrams contributing to the Higgs pair production process through gluon fusion (an additional diagram
obtained by crossing the box one is not shown).
}\label{fig:diagr_ggHH}
\end{figure}
The box-type diagrams, which depend on the top Yukawa couplings, and the triangle-type one, which in addition to the
top Yukawa also includes the trilinear Higgs self-interaction.
In the SM a partial cancellation between these two kinds of diagrams is present, which leads
to a $\sim 50\%$ suppression of the total cross section.
The behavior of the box and the triangle diagrams at high $\sqrt{\hat s} = m_{hh} \gg m_t, m_h$ is quite different however.
The corresponding amplitudes scale as
\begin{equation}
{\cal A}_\square \sim \frac{\alpha_s}{4 \pi} y_t^2\,,
\qquad
{\cal A}_\triangle \sim \lambda_3 \frac{\alpha_s}{4 \pi} y_t^2 \frac{m_h^2}{\hat s}\left(\log \frac{m_t^2}{\hat s} + i \pi\right)^2\,.
\end{equation}
From these equations it is apparent that, due to the presence of the off-shell Higgs propagator, the triangle diagram
is suppressed for high $\hat s$. This implies that the Higgs trilinear coupling affects the $m_{hh}$ distribution mostly at threshold,
while  the tail at  large invariant mass is mostly determined by the box contribution.

The shape of the Higgs pair invariant mass distribution for the SM signal is shown in Fig.~\ref{dsdQ100}~\cite{deFlorian:2015moa}.
%
\begin{figure}[t!]
\centering
\includegraphics[width=.55\textwidth]{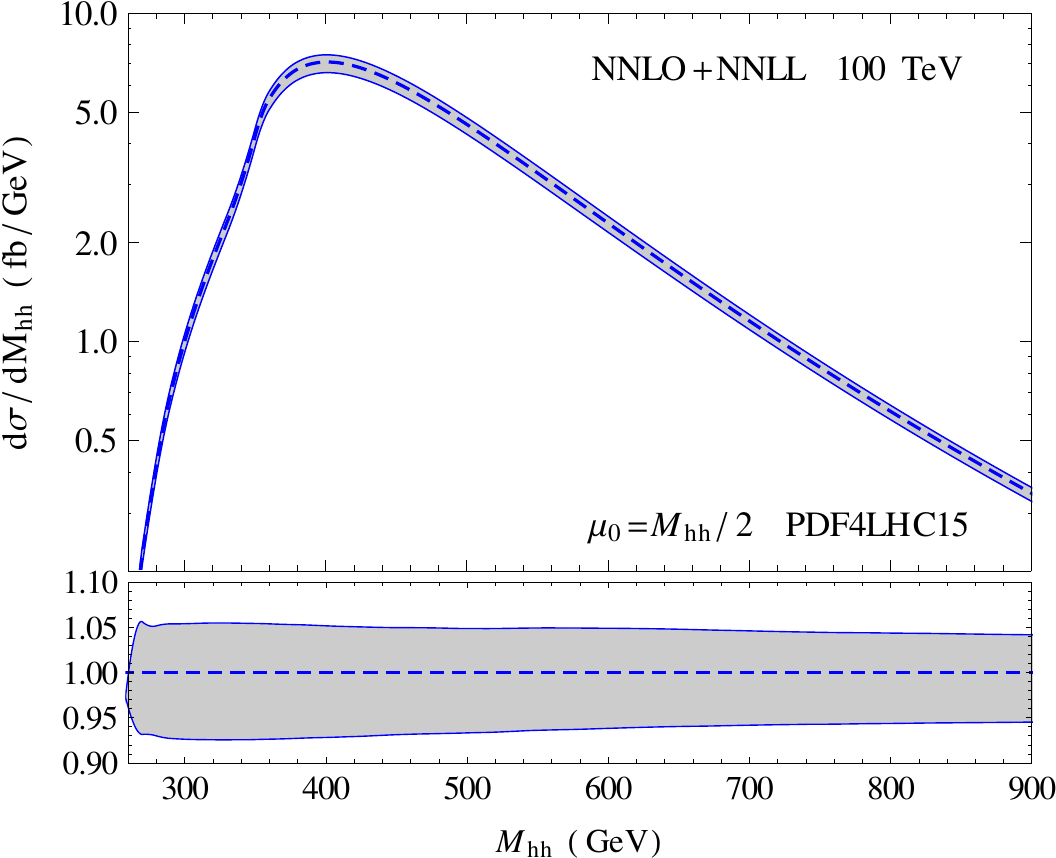}
\caption{\label{dsdQ100}
The invariant mass distribution at NNLO+NNLL \cite{deFlorian:2015moa} for a $100\text{ TeV}$ collider, with the corresponding scale uncertainty.
The lower panel shows the ratio with respect to the central prediction.
}
\end{figure}
%
The central line corresponds to the choice $\mu_F=\mu_R=M_{hh}/2$ for the factorization and renormalization scales, and
the band illustrates the scale uncertainty, evaluated by varying independently the above scales in the
range $\mu_0/2\leq\mu_R,\mu_F\leq 2\mu_0$ with the constraint $1/2\leq\mu_R/\mu_F<2$, where $\mu_0$ is the
central scale. The lower panel shows the ratio with respect to the central value, and it can be seen that the scale uncertainty
is roughly constant in the whole range, being of the order of $\pm 5\%$.
One can see that the peak of the distribution is at $m_{hh} \sim 400$~GeV and some suppression is present close to threshold.
The suppression is a consequence of the partial cancellation between the box and triangle diagrams that, as we already
mentioned, is present in the SM.

The invariant mass distribution at a $14$~TeV collider is similar to the one at $100$~TeV.
The comparison between the two distributions is shown in Fig.~\ref{fig:mhh_comparison}. 
%
\begin{figure}[t!]
\centering
\includegraphics[width=.45\textwidth]{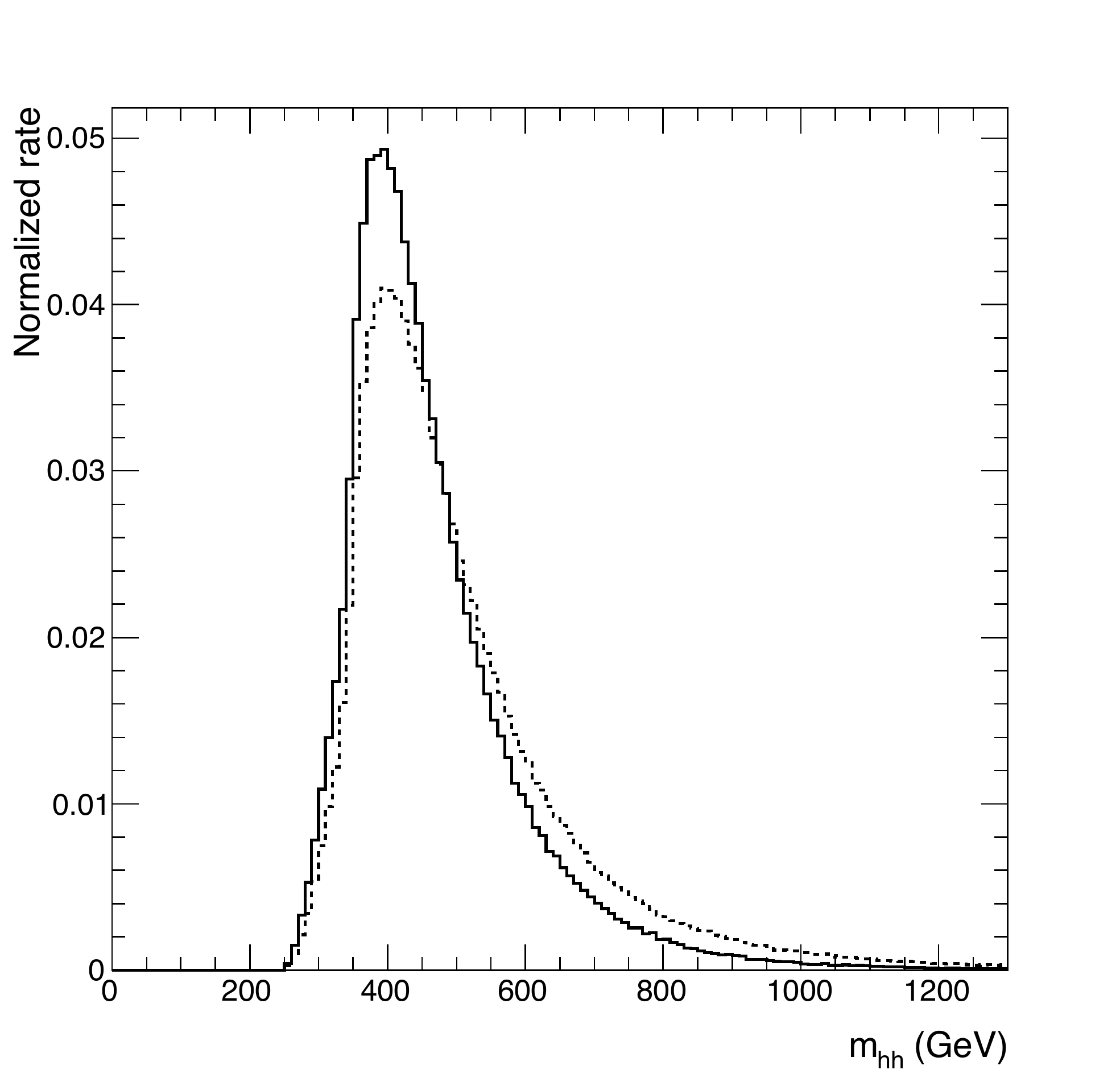}
\hspace{1em}
\includegraphics[width=.45\textwidth]{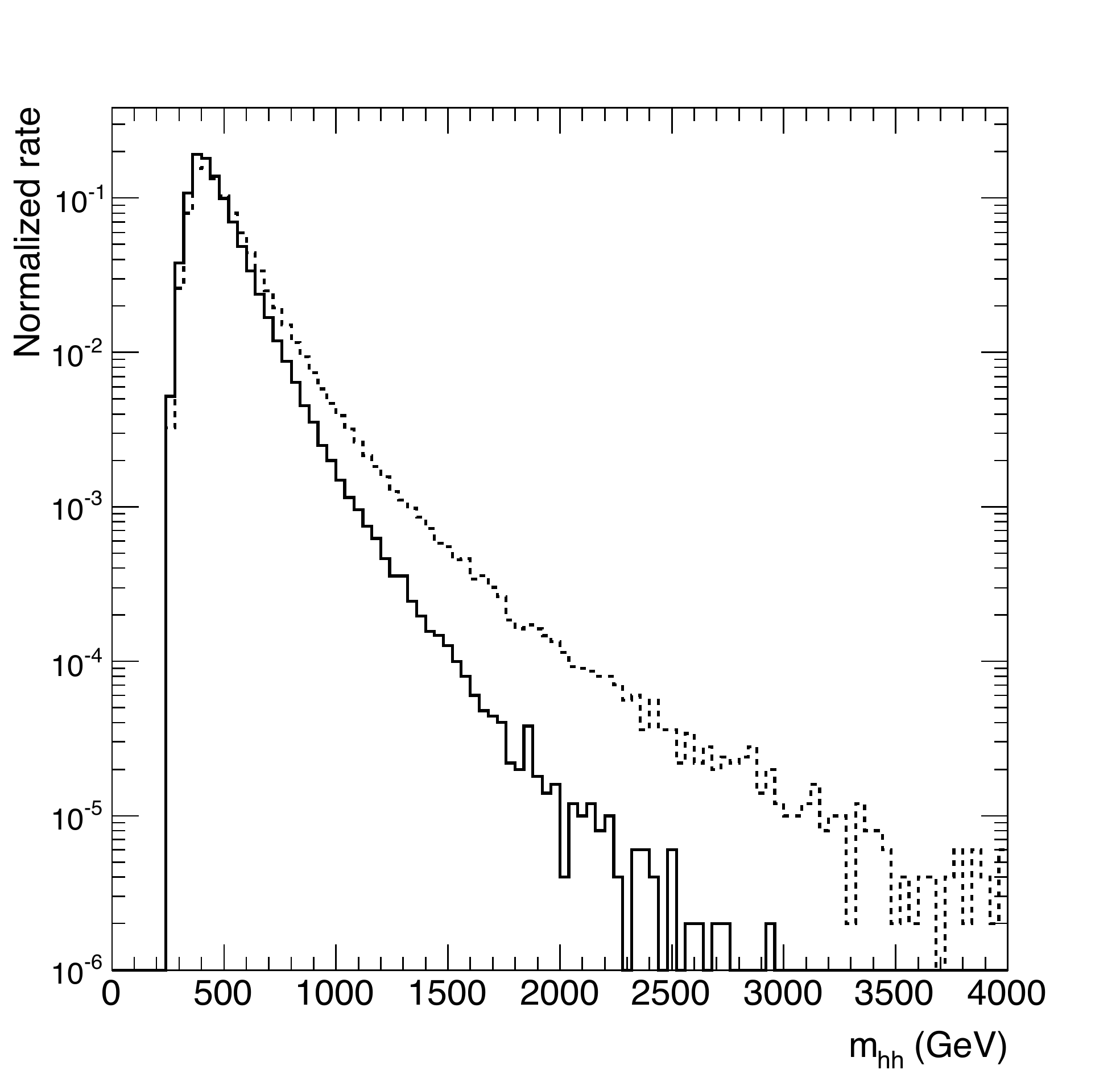}
\caption{Comparison between the normalized $m_{hh}$ distributions for the SM signal.
The solid and dotted lines correspond to $\sqrt{s} = 14$~TeV and $\sqrt{s} = 100$~TeV respectively.
The distributions have been computed at LO retaining the full top mass dependence. The plots are taken
from Ref.~\cite{Azatov:2015oxa}.}\label{fig:mhh_comparison}
\end{figure}
%
The position of the peak and the threshold behavior is unchanged.
The  tail of the distribution, on the other hand, is significantly larger at a $100$~TeV collider, starting from $m_{hh} \gtrsim 700$~GeV.
This modification of the tail, however, has only a small impact on the total production rate,
which is still dominated by the peak region $300~ \textrm{GeV} \lesssim m_{hh} \lesssim 600\ \textrm{GeV}$.

Non-standard Higgs interactions, in particular the couplings with the top (either a modified Yukawa or the
non-renormalizable interaction $hht\overline t$) and the contact interactions with the gluons (see Eq.~(\ref{eq:nonlin_lagr})),
lead to corrections that are not suppressed at high $m_{hh}$. Therefore they can significantly change the tail
of the kinematic distribution at large invariant mass. An analysis exploiting the differential $m_{hh}$ distribution can thus be helpful
to disentangle possible corrections to the various Higgs couplings~\cite{Azatov:2015oxa}. This kind of analysis goes however
beyond the scope of the present report. Here we will  analyze only the inclusive total signal rate and focus on
scenarios in which  the Higgs trilinear coupling is significantly modified, while the corrections to the other couplings
are negligible.

As already mentioned, the sizable production cross section via gluon fusion allows one to consider various decay channels for an experimental search.
In the following we will describe some
preliminary analyses that focus on the most relevant final states with the aim of determining the precision with which
the SM signal can be extracted. In particular the $b\overline b \gamma\gamma$ channel will be presented in
Subsection~\ref{sec:bbaa}, the $b\overline b b\overline b$ channel in Subsection~\ref{sec:bbbb} and finally
the rare final states containing leptons in Subsection~\ref{sec:hh_leptons}.

Before discussing the details of the analyses we briefly discuss in the next subsection the present status of the
computation of the SM Higgs pair production in gluon fusion, pointing out, in particular, the various sources of
theoretical uncertainty.

\subsubsection{Status of SM gluon fusion cross section computation}
\label{sec:status_ggHH}

In the last years, a lot of effort has been devoted towards the improvement of the theoretical prediction of the SM Higgs pair production cross section via gluon fusion.
Even though this loop-induced process is known in an exact way only at LO in the QCD perturbative expansion \cite{Glover:1987nx,Eboli:1987dy,Plehn:1996wb}, very useful approximations are available for the higher-order corrections.

One main approach, exploited extensively for the calculation of the single-Higgs production cross section, consists in working in the large top-mass approximation, in which the Higgs has a direct effective coupling to gluons.
Within this approximation, the LO becomes a tree-level contribution, and higher-order corrections can be computed.
In this way, both the NLO \cite{Dawson:1998py} and 
NNLO \cite{deFlorian:2013jea} corrections have been obtained, together with threshold resummation effects at NNLL accuracy \cite{Shao:2013bz,deFlorian:2015moa}.
The QCD corrections were found to be large, resulting in about a $50\%$ increase from LO to NLO, and a still sizeable $\sim 20\%$ increment from NLO to NNLO at a collider center-of-mass energy of $100\text{ TeV}$ (corrections are even larger for lower energies).
Of course, in order to use these results, an estimate of the accuracy of the approximation is needed.

Two different approaches have been used to estimate the finite top-mass effects at NLO.
In Refs.~\cite{Grigo:2013rya,Grigo:2015dia} the analysis was performed through the computation of subleading terms in the $1/m_t$ expansion.
By evaluating the deviation of the results containing powers of $1/m_t$ from the infinite top-mass prediction, the authors estimate that the effective theory is accurate to $\pm 10\%$ at NLO \cite{Grigo:2015dia}.
On the other hand, in Ref. \cite{Maltoni:2014eza} the exact one-loop real emission contributions were included via a reweighting technique, 
finding in this way that the  NLO total cross section decreases by about $10\%$.
The pure EFT result reproduces well the shape of the Higgs pair invariant mass distribution obtained retaining the exact real contributions.
It is worth mentioning that this is not the case for all the distributions, and for instance the EFT fails to reproduce the region where the Higgs pair system has a large transverse momentum.
Based on the two studies described above, it is possible to  estimate the current accuracy of the large top-mass approximation to be $\pm 10\%$ for the total cross section.\footnote{Recently the complete computation of the NLO cross section including
the finite-top-mass corrections has been performed~\cite{Borowka:2016ehy}. The results are at present available only for the
$14$~TeV LHC and show a $\sim 10\%$ reduction of the cross section with respect to the approximate results.}

The final result for the SM cross section for hadron colliders with center-of-mass energy $E_{cm}=14~\text{TeV}$ and $E_{cm}=100~\text{TeV}$ are listed in 
Table~\ref{table:Higgspair} together with the size of the different theoretical uncertainty.
These results are computed at NNLO+NNLL accuracy using the PDF4LHC recommendation for the parton flux~\cite{Butterworth:2015oua}, and the
values $m_h=125\text{ GeV}$ and $m_t=172.5\text{ GeV}$ for the Higgs and top quark masses.
The scale uncertainty at NNLO+NNLL is quite small, as it is also the case for the PDF and $\alpha_S$ uncertainty.
Therefore, theoretical uncertainties are currently driven by the use of the large top-mass approximation.
It is worth noticing that, once the exact NLO becomes available, the remaining EFT uncertainty at NNLO is expected to be at
most $\pm 5\%$ \cite{Grigo:2015dia}.

\subsubsection{The $HH\to b\bar{b}\gamma\gamma$ channel}
\label{sec:bbaa}

In this section we analyze the $HH\to b\bar{b}\gamma\gamma$ channel, which has been singled out in the literature
as the process that can lead to the highest SM signal significance and highest precision in the measurement of the
Higgs trilinear coupling. This channel has a relatively small branching ratio ($BR\simeq 0.264\%$ in the SM), which somewhat
limits the signal yield (in the SM the total rate for this channel at a $100\,$TeV pp collider is $\simeq 4.6\,$fb).
The presence of two photons, however, allows one to efficiently keep the background under control while preserving a fair fraction of the signal events.

Various studies included an analysis of this channel at future high-energy hadron
colliders, focusing mainly on the extraction of the Higgs trilinear coupling~\cite{Yao:2013ika,Barr:2014sga,He:2015spf},
or performing a global analysis of the impact of the modifications of the various Higgs couplings~\cite{Azatov:2015oxa}.
The differences among these analyses stem mainly from different assumptions about the detector performance, a different treatment of the backgrounds 
and the choice of benchmark integrated luminosity.
In the following we will present (in a summarized form) the results of a new analysis of the $HH\to b\bar{b}\gamma\gamma$
final state, specifically focused on the extraction of the trilinear Higgs coupling in a SM-like scenario~\cite{Contino:2016}.
Differently from most of the previous ones, this new analysis is tailored specifically on a $100$~TeV future hadron collider,
with the primary purpose of estimating how much the achievable precision is influenced by the detector performance.

\paragraph{Simulation setup}

The parton-level generation of the signal and the backgrounds is performed by using \texttt{MadGraph5\_aMC@NLO}
(version 2.3.3)~\cite{Alwall:2011uj} and the parton density functions \texttt{cteq6l1}~\cite{Stump:2003yu}. 
The signal is generated at LO retaining the finite-top-mass effects and 
afterwards rescaled by a k-factor in order to match the NNLL gluon-fusion SM cross section (see Table~\ref{table:Higgspair}).
The analysis includes the following main backgrounds: the non-resonant processes $b\bar b\gamma \gamma$,
$b\bar bj \gamma$ (with one fake photon), $bj\gamma\gamma$ and $jj\gamma\gamma$ (respectively with one and two fake $b$-jets),~\footnote{Here $j$ 
denotes a jet initiated by a gluon or a light quark $u$, $d$, $s$, $c$. For simplicity $bj\gamma\gamma$ denotes the sum of the processes 
where the $b$-jet is initiated by either a $b$ quark or an anti-$b$ quark.} and the resonant processes $b\bar b h$ and $t\bar t h$.
The cross sections for each background process after the acceptance cuts of Table~\ref{tab:HH_bbgaga_cuts} are given in Table~\ref{tab:HH_bbgaga_crosssections}.
The $b\bar b\gamma\gamma$ and $bj\gamma\gamma$ samples are generated by matching up to one extra parton at the matrix-element 
level.~\footnote{The $k_T$-MLM matching scheme has been used with matching scale $35$~GeV and matching parameter xqcut\;$= 25\,$GeV.} 
In the case of $b\bar b\gamma\gamma$, matching accounts for the bulk of the NLO correction to the
cross section, as virtual effects are small for this process, see Ref.~\cite{Azatov:2015oxa}.
The remaining backgrounds are instead  generated at LO without matching and rescaled by the following k-factors 
to take into account higher-order effects: $k = 1.08$ for $b\bar bj \gamma$, $k = 1.3$ for $t\bar t h$,
$k = 0.87$ for $b\bar bh$ and $k = 1.43$ for $\gamma\gamma jj$.

Showering and hadronization effects are included for the signal and background samples by using the \texttt{pythia6}
package~\cite{Sjostrand:2006za}. The simulation of the underlying event  has been found to have a minor impact on the analysis
and, therefore, has been omitted for simplicity.
Detector simulation effects are included by using the \texttt{Delphes} package (version 3.3.1)~\cite{deFavereau:2013fsa}
with a custom card that describes the FCC-hh detector parametrization. A more detailed discussion of the benchmarks used
for the calorimeters performance parametrization will be given in the next subsection.

The tagging of $b$-jets and photons is performed by using the \texttt{Delphes} flavor tagging information
that associates each jet with a parton after showering.
Events are then re-weighted according to the $b$- and photon-tagging probabilities.
The following benchmark efficiencies are considered:
The $b$-tagging probabilities are chosen to be constant throughout the detector and independent of the transverse momenta,
with values $p_{b\rightarrow b} = 0.75$, $p_{c\rightarrow b} = 0.1$
and $p_{j\rightarrow b} = 0.01$, for $b$, $c$ and light jet tagging to $b$-jets respectively.
The light-jet-to-photon mis-tagging probability is parametrized via the function
\begin{equation}\label{eq:gamistag}
p_{j\rightarrow \gamma} = \alpha \exp ( - p_{T,j} / \beta )\,,
\end{equation}
where $\alpha$ and $\beta$ are parameters whose
benchmark values are set to $\alpha = 0.01$ and $\beta = 30$~GeV. Photons are assumed to be reconstructed
with an efficiency that depends on $\eta$, namely
\begin{equation}
\left\{
\begin{array}{c@{\hspace{1.5em}}l}
95\% & {\rm for}\quad |\eta| \leq 1.5\\
\rule{0pt}{1.15em}90\% & {\rm for}\quad 1.5 < |\eta| \leq 4\\
\rule{0pt}{1.15em}80\% & {\rm for}\quad 4 < |\eta| \leq 6
\end{array}
\right.\,,
\end{equation}
provided that they have a transverse momentum $p_T(\gamma) > 10$~GeV.

\begin{table}
\centering
\def\arraystretch{1.5}
\begin{tabular}{c|c}
Acceptance cuts & Final selection\\
\hline
\hline
\rule{0pt}{1.8em} \parbox{11em}{\centering $\gamma$ isolation $R = 0.4$\\ ($p_T(had)/p_T(\gamma) < 0.15$)} & \parbox{11em}{\centering $\gamma$ isolation $R = 0.4$\\ ($p_T(had)/p_T(\gamma) < 0.15$)}\\
\rule{0pt}{1.1em} jets: anti-$k_T$, parameter $R = 0.4$ & jets: anti-$k_T$, parameter $R = 0.4$\\
\rule{0pt}{1.1em} $|\eta_{b,\gamma,j}|< 6$ & $|\eta_{b,\gamma}|< 4.5$\\
\rule{0pt}{1.35em} $p_T(b), p_T(\gamma), p_T(j) > 35\ \mathrm{GeV}$ & $ p_T(b_{1}), p_T(\gamma_{1}) > 60\ \mathrm{GeV}$  \\
& $ p_T(b_{2}), p_T(\gamma_{2}) > 35\ \mathrm{GeV}$ \\
\rule{0pt}{1.1em} $m_{bb} \in [60, 200]\ \mathrm{GeV}$ & $m_{bb} \in [100, 150]\ \mathrm{GeV}$\\
\rule{0pt}{1.1em} $m_{\gamma\gamma} \in [100,150]\ \mathrm{GeV}$ & $|m_{\gamma\gamma} - m_{h}| < 2.0, 2.5, 4.5\ \mathrm{GeV}$\\
& $p_T(bb), p_T(\gamma\gamma) > 100\ \mathrm{GeV}$ \\
& $\Delta R(bb), \Delta R(\gamma\gamma) < 3.5$\\
& no isolated leptons with $p_T > 25\ \mathrm{GeV}$
\end{tabular}
\caption{List of cuts at the acceptance level (left column) and final cuts (right column) used for the analysis.
The final cuts are optimized to increase the precision on the trilinear Higgs self-coupling. The three values listed
for the final cuts on the $m_{\gamma\gamma}$ invariant mass are used in the ``low'', ``medium'' and ``high''
detector performance scenarios.}\label{tab:HH_bbgaga_cuts}
\end{table}

\paragraph{Benchmark scenarios for the detector performance}\label{sec:ggHH_detector}

In the analysis three benchmark scenarios for the detector performance are considered, denoted in the following as
``Low'', ``Medium'' and ``High'' performance scenarios.
These three benchmarks are simulated through
\texttt{Delphes} by implementing different choices for the energy resolution in the electromagnetic and hadronic calorimeters cells.
For this purpose the variance of the energy distribution in a single cell is parametrized by the following formula
\begin{equation}
\Delta E = \sqrt{a^2 E^2 + b^2 E}\,,
\end{equation}
where $E$ is measured in GeV  and the values of the $a$ and $b$ parameters are listed in Table~\ref{tab:calor_param}.
The energy output is then assumed to follow a log-normal distribution, namely the $\log E$ variable follows
a Gaussian distribution. The advantage of this distribution lies in the fact that it tends asymptotically
to the usual normal distribution for large $E$ and provides only positive values for the energy.
For a more detailed discussion about the setup used for the \texttt{Delphes} simulation we refer the reader to Ref.~\cite{Contino:2016}.

\begin{table}
\centering
\def\arraystretch{1.05}
\begin{tabular}{c@{\hspace{1.em}}||@{\hspace{1.em}}cc@{\hspace{1.em}}|@{\hspace{1.em}}cc@{\hspace{1.em}}||@{\hspace{1.em}}cc@{\hspace{1.em}}|@{\hspace{1.em}}cc}
& \multicolumn{4}{c@{\hspace{1.em}}||@{\hspace{1.em}}}{ECAL} & \multicolumn{4}{c}{HCAL}\\
& \multicolumn{2}{c@{\hspace{1.em}}|@{\hspace{1.em}}}{$|\eta| \leq 4$} & \multicolumn{2}{c@{\hspace{1.em}}||@{\hspace{1.em}}}{$4  < |\eta| \leq 6$}
& \multicolumn{2}{c@{\hspace{1.em}}|@{\hspace{1.em}}}{$|\eta| \leq 4$} & \multicolumn{2}{c}{$4  < |\eta| \leq 6$}\\
& $a$ & $b$ & $a$ & $b$ & $a$ & $b$ & $a$ & $b$\\
\hline
\hline
\rule[-.5em]{0pt}{1.6em}Low & $0.02$ & $0.2$ & $0.01$ & $0.1$ & $0.05$ & $1.0$ & $0.05$ & $1.0$\\
\hline
\rule[-.5em]{0pt}{1.6em}Medium & $0.01$ & $0.1$ & $0.01$ & $0.1$ & $0.03$ & $0.5$ & $0.05$ & $1.0$\\
\hline
\rule[-.5em]{0pt}{1.6em}High & $0.007$ & $0.06$ & $0.01$ & $0.1$ & $0.01$ & $0.3$ & $0.03$ & $0.5$
\end{tabular}
\caption{Parameters defining the energy resolution in the electromagnetic (ECAL) and hadronic (HCAL) calorimeter cells for the
``Low'', ``Medium'' and ``High'' detector performance benchmarks.}\label{tab:calor_param}
\end{table}

\begin{figure}[t]
\centering
\includegraphics[width=0.47\textwidth]{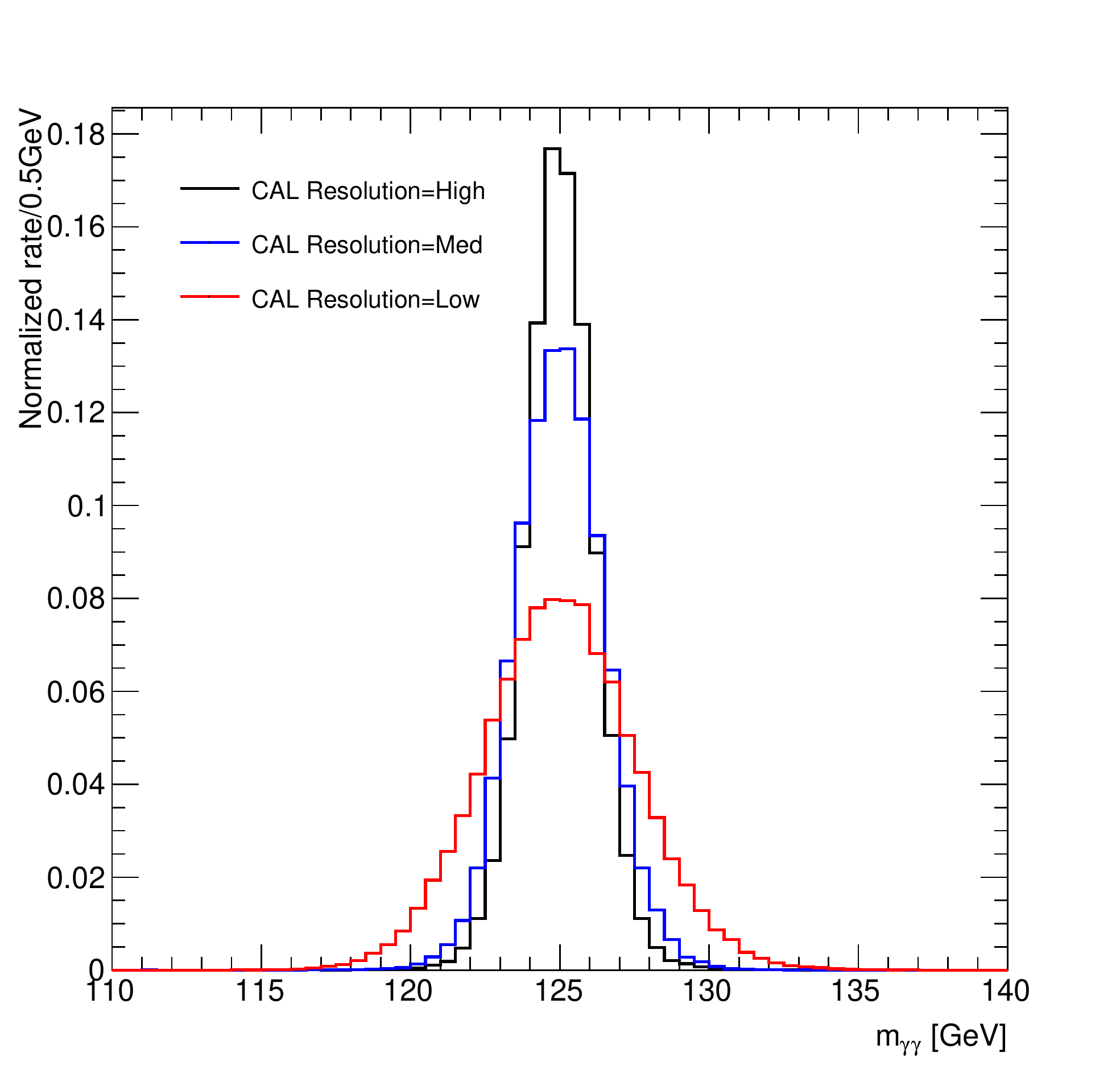}
\hfill
\includegraphics[width=0.47\textwidth]{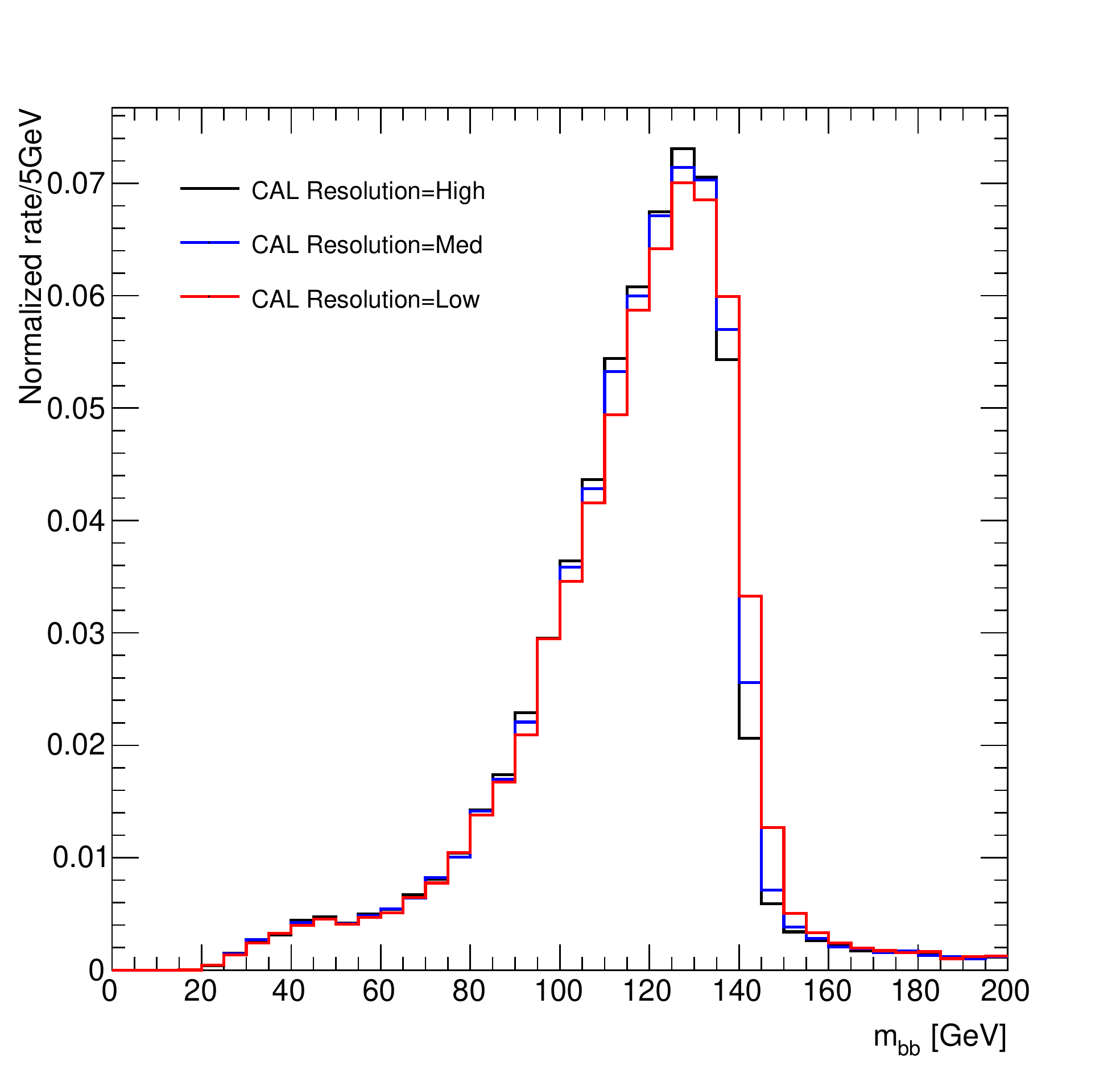}
\caption{Distribution of the reconstructed invariant mass of the photon pair (left panel) and bottom pair (right panel) for the signal.
The plots show how the distributions vary in the ``Low'' (red curve), ``Medium'' (blue curve) and ``High'' (black curve) detector
performance benchmarks.
}
\label{fig:ggHH_massres}
\end{figure}

\begin{figure}[t]
\centering
\includegraphics[width=0.47\textwidth]{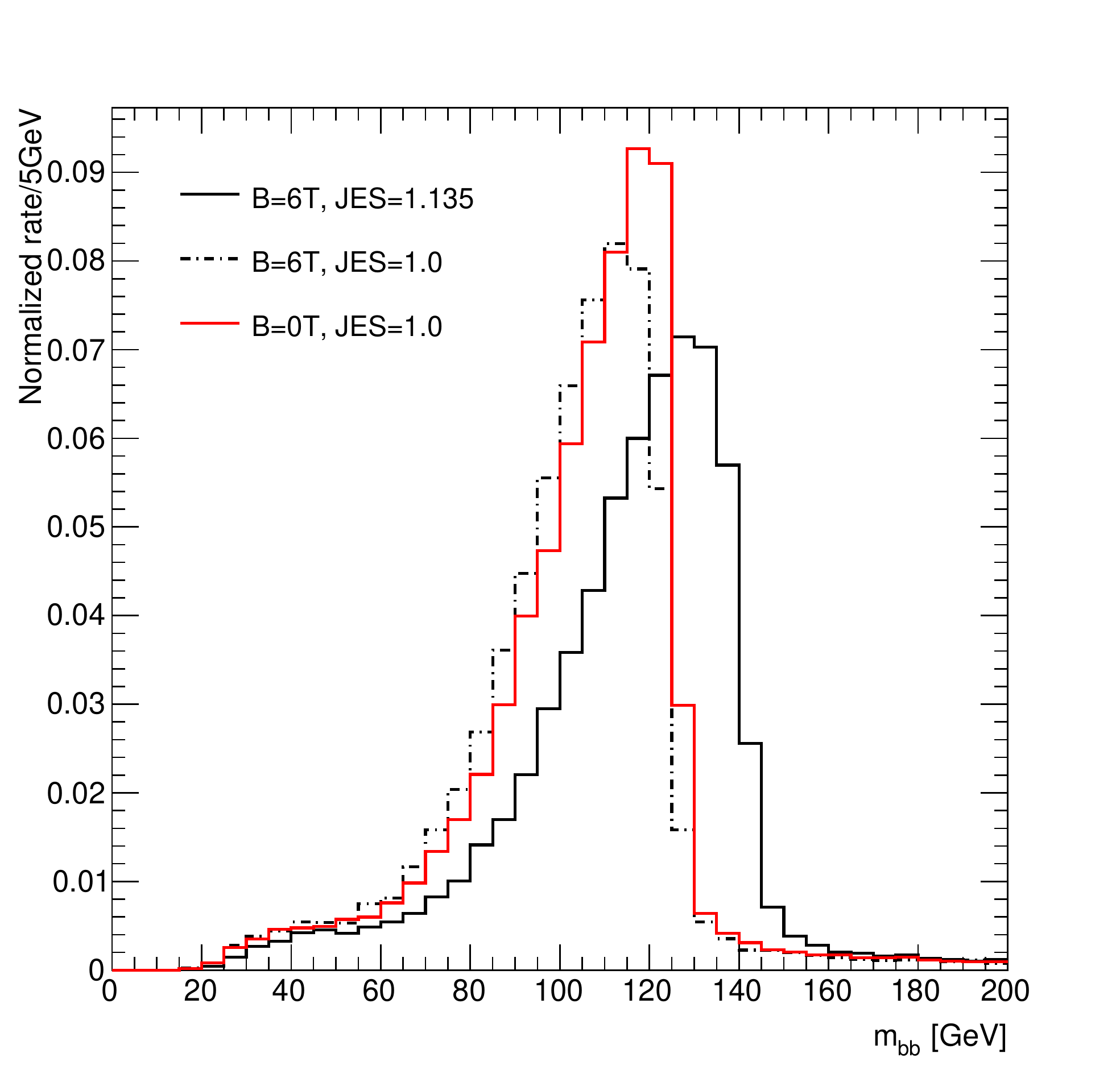}
\caption{Distribution of the reconstructed bottom pair invariant mass for different values of the detector magnetic field $B$ and
of the JES correction. The red curve corresponds to the case with no magnetic field $B=0$~T and no JES correction $r_\text{JES} = 1$,
the dot-dashed black curve corresponds to $B= 6$~T and $r_\text{JES} = 1$, and the solid black curve corresponds to $B = 6$~T
and $r_\text{JES} = 1.135$.
}
\label{fig:ggHH_Bfield}
\end{figure}

As it will be discussed later on, the detector performance can have a significant impact on the analysis and on the achievable precision
in the measurements of the signal cross section and Higgs self-couplings. The finite  energy resolution of the calorimeter induces a smearing in
the reconstruction of the photon- and bottom-pair invariant masses, which are crucial observables for differentiating signal
and background events. The distributions of the reconstructed invariant masses $m_{\gamma\gamma}$ and $m_{bb}$ are shown
in Fig.~\ref{fig:ggHH_massres}, for the three detector performance benchmarks. One can see that the impact on the photon
invariant mass can be sizable. In the ``Low'' performance benchmark the width of the distribution is
$\Delta m_{\gamma\gamma} \simeq 3\,$GeV, while it decreases to $\simeq 2\,$GeV and $\simeq 1.5\,$GeV in the ``Medium''
and ``High'' benchmarks respectively. This means that a cut on the $m_{\gamma\gamma}$ invariant mass can be twice more effective
in the ``High'' benchmark than in the ``Low'' one in reducing backgrounds containing non-resonant photons (as for instance the
non-resonant $b\overline b \gamma\gamma$ background).

The reconstruction of the bottom pair invariant mass $m_{bb}$ is instead only marginally affected by the calorimeters
energy resolution, as can be seen from the right panel of Fig.~\ref{fig:ggHH_massres}. In all cases the distribution
is peaked at the Higgs mass and at half-height it is contained in the region $\sim [100, 145]\,$GeV.
The reconstruction of the $m_{bb}$ invariant mass, on the other hand, is highly affected in the presence of a strong magnetic
field in the detector. In order to be able to bend very energetic charge particles, indeed, a very intense magnetic field
is needed. As a benchmark value, $B = 6\,$T has been chosen in the analysis. The problem with such a magnetic
field is the fact that low-energy charged particles (with a transverse momentum $p_T \lesssim 5\,$GeV) do not reach the
electromagnetic calorimeter, so that it is difficult to reconstruct their energy. This effect can be significant in processes
that happen dominantly at ``low-energy'', as in the case of double Higgs production where the bulk of the cross section comes from
the threshold region.
Figure~\ref{fig:ggHH_Bfield} shows that a strong magnetic field can distort the $m_{bb}$ distribution and shift its peak (by roughly $5$~GeV
for $B = 6$~T).
Such shift, together with the energy loss in the reconstruction of the $b$ momenta, can be partially compensated by
a rescaling of the jet energy scale (JES).\footnote{Actual experimental analyses adopt more sophisticated jet calibration procedures
and particle flow techniques are usually used to cope with these effects. The simple rescaling applied in the analysis discussed here
should be thus considered as a rough approximation of a more accurate experimental procedure.} 
As shown in the figure, a rescaling of the jets' four-momentum by a factor $r_\text{JES} = 1.135$ is sufficient to
shift back the distribution and move its peak to the value $m_{bb} \simeq m_h$.

\paragraph{Analysis strategy}

As discussed in the introduction, the main aim of the analysis described here is to determine the achievable precision on the SM signal
cross section and on the Higgs trilinear coupling. For this purpose a simple cut-and-count strategy focused on the inclusive event rate
can be used. More sophisticated analysis strategies, as for instance an exclusive one that also takes into account the differential
distributions, could be useful in disentangling different new-physics effects (see for instance Ref.~\cite{Azatov:2015oxa}).

In optimizing the selection cuts two different strategies can in principle be used: one can either maximize the precision on the SM signal 
or that on the Higgs trilinear coupling.
The two procedures lead to significantly different sets of cuts. The effect of a change
in the Higgs self-coupling mostly affects the threshold behavior, thus ``looser'' cuts aimed at preserving a large fraction of the
threshold events improve the precision on $\lambda_3$. On the other hand, the threshold region is also the one that has a
larger background, hence harder cuts might be convenient to improve the precision on the SM signal rate.

For the analysis presented here,  cuts have been optimized to maximize the precision on $\lambda_3$.
This choice is motivated by the fact that the SM signal significance is always quite high,
and optimizing for the extraction of the  Higgs trilinear coupling does not degrade significantly the precision
on the signal cross section. In fact, the sensitivity to the signal cross section depends mildly on the
choice of the cuts (provided they do not vary dramatically). On the contrary, the precision on the Higgs trilinear is much
more sensitive to variations of the cuts.

The set of benchmark cuts are listed in Table~\ref{tab:HH_bbgaga_cuts}. Mild $p_T$ cuts are imposed on the photons and
$b$-quarks, namely $p_{T}(b_1), p_T(\gamma_1) > 60$~GeV and $p_{T}(b_2), p_T(\gamma_2) > 35$~GeV,
where $b_{1,2}$, $\gamma_{1,2}$ denote the hardest/softest $b$-quark and photon.
Stronger cuts are imposed on the transverse momentum of the photon and $b$ pair,
$p_T(bb), p_T(\gamma\gamma) > 100$~GeV. All the reconstructed objects are required to be within a rapidity $|\eta_{b,\gamma}|<4.5$,
while the separation between the two photons and the two $b$-quarks is required
to be not too large, namely $\Delta R(bb), \Delta R(\gamma\gamma) < 3.5$. Notice that the angular cuts imposed on $\Delta R$
are  milder than the ones typically used in the previous literature~\cite{Yao:2013ika,Barr:2014sga,Azatov:2015oxa,He:2015spf}.
The invariant mass of the $b$ pair is required to be in the window $m_{bb} \in [100, 150]$~GeV.
For the invariant mass of the photon pair, three different windows are used optimized for each detector performance
scenario: $|m_{\gamma\gamma} - m_h| < 2.0, 2.5, 4.5$~GeV for the ``High'', ``Medium'' and ``Low'' performance benchmarks
respectively. As discussed in Subsection~\ref{sec:ggHH_detector}, these choices of invariant mass windows allow one to retain a sufficiently
large fraction of the signal events.

\begin{table}
\centering
\def\arraystretch{1.25}
\begin{tabular}{c|c|c|c}
Process & Acceptance cuts [fb] & Final selection [fb] & Events ($L=30$~ab$^{-1}$)\\
\hline
\hline
\rule{0pt}{1.05em}$h(b\bar b) h(\gamma\gamma)$ (SM) & $0.73$ & $0.40$ & $12061$\\
\hline
\rule{0pt}{1.05em}$bbj\gamma$ & $132$ & $0.467$ & $13996$\\
\rule{0pt}{1.05em}$jj\gamma\gamma$ & $30.1$ & $0.164$ & $4909$\\
\rule{0pt}{1.05em}$t\bar t h(\gamma\gamma)$ & $1.85$ & $0.163$ & $4883$ \\
\rule{0pt}{1.05em}$b\bar b\gamma\gamma$ & $47.6$ & $0.098$ & $2947$\\
\rule{0pt}{1.05em}$b\bar b h(\gamma\gamma)$ & $0.098$ & $7.6 \times 10^{-3}$ & $227$\\
\rule{0pt}{1.05em}$bj\gamma\gamma$ & $3.14$ & $5.2\times 10^{-3}$ & $155$ \\
\hline
\rule{0pt}{1.05em}Total background & $212$ & $1.30$ & $27118$
\end{tabular}
\caption{Cross section for SM signal and main backgrounds after the acceptance and final cuts of Table~\ref{tab:HH_bbgaga_cuts}. 
The last column shows the number of signal and background events after the final cuts for an integrated luminosity of $30$~ab$^{-1}$.}
\label{tab:HH_bbgaga_crosssections}
\end{table}

The signal and background cross sections after the acceptance and final selection cuts are given in
Table~\ref{tab:HH_bbgaga_crosssections}. One can see that the most significant background after all cuts is
$b\bar b j \gamma$, followed by $jj\gamma\gamma$ and $t\bar th$. 
Another non-negligible contribution comes from the irreducible process $b\bar b\gamma\gamma$. 
The backgrounds $b\bar b h$ and $bj\gamma\gamma$  turn out to be  negligible instead.

Notice that other double Higgs production channels, in particular $t\bar t HH$ and VBF, provide an additional contribution to the signal
of the order of $10\%$. Given the high precision on the SM signal rate and on the Higgs trilinear coupling,
these contributions should be taken into account in a full experimental analysis. However,  the inclusion of these
effects is not expected to  change significantly the estimated precision on the Higgs trilinear coupling presented in the following.

\paragraph{Results}

The results in Table~\ref{tab:HH_bbgaga_crosssections} suggest that, with an integrated luminosity of $L = 30$~ab$^{-1}$,
a precision on the SM signal of the order of $1.6\%$ can be obtained, corresponding to $S/\sqrt{S+B} \simeq 61$. By a simple
rescaling, one can see that already with $L = 500$~fb$^{-1}$ a $13\%$ determination of the cross section is possible.
Notice that these results, as well as those presented in the following, include only the statistical uncertainties and are obtained by neglecting the theoretical error 
on the prediction of the signal and the systematic uncertainty on the overall determination of the background rates.
Anticipating the size of these effects for a future $100\,$TeV collider is a difficult task.
An estimate of the impact of some of the possible systematic errors and of the geometry and performances of the detector is  provided in the following.

\begin{figure}[t]
\centering
\includegraphics[width=0.493\textwidth]{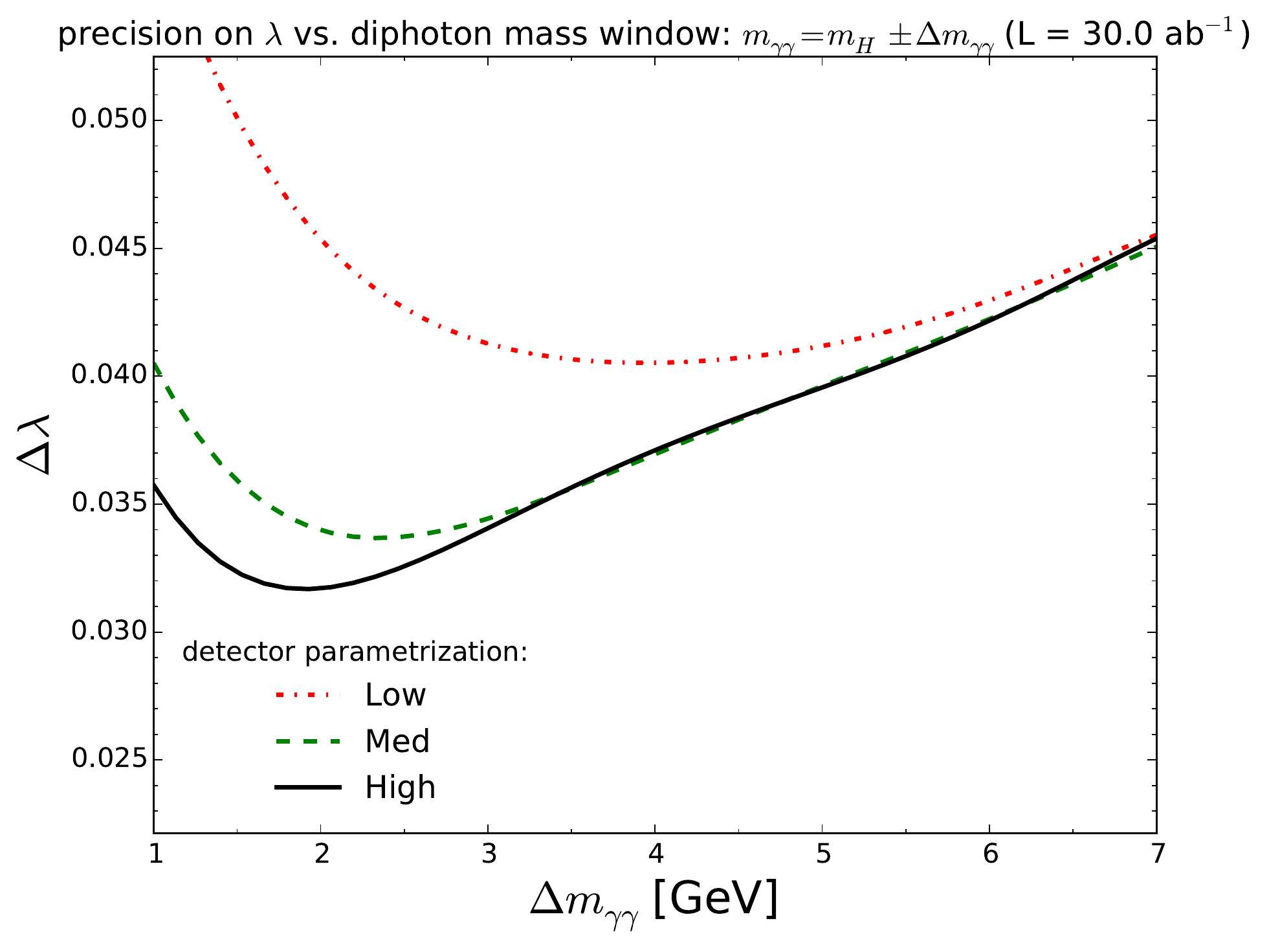}
\includegraphics[width=0.495\textwidth]{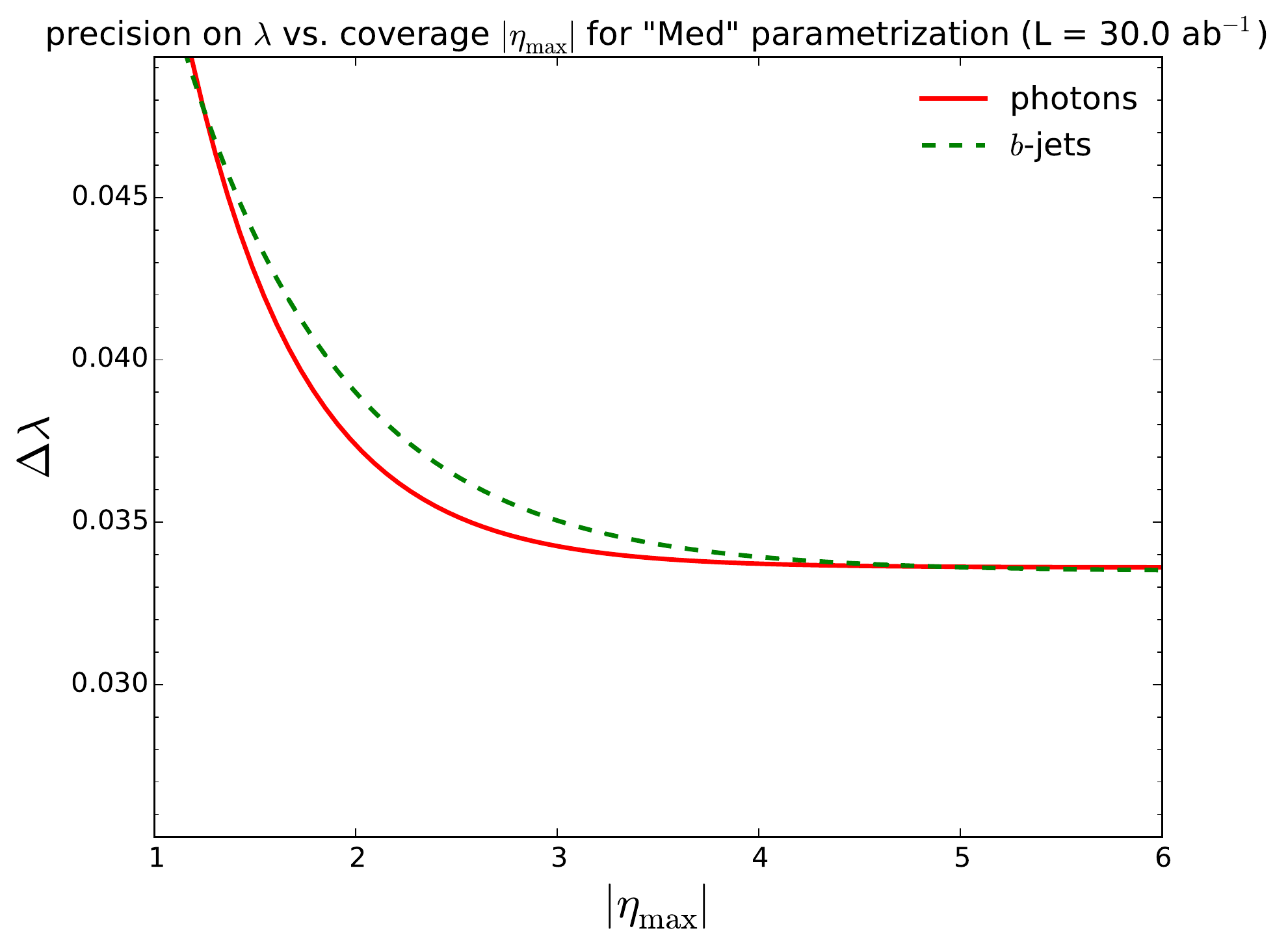}
\caption{Estimated precision on the measurement of the Higgs trilinear self-coupling. The left panel shows
to result as a function of the cut on the invariant mass os the photon pair $\Delta m_{\gamma\gamma}$ for the three
detector benchmark scenarios, ``Low'' (dot-dashed red), ``Medium'' (dashed green) and ``High'' (solid black). In the right panel the result
is shown as a function of the cut on the maximal rapidity of the reconstructed objects $\eta_{max}$ assuming the
``Medium'' detector benchmark (the solid red and dashed green curves correspond to a variation of the photon and
$b$-jets acceptances respectively). All the results have been obtained for an integrated luminosity of $30$~ab$^{-1}$.
}
\label{fig:deltaL1}
\end{figure}

As illustrated by the left panel of Fig.~\ref{fig:deltaL1},  a maximal precision on the Higgs trilinear coupling
of the order of $3.4\%$ can be obtained in the ``Medium'' detector performance benchmark. 
The figure shows that this value crucially depends on the size of the photon invariant mass window
used to select the events. If the size of the window is modified, the precision on the Higgs trilinear can be substantially degraded
(especially for smaller sizes of the window, which reduce the amount of reconstructed signal events).
In the ``Low'' and ``High'' scenarios a precision of respectively  $4.1\%$ and $3.2\%$ seems to be achievable.

Let us now discuss how the precision changes by varying the most important parameters related to the detector geometry and performances,
namely the $\eta$ coverage, the $b$-tagging efficiencies and the photon mis-tagging rate. For this purpose the ``Medium''  performance scenario 
will be taken as a reference and each parameter varied separately.

The precision on $\lambda_3$ is shown in the right panel of Fig.~\ref{fig:deltaL1} as a function of the maximal rapidity coverage $\eta_\text{max}$
for photons and b-jets. One can see that
extending the coverage beyond $|\eta_\text{max}| \sim 3.5$ does not lead to any substantial improvement. In other words, having a larger coverage
in rapidity does not seem a crucial feature for the extraction of the Higgs self coupling, and a reach up to $|\eta_\text{max}| = 2.5-3$ could be
considered to be an acceptable compromise.

\begin{figure}[t]
\centering
\includegraphics[width=0.99\textwidth]{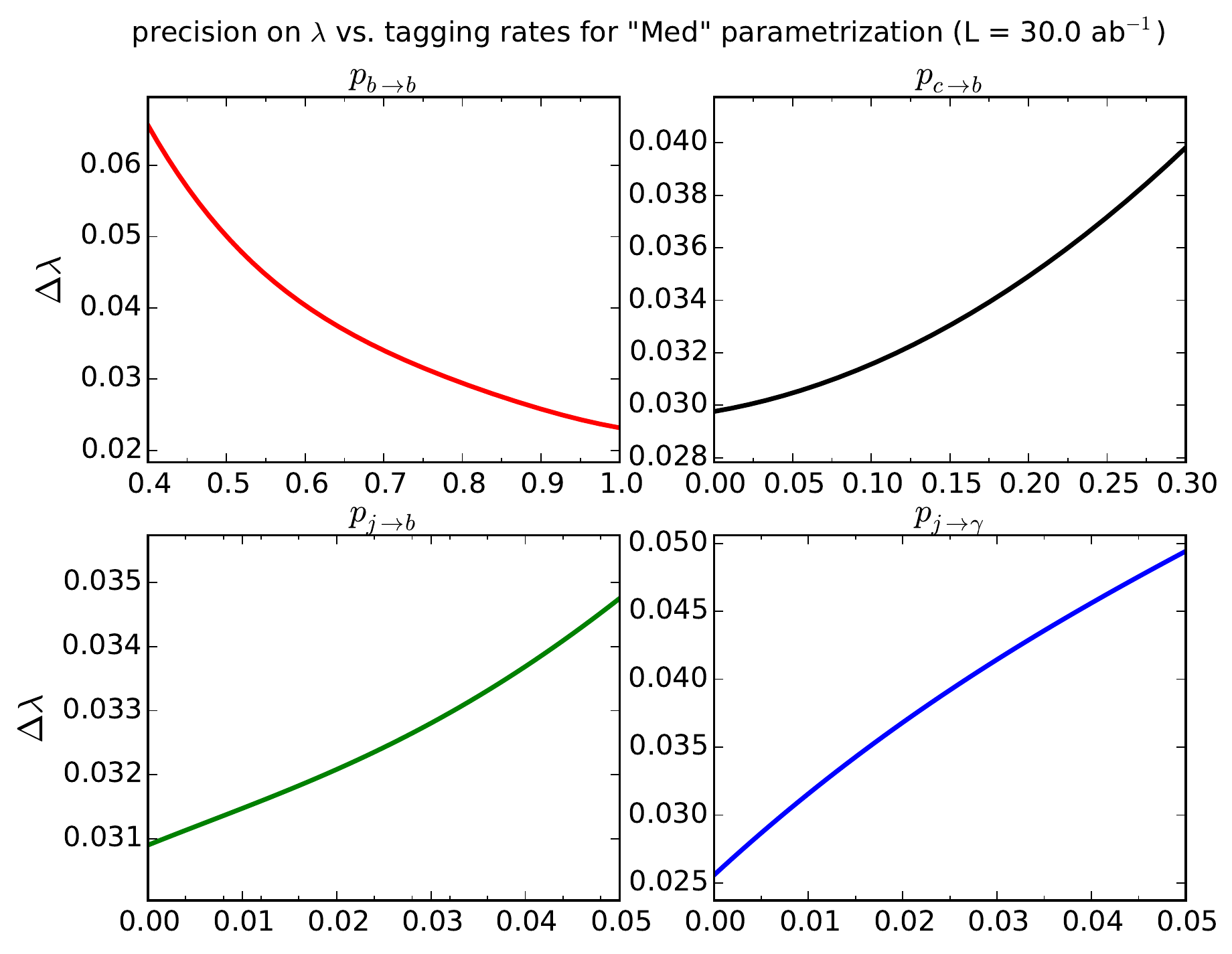}
\caption{Estimated precision on the measurement of the Higgs trilinear self-coupling as a function of the $b$-tagging
efficiency. Each plot shows how the precision changes by varying only one parameter, namely the $b$ reconstruction
efficiency $p_{b \to b}$ (upper left), the $c \to b$ mistag rate (upper right), the $j \to b$ mistag rate (lower left)
and the $j \to \gamma$ mistag rate (lower right). In the case of the $j \to \gamma$ mistag, on the horizontal axis we
give the coefficient $\alpha$ of the mistag function in Eq.~(\ref{eq:gamistag}).
All the results have been obtained in the ``Medium'' detector performance scenario with an integrated luminosity of $30$~ab$^{-1}$.
}
\label{fig:deltaL2}
\end{figure}

A  more crucial role is instead played by the $b$-tagging efficiencies and rejection rates, as shown in Fig.~\ref{fig:deltaL2}.
The reconstruction efficiency for the $b$-jets is the most important parameter, since it directly controls the signal
reconstruction rate. A minimal efficiency $p_{b \to b} \simeq 0.75$ is necessary to achieve a good precision
on the Higgs trilinear coupling. A value $p_{b \to b} \simeq 0.6$ already degrades the achievable precision
to $\simeq 4.0\%$. The mistag rates for charm-jets $p_{c \to b}$ plays a marginal role and does not affect too much the precision on $\lambda_3$
as long as $p_{c \to b} \lesssim 0.2$. The impact of the light-quark and gluon jets mistag rate $p_{j \to b}$ is even milder
and does not influence the result as long as $p_{j \to b} \lesssim 0.05$.
Finally, the lower right panel of Fig.~\ref{fig:deltaL2} shows how the precision on $\lambda_3$ changes when the mistag rate of fake photons from jets
is modified.  The curve is obtained by varying the overall coefficient $\alpha$ in Eq.~(\ref{eq:gamistag}) (values on the horizontal axis) 
and keeping fixed the functional dependence on $p_{T,j}$ with $\beta = 30$~GeV.
One can see that high mistag rates ($\alpha \sim 0.05$)
can significantly affect the achievable precision. This is a consequence of the fact that the main background, $b\bar b j \gamma$, 
contains one fake photon from jet mis-tagging. Keeping $\alpha$ below $0.02$ is enough not to
affect significantly the precision on $\lambda_3$.

\begin{table}
\centering
\def\arraystretch{1.25}
\begin{tabular}{l||c|c|c|c|c}
& $\Delta_S = 0.00$ & $\Delta_S = 0.01$ & $\Delta_S= 0.015$ & $\Delta_S = 0.02$ & $\Delta_S = 0.025$\\
\hline
\hline
$r_B = 0.5$ & $2.7\%$ & $3.4\%$ & $4.1\%$ & $4.9\%$ & $5.8\%$\\
\hline
$r_B = 1.0$ & $3.4\%$ & $3.9\%$ & $4.6\%$ & $5.3\%$ & $6.1\%$\\
\hline
$r_B = 1.5$ & $3.9\%$ & $4.4\%$ & $5.0\%$ & $5.7\%$ & $6.4\%$\\
\hline
$r_B = 2.0$ & $4.4\%$ & $4.8\%$ & $5.4\%$ & $6.0\%$ & $6.8\%$\\
\hline
$r_B = 3.0$ & $5.2\%$ & $5.6\%$ & $6.0\%$ & $6.6\%$ & $7.3\%$\\
\end{tabular}
\caption{Impact of the systematic uncertainties on the precision on the trilinear Higgs coupling.
The precision on $\lambda_3$ is shown for different values of the systematic uncertainty on the signal, $\Delta_S$, and of the
rescaling factor for the total background rate $r_B$.
The ``Medium'' detector performance scenario and an integrated luminosity of $30$~ab$^{-1}$ have been assumed.
}
\label{tab:bbaa_systematics}
\end{table}

To conclude, we briefly comment on the possible impact of the theoretical error on the signal cross section and of the systematic uncertainties 
on the overall background rate. 
Table~\ref{tab:bbaa_systematics} shows how the precision on the Higgs trilinear coupling varies as a function of the relative
error on the signal cross section, $\Delta_S \equiv \Delta \sigma(pp\to hh)/\sigma(pp\to hh)$, and of an overall rescaling of the total background by a factor $r_B$.
Notice that an actual experimental analysis  will most likely extract the background rate directly from the data
(by fitting for instance the $m_{\gamma\gamma}$ distribution away from the Higgs peak, as done for the diphoton channel in single-Higgs production).
The rescaling factor $r_B$  should be thus considered as a way to assess the impact of the error associated with the MonteCarlo calculation of the background rate
in Table~\ref{tab:HH_bbgaga_crosssections}.
The actual systematic uncertainty on the background rate in an experimental analysis will likely be much smaller,
and possibly negligible.
In the limit in which the systematic uncertainty (theory error + pdfs uncertainty) on the signal cross section becomes larger than its statistical error,
the precision on the Higgs trilinear measurement saturates to $\simeq 2 \Delta_S$. Since the statistical error on the signal rate
is expected to be small (of the order of $3 - 4\%$), the systematic uncertainty can easily become the main limitation in the extraction of $\lambda_3$.
At present, as already discussed, the computation of the signal has a $\sim 10\%$ uncertainty due to the use of the infinite top mass
approximation. It is highly probable that finite-mass computations will become available in the near future. The remaining
uncertainty from scale variation at NNLL order is still $\sim 5\%$, while the pdf error is $\sim 3\%$. Without further improvements
on these two issues, the systematic uncertainty will be the main limiting factor in the  determination of $\lambda_3$ and the
maximal precision would be limited to $\delta \lambda_3/\lambda_3 \sim 10\%$.

\subsubsection{The $HH\to b\bar{b}b\bar{b}$ channel}
\label{sec:bbbb}

In the analysis of the $b\bar{b} \gamma\gamma$ final state presented in the previous subsection,  a large fraction
of the double Higgs production cross section was sacrificed in order to select a clean final state, for which the background levels can
be easily kept under control. In this subsection a different strategy is considered which makes use of the final state with the largest branching ratio, 
namely $b\bar{b}b\bar{b}$.
The total cross section for this final state is $580$~fb at a hadronic $100$~TeV collider, which is two order of magnitude
larger than the $b\bar{b} \gamma\gamma$ one. The level of backgrounds one needs to cope with, however, is much larger
thus severely complicating the signal extraction.

One of the possible advantages of the $b\bar{b}b\bar{b}$ final state is the fact that it provides a reasonable number
of events in the tail at large invariant masses of the Higgs pair. This, in principle, allows one to analyse the high-energy kinematic
regime much better than  other final states with smaller cross sections. As we discussed before,
the tail of the $m_{hh}$ distribution is not particularly sensitive to the change of the trilinear Higgs coupling, which mostly affects
the kinematic distribution at threshold. However it can be more sensitive to other new-physics effects, such as deviations
induced by dimension-$6$ and dimension-$8$ effective operators that induce a contact interaction between the Higgs
and the gluons (see for instance the discussion in Ref.~\cite{Azatov:2015oxa}). The analysis of these effects, although interesting
and worth studying further, goes beyond the scope of the present report. In the following we will concentrate only on the
SM case and on the extraction of the Higgs trilinear coupling and we will discuss an analysis based on a recent feasibility
study at the $14$~TeV LHC~\cite{Behr:2015oqq},\footnote{Other studies of Higgs pair production in the same
final state at the LHC can be found in Refs.~\cite{Wardrope:2014kya,deLima:2014dta}.} with suitable modifications for the $100$~TeV case.

\paragraph{Monte Carlo samples generation}

Higgs pair production in the gluon-fusion channel is simulated at LO thorugh {\tt MadGraph5\_aMC@NLO}~\cite{Alwall:2014hca,Maltoni:2014eza}
by using the recently developed functionalities for loop-induced processes~\cite{Hirschi:2015iia}.
The calculation is performed in the $n_f = 4$ scheme and the renormalization and factorization
scales are taken to be $\mu_F=\mu_R=H_T/2$.
The NNPDF 3.0 $n_f=4$ LO set~\cite{Ball:2014uwa} is adopted with
$\alpha_s(m_Z^2)=0.118$, interfaced via {\tt LHAPDF6}~\cite{Buckley:2014ana}.
To achieve the correct higher-order value of the
integrated cross-section, the LO signal sample is rescaled to match the NNLO+NNLL
inclusive calculation~\cite{deFlorian:2013jea,deFlorian:2015moa}.
Parton level signal events are then showered with {\tt Pythia8}~\cite{Sjostrand:2007gs,Sjostrand:2014zea} (version {\tt v8.201})
using the  Monash 2013 tune~\cite{Skands:2014pea}, based on the NNPDF2.3LO PDF set~\cite{Ball:2012cx,Ball:2013hta}.
Background samples are generated at LO with {\tt SHERPA}~\cite{Gleisberg:2008ta} ({\tt v2.1.1})
and rescaled to known higher-order results, using the same $K$-factors as in~\cite{Behr:2015oqq}. 
The input PDFs and scales are the same as for the signal samples.
In order to keep the analysis simple enough, only the irreducible QCD $4b$ background is included.
This background is one of the most important at the LHC, together with $b\bar{b} jj$, and is thus
expected to provide a rough estimate of the total background also at $100$~TeV.
Single Higgs production processes and electroweak backgrounds are much smaller and are also neglected.

\paragraph{Analysis strategy}

After the parton shower, final state particles are clustered using the jet reconstruction algorithms of
{\tt FastJet}~\cite{Cacciari:2011ma,Cacciari:2005hq} ({\tt v3.1.0}).
First of all, {\it small-$R$ jets} are reconstructed with the anti-$k_T$ algorithm~\cite{Cacciari:2008gp} with $R=0.4$,
and required to have transverse momentum $p_T \ge 40$~GeV and pseudo-rapidity $|\eta|<2.5$.
In addition {\it large-$R$ jets} are defined, reconstructed with  anti-$k_T$ with $R=1.0$.
These are required to have $p_T \ge 200$~GeV, lie in the $|\eta|<2.0$ region and
satisfy the  BDRS mass-drop tagger (MDT)~\cite{Butterworth:2008iy}.
Finally, {\it small-$R$ subjets} are constructed by clustering all final-state particles with anti-$k_T$ with  $R=0.3$,
that are then ghost-associated  to the large-$R$ jets~\cite{Aad:2015uka}.
These are required to satisfy the condition $p_T > 50$~GeV and $|\eta|<2.5$.
For the boosted and intermediate categories, which involve large-$R$ jets,
a number of jet substructure variables~\cite{Salam:2009jx,Aad:2013gja} are used in the analysis:
the $k_T$-splitting scale~\cite{Butterworth:2002tt,Butterworth:2008iy}, the ratio of 2-to-1 subjettiness $\tau_{21}$~\cite{Thaler:2010tr,Thaler:2011gf}, and the ratios of energy correlation functions (ECFs)  $C^{(\beta)}_2$~\cite{Larkoski:2013eya} and $D_2^{(\beta)}$~\cite{Larkoski:2014gra}.

For each jet definition a different $b$-tagging strategy is adopted.
A {\it small-$R$ jet} is tagged as a $b$-jet with probability $f_b$ if it contains at least one $b$-quark among its constituents with 
$p_T \ge 15$ GeV~\cite{Aad:2015ydr}.
If no $b$-quarks are found among its constituents,  a jet with $p_T \ge 15$ GeV can be still be tagged as a $b$-jet
with a mistag rate of $f_l$ ($f_c$) in the case of a light (charm) jet constituent.
 {\it Large-$R$ jets} are $b$-tagged by ghost-associating anti-$k_T$ $R=0.3$ (AKT03) subjets to the original large-$R$
    jets~\cite{Cacciari:2007fd,Aad:2013gja,ATLAS-CONF-2014-004,Aad:2015uka}.
A large-$R$ jet is considered to be $b$-tagged if  the leading and subleading AKT03 subjets are both individually $b$-tagged,
with the same criteria as for the small-$R$ jets.
The treatment of the $b$-jet mis-identification from light and charm jets is the same as for the small-$R$ jets.
For the $b$-tagging probability $f_b$, along with the $b$-mistag probability of light ($f_l$) and charm ($f_c$) jets,
the following values are used: $f_b=0.8$, $f_l=0.01$ and  $f_c=0.1$.

The analysis strategy follows the scale-invariant resonance tagging method of Ref.~\cite{Gouzevitch:2013qca}.
Rather than restricting to a specific event topology, it consistently combines the information from
three possible topologies: boosted, intermediate and resolved, with the optimal cuts for each category being determined
separately.
The three categories are defined as follows.
Events are classified in the {\it boosted category} if they contain at least two large-$R$ jets, with the two leading jets
being $b$-tagged.
They are classified in the {\it intermediate category} if there is exactly one  $b$-tagged, large-$R$ jet, which
is assigned to be the leading Higgs candidate.
In addition, at least two $b$-tagged small-$R$ jets are required, which must be separated with respect to the large-$R$ jet
by  $\Delta R\ge 1.2$.
Finally, events are assigned to the {\it resolved category} if they contain at least four $b$-tagged small-$R$ jets.
The two Higgs candidates are reconstructed out of the leading four small-$R$ jets in the event by minimizing the relative difference of
dijet masses.
In all categories, once a Higgs boson candidate has been identified, its invariant mass is required to lie within a fixed window
of width $80$~GeV, symmetric around the nominal Higgs boson mass of $m_h= 125$~GeV.
The object and event selection are deliberately loose since their optimization is performed through a Multivariate Analysis (MVA)
strategy.

\paragraph{Results}

Following Ref.~\cite{Behr:2015oqq}, a preliminary cut-based analysis is performed,
followed by a MVA procedure aimed at the optimization of the separation between signal and backgrounds.
The specific type of  MVA that it is used is a multi-layer feed-forward artificial neural network (ANN),
known as {\it perceptron} or {\it deep neural network}.
The MVA inputs are the set of kinematic variables describing the signal and background events which satisfy the requirements of the
cut-based analysis, including the jet substructure variables.
The output of the trained ANNs allows for the identification, in a fully automated way,
of the most relevant variables for the discrimination between  signal and background.

\begin{figure}[t]
\centering
\includegraphics[width=0.48\textwidth]{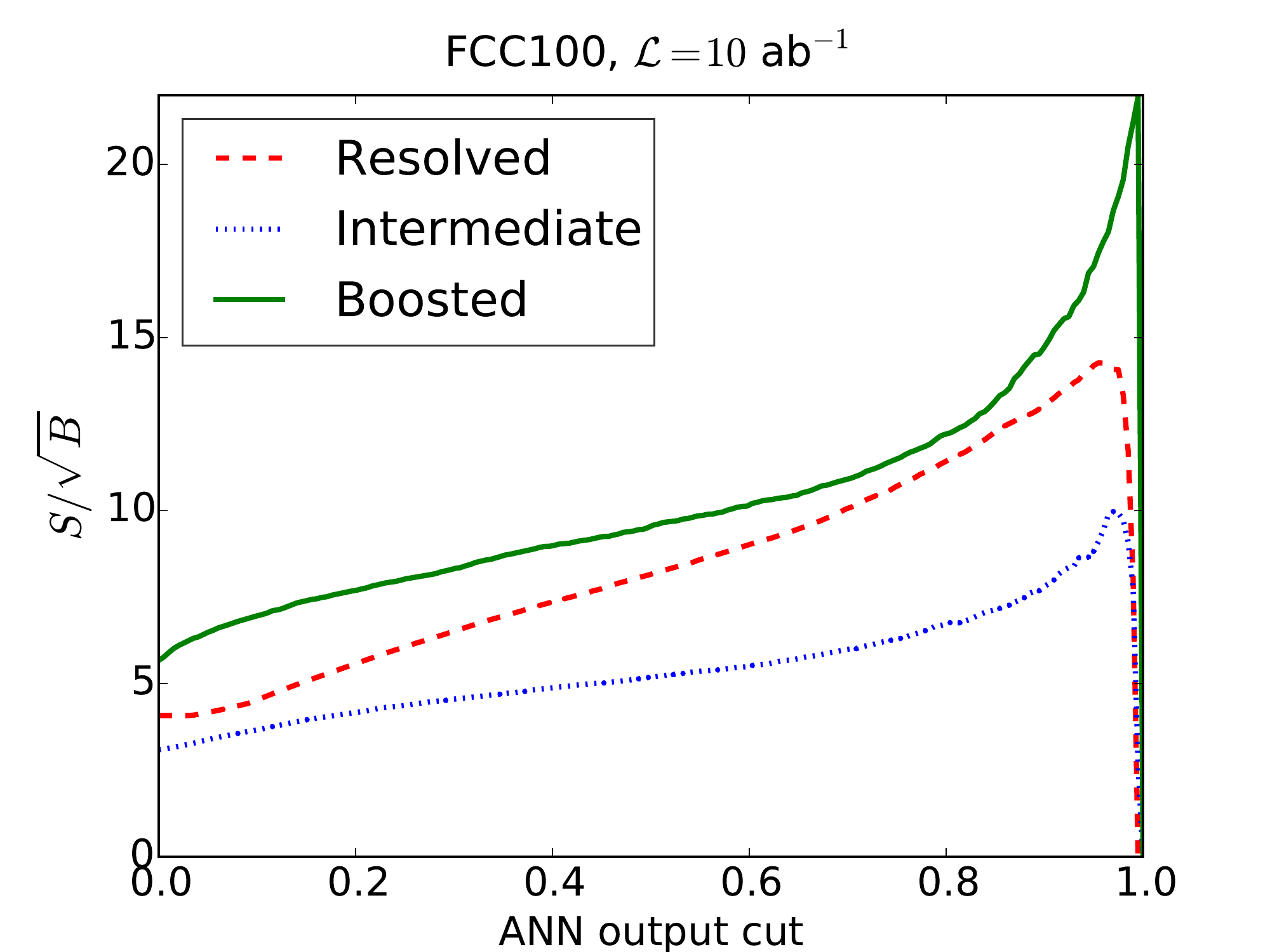}
\includegraphics[width=0.48\textwidth]{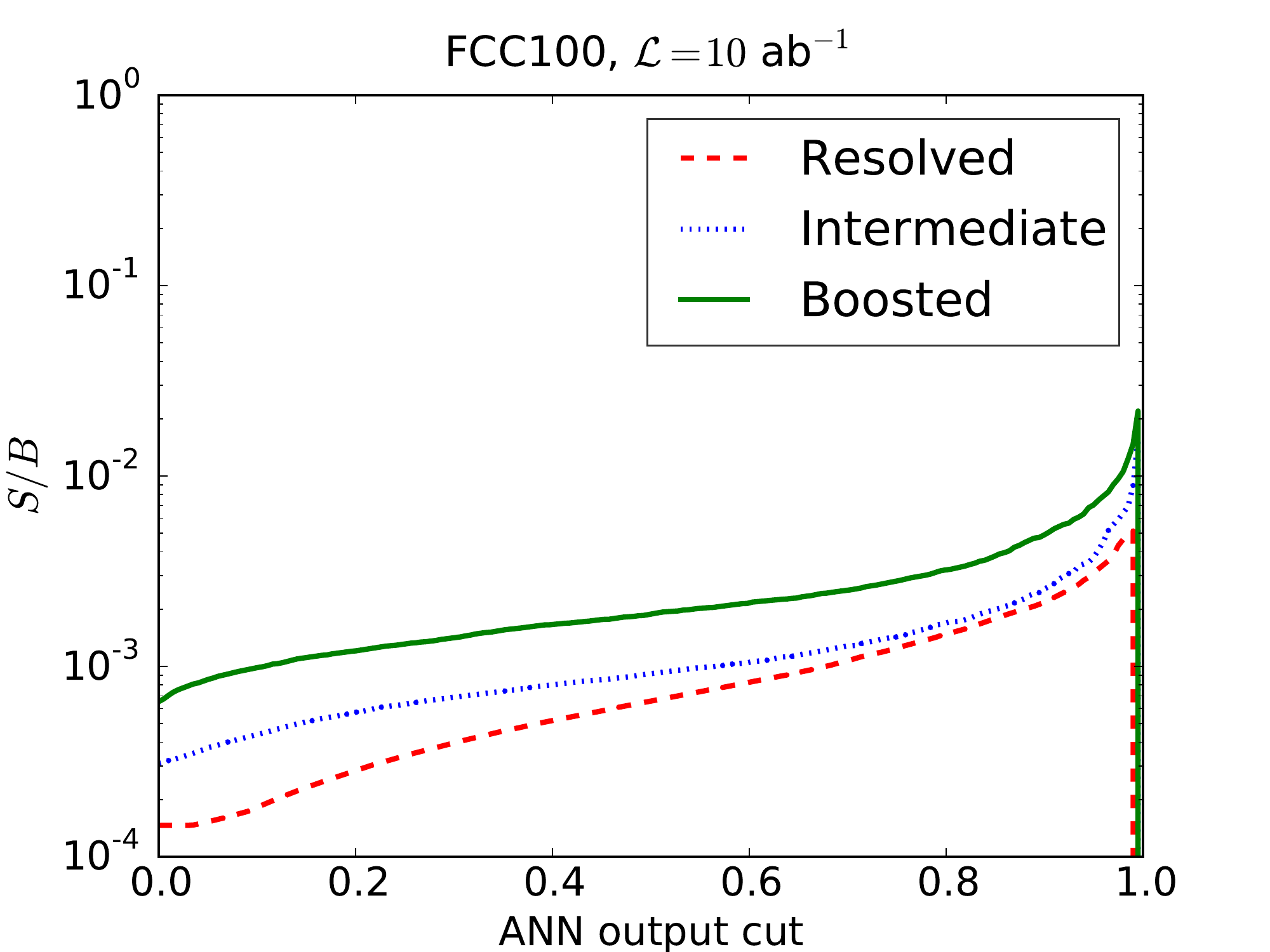}
\caption{ The values of the signal significance, $S/\sqrt{B}$, and of the
  signal over background ratio, $S/B$, for the boosted, intermediate
  and resolved categories as a function of the cut
  $y_{\rm cut}$ in the ANN output.
  Only the $4b$ QCD background is considered here.
  The $y_{\rm cut}=0$
  results are those at the end of the cut-based
  analysis. 
}
\label{fig:sb_mva}
\end{figure}

The results for the signal significance $S/\sqrt{B}$ and the signal-over-background ratio
$S/B$ at a $100\,$TeV collider are shown in Fig.~\ref{fig:sb_mva} as a function of the ANN output cut $y_{\rm cut}$ for the three categories.
A total integrated luminosity of $\mathcal{L}=10$ ab$^{-1}$ is assumed, and only the irreducible QCD $4b$ background is included.
The values  for $y_{\rm cut}=0$ correspond to those at the end of the loose cut-based analysis.
One can observe how in the three categories there is a marked  improvement both in signal
significance and in the signal over background ratio as compared to the pre-MVA results.
In Table~\ref{table:cutflowMVA} the post-MVA results are given for the optimal value of the
ANN discriminant $y_{\rm cut}$ in the three categories, compared with the corresponding pre-MVA results ($y_{\rm cut}=0$).
The number of signal and background events expected for an integrated luminosity of $\mathcal{L}=10$ ab$^{-1}$ is also quoted.

\begin{table}[t]
  \centering
  \begin{tabular}{c|l|c|c|c|c}
    \rule[-0.4em]{0pt}{1.6em} Category  &   &  $N_{\rm ev}$ signal &  $N_{\rm ev}$ back  &  $S/\sqrt{B}$ & $S/B$ \\ 
    \hline
    \hline
    \multirow{2}{*}{Boosted} &  \rule{0pt}{1.1em} $y_{\rm cut}=0$  &       $5\cdot 10^4$    &   $8\cdot 10^7$        &  $6$        &    $6\cdot 10^{-4}$         \\
    &  \rule{0pt}{1.1em} $y_{\rm cut}=0.99$ &   $2\cdot 10^4$    &  $1\cdot 10^6$         &      $22$    &    $2\cdot 10^{-2}$     \\
    \hline
    \multirow{2}{*}{Intermediate} &  \rule{0pt}{1.1em} $y_{\rm cut}=0$  &   $3\cdot 10^4$     & $1\cdot 10^8$  & $3$   &  $3\cdot 10^{-4}$ \\
       &  \rule{0pt}{1.1em} $y_{\rm cut}=0.98$ &         $2\cdot 10^4$       &  $2\cdot 10^6$   &  $10$   & $7\cdot 10^{-3}$        \\
    \hline
      \multirow{2}{*}{Resolved} &  \rule{0pt}{1.1em} $y_{\rm cut}=0$  &  $1\cdot 10^5$      & $8\cdot 10^8$  &   $4$  & $1\cdot 10^{-4}$  \\
    &  \rule{0pt}{1.1em} $y_{\rm cut}=0.95 $ &    $6\cdot 10^4$    &   $2\cdot 10^7$        &    $15$      &    $4\cdot 10^{-3}$  
      \end{tabular}
  \caption{Post-MVA results, for the optimal value of the
    ANN discriminant $y_{\rm cut}$ in the three categories, compared with the
    corresponding
    pre-MVA results ($y_{\rm cut}=0$).
    We quote the number of signal and
    background events expected for $\mathcal{L}=10$ ab$^{-1}$,
    the signal significance $S/\sqrt{B}$ and
    the signal over background ratio $S/B$.
    In this table, only the irreducible
    QCD $4b$ background has been considered.
    \label{table:cutflowMVA}
  }
\end{table}

From Fig.~\ref{fig:sb_mva} and Table~\ref{table:cutflowMVA} one can observe that the statistical
significance of the three categories is very large, with a post-MVA value of $S/\sqrt{B}\simeq 20$ in the boosted category.
However, one also finds that, as compared to the LHC case, the QCD $4b$ multijet background increases
more rapidly than the signal and thus $S/B$ is actually smaller than at $14$~TeV~\cite{Behr:2015oqq}.
Achieving  percent values in $S/B$ requires very hard cuts on the value of the ANN output $y_{\rm cut}$.
At $100$~TeV the boosted category is the most promising one: not only
it benefits from the highest signal significances, it also exhibits
the best signal over background ratio.
The result is analogous to the one found at $14$~TeV~\cite{Behr:2015oqq}, where the significance
of the three categories was quite similar, with the boosted one being the best without pile-up, and
the resolved one exhibiting the higher significance in the simulations with pile-up.
Unfortunately, as it was already mentioned and will be further discussed  below,
the boosted category is the less sensitive to the Higgs self-coupling,
and thus a measurement of the trilinear will depend to good extent on the resolved category.
The smallness of $S/B$ indicates that at a $100$~TeV collider, even more that at $14$~TeV, the feasibility of the measurement
of the $\sigma(hh\to b\bar{b}b\bar{b})$ cross-section will depend strongly on how small the systematic
uncertainties will be, in particular those associated to the background determination.

\paragraph{Extracting the Higgs self-coupling.}

The extraction of the trilinear coupling $\lambda_3$ from the
corresponding cross-section is complicated by the destructive interference
between diagrams that depend on $\lambda_3$ and those that do not.
Here a first estimate is provided of the accuracy on the Higgs self-coupling that can be obtained from the $b\bar{b}b\bar{b}$ final state.
A robust estimate  would require a careful study of the impact of
experimental systematic uncertainties, which is beyond the scope of this report.
Therefore, a number of simplifying assumptions will be used, in particular
a very simple estimate of the total systematic uncertainty in the cross-section
measurement, which is the limiting factor in the extraction of $\lambda_3$.

The sensitivity in the Higgs self-coupling is defined by the $\chi^2$ estimator
\begin{equation}
\label{eq:chi2profile}
\chi^2(\lambda_3) \equiv \frac{\left[ \sigma(hh,\lambda_3) - \sigma(hh,\lambda_{3\,\rm SM})
    \right]^2}{\left( \delta_{\rm stat}\sigma\right)^2+\left( \delta_{\rm sys}\sigma\right)^2 } \, ,
\end{equation}
where $\lambda_3$ is the Higgs self-coupling, $\lambda_{3\,\rm SM} = 1$ is its SM value, $\sigma(hh,\lambda_3)$ is the
post-MVA signal cross-section for a given value of $\lambda_3$, and $\delta_{\rm stat}\sigma$ and
$\delta_{\rm sys}\sigma$ are respectively the statistical and systematic uncertainties in the cross-section measurement.
Signal samples for a range of $\lambda_3$ values have been  processed by the same analysis chain, including the MVA (which is not
re-trained), as for the SM samples.
The $68\%$~CL range for the extraction of $\lambda_3$ is found using the usual parameter-fitting criterion to determine
the values $\pm \delta\lambda_3$ for which the cross-section satisfies
  \begin{equation}
  \label{eq:condition}
\chi^2(\lambda_{3\,\rm SM}~^{+\delta\lambda_3}_{-\delta\lambda_3})=\chi^2(\lambda_{3\,\rm SM})+1 \, .
\end{equation}
Figure~\ref{fig:chi2} shows the $\sigma(hh\to b\bar{b}b\bar{b})$ cross-section at
various steps of the cut-flow: generator level, after  kinematical cuts~\cite{Behr:2015oqq}, after $b$-tagging and finally after the MVA.
The results in the resolved (left plot)  and boosted (right plot) categories are shown
as a function of the Higgs self-coupling  $\lambda_3$.
For the MVA cut,  a representative value of $y_{\rm cut}\simeq 0.7$  has been used.
One finds that, although the MVA is only trained on the SM sample, the signal
selection efficiency of the MVA is relatively flat when $\lambda_3$ is varied, reflecting the fact
that the signal kinematics do not change dramatically.
From the plots one can also see that the resolved category (low and medium Higgs $p_T$) is more sensitive to variations
of $\lambda$ than the boosted one (large Higgs $p_T$), as indicated by the shallower minimum of the latter after  the $b$-tagging
and MVA cuts.
This reflects the fact that the triangle diagram (which depends on the Higgs self-coupling) is  dominant near the threshold region.

\begin{figure}[t]
\begin{center}
\includegraphics[width=0.43\textwidth]{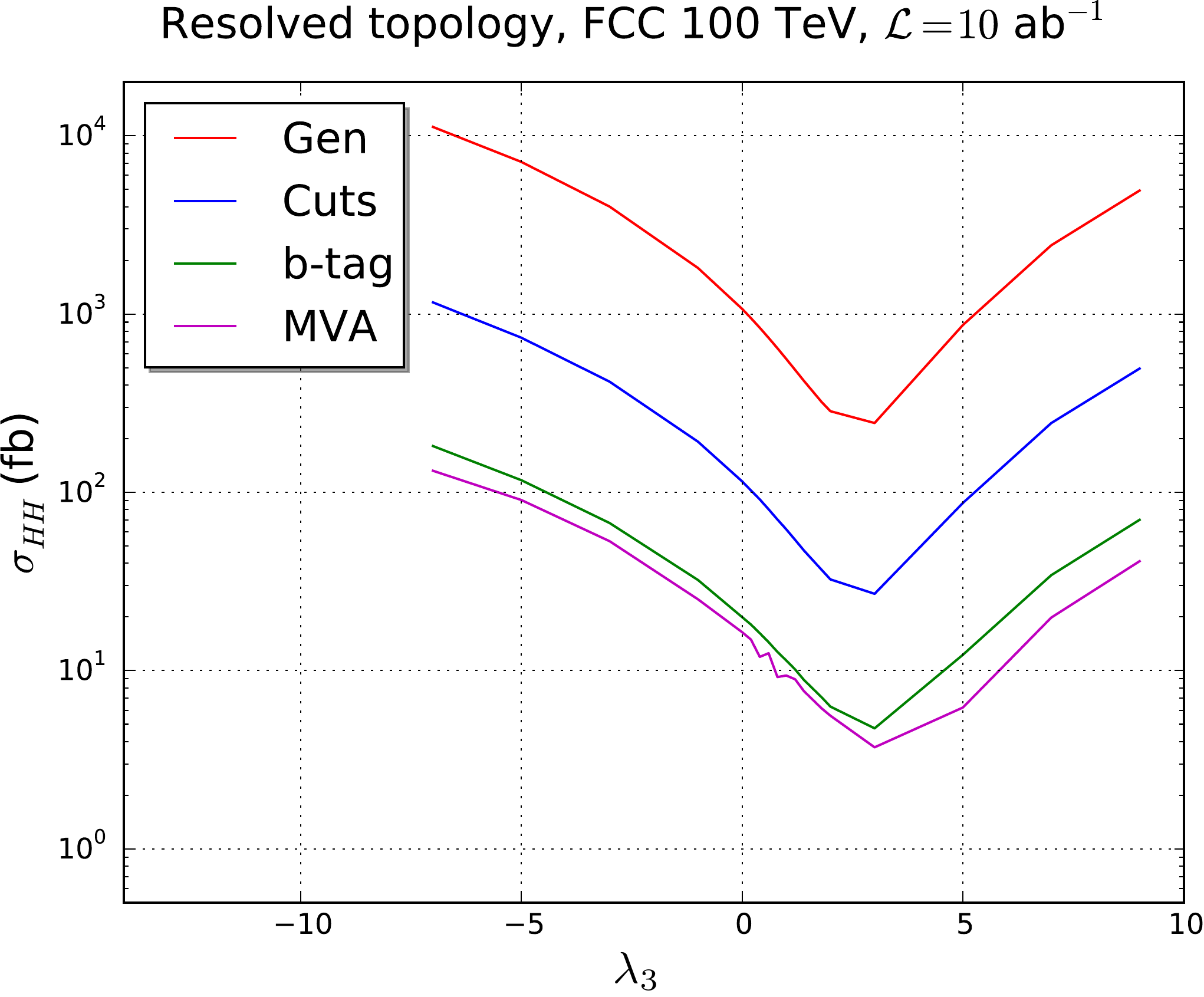}
\hspace{2em}
\includegraphics[width=0.43\textwidth]{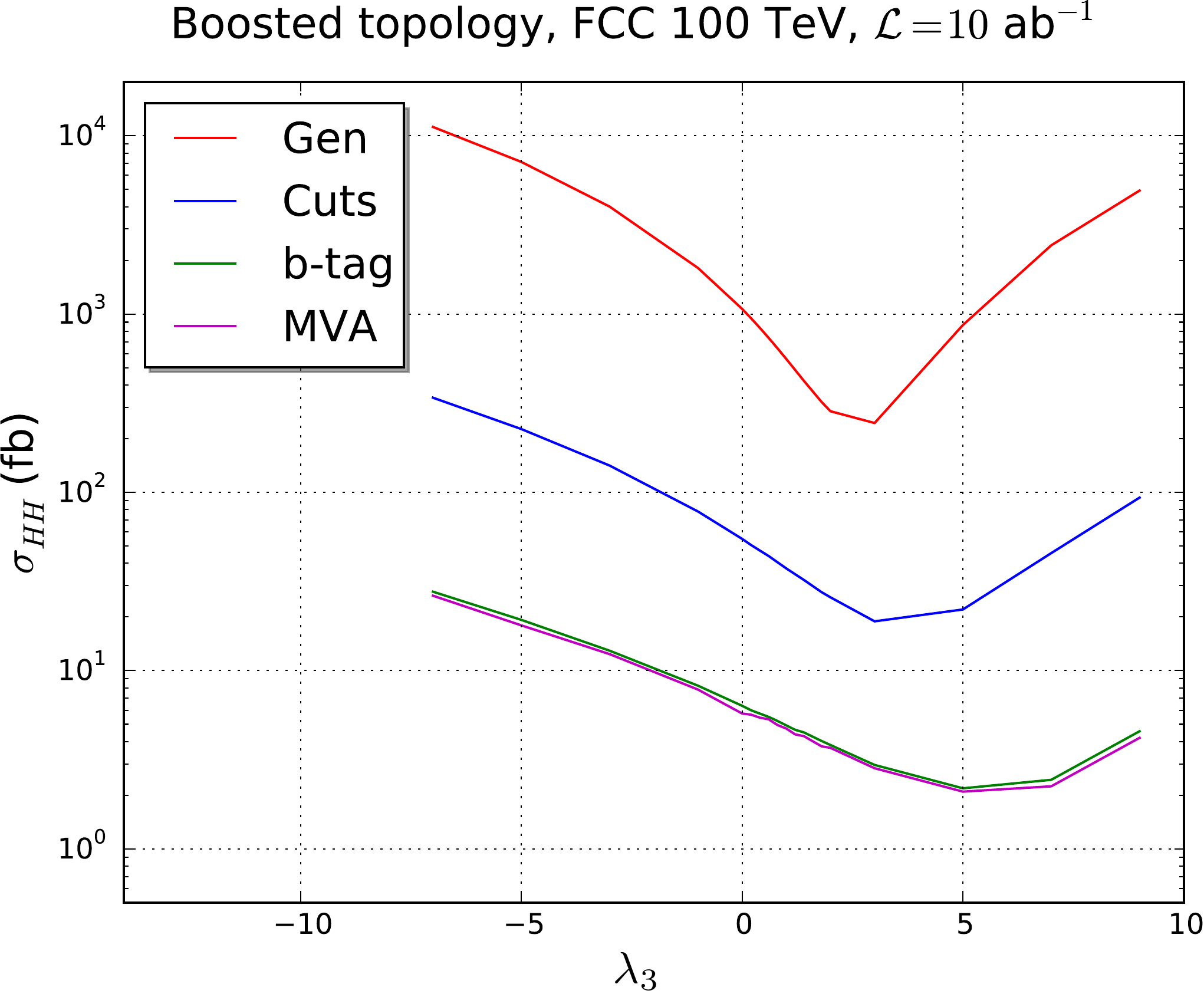}
\caption{The $\sigma(hh\to b\bar{b}b\bar{b})$ cross-section at various steps of the cut-flow: generator level, after
  kinematical cuts, after $b$-tagging and finally after the MVA.
  We show the results  in the resolved (left plot)  and boosted (right plot) categories,
  as  as a function of the Higgs self-coupling  $\lambda_3$.
  For the MVA, a representative value of the ANN output of $y_{\rm cut}\simeq 0.7$  has been used in this plot.
}
\label{fig:chi2}
\end{center}
\end{figure}

 %
  Imposing the condition Eq.~(\ref{eq:condition}), 68\% confidence level intervals are derived
  on the Higgs self-coupling $\lambda_{3}$ in the boosted, resolved and intermediate categories, for  different assumptions on the total systematic error in the measured cross-section.
  The ranges that are found are reported in Table~\ref{tab:chi2}.
  As expected, the results depend  strongly on the assumption for the systematic uncertainty  on the measured cross-section.
  In the optimistic scenario of a measurement with $\delta_{\rm sys}\sigma=25\%$,
  the best performance comes from the resolved category, where at the 68\% CL the trilinear
  can be determined to lie in the interval $\lambda_3 \in \left[ 0.9, 1.5\right]$.
  Looser constrains are derived from the intermediate and from the boosted category.
  On the other hand, for $\delta_{\rm sys}\sigma=100\%$, the constraints in all three categories
  degrade substantially, especially for $\lambda_3 \ge 1$, due to the negative interference effects.

\begin{table}[h]
  \centering
  \begin{tabular}{c|c|c}
\rule[-.6em]{0pt}{1em}  &  $\delta_{\rm sys}\sigma=25\%$ & $\delta_{\rm sys}\sigma=100\%$ \\
    \hline
    \hline
\rule[-.6em]{0pt}{1.7em} Boosted   & $\lambda_3 \in \left[ -0.1, 2.2\right]$  & $\lambda_3 \in \left[ -1.5, > 9\right]$  \\
\hline
\rule[-.6em]{0pt}{1.7em} Intermediate   & $\lambda_3 \in \left[ 0.7, 1.6\right]$  &  $\lambda_3 \in \left[ -0.4, > 9\right]$  \\
\hline
\rule[-.6em]{0pt}{1.7em} Resolved   & $\lambda_3 \in \left[ 0.9, 1.5\right]$  &  $\lambda_3 \in \left[ -0.1, 7\right]$ 
  \end{tabular}
  \caption{ \label{tab:chi2} The $68\%$ confidence level intervals
    on the Higgs self-coupling $\lambda_3$
    obtained from the condition
    Eq.~(\ref{eq:condition}) in the boosted, resolved and intermediate categories.
    We consider two different assumptions on the total systematic error in the measured
    cross-section,  $\delta_{\rm sys}=25\%$ and $\delta_{\rm sys}=100\%$.
  }
\end{table}

\subsubsection{Additional modes with leptons}
\label{sec:hh_leptons}

Due to the considerable increase of the Higgs pair production cross section at $100$~TeV, it is conceivable that rare,
but potentially cleaner, final states become accessible~\cite{Papaefstathiou:2015iba}. This is for instance the case for
decay channels including leptons. In the following we will examine the final states containing a pair of $b$-jets
and 2 or more leptons, namely $hh \rightarrow  (b\bar{b}) (ZZ^*) \rightarrow (b\bar{b}) (4\ell)$,
$hh \rightarrow  (b\bar{b}) (WW^*)/(\tau^+\tau^-) \rightarrow (b\bar{b}) (\ell^+\ell^-)$,
$hh \rightarrow  (b\bar{b}) (\mu^+\mu^-)$ and  $hh \rightarrow  (b\bar{b}) (Z\gamma) \rightarrow (b\bar{b})
(\ell^+ \ell^- \gamma)$.

\paragraph{Simulation setup and detector performance}

The signal events are generated at LO using the \texttt{Herwig++} event generator~\cite{Bahr:2008pv,Bellm:2015jjp} interfaced with the \texttt{OpenLoops} package for
the one-loop amplitudes~\cite{Cascioli:2011va,Maierhofer:2013sha}. The backgrounds are generated with the
\texttt{MadGraph 5/aMC@NLO} package~\cite{Frixione:2010ra, Frederix:2011zi, Alwall:2014hca},
at NLO QCD. The only exception is the $t\bar{t}$ background, which is
generated at LO and merged with the \texttt{Herwig++} parton shower using the MLM algorithm,
including $t\bar{t} + 1$~parton matrix elements. For the latter, the cross section is normalized to the total NLO result.
All simulations include modelling of hadronization as well as the underlying event through multiple parton
interactions, as they are available in \texttt{Herwig++}. No simulation of additional interacting protons (pile-up)
is included in this study. The \texttt{CT10nlo} pdf set~\cite{Lai:2010vv} is used for all simulations. 

The detector effects are included by smearing the momenta of all reconstructed objects and introducing suitable
reconstruction efficiencies. The smearing and efficiency parameters for jets and muons are taken from Ref.~\cite{ATL-PHYS-PUB-2013-009}, while for electrons follow Ref.~\cite{ATL-PHYS-PUB-2013-004}.
Jets are reconstructed by using the anti-$k_t$ algorithm available in the \texttt{FastJet} package~\cite{Cacciari:2011ma, Cacciari:2005hq}, with a radius parameter $R=0.4$. Only  jets with $p_T > 40$~GeV and  $|\eta| < 3$ are considered in the analysis. The jet-to-lepton mis-identification probability is taken to be $p_{j\rightarrow \ell} = 0.0048 \times
\mathrm{exp}({-0.035 p_{Tj}/\mathrm{GeV}})$, following Ref.~\cite{Barr:2014sga}.
A transverse momentum cut $p_T > 20$~GeV is applied on all the leptons, which are also required to lie in the
pseudorapidity range $|\eta| < 2.5$. An isolation criterion is also applied by considering a lepton isolated if it
has $\sum_i p_{T,i}$ less than $15\%$ of its transverse momentum in a cone of $\Delta R = 0.2$ around it.
The tagging of $b$ jets is simulated by looking for jets containing $B$-hadrons in the range $|\eta| < 2.5$.
The tagging efficiency is assumed to be $70\%$, with a mis-tagging probability of $1\%$ for light-flavor jets. Mis-tagged
$c$-jets are not included in the analysis, since their contributions is estimated to be negligible.
Finally no smearing is applied to the missing transverse energy.\footnote{Due to the large cross section, in order to generate
the $t\bar{t}$ samples the following generation-level cuts are applied on the final-state objects $(\ell^+ b \nu_\ell) (\ell'^-\bar{b} \bar{\nu}_{\ell'})$:
\begin{equation}\label{eq:ttcuts}
p_{T,b} > 40~\mathrm{GeV}\,,
\qquad
p_{T,\ell} > 30~\mathrm{GeV}\,,
\qquad
|\eta_\ell| < 2.5\,,
\qquad
0.1 < \Delta (b,b), \Delta(\ell,\ell) < 2.0\,.
\end{equation}
}

In addition to the previous parametrization of the detector effects (denoted as`LHC' parametrization in the following),
an `ideal' parametrization is also considered obtained by setting all efficiencies to $100\%$
(within the same acceptance regions for jets and leptons) and by removing all momentum smearing effects.
The mis-tagging rates for $b$-jets, leptons and photon are kept identical in both parametrizations.
However, additional backgrounds due to mis-tagging are not particularly important for the channels
considered here, provided that they remain at the levels estimated for the high-luminosity LHC.

\paragraph{The $hh \rightarrow (b\overline b)(4\ell)$ channel}\label{subsec:HH->bb4l}

At a $100$~TeV collider, the cross section for the final state $hh \to (b \bar{b}) (4\ell)$ increases to about $0.26$~fb.
The analysis strategy is focused on the reconstruction of all the relevant objects in the hard process, namely the two $b$-jets and
the $4$ leptons.
The events are selected by demanding the presence of two pairs of leptons of opposite charge and same flavor, as well as two identified $b$-jets.
To simulate a possible $4$-lepton trigger,  the following staggered  cuts are imposed on the leptons:
$p_{T,\ell _{\{1,2,3,4\}}} > \{ 35, 30, 25, 20 \}$~GeV.

Since the signal is not expected to possess a large amount of
missing transverse energy, a further cut $\slashed{E}_T < 100$~GeV is imposed. To further reduce the background it is also useful to add
a cut on the lepton separations. Since the distance between all leptons
in the $hh$ signal is substantially smaller than in most of the background processes, a cut $\Delta R ( \ell _1, \ell_ j ) < 1.0$,
with $j = \{2,3,4\}$, is imposed. The backgrounds coming from processes with multiple $Z$ bosons can be reduced by
rejecting events with two on-shell $Z$'s, namely rejecting the event if there are two combinations of same-flavor opposite-sign
leptons with an invariant mass $m_{\ell^+\ell^-} \in [80, 100]$~GeV.
No single pair of same-flavor opposite-sign leptons is allowed with a mass above $120$~GeV.
Finally, the mass windows for the reconstruction of the two Higgs bosons are chosen as
\begin{eqnarray}
M_{bb} \in [100, 150]~\mathrm{GeV}\,,\qquad M_{4\ell} \in [110,140]~\mathrm{GeV},
\end{eqnarray} 
and no cut on the total invariant mass of all the reconstructed objects is imposed.

After the cuts, the most relevant backgrounds are the ones coming from top pair production in association with an Higgs or a $Z$ boson.
The cross section and event yield of all the analyzed backgrounds are listed, together with the signal, in Table~\ref{tb:bb4l_bkg}.
For simplicity, only the mis-tagging of a single lepton are considered, with the dominant process in this case being $W^\pm Zh$.
Processes with multiple mis-tagged leptons are estimated to be totally negligible.

\begin{table}[h]
\resizebox{\linewidth}{!}{%
\def\arraystretch{1.15}
\begin{tabular}{llll}
\hspace{4em} channel \hspace{12em} & \hspace{-2em}$\sigma(100~\mathrm{TeV})$~(fb)\hspace{3.em} &  \hspace{-1.5em}$N_{30~\mathrm{ab}^{-1}}(\mathrm{ideal})$\hspace{3.em} & \hspace{-1.5em} $N_{30~\mathrm{ab}^{-1}}(\mathrm{LHC})$\\
\hline
\hline
$\mathbf{hh} \rightarrow (b\bar{b})  (\ell^+ \ell^- \ell^{'+} \ell^{'-})$ &  $0.26$ & $130$ & $41$ \\
\hline
$\mathbf{t\bar{t} h} \rightarrow (\ell^+ b \nu_\ell) (\ell'^- \bar{b} \bar{\nu}_{\ell'}) (2 \ell) $ & $193.6$ & $304$ & $109$ \\
$\mathbf{t\bar{t} Z} \rightarrow (\ell^+ b \nu_\ell) (\ell'^- \bar{b}
  \bar{\nu}_{\ell'}) (2 \ell) $ &  $256.7$ & $66$ & $25$ \\
$\mathbf{Zh} \rightarrow (b\bar{b})(4 \ell)$ &  $2.29$  &
                                                        $\mathcal{O}(1)$ & $\mathcal{O}(1)$ \\
$\mathbf{ZZZ} \rightarrow (4 \ell)(b\bar{b}) $ & $0.53$ & $\mathcal{O}(1)$  & $\mathcal{O}(1)$  \\
$\mathbf{b\bar{b} h} \rightarrow b\bar{b}(4 \ell)$\hspace{1em} {\small ($p_{T,b} > 15$~GeV)}
        &  $0.26$  & $\mathcal{O}(10)$  & $\mathcal{O}(1)$ \\
$\mathbf{ZZh} \rightarrow (4 \ell)(b\bar{b}) $ & $0.12$& $\mathcal{O}(10^{-2})$  & $\mathcal{O}(10^{-2})$ \\
\hline
$\mathbf{ZZjj} \rightarrow (4 \ell)\, +$ fake $b\bar{b}$ & $781.4$ & $\mathcal{O}(10^{-1})$  & $\mathcal{O}(10^{-1})$  \\
$\mathbf{hZjj} \rightarrow (4 \ell)\, +$ fake $b\bar{b}$ & $68.2$ & $\mathcal{O}(10^{-2})$  & $\mathcal{O}(10^{-2})$  \\
$\mathbf{W^\pm ZZj} \rightarrow (\ell \nu_\ell) (\ell^+\ell^-) (b\bar{b})\,+$ fake $\ell$ &  $7.5$ & $\mathcal{O}(10^{-1})$  &  $\mathcal{O}(10^{-1})$ \\
$\mathbf{W^\pm Zhj} \rightarrow (\ell \nu_\ell) (\ell^+\ell^-) (b\bar{b})\,+$ fake $\ell$ & $1.4$ & $\mathcal{O}(10^{-1})$  &$\mathcal{O}(10^{-2})$  \\
\end{tabular}
}
\caption{Signal and relevant backgrounds for the $(b\bar{b})(\ell^+ \ell^- \ell^{'+} \ell^{'-})$ channel.
The second column reports the cross section after the generation cuts (for the $b\overline b h$ channel the additional cut
listed in the table is imposed at generation level).
The third and fourth columns show the number of events, $N_{30~\mathrm{ab}^{-1}}$, after the cuts
for the `ideal' and `LHC' detector parametrizations obtained by assuming an integrated luminosity of $30$~ab$^{-1}$.
}
\label{tb:bb4l_bkg}
\end{table}

As a result of the analysis one gets that, for the SM signal in the ideal detector parametrization, $S/\sqrt{B + S} \simeq 5.8$
(with $S/B \simeq 0.35$). This corresponds to an estimated precision of ${\mathcal O}(20\%)$ on the SM cross section, which
roughly corresponds to a precision of ${\mathcal O}(30\%)$ on the Higgs trilinear coupling.
In the case of the LHC parametrization one instead has $S/\sqrt{B + S} \simeq 3.1$ (with $S/B \simeq 0.31$), which corresponds
to a precision of ${\mathcal O}(30\%)$ on the SM cross section and of ${\mathcal O}(40\%)$ on the Higgs trilinear.\footnote{In order to
estimate the precision on the trilinear Higgs coupling $\lambda_3$ it is assumed that the dependence of the total cross section
on $\lambda_3$ is the same as the one before the cuts.}

\paragraph{The $hh \rightarrow (b\overline b)(\ell^+ \ell^-)(+ \slashed E_T)$ channel}

As a second channel we consider the final state that includes a $b\bar{b}$ and two oppositely-charged leptons.
This final state receives contributions from three different $hh$ decay modes.
The largest one comes from $hh \rightarrow (b\bar{b}) (W^+ W^-)$ with the $W$'s decaying (either directly, or indirectly
through taus) to electrons or muons. The second-largest contribution comes from $hh \rightarrow (b\bar{b}) (\tau^+\tau^-)$,
with both taus decaying to electrons or muons. Both these channels include final-state neutrinos, and hence are associated
with large missing energy. A third, smaller contribution comes from $hh \rightarrow (b\bar{b}) (\mu^+\mu^-)$,
i.e.~through the direct decay of one Higgs boson to muons. 

Due to the different origin of the leptons in the three processes, the kinematics varies substantially. As
already mentioned, in $(b\bar{b}) (W^+ W^-)$  and $(b\bar{b}) (\tau^+\tau^-)$ large missing energy is expected.
In the $(b\bar{b}) (\tau^+\tau^-)$ channel, the $\tau$ leptons are light compared to the Higgs boson,
hence the leptons and the neutrinos in their decays are expected to be collimated.
On the contrary, in $(b\bar{b}) (W^+ W^-)$ both $W$'s are heavy, one being most of the time on-shell
and the other off-shell with $M_{W^*}$ peaking at $\sim 40$~GeV.
In order to take into account the different kinematics of the various sub-processes,  two separate signal regions are constructed.
The first aims at capturing events containing rather large missing energy, targeting the $(b\bar{b}) (W^+ W^-)$
and $(b\bar{b}) (\tau^+\tau^-)$ channels, whereas the second is aimed towards events with minimal
missing energy that are expected to characterize the $(b\bar{b}) (\mu^+\mu^-)$ channel.

The object reconstruction strategy in the two signal regions is similar.
Events with two tagged $b$-jets and two isolated leptons are considered, with isolation criteria equal to those used in the
$hh \rightarrow (b\overline b)(4\ell)$ analysis of subsection~\ref{subsec:HH->bb4l}.
In addition to the standard observables characterizing the final state objects, it is useful to introduce a further quantity, $M_\mathrm{reco.}$,
aimed at reconstructing the invariant mass of the Higgs decaying into leptons and neutrinos.
$M_\mathrm{reco.}$ is constructed by assuming that the missing energy arising from neutrinos in the decays of the $\tau$
leptons is collinear to the observed leptons:
\begin{equation}
M_\mathrm{reco.} = \left[ p_{b_1} + p_{b_2} + (1+f_1) p_{\ell_1}  + (1 + f_2)  p_{\ell_2}\right]^2 \;,
\end{equation}
where $p_{b_i}$, $p_{\ell_i}$ are the observed momenta of the $i$-th $b$-jet and $i$-th lepton and $f_{1,2}$ are constants of
proportionality between the neutrino and lepton momenta from the decay of the two $\tau$ leptons, namely $p_{\nu_i} = f_i p_{\ell_i}$.
The latter can be calculated from the observed missing transverse energy by inverting the missing transverse momentum
balance relation $L \cdot \mathbf{f} = \mathbf{\slashed{E}}$, where $L$ is the matrix $L_i^j = p_{{\ell_i}}^j$, in which the superscript
denotes the component of the $i$-th lepton momentum, $j = \{x,y\}$ and $E$ and $\mathbf{f}$ are the vectors
$\mathbf{\slashed{E}} = ( \slashed{E}^x, \slashed{E}^y )$ and $\mathbf{f} = (f_1, f_2)$. 

The two signal regions are denoted by SR$_{\slashed{E}}$ and SR$_\mu$. The former is optimized for a signal with significant missing
transverse energy and is aimed at the decay modes $(b\bar{b}) (W^+ W^-)$, $(b\bar{b}) (\tau^+\tau^-)$. The second region
SR$_\mu$ is instead focused on events with minimal missing energy, as in the $(b\bar{b}) (\mu^+\mu^-)$ channel.
The cuts defining the two signal regions are listed in Table~\ref{tb:srs}.

\begin{table*}[!t]
\centering
\def\arraystretch{1.15}
\begin{tabular}{l@{\hspace{2.75em}}l@{\hspace{2.25em}}l}
observable & \hspace{1.75em} SR$_\slashed{E}$ &  \hspace{1.75em} SR$_\mu$ \\
\hline
\hline
$\slashed{E}_T$  & $>100$~GeV & $<40$~GeV\\
$p_{T,\ell _1}$ & $>60$~GeV & $>90$~GeV \\ 
$p_{T,\ell _2}$ &$>55$~GeV  &$>60$~GeV \\ 
$\Delta R(\ell_1, \ell_2)$ & $<0.9$& $\in (1.0, 1.8)$\\ 
$M_{\ell\ell}$ & $\in (50, 80)$~GeV  &$\in (120, 130)$~GeV    \\ 
\hline
$p_{T,b_1}$ & $>90$~GeV &$>90$~GeV \\ 
$p_{T,b_2}$ & $>80$~GeV& $>80$~GeV\\ 
$\Delta R(b_1,b_2)$ &$\in (0.5, 1.3)$ & $\in (0.5, 1.5)$ \\ 
$M_{bb}$ & $\in (110,140)$~GeV& $\in (110,140)$~GeV\\
\hline
$M_{bb\ell\ell}$ & $>350$~GeV&$>350$~GeV \\
$M_\mathrm{reco.}$ & $>600$~GeV & none \\
\end{tabular}
\caption{Cuts defining the two signal regions constructed in the analysis of the $hh \to (\bar{b}b) (\ell^+ \ell^-) (+ \slashed{E}_T)$ channel. The signal regions SR$_\slashed{E}$ and SR$_\mu$ are optimized for the channels with and without missing transverse energy
respectively.}
\label{tb:srs}
\end{table*}

The main irreducible backgrounds for the $(\bar{b}b) (\ell^+ \ell^-) (+ \slashed{E}_T)$ final state include the following processes:
$t\bar{t}$ with subsequent semi-leptonic decays of both top quarks; $b\bar{b}Z$ with decays of the $Z$ boson to leptons;
$b\bar{b}h$ with subsequent decays of the Higgs boson to two leptons; and the resonant $hZ$ and $ZZ$ backgrounds.
The two largest reducible backgrounds are also considered, coming from the mis-tagging of a jet to a single lepton in
the $b\bar{b}W^\pm$ channel and the mis-tagging of a $b\bar{b}$ pair in the $\ell^+ \ell^-\!+$ jets
background.~\footnote{This last background has been estimated by simulating $\ell^+ \ell^-\!\!+ 1$ parton at NLO.}
As before, no mis-identification of $c$-jets to $b$-jets is included. 

\begin{table*}[!t]
\resizebox{\linewidth}{!}{%
\def\arraystretch{1.15}
\begin{tabular}{llll}
\hspace{4em} channel \hspace{12.5em} & \hspace{-1em} $\sigma(100~\mathrm{TeV})$~(fb) \hspace{1em} & \hspace{-1em} $N_{30~\mathrm{ab}^{-1}}(\mathrm{ideal})$ \hspace{1em} & \hspace{-1em} $N_{30~\mathrm{ab}^{-1}}(\mathrm{LHC})$\\
\hline
\hline
$\mathbf{hh} \rightarrow (b\bar{b}) (W^+ W^-) \rightarrow (b\bar{b}) (\ell'^+ \nu_{\ell'} \ell^- \bar{\nu}_\ell)$ & $27.16$ & $209$ & $199$\\ 
$\mathbf{hh} \rightarrow (b\bar{b}) (\tau^+ \tau^-) \rightarrow (b\bar{b})  (\ell'^+ \nu_{\ell'} \bar{\nu}_\tau \ell^- \bar{\nu}_\ell\nu_\tau)$ & $14.63$ & $385$ & $243$ \\ 
\hline
$\mathbf{t\bar{t}} \rightarrow  (\ell^+ b \nu_\ell) (\ell'^-\bar{b} \bar{\nu}_{\ell'})$\hspace{1em} {\small (cuts as in Eq.~\ref{eq:ttcuts})}
& $25.08 \times10^3$ & $343^{+232}_{-94}$  & $158^{+153}_{-48}$ \\
$\mathbf{b\bar{b} Z} \rightarrow b\bar{b}(\ell^+ \ell^-)$\hspace{1em} {\small ($p_{T,b} >30$~GeV)}  &  $107.36 \times 10^3$ & $2580^{+2040}_{-750}$ & $4940^{+2250}_{-1130}$  \\ 
$\mathbf{Z Z} \rightarrow b\bar{b}(\ell^+ \ell^-)$& $356.0$ &  $\mathcal{O}(1)$ &  $\mathcal{O}(1)$ \\ 
$\mathbf{h Z} \rightarrow b\bar{b}(\ell^+ \ell^-)$& $99.79$ & $498$ & $404$    \\ 
$\mathbf{b\bar{b}h} \rightarrow b\bar{b} (\ell^+ \ell^-)$\hspace{1em} {\small ($p_{T,b} > 30$~GeV)}  & $26.81$ & $\mathcal{O}(10)$  &$\mathcal{O}(10)$  \\
\hline
$\mathbf{b\bar{b}W^\pm }\rightarrow b\bar{b} (\ell^\pm\nu_\ell)\, + $ fake $\ell$\hspace{1em} {\small ($p_{T,b} > 30$~GeV)} & $1032.6$ & $\mathcal{O}(10^{-1})$  & $\mathcal{O}(10^{-1})$ \\
$\mathbf{\ell^+ \ell^-\!\!+\!\mathrm{jets}}\rightarrow (\ell^+ \ell^-)\, +$ fake $b\bar{b}$ & $2.14 \times 10^3$ &  $\mathcal{O}(10^{-1})$ & $\mathcal{O}(10^{-1})$
\end{tabular}
}
\caption{Signal and background cross sections for the $(b\bar{b}) (\ell^+ \ell^-+ \slashed{E})$ channel.
Due to the limited MonteCarlo statistics, the estimated number of events for the $t\bar{t}$ and $b\bar{b} Z$ backgrounds
has a rather limited precision (the $1\sigma$ interval is given in the table together with the central value).}
\label{tb:bb2ell_bkg}
\end{table*}

The signal and background cross section are shown in Table~\ref{tb:bb2ell_bkg}, where
the expected number of signal and background events in the signal region SR$_\slashed{E}$
is also reported for the `LHC' and the `ideal' detector parametrizations. Notice that, since the same set of cuts is used for both
parametrizations, the `ideal' case does not necessarily provide a substantial improvement in the signal efficiency.
This happens in particular for the $(b\bar{b}) (W^+ W^-)$ sample.
The high signal yield in this channel allows one to determine the SM-like $hh$ production with fair accuracy.
In the `ideal' case, with $30$~ab$^{-1}$ integrated luminosity, a large statistical significance $S/\sqrt{B+S} \sim 9.4$
is expected with $S/B \sim 0.17$, allowing a determination of the total SM cross section with a precision of ${\cal O}(10\%)$.
This corresponds to an estimated precision on the Higgs trilinear coupling of ${\cal O}(10\%)$.
In the `LHC' parametrization, the statistical significance remains fairly high, $S/\sqrt{B+S} \sim 5.7$ with
$S/B \sim 0.08$, leading to a precision of ${\cal O}(20\%)$ on the SM cross section and ${\cal O}(20\%)$
on the Higgs trilinear.

On the other hand, the prospects for the $(b\bar{b}) (\mu^+\mu^-)$ channel after the SR$_\mu$ cuts are applied are
rather bleak: with $30$~ab$^{-1}$ of integrated luminosity, only a handful of events are expected with
the `LHC' detector parametrization with a few hundred background events, even imposing hard transverse momentum cuts on the
muons and a tight mass window on the di-muon invariant mass around the Higgs boson mass.
Because of the latter cut, turning to the `ideal' situation improves the signal efficiency substantially,
since the smearing of the muon momenta is absent. Despite this, only $\mathcal{O}(80)$ events
would be obtained with $30$~ab$^{-1}$ integrated luminosity with a similar number of background events as for the
`LHC' parametrization. Hence, barring any significant enhancements of the rate due to new physics,
the $(b\bar{b}) (\mu^+\mu^-)$ contribution to the $hh \to (\bar{b}b) (\ell^+ \ell^-)$ final state is not expected to provide
significant information.

\begin{table*}[!t]
\resizebox{\linewidth}{!}{%
\def\arraystretch{1.15}
\begin{tabular}{llll}
\hspace{4em} channel \hspace{12em} & \hspace{-1em} $\sigma(100~\mathrm{TeV})$~(fb) \hspace{1em} & \hspace{-1em} $N_{30~\mathrm{ab}^{-1}}(\mathrm{ideal})$ \hspace{1em} & \hspace{-1em}$N_{30~\mathrm{ab}^{-1}}(\mathrm{LHC})$\\
\hline
\hline
$\mathbf{hh} \rightarrow (b\bar{b})  (\mu^+ \mu^-)$ & $0.42$ & $86$ & $18$\\
\hline
$\mathbf{t\bar{t}} \rightarrow  (\ell^+ b \nu_\ell) (\ell'^-  \bar{b} \bar{\nu}_{\ell'})$\hspace{1em} {\small (cuts as in Eq.~\ref{eq:ttcuts})} & $25.08 \times10^3$ & $480^{+1100}_{-140}$ & $158^{+150}_{-48}$  \\
$\mathbf{b\bar{b} Z} \rightarrow b\bar{b}(\ell^+ \ell^-)$\hspace{1em} {\small ($p_{T,b} > 30$~GeV)}  &  $107.36 \times 10^3$ & $< 740$ & $490^{+1130}_{-140}$  \\ 
$\mathbf{Z Z} \rightarrow b\bar{b}(\ell^+ \ell^-)$& $356.0$ & $\mathcal{O}(1)$ & $\mathcal{O}(1)$  \\ 
$\mathbf{h Z} \rightarrow b\bar{b}(\ell^+ \ell^-)$& $99.79$ &$\mathcal{O}(1)$  &  $25$ \\ 
$\mathbf{b\bar{b}h} \rightarrow b\bar{b} (\ell^+ \ell^-)$\hspace{1em} {\small ($p_{T,b} > 30$~GeV)}  & $26.81$  &$\mathcal{O}(10)$  & $\mathcal{O}(10)$ \\\hline
$\mathbf{b\bar{b}W^\pm }\rightarrow b\bar{b} (\ell^\pm\nu_\ell)\,+$ fake $\ell$\hspace{1em} {\small ($p_{T,b} > 30$~GeV)} & $1032.6$ & $\mathcal{O}(10^{-1})$ & $\mathcal{O}(10^{-1})$ \\
$\mathbf{\ell^+ \ell^-\!\!+\!\mathrm{jets}}\rightarrow (\ell^+ \ell^-)\,+$ fake $b\bar{b}$ & $2.14 \times 10^3$ & $\mathcal{O}(10^{-1})$& $\mathcal{O}(10^{-1})$
\end{tabular}
}
\caption{Signal and background cross sections for the $(b\bar{b}) (\mu^+ \mu^-)$ channel.
Due to the limited MonteCarlo statistics, the estimated number of events for the $t\bar{t}$ and $b\bar{b} Z$ backgrounds
has a rather limited precision. The $1\sigma$ interval is given in the table together with the central value.
For the case of $b\bar{b}Z$ in the `ideal' parametrization, we list the $1\sigma$-equivalent region, since no events
were left after the cuts.}
\label{tb:bb2mu_bkg}
\end{table*}

\paragraph{The $hh \rightarrow (b\overline b)(\ell^+ \ell^- \gamma)$ channel}

The $hh \to (\bar{b}b) (\ell^+ \ell^- \gamma)$ channel in the SM
has a cross section $\sigma_{SM} \simeq 0.21\ \mathrm{fb}$, only slightly lower than the $hh \to (b \bar{b}) (4\ell)$ one.
The backgrounds are, however, substantially larger. An estimate  of the relevance of this channel can be obtained by
including only the most significant irreducible backgrounds, namely those from ${b\bar{b} Z\gamma}$,
$t\bar{t}\gamma$, and $hZ \gamma$, as well as the dominant reducible ones, where a photon is mis-tagged
in $b\bar{b} Z$ or $t\bar{t}$ production.

Events are selected by requiring two leptons of the same flavor with $p_{T,\ell_{\{1,2\}}} > \{ 40, 35\}$~GeV,
two anti-$k_t$ $R=0.4$ $b$-jets with $p_{T,b_{\{1,2\}}} > \{ 60, 40\}$~GeV, $\slashed{E}_T < 80$~GeV and a
photon with $p_{T,\gamma} > 40$~GeV.
No isolation requirements are imposed on the photon. The additional cuts  are imposed: $\Delta R ( \ell _1, \ell_ 2 ) < 1.8$,
$\Delta R ( \ell _1, \gamma) < 1.5$ and $ 0.5 < \Delta R ( b_1, b_2 ) < 2.0$. The invariant mass of
the $b$-jet pair is required to be in the range $100 < M_{b\bar{b}} < 150$~GeV,
while the  system of the two leptons and the photon must have an invariant mass lying in the rage $100$~GeV~$< M_{\ell^+ \ell^- \gamma} < 150$~GeV.

Even after these cuts, the $b\bar{b} Z\gamma$ background dominates the final sample, giving a signal-to-background
ratio of $\mathcal{O}(0.02-0.03)$ with only $\mathcal{O}(100)$ signal events with $30$~ab$^{-1}$ integrated luminosity.
Therefore, this channel is not expected to provide significant information on the double Higgs production process
at a 100~TeV pp collider, unless a significant alteration of the $hh$ channel is present due to new physics effects.

\subsection{Triple Higgs production and the quartic Higgs self-coupling}
\label{sec:HH_triple}

In this section we discuss the prospects for the measurement of the triple-Higgs production process.
The main relevance of this channel lies in the possibility of directly accessing the quadrilinear Higgs self-coupling.
The very small production cross section, however, makes the measurement of $\lambda_4$ extremely challenging.

Early work on triple-Higgs production showed that lepton colliders can not access this channel.
For instance, at an $e^+e^-$ machine with a center-of-mass energy of $\sqrt{s}=1$~TeV,
the cross section of the process $e^+e^-\rightarrow ZHHH$ is only $0.4$~ab~\cite{Djouadi:1999gv}, leading to just $1.2$
signal events when assuming the designed integrated luminosity of $3$~ab$^{-1}$.

The situation can instead be more favorable at high-energy hadron colliders.
In this case the main production channel is gluon fusion, while production modes in association with a
gauge bosons, namely $WHHH + X$ and $ZHHH + X$, have a negligible cross section~\cite{Dicus:2016rpf}.
At the $14$~TeV LHC the total SM production cross section
is of the order of $0.1$~fb~\cite{Plehn:2005nk,Binoth:2006ym,Maltoni:2014eza}, which is too
small to be observed with the current designed luminosity.
On the other hand, at a $100$~TeV hadron collider, similarly to what happens for double-Higgs production,
the gluon fusion cross section increases by almost two orders of magnitude with respect to the LHC value,
reaching about $5$~fb (see Table~\ref{table:Higgspair}). This leads to a reasonable amount of signal events
to perform a dedicated analysis.

\begin{figure*}[!htp]
  \centering
\includegraphics[width=0.245\textwidth]{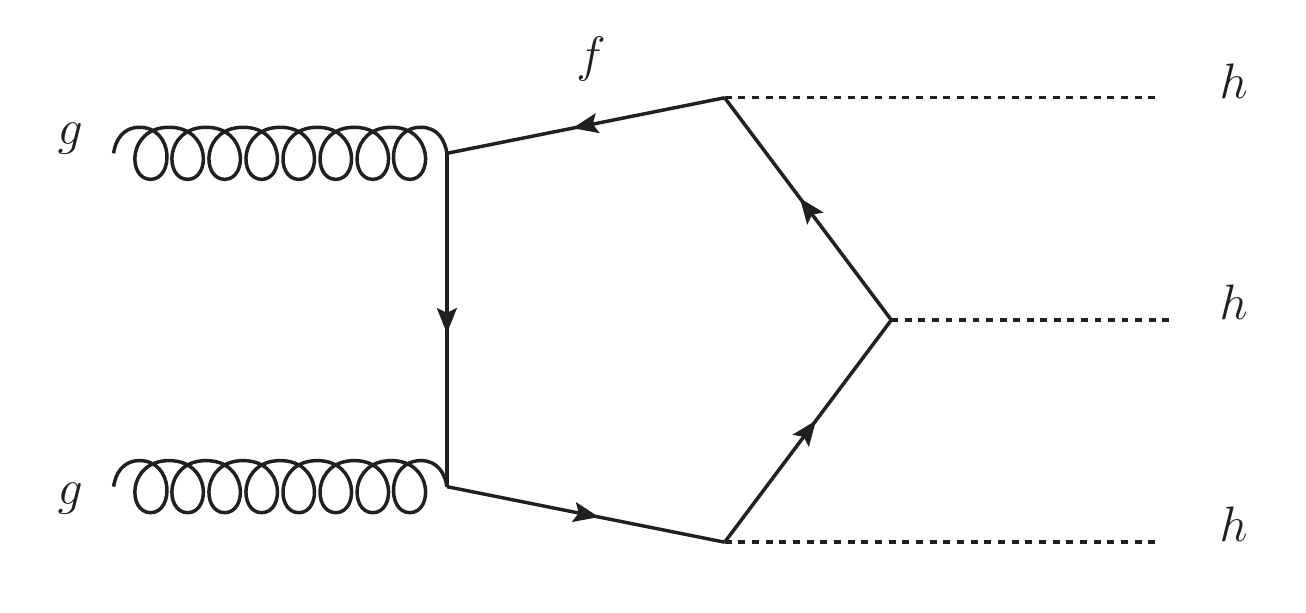}
\includegraphics[width=0.245\textwidth]{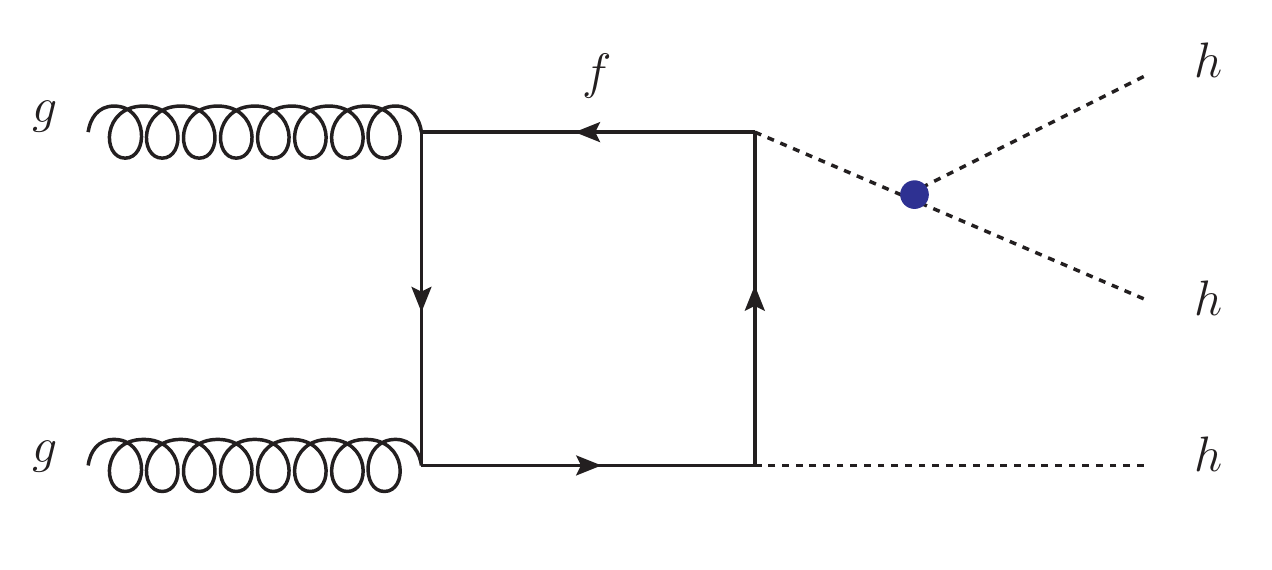}
\includegraphics[width=0.245\textwidth]{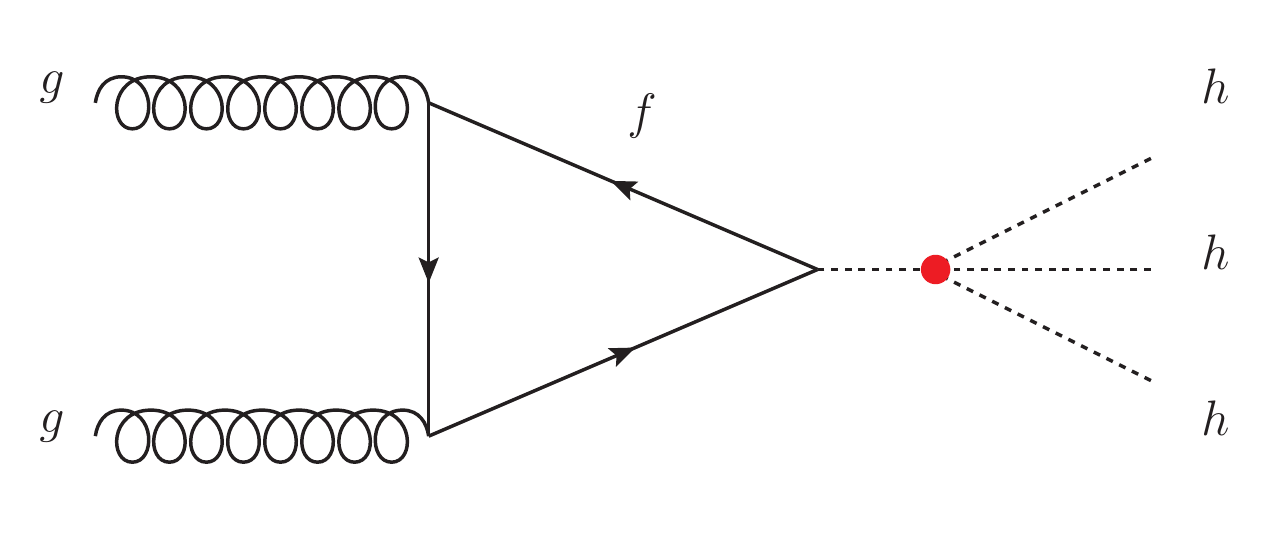}
\includegraphics[width=0.245\textwidth]{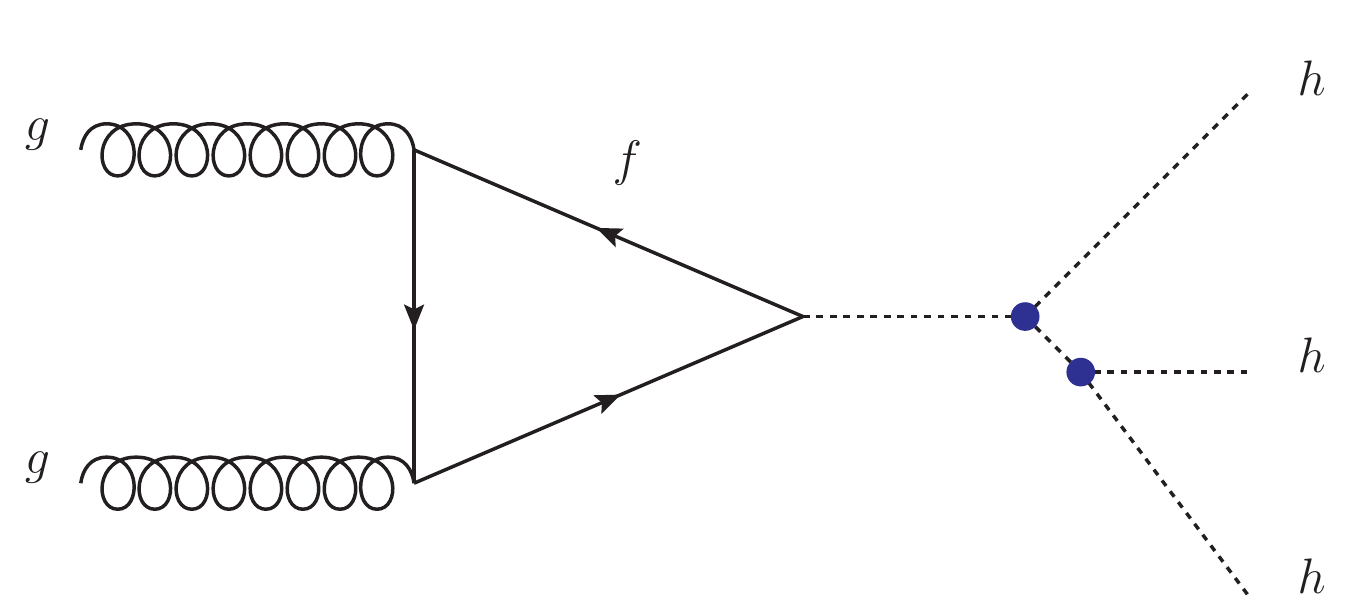}
\caption{Example Feynman diagrams contributing
to Higgs boson triple production via gluon fusion in the Standard Model. The vertices
highlighted with blobs indicate either triple (blue) or quartic (red) self-coupling contributions.}\label{fig:feyndiags}
\end{figure*}

The main diagrams contributing to the gluon fusion channel are shown in Fig.~\ref{fig:feyndiags}.
It turns out that, exactly as in the double-Higgs process, the main contribution to the amplitude comes from
the diagrams that do not contain the multi-Higgs interactions, namely the pentagon ones.
The diagrams with a trilinear and a quadrilinear Higgs coupling, on the other hand, are significantly suppressed.
The dependence of the total cross section on the Higgs self couplings is thus expected to be quite mild.
This expectation is indeed confirmed by Fig.~\ref{LI:cross_lambda4}, which shows the total cross
section as a function of the Higgs quartic coupling. A modification of the Higgs quartic self-coupling has also a
marginal impact on the kinematic distributions, as shown in Fig.~\ref{LI:diff_lambda4}.
These results suggest that the extraction of the $\lambda_4$ coupling is a very challenging task,
and can be problematic unless the triple-Higgs production channel can be measured with quite good accuracy.

\begin{figure*}[!htp]
\centering
\includegraphics[width=0.57\textwidth]{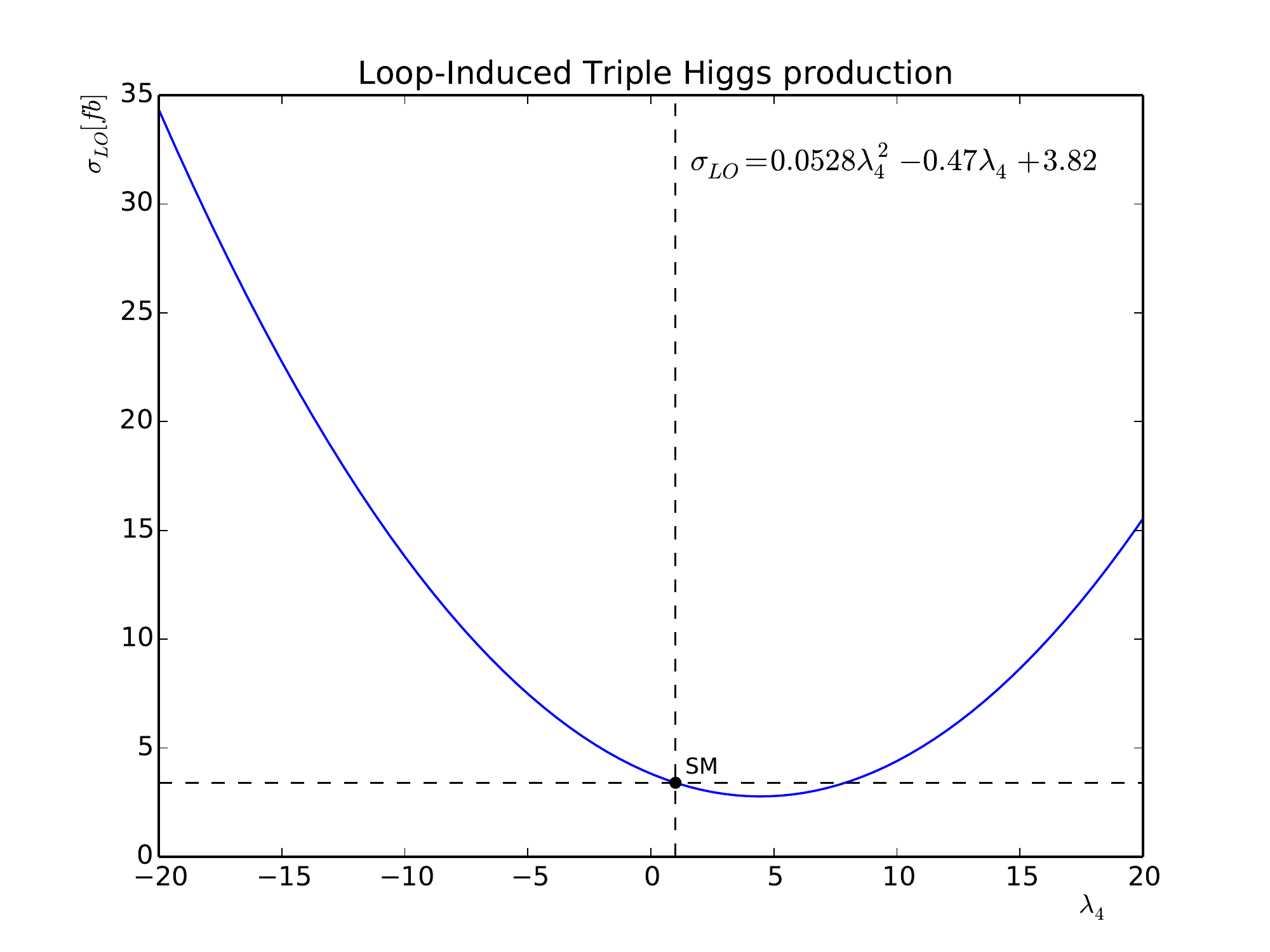}
\caption{Inclusive LO cross-section for $gg \to HHH$ as a function of the $\lambda_4$ parameter.
Details on the computation can be found in Ref.~\cite{Chen:2015gva}.}\label{LI:cross_lambda4}
\end{figure*}

\begin{figure*}[!htp]
  \centering
\includegraphics[width=0.49\textwidth]{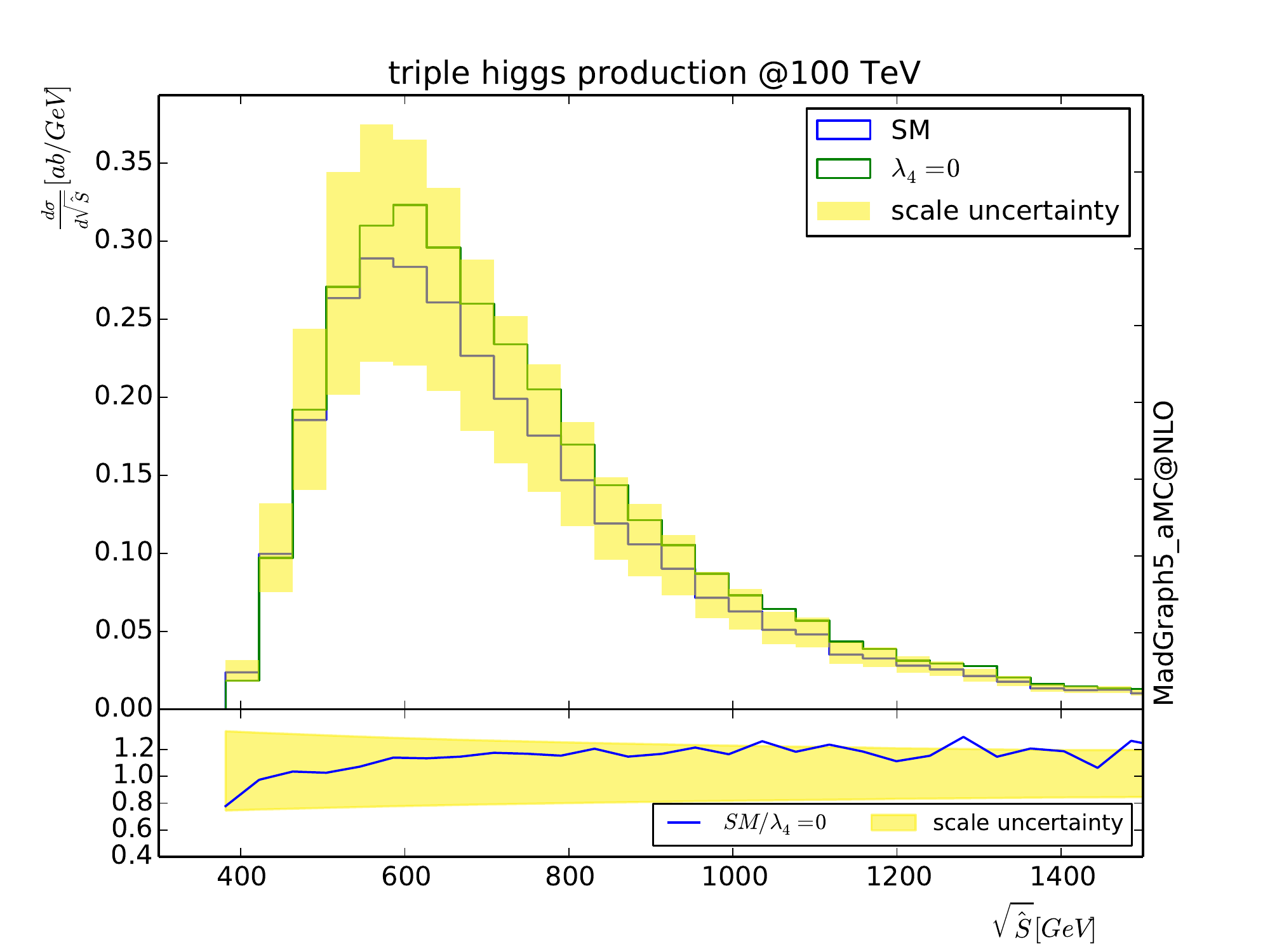}
\includegraphics[width=0.49\textwidth]{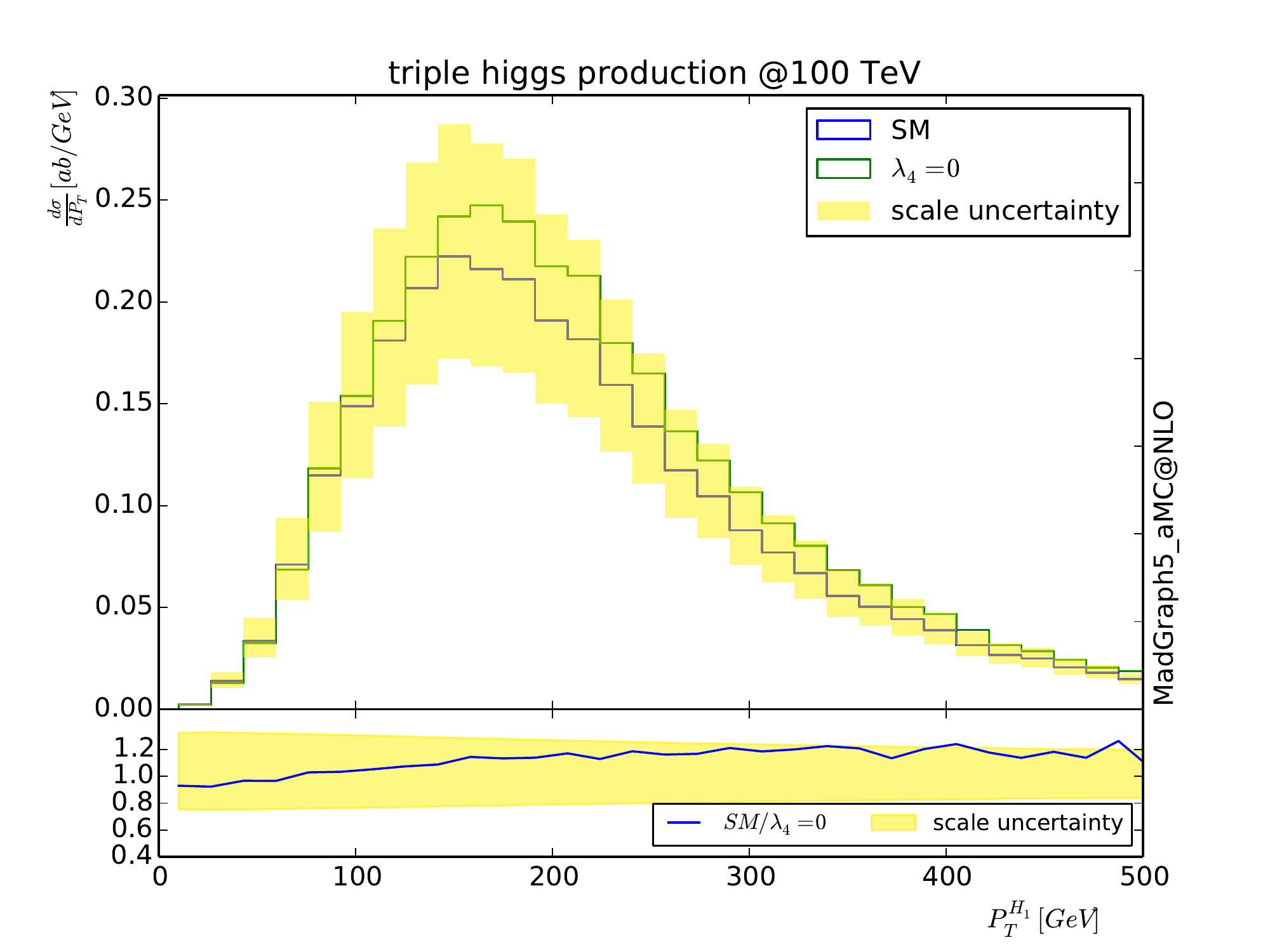}
  \caption{Dependence of the differential cross-section for the triple Higgs production channel on the Higgs
  quartic self-coupling. The left and right panels show the $\sqrt{\hat s}$ and $p_T^{H_1}$ distributions
  for the benchmark points $\lambda_4=1$ (SM) and $\lambda_4=0$.}
  \label{LI:diff_lambda4}
\end{figure*}

One of the most promising decay channel to observe the triple-Higgs production process is
$p p \to HHH\to b\bar{b}b\bar{b}\gamma\gamma$.
This channel combines a clear enough final state, which can be used discriminate the signal against the various backgrounds,
and a relatively large cross section. In the following subsections we will describe the three
analyses of Refs.~\cite{Papaefstathiou:2015paa,Fuks:2015hna,Chen:2015gva}, which focus on scenarios with
different $b$-tagging efficiency, namely $60\%$, $70\%$ and $80\%$.
The $60\%$ $b$-tagging benchmark can be considered as a pessimistic scenario since
it assumes the current $b$-tagging working point at the LHC. The other two analyses, on the other hand, give an
idea of how much the prospects for measuring triple-Higgs production can improve with a higher detector performance.

\subsubsection{Pessimistic hypothesis: \boldmath $60\%$ $b$-tagging
  efficiency}

Let us first consider a ``pessimistic'' scenario in which the
$b$-tagging efficiency is $60\%$~\cite{Chen:2015gva}.
The signal events are generated at LO (using the MadLoop/aMC@NLO~\cite{Hirschi:2011pa} and GoSam~\cite{Cullen:2014yla}
packages) and the NLO effects are taken into account through a rescaling by a $k$-factor $k=2$.
Two types of backgrounds are considered, namely $pp\rightarrow b\bar{b}jj\gamma\gamma$ and $pp\rightarrow H t\bar{t}$.
Parton shower and hadronization effects for the signal events are also included by using
{\sc\small PYTHIA}~\cite{Sjostrand:2006za,Sjostrand:2014zea}, while detector effects are taken into account through
{\sc\small DELPHES 3.0}~\cite{Ovyn:2009tx,deFavereau:2013fsa}.  In all the simulations the PDF
set  CTEQ6L1~\cite{Pumplin:2002vw} is used.

Jets are clustered using the anti-$k_t$ algorithm~\cite{Cacciari:2008gp} as implemented
in {\sc\small FASTJET}~\cite{Cacciari:2011ma} with a cone of radius $R=0.5$ and minimum $p_T (j)=30$ GeV. 
For photon identification, the maximum reconstruction efficiency is assumed to be $95\%$, for transverse momentum
$p_T(\gamma)>10$ GeV and rapidity $|\eta(\gamma)|\leq 2.5$, whereas
it decreases to $85\%$ for $2.5<|\eta(\gamma)|\leq 5.0$. Pile-up effects are neglected.
The $b$-tagging procedure is implemented by mimicking the $60\%$ $b$-jet efficiency LHC working point.
The (mis-)tagging efficiencies vary as a function of the transverse momentum $p_T$ and rapidity $\eta$ of the jets.
For a transverse momentum of $p_T(j) =120$ GeV, the $b$-tagging efficiencies for ($b$, $c$, light) jets are $(0.6, 0.1, 0.001)$,
while they drop to $(0.28, 0.046, 0.001)$ at $p_T(j)=30$ GeV. 

\begin{figure}[!t]
  \centering
  \includegraphics[width=0.4\textwidth]{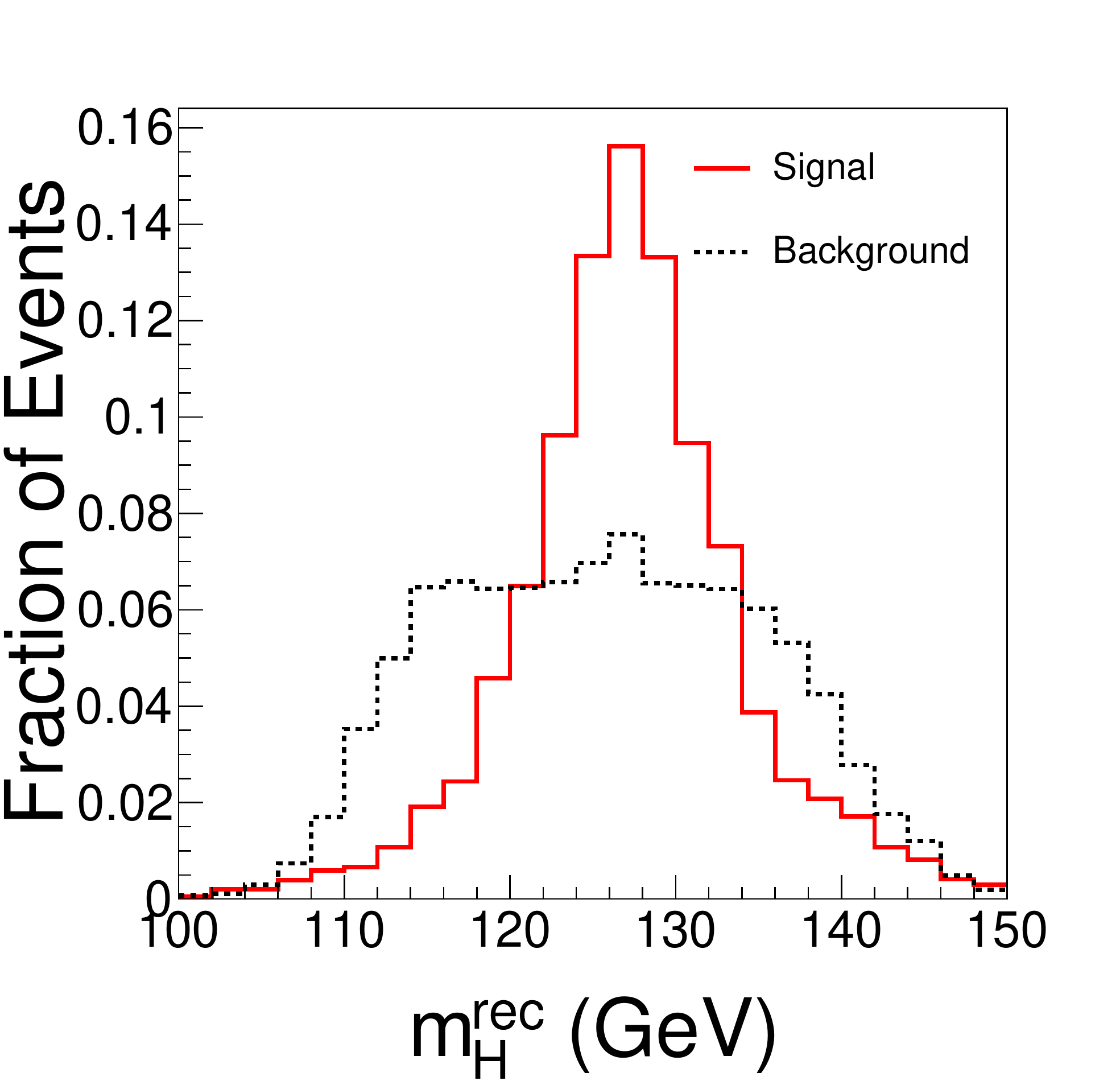}
  \includegraphics[width=0.4\textwidth]{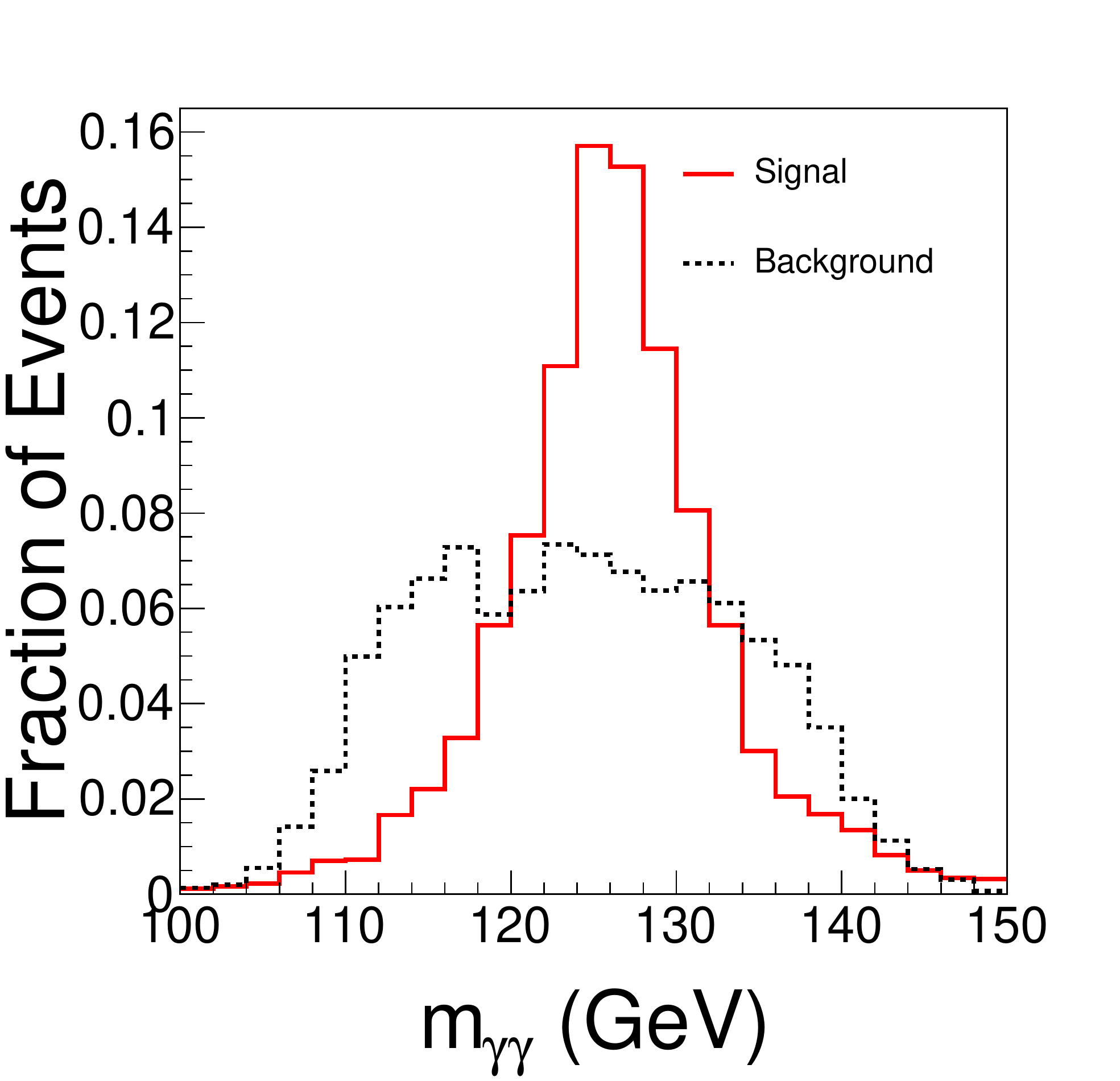}
  \caption{Distributions of the reconstructed Higgs mass $m_H^{rec}$ and invariant mass of the di-photon system
  $m_{\gamma\gamma}$ for the signal and background events.\label{LI_MREC}}
\end{figure}

In order to suppress the large background and select the most relevant events, several preselection cuts are introduced.
\begin{enumerate}
  \item Only events with $4$ or $5$ jets are considered, including at least $2$ tagged $b$-jets. The transverse momenta
  of the jets are required to be  $p_T(j)>30$ GeV. 
  \item The events are required to contain exactly $2$ isolated photons with $p_T(\gamma)>30$~GeV. 
  \item The total number of jets reconstructed by the detector is required to be $\leq 5$
  (this cut aims at suppressing the $p p \to t \bar{t} H$ background with fully hadronic $t\bar{t}$ decays, where $t \rightarrow b W^+$).
  \item Events with MET $>50$ GeV are vetoed (this cut aims at decreasing the $ p p \to t \bar{t} H$ background).
\end{enumerate}
In order to cope with the combinatoric issues, a ``Higgs reconstructed mass'' $m_H^{rec}$ is constructed
by considering all possible jet pairings and selecting the permutation that minimizes the $\chi$-squared fit
to three decaying particles with a mass $m_H^{rec}$. Events with large $\chi$-square ($\chi^2>6.1$) are discarded.
The reconstructed Higgs mass is then required to be in the window
$|m_H^{rec} - 126$~GeV~$|<5.1$~GeV.
A similar chi-square technique~\cite{Yang:2011jk} is used to look for the presence of a top final state and reject those type
of events. 
The distribution of the reconstructed mass $m_H^{rec}$ and of the photons invariant mass $m_{\gamma\gamma}$
is shown in Fig.~\ref{LI_MREC} for the signal and backgrounds events.

The impact of the various cuts listed before is shown in Table~\ref{tab:table2}. The ratio $S/B$ can be enhanced
by almost one order of magnitude, but the signal significance $S/\sqrt{S+B} \simeq 0.15$ remains quite poor.
The small size of the signal cross section, combined with the smearing induced by the finite detector resolution, prevents
an efficient suppression of the background. More sophisticated analysis strategies (multivariate approaches
and boosted decision trees) or the variation of the number of tagged $b$-jets do not substantially modify the results
and only allow a marginal improvement in the signal significance.
With such low sensitivity and weak dependence of the cross-section on the Higgs quartic self-coupling,
the analysis considered in this subsection is expected to lead to a determination of $\lambda_4$ in the range $[-20, 30]$
with an integrated luminosity of $30$~ab$^{-1}$.

\begin{table}
  \begin{center}
  \def\arraystretch{1.25}
  \begin{tabular}{c|c|c|c|c|c}
                        &  Signal          &  $b\bar{b}jj\gamma\gamma$   &  $Ht\bar{t}$        &  $S/B$               &  $S/\sqrt{B}$ \\
  \hline
  \hline
  preselection          &  $50$                    &  $2.3\times 10^{5}$               &  $2.2\times 10^{4}$ &  $2.5\times 10^{-4}$ &  $0.14$       \\
  \hline
  $\chi_{H,min}^2<6.1$  &  $26$                    &  $4.6\times 10^{4}$               &  $9.9\times 10^{3}$ &  $5.0\times 10^{-4}$ &  $0.14$       \\
  \hline
  $|m^{rec}_H-126$~GeV$|<5.1$~GeV  &  $20$         &  $1.7\times 10^{4}$               &  $7.0\times 10^{3}$ &  $8.1\times 10^{-4}$ &  $0.15$        \\
  \end{tabular}
  \end{center}
  \caption{Signal sensitivity in the ``pessimistic'' scenario after each group of selection cuts is imposed. The integrated luminosity
  is assumed to be $30$~ab$^{-1}$.}\label{tab:table2}
\end{table}

\subsubsection{Intermediate hypothesis: \boldmath $70\%$ $b$-tagging
  efficiency}

The second scenario considered is an ``intermediate'' one in which the $b$-tagging efficiency is assumed to be
$70\%$~\cite{Fuks:2015hna}.
For this analysis the signal and background events are generated by convoluting the LO
hard-scattering matrix elements (calculated by {\sc MadGraph5\_aMC@NLO}) with
the NNPDF~2.3 set of parton densities~\cite{Ball:2012cx}. To take into account NLO effects
the background samples are conservatively rescaled by including a $k$-factor of $2$.
All generated final-state particles are required to have a transverse-momentum
\mbox{$p_T>15$~GeV}, a pseudorapidity \mbox{$|\eta|<5$} and to be isolated from
each other by an angular distance, in the transverse plane, of \mbox{$\Delta R>0.4$}. Parton shower,
hadronization, underlying event and pile-up effects are not included. Detector effects are taken into account by
including generic reconstruction features based on the ATLAS detector performances,
namely a smearing of the momentum and energy of the photons and jets~\cite{Aad:2014nim,ATL-PHYS-PUB-2013-004}.
In particular, photons can be remarkably well reconstructed with a resolution that only weakly
depends on the energy. One consequently expects that a relatively narrow peak,
centered on the true Higgs-boson mass value, will emerge in the diphoton
invariant-mass spectrum.
As already mentioned, the $b$-tagging efficiency is assumed to be $70\%$ and the related mistagging rates
for a $c$-jet and a light jet are taken to be $18\%$ and $1\%$ respectively~\cite{ATL-PHYS-PUB-2014-014}.

Events are preselected by demanding that they contain at least four central jets and exactly two central photons with
\mbox{$|\eta|<2.5$}. The transverse momenta of the four
leading jets are required to be greater than $50$, $30$, $20$ and $15$~GeV, and the
ones of the two photons to satisfy \mbox{$p_T^{\gamma_1}>35$~GeV} and $p_T^{\gamma_2}>15$~GeV.
In order to reduce the signal contamination from jets misidentified as photons, isolation
constraints on the photons are imposed, namely the transverse energy
lying in a cone of radius \mbox{$R=0.3$} centered on each photon is required to be smaller than $6$~GeV~\cite{ATLAS:2014iha}.
After the preselection, the two Higgs bosons originating
from the four jets are reconstructed and their invariant masses $m_{\rm jj_k}$ (with
$k=1,2$) are required to satisfy \mbox{$|m_h - m_{\rm jj_k}|<15$~GeV}. In order to solve
possible combinatorics issues, the correct two dijet systems are selected as the
combination of jets that minimizes the mass asymmetry
\begin{equation}
  \Delta_{\rm j j_1, j j_2} = \frac{m_{\rm j j_1} - m_{\rm j j_2}}
     { m_{\rm j j_1} + m_{\rm j j_2} } \ .
\end{equation}
The third Higgs boson is reconstructed from the diphoton system whose
invariant mass $m_{\gamma\gamma}$ is required to be in a window \mbox{$|m_h-m_{\gamma\gamma}|<M$}
with \mbox{$M\in[1,5]$~GeV}. For the minimal number of $b$-tagged jets, $N_b^{\rm min}$, different options are
considered, namely \mbox{$N_b^{\rm min}\in[2,4]$}.
Finally the invariant mass of the four-jet system is required to be smaller than $600$~GeV. This cut has the
advantage of significantly reducing the $\gamma\gamma t\bar{t}$ background
without affecting the signal since jets arising from a top quark decay are generally harder.
As a result, the dominant sources of background consist of $\gamma\gamma b\bar{b}jj$, $\gamma\gamma Z(\rightarrow  bb)jj$
and $\gamma\gamma t \bar t$.\footnote{The additional backgrounds
$h(\rightarrow{\gamma\gamma})h(\rightarrow{bb})Z(\rightarrow{bb})$ and $h(\rightarrow{\gamma\gamma}) b\bar{b}b\bar{b}$
are also considered in the analysis, but are found to give a negligible contribution.}

\begin{table}
\setlength{\tabcolsep}{2.5mm}
\renewcommand{\arraystretch}{1.2}
\centering
\begin{tabular}{c|cccc|c}
  Selection step & Signal & $\gamma\gamma b\bar{b}jj$ & $\gamma\gamma Z_{bb}jj$
     & $\gamma\gamma t\bar{t}$ & Significance\\
  \hline \hline
  Preselection  & $2.6$~ab & $4.2 \times 10^{6}$~ab & $5.3 \times 10^{4}$~ab
      & $1.1 \times 10^{5}$~ab & $6 \times 10^{-3}\,\sigma$\\ 
  $| m_h - m_{\rm j j_{1,2}}| < 15$~GeV\! & $2.0$~ab & $1.7 \times 10^{5}$~ab
      & $1.8 \times 10^{3}$~ab & $1.1 \times 10^{4}$~ab & $0.021\,\sigma$\\ 
  \hline
  $| m_h - m_{ \gamma \gamma}| < 5$~GeV & $2.0$~ab & $6.9 \times 10^{3}$~ab & $68$~ab
      & $500$~ab & $0.1\,\sigma$\\ 
  $m_{\rm jjjj} < 600$~GeV & $1.7$~ab & $6.9 \times 10^{3}$~ab & $68$~ab & $280$~ab & $0.089\,\sigma$\\
  $N_b^{\rm min}=2$ & $1.4$~ab & $1.3 \times 10^{3}$~ab & $27$~ab & $74$~ab & $0.17\,\sigma$\\
  $N_b^{\rm min}=3$ & $1.1$~ab  & $160$~ab & $3.5$~ab& $12$~ab & $0.37\,\sigma$\\ 
  $N_b^{\rm min}=4$ & $0.42$~ab & $1.3$~ab & $0.27$~ab & $0.26$~ab & $1.3\,\sigma$\\ 
  \hline
  $| m_h - m_{ \gamma \gamma}| < 2$~GeV & $2.0$~ab & $2.9 \times 10^{3}$~ab & $34$~ab
      & $210$~ab & $0.16\,\sigma$\\ 
  $m_{\rm jjjj} < 600$~GeV & $1.7$~ab & $2.9 \times 10^{3}$~ab & $34$~ab & $120$~ab & $0.14\,\sigma$\\
  $N_b^{\rm min}=2$ & $1.3$~ab & $890$~ab & $17$~ab& $25$~ab & $0.19\,\sigma$\\ 
  $N_b^{\rm min}=3$ & $1.1$~ab & $76$~ab & $0.33$~ab & $5.2$~ab & $0.54\,\sigma$\\ 
  $N_b^{\rm min}=4$ & $0.40$~ab & $0.62$~ab & $1.7 \times 10^{-3}$~ab 
      & $0.15$~ab & $1.7\,\sigma$\\ 
   \hline
  $| m_h - m_{ \gamma \gamma}| < 1$~GeV & $1.5$~ab &$1.2 \times 10^{3}$~ab & $34$~ab
      & $94$~ab & $0.18\,\sigma$\\ 
  $m_{\rm jjjj} < 600$~GeV& $1.3$~ab &$1.2 \times 10^{3}$~ab & $34$~ab & $54$~ab & $0.16\,\sigma$\\
   $N_b^{\rm min}=2$& $1.0$~ab & $590$~ab & $17$~ab & $17$~ab & $0.18\,\sigma$\\ 
   $N_b^{\rm min}=3$& $0.84$~ab & $59$~ab & $0.33$~ab & $1.7$~ab & $0.48\,\sigma$\\ 
   $N_b^{\rm min}=4$& $0.31$~ab & $0.54$~ab & $1.7 \times 10^{-3}$~ab
     & $0.065$~ab & $1.5\,\sigma$ 
\end{tabular}
\caption{Effects of the selection strategy in the ``intermediate'' analysis for the SM case. The
signal and background cross sections after each of the selection steps is shown. The last column shows the signal
significance computed for a luminosity $20$~ab$^{-1}$.}
\label{tab:hhh_fkl:Cutflow}
\end{table}

The effects of the selection strategy for the SM case are shown in Table~\ref{tab:hhh_fkl:Cutflow}.
In the table the results are given for different choices of the diphoton invariant-mass resolution $M$ and
of the minimum number of required $b$-tagged jets $N_b^{\rm min}$.
For all $M$ values, a requirement of four $b$-tagged jets is needed in order to achieve
a fair sensitivity to the signal. With this choice a significance around $1.7\,\sigma$ can be obtained.
The signal significance $\sigma$ as a function of the Higgs trilinear and quartic self-couplings
is shown in Fig.~\ref{fig:hhh_fkl:ExclusionMaps} for a luminosity of $20$~ab$^{-1}$.
In the figure two benchmarks with $M = 5$~GeV and $M = 2$~GeV are shown. In both benchmarks
$4$ tagged $b$-jets are required, since this choice maximizes the significance.\footnote{Analogous
plots for $M = 1$~GeV and for a different number of tagged $b$-jets can be found in the original paper~\cite{Fuks:2015hna}.}

A strong dependence on the Higgs trilinear coupling $\kappa_3 \equiv \lambda_3 - 1$ is found due to the jet and photon $p_T$
distributions that are harder when $\kappa_3$ is large and positive, while
the analysis turns out to be less sensitive to the quartic Higgs coupling $\kappa_4 \equiv \lambda_4 - 1$.
For the benchmark value \mbox{$M=2$~GeV}, $N_b^{\rm min} = 4$ (left panel of Fig.~\ref{fig:hhh_fkl:ExclusionMaps}),
a good fraction of the parameter space is covered at the $3\sigma$ level for negative $\kappa_3$ values.
On the other hand, in the SM case ($\kappa_3=\kappa_4=0$) a signal significance of at most $\sim 2 \sigma$
can be obtained.
In all studied setups, the significance isolines closely follow the total cross section and thus it is very challenging to get any
sensitivity for positive shifts in the Higgs self couplings ($\kappa_{3,4} > 0$).

\begin{figure}
\includegraphics[scale=0.77]{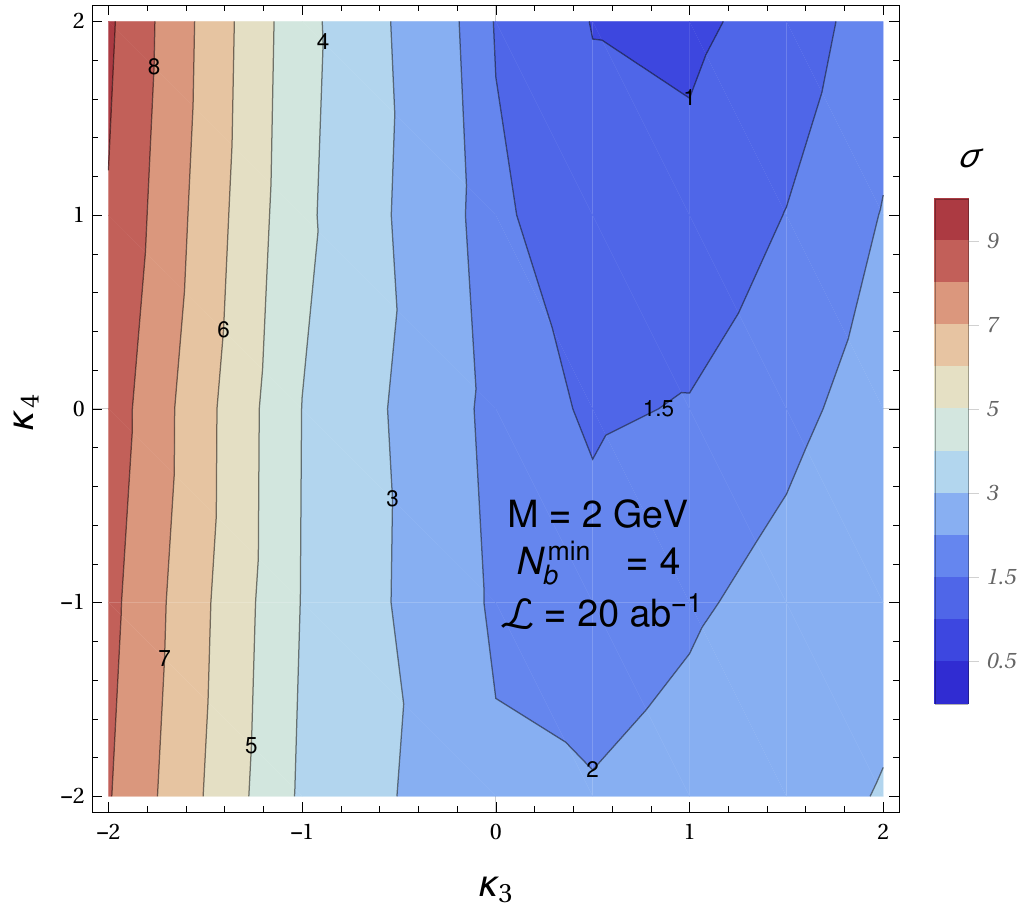}
\hfill
\includegraphics[scale=0.77]{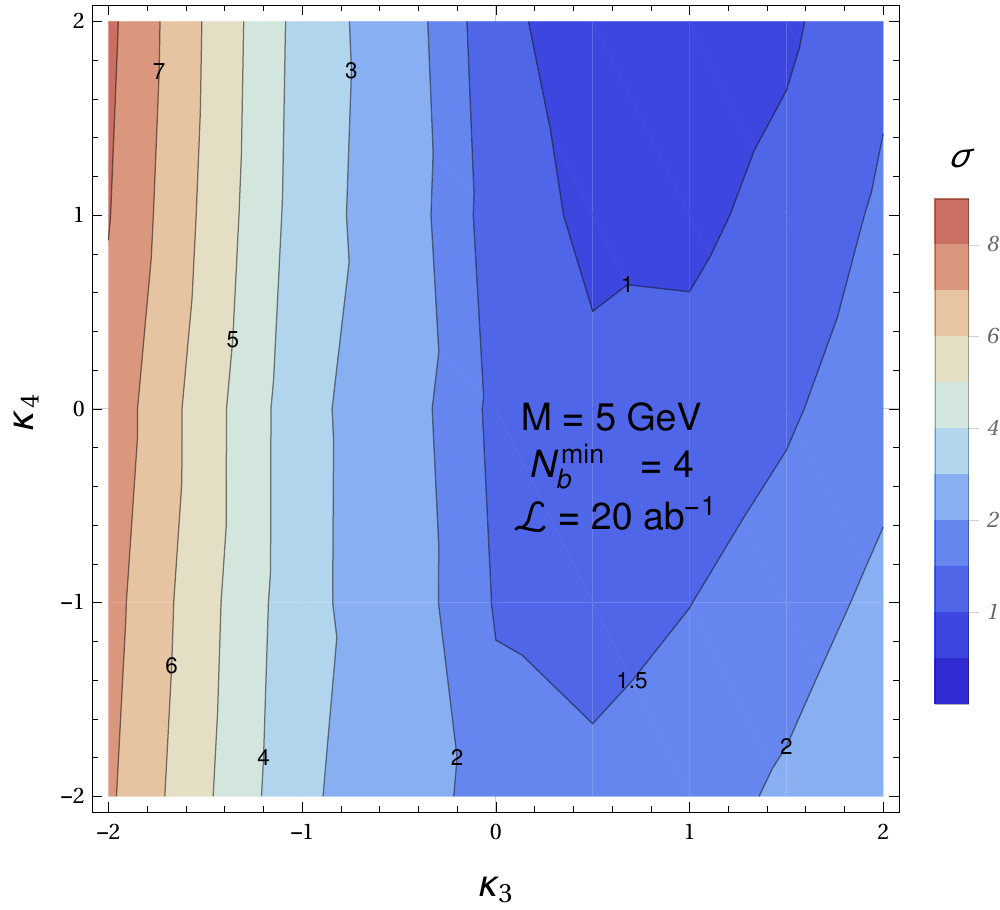}
\caption{Sensitivity of the FCC to the production of a triple-Higgs system
decaying into a $\gamma \gamma b \bar{b} b \bar{b}$ final state
for two different choices of the $M$ in the ``intermediate'' scenario.} \label{fig:hhh_fkl:ExclusionMaps}
\end{figure}

\subsubsection{Optimistic hypothesis: \boldmath $80\%$ $b$-tagging
  efficiency}

As a final case consider the ``optimistic'' scenario with a $b$-tagging efficiency of $80\%$~\cite{Papaefstathiou:2015paa}.
In this analysis a cut of $|\eta| < 5$ and $p_T > 400$~MeV is introduced on all particles of all event samples considered.
Jets are reconstructed from hadrons using the anti-$k_t$ algorithm~\cite{Cacciari:2011ma, Cacciari:2005hq},
with a radius parameter of $R=0.4$. Only jets with $p_T > 40$~GeV and $|\eta| < 3.0$
are kept (this includes $b$-jets). An ideal reconstruction efficiency of $100\%$ is assumed for the photons,
which are required to satisfy the conditions $|\eta| < 3.5$ and $p_T > 40$~GeV. A crude estimate of the detector effects
is included through jet-to-photon mis-identification probability and an heavy flavour (mis-)tagging efficiency.
The jet-to-photon mis-identification probability is set to $p_{j\rightarrow \gamma} = 10^{-3}$ and
is assumed to be constant in the whole kinematic range considered.
A flat $b$-jet identification rate of $80\%$ is assumed, while
the mis-tagging of a light jet to bottom-quark-initiated jet is set to be $1\%$.
All photons are required to be isolated, namely in a cone of radius $\Delta R = 0.2$ around the photon,
the sum of the transverse momenta of particles, \emph{i.e.}~$\sum_i p_{T,i}$, should be less than $15\%$
of the photon transverse momentum. The jet momenta are not smeared since a large
($60-80$~GeV) invariant $b \bar{b}$ mass window is considered in the analysis. The photon momenta are also not smeared, reflecting the fact that the photon momentum resolution at LHC is already at the $\sim 1\%$ level.

Given the large cross sections of processes with high-multiplicity final states at a
$100$~TeV collider, the only processes fully generated at parton level are those that include true photons
and true $b$-quarks. This implies that light extra jets are generated only at
the parton shower level, which is included via \texttt{HERWIG++}~\cite{Bahr:2008pv, Gieseke:2011na, Arnold:2012fq,
Bellm:2013lba, Bellm:2015jjp}.\footnote{\texttt{HERWIG++} is also used to simulate hadronization and
underlying event effects. Pile-up events, instead, are not considered.}
Additional phase-space cuts applied to the samples $b\bar{b} b\bar{b}$, $b\bar{b} b\bar{b} \gamma$, $b\bar{b} b\bar{b}\gamma\gamma$ and $b\bar{b} \gamma\gamma$ are shown in Table~\ref{tab:pscuts}.
\begin{table}[!t]
\centering
\renewcommand{\arraystretch}{1.2}
\begin{tabular}{l@{\hspace{2.5em}}l}
 observable & \hspace{4em} PS cut \\
 \hline
 \hspace{1em} $p_{T,b}$& $>35~\mathrm{GeV}$, at least one $> 70~\mathrm{GeV}$\\
 \hspace{1em} $|\eta _b|$ & $< 3.2$\\ 
 \hspace{1em} $p_{T,\gamma}$ & $>35~\mathrm{GeV}$, at least one $> 70~\mathrm{GeV}$\\
 \hspace{1em} $|\eta _\gamma|$ & $<3.5$\\
 \hspace{1em} $\Delta R_{\gamma\gamma}$ & $>0.2$  \\
 \hspace{1em} $m_{\gamma \gamma}$ & $\in [90, 160]~\mathrm{GeV}$
\end{tabular}
\caption{The phase-space cuts imposed on the background samples $b\bar{b} b\bar{b}$, $b\bar{b} b\bar{b} \gamma$, $b\bar{b}
b\bar{b}\gamma\gamma$, $b\bar{b} \gamma\gamma$ in the ``optimistic'' scenario.}
\label{tab:pscuts}
\end{table}

As a requirement for the signal sample, four $b$-jets, or light jets mis-identified as $b$-jets,
are required with $|\eta | < 3.0$, with transverse momenta $p_{T,b _{\{1,2,3,4\}}} > \{ 80, 50, 40, 40
\}$~GeV, where the subscripts $1$, $2$, $3$, $4$ denote the first, second,
third and fourth hardest $b$-jets respectively. Two photons, or
jets mis-identified as photons, are required with $|\eta| < 3.0$ and $p_{T,\gamma_{\{1,2\}}} > \{ 70, 40\}$~GeV.
Due to the fact that it is impossible for the majority of $b$-jets to identify whether they originated
from a $b$-quark or an anti-$b$-quark, there exists a $3$-fold combinatorial ambiguity in combining
$b$-jets into the two Higgs boson candidates.
As a simple choice, the highest-$p_T$ $b$-jet is paired with its closest
$b$-jet in $\Delta R = \sqrt{ \Delta \eta^2 + \Delta \phi^2} $, while other two remaining
$b$-jets are paired together.\footnote{An alternative method based on the minimization of the squared sum
of $(m_{bb}-m_h)$ from each $b$-jet combination yields results that differ by only $\mathcal{O}(1\%)$
compared to the simpler $\Delta R$ method.}
The invariant masses of the $b$-jet paired are constructed, $m_{bb}^{\mathrm{close},1}$ and
$m_{bb}^{\mathrm{close},2}$ respectively, which are required to be $m_{bb}^{\mathrm{close},1} \in [100, 160]$~GeV
and $m_{bb}^{\mathrm{close},2} \in [90, 170]$~GeV.
The rather large mass windows are chosen so as to retain most of the already rare signal. Moreover, the distance between
the highest-$p_T$ $b$-jet and the corresponding paired one is constructed and restricted to be within
$\Delta R_{bb}^{\mathrm{close}, 1} \in [ 0.2,1.6 ]$.\footnote{The distance between the other paired $b$-jets
was not found to have significant discriminating power.}
The invariant mass of the photon pair is required to be in a small window around the measured
Higgs boson mass $m_{\gamma\gamma} \in [ 124, 126 ]~$GeV.\footnote{This cut implies that the di-photon
resolution must be better than $\sim 1$~GeV at the FCC-hh. The current resolution at the LHC is
$1-2$~GeV,~\cite{Chatrchyan:2012twa, Aad:2014eha} and thus it is not unreasonable to expect a marginally improved
resolution at the detectors of the future collider.} Finally, the distance between the two photons
is required to be within $\Delta R_{\gamma \gamma} \in [ 0.2, 4.0 ]$. The selection cuts are summarized in
Table~\ref{tab:cuts}.

\begin{table}[!htp]
\centering
\renewcommand{\arraystretch}{1.2}
\begin{tabular}{l@{\hspace{2em}}l}
 observable & \hspace{2em} selection cut \\\hline
 $p_{T,b _{\{1,2,3,4\}}}$ & $> \{ 80, 50, 40, 40\}$~GeV\\
 $|\eta _b|$ & $< 3.0$\\ 
$m_{bb}^{\mathrm{close},1}$ & $\in [100, 160]$~GeV \\
$m_{bb}^{\mathrm{close},2}$ & $\in [90, 170]$~GeV \\
$\Delta R_{bb}^{\mathrm{close}, 1}$ &  $\in [ 0.2, 1.6 ]$ \\
$\Delta R_{bb}^{\mathrm{close}, 2}$ &  no cut\\
$p_{T,\gamma_{\{1,2\}}}$ & $> \{ 70, 40\}$~GeV\\
$|\eta _\gamma|$ & $<3.5$\\
$\Delta R_{\gamma\gamma}$ & $\in [ 0.2, 4.0 ]$  \\
$m_{\gamma \gamma}$ & $\in [124, 126]~\mathrm{GeV}$\\
\end{tabular}
\caption{Final selection cuts imposed in the ``optimistic'' analysis of the $(b \bar{b})  (b \bar{b}) (\gamma\gamma)$ final state.}
\label{tab:cuts}
\end{table}%

\begin{table*}[!htp]
\renewcommand{\arraystretch}{1.2}
\resizebox{\linewidth}{!}{
\begin{tabular}{lllll}
\rule[-.75em]{0pt}{1em}\hspace{3.5em} process & $\sigma_\mathrm{LO}$ (fb) & $\sigma_\mathrm{NLO}  \times \mathrm{BR} \times \mathcal{P}_\mathrm{tag} $ (ab) & \hspace{.5em} $\epsilon_\mathrm{analysis}$ & $N^{\mathrm{cuts}}_{30~\mathrm{ab}^{-1}}$  \\
\hline
\hline
$hhh \to (b \bar{b})  (b \bar{b}) (\gamma\gamma)$, SM &  $2.89$  &  \hspace{3em}  $5.4$ & $0.06$ & \hspace{.5em} $9.7$\\
\hline
$b \bar{b} b \bar{b} \gamma\gamma$ &  $1.28$  &  \hspace{3em} $1050$ &  $2.6\times10^{-4}$ &  \hspace{.5em} $8.2$\\
$hZZ$, (NLO) $(ZZ\rightarrow (b\bar{b})(b\bar{b}))$ &  $0.817$  & \hspace{3em} $0.8$ & $0.002$ &  \hspace{.5em} $\ll 1$\\
$hhZ$, (NLO)$(Z \rightarrow (b\bar{b}))$  &  $0.754$ & \hspace{3em} $0.8$ & $0.007$  &  \hspace{.5em} $\ll1$\\
$hZ$, (NLO) $(Z \rightarrow (b\bar{b}))$ & $8.02\times 10^3$& \hspace{3em} $1130$& $\mathcal{O}(10^{-5})$&  \hspace{.5em} $\ll 1$\\ 
$b \bar{b} b \bar{b} \gamma$ + jets &  $2.95 \times 10^3$  &  \hspace{3em} $2420$ & $\mathcal{O}(10^{-5})$ &  \hspace{.5em} $\mathcal{O}(1)$\\
$b \bar{b} b \bar{b}$ + jets &  $5.45 \times 10^3$ & \hspace{3em} $4460$ & $\mathcal{O}(10^{-6})$&  \hspace{.5em} $\ll 1$\\
$b \bar{b} \gamma\gamma$ + jets &  $98.7$  &  \hspace{3em} $4.0$ & $\mathcal{O}(10^{-5})$ &  \hspace{.5em} $\ll 1$ \\
$hh$ + jets, SM & $275$ &  \hspace{3em} $593$ & $7\times 10^{-4}$  &  \hspace{.5em} $12.4$
\end{tabular}
}
\caption{List of the various processes considered in the ``optimistic'' analysis of the $(b \bar{b})
  (b \bar{b}) (\gamma\gamma)$ final state. The parton-level cross
  section, including the cuts given in the main text is presented as well as the analysis efficiency and the
  expected number of events at $30$~ab$^{-1}$. A flat
  $k$-factor of $k=2.0$ has been applied to all tree-level processes (including
  $hh$+jets) as an estimate of the expected increase in cross section
  from LO to NLO. The $hZZ$, $hhZ$ and $hZ$ processes have been produced at NLO and hence no $k$-factor is applied.}
\label{tab:procs}
\end{table*}

The signal cross section after the cuts as well as the list of background processes considered in the analysis
is given in Table~\ref{tab:procs}.
The most significant backgrounds are the QCD production of $b \bar{b} b \bar{b} \gamma\gamma$, along with
all processes involving the production of only two Higgs boson in association with extra jets of QCD origin.
More precisely, the latter class of processes closely reproduces the kinematic distribution of the signal,
since in this case the tight di-photon mass window is of no help. Moreover, the Higgs bosons in di-Higgs production
processes are harder on average than those in $hhh$, thus passing transverse momentum cuts easily.
This background could be tackled in future studies with a $h\rightarrow b\bar{b}$ tagging algorithm based
on the jet substructure analysis techniques that exploit the differences between the energy spread of fat $b$-jets originating
from Higgs decay and the one of fat $b$-jets from QCD gluon splitting~\footnote{Note that the additional two $b$-jets
in $hh$+jets and $hZ$+jets have been generated by gluon splitting into $b\bar{b}$ performed by the shower
Monte Carlo program.}. The expected sensitivity to triple Higgs production in the SM obtained with this analysis is
$S/\sqrt{B} \sim 2.1$ for $30$~ab$^{-1}$, with $S/B \sim 0.5$.
Finally, assuming that the trilinear Higgs coupling is not modified from the SM value,
it can be estimated that the $\lambda_4$ parameter can be constrained
to the range $\lambda_4 \in [\sim -4, \sim +16]$ at $95\%$ confidence level with an integrated luminosity of
$30$~ab$^{-1}$ .

\subsubsection{Prospects of measuring the Higgs quartic self-coupling}

The comparison among the three analyses presented in this section allows one to draw some general conclusions
about the possibility to measure the triple-Higgs production cross section and to extract the Higgs quadrilinear self-coupling.

A crucial element that determines the experimental sensitivity are the efficiency and the fake rejection rates
of the $b$-tagging procedure. The ``pessimistic'' and ''intermediate'' analyses indeed show that two and even three
$b$-tags are not sufficient to efficiently suppress the large backgrounds. In particular, as it can be seen
from Table~\ref{tab:hhh_fkl:Cutflow}, the $\gamma\gamma b\overline b jj$ background can only be kept under control
with $4$ $b$-tags, a choice that allows one to reduce it to a level comparable with the SM signal yield. In this situation
the overall efficiency of the $b$-tagging procedure becomes an essential ingredient to determine the sensitivity of the
search. An increase in the reconstruction efficiency from the $60\%$ level assumed in the ``pessimistic'' analysis
to the $70\%$ used in the ``intermediate'' one already implies an enhancement of the signal by almost a factor two.
An $80\%$ efficiency would instead increase the number of reconstructed signal events by a factor three.
A minimal $b$-tagging efficiency of $70\%$ seems thus necessary to achieve some sensitivity to the signal.

It must be however noticed that an increase in the $b$-tagging efficiency can be effective only if it can be achieved by keeping
the fake rejection rates to an acceptable level (namely at most $\sim1\%$ for the light jets and $10 - 20\%$ for $c$-jets).
Indeed, in all the analyses the main backgrounds include some with fake $b$-jets
(e.g. $\gamma\gamma b\overline b jj$ in the ``pessimistic'' and ``intermediate'' analyses and  $hh jj$ in the ``optimistic'' one).

Another element that can significantly affect the analysis is the experimental resolution in the reconstruction
of the invariant mass of the di-photon system, $m_{\gamma\gamma}$.
This resolution affects linearly the size of the backgrounds containing
two photons not coming from an Higgs boson decay, as in the case of the $\gamma\gamma b\overline b jj$
and $\gamma\gamma b\overline b b \overline b$ non-resonant
backgrounds. As it can be seen from the ``pessimistic'' and ''intermediate'' analyses (see in particular the cut-flow in
Table~\ref{tab:hhh_fkl:Cutflow}) a resolution around $1 - 2$~GeV is a minimal requirement for the analysis to be effective.

An important ingredient to be further investigated is the relevance of the various backgrounds in the analysis when showering
and detector effects are included. From the available analyses, it seems that different choices for the cut strategy can
be used, which lead to comparable sensitivity to the signal, but significantly change the relevance of the various
backgrounds. 

Summarizing the results of the ``intermediate'' and ``optimistic'' analyses, it seems that in a realistic situation
a signal comparable to the SM one would lead to a significance around $2\sigma$. This would allow an ${\cal O}(1)$
determination of the production cross section. The situation could get better if a modification of the trilinear Higgs coupling
is present. In particular a negative shift in $\lambda_3$ would lead to a significant increase in the signal cross section
allowing for a higher significance as shown in Fig.~\ref{fig:hhh_fkl:ExclusionMaps}.

The prospects for extracting the quartic Higgs self-coupling, unfortunately, are not very promising. As mentioned before,
the dependence of the $HHH$ production cross section on $\lambda_4$ is very mild (see Fig.~\ref{LI:diff_lambda4})
and an order-one change in the signal can only be obtained for large deviations ($|\delta \lambda_4| \gtrsim 5$)
with respect to the SM value. As a reference result we can quote the one obtained in the ``optimistic'' scenario,
in which the quartic self-coupling, in the absence of modifications of the other Higgs couplings, is expected to be
constrained in the range $\lambda_4 \in [\sim -4, \sim +16]$ with an integrated luminosity of $30$~ab$^{-1}$.

\clearpage
\section{BSM aspects of Higgs physics and EWSB}
\label{sec:BSM}
\newcommand{\nc}{\newcommand}

\newcommand{\fref}[1]{Fig.~\ref{f.#1}}
\newcommand{\eref}[1]{Eq.~(\ref{e.#1})}
\newcommand{\erefn}[1]{ (\ref{e.#1})}
\newcommand{\erefs}[2]{Eqs.\ (\ref{e.#1}) - (\ref{e.#2}) }
\newcommand{\aref}[1]{Appendix~\ref{a.#1}}
\newcommand{\sref}[1]{Section~\ref{s.#1}}
\newcommand{\ssref}[1]{Section~\ref{ss.#1}}
\newcommand{\sssref}[1]{Section~\ref{sss.#1}}
\newcommand{\cref}[1]{Chapter~\ref{c.#1}}
\newcommand{\tref}[1]{Table~\ref{t.#1}}

\def\ifb{{\ \rm fb}^{-1}}
\def\ipb{{\ \rm pb}^{-1}}

\nc{\beq}{\begin{equation}}
\nc{\eeq}{\end{equation}}
\nc{\barray}{\begin{eqnarray}}
\nc{\earray}{\end{eqnarray}}
\nc{\barrayn}{\begin{eqnarray*}}
\nc{\earrayn}{\end{eqnarray*}}
\nc{\bcenter}{\begin{center}}
\nc{\ecenter}{\end{center}}

\def\AUTHORS#1{{\it  \bf \textcolor{red}{[AUTHORS: {#1}]}}}

\def\be{\begin{equation}}
\def\ee{\end{equation}}
\def\ba{\begin{array}}
\def\ea{\end{array}}
\def\beqn{\begin{eqnarray}}
\def\eeqn{\end{eqnarray}}
\def\beqs{\begin{subequations}}
\def\eeqs{\end{subequations}}
\def\non{\nonumber\\}
\def\beqa{\begin{eqnarray}}
\def\eeqa{\end{eqnarray}}

\def\bes{\begin{subequations}}
\def\ees{\end{subequations}}
\def\bea{\begin{eqnarray}}
\def\eea{\end{eqnarray}}
\def\bry{\begin{array}}
\def\ery{\end{array}}
\def\bit{\begin{itemize}}
\def\eit{\end{itemize}}
\def\nn{\nonumber}
\def\mc{\mathcal}
\def\tr{\textrm{Tr}}
\def\I{\mathbb{1}}
\def\t{\widetilde}
\def\o{\omega}
\def\ob{{\overline\omega}}
\def\ot{\widetilde\omega}
\def\obt{\widetilde{\overline\omega}}
\def\p{{\bf{p}}}
\def\q{{\bf{q}}}
\def\pit{\widetilde\pi}
\def\wt{\widetilde{w}}
\def\mt{\widetilde{m}}
\def\W{{\mathcal{W}}}
\def\dst{\displaystyle}
\def\f{\frac}
\def\gst{g_{V}}
\def\eq{Eq.~\eqref}
\newcommand{\eqs}[2]{Eqs.~\eqref{#1} and \eqref{#2}}

\def\ih{{\overline{I}}}
\def\jh{{\overline{J}}}

\def\O{{\mathcal{O}}}

\newcommand\dW[1]{{\frac{\delta\W}{\delta{#1}}}}
\newcommand\ddW[2]{{\frac{\delta^{2}\W}{\delta{#1}\delta{#2}}}}

\def\DC#1{{\it  \bf \textcolor{magenta}{[DC: {#1}]}}}
\def\AK#1{{\it  \bf \textcolor{blue}{[AK: {#1}]}}}
\def\MRM#1{{\it  \bf \textcolor{orange}{[MRM: {#1}]}}}

\subsection{Overview}
\label{s.overview}

\subsubsection{Motivations for BSM Higgs Sectors}

Here we briefly summarize the most important motivations for the
existence of BSM Higgs sectors, both experimentally and
theoretically. This is explored in more detail in
Sections~\ref{ss.EWPT}
-~\ref{ss.naturalness}. Additional details on the cosmological implications of BSM Higgs sectors and their phenomenological consequences may be found in Refs.~\cite{Assamagan:2016azc,ACFI15}.

\paragraph{The Electroweak Phase Transition and Baryogenesis}
It is well-known that the SM cannot explain the origin of the
matter-antimatter asymmetry, characterized by the tiny
baryon-to-photon ratio~\cite{Ade:2013zuv}: 
\begin{eqnarray}
Y_B=\frac{n_B}{s} = (8.59 \pm 0.11) \times 10^{-11} \quad \mathrm{(Planck)}
\end{eqnarray}
where $n_B$ ($s$) is the baryon number (entropy) density. This tiny
baryon asymmetry of the universe (BAU) nevertheless comprises roughly
5\% of the present cosmic energy density. Explaining why it is not
significantly smaller is a key challenge for BSM physics. While it is
possible that the universe began with a non-zero BAU, the inflationary
paradigm implies that the survival of any appreciable BAU at the end
of inflation would have been highly unlikely. Thus, one looks to the
particle physics of the post-inflationary universe to account for the
observed value of $Y_B$.

The necessary particle physics ingredients, identified by
Sakharov~\cite{Sakharov:1967dj}  nearly half a century ago, include:
\begin{enumerate}
\item baryon number violation; 
\item  C- and CP-violation; and 
\item either departure from equilibrium dynamics or CPT-violation. 
\end{enumerate}
While the SM
contains the first ingredient in the guise of electroweak sphalerons,
it fails with respect the second and third. The known CP-violation
associated with the SM CKM matrix is too feeble to have produced an
appreciable $Y_B$, and the SM universe was never sufficiently
out-of-equilibrium to have preserved it, even if 
 it had been sufficiently large. Consequently, BSM interactions are
 needed to remedy these shortcomings.  

A number of theoretically attractive BSM baryogenesis scenarios have
been developed over the years, with greater or lesser degrees of
testability. From the high energy collider standpoint one of the most
interesting possibilities is electroweak baryogenesis (EWBG) (for a
recent review, see Ref.~\cite{Morrissey:2012db}). EWBG requires that
electroweak symmetry breaking (EWSB) occur {\em via} a strong, first
order electroweak phase transition (EWPT). For the SM universe,
lattice studies indicate EWSB takes place through a cross over
transition, as the observed mass the of the Higgs-like scalar is to
heavy to allow for a first order transition. Nevertheless, the
presence of additional scalar fields in BSM Higgs scenarios could
allow for a first order transition in a variety of ways, as discussed
in detail in \ssref{EWPT}.  

Present measurements of Higgs properties imply that these new scalars 
are unlikely to be charged under SU(3$)_C$. Since they must couple to
the SM Higgs scalar in order to affect the thermal history of EWSB,
they necessarily also contribute to the Higgs production cross section
in gluon-gluon fusion if they are charged under SU(3$)_C$. Recent
model-independent studies as well as analyses of the \lq\lq light
stop" catalyzed EWPT in the MSSM and simple  
extensions~\cite{Curtin:2012aa,Cohen:2012zza,Katz:2015uja}, indicate
that a strong first order EWPT is usually incompatible  
with Higgs signal data if the new scalar is colored. 
Consequently, any new scalars that enable a
first order EWPT are likely to be SU(3$)_C$  
singlets.  

Present data still allow for first order EWPT-viable scalar sector
extensions to contain $SU(2)_L\times U(1)_Y$  
non-singlets as 
well as scalars that carry no SM charges and that interact with the
Higgs solely via Higgs portal interactions.  
In either case, discovery at the LHC and possibly future $e^+e^-$ is
possible, but probing the landscape  
of possible scenarios will require a higher energy pp collider, given
the generically small production cross sections.  
A detailed discussion appears in~\ssref{EWPT}.

In considering the EWPT, several additional observations are worth
bearing in mind. First, the existence of a strong first order EWPT is
a necessary condition for successful EWBG, but not a guarantee that
the BAU was produced during the era of EWSB. The CP-violation
associated with the BSM Higgs sectors may still have been too
feeble. In this respect, probes of BSM CP-violation with searches for
permanent electric dipole moments of atoms, molecules, nucleons and
nuclei provide a powerful probe, as do studies of CP-violating
observables in heavy flavor systems under certain
circumstances. Second, there exist well-motivated weak scale
baryogenesis scenarios that do not rely on a first order EWPT, such as
the recently introduced \lq\lq WIMPY baryogenesis"
paradigm~\cite{Cui:2011ab}. Nevertheless, since our emphasis in this section falls on
BSM Higgs, we will concentrate on the implications for EWBG. 

Against this backdrop, understanding the thermal history of EWSB is
interesting in its own right, independent from the EWBG
implications. For example, it could have implications for the generation of primordial gravitational
waves~\cite{Caprini:2015zlo}. 
By analogy, we note that exploring the phase diagram of
QCD has been a topic of intense theoretical and experimental effort
for several decades, involving a combination of lattice QCD studies
and experiments at the Relativistic Heavy Ion Collider and ALICE
detector at the LHC. With the additional motivations for BSM Higgs
sectors to be discussed below, it is interesting to ask about the
implications of these SM extensions for the phase diagram of the
electroweak sector of the more complete theory that enfolds the
SM. Both discovery of new scalars as well as detailed probes of the
scalar potential will in principle allow us to flesh out the thermal
history of EWSB.  

\paragraph{Dark Matter}

The astrophysical and cosmological evidence for the existence of Dark
Matter is overwhelming. However, these observations almost entirely
rely on DM's gravitational interactions, revealing little information
about its identity or interactions with the SM. Determining just what
constitutes the DM, how it interacts, and why it comprises roughly
27\% of the energy density of the present universe is one of the key
challenges at the interface of particle physics and cosmology. From a
particle physics standpoint, many BSM theories admit a variety of DM
candidates of various types (cold DM, axions, etc.) but the lack of a
signal to date leaves all possibilities open.  

Experimentally, a host of dark matter direct detection experiments
 are looking for nuclear recoils from collisions with ambient
DM, under the assumption that interactions are not too weak. The
sensitivity of  these searches has increased tremendously, and one
expects that they will enter the neutrino-background dominated regime
in the next few decades.  If a signal is detected, additional
measurements from either astrophysics or collider experiments would
not only corroborate the existing evidence for dark matter but would
help reveal its particle nature and  interactions. Indeed, recent
astrophysical observations, such as the positron excess observed by
the Pamela, Fermi-LAT, and AMS-02, or the excess of gamma rays at 2.2
GeV from the galactic center, provide tantalizing clues and have
inspired a flurry of particle physics model-building. Even if these
indirect signatures prove to be entirely of astrophysical origin and
direct detection experiments continue yield null
results,\footnote{While a few direct detection experiments have
  reported positive signals, they remain inconclusive} cosmological
constraints like the DM relic abundance  and its effect on large-scale
structure formation  could still point towards certain
interactions with the SM.

Colliders can probe dark matter through direct production of pairs of
dark matter particles, $\chi$. Since dark matter is stable on
cosmological timescales, the process $pp\to \chi\chi$ is unobservable
since $\chi$ leaves no visible trace in the detector. Consequently,
one must look for a visible signature arising from production of
additional visible particles in association with $\chi$ pairs. The
standard collider search strategy involves $pp\to \chi\chi + X$, where
$X$ may be a jet, SM gauge boson, or even the SM Higgs boson -- the
so-called \lq\lq mono-X plus missing energy" signature. Depending on
the specific dark matter model, the reach in $M_\chi$ anticipated by
the end of the LHC high luminosity phase is projected to be a few TeV. 
%

Dark matter can either be a part of the scalar sector, or connected to the SM via a scalar portal. Furthermore, it is  possible that the dark matter mass and
nature of its interactions would make observation in the mono-X plus
MET channel inaccessible at the LHC. 
For example, if the Higgs sector is extended in inert-doublet models~\cite{Barbieri:2006dq} 
to include a dark matter candidate, which is very hard to detect at the LHC for masses heavier than a few hundred GeV.
The situation is even more severe for scalars in non-doublet electroweak representations, which  typically requires a mass in the 2-3 TeV range in order to saturate the observed
relic abundance under a thermal dark matter scenario (lighter states
would annihilate away too quickly due to gauge interactions). The
corresponding production cross section can be far too small to be LHC
accessible, but with the higher energy and associated parton
luminosity, a 100 TeV $pp$ collider could allow for discovery. 
The extended scalar sector associated with such a scenario would also
provide rich opportunities for discovery and characterization.  

The conclusions are similar for scenarios where  scalars are responsible for the
interactions of DM with the SM. For example, if the DM is part of a
hidden sector, it may communicate solely to the SM Higgs sector
through the exchange of a SM gauge singlet scalar, making the Higgs
sector our only portal to the dark sector.  
Direct searches for this mediator or  for dark matter production
through the mediator are complementary to direct detection
experiments.  
The high mass reach and luminosity of a 100 TeV collider allows it to
probe  TeV-scale mediator and dark matter masses~\cite{Harris:2015kda, Khoze:2015sra}, a feat
which is very difficult to accomplish for most scenarios at the
LHC.
 This will be discussed in more detail in~\ssref{dm}. 

\paragraph{Origins of Neutrino Mass}

Explaining the origin of the small scale of neutrino masses, relative
to the masses of the other known elementary fermions, remains a
forefront challenge for particle and nuclear physics. While the Higgs
mechanism thus far appears to account for the non-vanishing masses of
the charged fermions and electroweak gauge bosons, the significantly
smaller scale of the active neutrino masses, as inferred from neutrino
oscillation and nuclear $\beta$-decay, suggests that an alternate
mechanism may be responsible in this sector. The longstanding,
theoretically favored explanation, the see saw mechanism, postulates
the existence of additional fields whose interactions with the SM
lepton doublets violates total lepton number. The fields may be
right-handed, electroweak gauge singlet  neutrinos; scalars that
transform as triplets under SU(2$)_L$; or fermions that transform as
electroweak triplets. These three possibilities correspond to the Type
I, II, and III see-saw mechanisms, respectively (for a recent review,
see, {\em e.g.}, Ref.~\cite{Perez:2015rza}).\footnote{Note that in
  left-right symmetric models, the Type II see-saw mechanism contains
  a parallel structure involving SU(2$)_R$ scalars.} The light
neutrino masses are inversely proportional to the mass scale $\Lambda$
of the new fields, with 
\beq
m_\nu \sim \frac{C v^2}{\Lambda}
\eeq
where $v=246$~GeV is the weak scale and C involves one or more of the
coupling constants in the specific realization.  

For $C\sim \mathcal{O}(1)$ one has
$\Lambda\sim\mathcal{O}(10^{14}-10^{15})$ GeV, a scale that is clearly
inaccessible to terrestrial experiments. However, there exist a
well-motivated variations on the conventional see-saw mechanism in
which $\Lambda$ may be at the TeV scale, such as low-scale see-saw
scenarios or radiative neutrino mass models. In this case, direct
production of the new fields may be accessible in high-energy $pp$
collisions. Of particular interest to this chapter are situations
involving new scalars, as in the Type II see-saw mechanism or
radiative neutrino mass models. A discussion of the opportunities for
discovery of these new scalars and their neutrino mass-related
properties with a 100 TeV $pp$ collider appears in \ssref{neutrinos}. 

\paragraph{Naturalness}

The SM can be viewed as a Wilsonian effective field theory with a
finite momentum cutoff $\Lambda$, parameterizing the scale at which
new degrees of freedom appear. In that case, the EW scale is not
stable with respect to quantum corrections from the UV. This
\emph{Hierarchy Problem} is most transparent in the expression for the
physical Higgs mass parameter in the SM Lagrangian, which sets the
electroweak scale: 
\beq
\mu^2 = \mu_{0}^2 + \frac{3 y_t^2}{4 \pi^2} \Lambda^2 + \ldots \ ,
\eeq
where the first term is the bare Higgs mass term and the second term is the dominant radiative contribution, 
which arises from top quarks at one-loop. If $\Lambda$ is much
higher than a TeV these loop corrections are much larger than the
physical Higgs mass, meaning the EW scale is
\emph{tuned}. Naturalness, as a guiding principle for BSM model
building, suggests one of the following:\footnote{The only known counterexample 
to this reasoning is the so-called ``dynamical solution" to the hierarchy problem, 
the relaxion~\cite{Graham:2015cka}. The minimal relaxion scenario has no interesting 
predictions for the collider experiments. However, it is not yet clear that a full consistent 
relaxion model, consistent with the cosmological constraints, exists.}
\begin{itemize}
\item[(a)] New degrees of freedom appear at the TeV scale to regulate
  this quadratic divergence, thereby protecting the Higgs mass from
  large UV contributions. This motivates supersymmetric solutions to
  the hierarchy problem, where stops below a TeV cancel the top loop. 
\item[(b)] The Higgs ceases to be a sensible degree of freedom above
  the TeV scale. This is the case for techni-color, or more recently
  Composite Higgs type solutions to the Hierarchy Problem. 
\end{itemize}
There is, at least naively, some tension between the expectations of
Naturalness and null results from recent LHC searches. Even so,
standard supersymmetric or composite theories could still show up at
LHC run 2. Some scenarios, including more exotic theoretical
realizations of naturalness, could escape detection at the LHC
all-together. This makes it vital to study their signatures at a 100~TeV collider. 

Given the direct connection between the Higgs boson and the Hierarchy
Problem, it is not surprising that Naturalness is strongly connected
to BSM Higgs sectors. This is explored in detail in
\ssref{naturalness}. 

For example, supersymmetric theories feature larger Higgs sectors than
the single SM doublet in order to cancel anomalies; the MSSM realizes
a particular subset of Type 2 Two-Higgs Doublet models. Direct
production of additional TeV-scale Higgs states will therefore be an
important physics goal of a 100 TeV collider. The existence of top
partners or the composite nature of the Higgs can also change the
Higgs couplings to SM particles at loop- and tree-level
respectively. Additional light states can be part of these extended
Higgs sectors. This makes exotic Higgs decays an attractive signature,
given the huge production rates for the SM-like Higgs boson at a
100~TeV collider. Finally, the full structure of the natural theory 
implies the existence of additional EW-charged states at or near the
TeV scale, such as vector resonances in Composite Higgs theories or
EWinos in supersymmetry.

An especially enticing scenario for the 100 TeV collider are theories
of Neutral Naturalness, like the Twin Higgs~\cite{Chacko:2005pe} or 
Folded SUSY~\cite{Burdman:2006tz}.  
In these theories, the quadratically divergent top contribution to the
Higgs mass is cancelled at one-loop by \emph{colorless} top partner
states, which only carry EW quantum numbers or can even be SM
singlets.  
While there are some very attractive discovery avenues for the LHC,
many cases can only be probed at future lepton and hadron
colliders. This can lead to a plethora of signatures, including exotic
Higgs decays to long-lived particles, direct production of uncolored
top partners through the Higgs portal, and Higgs coupling deviations
or direct production of new singlet states due to mixing effects.  
Perhaps the most exciting possibility is discovering many states carrying SM charges with masses in the 5 - 10 TeV range, which are predicted by all known UV completions of Neutral Naturalness. The 100 TeV collider is the only machine that would allow us to explore the full symmetry underlying these theories. 

\paragraph{BSM Scalar Sectors}


We discussed above several theories or frameworks which address the strength of the electroweak phase transition, the hierarchy problem, dark matter, and the origin of neutrino masses. All of them involve extended scalar sectors. However, the list of scenarios we study is not exhaustive, and no matter where we turn, BSM model building often involves modifying the Higgs sector. 
For example, {\it Flavor} is usually seen as a problem in theories with more elaborate Higgs sectors because heavy Higgs mass eigenstates are not necessarily aligned in flavor space, leading to unacceptable tree-level contributions to FCNC processes. This problem could turn into a virtue, as the richer Higgs-induced flavor structure may be used to explain the pattern of quark and lepton masses. Examples are the extension of the flavor symmetry to the Higgs sector~\cite{Weinberg:1991ws}, the gauging of the flavor group~\cite{Grinstein:2010ve}, the Froggatt-Nielsen scheme~\cite{Froggatt:1978nt}, or higher-dimensional Yukawa couplings~\cite{Babu:1999me,Giudice:2008uua}. 
The purported {\it instability of the SM scalar potential}~\cite{Degrassi:2012ry,Buttazzo:2013uya,Bednyakov:2015sca} may also motivate extended scalar sectors. It can be interpreted as indicating the existence of new particles, and introducing new Higgs bosons is the most economical  solution~\cite{Gonderinger:2009jp,Gonderinger:2012rd,Lebedev:2012zw,EliasMiro:2012ay,Chao:2012mx,Cheung:2012nb,Ferreira:2015rha}.

These considerations suggest that maybe the right question is not why additional Higgs bosons should exist, but rather why they shouldn't exist.
The argument in favor of the SM Higgs structure is usually based on an Occam's razor criterion: simplicity and  minimality. But are simplicity and minimality really conceptual ingredients of the SM? Wouldn't logical simplicity prefer a gauge  structure for EW breaking (such as technicolor) rather than the introduction of scalar particles with new non-gauge interactions?  Are {\it three} generations of quarks and leptons the choice preferred by minimality? 
 Indeed, the existence of an enlarged Higgs sector is an almost inescapable consequence of theoretical constructions that address the naturalness problem with new dynamics. As an extreme conclusion, one could say that an enlarged scalar sector is a good discriminator between theories with or without dynamical explanations of the hierarchy at the weak scale. The discovery of new Higgs bosons would strike a mortal blow to the logical arguments in favor of an anthropic explanation of the hierarchy. Also cosmological relaxation mechanisms~\cite{Graham:2015cka} would be disfavored by the uncovering of an extended Higgs sector.

All of this provides ample motivation to study extended scalar sectors as  theoretical structures in their own right. The possibilities include EW sterile scalars (e.g. in models motivated by EWSB or Neutral Naturalness), or more complicated landscape of the scalar particles, like, for example,  in theories of partial compositeness. 
Probably one of most famous scenarios of the extended higgs sector is a so-called two-higgs doublets model (2HDM).  A particular version of the 2HDM is inevitably a part of the supersymmetrized version  of the SM, motivated by naturalness. However, 2HDM is an attractive scenario of its  own, not necessarily motivated by naturalness and any other BSM scenario.  

For this reason, we pay special attention to the direct production reach of a 100~TeV collider for new scalar states. Previous studies \cite{Hajer:2015gka} indicate a multi-TeV reach for new Higgs doublets and TeV-reach for new singlet scalar states in most scenarios. We discuss this in more detail,  in~\ssref{bsmhiggs}.  We also provide a theoretical  overview of the 2HDM and explain the parts of parameter space that are still relevant in light of LHC results.

\subsubsection{Unique Opportunities at 100 TeV}

Here we summarize the most important Higgs-related measurements 
for which a 100 TeV collider is uniquely suited, compared to the LHC
or planned future lepton colliders.

\paragraph{Measurement of the Higgs self-coupling}

The Higgs cubic coupling reveals direct information about the shape of the Higgs potential near our vacuum, and can be determined from measurements of non-resonant di-Higgs production. Unfortunately, such a measurement is extremely difficult. 
The HL-LHC can only determine this cubic coupling with $\mathcal{O}(100\%)$ precision
(see~\cite{ATL-PHYS-PUB-2014-019,ATL-PHYS-PUB-2015-046,CMS:2015nat} and~\cite{Yao:2013ika,Azatov:2015oxa,Barr:2013tda,Goertz:2013kp}), and
proposed lepton colliders with sub-TeV center-of-mass energies are expected to have similar or slightly better precisions (see~\cite{Baer:2013cma,Fujii:2015jha}).
A~$1\,$TeV ILC program with 2~$\mathrm{ab}^{-1}$ of luminosity could yield one-sigma precision of 16\%~\cite{Tian:2013,Kurata:2013,Fujii:2015jha}. 

Fortunately, the 100 TeV collider  is the ultimate machine for measuring the self-interaction of the Higgs boson. 
The study in Section~\ref{sec:HH_intro} found that $3-4\%$ statistical precision is achievable with $30 \mathrm{ab}^{-1}$ of luminosity, 
see Table~\ref{table:HHprecision}.  Inclusion of systematic errors could lead to a precision of about $5-6$\% with the same luminosity, 
see Table~\ref{tab:bbaa_systematics} and the discussions in~\cite{Yao:2013ika,Azatov:2015oxa,Barr:2014sga}. 

Higgs self-coupling measurements with $\lesssim 5\%$ precision are required to exclude a strong electroweak phase transition in $\mathbb{Z}_2$ symmetric singlet scalar extensions of the SM, and provide an important probe of mixed singlet scenarios that is orthogonal to precision Higgs coupling measurements (to other SM particles) at lepton colliders.  Such precision is also required to exclude certain neutral top partner scenarios, which can provide additional probes of neutral naturalness \cite{Curtin:2015bka}, and may also provide sensitivity for low-scale neutrino see-saws.

Regardless of any BSM motivation, measuring the shape of the Higgs potential around our vacuum is a worthy goal of precision Higgs physics in itself, and one which can only be carried out with any real precision at a 100 TeV machine.

\paragraph{Direct production of new electroweak states}
A 100 TeV collider would be sensitive to EW-charged BSM states with masses of $5-10$ TeV. This chapter contains or summarizes several studies of direct Higgs production in scenarios with extended scalar sectors that demonstrate this point, but even for squeezed fermion spectra like EWinos in split SUSY, a 100 TeV collider can have sensitivity to multi-TeV masses \cite{Low:2014cba, Gori:2014oua, Bramante:2014tba}. 

This has crucial implications for many fundamental questions in particle physics. Many models of new physics contain additional Higgs doublets or triplets, including supersymmetry, or possible mechanisms for generating the neutrino mass or inducing a strong EWPT. A reach for multi-TeV EW states also has relevance for naturalness: it could allow EWinos in split-SUSY to be detected \cite{Gori:2014oua}, and it would allow direct production of states that are part of the UV completion for theories of neutral naturalness. Most such UV completions also contain heavy colored states around 5-10 TeV, which are an obvious target for a 100 TeV collider, but even in models where such colored states might be avoided, it seems difficult to avoid new EW states. Therefore, the capability of probing heavy EW states allows the 100 TeV machine to probe theories of naturalness, possibly exhaustively \cite{Curtin:2015bka}.

\paragraph{Direct production of new singlet states}

A 100 TeV collider allows us to probe the most challenging aspect of the TeV-scale: SM singlets. Singlets with sub-TeV masses occur in many BSM extensions, motivated by Neutral Naturalness, a strong EWPT, the NMSSM, etc. These states are notoriously hard to probe at the LHC, and too heavy to produce at most proposed lepton colliders. Searches for di-higgs or di-$Z$ final states at 100 TeV are sensitive to singlet scalar masses up to about a TeV, and in some cases significantly above a TeV, depending on the model. 
A $pp$ center-of-mass energy of 100 TeV with $30 \mathrm{ab}^{-1}$ is needed to exhaustively probe a sub-TeV singlet that induces a strong phase transition, both with and without Higgs mixing.
For heavier mixed singlets, the FCC-hh could extend the discovery reach up to several TeV and down to Higgs-singlet mixing angles below 0.001. 

\paragraph{Ultra-rare exotic Higgs decays}

A 100 TeV collider with $\mathcal{O}(10 \ab^{-1}$) of luminosity produces $\sim 10^{10}$ SM-like Higgs bosons. Even when accounting for triggering requirements (which may well be absent at such a machine) this enormous rate allows for the detection of exotic Higgs decays with tiny branching fractions smaller than $10^{-8}$, as long as the final states are conspicuous enough to stand out from the SM background. Exotic Higgs decays are motivated for a myriad of reasons in many BSM scenarios (see e.g.~\cite{Curtin:2013fra}). For example, the Higgs portal is the lowest-dimensional interaction one  can add between the SM sector and a hidden sector. This makes exotic Higgs decays a prime discovery channel of new physics, as long as the new states are relatively light. This was demonstrated for dark photons in \cite{Curtin:2014cca} and for displaced decays in the context of Neutral Naturalness by \cite{Curtin:2015fna}, but applies to any theory which produces ultra-rare exotic Higgs decays with conspicuous final states.

\paragraph{High-Precision High-Energy Measurements}
A somewhat under-appreciated capability of the 100 TeV collider is the potential for high-energy high-precision measurements. A good example was studied by \cite{Alves:2014cda}, which showed that dilepton measurements of Drell-Yan production at 100 TeV can be sensitive to new states with EW charges around a TeV. This is especially important since such a measurement is almost completely model-independent, depending only on the masses and gauge charges of new particles, and being completely independent of decay modes etc, which could in some scenarios serve to hide the signatures of new states. 
This measurement is complementary to another high-precision measurement that is possible at the 100 TeV collider: the determination of the $h\gamma \gamma$ coupling with percent-level precision, see Section~\ref{sss.htogammagamma}.

An important application of this model-independent measurement is to theories of Neutral Naturalness, which includes scenarios with electroweak-charged top partners that are neutral under SM color. The results of \cite{Alves:2014cda} were used in \cite{Curtin:2015bka} to argue that such EW-charged top partners would be detectable at a 100 TeV collider with masses of at least 2 TeV or more, depending on multiplicity. This exhausts the natural range of top partner masses in an untuned theory, and essentially guarantees discovery if the hierarchy problem is solved by such states. 

Another application are theories of a strong EWPT, which can be induced by new light bosons with masses around a few hundred GeV. If these new degrees of freedom carry SM gauge charges, detecting their effect in the DY spectrum would be an orthogonal, model-independent way to guarantee their discovery.

\paragraph{Searches for invisible states}

There is ample motivation to search for new invisible (stable) states at colliders, the most obvious one being Dark Matter. Such searches are notoriously difficult at the LHC, but the 100 TeV collider will be sensitive to scalar mediators as well as dark matter masses of more than a TeV using for mono-X + MET or dijet+MET searches. 

Another important motivation is the $\mathbb{Z}_2$ symmetric singlet extension of the SM, which can induce a strong EWPT. A VBF jets + MET search, exploiting pair production of the singlets, is vital in excluding the entire EWBG-viable parameter space of this model, and requires $30 \ab^{-1}$ of luminosity at a 100 TeV collider.

\subsubsection{Probing the Intensity Frontier at 100 TeV}

The study of new physics opportunities in general, and BSM Higgs physics in particular, leads us to an important complementary perspective on the role of a 100 TeV collider. Of course, one of the most important reasons for increasing the center-of-mass energy is to increase the reach for direct production of heavy new states. However, an equally important reason is the huge increase in the production rate of light states like the 125 GeV Higgs. In producing $\sim 10^{10}$ Higgs bosons the 100 TeV collider has no equal in measuring certain Higgs-related processes, such as rare decays and the self-coupling. Similarly, relatively light states near a TeV (compared to $\sqrt{s}$) with sufficiently weak interactions  can only be discovered at a 100 TeV collider. This includes some  singlet scalar states, or an electroweak multiplet whose neutral component contributes substantially to the dark matter relic density.

In that sense, the 100 TeV collider acts as an \emph{intensity frontier experiment} for uncolored physics near the TeV-scale. This has several important implications for detector design. It is vital to maintain sensitivity for relatively soft final states, which may arise from the decay of e.g. the SM-like Higgs. The ability to reconstruct soft $b$-jets with $p_T \sim \mathcal{O}(20 \gev)$ and long-lived particle decays that decay (ideally) only $\mathcal{O}(50 \mu m)$ from the interaction point are important for realizing the full discovery potential of such a machine. Triggers might also be a concern: at such high center-of-mass energies, trigger strategies analogous to current LHC operation would miss many important low-mass processes. This provides powerful motivation to realize \emph{trigger-less} operation (or at least, full  event reconstruction at low trigger level so that interesting soft physics can be directly selected for).

\subsection{Electroweak Phase Transition and Baryogenesis}
\label{ss.EWPT}

As indicated in the Overview, determining the thermal history of EWSB is both interesting in its own right and relevant to the possibility of electroweak baryogenesis (EWBG). The latter requires a strong first order EWPT. From this standpoint, the object of interest is the finite-temperature effective action $S_\mathrm{eff}$ whose space-time independent component is the effective potential $V_\mathrm{eff}(T, \varphi)$. Here, $\varphi$ denotes the vevs of the scalar fields in the theory and $T$ is the temperature. A first order EWPT can arise when $V_\mathrm{eff}(T,\varphi)$ contains a barrier between the electroweak symmetric minimum (vanishing vevs for all fields that carry SM gauge charges) and the EWSB minimum. 
In principle, such a situation could have pertained to the SM universe, as thermal gauge boson loops induce a barrier between the broken and symmetric phases. In practice, the effect is too feeble to lead to a first order EWPT. 
More specifically, the character of the SM EWSB transition depends critically on the Higgs quartic self-coupling, $\lambda\propto m_h^2/v^2$. The maximum value for this coupling that is compatible with a first order phase transition corresponds to an upper bound on $m_h$ between 70 and 80 GeV~\cite{Gurtler:1997hr,Laine:1998jb,Csikor:1998eu}, 
clearly well below the experimental value. For a 125 GeV Higgs, lattice studies indicate that the EWSB  transition in the SM is of a cross-over type with no potential for baryon number generation. 

Nevertheless, well-motivated BSM scenarios can lead to a first order phase transition that may provide the necessary conditions for EWBG. The barrier between the two phases can arise from a number of effects, either singly or in combination: 
\begin{enumerate}
\item  finite-temperature loops involving BSM degrees of freedom; 
\item large zero-temperature loop effects from new BSM states with sizable Higgs couplings;
\item new tree-level interactions; 
\item additional contributions to $m_h$ that allow $\lambda$ to be smaller than its SM value. 
\end{enumerate}
In addition, the presence of such effects may lead to a richer thermal history than in a purely SM universe. One of the most compelling opportunities for the FCC-hh is to explore as fully as possible the set of possibilities for the finite-temperature EWSB dynamics. In what follows, we briefly review the present theoretical situation, followed by a discussion of representative scenarios that may be particularly interesting for a 100 TeV proton-proton collider. 
More detailed discussions may be found in two white papers~\cite{Assamagan:2016azc,ACFI15}. 

\subsubsection{Theoretical Studies}

Here we classify BSM scenarios according to the dynamics by which they generate a strong, first order EWPT. 

\vskip 0.1in

\noindent{\em First order transitions induced by BSM thermal loops}. The MSSM represents the most widely-considered BSM scenario that, in principle, could give rise to a loop-induced, strong first order EWPT (SFOEWPT). The effect relies on contributions to $V_\mathrm{eff}(T,\varphi)$ from stops, whose coupling to the Higgs field is $\mathcal{O}(1)$ and whose contribution is $N_C$-enhanced. Generation of a SFOEWPT requires that at least one of the stop mass eigenstates be relatively light, with mass $m_{\tilde t}\sim \mathcal{O}(100\, \mathrm{GeV})$ in our vacuum \cite{Carena:1997ki}. Such a light stop requires a tachyonic soft mass-squared parameter, and allowing for the possibility of a stable color- and charge-breaking vacuum associated with a non-vanishing stop vev. Recent theoretical work  indicates that for the \lq\lq light stop scenario" the universe may undergo a SFOEWPT transition to the color-symmetric Higgs phase and that the latter is metastable with respect to a deeper color-breaking phase. However, the lifetime of the Higgs phase is longer than the age of the universe, so once the universe lands there at high-$T$, it stays there\cite{Carena:2008rt}. 

Unfortunately, LHC Higgs data now preclude this interesting possibility within the
 MSSM~\cite{Curtin:2012aa, Cohen:2012zza}, even if one augments the scalar potential by \lq\lq hard" SUSY-breaking 
 operators beyond the MSSM~\cite{Katz:2015uja}. On the other hand, in a more general framework, loop-induced 
 SFOEWPT remains a viable possibility. The reason is that supersymmetry rigidly relates the stop-Higgs coupling to the SM top Yukawa; without this assumption, new scalar fields may have stronger couplings to the Higgs, and hence have a stronger effect on the EWPT dynamics through loops. If the new scalar field responsible for the SFOEWPT is colored, the deviations in the Higgs coupling to gluons induced by its loops will be sufficiently large to be discovered in the upcoming runs of the LHC and HL-LHC, unless some cancellation mechanism is operational~\cite{Katz:2014bha}. However, the scalar responsible for the SFOEWPT may also be charged only under electroweak interactions, or in fact be a complete SM-gauge singlet. (In the latter case, both tree-level and loop-level modifications of the potential may be important; see below.) In these scenarios, the LHC Higgs program alone will not be sufficient to conclusively probe the parameter space where the SFOEWPT occurs, leaving this important task for future colliders. Among the most important measurements that will constrain such scenarios are the precision measurements of the $h\gamma\gamma$ and $hZZ$ couplings at the HL-LHC and electron-positron Higgs factories, and the measurement of the Higgs cubic coupling at the 100 TeV proton-proton collider~\cite{Katz:2014bha,Noble:2007kk}.

\vskip 0.1in

\noindent{\em Zero-temperature loop effects}. 
It is possible for thermal loops from $W$ and $Z$ bosons to generate a SFOEWPT even for a 125 GeV Higgs mass, so long as the shape of the potential differs from the SM case. 
This can be realized if there are relatively large (but still perturbative) couplings between the Higgs and some new degrees of freedom. Non-analytical zero-temperature loop corrections can then lift the EWSB minimum, effectively reducing the depth of the potential well and allowing SM thermal contributions to generate the potential energy barrier required for a strong first order transition. Note that unlike in the above scenario of BSM thermal loops, the new degrees of freedom generating these zero-temperature loop corrections do not have to be so light as to be in thermal contact with the plasma during the phase transition. This has been studied in many contexts, most recently in \cite{Curtin:2014jma} with a focus on 100 TeV signatures, most importantly $\mathcal{O}(10\%)$ deviations in the Higgs cubic coupling.

\vskip 0.1in

\noindent{\em Tree-level barriers}. A  promising avenue  appears to entail BSM scenarios that contain gauge singlet scalars or scalars carrying only electroweak gauge charges. The former class has received the most attention in recent years, both in the context of the NMSSM and in non-supersymmetric singlet extensions.  For these scenarios the phase transition dynamics may rely on a tree-level barrier between the electroweak symmetric and broken phases.  (Tree-level effects can also generate a strong phase transition in the 2HDM, though this mechanism has not yet been fully explored.\footnote{Private Communication with Jose Miguel No.}) Thermal loops, of course, also contribute to $V_\mathrm{eff}(T,\varphi)$, and they are essential for symmetry restoration at high $T$. It is important to note that both the electroweak symmetric  and broken phases may involve non-vanishing vevs for the singlet fields. The transition to the EWSB phase may, thus, proceed first through a \lq\lq singlet phase"\cite{Profumo:2007wc,Curtin:2014jma,Jiang:2015cwa}, a possibility that can lead to a stronger first order EWPT than if the transition to a joint Higgs-singlet phase occurs in a single step\cite{Profumo:2007wc}. The possibilities of a SFOEWPT associated with a tree-level barrier in singlet extensions have been studied extensively in both supersymmetric and non-supersymmetric contexts. As we discuss below, work completed to date indicates that there exist interesting opportunities to probe this class of scenarios with a 100 TeV $pp$ collider.

\vskip 0.1in

\noindent{\em Combinations}. Within the context of the singlet extension, one may also encounter a SFOEWPT even in the absence of a tree-level barrier. The presence of a quartic singlet-Higgs operator may reduce the effective quartic coupling at high-$T$. In conduction with the gauge loop-induced barrier, a SFOEWPT may arise\cite{Profumo:2007wc}. A possibility of more recent interest is multi-step EWSB that involves a combination of thermal loop- and tree-level dynamics\cite{Patel:2012pi,Blinov:2015sna,Inoue:2015pza}. Multi-step transitions may arise in BSM scenarios involving new electroweak scalar multiplets, generically denoted here as $\phi$. For non-doublet representations, a SFOEWPT to a phase of non-vanishing $\langle \phi \rangle$ may occur as a result of a loop-induced barrier, followed by a first order transition to the Higgs phase associated with a tree-level barrier generated by a $\phi^\dag\phi H^\dag H$ interaction.  The baryon asymmetry may be produced during the first step, assuming the presence of appropriate sources of CP-violation\cite{Inoue:2015pza}, and transferred to the Higgs phase during the second step provided that electroweak sphalerons are not re-excited and that the entropy injection associated with the second transition is sufficiently modest. Measurements of Higgs diphoton decay signal strength provide an important probe of this possibility if the new scalar masses are relatively light. For heavier new scalars, direct production may provide an interesting avenue for a 100 TeV collider.

\subsubsection{Representative Scenarios}
\label{sec:ewptrep}

Here, we concentrate in more detail on those scenarios for which dedicated studies have been performed for a 100 TeV $pp$ collider. In doing so, we emphasize that exploration of the EWPT with a 100 TeV collider is a relatively new area of investigation and that there exists considerable room for additional theoretical work. Thus, our choice of representative scenarios is not intended to be exhaustive but rather is dictated by the presence of existing, quantitative studies. For purely organizational purposes, we group these scenarios according to the transformation properties of the BSM scalars under SM gauge symmetries.

\noindent{\em  Scalar singlet extensions}. The simplest extension of the SM scalar sector entails the addition of a single, real gauge singlet $S$. In the NMSSM, of course, the new singlet must be complex, but many of the generic EWPT features of well-motivated singlet extensions can be studied using the real singlet extension, the  \lq\lq xSM". The most general, renormalizable potential has the form\footnote{We eliminate a term linear in $S$ by a linear shift in the field by a constant.}
\begin{eqnarray}
\label{eq:vhs}
V(H,S) &=&  -\mu^2 \left( H^\dagger H \right) + \lambda \left( H^\dagger H \right)^2 + \frac{a_1}{2} \left( H^\dagger H \right) S  \\
\nonumber
&& + \frac{a_2}{2} \left( H^\dagger H \right) S^2 + \frac{b_2}{2} S^2 + \frac{b_3}{3} S^3 + \frac{b_4}{4} S^4 \ \ \ .
\end{eqnarray}
The presence of the cubic operators implies that $S$ will have a non-vanishing vev at $T=0$. Diagonalizing the resulting mass-squared matrix for the two neutral scalars leads to the mass eigenstates
\be
\label{eq:higgstheta}
\left(\begin{array}{c}
h_1\\
h_2
\end{array}\right)
= 
\left(\begin{array}{cc}
 \cos \theta &  \sin \theta  \\
-  \sin \theta &  \cos \theta 
\end{array}\right)
\left(\begin{array}{c}
h\\
s
\end{array}\right)
\ee
The mixing angle $\theta$ and $h_{1,2}$ masses $m_{1,2}$ are functions of the parameters in Eq.~(\ref{eq:vhs}) and of the doublet and singlet vevs, once the minimization conditions are imposed. 

For positive $b_2$, the cubic operator $H^\dag HS$ will induce a barrier between the origin and the EWSB minimum wherein both $\langle H^0\rangle$ and $\langle S\rangle$ are non-vanishing\footnote{In the region where $S>0$ one must have $a_1<0$ for this to occur.}. For an appropriate range of the potential parameters the transition to the EWSB can be strongly first order. For $b_2<0$, a minimum along the $S$-direction will occur with singlet vev $\langle S\rangle = x_0$. It is possible that the Higgs portal operator $H^\dag H S^2$ can generate a barrier between the $(\langle H^0\rangle$,$\langle S\rangle= (0, x_0)$ minimum and an EWSB minimum wherein $\langle H^0\rangle\not=0$, even in the absence of cubic terms in Eq.~(\ref{eq:vhs}) . The thermal history in the latter case involves a two-step transition to the EWSB vacuum, with a first step to the $(0, x_0)$, followed by a second transition to the EWSB vacuum\cite{Profumo:2007wc,Curtin:2014jma}. Under suitable conditions, the latter transition may also be strongly first order. Studies carried out to date indicate\cite{Profumo:2007wc,Curtin:2014jma,Profumo:2014opa} that a SFOEWPT can arises when the mass $m_2$ of the singlet-like scalar is less than one TeV for perturbative values of the couplings in in Eq.~(\ref{eq:vhs}). The phenomenological probes for this scenario are discussed in Section \ref{sec:ewptposter} below.

For much larger masses, it is appropriate to integrate the singlet out of the theory, leading to additional terms in the effective Higgs Lagrangian of the form\cite{Henning:2014gca}
\be
\mathcal{L}_\mathrm{eff}\supset \frac{a_1^2}{m_S^4}\mathcal{O}_H-\left( \frac{a_1^2 a_2}{m_S^4}-\frac{2 a_2^3 b_3}{m_S^6}\right)\mathcal{O}_6
\ee
where 
\beqn
\mathcal{O}_H & = & \frac{1}{2}\left(\partial_\mu H^\dag H\right)^2 \\
\mathcal{O}_6 & = &  (H^\dag H)^3\ \ \ .
\eeqn
A SFOEWPT can arise if
\begin{equation}
\label{eq:heavyewpt}
\frac{2 v^4}{m_H^2} < \frac{m_S^2}{a_1^2 a_2} < \frac{6 v^4}{m_H^2}\ \ \ .
\end{equation} 
Precision Higgs studies, such as a measurement of $\sigma(e^+e^-\to Zh)$ or the Higgs cubic coupling, could probe this regime.

An instructive special case of the xSM is obtained by imposing a $\mathbb{Z}_2$ symmetry on the potential in Eq.~(\ref{eq:vhs}). 
The number of free parameters in this scenario, which was studied in detail by the authors of ref.~\cite{Curtin:2014jma}, is reduced to just three (singlet mass, quartic coupling and Higgs portal coupling), making it amenable to full exploration via analytical methods. It also serves as a useful ``experimental worst-case'' benchmark scenario of a SFOEWPT, since the the $\mathbb{Z}_2$ symmetry turns off most of the signatures of generic singlet extensions by precluding doublet-singlet mixing.

In the $\mathbb{Z}_2$-symmetric xSM, a SFOEWPT can occur in two ways. For $b_2 < 0$, a two-step transition via the vacuum with a singlet vev can be made very strong for some range of self-couplings $b_4$. For $b_2 > 0$ and large Higgs-portal couplings, zero-temperature loop effects lift the EWSB vacuum, allowing SM thermal loops to generate the necessary potential barrier. This is illustrated as the red and orange shaded/outlined regions in \fref{Z2singlet}.

This scenario is almost completely invisible at the LHC, and only part of the relevant parameter space can be probed at lepton colliders. However, as we will review in Section \ref{sec:ewptposter} below, a 100 TeV collider can probe the entire EWBG-viable parameter space in this scenario, via either direct singlet pair production or measurements of the Higgs cubic coupling. This demonstrates the tremendous discovery potential for EWBG contributed by such a machine.

\begin{figure}
\centering
\includegraphics[width=10cm]{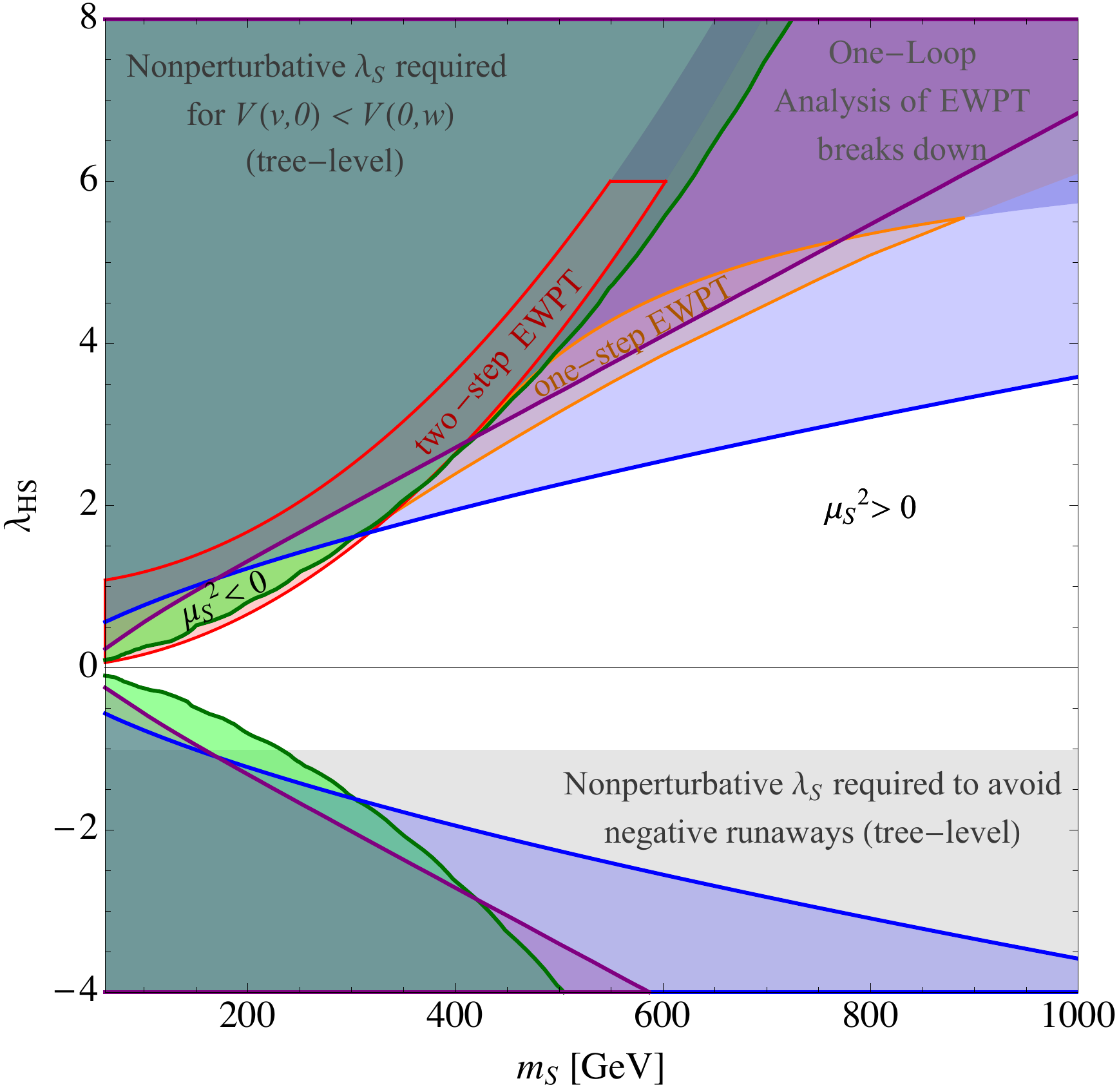}
\caption{
Summary of the $\mathbb{Z}_2$-symmetric singlet's parameter space for a strong EWPT, from~\cite{Curtin:2014jma}. $m_S$ is the physical singlet mass at the EWSB vacuum, while $\lambda_{HS} = a_2/2$, $\mu_S^2 = b_2$ and $\lambda_S = b_4$ in the notation of Eq. (\ref{eq:vhs}).
 All 100 TeV sensitivity projections assume $30 \mathrm{ab}^{-1}$ of luminosity.
\textbf{Gray} shaded regions require non-perturbative $\lambda_S > 8$ and are not under theoretical control. \textbf{Red} shaded region with red boundary: a strong two-step PT from tree-effects is possible for some choice of $\lambda_S$. \textbf{Orange} shaded region with orange boundary: a strong one-step PT from zero-temperature loop-effects is possible. Gray-Blue shading in top-right corner indicates the one-loop analysis becomes unreliable for $\lambda_{HS} \gtrsim 5 (6)$ in the one-step (two-step) region.
In the \textbf{blue} shaded region,
higgs triple coupling is modified by more than 10\% compared to the SM, which could be excluded at the $2\sigma$ \cite{Barr:2014sga} or better, see Table \ref{table:HHprecision}.
In the \textbf{green} shaded region, a simple collider analysis yields $S/\sqrt{B} \geq 2$
for VBF production of $h^*\to S S$. (Confirmed in later collider study by \cite{Craig:2014lda}.)
In the \textbf{purple} shaded region, $\delta \sigma_{Zh}$ is shifted by more than 0.6\%, which can be excluded by TLEP.
Note that both EWBG preferred regions are excludable by XENON1T if $S$ is a thermal relic.
}
\label{f.Z2singlet}
\end{figure}

\vskip 0.1in
\noindent{\em Electroweak scalar multiplets}.
Extensions of the SM scalar sector containing new color neutral, electroweak multiplets arise in a  variety of contexts, including type-II see-saw models, GUTs, and simple dark matter scenarios. The most widely-considered possibility is likely the two-Higgs doublet model (2HDM). In the general case where the origins of the 2HDM operators are not constrained by SUSY, it has been shown that a SFOEWPT can arise through a suitable choice of potential parameters. The precise dynamics responsible are not yet fully understood, but one likely candidate are tree-level effects that generate a barrier\footnote{Private Communication with Jose Miguel No.}. However, it has been found \cite{Dorsch:2013wja,Dorsch:2014qja} that a phenomenological consequence is the existence of the exotic decay channel for the CP-odd neutral scalar: $A^0\to Z H^0$. It appears likely that this scenario will be well-probed through LHC $A^0$ searches using this decay mode. Consequently, we will not consider it further here.  

For non-doublet electroweak multiplets, denoted here $\phi$, the $\rho$-parameter constrains the $T=0$ neutral vev to be rather small. As a result, the tree-level barriers associated with cubic operators are not pronounced. On the other hand, it is possible the EWSB occurs twice: first along the $\phi^0$ direction with vanishing $H^0$, and subsequently to the non-zero Higgs vacuum with small or vanishing $\phi^0$ vev. The first transition may be strongly first order, leading to the conditions needed for EWBG. The resulting baryon asymmetry will be transferred to the Higgs phase during the second step if the entropy injection is not too large. Studies of the phase transition dynamics and phenomenological tests have been reported in Ref.~\cite{Patel:2012pi} for a concrete illustration with a real triplet, and general considerations outlined in the subsequent work of Ref.~\cite{Blinov:2015sna}  The corresponding CP-violating dynamics needed for baryon asymmetry generation have been discussed in general terms in Ref.~\cite{Inoue:2015pza}  along with a concrete illustration of its viability. To date, no work has been completed on the probes using a 100 TeV $pp$ collider. However, the new electroweak states must generally be pair produced. The corresponding phase space considerations, along with the electroweak scale cross sections, make this class of scenarios an interesting opportunity for a next generation hadronic collider.

While the  signal associated with direct production is  highly model dependent, 
the deviations of the Higgs boson couplings from the SM values more  generic in the presence of the EW scalar.   One of the
most important observables in this case is $h \gamma \gamma $ coupling, which is necessarily affected due to the new light 
EW charged states running in the loop. In the next subsection we will estimate the deviations and comment on the 
prospects of the 100~TeV machine.

\subsubsection{Prospective signatures}
\label{sec:ewptposter}
While there exist a number of studies analyzing the prospects for LHC probes of the EWPT (for a review and references, see, {\em e.g.} Ref.~\cite{Morrissey:2012db}), relatively few have focused on the prospects for a next generation high energy $pp$ collider. Here, we review work completed to date, following the same organization as in Section~\ref{sec:ewptrep}.


\noindent{\em Gauge Singlets}. We start by considering the $\mathbb{Z}_2$-symmetric xSM, which was studied in detail by the authors of ref.~\cite{Curtin:2014jma}. Remarkably, despite the fact that this model represents an experimental worst-case scenario for EWBG, \emph{all parameter regions with a SFOEWPT can be probed at a 100 TeV collider}. 

Unlike the general xSM, this scenario has only a handful of signatures. The singlet can only be pair-produced via the $H^\dag H S^2$ operator through the processes $pp\to h^\ast \to SS$ and $pp\to h^\ast \to SS jj$, where the former corresponds to gluon fusion production of the off-shell Higgs and the latter to VBF production. A search for VBF-tagged dijets + MET can be sensitive to $SS$ production, though mono-jet analyses are also worth exploring in more detail. Singlet loops will modify the Higgs cubic coupling and $Z h$ coupling at the $\sim 10\%$ and $\sim 0.5\%$ level respectively. The former are best measured at the 100 TeV collider, see \cite{Barr:2014sga} and Table \ref{table:HHprecision},
while the latter can be detected at lepton colliders like FCC-ee \cite{Dawson:2013bba, Peskin:2012we,Blondel:2012ey,Klute:2013cx}. As \fref{Z2singlet} shows, direct singlet pair production (green region) is sensitive to the two-step phase transition, while measurements of the Higgs cubic coupling (blue region) are sensitive to the one-step region. This allows the 100 TeV collider to achieve full coverage of the parameter space viable for EWBG.

The authors of Ref.~\cite{Curtin:2014jma} observe that since $S$ is stable, it constitutes a dark matter candidate, a possibility that has been considered widely by other studies, most recently \cite{Cline:2013gha}. The XENON 1T direct detection search could probe the entire SFOEWPT-viable region, well in advance of the initiation of the FCC-hh program. Non-observation of a direct detection signal, however, would not preclude this scenario. In principle, introduction of small $Z_2$-breaking terms would render the singlet-like state unstable, thereby evading DM direct detection searches. For a sufficiently long decay length, $SS$ would nevertheless appear as MET, leaving the VBF channel as the only viable probe under these conditions. Furthermore, even if a dark matter signal is detected, collider studies will be necessary to determine the nature of the new particles, and their possible connection to the EWPT.



The general xSM has many more signatures, since the presence of $\mathbb{Z}_2$ breaking operators in the potential can lead to non-negligible doublet-singlet mixing. In this case, one may directly produce the singlet-like state $h_2$, with reduction in production cross section by $\sin^2\theta$ compared to the SM Higgs production cross section. For a given $m_2$, it will decay the same final states as would a pure SM Higgs of that mass. However, for $m_2> 2 m_1$, the decay $h_2\to h_1 h_1$ becomes kinematically allowed, leading to the possibility of resonant di-Higgs production. Studies of this possibility have been carried out for the LHC\cite{Dolan:2012ac,No:2013wsa,Chen:2014ask,Barger:2014taa}, and there exist promising possibilities for both discovery and exclusion if $h_2$ is relatively light. The resonant di-Higgs cross section can be significantly larger than the non-resonant SM di-Higgs cross section, so observation of this process could occur as early as Run II of the LHC.

Nonetheless, we are again led to the conclusion that a full probe of the SFOEWPT-viable xSM {\em via} resonant di-Higgs production will likely require a 100 TeV $pp$ collider. Recently, the authors of Ref.~\cite{Kotwal:2016tex} have investigated the discovery reach for the LHC and future $pp$ colliders for the SFOEWPT-viable parameter space of the xSM. After scanning over the parameters in the potential (\ref{eq:vhs}) and identifying choices that lead to a SFOEWPT, the authors selected points yielding the maximum and minimum $\sigma(pp\to h_2)\times \mathrm{BR}(h_2\to h_1 h_1)$. Results were grouped by $m_{h_2}$ in bins 50 GeV-wide, and a set of 22 benchmark parameter sets chosen  (11 each for the minimum and maximum resonant di-Higgs signal strength). Two sets of final states were considered: $b{\bar b}\gamma\gamma$ and $4\tau$. After taking into account SM backgrounds and combining the prospective reach for the two channels, the significance $N_\sigma$ for each benchmark point was computed using a boosted decision tree analysis. Results are shown in Fig.~\ref{fig:dihiggs}. The left panel compares the reach of the high-luminosity phase of the LHC with that of a 100 TeV $pp$ collider with 3 and 30 ab$^{-1}$, respectively. It is apparent that under this \lq\lq best case" study, wherein no pile up or detector effects have been included, \emph{the discovery reach of a 100 TeV $pp$ collider could cover nearly all of the SFOEWPT-viable parameter space with 30 ab$^{-1}$}. For the LHC, the reach with these two channels is more limited. We note that inclusion of pile up an detector effects will likely degrade the discovery potential. However, for this best case analysis, the significance lies well above $5\sigma$ for nearly all of the SFOEWPT-viable parameter space. Thus, we expect this discovery potential to persist even with a more realistic analysis.

\begin{figure}[ht!]
\begin{center}
\begin{tabular}{cc}
\includegraphics[width=0.45\linewidth]{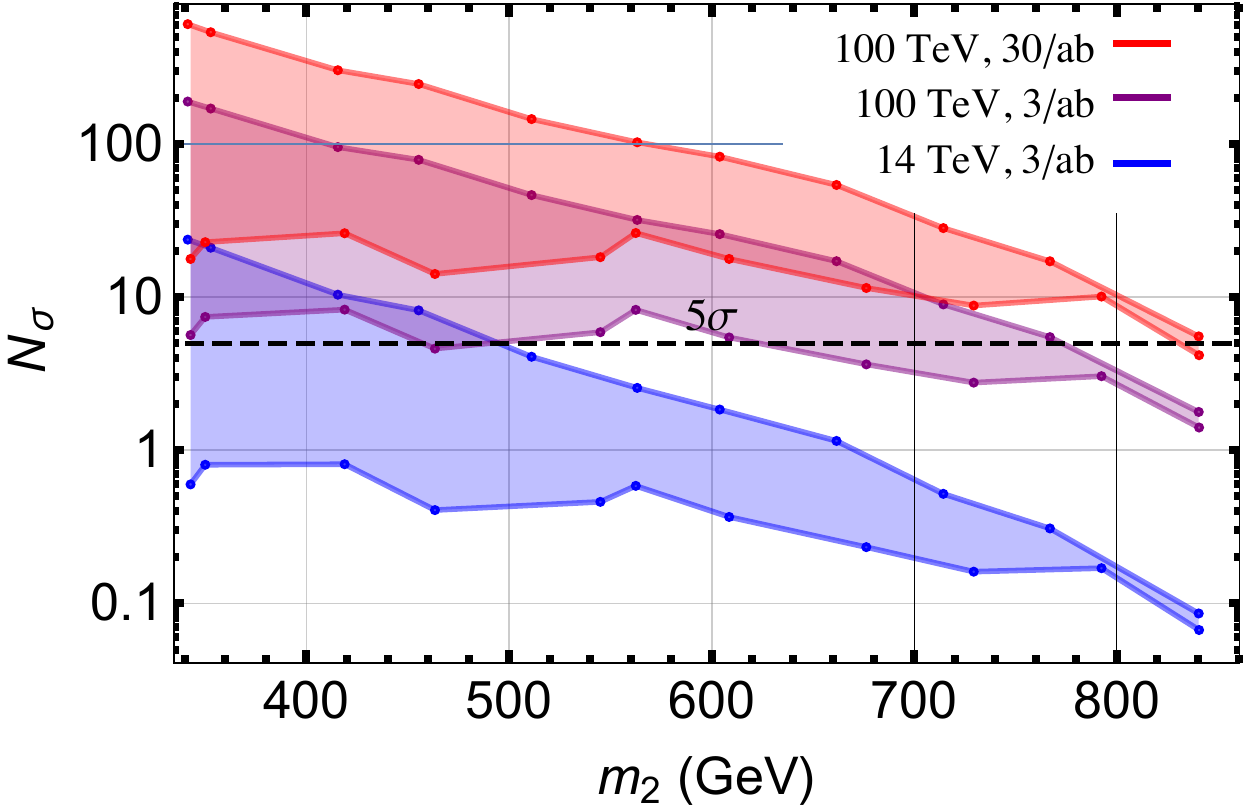}
&
\includegraphics[width=0.45\linewidth]{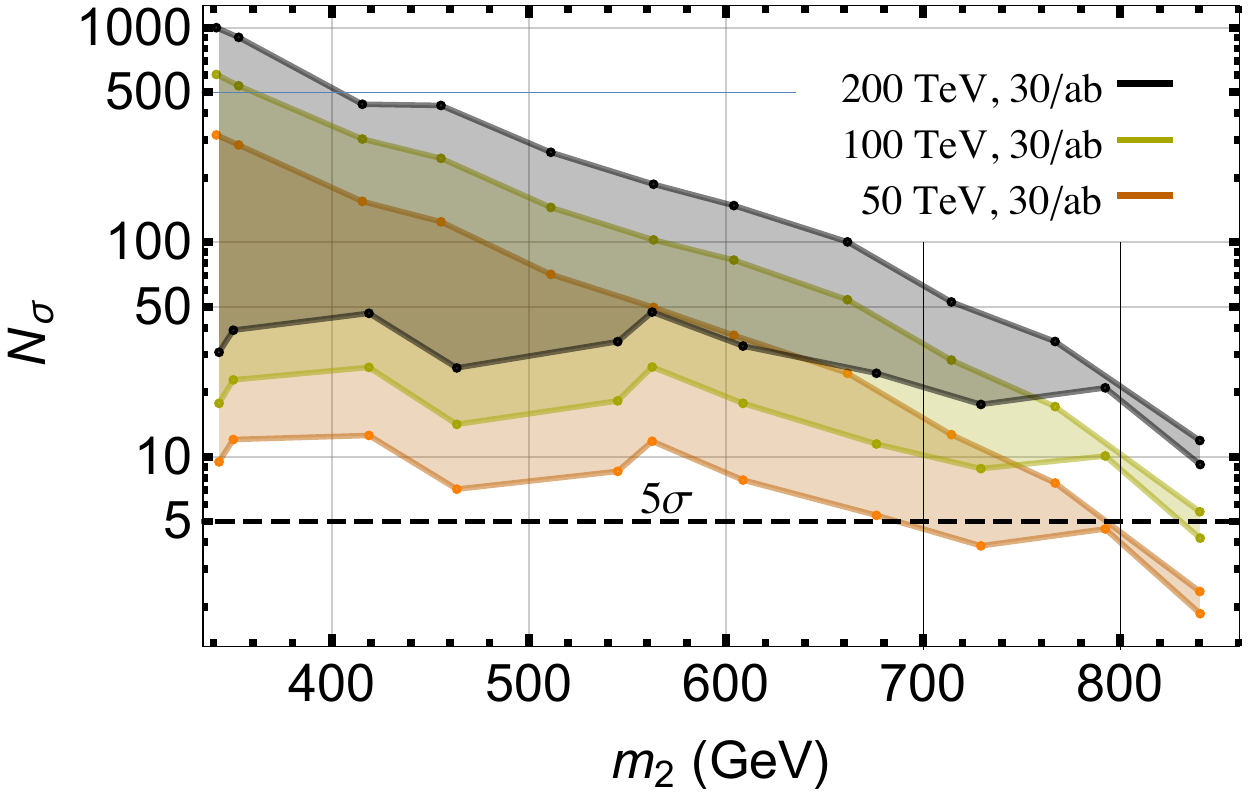}
\end{tabular}
\end{center}
 \caption{\label{fig:dihiggs} Physics reach for a SFOEWPT in the xSM with the LHC and a higher energy $pp$ collider, considering resonant di-Higgs production in the $b{\bar b}\gamma\gamma$ and $4\tau$ channels\cite{Kotwal:2016tex}. For each panel, vertical axis gives the significance $N_\sigma$ for each of the 22 SFOEWPT-viable benchmark points (see text), combining the significance of the two channels. For a given colored band, the upper (lower) edges give the maximum (minimum) $\sigma(pp\to h_2)\times \mathrm{BR}(h_2\to h_1 h_1)$. Left panel: comparison of the reach for the high luminosity phase of the LHC with a 100 TeV $pp$ collider at two different integrated luminosities. Right panel: comparison of the reach with 30 ab$^{-1}$ at three different center of mass energies.}
\end{figure}

In this context, it is also interesting to ask whether 100 TeV is the optimal energy for this probe. To address this question, the authors of Ref.~\cite{Kotwal:2016tex} performed a similar study for $\sqrt{s}=50$ and 200 TeV as well. The results are given in the right panel of Fig.~\ref{fig:dihiggs}. Unsurprisingly, the reach of a 200 TeV collider would exceed that of a 100 TeV machine, with the advantage being particularly pronounced for the higher mass region. On the other hand, for lower $\sqrt{s}$, one would begin to lose discovery potential in the high mass region and face less room for degradation of the significance once pile up and detector performance are considered.

Measurements of the Higgs trilinear self-coupling provide an alternate probe of the EWPT~\cite{Noble:2007kk}. In the absence of a $Z_2$ symmetry, this coupling will be modified by a combination of the parameters in the potential and the non-zero mixing angle. The opportunities for probing this effect at the LHC and various prospective future colliders are illustrated in Fig.~\ref{fig:selftc}, where the the critical temperature for the EWPT is plotted vs. the trilinear self coupling $g_{111}$. The SM value corresponds to the solid vertical black line. The colored vertical bands indicate the prospective sensitivities of the LHC and future colliders. The black dots indicate results of a scan over the parameters in Eq.~(\ref{eq:vhs}) that lead to a SFOEWPT, taking into account present LHC, electroweak precision, and LEP Higgs search constraints. It is clear that significant modifications of the self-coupling can occur. Moreover, even in the absence of an observed deviation at the LHC or future $e^+e^-$ colliders, there exists significant opportunities for discovery with a next generation $pp$ collider.

\begin{figure}[ht!]
  \centering
\includegraphics[width=0.5\linewidth]{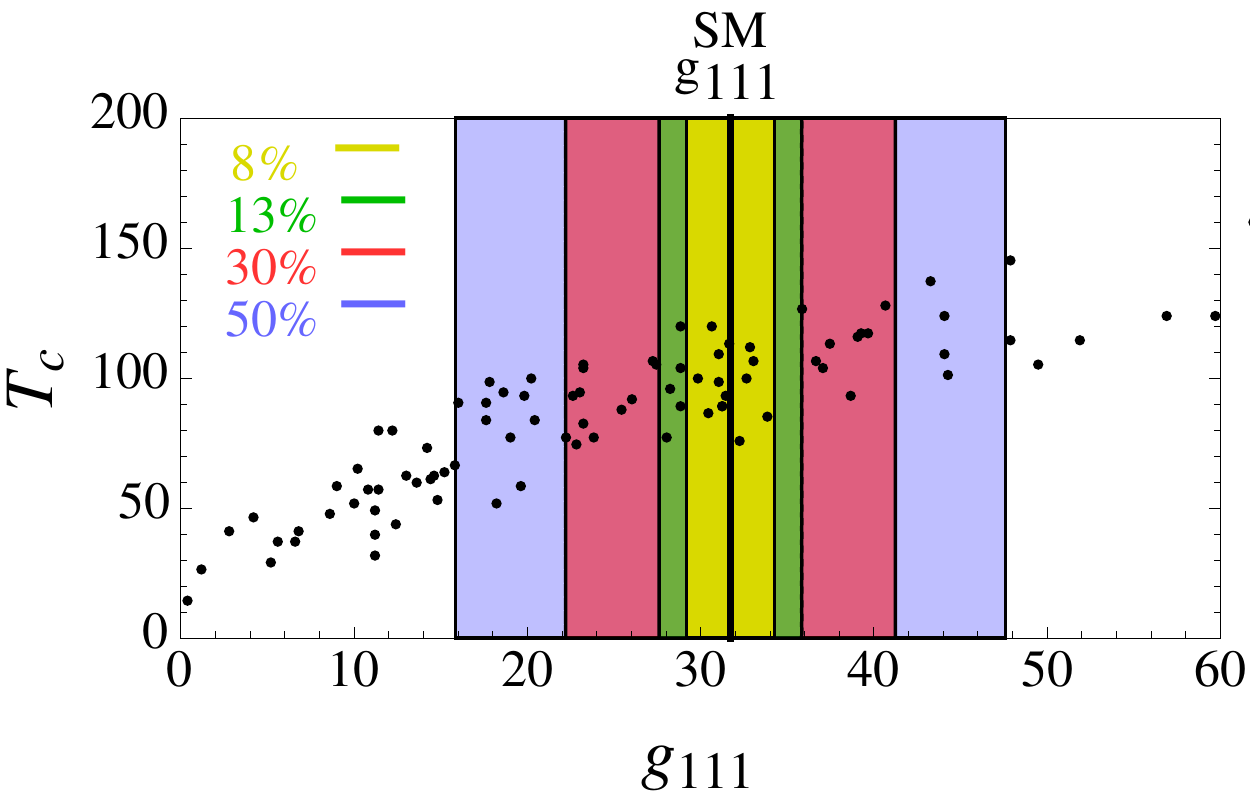}
  \caption{\label{fig:selftc} Correlation between the critical temperature and SM-like Higgs scalar self-coupling in the singlet-extended SM with a strong first-order electroweak phase transition, adapted from Ref.~\cite{Profumo:2014opa}. Colors indicate prospective sensitivities of the HL-LHC (purple), CEPC/FCC-ee (red), ILC (green), and SPPC/FCC-hh (yellow). The latter was assumed to be 8\%, but the precision may be as good as 3\%, see Table \ref{table:HHprecision}.
 }
\end{figure}


\vskip 0.1in

\noindent{\em Non-trivial representations of the SM}. In this case, where the strong EWPT is induced by  thermal loops of the new degrees of freedom, we expect strong deviations of the higgs couplings to 
$gg$, $\gamma \gamma $ and $\gamma Z$ are expected. These couplings are the most important, because at the 
SM these couplings show up at one-loop at the LO, and therefore any new light state might potentially lead to 
strong deviations from the SM. We will focus here on the first two coyplings, namely $gg$ and $\gamma \gamma$. While 
the latter can be relatively precisely probed at hadron colliders via appropriate decay mode of the higgs, the former 
affects the dominant higgs production mode. 

The expected deviations of the coupling to the $gg$ is the case of the colored scalars is appreciable. We illustrate this 
on an example of the $SU(3)_c$ triplet with the EW quantum numbers $1_{-4/3}$ on Fig.~\ref{fig:diquark}. In this 
example we overlay the contours of the deviations from the SM higgs couplings on the strength of the EWPT, which 
we parametrize as 
\beq
\xi \equiv \frac{v_c}{T_c}
\eeq
where $v_c$ stands for the higgs VEV at the temperature of the PT, and $T_c$ is the temperature of the 
PT.\footnote{The quantity $\xi$ is not, in general, gauge invariant\cite{Patel:2011th}. A more rigorous, gauge invariant characterization of the strength of the EWPT requires computation of the sphaleron energy and a careful treatment of $T_c$. These computations are also subject to additional theoretical uncertainties. For a detailed, discussion, see \cite{Patel:2011th}. In what follows, we will treat it as a rough \lq\lq rule of thumb", deferring a gauge invariant analysis to future studies.}
The value of $\xi$ is calculated in the one-loop approximation. In principle 
one demands $\xi \gtrsim 1$ for the strong 1st order PT, however, given the order one uncertainties one 
usually gets in thermal
loops calculation, even nominally smaller values of $\xi$ in this approximation might be viable. As we clearly
see from here most of the valid parameter space for the triplet has either been already probed at the LHC, are 
will be probed in LHC13 or VLHC. 

\begin{figure}[ht!]
\centering
\includegraphics[width=.49\textwidth]{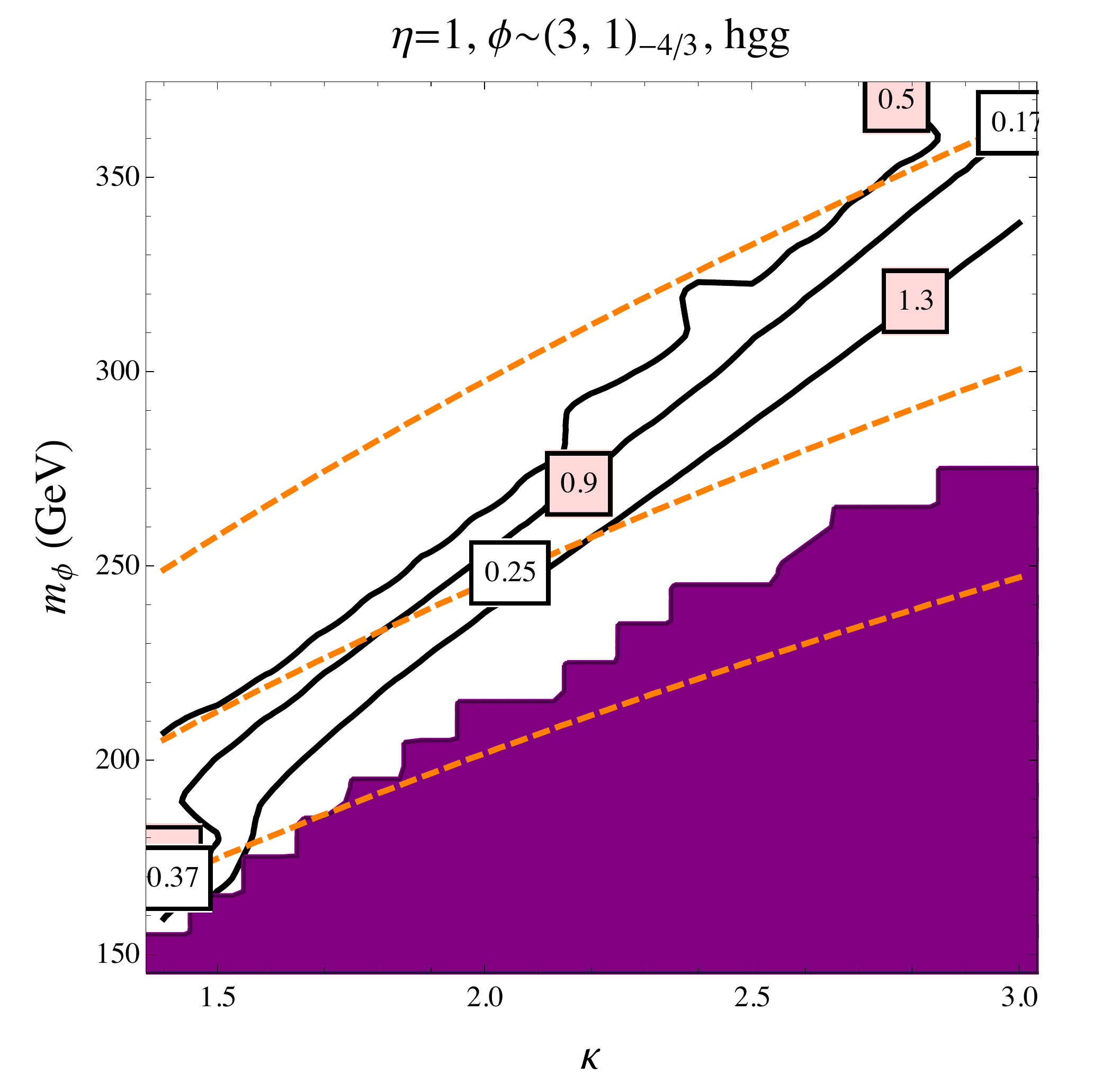}
\includegraphics[width=.49\textwidth]{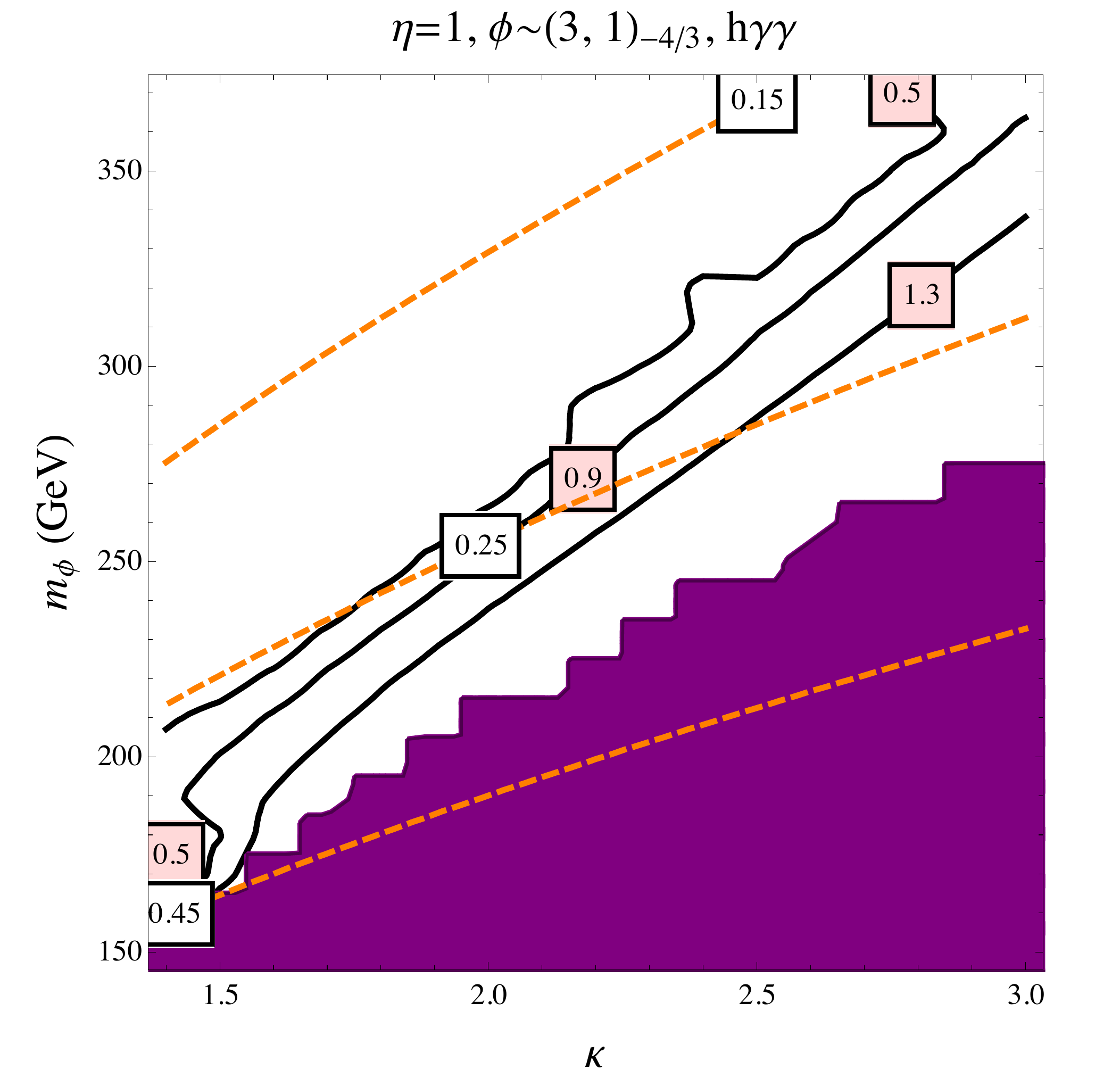}
\caption{ Orange contours: deviations (in percent from SM) of the $hgg$ and $h\gamma \gamma$ couplings from the SM values in the presence of the diquark with the quantum numbers~$1_{-4/3}$. In the shaded region there is no one-step transition to the EW vacuum. The black solid lines stand for the strength of the EW phase transition $\xi$. 
The plots are from Ref.~\cite{Katz:2014bha}}
\label{fig:diquark}
\end{figure} 

The situation is different one we consider EW-charged colorless states. Here the main deviations are in 
$\gamma \gamma $ and $\gamma Z$ channels. We will show the deviations in the first channel, to the best 
of our knowledge the deviations in $\gamma Z$ have not yet been explored in the context.  On the other hand, 
deviations of $h\gamma \gamma$ couplings can be as small as 5\% in the relevant part of parameter space. 
While the HL-LHC is may be sensitivity to diphoton decay branching ratio at the few percent level,  
the FCC-pp will be able to make important gains in 
addressing this option, see Section \ref{sss.htogammagamma}. We illustrate this point on Fig.~\ref{fig:EWscalars}.  Although it is not clear whether 
the HL-LHC will be able to achieve the absolute sensitivity required to completely probe this possibility (better than
5\%), substantial gains can be made by measuring the ratios of the various cross sections, for example the BRs of 
$\gamma \gamma$ relative to $ZZ^*$~\cite{Almeida:2013jfa, Djouadi:2015aba}

\begin{figure}[ht!]
\centering 
\includegraphics[width = .49\textwidth]{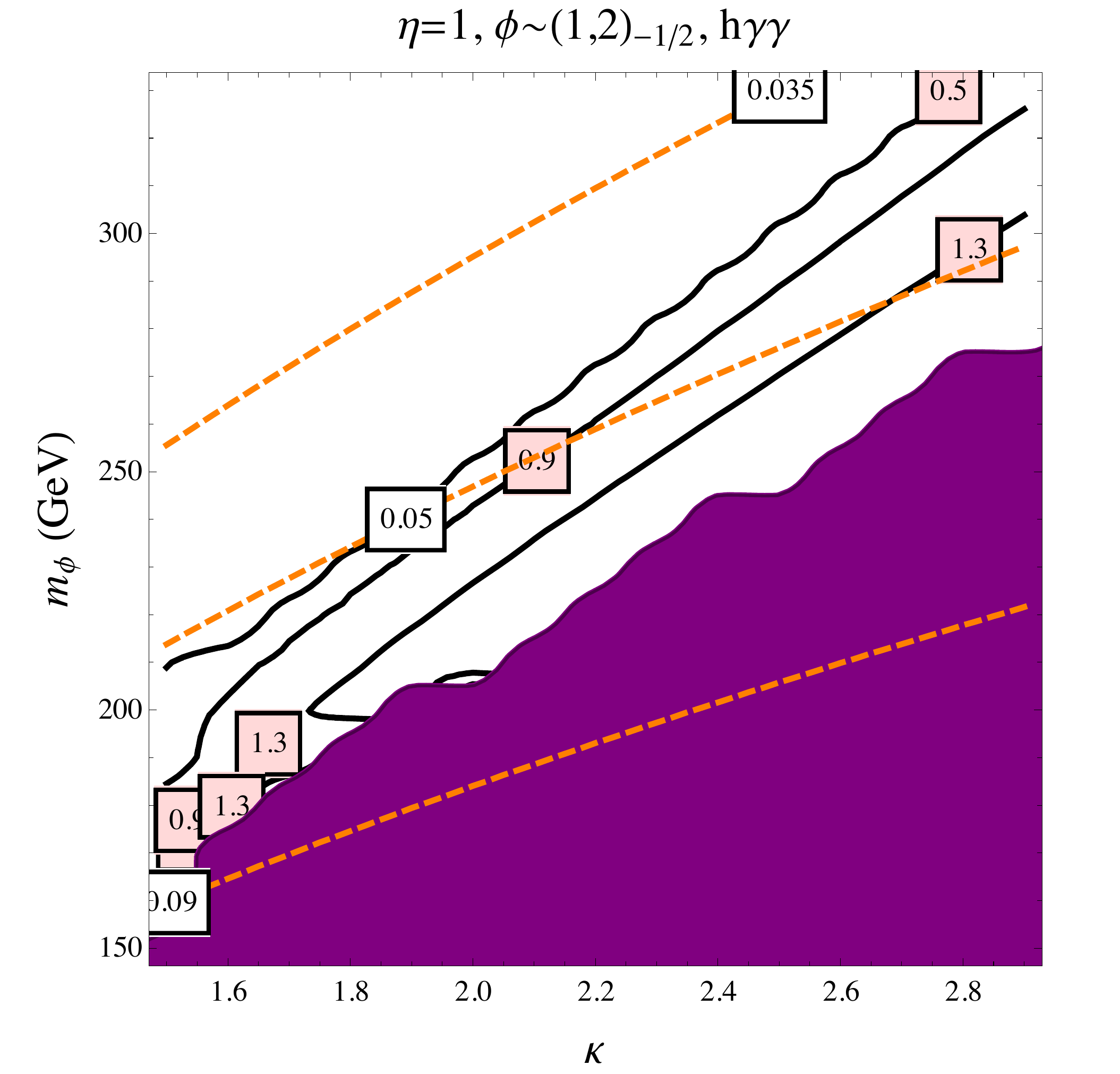}
\includegraphics[width= .49\textwidth]{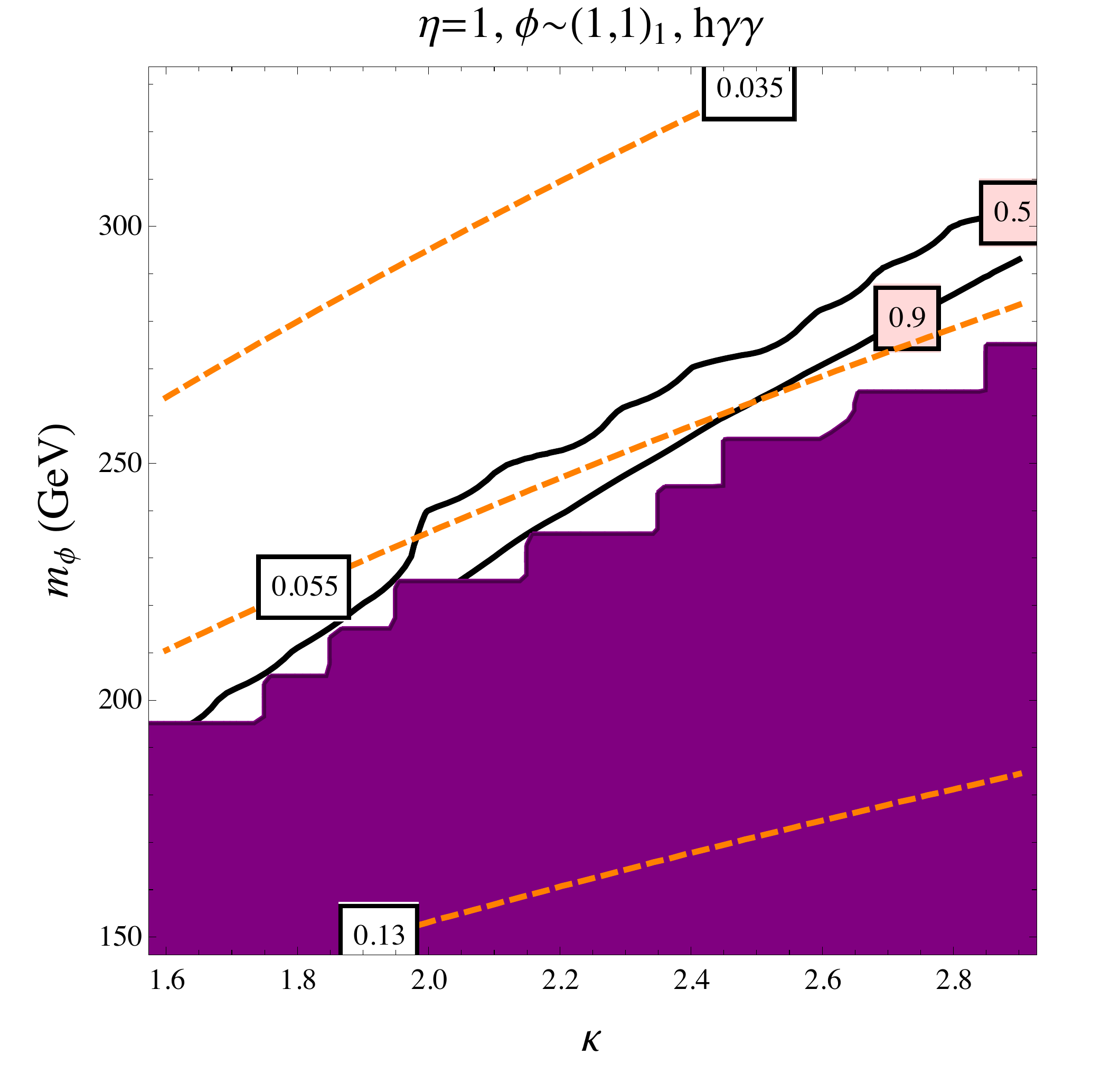}
\caption{Expected deviation of $h \gamma \gamma $ couplings in case of EW-charged scalars with quantum 
numbers $1, 2, -1/2$ (left panel)  and $1, 1, 1$ (right panel) from Ref.~\cite{Katz:2014bha}. 
Same labeling as  Fig.~\ref{fig:diquark}.} 
\label{fig:EWscalars}
\end{figure}

\subsubsection{EWPT: The discovery landscape}

Given the relatively small set of studies dedicated to probes of the EWPT at a 100~TeV collider, it would be 
premature to draw far-reaching conclusions about the range of opportunities for the FCC-hh. Indeed, the importance of engaging the community in performing these investigations was one of the key conclusions to the recent ACFI workshop that focused on this physics\cite{ACFI15}. Nonetheless, the work performed to date points to what is likely a rich opportunity. As indicated in Fig.~\ref{fig:dihiggs}, for the simplest BSM scenario yielding a SFOEWPT, the LHC will begin to \lq\lq scratch the surface", whereas the FCC-hh would provide an essentially exhaustive probe, which is also illustrated by~\fref{Z2singlet}. Moreover, 
$\sqrt{s}=100$~TeV appears to be close to the minimum needed for discovery.

Of course, it is possible that if this scenario is realized in nature, the parameters will put it in an LHC-accessible region. In this case, the FCC-hh  could provide confirmation relatively early in its operation, and could then be used to explore additional signatures, such as small deviations of the Higgs trilinear coupling from its SM value (see Fig.~\ref{fig:selftc}). Importantly, these observables provide an orthogonal probe of EWBG compared to, for example, measurements of Higgs mixing at lepton colliders through Higgs coupling measurements, since the Higgs self coupling and mixing angle are not correlated in xSM scenarios with a 
SFOEWPT~\cite{Profumo:2014opa}. While the 100 TeV collider may be able to probe EWBG exhaustively, it seems especially unlikely that such new physics could escape detection at both the 100 TeV \emph{and} a lepton collider. This complementarity provides a strong argument for the construction of both machines.

\subsection{Dark Matter}
\label{ss.dm}

An extended Higgs sector offers new possibilities for dark matter candidates and new avenues for the dark matter to communicate with the Standard Model states.  In general, there is a wide variety of theoretical
constructs exhibiting this feature, including models where the dark matter is a scalar, fermion, or vector particle, and constructions where it is either an electroweak
singlet or part of an SU(2) multiplet, charged under the electroweak interaction.  Similarly, there are a variety of possibilities for the SU(2) representations of the extended
Higgs sector.  If the couplings are large enough, this class of theories results in potentially visible phenomena resulting from dark matter annihilation, scattering with
heavy nuclei, and production at high energy colliders.  Colliders offer a particular opportunity when the interactions between the dark matter and the standard model
are suppressed at low momentum transfer, which suppresses its annihilation and/or scattering with heavy nuclei, because the ambient dark matter in the galaxy
is highly non-relativistic, with $v \sim 10^{-4}$.


\subsubsection{Landscape of Current Models}


In the limit in which the mediator particles are heavy compared to all energies of interest, all theories flow to a universal effective field theory (EFT) consisting of the Standard Model
plus the dark matter, and residual non-renormalizable terms in the form of contact interactions which connect 
them \cite{Beltran:2008xg,Cao:2009uw,Beltran:2010ww,Goodman:2010ku,Bai:2010hh,Goodman:2010yf}.  The EFT limit has been widely studied using data from run I of the LHC.
At the same time, it is recognized that theories in which the mediators are light enough to play an active role in collider phenomenology are of great interest, and
simplified model descriptions including such particles have been widely discussed \cite{Abdallah:2014hon,Malik:2014ggr,Boveia:2016mrp}.

We can discriminate between various classes of simplified models using scalar particles to communicate with a secluded sector:

\paragraph{Inert multiplet models}
\label{sec:inert}

In inert models, the Standard Model is extended by a scalar multiplet in a certain electroweak representation:
\begin{equation}
\label{eq:portal}
\mathcal{L} = \mathcal{L}_{\mathrm{SM}} + \frac{1}{2}\partial_\mu \phi \partial^\mu \phi - \frac{1}{2}m_\phi^2 \phi^2 - c_\phi |H^2| |\phi^2| - \lambda_\phi |\phi^2|^2,
\end{equation}
where $H$ is the SM-like Higgs doublet, and $\phi$ is the additional scalar field that may be a SM gauge singlet or charged under SM electroweak symmetry.  Note that this construction contains a $\mathbb{Z}_2$ symmetry $\phi \rightarrow - \phi$, such 
that (provided $\phi$ does not develop a vacuum expectation value, and thus mix with the SM Higgs) its lightest component is stable.  Cases in which $\phi$
is an even-dimensional SU(2) representation are generically in tension with null searches for scattering with nuclei, but odd-dimensional SU(2) representations
remain relatively unconstrained \cite{Burgess:2000yq,Cirelli:2005uq}. For recent studies for the case when $\phi$ transforms non-trivially under SM electroweak symmetry, see, {\em e.g.} \cite{FileviezPerez:2008bj,Hambye:2009pw}. Scenarios wherein $\phi$ is a gauge singlet (real or complex) correspond to setting the $\mathbb{Z}_2$-breaking coefficients $a_1=b_3=0$ in Eq.~(\ref{eq:vhs}) and identifying $b_2\to m_\phi^2$, $a_2\to 2 c_\phi$, and $b_4\to 4 \lambda_\phi$. This scenario has been studied extensively in Refs.~\cite{Barger:2007im,He:2008qm,Barger:2008jx,Gonderinger:2009jp,Gonderinger:2012rd,He:2013suk,Cline:2013gha,Feng:2014vea,Curtin:2014jma,Jiang:2015cwa,Kahlhoefer:2015jma}. An extension to the 2HDM plus a real singlet has been considered in Refs.~\cite{He:2013suk,Drozd:2014yla}.

\paragraph{Higgs-multiplet mixing models }
\label{sec:mix}

If the $\mathbb{Z}_2$ symmetry is broken, either explicitly by including a trilinear interaction such as $\phi |H|^2$, or spontaneously by engineering a potential for $\phi$ which
results in it obtaining a vacuum expectation value, it will mix with the SM Higgs.  In general, this removes the possibility that $\phi$ itself will play the role of dark matter, but
it may nonetheless serve as the portal to the dark sector if it couples to the dark matter.  For example, if $\phi$ and the dark matter $\chi$ are both electroweak
singlets, the only renormalizable interactions of $\chi$ respecting a $\mathbb{Z}_2$ are with $\phi$.  Through mixing with the Higgs, $\phi$ picks up coupling to the Standard Model,
and thus serves as the bridge between the two sectors (as does the SM Higgs) \cite{Barger:2007im,Buckley:2014fba,Baek:2015lna}.

An special case occurs when $\phi$ is a complex singlet\cite{Barger:2008jx,Gonderinger:2012rd,Jiang:2015cwa}. In this scenario, if the potential contains a global U(1) symmetry that is both spontaneously and softly broken, the massive Goldstone mode can serve as a dark matter candidate while the remaining degree of freedom mixes with the SM Higgs boson. 

\paragraph{Vector mediators}
\label{sec:vec}
While scalar mediators can be directly related to Higgs phenomenology, either by mixing additional scalar degrees of freedom with the Standard Model Higgs field, or by involving the Higgs boson in the production of new scalar particles, spin-1 mediators connecting the visible and dark sector is an interesting alternative. 
A vector mediator can arise from extended or additional gauge sectors to the Standard Model gauge group. Often a second Higgs boson is needed for the vector mediator to acquire a mass in a gauge invariant way. Such scenarios can for example arise from radiative symmetry breaking in the dark sector \cite{Foot:2007iy,Englert:2013gz}. 

\paragraph{Fermionic Dark Matter}
\label{fermion}

Fermionic dark matter can communicate with the Standard Model through the Higgs portal provided the dark matter is charged under the electroweak group.  Coupling
to the SM Higgs requires a combination of a $n$-dimensional representation with an $n+1$-dimensional one, and an appropriate choice of hypercharge.
Given current constraints, this is a region of particular interest in the MSSM, and can also be represented by simplified models, including the
``singlet-doublet" \cite{Cohen:2011ec,Calibbi:2015nha,Banerjee:2016hsk}, ``doublet-triplet" \cite{Dedes:2014hga}, and 
``triplet-quadruplet" \cite{Tait:2016qbg} implementations.
The generic feature in such models is electroweak-charged matter, which the relic abundance suggests typically has TeV scale masses.  In the absence of additional
ingredients, this is a regime which is difficult or impossible to probe effectively at LHC energies, but is typically within reach of a 100 TeV future collider.

\subsubsection{Signatures}
\label{sec:signatures}

When the mediators are heavy compared to the typical parton energies, all theories flow to a universal set of effective field theories, and lead to signatures
in which the dark matter is produced directly (with additional radiation to trigger) through contact interactions.  Projections for the limits on such interactions
at 100 TeV were derived in \cite{Zhou:2013raa}.

For models discussed in Sec.~\ref{sec:inert}, a neutral scalar of the multiplet $\phi$ could act as DM candidate. If then $m_\phi < m_H/2$ the decay $H\to \phi \phi$ contributes to the total Higgs width $\Gamma_H$ and can be probed in searches for invisible Higgs decays \cite{Djouadi:2011aa}. If realised within the Higgs portal model of Eq.~\ref{eq:portal}, with only two free parameters, this scenario is very predictive. We show the branching ratio of the Higgs boson into the stable particle $\phi$ in Fig.~\ref{fig:hportdm} (left). Current LHC limits \cite{Chatrchyan:2014tja, Aad:2015pla,CMS-PAS-HIG-16-009} reach as low as BR$(H\to \mathrm{inv}) \lesssim 30\%$, while an extrapolation to 3000 $\mathrm{fb}^{-1}$ yields BR$(H\to \mathrm{inv}) \lesssim 5\%$ \cite{Brooke:2016vlw} if systematic uncertainties scale with $1/\sqrt{\mathcal{L}}$.

It has been pointed out that off-shell Higgs measurements can set an indirect limit on the total Higgs width \cite{Caola:2013yja}, which could in turn result in a limit on Higgs decays into dark matter candidates, however such an interpretation is highly model-dependent \cite{Englert:2014aca} and can only be invoked on a case-by-case basis \cite{Englert:2014ffa}. 

If $m_\phi > m_H/2$ 2-$\phi$-production in association with one or two jets or a pair of heavy quarks can be probed at future hadron colliders. The authors of \cite{Craig:2014lda, Endo:2014cca} find the VBF configuration to be most promising to limit $m_\phi$ and $c_\phi$ of Eq.~\ref{eq:portal}. For a combined limit on $m_\phi$ and $c_\phi$ in the mono-jet, $t\bar{t}h$ and VBF channel see Figure~\ref{fig:hportdm} (middle). Increasing the collision energy from $\sqrt{s}=14$ TeV to $\sqrt{s}=100$ TeV and the integrated luminosity from $3~\mathrm{ab}^{-1}$ to $30~\mathrm{ab}^{-1}$ extends the testable parameter range significantly, e.g. for $m_\phi=200$ GeV from $c_\phi=2.7$ to $c_\phi \leq 0.7$ at $95\%$ C.L.  Requesting $\phi$ to contribute to a certain fraction of the relic dark matter density results in the contours of Fig.~\ref{fig:hportdm} (right).

\begin{figure}[t]
\includegraphics[width=5cm]{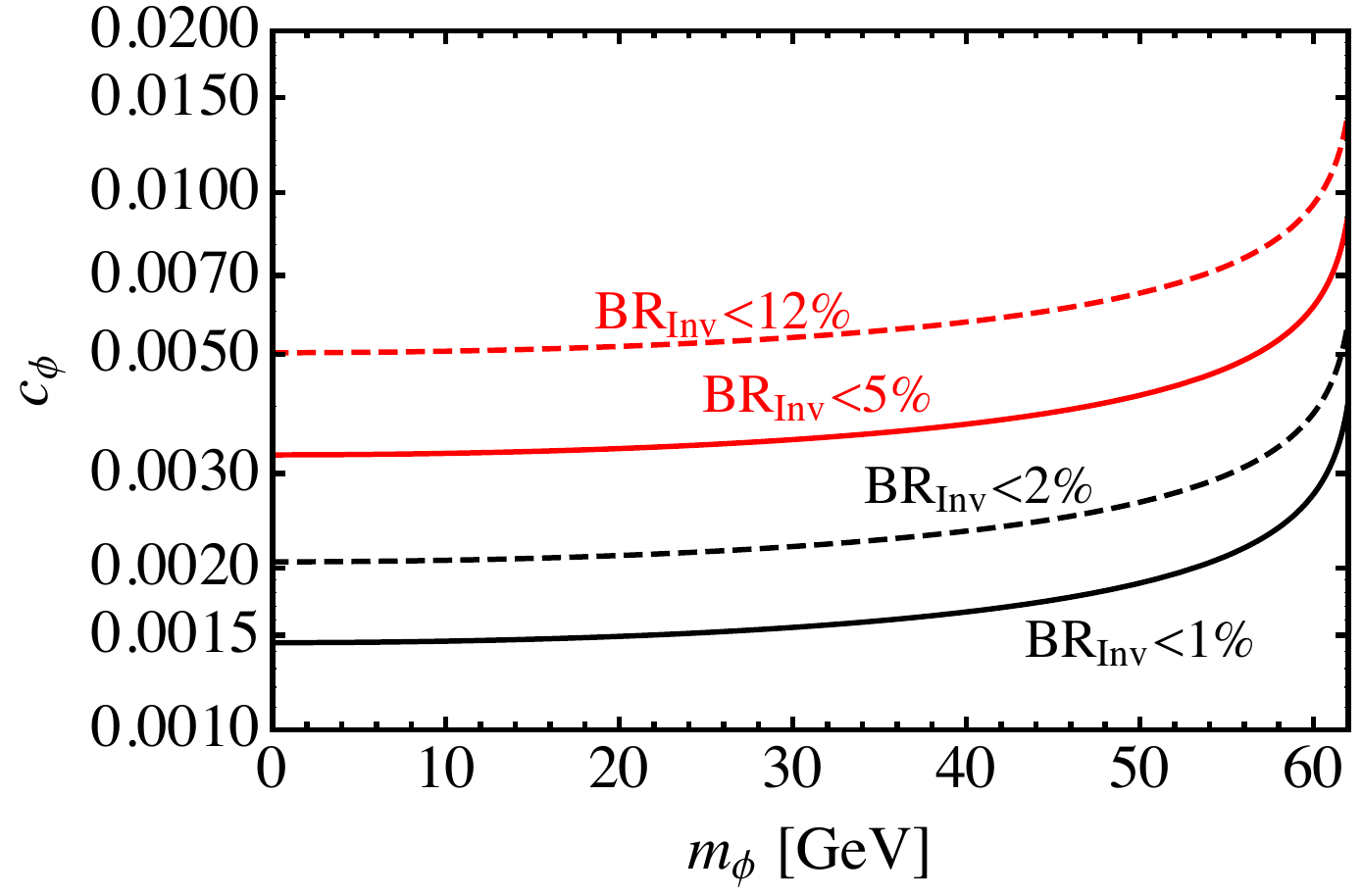} 
\includegraphics[width=5.2cm]{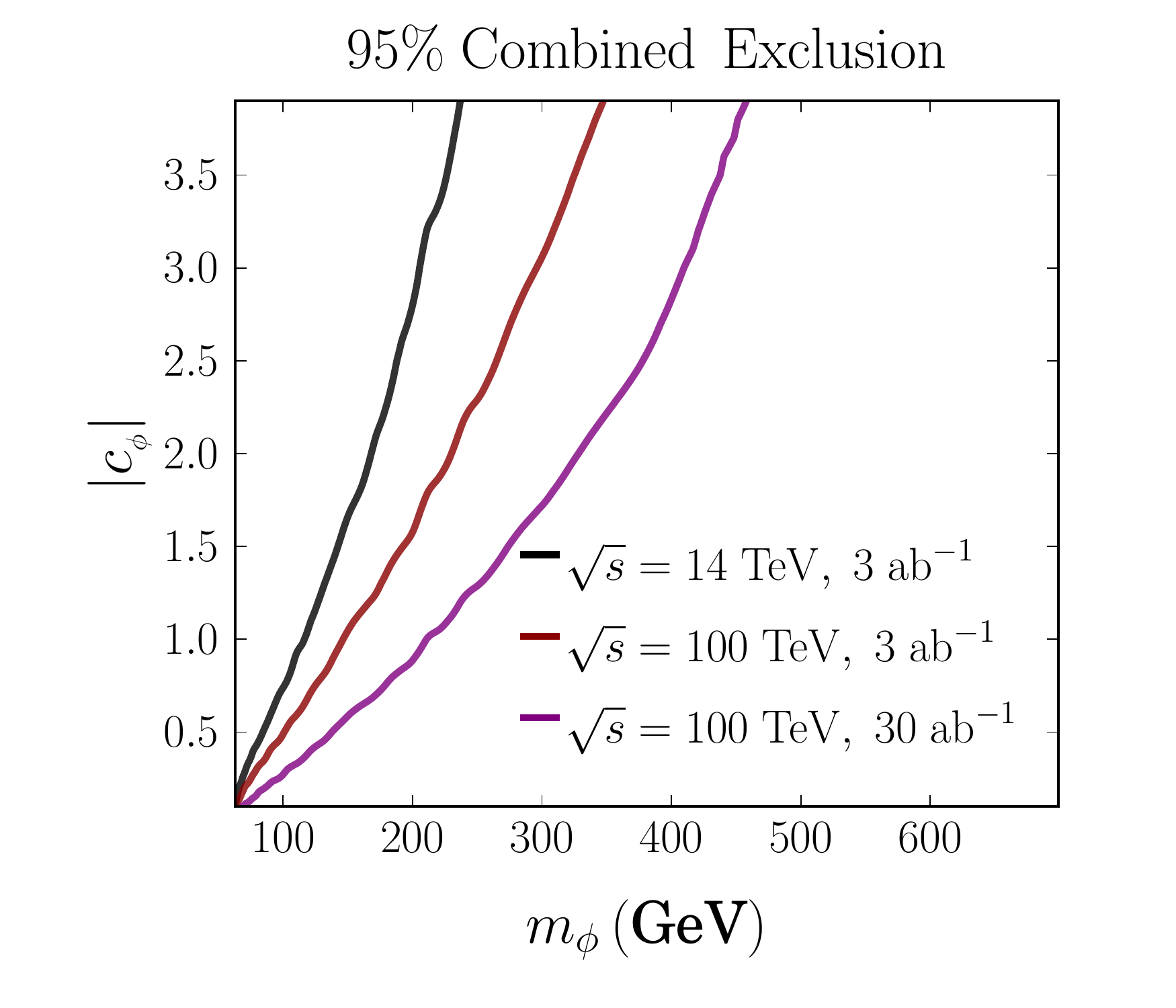}
\includegraphics[width=5cm]{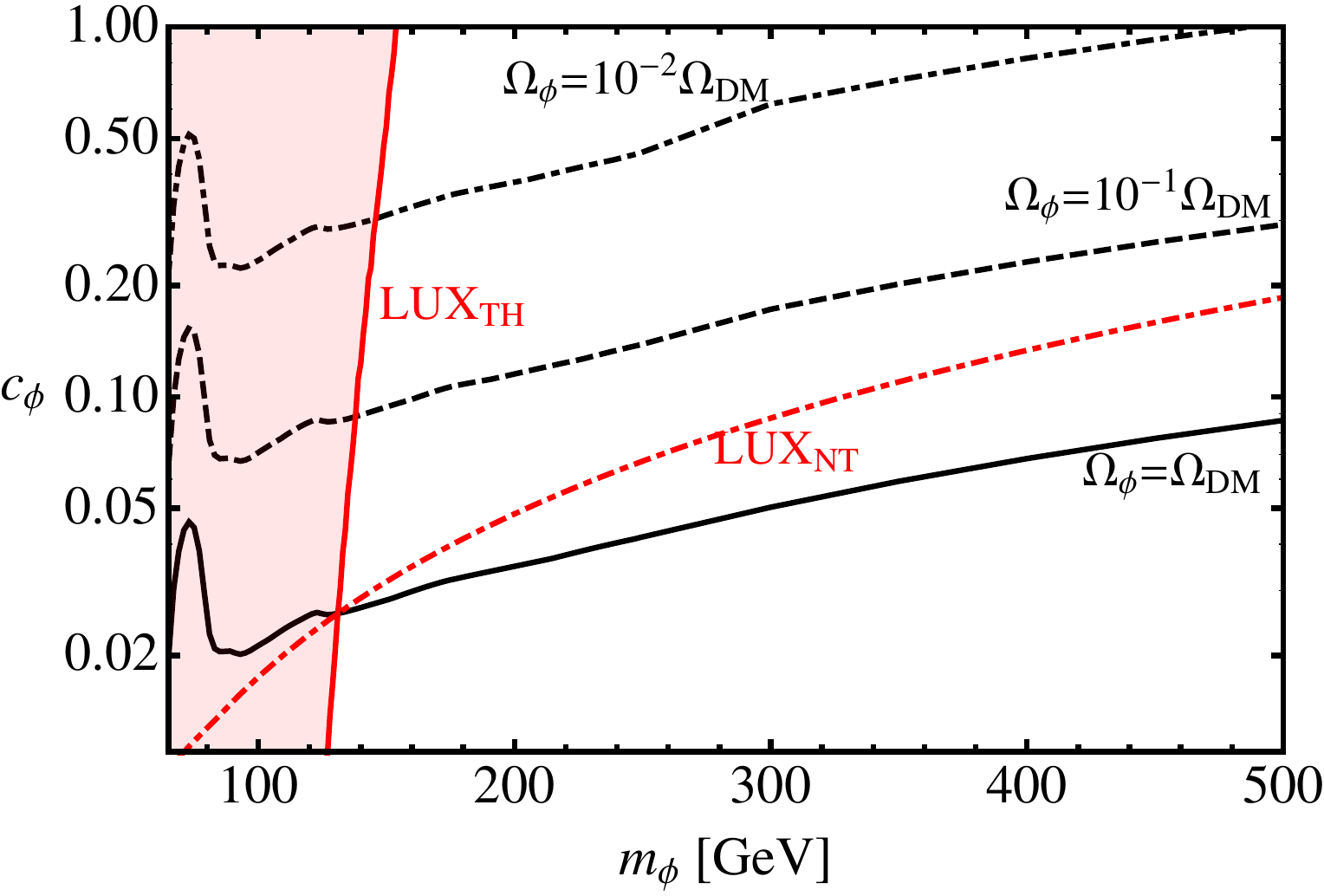}
\caption{
The left figure shows the branching ratio of the decay $H\to \phi \phi$ \cite{talkMcCullough} for $m_\phi \leq m_H/2$. The figure in the middle shows the expected improvement on the Higgs portal coupling $c_\phi$ against $m_\phi$ when increasing the collision energy from $\sqrt{s}=14$ TeV to $\sqrt{s}=100$ TeV and the integrated luminosity from $3~\mathrm{ab}^{-1}$  to $30~\mathrm{ab}^{-1}$. The fraction of the relic dark matter density $\Omega_\mathrm{DM}$ is shown in the right figure. More information on middle and right figures can be found in \cite{Craig:2014lda}.
}
\label{fig:hportdm}
\end{figure}

In the context of models of Secs.~\ref{sec:mix} and \ref{sec:vec} possible signatures at future colliders can be far more diverse than modified Higgs branching ratios or final states with missing energy. Depending on the particle content and their representations mixing between scalars and gauge bosons, e.g. via kinetic mixing, can result in a rich phenomenology. Not only can $\phi$ be probed effectively in an indirect way by global Higgs fits \cite{Lopez-Val:2013yba,Pruna:2013bma} but also in direct searches without involvement of the Higgs boson. In addition, for models wherein $\phi$ transformers nontrivially under SM electroweak symmetry, Drell-Yan pair production that includes at least one electrically charged component of the multiplet may lead to the appearance of a disappearing charged track, providing an additional probe of this class of scenarios\cite{FileviezPerez:2008bj}.

In \cite{Harris:2015kda} predictions for searches for scalar and vector mediators at a possible 100 TeV have been obtained, see Fig.~\ref{fig:hportdm2}. They show strong complementarity between the reach of hadron colliders, indirect and direct detection experiments. Further, it has been shown that the mediator mass and CP property can be inferred from jet distributions in VBF topologies \cite{Khoze:2015sra}.

\begin{figure}[t]
  \begin{center}
    \includegraphics[width=0.4\textwidth]{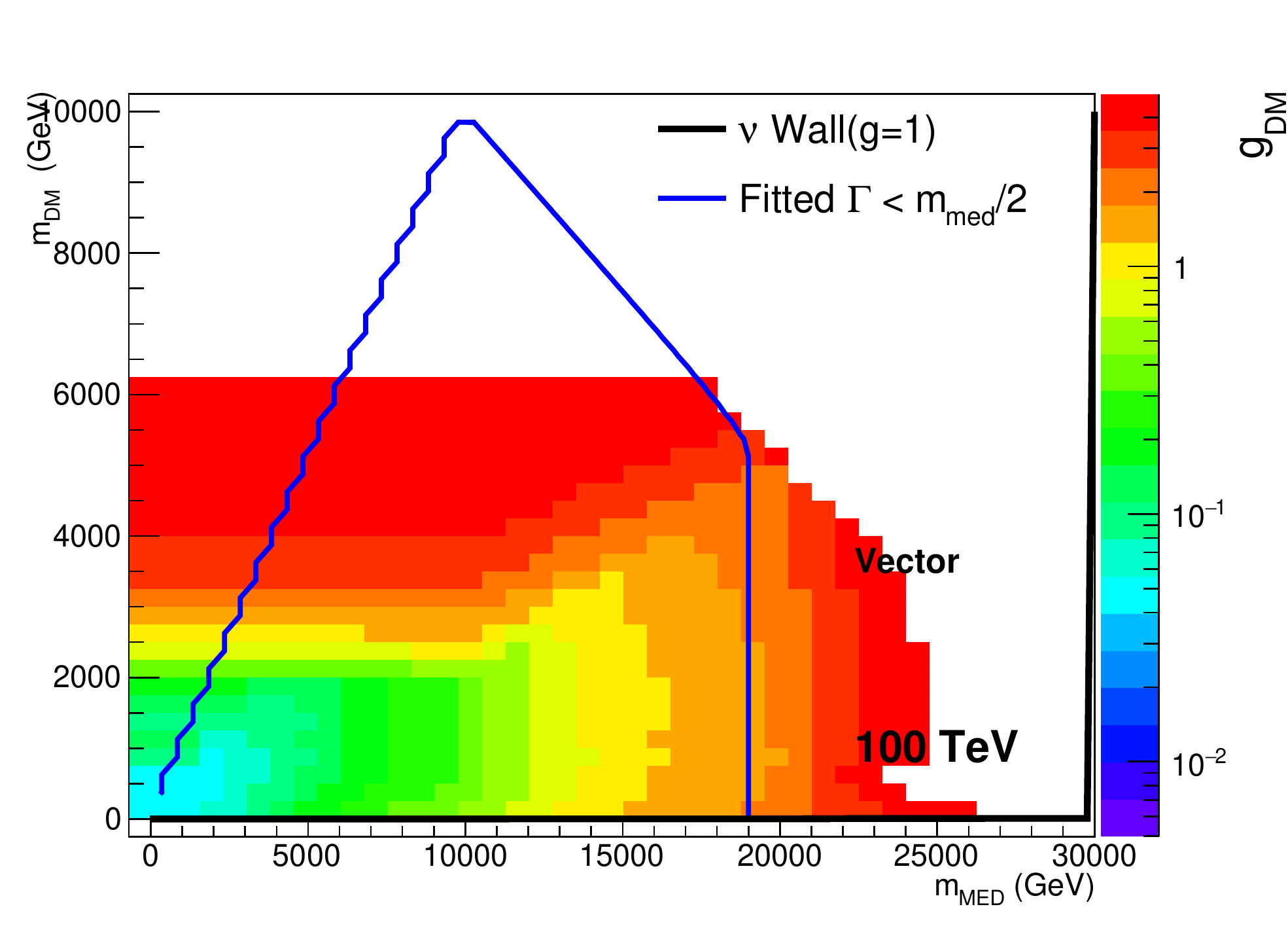} 
    \includegraphics[width=0.4\textwidth]{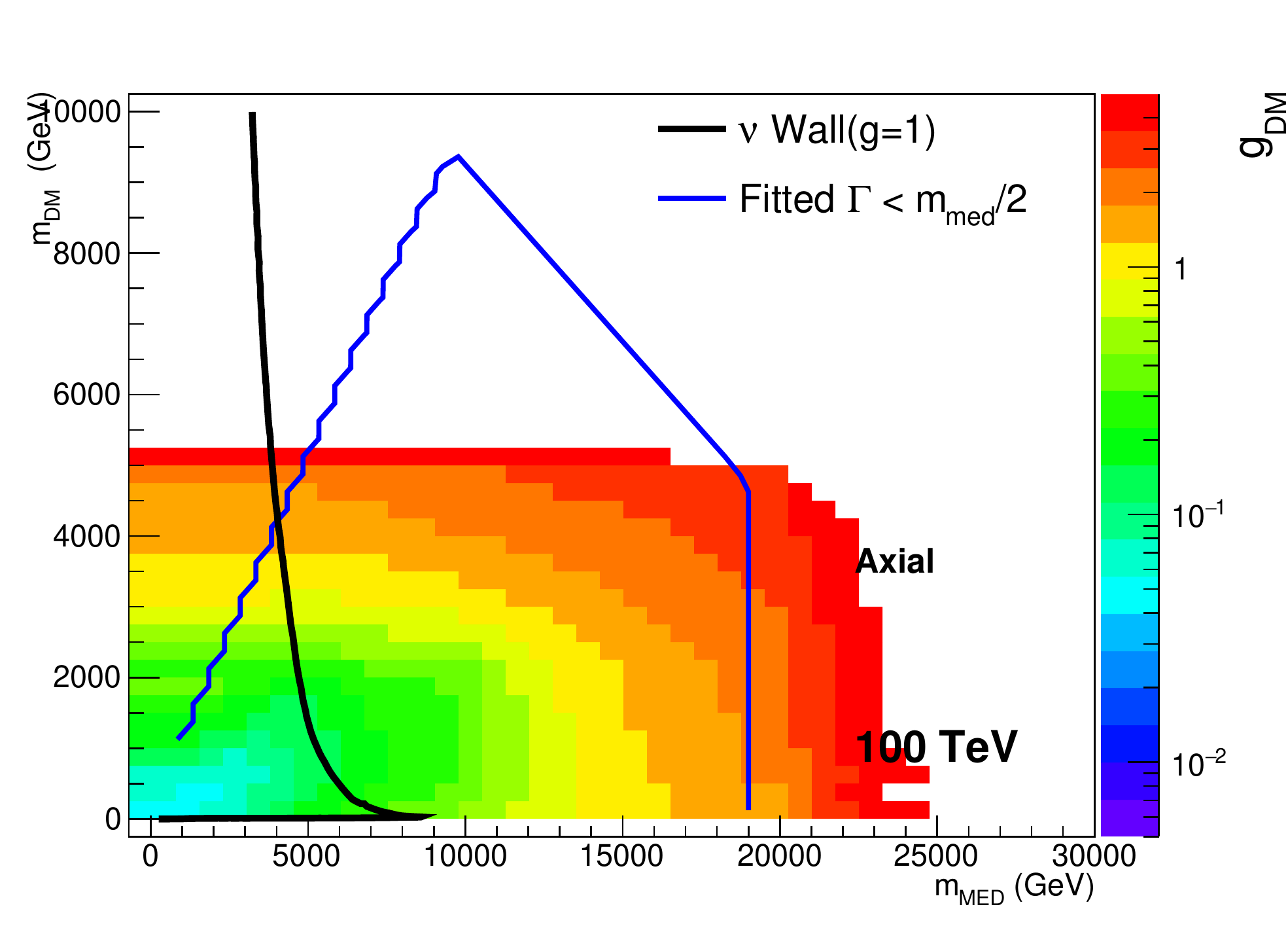}
    \\
    \includegraphics[width=0.4\textwidth]{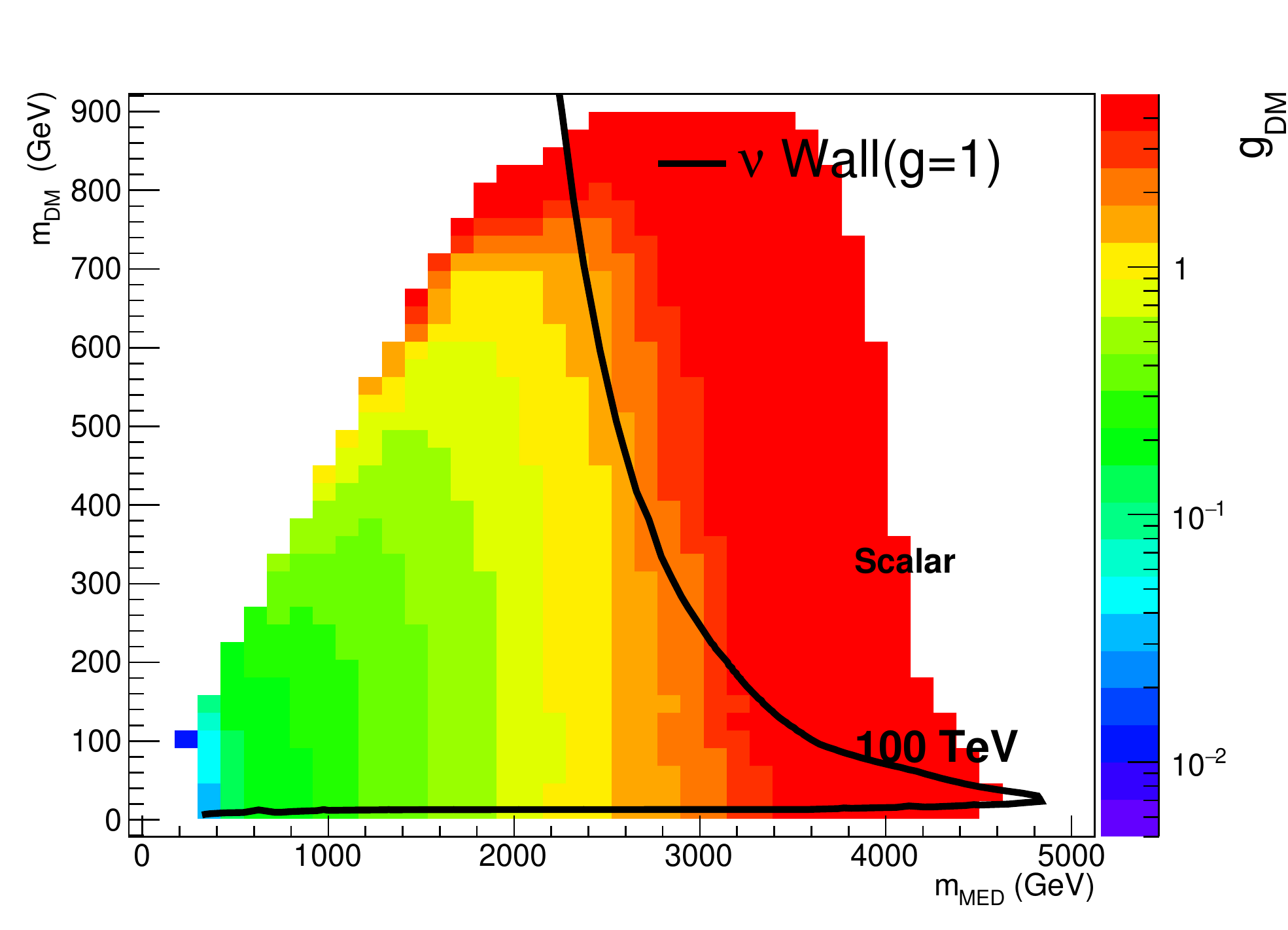}
    \includegraphics[width=0.4\textwidth]{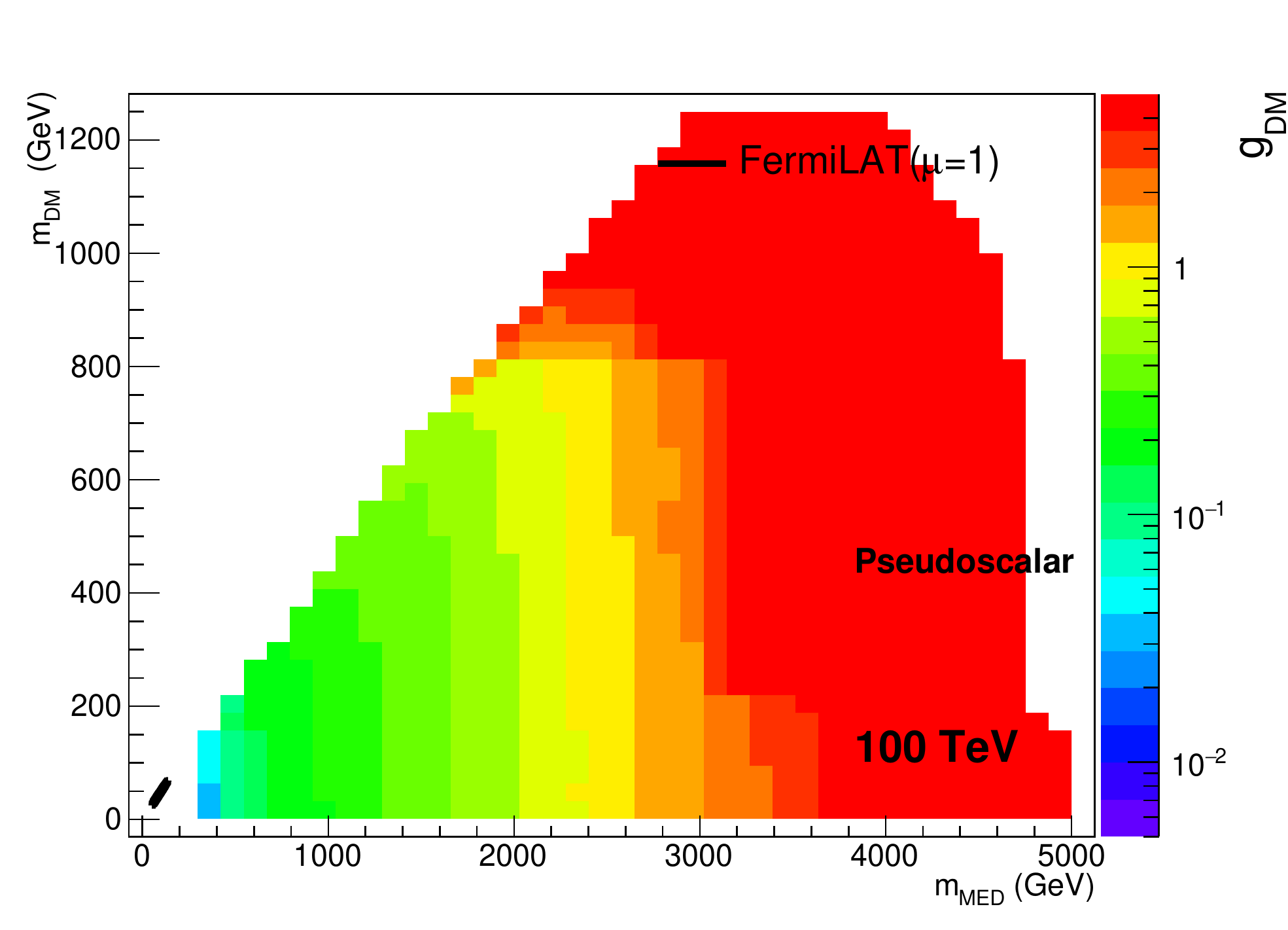}
\caption{
100 TeV limits for vector (upper left), axial-vector (upper right),
scalar (lower left) and pseudoscalar (upper right) mediators. The blue curves on the figures for the vector and axial-vector depict the regions inside which the width of the mediator is smaller than its mass, i.e. $\Gamma_\mathrm{MED} < m_{\mathrm{MED}}$. The black lines for the vector, axial-vector and scalar cases show the limits obtained if cross sections down to the neutrino wall \cite{Cushman:2013zza} can be probed. For the pseudoscalar the black line shows the limit from FermiLAT \cite{Ackermann:2011wa}. More information on these figures can be found in \cite{Khoze:2015sra}.
}\label{fig:hportdm2}
  \end{center}
  
\end{figure}

Striking signatures, with little Standard Model background, are displaced vertices or even displaced jets. They can arise if the mediator has a sufficiently long lifetime and decays back into electrically charged Standard Model particles \cite{Strassler:2006ri,Curtin:2013fra} or mesons of a dark sector which in turn decay into Standard Model mesons, e.g. if the Standard Model gauge group is extended by a dark $SU(N_d)$ \cite{Schwaller:2015gea}. In the latter case, if the mediator is pair-produced, resulting in more than one so-called "emerging jets", the QCD background can be rejected completely. All scenarios with rare but rather clean Higgs decays benefit greatly from the enhanced Higgs production rate and increased integrated luminosity of a 100 TeV collider.

\subsection{The Origins of Neutrino Mass and Left-right symmetric
  model}
\label{ss.neutrinos}

The neutrino oscillation data have unambiguously established that neutrinos have tiny but non-zero masses, as well as mixing between different flavors. Understanding these necessarily requires physics beyond the Standard Model (SM). Since the origin of masses for all the SM charged fermions has now been clarified by the discovery of the Higgs boson~\cite{Aad:2012tfa,Chatrchyan:2012xdj}, an important question is what physics is behind neutrino masses. If we simply add three right-handed (RH) neutrinos $N_R$ to the SM, one can write Yukawa couplings of the form ${\cal L}_{\nu,Y}=h_\nu \bar{L}H N_R$, where $H$ is the SM Higgs doublet and $L$ the lepton doublet.  After electroweak symmetry breaking by the vacuum expectation value (VEV) of the neutral component of the SM Higgs, i.e. $\langle H^0\rangle=v_{\rm ew}$, this gives a Dirac mass to neutrinos of magnitude  $m_D = h_\nu v_{\rm ew} $. To get sub-eV neutrino masses, however, we need $h_\nu \lesssim 10^{-12}$, which is an ``unnaturally" small number, as compared to the Yukawa couplings involving other SM fermions. So the strong suspicion is that there is some new physics beyond just the addition of RH neutrinos,  as well as new Higgs bosons associated with this, which is responsible for neutrino mass generation.

A simple paradigm is the (type-I) seesaw mechanism~\cite{Minkowski:1977sc, Mohapatra:1979ia, GellMann:1980vs, Yanagida:1979as} where the RH neutrinos alluded to above have Majorana masses, in addition to having Yukawa couplings like all charged fermions. Neutrinos being electrically neutral allows for this possibility, making them different from the charged fermions and suggesting that this might be at the root of such diverse mass and mixing patterns for leptons compared to quarks.  The crux of this physics is the seesaw matrix with the generic form in the $ (\nu_L, N_R)$ space:
\begin{eqnarray}
{\cal M}_\nu \ = \ \left(\begin{array}{cc}0 & m_D\\ m^T_D & M_N\end{array}\right)
\end{eqnarray}
where $M_N$ is the Majorana mass for $N_R$ which embodies the new neutrino mass physics, along with the mixing between the light ($\nu_L$) and heavy ($N_R$) neutrinos governed by the parameter $V_{\ell N}\sim m_D M_N^{-1}$.   The mass of light neutrinos is given by the seesaw formula
\begin{eqnarray}
M_\nu \ \simeq \ -m_DM_N^{-1}m_D^T .
\label{eqn:seesaw}
\end{eqnarray}
The question that one is led to ask is what is the origin of $N_R$ and the associated Majorana masses which represents the seesaw scale. We require that the new physics should naturally provide the key ingredients of the seesaw mechanism, i.e. the RH neutrinos and a symmetry origin of their masses $M_N$. This will necessarily involve new Higgs bosons, whose collider signals are discussed in this article for a future collider with center-of-mass $\sqrt s=100$ TeV.  Clearly for the seesaw scale to be accessible at such colliders, it must be below the multi-TeV regime, which implies that there will exist new Higgs bosons with TeV masses. A look at the seesaw formula makes it clear that with a multi-TeV seesaw scale, a sub-eV neutrino mass is quite compatible with Yukawa couplings similar to electron Yukawa of the SM (i.e. $h_\nu \sim h_e$), thus obviating the need for any ultra-small Yukawas (as for example in the pure Dirac case).

\subsubsection{Left-Right Symmetric models (LRSM)}
An appealing UV-complete model for the TeV-scale seesaw is the left-right symmetric model (LRSM) which extends the SM electroweak gauge group to $SU(2)_L\times SU(2)_R\times U(1)_{B-L}$ ~\cite{Mohapatra:1974hk, Mohapatra:1974gc, Senjanovic:1975rk}. The  fermions are assigned to the LR gauge group as follows: denoting $Q \equiv (u \; d)^T$ and $\psi \equiv (\nu \; e)^T$
as the quark and lepton doublets respectively, we assign $Q_L$ and $\psi_L$ as the doublets under the $SU(2)_L$ group and $Q_R$ and $\psi_R$ as the doublets under the $SU(2)_R$ group. The Higgs sector of the model consists of one or several of the following multiplets, that go the beyond the SM Higgs doublet:
\begin{eqnarray}
 \phi\equiv\left(\begin{array}{cc}\phi^0_1 & \phi^+_2\\\phi^-_1 & \phi^0_2\end{array}\right), \quad \Delta_L\equiv\left(\begin{array}{cc}\Delta^+_L/\sqrt{2} & \Delta^{++}_L\\\Delta^0_L & -\Delta^+_L/\sqrt{2}\end{array}\right), \quad \Delta_R\equiv\left(\begin{array}{cc}\Delta^+_R/\sqrt{2} & \Delta^{++}_R\\\Delta^0_R & -\Delta^+_R/\sqrt{2}\end{array}\right).
\label{eqn:Higgs}
\end{eqnarray}
There are versions of the model where parity and $SU(2)_R$ gauge symmetry scales are decoupled so that $\Delta_L$ fields are absent from the low energy theory~\cite{Chang:1983fu}. An important practical implication of the parity decoupling is that it suppresses the type-II seesaw contribution to neutrino masses and thus provides a natural way to realize the TeV-scale type-I seesaw mechanism, as in Eq.~\eqref{eqn:seesaw}.

It has also been pointed out recently that the class of minimal left-right models discussed here provide a natural setting for new fermions or scalars that are stable without the need for extra symmetries and therefore become candidates for dark matter of the universe~\cite{Heeck:2015qra, Ko:2015uma, Garcia-Cely:2015quu, Nagata:2015dma}. We do not elaborate on these issues here since they do not affect our considerations reported here.

The gauge symmetry $SU(2)_R\times U(1)_{B-L}$ is broken by the VEV $\langle \Delta^0_R\rangle = v_R$ to the group $U(1)_Y$ of the SM. So $v_R$ will be the seesaw scale as we see below and is chosen to be in the multi-TeV range.
The VEV of the $\phi$ field, $\langle \phi\rangle={\rm diag}(\kappa, \kappa' e^{i\alpha})$, breaks the SM gauge group to $U(1)_{\rm em}$, with $\alpha$ being a CP-violating phase. We will work in the limit that $\kappa'\ll \kappa$, so that $\kappa \simeq v_{\rm ew}$.

To see how the fermions pick up mass and how the seesaw mechanism arises, we write down the Yukawa Lagrangian of the model:
\begin{align}
{\cal L}_Y \ = \ & h^{\ell,a}_{ij}\bar{\psi}_{L_i}\phi_a\psi_{R_j}+\tilde{h}^{\ell,a}_{ij}\bar{\psi}_{L_i}\tilde{\phi}_a\psi_{R_j}+h^{q,a}_{ij}\bar{Q}_{L_i}\phi_aQ_{R_j} +\tilde{h}^{q,a}_{ij}\bar{Q}_{L_i}\tilde{\phi}_aQ_{R_j}\nonumber \\
& \qquad +f (\psi_{R_i}\Delta_R\psi_{R_j} +\psi_{L_i}\Delta_L\psi_{L_j})~+~{\rm H.c.}
\label{eq:yuk}
\end{align}
where $i,j$ stand for generations and $a$ for labeling Higgs bi-doublets, and $\tilde{\phi}=\tau_2\phi^*\tau_2$ ($\tau_2$ being the second Pauli matrix). After symmetry breaking, the quark and charged lepton masses are given by the generic formula $M_f~=~h^f\kappa + \tilde{h}^f\kappa' e^{-i\alpha}$ for up-type quarks, while for down-type quarks and charged leptons, it is the same formula with $\kappa$ and $\kappa'$ interchanged and $\alpha \to -\alpha$.  The above Yukawa Lagrangian leads to the Dirac mass matrix  for neutrinos, $m_D = h^{\ell}\kappa + \tilde{h}^{\ell}\kappa' e^{-i\alpha}$, and the Majorana mass matrix for the heavy RH neutrinos, $M_N=fv_R$, which go into the seesaw formula (\ref{eqn:seesaw}) for calculating the light neutrino masses.

\subsubsection{Scalar Potential}
The most general renormalizable scalar potential for the bidoublet and triplet fields, which is invariant under parity, is given by
\begin{eqnarray}
\label{eqn:potential}
\mathcal{V} & \ = \ & - \mu_1^2 \: {\rm Tr} (\phi^{\dag} \phi) - \mu_2^2
\left[ {\rm Tr} (\tilde{\phi} \phi^{\dag}) + {\rm Tr} (\tilde{\phi}^{\dag} \phi) \right]
- \mu_3^2 \:  {\rm Tr} (\Delta_R
\Delta_R^{\dag}) \nonumber
\\
&&+ \lambda_1 \left[ {\rm Tr} (\phi^{\dag} \phi) \right]^2 + \lambda_2 \left\{ \left[
{\rm Tr} (\tilde{\phi} \phi^{\dag}) \right]^2 + \left[ {\rm Tr}
(\tilde{\phi}^{\dag} \phi) \right]^2 \right\} \nonumber \\
&&+ \lambda_3 \: {\rm Tr} (\tilde{\phi} \phi^{\dag}) {\rm Tr} (\tilde{\phi}^{\dag} \phi) +
\lambda_4 \: {\rm Tr} (\phi^{\dag} \phi) \left[ {\rm Tr} (\tilde{\phi} \phi^{\dag}) + {\rm Tr}
(\tilde{\phi}^{\dag} \phi) \right]  \\
&& + \rho_1  \left[ {\rm
Tr} (\Delta_R \Delta_R^{\dag}) \right]^2 
+ \rho_2 \: {\rm Tr} (\Delta_R
\Delta_R) {\rm Tr} (\Delta_R^{\dag} \Delta_R^{\dag}) \nonumber
\\
&&+ \alpha_1 \: {\rm Tr} (\phi^{\dag} \phi) {\rm Tr} (\Delta_R \Delta_R^{\dag})
+ \left[  \alpha_2 e^{i \delta_2}  {\rm Tr} (\tilde{\phi}^{\dag} \phi) {\rm Tr} (\Delta_R
\Delta_R^{\dag}) + {\rm H.c.} \right]
+ \alpha_3 \: {\rm
Tr}(\phi^{\dag} \phi \Delta_R \Delta_R^{\dag}) \,.  \nonumber
\end{eqnarray}
Due to the left-right symmetry, all 12 parameters $\mu^2_{1,2,3}$, $\lambda_{1,2,3,4}$, $\rho_{1,2}$, $\alpha_{1,2,3}$ are real, except for the CP violating phase $\delta_2$, as explicitly stated in Eq.~\eqref{eqn:potential}. If the $v_R$ is in the multi-TeV range, the parity symmetric theory above leads to an unacceptably large contribution to neutrino masses from the $\Delta_L$ VEV (the so-called type-II seesaw contribution). In  order to make the TeV-scale LRSM an acceptable and natural theory for small neutrino masses (without invoking any fine-tuning or cancellations between the type-I and type-II terms), one needs to suppress the type-II contribution. This can be done simply by decoupling parity and $SU(2)_R$ breaking scales, in which case in the low energy spectrum (and hence, in the scalar potential), the $\Delta_L$ field is absent~\cite{Chang:1983fu}.  In this section, we will consider this class of TeV-scale LRSM (unless otherwise specified)  and study its implications in the Higgs sector.

\subsubsection{New Higgs bosons in LRSM}
In the minimal LRSM with the left-handed triplet $\Delta_L$ decoupled, there are 14 degrees of freedom in the scalar sector, of which two neutral components and two pairs of singly-charged states are eaten by the six massive gauge bosons ($W^\pm,~W_R^\pm,~Z,~Z_R$), thus leaving 8 physical scalar fields, namely, three CP-even ($h,H^0_{1,3}$), one $CP$-odd ($A^0_{1}$), two singly-charged ($H^\pm_{1}$) and RH doubly-charged fields ($H^{\pm\pm}_{2}$) ($h$ being the SM Higgs boson).\footnote{The physical scalars from $\Delta_L$ are labeled respectively as $H_2^0$, $A_2^0$, $H_2^\pm$ and $H_1^{\pm\pm}$, and are decoupled from the low-energy spectrum.} Their mass eigenvalues are given by (with $\xi \equiv \kappa'/\kappa$)
\begin{eqnarray}
\label{eqn:hmass}
M_h^2 &=& \left(  4 \lambda _1-\frac{\alpha _1^2}{\rho _1} \right) \kappa^2 \,, \\
M_{H_1^0}^2 &=& \alpha _3 ( 1 + 2 \xi ^2 ) v_R^2 + 4 \left( 2 \lambda
   _2+\lambda _3 + \frac{4 \alpha _2^2 }{\alpha _3-4 \rho _1} \right) \kappa^2 \,, \\
\label{eqn:H3mass}
M_{H_3^0}^2 &=& 4 \rho_1 v_R^2 + \left( \frac{\alpha_1^2}{\rho_1} - \frac{16 \alpha_2^2}{\alpha_3 -4\rho_1} \right) \kappa^2 \,, \\
M_{A_1^0}^2 &=& \alpha _3 ( 1 + 2 \xi ^2 ) v_R^2 +4 \left(\lambda _3-2 \lambda _2\right) \kappa^2 \,, \\
\label{eqn:Hpmass}
M^2_{H_1^\pm} &=& \alpha _3 \left((1 +2 \xi ^2) v_R^2 + \frac12 \kappa^2\right) \,, \\
\label{eqn:Hppmass}
M^2_{H_2^{\pm\pm}} &=& 4 \rho_2 v_R^2 + \alpha_3 \kappa^2 \,.
\end{eqnarray}

Note that prior to symmetry breaking, there are two distinct types of Higgs bosons in the minimal version of the model [cf. Eq.~\eqref{eqn:Higgs}]:  the bi-fundamental Higgs field $\phi({\bf 2},{\bf 2},0)$ that is responsible for breaking the SM electroweak gauge symmetry and generating Dirac fermion masses, and the triplet field $\Delta_R({\bf 1},{\bf 3},2)$ that is responsible for breaking the $SU(2)_R\times U(1)_{B-L}$ symmetry and generating the seesaw scale. Apart from their interactions with the gauge bosons and the bi-doublet fields, the triplet fields are hadrophobic, i.e. couple exclusively to leptons in the limit of $\kappa \ll  v_R$. After symmetry breaking,  these Higgs fields mix among themselves, but in the limit $\epsilon \equiv \kappa/v_R, \xi\equiv \kappa'/\kappa \ll 1$, they can be considered almost pure states. With this approximation, we find the predominantly bi-fundamental Higgs mass eigenstates at the TeV-scale to be
\begin{eqnarray}
H_1^0 \  &=&  \ {\rm Re}\: \phi^0_2 - \xi \, {\rm Re}\: \phi^0_1 - \beta \epsilon \, {\rm Re}\: \Delta_R^0 \; , \nonumber \\
A_1^0 \ &=& \ {\rm Im}\: \phi^0_2 + \xi \, {\rm Im}\: \phi^0_1 \; , \nonumber \\
H_1^{\pm} \ &=& \ \phi^\pm_2 + \xi \, \phi^\pm_1 + \frac{\epsilon}{\sqrt{2}}\Delta_R^\pm \; .
\end{eqnarray}
Similarly, the predominantly hadrophobic Higgs mass eigenstates at the TeV-scale are
\begin{eqnarray}
H^0_3 \ &=& \ {\rm Re} \: \Delta_R^0 + \beta \epsilon \, {\rm Re}\: \phi^0_1 + \beta' \epsilon \, {\rm Re}\: \phi^0_2  \; , \nonumber \\
H_2^{\pm\pm} \ &=& \ \Delta^{\pm\pm}_R \; ,
\end{eqnarray}
where $\beta,\, \beta^\prime$ are some combinations of the scalar couplings in the Higgs potential and are expected to be of order $\sim 1$. The hadrophobic Higgs masses are typically of order $\beta v_R$.  Since our goal is to explore the Higgs sector of the minimal LRSM at the $\sqrt s=100$ TeV collider, we will assume that the $SU(2)_R$-symmetry breaking scale is in the multi-TeV range, which generally means that the new Higgs fields are also in the multi-TeV range. For an earlier discussion of the Higgs  mass spectrum in this model, see Refs.~\cite{Gunion:1986im,Gunion:1989in}. A recent detailed study at the future 100 TeV collider, including the relevant couplings, production and decay modes of these new Higgs bosons, can be found in Ref.~\cite{Dev:2016dja}.

\paragraph{Bidoublet Higgs Sector}
We identify the lightest CP-even Higgs boson (denoted by $h$) as the SM-like Higgs field and fix its mass to be 125 GeV by appropriately choosing the parameters of the scalar potential. Its trilinear coupling is then related in the same way as in the SM  in the limit of  $\xi\ll 1$:
\begin{eqnarray}
\label{eqn:lambdahhh}
\lambda_{hhh} = \frac{1}{2\sqrt2} \left( 4\lambda_1 - \frac{\alpha_1^2}{\rho_1} \right)\kappa + \sqrt2 \left( 4\lambda_4 - \frac{\alpha_1\alpha_2}{\rho_1} \right)\xi \kappa \,,
\end{eqnarray}
but differs from this prediction once $\kappa'$ becomes comparable to $\kappa$.  So any observed deviation of the $m_h-\lambda_{hhh}$ relation of the SM would be a measure of the ratio $\kappa'/\kappa$ in the LRSM.

Turing now to the heavier fields, namely $H^0_1$, $A_1^0$ and $H_1^\pm$, being in the same bidoublet, they are expected to have similar masses. The scale of their masses is severely constrained in the minimal version of the model by low energy flavor changing neutral current (FCNC) processes, such as $K_L-K_S$ mass difference, $B-\bar{B}$ mixing etc~\cite{Beall:1981ze, Zhang:2007da, Maiezza:2010ic, Bertolini:2014sua} and is known to imply $M_{H^0}\geq 8-10$ TeV. These fields are therefore not accessible at the LHC but ripe for searches at the 100 TeV collider. 

\paragraph{Hadrophobic Higgs Sector}
The second set of Higgs fields in this model consists of the hadrophobic scalars $H^0_3$ and $H_2^{\pm\pm}$ that are part of $\Delta_R({\bf 1},{\bf 3},2)$ which is responsible for giving Majorana mass to the RH neutrinos. Prior to symmetry breaking, they do not couple to quarks, as is evident from Eq.~\eqref{eq:yuk}.  

\subsubsection{Production of the Heavy Higgs Bosons}\label{sec:prod}
Using the relevant couplings given in~\cite{Dev:2016dja}, we can read off the main collider signals of the heavy Higgs sector in the minimal LRSM. As noted above, we require $M_{H^0_1}, M_{A^0_1}, M_{H^\pm_1} \gtrsim 10$ TeV to satisfy the FCNC constraints, whereas  $M_{H^0_3}, M_{H_2^{\pm\pm}}$ can be much lighter, since there are no such stringent low energy flavor constraints on them. The doubly-charged scalars must be above a few hundred GeV to satisfy the existing LHC constraints~\cite{ATLAS:2014kca}.

The productions of the heavy CP-even/odd Higgs fields $H_1^0/A_1^0$ are mainly through the $b$-parton content of the proton, i.e. $b \bar{b} \to H_1^0/A_1^0$. This is due to the fact that the couplings of $H_1^0$ and $A_1^0$ to light quarks are Yukawa-suppressed and to top-quark is suppressed by $\kappa'/\kappa$, while the gluon fusion channel is highly suppressed by the loop factor in the chiral limit of small $m_b^2 / M_{H_1^0}^2\to 0$. The parton-level cross sections for $pp\to H_1^0/A_1^0$ and other relevant sub-dominant processes at $\sqrt s=100$ TeV are shown in Figure~\ref{fig:prod_Higgs1} (left). Here we have computed the leading order (LO) cross sections using {\tt CalcHEP3.6.25} event generator~\cite{Belyaev:2012qa} and {\tt CT14}~\cite{Dulat:2015mca} parton distribution functions (PDFs).
We also include the NLO and NNLO QCD corrections estimated using an appropriately modified version of {\tt SuSHi}~\cite{Harlander:2012pb} and find that the NNLO $K$-factor is sizable $\sim 2.6-2.8$.


For the singly-charged Higgs field $H_1^\pm$, the dominant production process is via associated production with a highly boosted top quark jet, e.g. $\bar{b} g \to H_1^+ \bar{t}$. This is mainly due to the large gluon-content (and sizable bottom content) of the proton and the large Yukawa coupling of $H_1^{\pm}$ to third-generation fermions.
The NLO corrections, e.g. the process with an extra b-quark jet, are found to be about 1.6.
The associated production with two light quark jets is also important, which is predominantly via the SM $W$ boson: $pp \to H_1^\pm W^\mp \to H_1^\pm jj$, with subleading contribution from heavy $W_R$ vector boson fusion (VBF) process.
Without imposing any specialized selection cuts on the light and heavy quark jets and just using the basic trigger cuts $p_{T_j}>50$ GeV and $\Delta R_{jj} > 0.4$, we show the parton-level cross sections in these three channels for $H_1^\pm$ production as a function of its mass in Figure~\ref{fig:prod_Higgs1} (left).

\begin{figure}[t!]
\centering
\includegraphics[width=7.3cm]{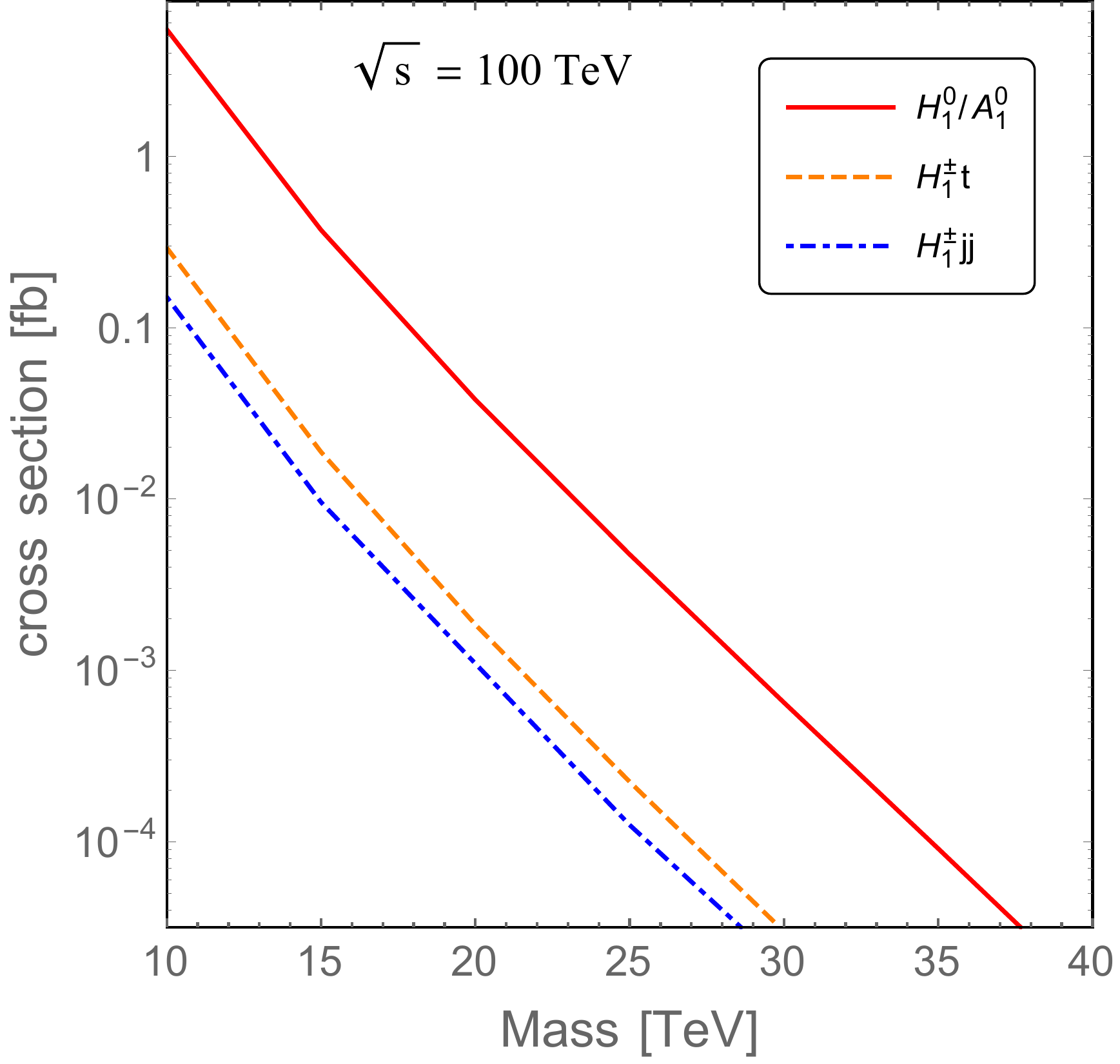}
\includegraphics[width=7.3cm]{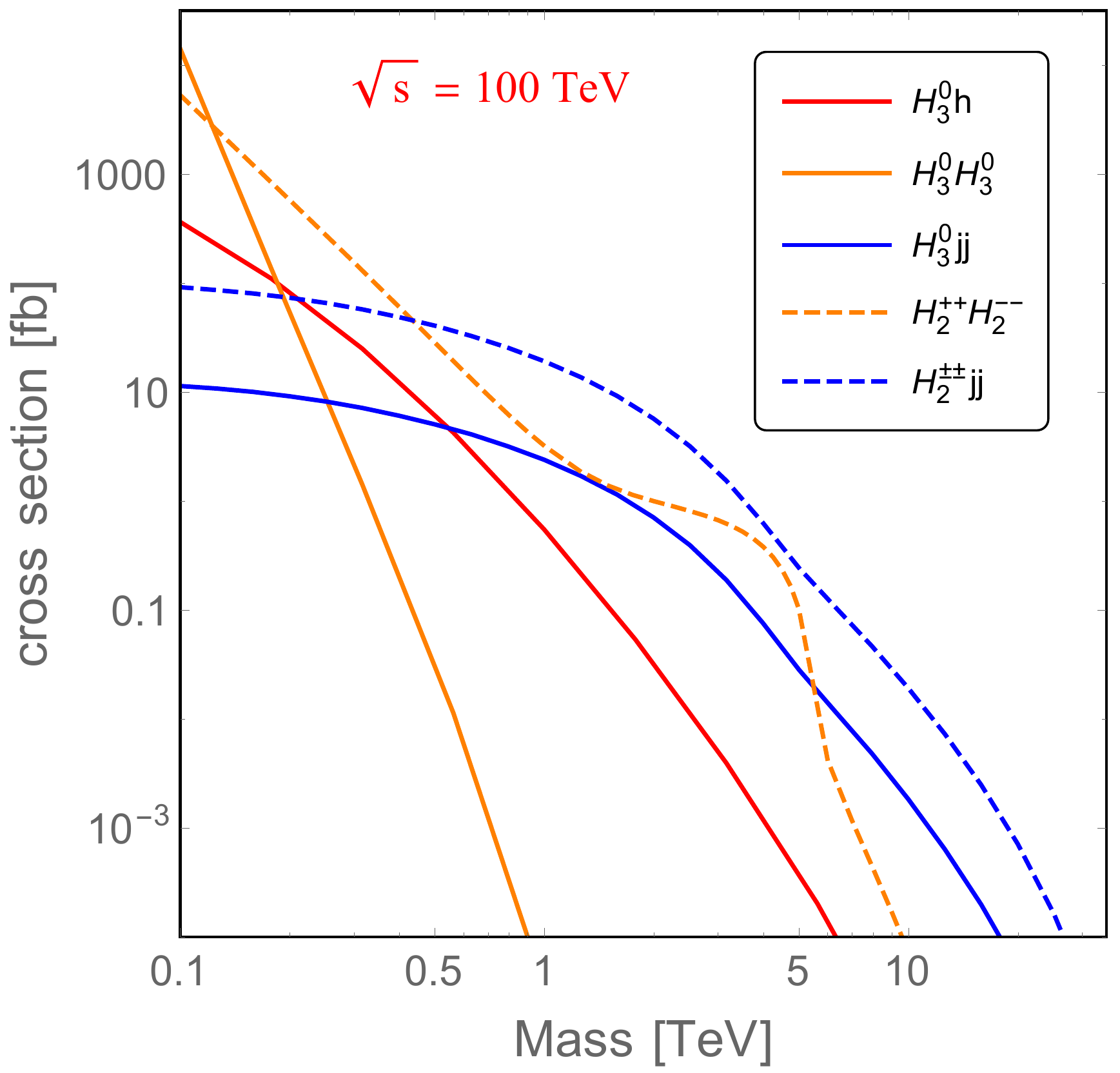}
  \caption{Dominant production cross sections for the heavy bidoublet Higgs bosons $H_1^0$, $A_1^0$ and $H_1^\pm$ (left) and hadrophobic Higgs bosons $H_3^0$ and $H_2^{\pm\pm}$ (right) in the minimal LRSM at a $\sqrt s= 100$ TeV FCC-hh. Reproduced from~\cite{Dev:2016dja}.}
\label{fig:prod_Higgs1}
\end{figure}

For the hadrophobic Higgs sector, the dominant production mode for $H_3^0$ is either via the VBF process involving RH gauge bosons in $t$-channel: $pp\to H_3^0 jj$ (with potentially important contribution from Higgsstrahlung processes $pp \to H_3^0 V_R \to H_3^0 jj$ where $V_R = W_R,\, Z_R$) or via associated production with the SM Higgs or pair-production of $H_3^0$: $pp\to h^*/H_1^{0\,(\ast)} \to H_3^0 h/H_3^0 H_3^0$, depending on the mass spectrum. The VBF processes are guaranteed by the gauge couplings, while the other two channels depend on the quartic couplings, mainly $\alpha_1$ and $\alpha_2$. The $H_1^0$ portal in the $H_3^0 h$ and $H_3^0 H_3^0$ channels is highly suppressed by the large bidoublet mass in most of the parameter space of interest, and we switch it off by setting $\alpha_2 = 0$. Regarding the SM Higgs portal, from the masses of the SM Higgs $h$ and $H_3^0$ [cf.~Eq.~(\ref{eqn:hmass}) and (\ref{eqn:H3mass})], one can easily obtain that $\lambda_1 = {M_h^2}/{4 \kappa^2} + {\alpha_1^2 v_R^2}/{M_{H_3^0}^2}$.
To prevent an unacceptably large $\lambda_1$ when $H_3^0$ is light below the TeV scale, we set a small value of $\alpha_1 = 0.01$. For the VBF channel, we set explicitly the gauge coupling $g_R = g_L$\footnote{Note that the parameter $g_R/g_L$ has significant effect on the $H_3^0$ production in the VBF channel~\cite{Dev:2016dja}.} and the RH scale $v_R = 10$ TeV to fix the masses of heavy gauge bosons, and apply the same basic cuts on the light quark jets as above.  The corresponding production cross sections in the three dominate channels are shown in Figure~\ref{fig:prod_Higgs1} (right). For the Higgs portal, we include the NLO QCD $k$-factor, which is known to be large $\sim 2$ for the top-quark loop~\cite{Heinemeyer:2013tqa}. It is obvious that when $H_3^0$ is light, say $M_{H_3^0} \lesssim 500$ GeV, the Higgs portal dominates, otherwise the VBF process takes over as the dominant channel.

For the doubly-charged Higgs sector, the dominant production mode is either via the Drell-Yan (DY) mechanism: $pp \to \gamma^\ast / Z^\ast /Z_R^{(\ast)}\to H_2^{++}H_2^{--}$ (with potentially sub-leading contribution from the SM Higgs or resonant enhancement from the heavy $H_1^0$ or $H_3^0$) or via the VBF process $pp\to H_2^{\pm\pm} jj$ mediated by RH gauge bosons $W_R^\pm$ in the $t$-channel (with potentially important contribution from the Higgsstrahlung process $pp \to H_2^{\pm\pm} W_R^\mp \to H_2^{\pm\pm} jj$). The LO cross sections are shown in Figure~\ref{fig:prod_Higgs1} (right), where we have chosen the same set of parameters and cuts given above, as well as $M_{H_3^0} = 5$ TeV to completely fix the coupling $h H_2^{++} H_2^{--}$ [cf.~Table 5 in Ref.~\cite{Dev:2016dja}]. We find that  for $M_{H_2^{\pm\pm}} \lesssim 400$ GeV, the DY process is dominant, whereas for relatively larger $M_{H_2^{\pm\pm}}$, this is suppressed, compared to the VBF process, due to kinematic reasons. The bump in the DY channel around 5 TeV is due to the resonant $Z_R$ contribution with $M_{Z_R} \simeq 2M_{H_2^{\pm\pm}}$.

\subsubsection{Decays of the heavy Higgs bosons}



For the bidoublet scalar $H_1^0$, the dominant decay channels are $b\bar{b}$, $hH_3^0$ and $WW_R$ (if kinematically allowed) which almost saturate  the total decay width. The branching fractions are comparable, depending on the top Yukawa coupling $y_t$ and the quartic couplings $\alpha_2$ and $\alpha_3$ (relevant to the mass of $H_1^0$). It is remarkable that for all the heavy Higgs bosons in the LRSM at LO all the dependence on the gauge coupling $g_R$ is cancelled out (except for the dependence through the heavy gauge bosons), and the decay widths are proportional to the RH scale $v_R$, as that is the only relevant energy scale in the high-energy limit. The other channels are suppressed by the relatively smaller couplings (e.g. $hh$ and $t\bar{t}$) or the phase space (e.g. $W_RW_R$ and $Z_RZ_R$). Given the three dominant channels with large couplings, the total decay width of $H_1^0$ is generally very large in most of the parameter space, up to a TeV or even larger.

\begin{table}[t!]
  \centering
  \caption[]{Dominant decay channels of the heavy bidoublet and hadrophobic Higgs bosons in the minimal LRSM and their corresponding branching fractions.
  See text and Ref.~\cite{Dev:2016dja} for more details.}
  \label{tab:decay}
  \begin{tabular}{lll}
  \hline\hline
  scalar & channels & BR / comments \\ \hline
  $H_1^0$  & $b\bar{b}$ & The BRs of the three channels are comparable in most  \\
  & $hH_3^0 \rightarrow hhh \rightarrow 6b / 4b2\gamma$ &  of the parameter space of interest, with the exact values \\
  & $WW_R \rightarrow 4j / \ell^\pm \ell^\pm 4j $ & depending on the parameters in LR model.
\\ \hline
  $A_1^0$ & $b\bar{b}$ &  The two channels  are comparable, depending on the parameters. \\
  & $WW_R \rightarrow 4j / \ell^\pm \ell^\pm 4j $ & $\Gamma(A_1^0 \rightarrow b\bar{b}) \simeq \Gamma(H_1^0 \rightarrow b\bar{b})$ \\
  && $\Gamma(A_1^0 \rightarrow WW_R) \simeq \Gamma(H_1^0 \rightarrow WW_R)$ \\ \hline
  $H_1^\pm$ & $t\bar{b} (\bar{t}b) \rightarrow bbjj / bb\ell\nu$ &  The three channels are comparable, depending on the parameters. \\
  & $ZW_R \rightarrow 4j / \ell^\pm \ell^\mp\ell^\pm \ell^\pm jj$ & $\Gamma(H_1^+ \rightarrow t\bar{b}) \simeq \frac12 \Gamma(H_1^0 \rightarrow b\bar{b})$ \\
  & $hW_R \rightarrow bbjj/ \ell^\pm \ell^\pm bb jj$ & $\Gamma(H_1^+ \rightarrow ZW_R) \simeq \Gamma(H_1^+ \rightarrow hW_R) \simeq \frac12 \Gamma(H_1^0 \rightarrow WW_R)$ \\ \hline
  $H_3^0$ & $hh \rightarrow 4b / 2b2\gamma$ &  $\sim 100\%$ (if the heavy particle channels are not open ) \\
  & $N_R N_R \rightarrow \ell^\pm\ell^\pm 4j$ & sizable if the four heavy particle channels are open   \\
  & $W_R^\pm W_R^\mp \rightarrow 4j / \ell^\pm \ell^\pm 4j$ \\
  & $Z_RZ_R \rightarrow 4j / \ell^\pm \ell^\mp jj$ \\
  & $H_2^{++}H_2^{--} \rightarrow \ell^+ \ell^+ \ell^- \ell^-$ &
  \\  \hline
  $H_2^{\pm\pm}$ & $\ell^\pm \ell^\pm$ &  $\sim 100\%$ (if $W_R W_R$ channel is not open ) \\
  & $W_R^\pm W_R^\pm \rightarrow 4j / \ell^\pm \ell^\pm 4j$ & sizable if kinematically allowed \\
  \hline\hline
  \end{tabular}
\end{table}

The decay of $A_1^0$ is somewhat similar to $H_1^0$, and is dominated by $b\bar{b}$ and $WW_R$, with the partial decay widths the same as that for $H_1^0$ at the leading order.  This implies that the bi-doublet CP-even and odd scalars in the LRSM will appear as wide resonances at the FCC-hh.

For the singly-charged sector, since $H_1^\pm$ comes from the same doublet as $H_1^0$ and $A_1^0$, its decay is closely related to the two neutral scalars. From the couplings in Table 3 of Ref.~\cite{Dev:2016dja}, it is easily found that $H_1^\pm$ decays dominantly to $t\bar{b}$ ($\bar{t}b$), $ZW_R^\pm$ and $hW_R^\pm$, with the partial widths half of those corresponding to the two neutral scalars, cf. Table~\ref{tab:decay}. The latter two decay modes are related via the Goldstone equivalence theorem before symmetry breaking at the RH scale. These partial decay width relations among $H_1^0$, $A_1^0$ and $H_1^\pm$ could be used as a way to distinguish the LRSM Higgs sector from other beyond SM Higgs sectors, such as in the MSSM.

For the hadrophobic scalar $H_3^0$, if it is not heavy enough to produce $N_R N_R$, $W_RW_R$, $Z_RZ_R$ or $H_2^{++}H_2^{--}$, it can decay only to $hh$, since the $t\bar{t}$ and $b\bar{b}$ channels are suppressed by the small mixing parameter $\epsilon$. In this case the width depends on the quartic parameter $\alpha_1$ which is directly related to the SM Higgs mass and trilinear coupling $\lambda_{hhh}$ [cf.~Eqs.~(\ref{eqn:hmass}) and (\ref{eqn:lambdahhh})]. Due to the theoretical and experimental constraints, the decay width of $H_3^0$ in this case could possibly be very small, around 10 GeV scale. If the decays to heavy particles are open, the width would be largely enhanced, as none of those couplings are suppressed; see Table 4 of Ref.~\cite{Dev:2016dja}. An interesting case is the decay into a pair of doubly-charged Higgs, which decays further into four leptons. In this case we can study the two scalars simultaneously in one chain of production and decay processes. Note that in this channel, the trilinear coupling for the vertex $H_3^0 H_2^{++} H_2^{--}$ is directly related to the masses of the two particles [cf. Eqs.~(\ref{eqn:Hpmass}) and (\ref{eqn:Hppmass}) and Table 5 of Ref.~\cite{Dev:2016dja}].

For the doubly-charged scalar $H_2^{\pm\pm}$, the dominant decay channel is to two same-sign leptons. If its mass is larger than twice the $W_R$ mass, the $W_RW_R$ channel is also open and contributes sizably to the total width. As stated above, the $W_R$ channel depends on the gauge coupling $g_R$ only through the $W_R$ boson mass.

More details of the dominant decay channels can be found in Ref.~\cite{Dev:2016dja}, including the analytic formulae for all the partial decay widths at LO. There are also rare lepton number violating Higgs decays that could provide additional signals for the LRSM at colliders~\cite{Maiezza:2015lza}.


\subsubsection{Key discovery channels at the 100 TeV collider}

Given the dominant production and decay modes of heavy Higgs states in the minimal LRSM demonstrated above, we list here the key discovery channels at the FCC-hh. 
For concreteness, we mainly focus on the channels with least dependence on the hitherto unknown model parameters.

Since the production of bidoublet Higgs bosons is solely determined by their Yukawa couplings to the third generation quarks, their signal sensitivities depend only their masses but not on the RH scale $v_R$ or the gauge coupling $g_R$.
For the bidoublet neutral scalars $H_1^0/A_1^0$, the main discovery channel is $pp\to H_1^0/A_1^0 \to b\bar{b}$. Due to the high center-of-mass energy and large masses of $H_1^0/A_1^0$, as required by FCNC constraints, the $b$-jets are highly boosted, which could help to distinguish them to some extent from the SM $bb$ background, for instance with a large invariant mass cut of $M_{bb} > 10$ TeV. With the additional basic transverse momentum and jet separation cuts, it is found that the neutral bi-doublet scalars $H_1^0 / A_1^0$ can be probed in the $bb$ channel up to 15.2 TeV at $3\sigma$ C.L., assuming an optimistic integrated luminosity of 30 ab$^{-1}$~\cite{Dev:2016dja}.


For the CP-even $H_1^0$, there is an additional key channel, i.e. $pp\to H_1^0\to hH_3^0\to hhh$ [cf. Table~\ref{tab:decay}]. If $H_3^0$ is not very heavy, e.g. at the TeV scale, this is a viable channel for both $H_1^0$ and $H_3^0$ discovery, by examining the triple Higgs production, for instance with the distinct final state of $6b$ or $4b+\gamma\gamma$. The LO $gg\to hhh$ production cross section in the SM is 3 fb at $\sqrt s=100$ TeV, with a large NLO $K$-factor of $\sim 2$~\cite{Papaefstathiou:2015paa,Chen:2015gva}. However, this large background can be suppressed effectively by applying $M_{bb} > 10$ TeV. Assuming a branching ratio of 10\% for $H_1^0$ decaying into $h H_3^0$, it is found that the sensitivity in this channel is comparable to the $bb$ mode, reaching up to 14.7 TeV for $H_1^0$~\cite{Dev:2016dja}.


For the singly charged $H_1^{\pm}$, the key discovery channel is $pp\to H_1^\pm t\to ttb$. Again, due to the large mass of $H_1^\pm$, both $t$ and $b$-jets will be highly boosted, which will be a key feature to extract the signal from the irreducible QCD $ttb$ background. In particular, jet substructure analysis of the heavy quark jets and the kinematic observables could help to suppress the SM background and also to distinguish the LRSM model from other scenarios such as the MSSM. With solely a simple cut on the bottom jets $M_{bb} > 5$ TeV, as well as the basic cuts, it is found that the the singly-charged scalar $H_1^\pm$ can be probed only up to 7.1 TeV at the C.L. level of $3\sigma$, mainly due to the small production cross section and the large QCD background~\cite{Dev:2016dja}.


The situation is more intricate for the hadrophobic scalars, as the dominant  production channels depend on the RH scale $v_R$, either through the vertices or through the RH gauge boson propagators, as well as the gauge coupling $g_R$.
For $H_3^0$, the key channel is $pp\to H_3^0 jj\to hhjj$, which can be searched for in either $4b+jj$ or $bb\gamma\gamma jj$ channels. The dominant SM backgrounds are from VBF production of SM Higgs pair and $ZZW$ processes~\cite{Baglio:2015wcg,Torrielli:2014rqa}.
For smaller $H_3^0$ masses, the triple Higgs channel $pp\to H_3^0h\to hhh$ becomes  important, which can be readily distinguished from the same final state due to $H_1^0$ decay, because of the different invariant masses and due to the fact that the $H_3^0$ resonance width is rather small, say $\sim$ few times 10 GeV, as compared to the broad resonance of order TeV for $H_1^0$.


Regarding the doubly-charged scalars $H_2^{\pm\pm}$, the key channels are (i) for low masses, the DY process $pp\to H_2^{++}H_2^{--}\to \ell^+ \ell^+ \ell^- \ell^-$, where some of the leptons could in principle be of different flavor, thus probing lepton flavor violation, with the dominant SM $ZZ$ background~\cite{ZZ}, and (ii) for high masses, the VBF process $pp\to H_2^{\pm\pm}jj\to \ell^\pm\ell^\pm jj$, which is a high-energy analog of the neutrinoless double beta decay, thus probing lepton number violation at FCC-hh. The leptonic channels are rather clean and the backgrounds are mostly from the SM $ZZ$ and $WZ$ leptonic decays with one of the signs of leptons wrongly reconstructed.
As demonstrated in Section~\ref{sec:DCS} the VBF process $H_2^{++} H_2^{--} jj$ is also promising at the FCC-hh. It is interesting to note that this channel could also stem from $pp \to H_3^0 jj \to H_2^{++} H_2^{--} jj$ with on-shell VBF production of $H_3^0$, provided $M_{H_3^0}> 2M_{H_2^{\pm\pm}}$, which could significantly enhance this signal.


Adopting the benchmark values of parameters given in Section~\ref{sec:prod}, we show the projected sensitivities for $H_3^0$ and $H_2^{\pm\pm}$ in all the dominant channels in Figure \ref{fig:sensitivity} for an integrated luminosity of 30 ab$^{-1}$. In all the channels, we choose only the decay modes with the largest significance: for $H_3^0$, it is the decay chain $H_3^0 \to hh \to 4b$, while for the doubly-charged scalar $H_2^{\pm\pm}$ it is the final states of $\ell^\pm \ell^\pm$ with $\ell = e,\, \mu$. All the corresponding SM backgrounds have been taken into consideration in a conservative manner; see Ref.~\cite{Dev:2016dja} for more details.
The sensitivities in the SM Higgs portal $H_3^0$ production channels ($H_3^0 h$ and $H_3^0 H_3^0$) increase for a larger $v_R$ and in these two channels $H_3^0$ can be probed up to multi-TeV range. In the DY channel, $H_2^{\pm\pm}$ is produced predominately through the SM $\gamma/Z$ mediators and thus the sensitivity is almost independent of $v_R$, except a resonance-like enhancement due to a heavy $Z_R$ boson with mass $M_{Z_R} \simeq 2M_{H_2^{\pm\pm}}$. In the VBF channel, both $H_3^0$ and $H_2^{\pm\pm}$ can be probed up to the few TeV range when $v_R$ is small; when $v_R$ becomes larger, due to the increasing $W_R$ (and $Z_R$) mass, the sensitivities drop rapidly, especially when the heavy gauge bosons can not be pair produced on-shell. Even when $v_R$ is in the range of few times 10 TeV, a TeV-scale hadrophobic scalar in the minimal LRSM can still be seen at the 100 TeV collider. The Higgsstrahlung sensitivities are lower for both $H_3^0$ and $H_2^{\pm\pm}$ compared to the VBF channels, and are not shown in Figure \ref{fig:sensitivity}.

More details of the sensitivity studies can be found in Ref.~\cite{Dev:2016dja}. This parton-level analysis is intended to serve as a guideline for more sophisticated and accurate simulations in future, including optimized selection acceptance and cut efficiencies, and other experimental issues, such as jet energy calibration, boosted top and bottom quark tagging efficiencies, etc. For a full detector-level case study of the pair-production of doubly-charged scalars in association with two jets, see Section~\ref{sec:DCS}.

\begin{figure}[t!]
  \centering
  \includegraphics[width=7.5cm]{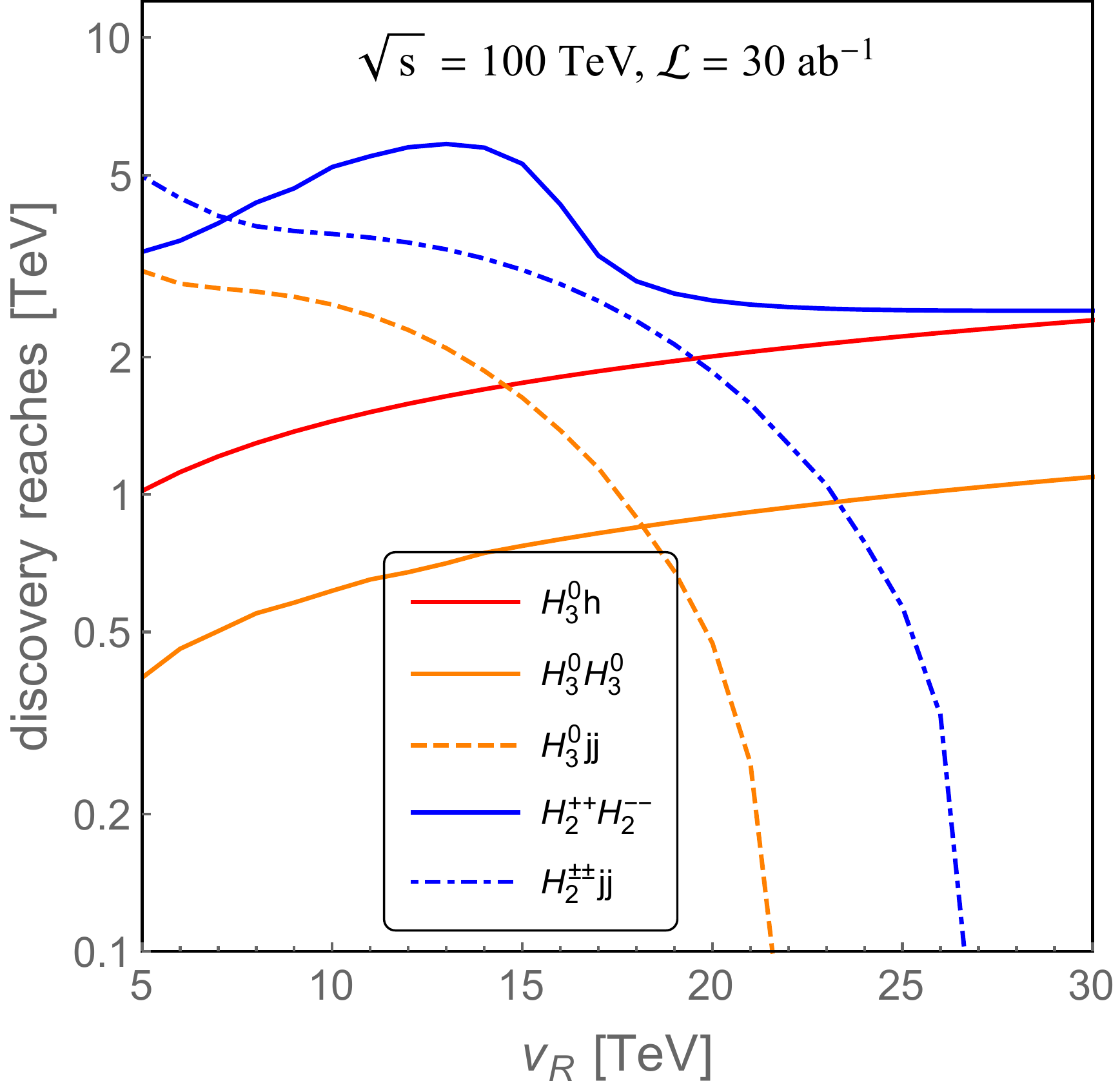}
  \caption{Projected sensitivities of the heavy hadrophobic Higgs bosons $H_3^0$ and $H_2^{\pm\pm}$ in the dominant channels in the minimal LRSM at $\sqrt s= 100$ TeV FCC-hh for an integrated luminosity of $30~{\rm ab}^{-1}$. Reproduced from~\cite{Dev:2016dja}. See text for more details.}
  \label{fig:sensitivity}
\end{figure}

\subsubsection{Case study:  $pp\to H_{1,2}^{++} H_{1,2}^{--} jj$}
\label{sec:DCS}

As stated in Section~\ref{sec:prod}, the dominant production channel for the RH doubly-charged scalars $H_2^{\pm\pm}$ is via the DY pair-production or VBF single production, depending on the model parameters. Another interesting possibility in the VBF scenario is the pair-production $H_{1,2}^{++} H_{1,2}^{--} jj$ (where $H_1^{\pm\pm}$ is the LH triplet counterpart of $H_2^{\pm\pm}$), which has been studied in great detail in Ref.~\cite{Bambhaniya:2015wna}. In this section, we summarize the main results for this case study. One should be aware that in presence of the left-handed triplet $\Delta_L$, not all charged scalars are {\it always} simultaneously light; however,  there are parameter domains where it is possible~\cite{Bambhaniya:2013wza,Bambhaniya:2014cia,Bambhaniya:2015wna}. 
In general, the doubly-charged scalars decay to either a pair of same-sign charged leptons or a pair of $SU(2)_{L,R}$ charged gauge bosons. The decay branching ratios are controlled by their respective VEVs. See Table 1 of \cite{Bambhaniya:2013wza} for more details. In this case study, it is assumed that the doubly-charged scalars dominantly decay to a pair of same- sign charged leptons, thus leading to the signal of four leptons associated with two forward jets, i.e., $pp\to H_{1,2}^{++} H_{1,2}^{--} jj \to 4\ell+2j$.

To perform the analysis, we have imported our own implemented minimal LRSM  files in {\tt{Madgraph}} \cite{Alwall:2014hca} using {\tt FeynRules}~\cite{Degrande:2011ua}.
In Ref.~\cite{Bambhaniya:2015wna}, two benchmark points consistent with experimental and theoretical constraints were shown which lead to two different sets of scalar spectra, where the common mass of the doubly-charged scalars was  500 GeV and 1000 GeV, respectively.  In this case, the LO
production cross-sections are 
\begin{eqnarray}\label{eq:xsection_1effH1}
\sigma(p p \rightarrow H_{1}^{\pm \pm} H_{1}^{\mp \mp} j j)
&& = 599.70\;[73.28]\times 10^{-2}~\; {\rm fb},\\
\sigma(p p \rightarrow H_{2}^{\pm \pm} H_{2}^{\mp \mp} j j )
&&= 401.40\;[37.43]\times 10^{-2}~\; {\rm fb},
 \nonumber
\end{eqnarray}
for  $\sqrt{s}$=100 TeV with $M_{H_{1,2}^{\pm \pm}} = 500\; [1000]$ GeV respectively \cite{Bambhaniya:2015wna}.
Then  we have allowed  $H^{\pm \pm}_{1,2}$ to decay leptonically within {\tt Madgraph} and that has been interfaced with {\tt DELPHES} \cite{deFavereau:2013fsa} to isolate the leptons and jets. For lepton/jet identifications and construction we have used default FCC-hh card in {\tt DELPHES}  which also includes the basic isolation and selection criteria. We have also incorporated the following VBF
cuts \cite{Dutta:2014dba, Bambhaniya:2015wna} within {\tt DELPHES-root} signal analysis code: ${p_T}_{j_1},{p_T}_{j_2}>50$ GeV, $|\eta_{j_1} - \eta_{j_2}|>4$, $m_{j_1j_2} >500$ GeV and $\eta_{j_1}*\eta_{j_2} <0$.
After implementing VBF cuts and hard $p_T$ cuts ($p_{T_{\ell_1}}>30\;{\rm GeV}$, $p_{T_{\ell_2}}>30\;{\rm GeV}$, $p_{T_{\ell_3}}>20\;{\rm GeV}$, $p_{T_{\ell_4}}>20\;{\rm GeV}$) for four leptons in {\tt DELPHES-root} code \cite{deFavereau:2013fsa}, we find signal cross section to be:
\begin{eqnarray}\label{eq:xsection_1effH2}
\sigma(p p \rightarrow 4l+ j j)_{\rm sig.}
&& = 48.92 \, [5.5146] \times 10^{-2}~\; {\rm fb},
\end{eqnarray}
for  $\sqrt{s}$=100 TeV with $M_{H_{1,2}^{\pm \pm}} = 500 \, [1000]$ GeV,  respectively. In the analysis without {\tt DELPHES} FCC-hh cards \cite{Bambhaniya:2015wna},  this cross section is
$37.01 \, [3.54]\times 10^{-2}$ fb. The departures in the signal cross sections are quite large -- around 32\% and 56 \% respectively for first and second benchmark points. It shows that the implementation of dedicated {\tt DELPHES} cards which take care of the lepton and jet reconstructions, and isolations, is promising and worth of further development for FCC-hh.

The dominant SM background comes from $ZZjj$ final state. We have computed and estimated this background using same set of
selection criteria, hard $p_T$  and VBF cuts for leptons and jets, at parton level using {\tt{Madgraph}}~\cite{Alwall:2011uj}, and at hadron level using {\tt PYTHIA} \cite{Sjostrand:2006za} after incorporating showering and hadronization:
\begin{eqnarray}\label{eq:xsection_1effH3}
\sigma(p p \rightarrow 4l+ j j)_{\rm bkg.}
&& = 479.4 \, [3.8] \times 10^{-2}~\; {\rm fb}.
\end{eqnarray}
\begin{figure}[t!]
  \centering
  \includegraphics[height=8.5cm]{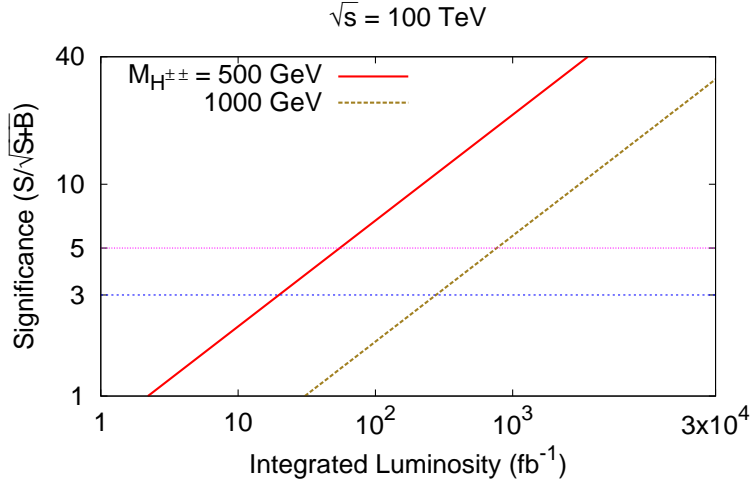}
  \vspace{-1.2cm}
  \caption{Significance vs integrated luminosity for $pp\to H_{1,2}^{++} H_{1,2}^{--} jj \to 4\ell+2j$ with doubly-charged scalar masses of 500 GeV (solid-red line) and 1000 GeV (dotted-yellow) and center of mass energy at $\sqrt{s}$=100 TeV. Here the left- and right-handed doubly-charged scalar contributions are summed up, and the two horizontal lines represent the significances at the level 5 and 3 respectively.}
  \label{fig:Significance_vs_Luminosity}
\end{figure}
For the suggested benchmark points with $M_{H^{\pm \pm}_{1,2}}$=500 and 1000 GeV we have also analyzed the significances  of the signal events for different set of integrated luminosities, see Fig.~\ref{fig:Significance_vs_Luminosity}. We have defined the significance as $S/\sqrt{S+B}$, where $S$ and $B$ are the signal and background events (cross section$\times$luminosity). It is interesting to note that below a luminosity of 100 ${\rm fb}^{-1}$ it is possible to adjudge the signal strength with significance level 5 (magenta dotted line) for $M_{H^{\pm \pm}_{1,2}}$=500 GeV. To make a definite comment on the other benchmark points with larger doubly-charged scalar masses we need to wait till we collect enough data with integrated luminosity $\sim \mathcal{O}(1000)\;{\rm fb}^{-1}$.

\subsubsection{Distinguishing from the MSSM Higgs Sector}

One of the key features which distinguishes the LRSM Higgs sector from other popular beyond SM Higgs sectors, such as the 2HDM, is the presence of the doubly-charged scalars. Thus, a positive signal for any of the doubly-charged scalars discussed here will be a strong evidence for the LRSM. Another distinction is due to the $H_3^0\to hh$ decay mode of the neutral hadrophobic scalar in LRSM, which is absent in generic 2HDM scenarios in the so-called {\it alignment limit}, since the $Hhh$ coupling identically vanishes~\cite{Gunion:2002zf, Carena:2013ooa, Dev:2014yca}.


As for the bidoublet Higgs sector in the LRSM, this is similar to the MSSM Higgs sector, which also contains two SM Higgs doublets. However there is a profound difference between the two models, since in the LR case, the second Higgs, in the limit of $\kappa'=0$ does not contribute to fermion masses and therefore the decay properties are very different, as illustrated in Table~\ref{tab:decay2}. In particular, the $\tau^+ \tau^-$ final state is suppressed by either the Dirac Yukawa coupling or the left-right mixing for the neutral bi-doublet scalars $H_1^0 / A_1^0$ in the LRSM [cf. Table 12 of Ref.~\cite{Dev:2016dja}], whereas this is one of the cleanest search channels for the MSSM heavy Higgs sector in the large $\tan\beta$ limit. Furthermore, due to the presence of extra gauge fields in our case i.e. $W^\pm_R, Z_R$, new modes appear, e.g. $H^0_1 \to W^\pm_RW^\mp $ and $H^\pm_1 \to W^\pm_R Z$, which have no MSSM analog. These modes can lead to distinguishing signals in leptonic channels e.g. $\ell^\pm\ell^\pm 4j$ with $\sim 5\%$ branching ratio. With 30 ab$^{-1}$ data, this can lead to about 1000 events before cuts, while the SM background for these sign-sign dilepton processes is expected to be very small. 
One can also use the relations between the various partial decay widths as shown in Table~\ref{tab:decay} to distinguish the LRSM Higgs sector from other scenarios.


\begin{table}[t]
  \centering
  \caption[]{A comparison of the dominant collider signals of neutral and charged scalars in the minimal LRSM and MSSM. }
  \label{tab:decay2}
  \begin{tabular}{lll}
  \hline\hline
  Field & MSSM & LRSM \\ \hline
  $H^0_1, A^0_1$  & $b\bar{b},~\tau^+\tau^-$ (high $\tan\beta$) &  $b\bar{b}$ \\ \hline
  & $t\bar{t}$ (low $\tan\beta$) & $WW_R \to \ell^\pm\ell^\pm 4j$ \\ \hline
  $H^+$ &  $t\bar{b}t\bar{b}$,  $t\bar{b}\tau^+\nu$  &  $\bar{t}_L b_R$ \\
  \hline\hline
  \end{tabular}
\end{table}

If a positive signal is observed, one can also construct various angular and kinematic observables to distinguish the LRSM scenario from other models giving similar signals~\cite{Han:2012vk, Chen:2013fna, Dev:2015kca}. For instance, we find from Table~\ref{tab:decay2} that $\bar{t}_Lb_R$ final states are preferred over the $\bar{t}_Rb_L$ final states for the $H_1^+$ production, which can be utilized to distinguish it from other 2HDM scenarios, including the MSSM.


\subsubsection{How would this fit into the discovery landscape?}
Discovery of any of the signals of the LRSM, and in particular its Higgs sector, would bring about a fundamental change in our thinking about neutrino masses and will change our perspective on supersymmetry and grand unification. This will also affect the discussion of the origin of matter via leptogenesis in a profound manner.  For instance, if the $W_R$ gauge boson is discovered below $9 \, (g_R / g_L)$ TeV, it will rule out the whole leptogenesis approach~\cite{Frere:2008ct, Deppisch:2013jxa, Dev:2014iva, Dev:2015vra, Dhuria:2015cfa}. Discussions of issues such as naturalness will have to assume the low energy group to be the left-right symmetric group rather than the SM gauge group. This has implications for the stability of the electroweak vacuum~\cite{Mohapatra:1986pj, Maiezza:2016bzp}.

\subsection{Naturalness}
\label{ss.naturalness}

\subsubsection{Supersymmetry}
\label{sec:SUSY}

In spite of the stringent bounds, which have been put on superpartners' masses by the ATLAS and CMS collaborations, supersymmetry (SUSY)
is still an attractive candidate for physics beyond the SM. It can successfully address the big hierarchy problem, 
although with some 
(potentially mild) residual fine tuning. 

The superpartners, and most importantly stops, gauginos and higgsinos play a crucial role in restoring the naturalness, and 
their masses are directly related to the fine tuning of the supersymmetric extension of the 
SM~\cite{Dimopoulos:1995mi,Cohen:1996vb,Papucci:2011wy,Brust:2011tb}. Summaries of the search program can be found, 
for example, in Ref.~\cite{Craig:2013cxa}.  
However, SUSY also necessarily  modifies the higgs sector of the SM, and we will mostly concentrate on these modifications here.

The modifications of the higgs sector in SUSY are twofold. First, low masses superpartners, required by naturalness 
might significantly affect the higgs couplings at the loop level. Given that the leading-order higgs coupling to the photons 
and gluons show up in the SM at the one-loop level, light stops and, to a lesser extent, light gauginos might affect 
these couplings appreciably. These effects have been extensively studied in  the literature, see for example
Refs.~\cite{Blum:2012ii,Espinosa:2012in}.    
In particular,  Ref.~\cite{Blum:2012ii} found that stops  with mass of order $\sim 250$~GeV imply a deviation in the higgs couplings of
order $r_{g} \equiv g_{hgg}/g_{hgg}^{SM}\approx 1.25$, triggering an order one deviation in the higgs gluon fusion production rate.
In general, the contributions of the stops to the gluon coupling in the small mixing limit 
is approximately given by the very well known formula
\beq
r_g - 1 \approx \frac{1}{4} \left( \frac{m_t^2}{m_{\tilde t_1}^2} + \frac{m_t^2}{m_{\tilde t_1}^2} -
\frac {X_t^2 m_t^2}{m_{\tilde t_1}^2 m_{\tilde t_2}^2}\right)~,
\eeq 
with $X_t \equiv  A_t - \mu \cot \beta$ being the left-right mixing between the stops. One can derive 
a very similar formula for $r_{\gamma}\equiv g_{h\gamma\gamma}/g_{h\gamma\gamma}^{SM}$. The modifications due to the stops are large as long as they are
light  (with mass around $\sim 200$~GeV) and have small mixing, while the effect rapidly decreases for larger masses and becomes  
negligible for $m_{\tilde t } \gtrsim 400$~GeV. 
Of course, exclusions based on these considerations are never 
completely robust because of the so-called ``funnel region" where the left-right mixing completely cancels out 
the contribution to the $hgg$ coupling, but this regime is also less interesting from the point of view of naturalness 
considerations. 

Although these modifications to the higgs couplings are interesting and helped until now to rule 
out certain SUSY scenarios, most of the parameter space with light stops has been already excluded
by the direct searches at the LHC. Higgs couplings are currently superior to direct searches only for the challenging case of compressed spectra. If such squeezed stops are the cause of higgs coupling deviations and escape detection at the LHC, the vastly superior sensitivity of a 100 TeV collider to weak scale colored states makes discovery extremely likely.  Precision high-energy measurements of DY production at a 100 TeV collider are also likely to model-independently detect such states via their effect on EW RG evolution~\cite{Alves:2014cda}. 

The charginos (winos and higgsinos) 
might also have an interesting effect on the $h \gamma \gamma$ coupling.
This effect is much more modest than that from stops, but it can still be important because of the challenges that the 
EWeakinos searches usually pose: small cross sections and relatively soft signatures.   This point has been first emphasized 
in~\cite{Djouadi:1996pb,Diaz:2004qt}. In practice, the effect becomes important only for relatively  small $\tan \beta$. 
This region of parameter space is somewhat disfavored by naturalness, at least if one restricts to more minimalistic
scenarios, but it can still be important if a larger degree of residual fine-tuning is tolerated. At $\tan \beta  \approx 1 $
the contribution to the higgs couplings can be approximated as
\beq
\label{eq:rgamma}
r_\gamma \approx  1 + 0.41 \frac{m_W^2}{M_2 \mu -m_W^2}
\eeq 
Practically for $\tan \beta \approx 1$ this value varies varies between $0.7$ and $1.13$, while for 
$\tan \beta > 2$ the allowed range further shrinks to $0.8 < r_\gamma < 1.1$.  The effect also decouples 
quickly with increasing gaugino mass. 
The precision of a $h\gamma \gamma$ coupling measurement might be below percent-level at a 100 TeV collider, 
see~\sssref{htogammagamma}, which corresponds to a limit $M_2 \mu \gtrsim (500 \ \gev)^2$ from Eq.~(\ref{eq:rgamma}). 
The generic EWino reach through direct production is in the TeV range or 
above~\cite{Low:2014cba, Gori:2014oua, Bramante:2014tba}, but it would be interesting to understand in which 
scenarios a $h\gamma \gamma$ coupling measurement could provide superior sensitivity.

Another effect, which has an important impact on the SUSY higgs sector, has to do with the fact that SUSY necessarily 
involves 2HDM of type~II. Moreover, if one insists on naturalness, the heavy higgses cannot be arbitrarily
heavy. Naturalness considerations imply upper bounds on their masses~\cite{Hall:2011aa,Gherghetta:2014xea,Katz:2014mba}, 
which are however much milder than those on stops or higgsinos.
In particular,  Ref.~\cite{Katz:2014mba} showed that in order not to exacerbate the fine-tuning of the supersymmetrized SM one 
would plausibly expect to see the heavy higgses of the 2HDM at masses of $1 - 3$~TeV. Needles to say, such a range of 
masses is far beyond the reach of the LHC, while it represents
a promising opportunity for a $100\,$TeV collider.
We will elaborate on the reach on these new states in Section~\ssref{bsmhiggs}, where
it will be also discussed in the more generic context of the 2HDM.    

\subsubsection{Composite Higgs}

\paragraph{Higgs  compositeness -- General Overview}
In the past decade a realistic framework has 
emerged~\cite{Contino:2003ve,Agashe:2004rs,Agashe:2005dk,Contino:2006qr,Panico:2005dh,Giudice:2007fh,
Barbieri:2007bh,Contino:2010rs}
(for a recent review see~\cite{Panico:2015jxa}) in which the Higgs
boson arises as a pseudo-Nambu-Goldstone Boson (pNGB) 
from the spontaneous breaking of a global symmetry $G\to G'$ of a new strongly interacting sector.
These theories have two crucial advantages over plain technicolor models. Firstly, the presence of a light Higgs boson allows 
a parametric separation between the $G\to G'$ breaking scale $f$ and the electroweak symmetry breaking scale $v$. This 
alleviates the tension of technicolor models with electroweak precision tests.
Secondly, the flavor problem of technicolor can be greatly improved by the implementation of partial compositeness~\cite{Kaplan:1991dc,Contino:2006nn}. 
The simplest realistic realization of the composite Higgs idea is represented by \sloppy\mbox{$G=SO(5)\times U(1)_X$} and $G'=SO(4)\times U(1)_X$ and called the Minimal Composite Higgs Model (MCHM)~\cite{Agashe:2005dk,Contino:2006qr}. The $U(1)_X$ factor is needed to obtain the correct hypercharge, $Y\equiv T_R^3+X$, for the SM fermions. This breaking pattern satisfies the conditions of a viable model: the SM vector bosons gauge a subgroup $SU(2)_L \times U(1)_Y\subset G$ and $G/G'$ contains an $SU(2)_L$ doublet which can be identified with the Higgs doublet. The coset space of the MCHM contains four NGBs transforming as a ${\bf 4}$ of $SO(4)$, three of which are eaten by the SM gauge bosons while the fourth is the physical Higgs boson.
Larger cosets give rise to more NGBs, including, for example, extra singlets and 
doublets~\cite{Gripaios:2009pe,Mrazek:2011iu,Chala:2012af}. 
Interestingly, an additional singlet could be interpreted as a Dark Matter candidate~\cite{Gripaios:2009pe,Frigerio:2012uc,Chala:2012af,Marzocca:2014msa,Fonseca:2015gva,Carmona:2015haa,Kim:2016jbz}.

At low energy, below the mass scale of the heavy resonances of the strong dynamics, the theory is described by an effective Lagrangian involving the 
composite Higgs doublet and the SM fields. Effects induced by the virtual exchange of the heavy modes are encoded by local operators whose relative importance 
can be estimated by assuming
a power counting. For example, under the assumption that the new strongly-coupled dynamics is described by a single mass scale $m_\ast$ and a single
coupling strength~$g_\ast$, the effective Lagrangian has the form~\cite{Giudice:2007fh}
\begin{equation} \label{LeffNDA}
\mathcal{L}_{\text{eff}} = \dst \frac{m_{\ast}^4}{g_{\ast}^{2}}\, \mathcal{L}\!\left(\frac{D_\mu}{m_{\ast}}, \frac{g_{\ast} H}{m_{\ast}}, \frac{\lambda \Psi}{m_{\ast}^{3/2}} \right)  \, .
\end{equation} 
One naively expects a typical coupling strength among the bound states of order $g \lesssim g_\ast \leq 4\pi$, where values $g_\ast < 4\pi$ allow a
perturbative expansion  in the effective theory.
The mass scale of the heavy resonances, $m_\ast$, represents the cutoff of the effective theory and sets its range of validity. 
Equation~(\ref{LeffNDA}) describes the low-energy dynamics of the light composite Higgs $H$ with elementary SM fermions $\Psi$, as first discussed in Ref.~\cite{Giudice:2007fh}.
The coupling $\lambda$ controls the strength of the interaction between the elementary and composite fermions, where the latter have been integrated out.
If the Higgs is strongly coupled, a simple yet crucial observation is that any additional power of $H$ costs a factor $g_\ast/m_\ast \equiv 1/f$,~\footnote{Note that a weakly-coupled, elementary Higgs would cost a factor $g/m_\ast$.} while any additional derivative is suppressed by a factor 
$1/m_\ast$. Note that extra powers of the gauge fields are also suppressed by $1/m_\ast$ as they can only appear through covariant derivatives in minimally coupled theories. If the light Higgs interacts strongly with the new dynamics, $g_\ast \gg 1$, then the leading corrections to low-energy observables arise from operators with extra powers of $H$ rather than derivatives. This remark greatly simplifies the list of important operators.

Composite Higgs models predict various new physics effects that can be probed at current and future colliders. In particular, new heavy vectors and fermions (the top partners) are expected and can be directly searched for. In addition, the composite nature of the pNGB Higgs implies deviations of the Higgs couplings from their SM value by an amount proportional to $\xi = v^2/f^2$, where $v$ is the scale of EWSB and $f$ the decay constant of the pNGB. In the MCHM, and more in general in theories
with coset $SO(5)/SO(4)$, the following prediction holds for the couplings of one and two Higgs bosons to gauge bosons:
\begin{equation}
\label{eq:pngbparameters}
\frac{g^{MCHM}_{hVV}}{g^{SM}_{hVV}}=\sqrt{1-\xi}\, , \qquad \frac{g^{MCHM}_{hhVV}}{g^{SM}_{hhVV}}=1-2\xi\, , 
\end{equation}
where $g^{SM}_{hVV}$ and $g^{SM}_{hhVV}$ represent the SM couplings,
while $g^{MCHM}_{hVV}$ and $g^{MCHM}_{hhVV}$ stand for the 
couplings in the MCHM. At low energy, virtual effects of the heavy resonances can be parametrized in terms of local operators, which
also lead to anomalous Higgs couplings (such as, for example, derivative interactions between the Higgs and gauge bosons). 
Precision measurements of the Higgs couplings thus constraint the compositeness scale and indirectly probe the heavy resonances.
In fact, direct and indirect measurements represent complementary strategies to test the parameter space of a composite Higgs models.

As discussed in detail in the Volume of this report dedicated to BSM physics~\cite{Golling:2016gvc},
the parameter space of the MCHM can be described by the mass of the heavy vectors, $m_\rho$, and their coupling strength, $g_\rho$ (to be identified with $m_\ast$ and $g_\ast$ respectively). These two parameters are related through the relation $\xi \sim g_\rho^2 v^2 / m_\rho^2$. In a recent study \cite{Thamm:2015zwa,Andreazza:2015bja}, the expected direct reach of a $100\,$TeV collider was compared to the indirect reach on $\xi$ of various lepton colliders. Indirect searches are sensitive to $\xi$ through precision measurements of the Higgs couplings: a high-luminosity upgrade of the LHC can probe values down to $\xi \ge 0.1$~\cite{CMS-NOTE-2012-006,ATL-PHYS-PUB-2014-016}, while lepton colliders like TLEP and CLIC are expected to reach the sub-percent level~\cite{Dawson:2013bba,Abramowicz:2013tzc}. Direct resonance searches for heavy vector particles at $100\,$TeV with $10\,$ab$^{-1}$, on the contrary, are sensitive to masses between $10$ and $20\,$TeV for coupling strengths between $g_\rho \sim 8$ and $2$. Masses up to $35\,$TeV become accessible for $g_\rho \lesssim 1.5$. Note that this corresponds to values of $\xi$ of the order $O(10^{-4})$. This illustrates the complementarity of the two searches strategies: indirect searches are more powerful for large couplings, while direct searches can access considerably larger masses for small coupling values.

\paragraph{Unnatural (or Split) Composite Higgs}
Composite Higgs models must satisfy a number of indirect constraints that arise from flavor and precision electroweak observables. While the precision electroweak constraints from the $T$ parameter are avoided with a custodial symmetry and those from the $S$ parameter are ameliorated with $g_\rho\gtrsim 3$, the most stringent constraints actually arise from flavor observables which gives rise to an approximate lower bound on the scale of spontaneous symmetry breaking $f \gtrsim 10$ TeV~\cite{Bellazzini:2014yua, Panico:2015jxa}. It is therefore clear that composite Higgs models require additional flavor structure in order to satisfy these constraints. Instead, if a more minimal approach is taken, one can simply assume that $f\gtrsim 10$ TeV. Of course this simplicity comes at the price of a tuning in the Higgs potential of order $v^2/f^2 \sim 10^{-4}$. This meso-tuning is still a many orders of magnitude improvement compared to that encountered in the Standard Model with a Planck scale cutoff and leads to an unnatural (or split) version of 
composite Higgs models.

Interestingly, even though the resonances are now very heavy these models can still give rise to distinctive experimental signals. 
The crucial requirement involves preserving gauge coupling unification due to a composite right-handed top quark~\cite{Agashe:2005vg}. The minimal coset preserving this one-loop result together with a discrete symmetry needed for proton and dark matter stability is $SU(7)/SU(6)\times U(1)$~\cite{Barnard:2014tla}. This coset contains twelve Nambu-Goldstone bosons, forming a complex $\bf 5$ containing the usual Higgs doublet, $H$, with a color triplet partner, $T$, and a complex singlet, $S$ which can be a stable dark matter candidate. In addition, the composite right-handed top quark, needed for gauge coupling unification, is part of a complete $SU(6)$ multiplet containing extra exotic states, $\chi^c$, that will be degenerate with the top quark. These states can be decoupled by pairing them with top companions, $\chi$ to form a Dirac mass of order $f$.

Interestingly the split compositeness can posses a rich non-minimal higgs sector. 
The unnatural or split spectrum will consist of the pseudo Nambu-Goldstone bosons, $H, T$ and $S$ with masses $\ll f$, which are split from the resonances with masses $> f$, while the top companions $\chi$ have Dirac masses of order $f$.  Thus for $f\gtrsim 10$ TeV, the color triplet partner $T$ of the Higgs doublet and the top companion states $\chi$, crucial for gauge coupling unification, will be accessible at a future 100 TeV collider.\footnote{In fact the top companions cannot be made arbitrarily heavy because this will worsen the unification and therefore there is an approximate upper bound $f\lesssim 500$ TeV~\cite{Barnard:2014tla}.}

In the model of Ref.~\cite{Barnard:2014tla}, the color triplet partner $T$ of the Higgs doublet will be the lightest colored state. 
Its dominant decay mode is $T\rightarrow t^c b^c S S$ which arises from a dimension-six term, where $t^c, b^c$ 
are Standard Model quarks and $S$ is the singlet scalar.  The decay length is given by
\beq
       c\tau = 0.6~\text{mm}\, \biggl( \frac{1}{c_3^T} \biggr)^2 \biggl( \frac{8}{g_\rho} \biggr)^3 \biggl( \frac{3~\text{TeV}}{m_T} \biggr)^5 \biggl( \frac{f}{10~\text{TeV}} \biggr)^4 \frac{1}{J(m_t, m_S)} \,,
\eeq
where $c_3^T$ is an order one constant, $m_T (m_S)$ is the color triplet (singlet scalar) mass and $J(m_t, m_S)$ is a phase 
space factor (see Ref.~\cite{Barnard:2015rba} for details). Thus, since the scale $f\gtrsim 10$ TeV, the color triplet is long-lived and can decay via displaced vertices or outside the detector.  This signal at a 100 TeV collider was analyzed in 
Ref.~\cite{Barnard:2015rba}. When $f=10$ TeV, displaced vertex and collider stable searches are sensitive to triplet masses in the range 3-10 TeV, while for heavier triplet masses, prompt decay searches are sensitive to color triplet masses in the range 4-7 TeV. For $f=100$ TeV there are no accessible prompt decays and the displaced vertex and collider stable searches can now cover color triplet masses up to 10 TeV. 
These results are depicted in Figure~\ref{fig:100TeV}.

\begin{figure}[h]
  \centering
  \begin{minipage}[b]{0.49\textwidth}
    \centering
  \includegraphics[width=\textwidth]{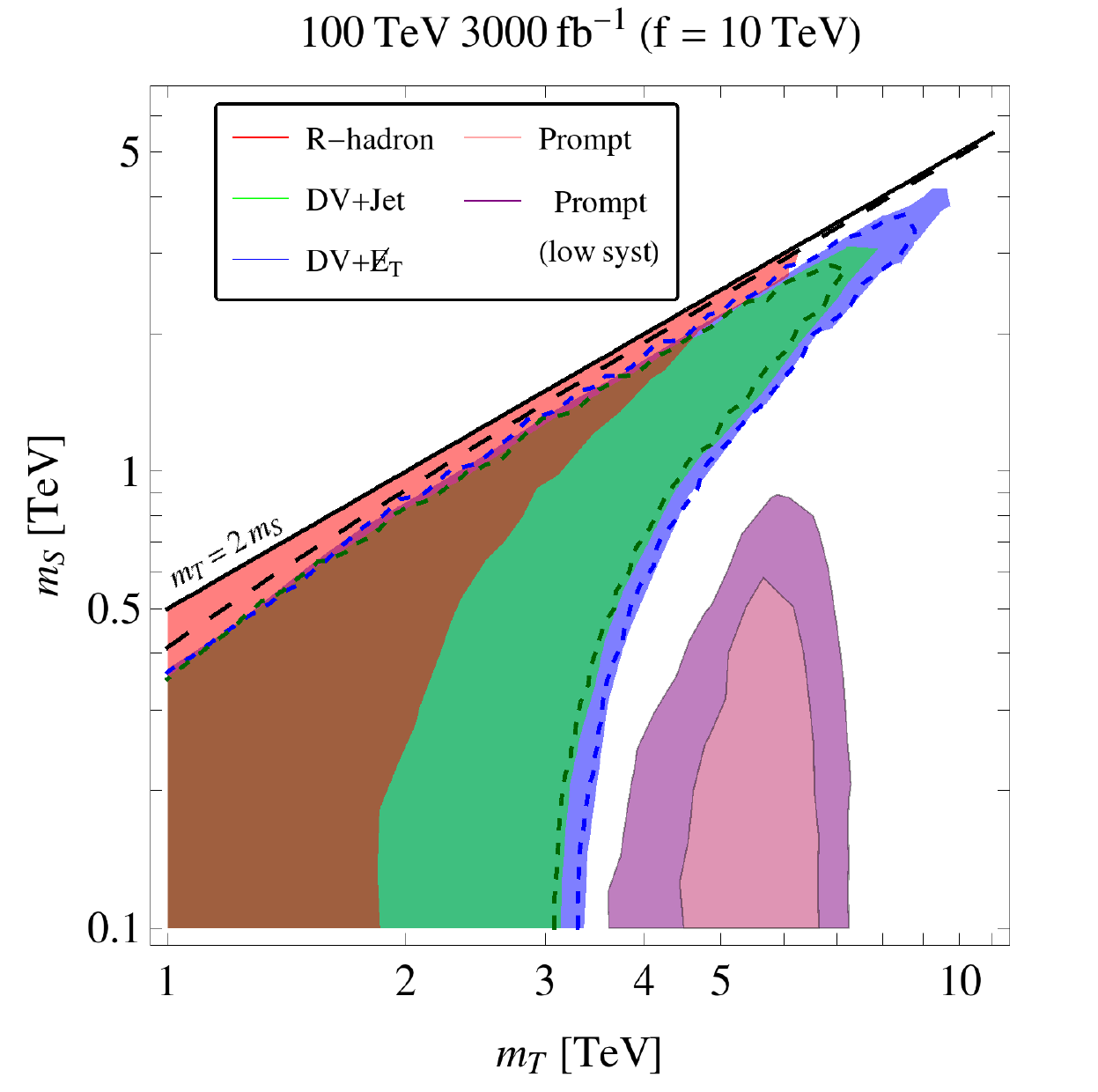}
  \end{minipage}
  \begin{minipage}[b]{0.49\textwidth}
    \centering
  \includegraphics[width=\textwidth]{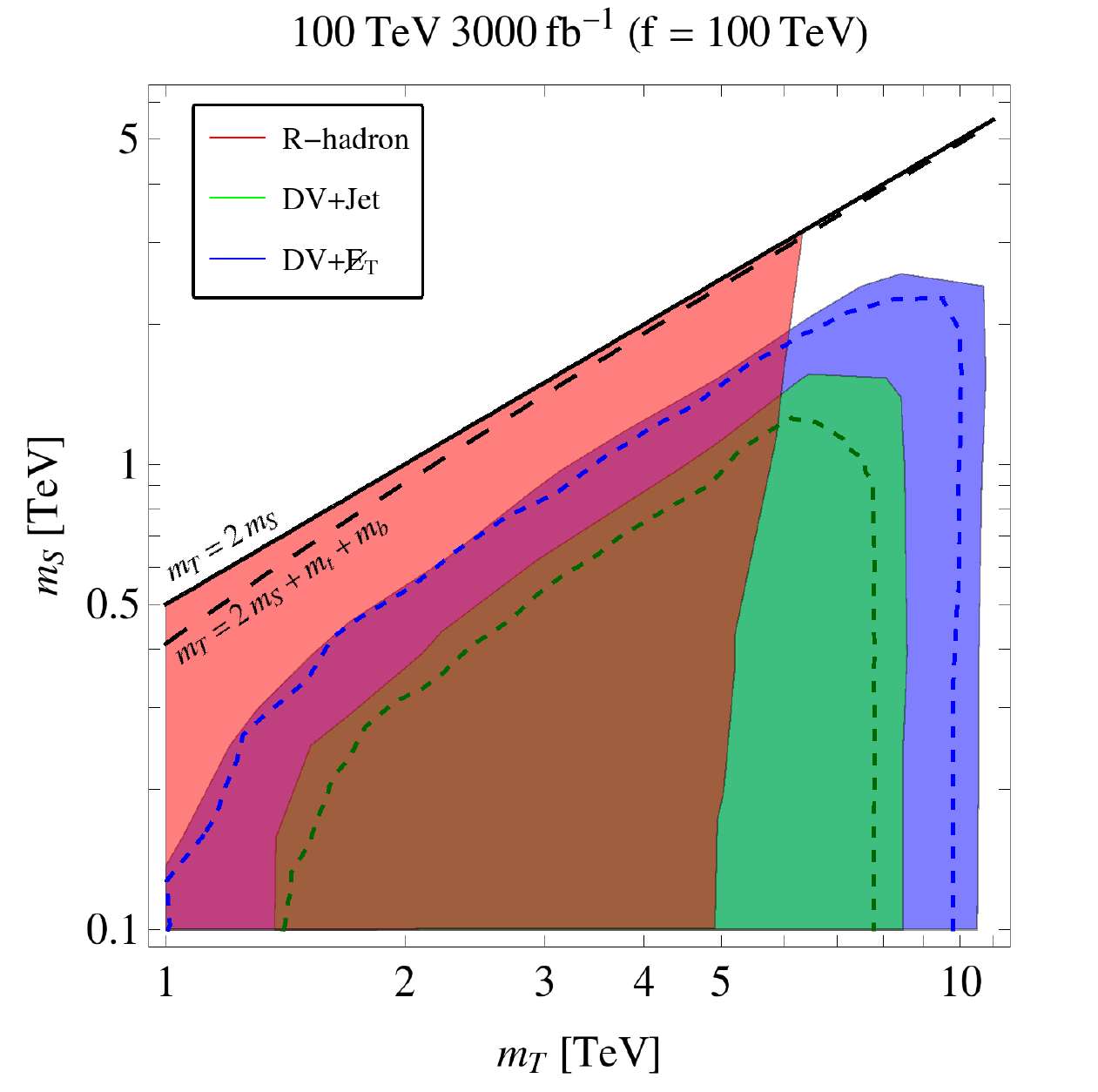}
  \end{minipage}
  \caption{Projections for a $100$~TeV collider with 3 ab$^{-1}$ of integrated luminosity as functions of the scalar mass $m_S$ and triplet mass $m_T$. The shaded regions show the $5\sigma$ discovery reach (95\% CLs exclusion limit) for the R-hadron/displaced (prompt) searches. The dashed lines include an additional factor of two reduction in the signal efficiency for DV searches to account for the impact of more stringent experimental cuts. The left and right panels correspond to $f=10$ and 100~TeV respectively. This figure is taken from Ref.~\cite{Barnard:2015rba}.} 
  \label{fig:100TeV}
\end{figure}

The top companions are the next heaviest states and have $SU(3)\times SU(2)\times U(1)_Y$ quantum numbers:
\beq
      \chi \equiv {\tilde q}^c\oplus {\tilde e}\oplus {\tilde d}^c\oplus {\tilde l} =  
      ({\bar{\bf 3}}, {\bf 2})_{-\frac{1}{6}}\oplus ({\bf 1},{\bf 1})_{-1}\oplus ({\bar{\bf 3}}, {\bf 1})_{\frac{1}{3}}\oplus ({\bf 1},{\bf 2})_{-\frac{1}{2}}
\eeq
These states are similar to excited Standard Model quarks and leptons and will decay promptly at a $100\,$TeV collider. Assuming a scale $f\gtrsim 10\,$TeV the top companions will have masses in the $10-20\,$TeV range. The colored top companions ${\tilde q}^c, {\tilde d}^c$ will have unsuppressed decays to quarks and the color triplet~$T$, whereas the SU(2) singlet, ${\tilde e}$ has a three-body decay into a bottom quark and two scalar triplets and the SU(2) doublet, ${\tilde l}$ decays to a quark, a color triplet and a scalar singlet~\cite{Barnard:2014tla}. A further study of top companion mass limits from these decays at a $100\,$TeV collider will be useful.

In summary, at a $100\,$TeV collider color triplet scalars give rise to distinctive signals and together with the top 
companions provide a smoking-gun signal for unnaturalness in composite Higgs models.

\subsubsection{Neutral Naturalness}
\label{sss.neutralnaturalness}
Here we briefly discuss the signatures of \emph{Neutral Naturalness}, with emphasis on the Higgs-related measurements most suited for a 100 TeV collider. We anchor the discussion by referring to two concrete benchmark models, \emph{Folded SUSY} (FSUSY) \cite{Burdman:2006tz} which features EW-charged scalar top partners, and the \emph{Twin Higgs} \cite{Chacko:2005pe} featuring SM-singlet fermionic top partners. However, some aspects of the phenomenology can also be derived more model-independently \cite{Craig:2013xia, Curtin:2015bka}. The phenomenology of Neutral Naturalness is rich, and includes potentially measurable Higgs coupling deviations, exotic Higgs decays, and direct production of partner states, additional Higgs states, or multi-TeV SM-charged states that are part of the more complete protective symmetry that ultimately underlies the model.

\paragraph{Theory overview}


In perturbative extensions of the SM, the hierarchy problem can be solved by introducing top partners. The coupling of these states to the Higgs is related to the top Yukawa by some symmetry (like supersymmetry, or a discrete symmetry in 
Little Higgs models~\cite{ArkaniHamed:2001nc,ArkaniHamed:2002pa,ArkaniHamed:2002qx,Schmaltz:2005ky}) 
such that the partner's quadratically divergent one-loop Higgs mass contribution cancels that of the top quark. In most 
theories, this symmetry ensures that the top partners carry SM color charge, allowing copious production at hadron colliders 
as long as their mass is in the natural $\lesssim \tev$ range. 

It is possible to devise concrete models in which the symmetry which protects the Higgs includes a discrete group 
like $\mathbb{Z}_2$, and does not commute with SM color. This leads to the possibility of color-neutral top partners. Such theories of \emph{Neutral Naturalness} feature very different phenomenology from  theories like the MSSM.

We anchor our discussion by referring to two archetypical theory examples of Neutral Naturalness. The first is 
\emph{Folded SUSY} (FSUSY)~\cite{Burdman:2006tz} which features a mirror sector of sparticles carrying SM 
EW quantum numbers. This mirror sector is charged under its own copy of QCD, which confines at a few -- 10 GeV. 
This leads to \emph{Hidden Valley} phenomenology~\cite{Strassler:2006im,Strassler:2006ri,Strassler:2006qa,Han:2007ae}: 
since LEP limits forbid light EW-charged particles below 
$\sim$ 100 GeV, the lightest new particles are mirror glueballs. Because the only interactions between the mirror sector
and the SM proceed via the EW-scale particles, the lifetimes of the glueballs are necessarily suppressed by powers 
of $\frac{\Lambda}{{\rm EW}}$, leading to the above mentioned hidden valley scenarios.  
Top partner loops couple the Higgs to mirror gluons, 
allowing for mirror glueball production in exotic Higgs decays, and displaced glueball decay via mixing with the SM-like Higgs. 

Our second theory benchmark is the \emph{Twin Higgs} \cite{Chacko:2005pe}, which features SM-singlet fermionic top partners. These are part of a mirror sector containing copies of all SM particles and gauge forces. The original mirror Twin Higgs model has several cosmological problems due to an  abundance of light invisible mirror states. One simple modification,
 which satisfies all cosmological constraints~\cite{Garcia:2015loa,Craig:2015xla}, 
 is the Fraternal Twin Higgs (FTH) ~\cite{Craig:2015pha}, which only duplicates the third generation in the mirror sector.  
 In that case, the hadrons of mirror QCD can be made up of mirror glueballs, mirror quarkonia, or a mixture of both.

A common feature of these theories is the existence of a \emph{mirror QCD} gauge group under which the top partners 
(and other fields in the mirror sector) are charged. From a top-down perspective this mirror QCD arises as a consequence 
of the discrete symmetry relating the SM and mirror sector. At some high scale, the mirror $g_S^B$ and $y_t^B$ are  
(almost) equal the SM $g_S^A, y_t^A$. From a bottom-up perspective~\cite{Craig:2015pha}, the existence of a mirror 
QCD force is expected, since otherwise $y_t^B$ would run very differently from $y_t^A$, ruining the cancellation between 
the top loop and the top partner loop in the Higgs mass at a scale of a few TeV. As we see below, this mirror QCD, and 
the associated low-energy hadron states, have important phenomenological consequences. Of course, the discrete 
symmetry usually has to be broken in some way (otherwise the two sectors would be identical), and it is possible to break 
mirror QCD as well~\cite{Poland:2008ev, Batell:2015aha}. In natural versions of these models, new SM-charged states 
appear at a few TeV, which is in line with the above bottom-up expectation.

There are several features of Neutral Naturalness that are even more model-independent. Top partners, by their very nature, 
have to couple to the SM through the Higgs-portal, which leads to loop-level deviations in the $Zh$ production cross section 
that is potentially detectable at future lepton colliders~\cite{Craig:2013xia}. Other possibilities include tree-level Higgs 
coupling deviations can also arise due to mixing effects, modifications of the Higgs cubic coupling due to top partner loops, electroweak precision observables~\cite{Curtin:2015bka}, and direct top partner production via off-shell Higgs 
bosons \cite{Curtin:2014jma, Craig:2014lda}. Crucially, it seems very challenging to construct a model of Neutral Naturalness which does not lead to detectable signatures at a 100 TeV collider, lepton colliders, or both~\cite{Curtin:2015bka}. This lends additional motivation for Higgs-related measurements at a 100 TeV collider.

Models of Neutral Naturalness solve the Little Hierarchy problem via a one-loop cancellation between top and top partner Higgs mass contributions. This cancellation is not enforced at two-loop order, necessitating a full solution to the hierarchy problem 
to become apparent in a UV completion at scales of $\sim 5-10 \tev$~\cite{
Batra:2008jy,Barbieri:2015lqa, Low:2015nqa, Geller:2014kta,
Craig:2013fga, Craig:2014fka, Chang:2006ra, Craig:2016kue, Harnik:2016koz}. In all known examples, these UV completions involve BSM states carrying 
SM charges, allowing for direct production at a 100 TeV collider. This will not be our focus here, but since the UV completion 
scale is connected to the degree of tuning in the theory, the ability of a 100 TeV collider to probe the full theory is 
complementary to Higgs-related measurements.


\paragraph{Higgs coupling deviations}

In the minimal Folded SUSY model, the electroweak and Higgs sectors are identical to the MSSM. Realistically there is significant uncertainty as to the exact structure of the scalar sector: the MSSM itself favors a light Higgs below 125 GeV, motivating extensions like non-decoupling $D$-terms \cite{Batra:2003nj}, while Folded SUSY needs some additional structure for viable electroweak symmetry breaking \cite{Cohen:2015gaa}. Even so, the required features of FSUSY imply the existence of additional Higgs bosons, leading to measurable Higgs coupling deviations if the decoupling limits is not satisfied. 
However, given that the naturalness limits, which are similar to the SUSY case, discussed in Sec.~\ref{sec:SUSY}, 
the deviations might potentially be too small not only for LHC, but also for the future leptonic collider. 


The electroweak charge of the top partners in FSUSY implies loop-corrections to  $\mathrm{Br}(h \to \gamma \gamma)$, which can be percent-level for top partner masses below 500 GeV \cite{Burdman:2014zta}. This represents a significant opportunity for lepton colliders \cite{Dawson:2013bba}, but the LHC or a 100 TeV collider is more likely to produce the EW-charged top partners directly than to see deviations in the diphoton rate.


In all known Twin Higgs models, a soft mass which breaks the discrete symmetry between SM- and mirror-Higgs has to be balanced against $f$, the vev of the mirror Higgs (or, more generally, the scale at which the higher symmetry which protects the light Higgs from quadratically divergent corrections is spontaneously broken) in order to achieve SM-like couplings of the 125 GeV Higgs to SM states. This leads to a tree-level tuning in the model, which is of order $v^2/f^2$, which is also the size of the mixing between the SM-like Higgs and the mirror Higgs. Therefore, in natural models where $v^2/f^2$ is not too small, there are detectable universal Higgs coupling deviations of the same order due to this tree-level mixing effect. These deviations can be detected at the percent-level at future lepton colliders \cite{Dawson:2013bba}, which are the smoking gun of Twin Higgs theories. Such a deviation, if detected at lepton colliders, would greatly motivate 100 TeV searches for additional signals of the Twin Higgs, such as SM-charged multi-TeV states.

In principle, it is possible to imagine Neutral Naturalness scenarios from a bottom-up perspective without measurable Higgs coupling deviations \cite{Curtin:2015bka}. However, avoiding this  smoking-gun-signature at lepton colliders comes at the cost of larger couplings in the hidden sector, reducing the required scale of the UV completion in the absence of strong tuning. The 100 TeV collider would then be able to produce the states of the UV completion directly, assuming they carry SM charge. This is the strongest demonstration of the important complementarity between the two types of possible future colliders.

\paragraph{Exotic Higgs decays}

Exotic Higgs decays are one of the best-motivated signatures of Neutral Naturalness. 
As outlined above,  most theories of Neutral Naturalness feature a mirror-QCD gauge symmetry under which the top partners are charged. The Higgs therefore couples to mirror-gluons via a top partner loop in analogy to its coupling to SM-gluons through the top loop. This means the Higgs can decay to mirror gluons, and therefore into light states in the mirror sector. The size of the exotic branching fraction is related to the top mass, and therefore to the naturalness of the theory itself, with less tuned scenarios giving higher exotic decay rates. The specific phenomenology of these exotic Higgs decays depends on the structure of the mirror QCD sector.

The mirror QCD sector could be SM-like, with quarks that are light compared to the confinement scale. This would allow the Higgs to decay to hidden pions and other hadrons, making it a classical Hidden Valley scenario \cite{Strassler:2006im, Strassler:2006ri, Strassler:2006qa, Han:2007ae}, realized for example by the original mirror Twin Higgs. If the exotic Higgs decays proceed through Yukawa couplings comparable to $y_b \sim 0.02$, exotic branching ratios could easily be as large as 10\%, which is detectable at the LHC and future colliders even if the decay products are detector-stable and hence invisible. On the other hand, if the decay proceeds through a top loop to mirror gluons only, the exotic branching fraction is of order $\lesssim 10^{-3}$ (see below). A high-luminosity lepton collider like TLEP/FCC-ee \cite{Gomez-Ceballos:2013zzn} is sensitive to exotic Higgs decays with branching fractions $\sim 10^{-5}$ ($10^{-3}$) if the decay products are very conspicuous with very little background (invisible or inconspicuous with sizable backgrounds). Therefore, depending on the detailed final state, discovering such a decay may be challenging even at lepton colliders.

In studying exotic Higgs decays, the clean environment of lepton colliders makes them superior if the final states  are not very distinctive, e.g. $\bar b b + \mathrm{MET}$ or only $\mathrm{MET}$. On the other hand, the huge Higgs production rates make the LHC and the 100 TeV collider vastly superior to lepton colliders when studying exotic Higgs decays with \emph{highly distinctive} final states, allowing access to much lower branching fractions. Few final states are more distinctive than long-lived particles that decay within the detector with measurable displacement from the interaction point. Neutral Naturalness strongly motivates exotic Higgs decays to displaced final states.

The simplest scenario is a mirror sector without any light matter charged under mirror QCD. This is guaranteed in FSUSY, where LEP limits constrain the mass of the EW-charged mirror sector. It can occur in Fraternal Twin Higgs scenarios, if the mirror $b$-quark is not too light. In that case, the mirror hadrons are \emph{glueballs}. There are $\sim 12$ stable glueball states in pure $SU(3)$ gauge theory \cite{Morningstar:1999rf}, with masses ranging from $m_0 \approx 7 \Lambda_\mathrm{QCD}^B$ for the lightest $G_0 \equiv 0^{++}$ state, to $\sim 2.7 m_0$ for the heaviest state. In the presence of top partners much 
heavier than $\Lambda_\mathrm{QCD}^B$, some of these states can decay via a top partner loop to SM particles via 
an off-shell Higgs boson~\cite{Juknevich:2009gg}. We concentrate on the $0^{++}$ state, since it has one of the shortest lifetimes and, as the lightest glueball, presumably produced commonly by mirror-hadronization.\footnote{Thermal estimates \cite{JuknevichPhD} suggest that roughly half of all produced glueballs are in the $0^{++}$ state, but given our ignorance about pure-gauge hadronization, this estimate is highly uncertain.}
For $m_0 \gtrsim 2 m_b$ (in the FTH case) 
and top partner mass $m_T$, the decay length of $G_0$ is approximately
\begin{eqnarray}
\label{e.glueballlifetime}
c \tau \ &\approx& \  \left( \frac{m_T}{400 \gev}\right)^4  \ \left(\frac{20 \gev}{m_0}\right)^7 \times
\left\{ 
\begin{array}{lll}
\displaystyle 
(35 \mathrm{cm}) & \ \ \ \ \ & \mbox{[FSUSY]}\\
\displaystyle (8.8 \mathrm{cm}) & \ \ \ \ \ & \mbox{[FTH]}
\end{array}
\right.
\end{eqnarray}
where we assume $m_T \gg m_t/2$ for FTH and degenerate unmixed stops for FSUSY. For natural theories, these decay lengths are in the $\mu$m - km decay range. This leads to  displaced signals at hadron colliders, with some glueballs decaying dominantly to $3^\mathrm{rd}$ generation SM fermion pairs.

\begin{figure}
\begin{center}
\includegraphics[width=9cm]{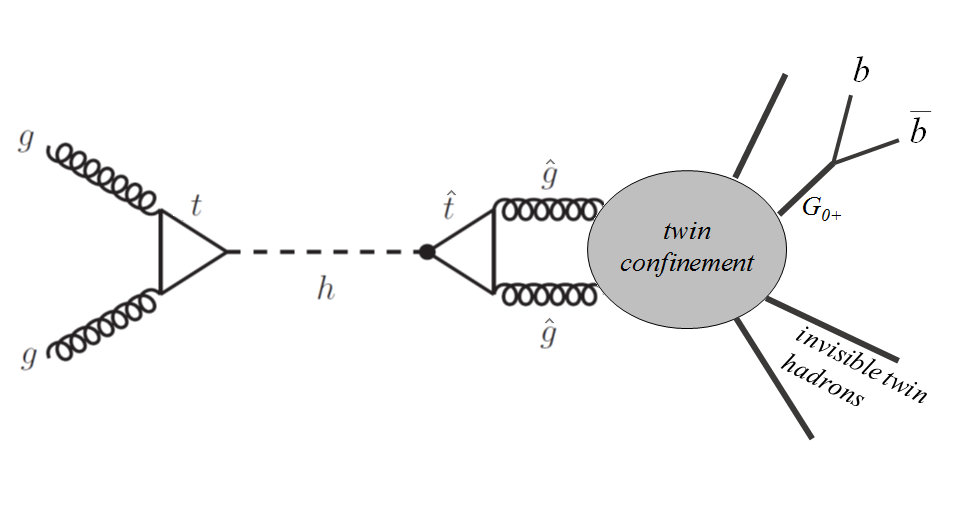}
\end{center}
\caption{Production of mirror hadrons in exotic Higgs decays, and their decay back to the SM, in the Fraternal Twin Higgs model. Figure from \cite{Craig:2015pha}.}
\label{f.mirrrorhadronproduction}
\end{figure}

\begin{figure}
\begin{center}
\hspace*{-10mm}
\begin{tabular}{ccc}
\includegraphics[height=6.6cm]{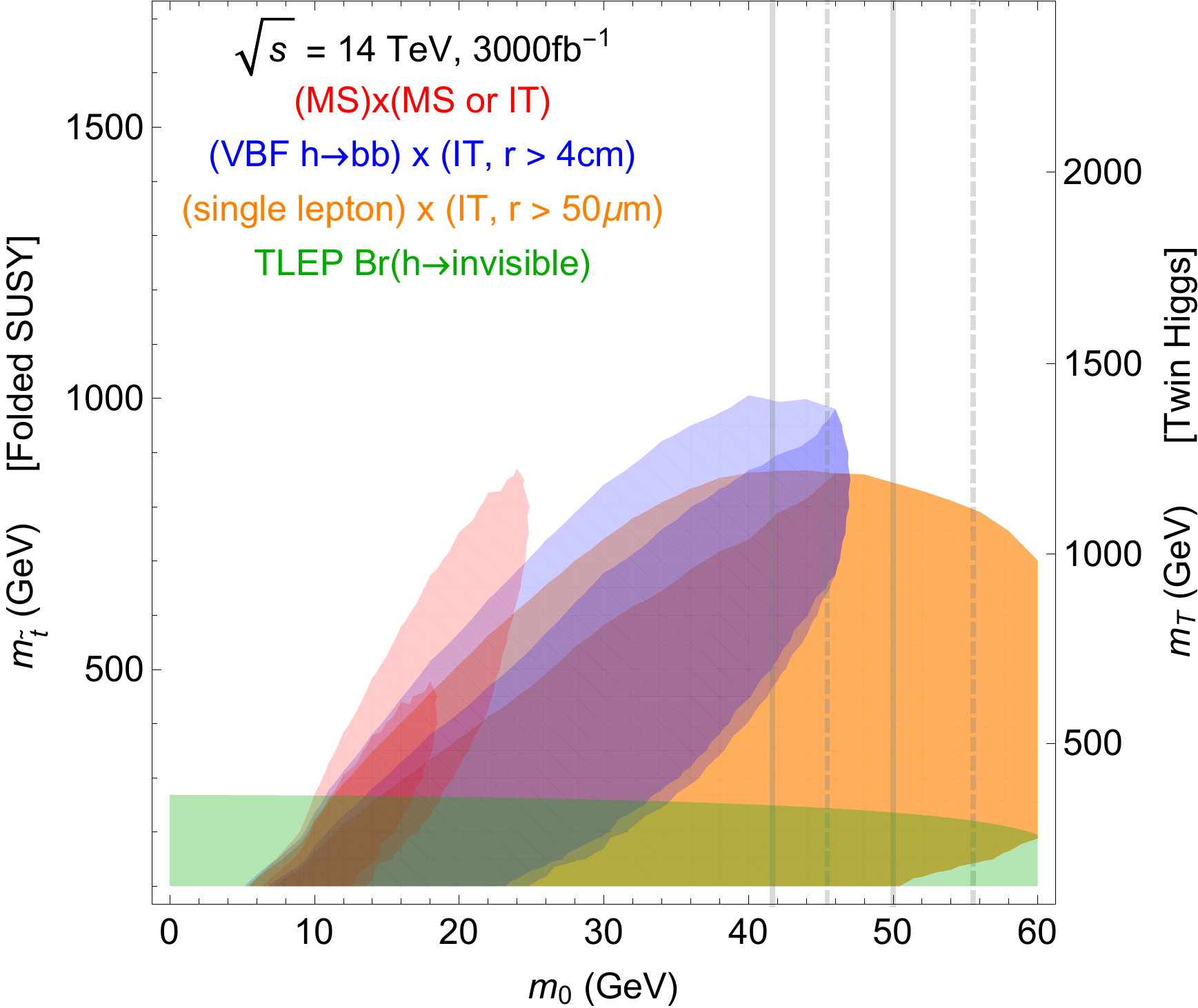} 
& &
\includegraphics[height=6.6cm]{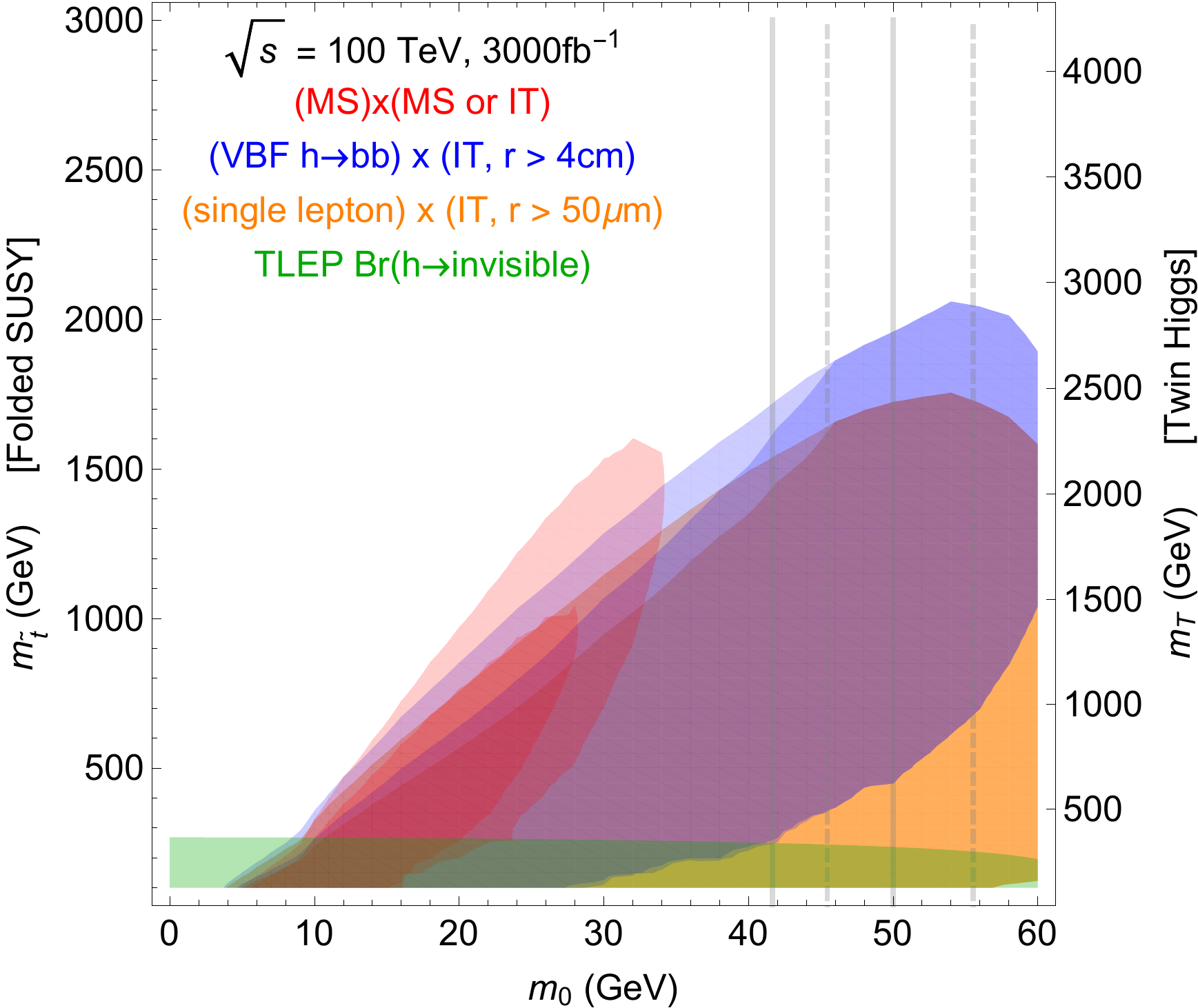} 
\end{tabular}
\end{center}
\caption{Summary of discovery potential for the simplified model of Neutral Naturalness with light mirror glueballs at LHC14 and 100 TeV collider with 3000$\ifb$, from looking for (i) one DV in the muon system (MS) and one additional DV in either the MS or the inner tracker (IT), (ii) one DV at least 4 cm from beam line and VBF jets (blue) and (iii) one DV with at least $50 \mu$m from beam line and a single lepton (orange). Assuming negligible backgrounds and 10 events for discovery. See \cite{Curtin:2015fna} for details. Note different scaling of vertical axes.  For comparison, the inclusive TLEP $h\to \mathrm{invisible}$ limit, as applied to the perturbative prediction for $\mathrm{Br}(h\to \mathrm{all\ glueballs})$, is shown for future searches as well, which serves as a pessimistic estimate of TLEP sensitivity. Lighter and darker shading correspond to the optimistic (pessimistic) estimates  of exclusive $0^{++}$ yield, under the assumption that $h$ decays dominantly to two glueballs. Effect of glueball lifetime uncertainty is small and not shown. $m_0$ is the mass of the lightest glueball $G_0$; the vertical axes correspond to mirror stop mass in FSUSY and mirror top mass in FTH and Quirky Little Higgs. Vertical solid (dashed) lines show where $\kappa$ might be enhanced (suppressed) due to non-perturbative mixing effects \cite{Craig:2015pha}.
}
\label{f.summaryplot}
\end{figure}

RG arguments favor masses for $G_0$ in the $\sim 10 - 60 \gev$ range (though model-dependent effects can easily shift that range) \cite{Curtin:2015fna}. The same top partner loop which allows glueballs to decay also allows the Higgs boson to decay to mirror gluons, which then hadronize to yield mirror glueballs, see \fref{mirrrorhadronproduction}. 
The rate for \emph{inclusive} production of mirror glueballs  from exotic Higgs decays can be estimated by rescaling $\mathrm{Br}(h \to \mathrm{gluons}) \sim 8\%$ in the SM:
\begin{equation}
\label{e.Brhgginclusive}
\mathrm{Br}(h \to \mathrm{mirror\ glue}) \  \approx \ 10^{-3} \ \left(\frac{400 \gev}{m_T}\right)^4 \  \times  \ 
\left\{ 
\begin{array}{lll}
\displaystyle 
1 & \ \ \ \ \ & \mbox{[FSUSY]}\\
\displaystyle 4 & \ \ \ \ \ & \mbox{[FTH]}
\end{array}
\right.
\end{equation}
Non-perturbative or RG effects on the mirror QCD coupling can change this by an order 1 factor~\cite{Craig:2015pha, Curtin:2015fna}. The \emph{exclusive} branching fraction to the unstable $0^{++}$ glueballs can be parameterized as the above inclusive branching fraction multiplied by a nuisance factor $\kappa$ which ranges from $\sim 0.1 - 1$ when there is phase space available to produce heavier glueballs. This simple approach was used by the authors of \cite{Curtin:2015fna}, who estimated the reach of various displaced searches at the LHC using ATLAS reconstruction efficiencies for displaced vertices (DV). A very conservative estimate of 100 TeV reach was derived by simply repeating the analysis for higher energy. Top partner mass reach, which is $\sim 1.5 \tev$ for TH top partners at the HL-LHC and $\sim 3 \tev$ at the 100 TeV collider with $3 \mathrm{ab}^{-1}$ of luminosity, is shown in \fref{summaryplot} as a function of glueball mass. Light shading indicates estimated uncertainties due to unknown details of mirror hadronization. Different search strategies are required to cover the entire range of glueball masses and lifetimes, most importantly searches for single displaced vertices together with VBF jets of leptons from Higgs production (see \cite{Curtin:2015fna} for details, and \cite{Csaki:2015fba} for other possibilities involving displaced triggers). This sensitivity projection is highly conservative, since the exotic Higgs decay was assumed to be 2-body, which underestimates the displaced vertex signal yield for light glueballs. Even so, the reach is impressive and provides good coverage for the natural regime of these theories. It also underlines that the detector of a 100 TeV collider needs to be able to reconstruct soft and displaced objects stemming from Higgs decays to maximize its potential for new physics discoveries. The reach could also be dramatically improved if a future detector could trigger on displaced decays, or indeed operate without a trigger.

Another possible scenario is a mirror QCD sector containing only light mirror bottom quarks $B$. This is one possible outcome of the Fraternal Twin Higgs (see \cite{Craig:2015pha} for a detailed discussion). If the mirror bottom quark mass satisfies $m_0 \lesssim  m_B < m_h/2$, then the Higgs can decay to $\bar B B$, which forms a mirror bottomonium state and annihilates into mirror glueballs. This enhances the inclusive twin glueball rate to 
\begin{equation}
\label{e.BrhBB}
\mathrm{Br}(h \to \mathrm{\mathrm{mirror} \ glue})  \approx \mathrm{Br}(h \to \bar B B) \  \approx  \ 0.15 \  \left( \frac{m_{B}}{12 \gev}\right)^2 \ \left( \frac{400 \gev}{m_T}\right)^4  
\ \ \ \   , \ \ (m_B<m_h/2) \ .
\end{equation}
which can be as large as $\sim 10\%$. 

Alternatively, in the FTH, exotic Higgs decays can produce long-lived mirror bottomonium states if they are the 
lightest mirror hadrons. The rate is equal to that shown in \eref{BrhBB}. The bottomonium spectrum also contains an unstable $0^{++}$ state that decays via the top partner loop.  The lifetime of this  state is 
\begin{eqnarray}
\label{e.GammaBottomonium}
\Gamma_{\chi\to YY} & \sim & 2\times 10^{-3} \left( \frac{v}{f}\right)^4 \frac{m_{\chi}^{11/3} m_0^{10/3}}{v^2 m_h (m_h^2 - m_{\chi}^2)^2} \Gamma_{h \to YY}(m_h) \ ,
\end{eqnarray}
assuming there are no light twin neutrinos which could short-curcuit this decay mode. This leads to similar phenomenology as the pure glueball scenario described above, however in this case the lifetime can depend very differently on the $0^{++}$ mass, which motivates the search for relatively short decay lengths $\sim 10 \mu\mathrm{m}$ even for unstable particle masses near 15 GeV.

Finally, it is important to point out that in all of the above scenarios, the lifetime can be shorter than $\sim 10~\mu$m (for 
very heavy glueballs with light top partners, or for bottomonia), which motivates searches for non-displaced $b$-rich final 
states of exotic Higgs decays.

\paragraph{Direct production of top partners}

The Higgs portal guarantees that neutral top partners can be produced at the 100 TeV colliders via an intermediate off-shell Higgs boson. Measurement of the top partner masses or couplings via direct production could serve as a powerful diagnostic of Neutral Naturalness to distinguish it from generic Hidden Valleys, which can also lead to displaced exotic Higgs decays.


Let us first consider the FTH scenario. In that case, any produced top partner pair will quickly mirror-beta-decay to mirror-bottoms, which then either decay to light glueballs or bottomonia, leading to displaced vertices in the detector. 
Higgs portal direct production, however, has a very low cross section, making direct production measurements unfeasible at the LHC for top partners heavier than a few 100 GeV \cite{Chacko:2015fbc}. The 100 TeV collider with $3 \mathrm{ab}^{-1}$, on the other hand, will produce hundreds of top partner pairs with potentially multiple displaced vertices, even if $m_T = 1 \tev$ \cite{Curtin:2015bka}.

Top partner direct production is most spectacular for EW-charged partners, as for FSUSY. In that case, DY-like pair production leads to large signal rates. Since there is no light mirror-QCD-charged matter, the top partner pairs form a quirky bound state  which de-excites via soft glueball and photon emission \cite{Kang:2008ea,Burdman:2008ek,Harnik:2008ax,Harnik:2011mv} before annihilating dominantly into mirror gluon jets, which hadronize to mirror glueballs \cite{Chacko:2015fbc}. This can lead to spectacular ``emerging jet'' \cite{Schwaller:2015gea} type signatures with many displaced vertices for top partner masses in the TeV  range at the LHC and multi-TeV range at the 100 TeV collider. This has been recently studied in \cite{Chacko:2015fbc}, which also addresses how to parameterize the unknown hadronization of the pure-glue mirror sector. It is shown that exotic Higgs decays and direct top partner pair production have complementary sensitivity to EW-charged partners in different regions of parameter space. At a 100 TeV collider, the mass reach provided by both channels extends to several TeV or more.

The most challenging scenario is a scenario with neutral top partners but without mirror QCD. In that case, top partner production proceeds through the Higgs portal but without producing displaced vertices in the final state. This has been studied by \cite{Curtin:2014jma, Craig:2014lda}, which found that $j j + \mathrm{MET}$ searches for VBF $h^* \to \bar T T$ production was the most promising channel. Even so, the production rate is so low that now meaningful bounds are derived at the LHC. A 100 TeV collider is necessary to achieve sensitivity to top partner masses of a few 100 GeV, which is the naturally preferred range. 

In all of the above cases, direct production of neutral top partners is complementary to direct production of SM-charged BSM states at the 5-10 TeV scale, which are expected in the UV completion of Neutral Naturalness theories. The latter was recently explored in \cite{Cheng:2015buv}, which found reach for masses up to about 10 TeV.  Both direct production measurements generally  require a 100 TeV collider (especially for the Twin Higgs) and would provide valuable information on the structure of the theory.





\subsection{BSM Higgs Sectors}
\label{ss.bsmhiggs}

In this section we will overview the direct reach for the new BSM higgs states at 100~TeV machine. 
We first review the two-Higgs doublet model (2HDM), which is relevant both as a standalone scenario and as an integral part of SUSY, models that explain the neutrino mass, etc. There have been a variety of studies studying the mass reach of a 100 TeV colliders to probe these new states by direct production. 
Singlet extensions are another highly relevant scenario, which occur in models of Neutral Naturalness, electroweak baryogenesis, and supersymmetry as the NMSSM and its derivatives, and we summarize recent work on the reach of future colliders for these states. 
Of course, this exploration of possibilities for BSM scalar sectors is not exhaustive, covering two of the most representative classes of scenarios. Particular examples of higgs triplet models are studied in the context of generating the neutrino mass in \ssref{neutrinos}, and more work on general extensions at 100 TeV is needed.

\subsubsection{Two-Higgs Doublet Models}
\label{ss.2hdm}

\paragraph{Higgs couplings}

Two-doublet models are one of the most common extensions of the SM Higgs sector and are naturally realised in 
supersymmetry. Besides the ordinary Goldstones eaten by the gauge bosons, such models describe two CP-even ($h^0$ and $H^0$), one CP-odd ($A^0$) and one charged ($H^\pm$) physical states. Let us consider, for simplicity, the type-II 
structure that arises in supersymmetry. 

The physical content of the models can be described in terms of two angles. The angle $\beta$, which defines the direction of the Higgs vev in the plane of the two neutral CP-even current eigenstates (usually denoted by $H_d^0$ and $H_u^0$). And the angle $\alpha$, which defines the direction of the lightest CP-even state ($h^0$) in the same plane. Following the usual convention for the definition of these angles, the directions of the Higgs vev and $h^0$ coincide when $\beta - \alpha =\pi /2$. In the literature, the condition that these two directions coincide is usually referred to as {\it alignment}. When alignment holds, $h^0$ behaves as the SM Higgs and the orthogonal state does not participate in the process of EW breaking.

Present LHC measurements tell us that the observed Higgs boson is SM-like at the level of about 20--30\% while the full LHC program will be able to make tests in the range between a few to 10\%. Assuming that no deviation is observed, it is a good starting point to take the Higgs as approximately SM-like. This situation is automatically obtained in the limit in which we take all new states ($H^0$, $A^0$, $H^\pm$) much heavier than $h^0$, which corresponds to $m_A\to \infty$ and which is usually referred to as the {\it decoupling limit}.

Since the Higgs couplings can be written in terms of the angles $\alpha$ and $\beta$, we can easily obtain their expression in the decoupling limit ($m_A\to \infty$)
\beqa
g_{hVV} &=& \sin (\beta-\alpha) \approx 1-\frac{2\, (1-t^{-2})^2}{t^2\, (1+t^{-2})^4}\left( \frac{m_Z}{m_A}\right)^4 
\label{coupH1}\\
g_{htt} &=& \frac{\cos\alpha}{\sin \beta} \approx 1-\frac{2\, (1-t^{-2})}{t^2\, (1+t^{-2})^2}\left( \frac{m_Z}{m_A}\right)^2 \label{coupH2}\\
g_{hbb} &=& -\frac{\sin\alpha}{\cos \beta} \approx 1+\frac{2\, (1-t^{-2})}{ (1+t^{-2})^2}\left( \frac{m_Z}{m_A}\right)^2\, ,
\label{coupH3}
\eeqa
where $t\equiv \tan\beta$ and $g_{hXX}$ denote the couplings of the SM-like Higgs to weak gauge bosons ($X=V$), top ($X=t$), and bottom ($X=b$), in units of the corresponding SM couplings. Equations~(\ref{coupH1})--(\ref{coupH3}) show explicitly how $h$ behaves exactly as the SM Higgs in the decoupling limit.
Note that the $h$ coupling to gauge bosons becomes SM-like very rapidly, since deviations scale as $(m_Z/m_A)^4$. The convergence becomes even more rapid for large $\tan\beta$, since deviations scale as $1/t^2$. 

To summarise: decoupling implies alignment, since $m_A\to \infty$ implies $\beta - \alpha \to\pi /2$. Moreover, decoupling implies a special pattern of deviations from the SM predictions
\beqa
\delta g_{hVV} &\approx& 0.02\% \, \left( \frac{10}{t}\right)^2 \left( \frac{300\, {\rm GeV}}{m_A}\right)^4 \\
\delta g_{htt} &\approx& 0.2\% \, \left( \frac{10}{t}\right)^2 \left( \frac{300\, {\rm GeV}}{m_A}\right)^2 \\
\delta g_{hbb} &\approx& 18\% \,  \left( \frac{300\, {\rm GeV}}{m_A}\right)^2 \, .
\eeqa

In the decoupling limit, the couplings of the heavy states are also simply determined. For instance, for the heavy CP-even 
state $H^0$, one finds
\beq
g_{HVV} = \cos (\beta-\alpha) \approx -\frac{2}{t}\left( \frac{m_Z}{m_A}\right)^2 \, , ~~~
g_{Htt}  \approx -\frac{1}{t} \, , ~~~
g_{Hbb} \approx t \, .
\label{coupH4}
\eeq
This means that the coupling of a single heavy Higgs to gauge bosons vanishes in the decoupling limit. 
Moreover, at large $\tan\beta$, the coupling to bottom quarks is enhanced. The production of heavy Higgses is dominated by $gg \to H^0/A^0$ via loops of top or bottom quarks, $b \bar b \to H^0/A^0$, or associated production with $H^0/A^0$ emitted from a top or bottom quark line.

If the observed Higgs boson is confirmed to be nearly SM-like, we must conclude that the alignment condition is approximately satisfied. We have shown that decoupling implies alignment. So one may now wonder: does alignment imply decoupling? The answer is no. 
In a general two-Higgs doublet model it is possible to satisfy $\beta -\alpha \approx \pi /2$, while still keeping light the new 
Higgs states~\cite{Gunion:2002zf,Ginzburg:2004vp,Craig:2013hca,Dev:2014yca}. Although this cannot be achieved in the most minimal version of supersymmetric models, alignment without decoupling can occur in supersymmetry with new singlet or triplet states~\cite{Delgado:2013zfa} or for extreme values of radiative corrections~\cite{Carena:2013ooa}.

As in the case of decoupling, also {\it alignment without decoupling} predicts a well-defined pattern of Higgs couplings. Using $\epsilon \equiv  \cos(\beta -\alpha)\tan\beta$ as expansion parameter, one finds
\beq
g_{hVV} \approx 1-\frac{\epsilon^2}{2t^2}\, , ~~~
g_{htt} \approx 1+\frac{\epsilon}{t^2}\, , ~~~
g_{hbb} \approx 1-\epsilon \, ,
\eeq
\beq
g_{HVV} \approx \frac{\epsilon}{t}\, , ~~~
g_{Htt} \approx -\frac{1}{t}(1-\epsilon)\, , ~~~
g_{Hbb} \approx t\left( 1+\frac{\epsilon}{t^2}\right) \, .
\eeq
Comparing this result with eqs.~(\ref{coupH1})--(\ref{coupH3}) and (\ref{coupH4}), one finds the same result as decoupling once we identify $\epsilon \to -2(m_Z/m_A)^2$. Yet, the two cases do not lead to the same phenomenological predictions, since in alignment without decoupling $\epsilon$ can take either sign.

To summarise, there are two cases in which a two-Higgs doublet model can predict a nearly SM-like Higgs boson, such that Higgs coupling measurements can be satisfied to an arbitrary degree of precision. The two cases are decoupling and alignment without decoupling. Each case leads to a well-defined pattern for the couplings of the light and heavy Higgses, which can be expressed in terms of two parameters: $m_A$ and $\tan\beta$ for decoupling, or $\cos(\beta -\alpha)$ and $\tan\beta$ for alignment without decoupling. The important difference, from the phenomenological point of view, is that the new Higgs bosons must be heavy for decoupling, but can be arbitrarily light for alignment without decoupling. Furthermore, while standard decay channels of heavy Higgs bosons, like $H/A \to WW, ZZ$, $A \to Zh$, $H \to hh$, $H^\pm \to Wh$ are important in the decoupling limit, they become strongly suppressed in the alignment limit, where decays to SM fermions or photons become dominant \cite{Craig:2015jba}. This significantly affects the strategy of direct searches, as discussed below.




\paragraph{Direct Searches for Heavy Higgs Bosons in 2HDM}

\begin{figure}[h!]
\centering
  \includegraphics[width=0.4\textwidth]{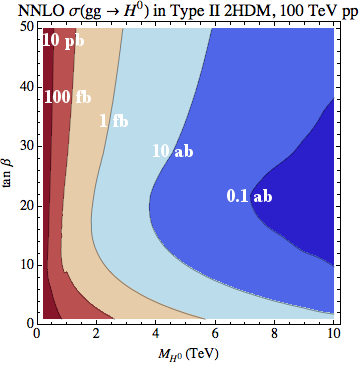}
  \includegraphics[width=0.4\textwidth]{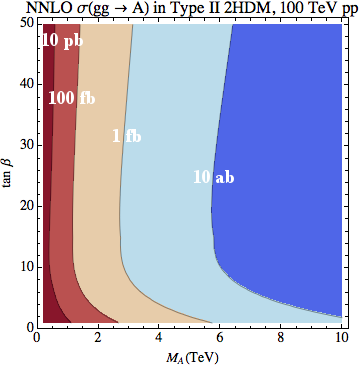}
  \includegraphics[width=0.4\textwidth]{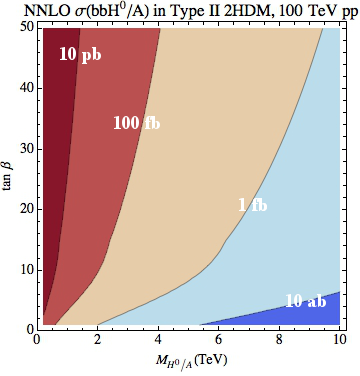}
  \includegraphics[width=0.4\textwidth]{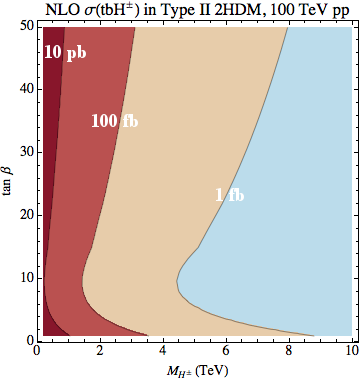}
 \caption{Dominant production cross sections for non-SM like Higgses in the Type II 2HDM at the 100 TeV $pp$ collider: NNLO cross section for $gg\rightarrow H^0$ or $A$ (top left and top right panel, calculated using HIGLU \cite{Spira:1995mt} with the NNPDF2.3 parton distribution functions \cite{Ball:2012cx}),   NNLO cross section for bottom-associated production $bbH^0/A$ (lower left panel, calculated using SusHi \cite{Harlander:2012pb, Harlander:2003ai, Harlander:2002wh}.  $bbH^0$ and $bbA$ cross sections are the same in the alignment limit),  NLO cross section for $tbH^\pm$ (lower right panel, calculated in Prospino \cite{Plehn:2002vy, Beenakker:1996ed}). }
\label{fig:CS}
\end{figure}

At the 100 TeV $pp$ collider the heavy Higgses of the 2HDM are dominantly produced via gluon fusion $gg \rightarrow H^0/A$, 
with dominant top and bottom (for large $\tan\beta$) loops, as well as $bbH^0/A$ associated production. 
$ttH^0/A$ associated production could be important as well. The dominant production process for the charged Higgses is 
$ tb H^\pm$ associated production. In Fig.~\ref{fig:CS} we show the production cross sections for $A$, $H^0$ and 
$H^\pm$ at 100 TeV $pp$ collider in the Type II 2HDM with $\cos (\beta - \alpha) = 0$. For neutral Higgses, gluon 
fusion production dominates at low $\tan\beta$ while $bb H^0/A$ associated production dominates at large $\tan\beta$.   
The $tb H^\pm $ production cross section   gets enhanced at both small and large $\tan\beta$. 

Comparing to the 14~TeV LHC, the production rates can be enhanced by about a factor of 30~$-$~50 for gluon fusion and 
$bb$ associated production, and about a factor of 90 for the charged Higgs for Higgs mass of about 500~GeV, 
and even more for heavier Higgses,  resulting in great discovery potential for heavy Higgses at a 100 TeV $pp$ colliders.

At the LHC, most of the current searches for non-SM neutral Higgs bosons  focus on the conventional Higgs search channels 
with a $WW$, $ZZ$, $\gamma\gamma$, $\tau\tau$ and $bb$ final 
state~\cite{Aad:2014vgg,CMS:2014cdp,Chatrchyan:2013qga,ATLAS-CONF-2013-027,Khachatryan:2014jya,ATLAS-CONF-2013-090,CMS-PAS-HIG-13-021}. These decays modes are characteristic to the 2HDM in the decoupling
limit, where we in general do not expect big splittings between the various heavy Higgses.
Typically, the production rate of the extra Higgses is suppressed, compared to that of the SM Higgs, either due to its 
larger mass or its suppressed couplings to the SM particles. The decay of the heavy neutral Higgses to the $WW$ and $ZZ$ 
final states, which provided a large sensitivity for SM Higgs searches, is absent for the CP-odd Higgs, and could be highly suppressed for the non-SM like CP-even Higgs, especially in the alignment limit \cite{Craig:2015jba}, in which case SM fermion and photon final states become more important. 
The decay modes into $\tau\tau$ or $bb$ suffer from either suppressed branching fraction once the top pair mode opens up or large SM backgrounds, and are therefore only relevant for regions of the parameter space with an enhanced $bb$ or $\tau\tau$ coupling.  If the non-SM neutral Higgs is heavy enough, the decay mode into top pairs becomes important. However, when the Higgs is produced in gluon fusion,
such decay suffers from large $t\overline{t}$ background and a possible destructive interference with the 
SM background~\cite{Dicus:1994bm, Craig:2015jba}.   
Direct searches for charged Higgs bosons are even more difficult at the LHC. For $m_{H^\pm}>m_t$, the cross section 
for the dominant production channel of $tbH^\pm$ is typically small. The dominant decay mode 
$H^\pm \rightarrow tb$  is hard to identify given the large $tt$ and $ttbb$ background, while the subdominant decay 
of $H^\pm \rightarrow \tau\nu$ has suppressed branching fraction. In the MSSM, even at the end of the LHC running, 
there is a ``wedge region''~\cite{Dawson:2013bba} in the $m_A-\tan\beta$ plane for $\tan\beta\sim 7$ and 
$m_A\gtrsim 300$ GeV in which only a SM-like Higgs can be discovered at the LHC. Similarly, the reach for the non-SM Higgses in other models with an extended Higgs sector is limited as well. 

The situation is very different at a 100 TeV collider.
Two recent studies~\cite{Craig:2016ygr, Hajer:2015gka} estimated the reach of Higgs production searches in the MSSM at 100 TeV, where the heavy bosons are produced in association with and decay into SM fermions. The reach, shown in Fig.~\ref{fig:exclusion_normal}, is impressive. Heavy Higgs masses up to 5-10 TeV will be probed with $3\ \mathrm{ab}^{-1}$ for all values of $\tan \beta$. This also shows that the the wedge region could be covered by making use of new kinematic features of such signal events at a 100 TeV pp collider, in this case top tagging in the boosted region. 
At low $\tan \beta$, the greatest sensitivity to neutral Higgs bosons is achieved with a same-sign dilepton search for Higgses produced in association with one or two top quarks, which subsequently decay to $\bar t t$. The associated production channel avoids the difficult interference issues of a $p p \to H^0 \to \bar t t$ search. 


\begin{figure}[h!]
\centering
  \includegraphics[width=0.45\textwidth]{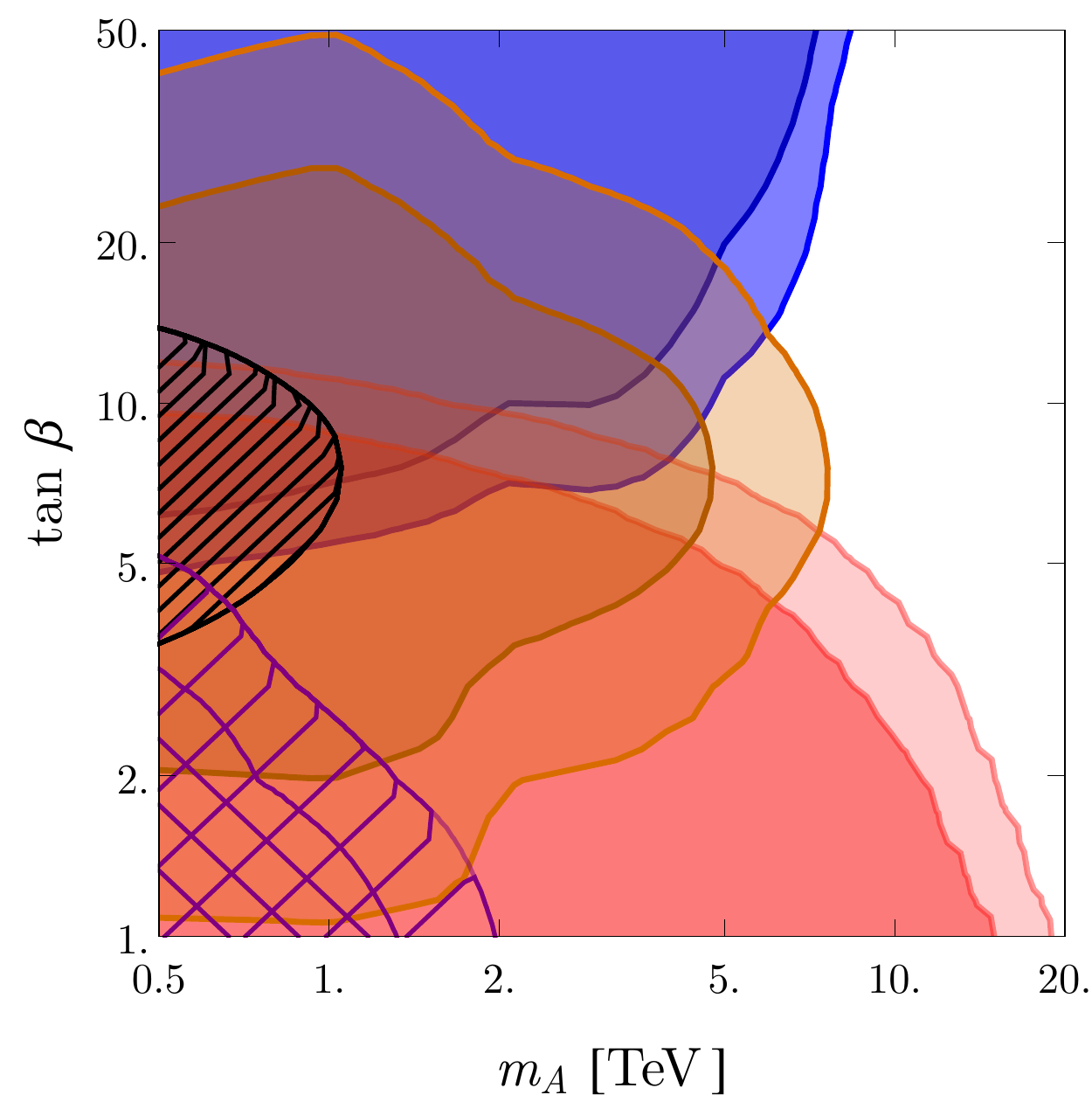}
  \includegraphics[width=0.45\textwidth]{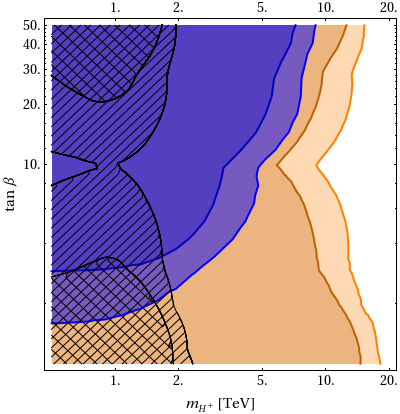}
  \caption{
  95\% C.L. exclusion bounds for neutral (left panel, from~\cite{Craig:2016ygr}) and charged (right panel, from~\cite{Hajer:2015gka}) Higgses of the MSSM at a 100 TeV collider.  %
  The blue and orange regions are probed   by the channels $pp\rightarrow bb H^0/A \rightarrow bb\tau\tau$ and  $pp\rightarrow bb H^0/A \rightarrow bb tt$ for the neutral Higgses and $pp\rightarrow tbH^\pm \rightarrow tb\tau\nu$ $pp\rightarrow tbH^\pm \rightarrow tbtb$  for the charged Higgses, respectively. 
  %
The red region is probed by heavy Higgs production in association with one or two top quarks, with subsequent decay to $\bar t t$, yielding a same-sign dilepton signature. 
   Given the same channel or the same color, the two different opacities indicate the sensitivities w.r.t. a luminosity of 3 ${\rm ab^{-1}}$ and 30 ${\rm ab^{-1}}$ at a 100 TeV $pp$ collider, respectively. 
   The cross-hatched and diagonally hatched regions are the predicted exclusion contours for associated Higgs production at the LHC for 0.3 ${\rm ab^{-1}}$, and 3 ${\rm ab^{-1}}$, respectively.}
 \label{fig:exclusion_normal}
 \end{figure}

 \begin{table}
 {\small 
 \begin{tabular}{l|l|l|l} \hline
Parent Higgs&Decay&Possible Final States&Channels in 2HDM  \\ \hline  
 &$HH$ type & $(bb/\tau\tau/WW/ZZ/\gamma\gamma)(bb/\tau\tau/WW/ZZ/\gamma\gamma)$ & $H^0 \rightarrow A A, h^0h^0$ \\   \cline{2-4}
Neutral Higgs&$HZ$ type & $(\ell\ell/qq/\nu\nu)(bb/\tau\tau/WW/ZZ/\gamma\gamma)$ & $H^0 \rightarrow A Z, A \rightarrow  H^0 Z, h^0Z$  \\   \cline{2-4}
$H^0$, $A$& $H^+H^-$ type & $(tb/\tau\nu/cs)(tb/\tau\nu/cs)$ & $H^0  \rightarrow H^+H^-$ \\  \cline{2-4}
&$H^\pm W^\mp$ type & $(\ell\nu/qq^\prime)(tb/\tau\nu/cs)$ & $H^0/A \rightarrow H^\pm W^\mp$\\  \hline
Charged Higgs&$ H W^\pm$ type & $(\ell\nu/qq^\prime)(bb/\tau\tau/WW/ZZ/\gamma\gamma)$ & $H^\pm \rightarrow  h^0W, H^0 W, A W$  \\  \hline
 \end{tabular}
}
 \caption{Summary of exotic decay modes for non-SM Higgs bosons. For each type of exotic decays (second column), we present possible final states (third column) and relevant channels in 2HDM. Note that $H$ in column two refers to any of the neutral Higgs, e.g. $h^0$, $H^0$ or $A$ in 2HDM.  }
 \label{tab:decay_exo}
 \end{table}
 
In addition to their decays to the SM particles, non-SM Higgses can decay via exotic modes, i.e., 
heavier Higgs decays into two light Higgses, or one light Higgs plus one SM gauge boson. Clearly this happens in the case when
the splitting between the various heavy higgses is not small. This can happen in the alignment limit of the 2HDM without 
decoupling. As outlined above, this limit is less generic than the decoupling limit, but still worth a detail study. 

Five main exotic decay categories for Higgses of the  2HDM are shown in Table~\ref{tab:decay_exo}. Once these decay modes are kinematically open, they typically dominate over the conventional decay channels. Recent studies on   exotic decays of heavy Higgs bosons  can be found in  Refs.~\cite{Coleppa:2013xfa,Coleppa:2014hxa,Brownson:2013lka, Coleppa:2014cca,Kling:2015uba,Li:2015lra,Maitra:2014qea,Basso:2012st, Dermisek:2013cxa,Mohn:833753,Assamagan:2000ud,Assamagan:2002ne}.   

Theoretical and experimental constraints restrict possible mass hierarchies in 2HDM. At high Higgs mass and close to 
the alignment limit, unitarity imposes a relation between the soft $Z_2$-breaking term and the heavy CP-even neutral 
Higgs mass $m_{12}^2=m_{H^0}^2 s_\beta c_\beta$\footnote{Note that this is automatically fulfilled in the MSSM.}. 
In this limit, the decay branching fraction $H^0 \to h^0 h^0, AA, H^+ H^-$ vanishes and vacuum stability further requires the CP-even non-SM Higgs $H^0$ to be the lightest non-SM like Higgs. In addition, electroweak precision measurements require the charge Higgs mass to be close to either of the neutral non-SM Higgs masses. This leaves us with only two possible mass hierarchies permitting exotic Higgs decays: $m_{H^0}\approx m_{H^+}<m_A$ and $m_{H^0}<m_A\approx m_{H^+}$.   At high Higgs masses, unitarity further requires the mass splitting between the non-SM Higgses to be small and therefore imposes an upper bound on the Higgs mass permitting exotic Higgs decays around $m_H\sim 1.5 - 2 $ TeV. These restrictions can be significantly relaxed at lower Higgs mass, allowing a larger spectrum of mass hierarchies including those permitting the decays $H^0 \to AA, H^+ H^-$. Considering the limited reach of the LHC around the ``wedge region'', extotic Higgs decay channels in the low Higgs mass region, $m_{H^0}\lsim500$ GeV, might still provide discovery potential at a 100 TeV $pp$ collider.  In Fig.~\ref{fig:decay} we show the branching fraction of non-SM Higgs bosons in Type II 2HDM for $\sin(\beta-\alpha)=1$ and a mass hierarchy containing a 1 TeV parent Higgs and a 850 GeV daughter Higgs. 

\begin{figure}[h!]
\centering
  \includegraphics[width=0.32\textwidth]{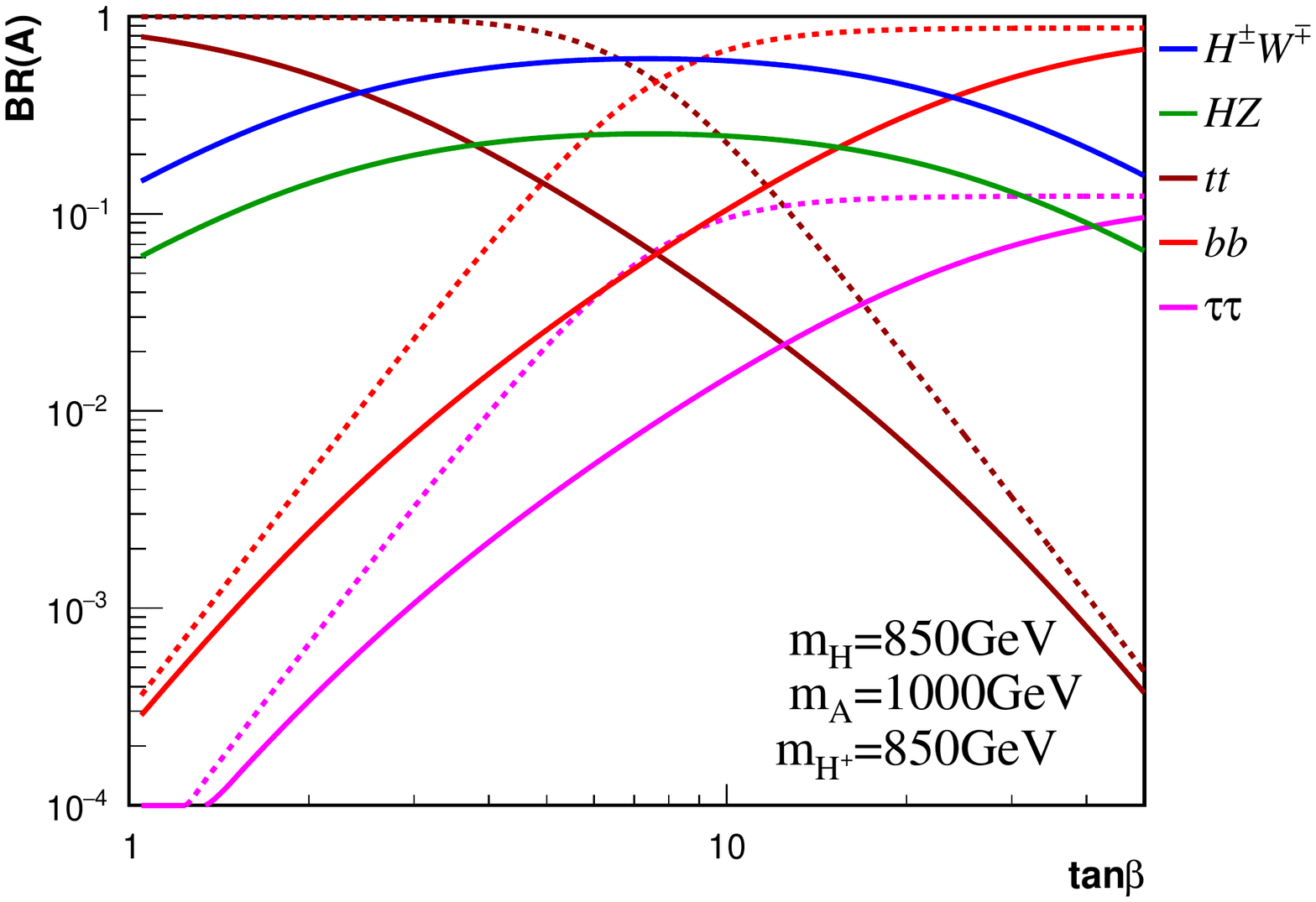}
  \includegraphics[width=0.32\textwidth]{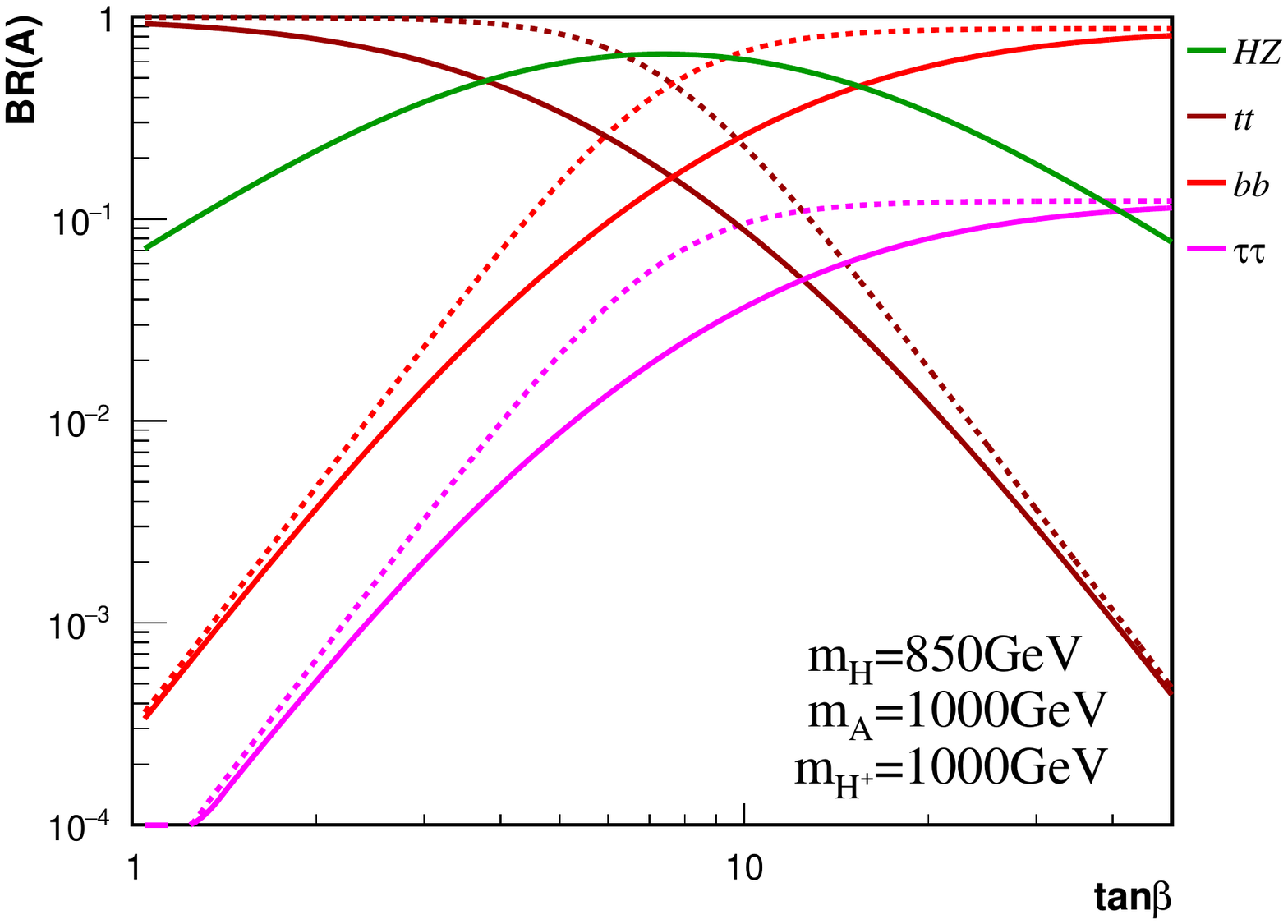}
  \includegraphics[width=0.32\textwidth]{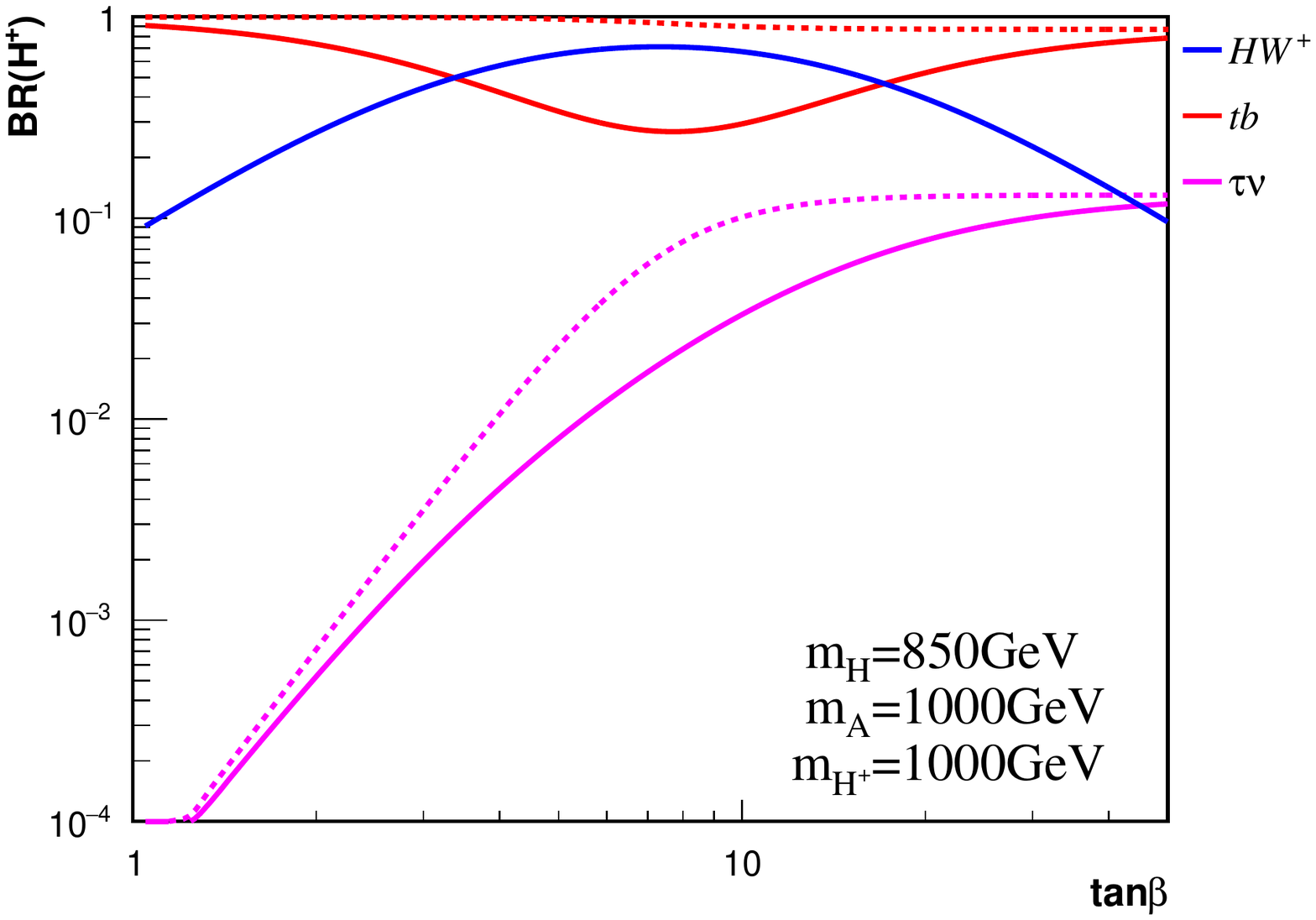}
 \caption{Branching fractions for $A$ (left and center panel) and $H^+$ (right panel) in the Type II 2HDM with $\sin(\beta-\alpha)=1$ and $m_{12}^2=m_{H^0}^2 s_\beta c_\beta$. We show the two allowed mass hierarchies $m_{H^0}=m_{H^+}<m_A$ (left panel) and $m_{H^0}<m_A=m_{H^+}$ (center and right panel). The parent and daughter Higgs masses are chosen to be 1 TeV and 850 GeV, respectively. Dashed curves are the branching fractions when exotic decay modes are kinematically forbidden.  All decay branching fractions are calculated using the program 2HDMC~\cite{Eriksson:2009ws}.}
 \label{fig:decay}
 \end{figure}
 
Note that most of the current experimental searches for the non-SM Higgs  assume the absence of exotic decay modes.  Once there are light Higgs states such that these exotic modes are kinematically open, the current search bounds can be  greatly relaxed given the suppressed decay branching fractions into SM final states~\cite{Coleppa:2014hxa,Coleppa:2014cca,Li:2015lra}.  Furthermore, exotic Higgs decays to final states with two light Higgses or one Higgs plus one SM gauge boson provide complementary search channels.  Here, we list such exotic Higgs decays and consider potential search strategies.

\begin{itemize}
\item{$H^0\rightarrow A A$ or $H^0\rightarrow h^0 h^0$}
\end{itemize}
With the final state Higgs decaying via $bb$, $\gamma\gamma$, $\tau\tau$, $WW^*$, final states of $bbbb$, $bb\tau\tau$, $bb\gamma\gamma$ and $\gamma\gamma WW^*$ can be used to search for resonant Higgs decay to two light neutral Higgses.    Current searches at the LHC 8 TeV with about 20 ${\rm fb}^{-1}$ luminosity gave  observed limits of   2.1 pb at 260 GeV and about 0.018 pb at 1000 GeV~\cite{Aad:2015xja}.  While $bb\gamma\gamma$ and $bb\tau\tau$ have comparable sensitivities at low mass, $bbbb$ mode dominates at high mass. 
 
\begin{itemize}
\item{$H^0\rightarrow A Z$ or $A \rightarrow H^0 Z $}
\end{itemize}
With $Z \rightarrow \ell\ell$ and $H^0 /A \rightarrow bb, \tau\tau$, the final states of $bb\ell\ell$, $\tau\tau\ell\ell$ can be obtained with gluon fusion production, or in the $bb$ associated production with two additional $b$ jets~\cite{Coleppa:2013xfa,Coleppa:2014hxa,Brownson:2013lka}.   Recent searches from ATLAS and CMS have shown certain sensitivity in this channel~\cite{CMS-PAS-HIG-14-011,CMS-PAS-HIG-13-025,Aad:2015wra,CMS-PAS-HIG-15-001}.   In parameter regions where ${\rm Br}(A \rightarrow H^0 Z)\times {\rm Br}(H^0 \rightarrow Z Z)$ is not   completely suppressed, $ZZZ$ final states with two $Z$ decaying leptonically and one $Z$ decaying hadronically can also be useful~\cite{Coleppa:2014hxa}.   Other channels with top final states could  be explored as well.

Note that the decay $A\to Z H^0$ has been identified as a particular signature of a SFOEWPT in the 2HDM\cite{Dorsch:2013wja}. As discussed below, the prospects for observing this channel in the $\ell\ell b{\bar b} $ and $\ell\ell W^+ W^-$ model have been analyzed in Ref.~\cite{Dorsch:2014qja}.

\begin{itemize}
\item{$H^0 \rightarrow H^+ H^- $}
\end{itemize}
With both $H^\pm$ decaying via $\tau\nu$ final states, the signal of $\tau\tau \nu \nu$ can be separated from the SM $W^+W^-$ background since the charged tau decay product   in the signal typically has a hard spectrum compared to that of the background~\cite{Li:2015lra}.      

\begin{itemize}
\item{$H^0/A \rightarrow  H^\pm W^\mp$ }
\end{itemize}

Similar to the $H^+H^-$ case, $H^\pm \rightarrow \tau \nu, tb$ and $W\rightarrow \ell \nu$ with $\ell \tau \nu \bar\nu$  or $tb\ell \nu$ could be used to search for $H^0/A \rightarrow  H^\pm W^\mp$.   Note that for the CP-even Higgs $H^0$, the branching fraction of $H^0 \rightarrow H^\pm W^\mp$ is mostly suppressed comparing to $H^0 \rightarrow H^+H^-$ as long as the latter process is kinematically open and not accidentally suppressed~\cite{Li:2015lra}.  However, for the CP-odd Higgs $A$, this is the only decay channel with a charged Higgs in the decay products.

\begin{itemize}
\item{$H^\pm \rightarrow H^0 W, A W $}
\end{itemize}
This is the only exotic decay channel for the charged Higgs in the 2HDM.  Given the associated production of $tbH^\pm$, and the decay of $H^0$, $A$ into the $bb$ or $\tau\tau$ channel, $\tau\tau bb WW$ or $bbbbWW$  can be used to probe this channel~\cite{Coleppa:2014cca}.  $H^0/A \rightarrow t\bar{t}$ could also be used given the boosted top in the high energy environment.

Note that while  $H^\pm \rightarrow WZ$ is absent in 2HDM type extension of the SM Higgs sector, it could appear, however,  in the real triplet models extension of the SM once the triplet obtains a vev~\cite{FileviezPerez:2008bj}.

\subsubsection{Singlet Extensions of the Higgs Sector\footnote{Contribution by Dario Buttazzo, Filippo Sala and Andrea Tesi}}
\label{sss.SingletExtensions}

The simplest example of an extended Higgs sector consists in the addition of a real scalar field, singlet under all the gauge groups, to the SM. Despite its great simplicity, this scenario arises in many of the most natural extensions of the SM -- e.g.\ the Next-to-Minimal Supersymmetric SM (NMSSM, see Section~\ref{sss.nmssm}), 
Twin Higgs (see Section~\ref{sss.twin}), some Composite Higgs models -- and is therefore of considerable physical interest.

In general, a CP-even scalar singlet will mix with the Higgs boson at a renormalizable level. As a consequence, both physical scalar states can be produced at colliders through their couplings to SM particles, and be observed by means of their visible decays.


Let us denote by $h$ and $\phi$ the physical mass eigenstates, and by $\gamma$ the mixing angle.
In a weakly interacting theory,
the couplings of $h$ and $\phi$ are just the ones of a SM Higgs, rescaled by a universal factor of $\cos\gamma$ or $\sin\gamma$, respectively. Hence, assuming no invisible decays, the signal strengths $\mu_{h,\phi}$ into SM particles are
\begin{eqnarray}
\mu_h &=& \mu_{\rm SM}(m_h)\times c_\gamma^2,\label{muh}\\
\mu_{\phi\to VV,ff} &=& \mu_{\rm SM}(m_\phi)\times s_\gamma^2\times \left(1 - {\rm BR}_{\phi\to hh}\right),\label{mus}\\
\mu_{\phi\to hh} &=& \sigma_{\rm SM}(m_\phi)\times s_\gamma^2\times {\rm BR}_{\phi\to hh},\label{mushh}
\end{eqnarray}
where $\mu_{\rm SM}(m)$ is the corresponding signal strength of a standard Higgs with mass $m$, and ${\rm BR}_{\phi\to hh}$ is the branching ratio of $\phi$ into two Higgses -- an independent parameter at this level.
The phenomenology of the Higgs sector is therefore completely described by three parameters: $m_\phi$, $s_\gamma$, and ${\rm BR}_{\phi\to hh}$. The second state $\phi$ behaves like a heavy SM Higgs boson, with reduced couplings and an additional decay width into $hh$.

The mixing angle $\gamma$ and $m_\phi$ are not independent quantities, but are related via
\begin{equation}
\label{e.singletmhh}
\sin^2\gamma = \frac{M_{hh}^2 - m_h^2}{m_\phi^2 - m_h^2},
\end{equation}
where $M_{hh}^2\propto v^2$ is the first diagonal entry of the Higgs squared mass matrix, in the gauge eigenstate basis. In the following we will often use $M_{hh}$ as a parameter, instead of $\gamma$, to avoid considering unnatural regions of the parameter space with high mass and large mixing angle.

\begin{figure}
\includegraphics[width=0.48\textwidth]{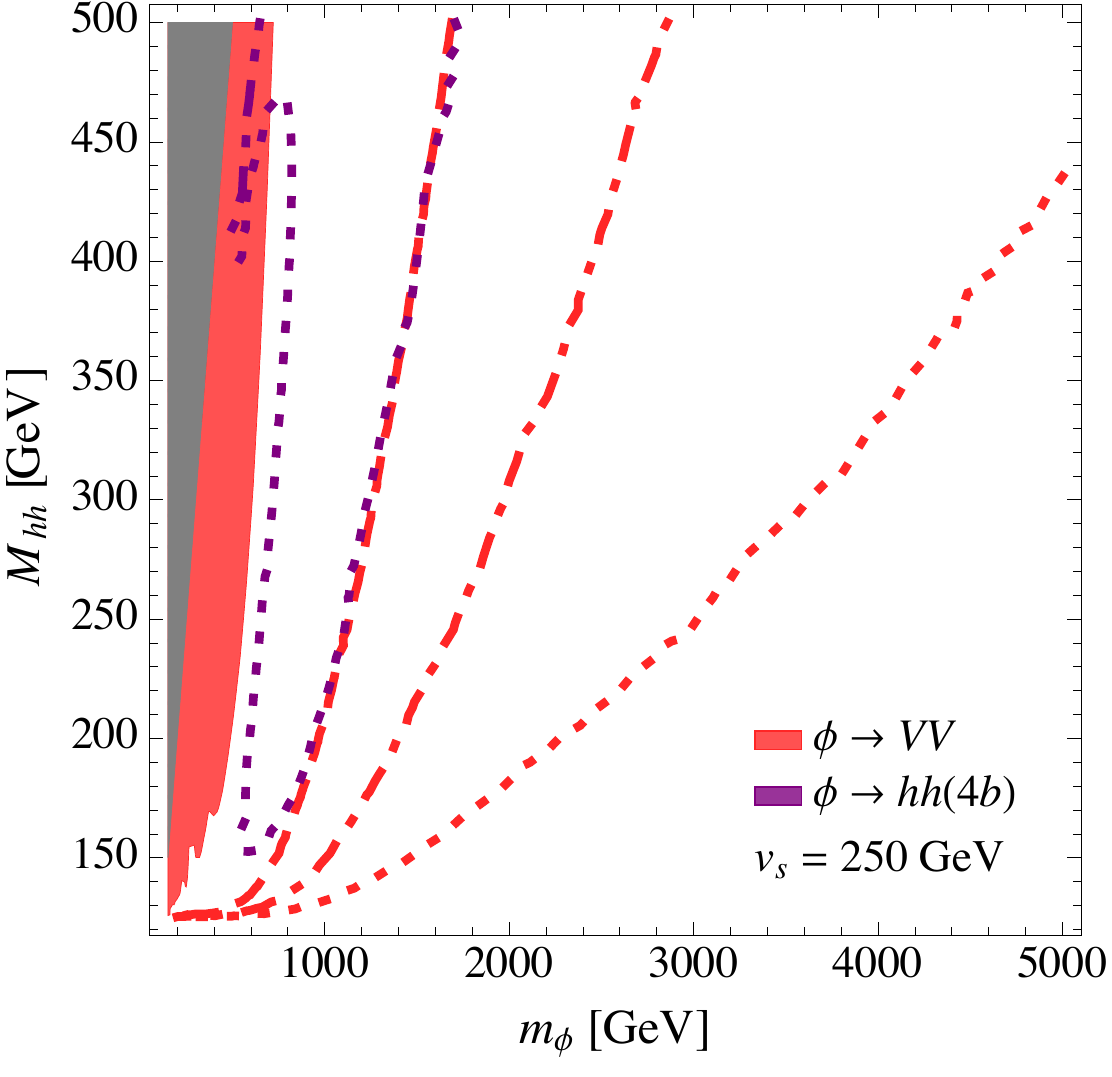}\hfill
\includegraphics[width=0.48\textwidth]{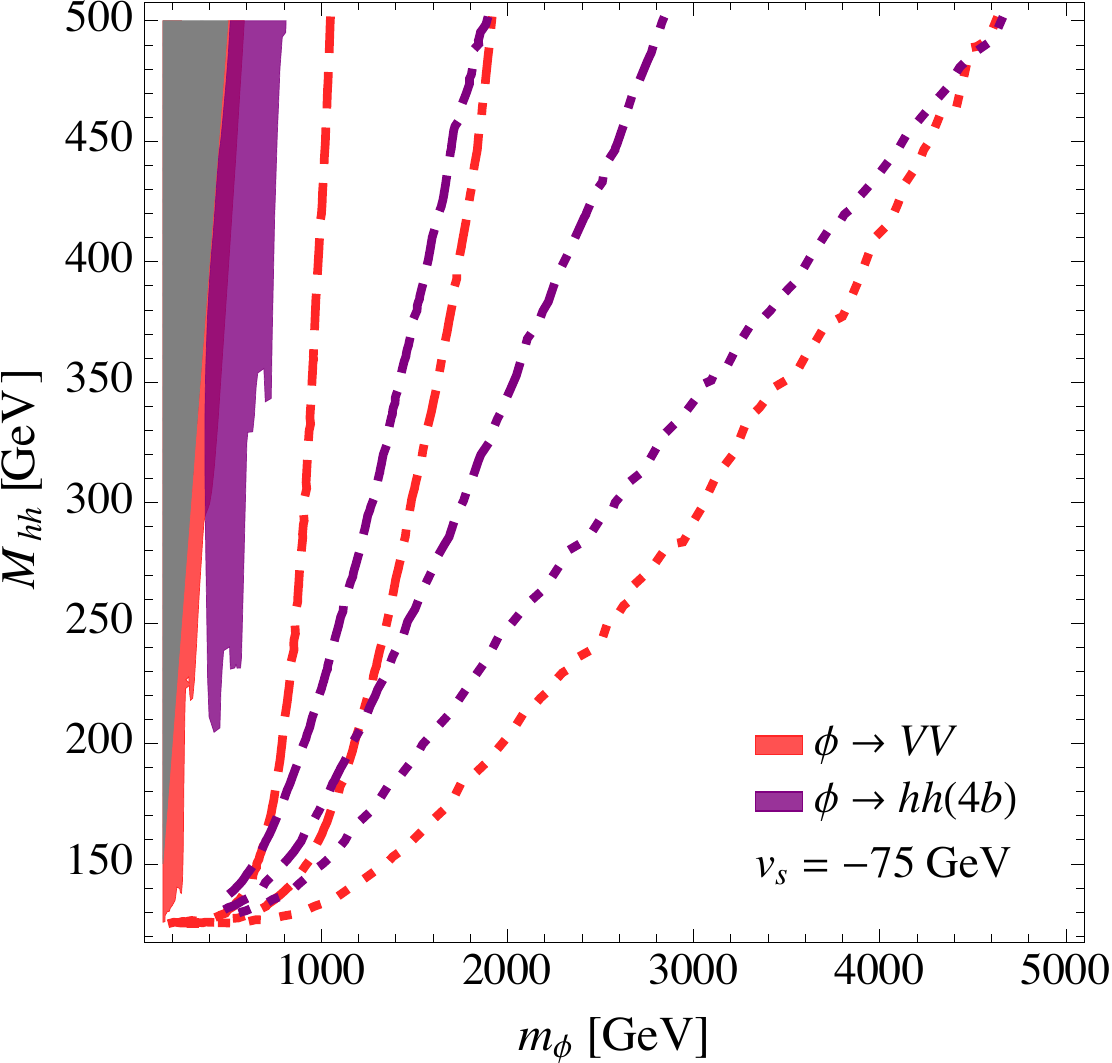}
\caption{Reach of direct searches for generic singlet extensions of the SM in the $m_\phi$--$M_{hh}$ plane, see \eref{singletmhh}. The singlet vev is  $v_s = 250$ GeV (left) and $v_s = -75$ GeV (right). 
 Searches using the $VV$ and $hh$ final state are shown in red and purple. The solid colored regions are excluded at 95\% CL by the 8 TeV LHC. 
 Lines represent the expected bounds for FCC-hh (dotted), high-luminosity LHC (dashed), and high-energy LHC (dot-dashed), all with a luminosity of 3~ab$^{-1}$.  Taken from ref.~\cite{Buttazzo:2015bka}.\label{fig:singlet_direct}}
\end{figure}


Given the singlet nature of $\phi$, its main decay channels are into pairs of $W$, $Z$, and Higgs bosons.
In the large-$m_\phi$ limit, the equivalence theorem implies that
\begin{equation}
{\rm BR}_{\phi\to hh} = {\rm BR}_{\phi\to ZZ} = \frac{1}{2}{\rm BR}_{\phi\to WW}.
\end{equation}
The leading corrections to this relation\footnote{The exact expressions for the triple couplings $g_{\phi hh}$ and $g_{hhh}$ depend on the details of the scalar potential, and can be found in ref. \cite{Buttazzo:2015bka}.} for finite masses depend only on the vacuum expectation value of the singlet, $v_s$ \cite{Buttazzo:2015bka}. Therefore, to a good approximation $m_\phi$, $M_{hh}$, and $v_s$ constitute a set of independent parameters that describe the phenomenology.

The FCC reach for a generic resonance in these channels has been discussed in Ref.~\cite{Buttazzo:2015bka}, where it has been obtained through a parton luminosity rescaling from the 8 TeV LHC results~\cite{Thamm:2015zwa}. 
Figure~\ref{fig:singlet_direct} shows the combined reach from all the $VV$ final states, compared with the one in the $hh$ channel from the $4b$ final state, again for two different values of $v_s$. The reach of the high-luminosity (14 TeV, 3 ab$^{-1}$) and high-energy (33 TeV, 3 ab$^{-1}$) upgrades of the LHC are also shown for comparison. It can be seen that the $VV$ searches are always dominant at FCC.  
Moreover, the detection of a $hh$ resonance in the multi-TeV range is possible only for values of $v_s$ related also to very large modifications (a factor of two or even larger) in the triple Higgs coupling $g_{hhh}$. Moderate to large deviations in such a coupling are a generic prediction of a singlet mixed with the Higgs, contrarily for example to the case of 2HDM, and the very high energy of the FCC-hh appear to provide a unique opportunity to test this possibility.

\begin{figure}
\includegraphics[width=0.48\textwidth]{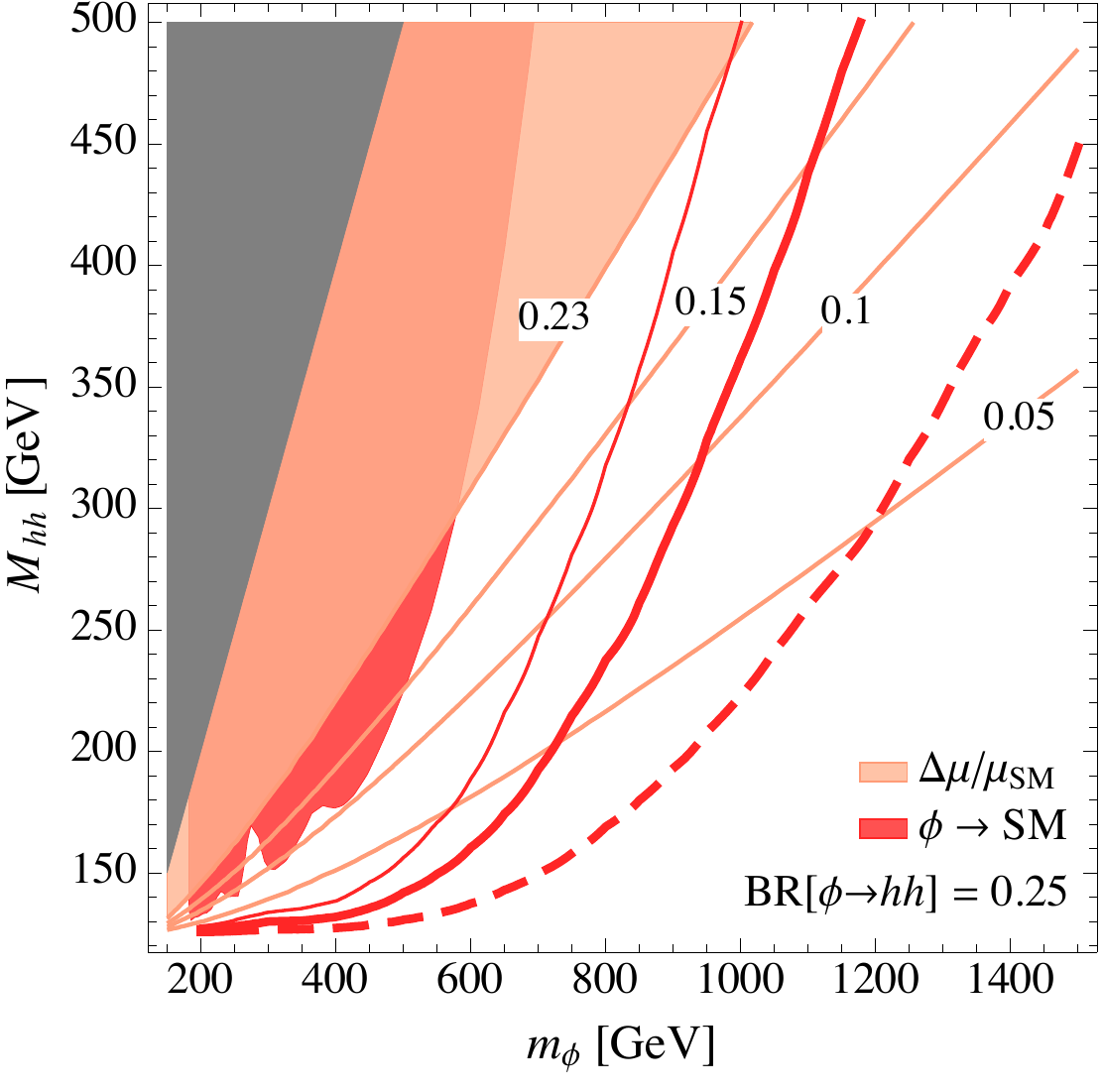}\hfill
\includegraphics[width=0.48\textwidth]{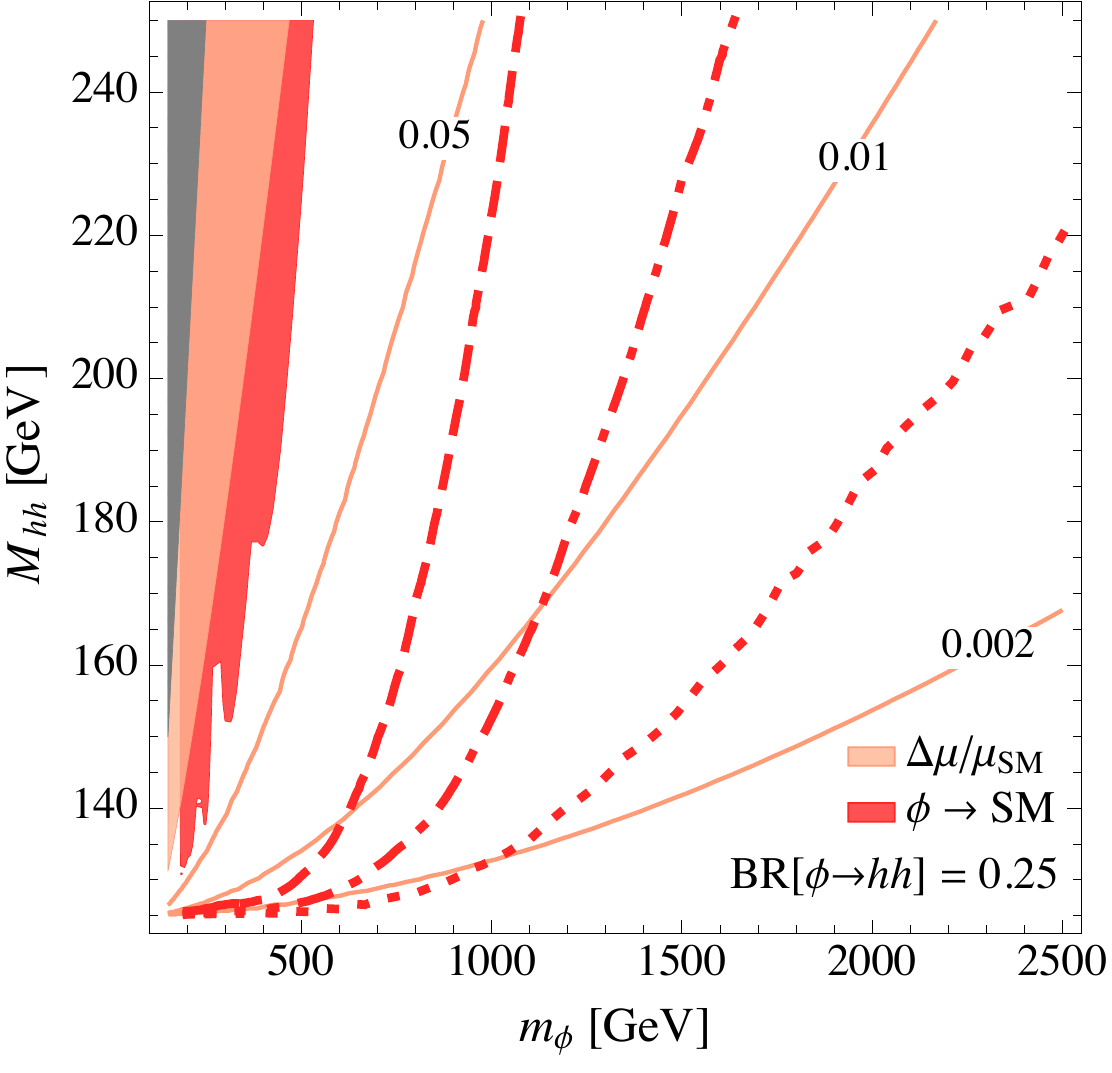}
\caption{Comparison between direct and indirect searches in the singlet model, and for ${\rm BR}_{\phi\to hh} = 0.25$. Left: region relevant for the LHC (thin lines: 13 TeV, thick lines: 14 TeV, dashed: high luminosity). Right: projections for future colliders (dot-dashed: 33 TeV LHC, dotted: FCC-hh with 3 ab$^{-1}$). Coloured isolines of $s_\gamma^2$. The colored region is excluded at 95\% CL. Taken from ref.~\cite{Buttazzo:2015bka}.\label{fig:singlet_comparison}}
\end{figure}

Finally, Fig.~\ref{fig:singlet_comparison} compares the reach of direct and indirect searches, for fixed ${\rm BR}_{\phi \to hh} = 0.25$, and for two regions relevant for the LHC and for FCC, respectively.
The deviation in Higgs signal strengths, shown as colored isolines, are proportional to $s_\gamma^2$.
Given the universal rescaling of all the couplings, the power of Higgs coupling measurements in the singlet case is rather limited, as compared e.g. to a 2HDM (see Section~\ref{ss.2hdm}).  
It is interesting to note that direct searches at FCC-hh are more powerful than indirect measurements at FCC-ee for resonance masses below about 1 TeV.

The results of the simple scenario presented in this Section apply in general to any singlet scalar, and can easily be applied to more concrete cases, as we discuss below for the NMSSM and the Twin Higgs.


\paragraph{NMSSM}
\label{sss.nmssm}
The NMSSM constitutes a particularly interesting physics case for several reasons. In particular, since the extra Higgs bosons could be the first new degrees of freedom to be detected, it is important to quantify the reach of LHC and future colliders for scalar states in this scenario.

The NMSSM consists in the MSSM with the addition of an extra gauge singlet $S$, so that the superpotential reads
\begin{equation}
W_{\rm NMSSM} = W_{\rm MSSM} + \lambda S H_u H_d + V(S),
\end{equation}
where $V$ is a polynomial up to order three in the new field $S$.
This addition is relevant form the point of view of naturalness for large enough $\lambda$, since the different dependence of the weak scale $v$ in the high energy parameters allows, for a given amount of tuning, to raise the stop and gluino masses by a factor $\sim \lambda/g$ with respect to the MSSM.  The Higgs mass value $m_h$ is also less constraining in the NMSSM than in the MSSM, thanks to the extra SUSY-preserving contribution of the form $\sin^22\beta\, \lambda\, v^2/2$. The drawback of this last feature is that, for a too large $\lambda$, a tuning is reintroduced to lower the Higgs mass to 125 GeV. These considerations have been made precise in the analysis of ref.~\cite{Gherghetta:2012gb}, that identified $\lambda \sim 1$ as a region of minimal tuning, capable of accommodating stops and gluinos as heavy as 1.2 and 2.5 TeV respectively.
The loss of perturbativity before the GUT scale, implied by a value of $\lambda \gtrsim 0.75$, is avoidable without spoiling unification, see \textit{e.g.} the model proposed in \cite{Barbieri:2013hxa} and references therein.

\medskip

The Higgs sector of the NMSSM consists of six particles: three CP-even ones, $h$, $H$ and $\phi$, two CP-odd ones $A$ and $A_s$, and a charged state $H^{\pm}$. We mostly concentrate on the CP-even ones, and identify $h$ with the 125 GeV state discovered at the LHC, $H$ with the mostly doublet mass eigenstate, and $\phi$ with the mostly singlet one. Three mixing angles, to be called $\delta$, $\gamma$, and $\sigma$, control the rotation between the mass eigenstates and the gauge eigenstates where one doublet takes all the vev.
%

It has been shown in ref. \cite{Barbieri:2013hxa} that the phenomenology of the three CP-even states, with the exception of the trilinear couplings among the Higgses, can be described in terms of only six parameters, which we find convenient to choose as:
\begin{equation}
\label{six_NMSSM_pars}
m_\phi^2, \quad m_H^2, \quad m_{H^\pm}^2, \quad \tan\beta, \quad \lambda, \quad \Delta^2,
\end{equation}
where with $\Delta^2$ we denote all the radiative contributions to the Higgs mass, which sums up the contributions from the rest of the superpartner spectrum.\footnote{We will choose values representative of TeV-scale stops, but the phenomenology does not sensitively depend on $\Delta^2$ unless the value is much larger than what we assume.}
Radiative corrections to other elements of the scalar mass matrix are assumed to be small.
The mixings angles
can be expressed in terms of the parameters in (\ref{six_NMSSM_pars}), see 
Refs.~\cite{Barbieri:2013hxa,Barbieri:2013nka} for the full analytical formulae.

Deviations in the Higgs couplings,
\begin{equation}
\frac{g_{h\bar{u}u}}{g_{h\bar{u}u}^{\rm SM}} = c_\gamma (c_\delta +\frac{s_\delta}{\tan\beta}), \qquad \frac{g_{h\bar{d}d}}{g^{\rm SM}_{h\bar{d}d}}= c_\gamma (c_\delta -s_\delta \tan\beta ), \qquad \frac{g_{hVV}}{g^{\rm SM}_{hVV}}=  c_\gamma c_\delta\,,
\label{Higgs_couplings}
\end{equation}
(where $s_\theta, c_\theta = \sin\theta, \cos\theta$) constrain only the two mixing angles $\delta$ and $\gamma$ that involve $h$.
While a fit to the Higgs couplings leaves space for a sizable mixing $\gamma$, at the level of $\sin\gamma \sim 0.45$, it leaves little space for two mixed doublets.
Perhaps more important than that,
the LHC14 with 300 fb$^{-1}$ is not expected to probe substantially the $h$-$S$ mixing $\gamma$, while the opposite is true for $\delta$, which will be constrained to the few percent level \cite{Barbieri:2013nka}.

Since the purpose of this document is to set a general strategy for searching for BSM Higgses, for convenience we now summarize the phenomenology in two limiting cases, identified by their relevant degrees of freedom\footnote{For a discussion with the same logic and more details, we refer the reader to the short review \cite{Sala:2015lza}.}. 
%
\begin{itemize}
\item[$\diamond$] $h$ and a singlet-like state $\phi$ in the low-energy spectrum, with $H$ decoupled and $\delta = 0$. The generic parameter $M_{hh}$ of Section \ref{sss.SingletExtensions} is now identified with the upper bound on the Higgs boson mass
\begin{equation}
M_{hh}^2 = c^2_{2 \beta} m_Z^2 + s^2_{2 \beta} \lambda^2 v^2/2 + \Delta^2,
\label{Mhh_NMSSM}
\end{equation}
and the total number of free parameters decreases from six to four, $m_\phi$, $\lambda$, $\tan\beta$ and $\Delta$.
Since $\Delta$ has little impact on the phenomenology of the model (unless it is very large),
it is convenient to fix it to some reference value, typical of stops in the TeV range.

%
\item[$\diamond$] $h$ and a second doublet $H$ in the low-energy spectrum, with $\phi$ decoupled and $\gamma = 0$. The Higgs sector now realizes a particular type-II 2HDM. 
Here the total number of free parameters is three, $m_H$, $\tan\beta$ and $\Delta$, and also the charged scalar $H_\pm$ is predicted with a mass close to that of the CP-even one $H$. 
The MSSM is a particular realization of this case, where $\lambda = 0$ and $\Delta$ is fixed to reproduce the correct value of the Higgs mass. However, for generic values of $\Delta$, $\lambda \neq 0$ and the realized 2HDM will be different from the MSSM.
\end{itemize}

Let us summarize the phenomenology of these two cases. In the $h$-singlet mixing scenario, Higgs coupling measurements will leave much of the parameter space unexplored, unless a per-mille precision is reached, as expected at the FCC-ee (or CEPC).
%
\begin{figure}[t]
\centering
\resizebox{0.48\textwidth}{!}{%
\includegraphics{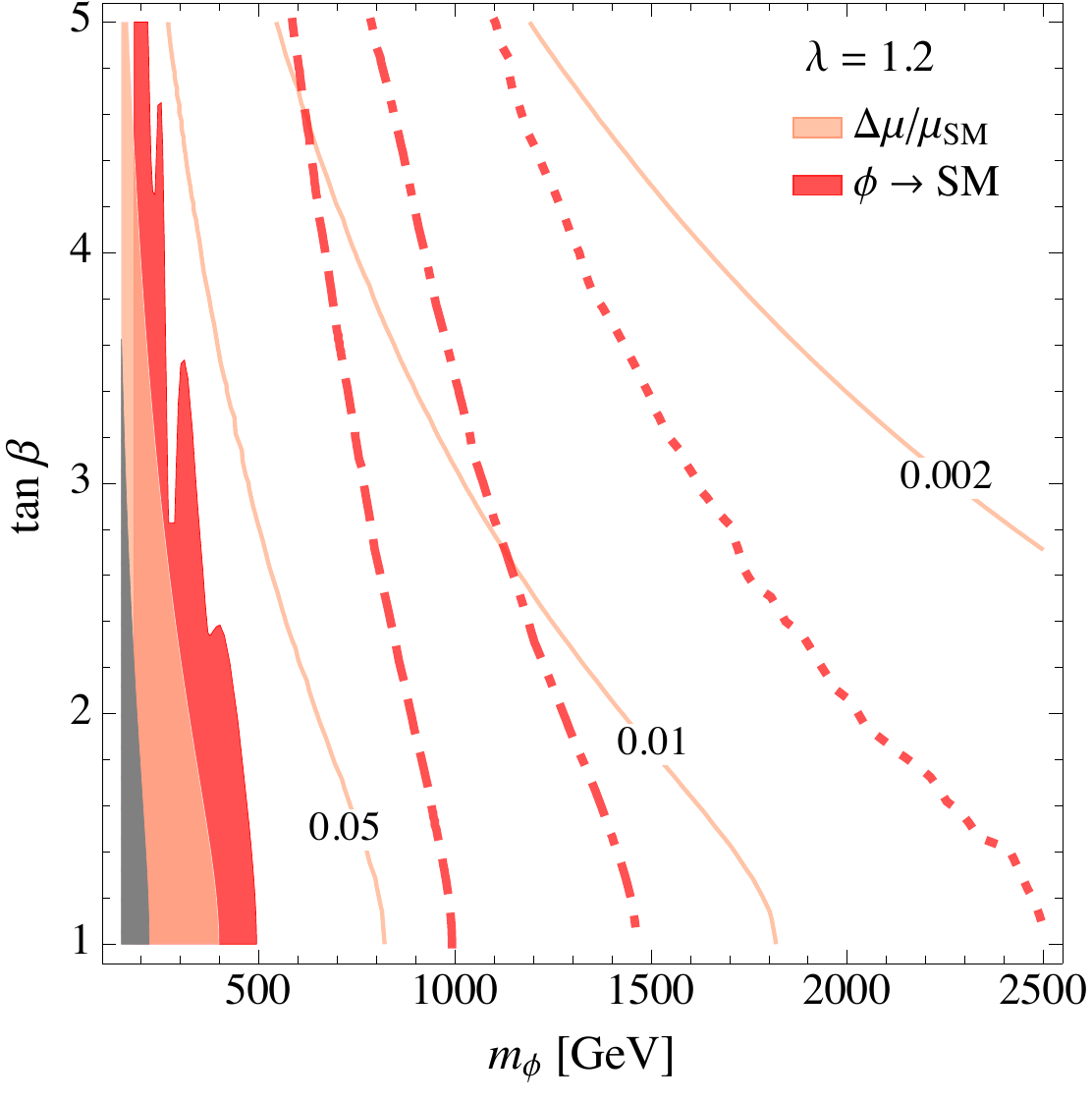}
}\hspace{.4 cm}
\resizebox{0.48\textwidth}{!}{%
\includegraphics{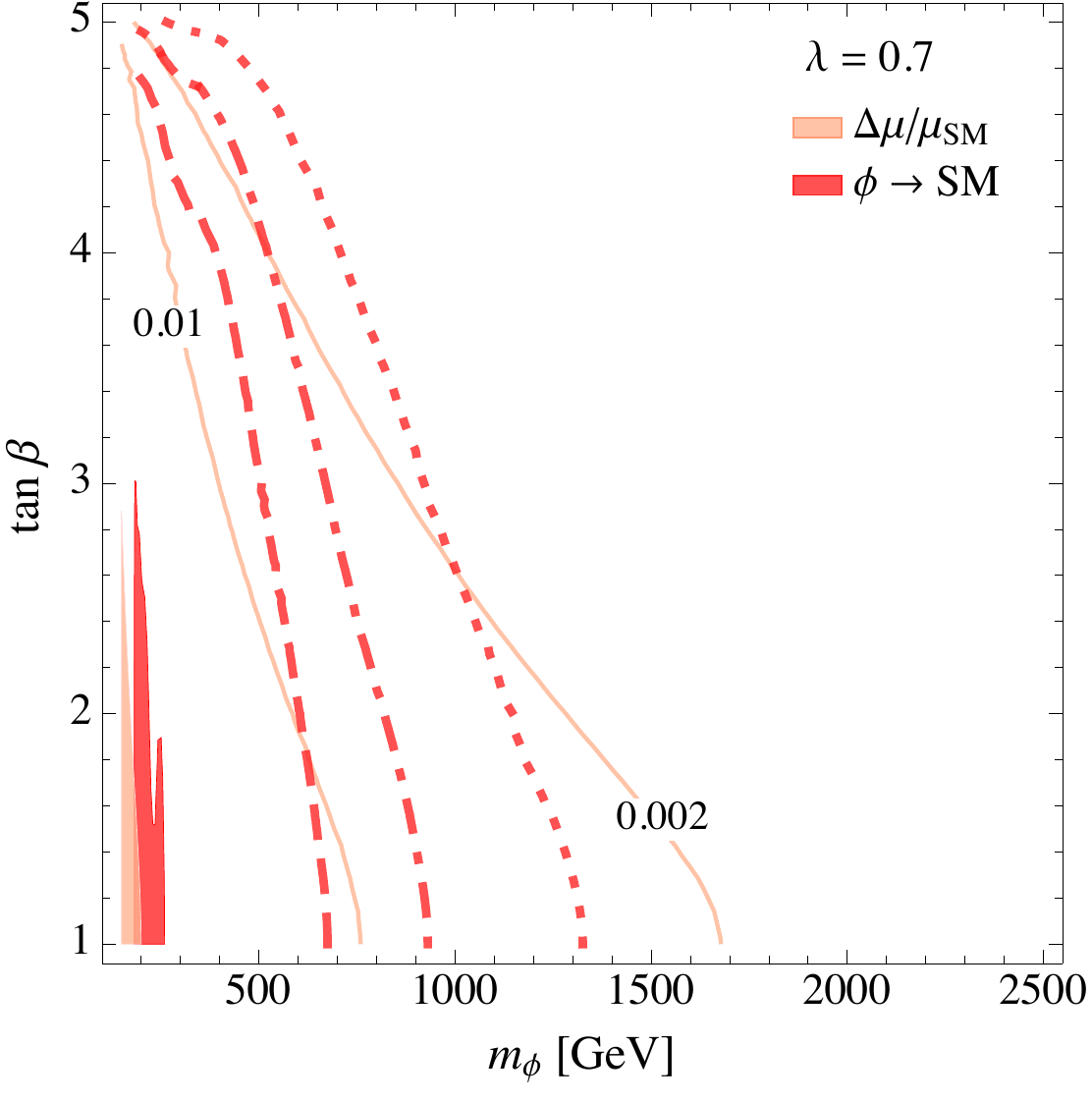}}
\caption{Parameter space of the NMSSM scenario in the limit of decoupled heavy doublet $H$.
Shaded areas: 95\% CL exclusions from Higgs signal strengths (pink) and direct searches for $\phi \to {\rm SM}$ (red) at LHC8. Lines: contours of $s_\gamma^2$ (pink), expected reach of direct searches at 
HL-LHC (dashed), HE-LHC (dot-dashed), and FCC-hh with 3~ab$^{-1}$ (dotted). Grey: unphysical regions. BR$_{\phi \to hh}$ fixed to its asymptotic value 1/4. Left: $\Delta = 70$ GeV and $\lambda = 1.2$; right: $\Delta = 80$ GeV and $\lambda = 0.7$. Taken from ref.~\cite{Buttazzo:2015bka}.}
\label{fig:NMSSMl12} 
\end{figure}
%
%
%
%
Direct searches for the extra Higgs is therefore the most powerful probe of the Higgs sector in this case.
The impact of current and future searches is shown in Figure~\ref{fig:NMSSMl12}, for the values of $\lambda = 1.2$ and $0.7$, respectively. For simplicity we have fixed BR$(\phi \to hh)$ to its asymptotic value of 1/4
, a case where $VV$ searches dominate over $hh$ ones\footnote{``Small'' values of $v_s$ can make resonant di-Higgs production more important, a case which we do not discuss here.}.
From Figure~\ref{fig:NMSSMl12} one reads that
direct searches are expected to dominate the reach in the parameter space of the model. At a 100 TeV $pp$ collider they will be complementary with Higgs coupling measurements of a leptonic collider like the FCC-ee or the CEPC. No matter how we look for BSM in the Higgs sector, the region of smaller $\lambda$ will be more difficult to probe than that of a larger one, giving more importance to other SUSY searches (like stop and gluino ones). In each scenario, singlet scalar masses in excess of a TeV can be probed, depending on the value of $\tan \beta$.

We now move to the second case, where the relevant degrees of freedom are $h$ and the doublet-like state $H$. This scenario is best probed via measurements of the Higgs signal strengths into SM particles, as evident from Figure~\ref{fig:NMSSM_hH}. A region of ``alignment without decoupling'' of the state $H$ survives for $\lambda \simeq 0.65$ and $\tan\beta \simeq 2.5$, corresponding to a zero of the mixing angle $\delta$.
That region -- which is already constrained by the bounds on $m_{H^\pm}$ coming from flavour measurements like BR$(B \to X_s\gamma)$ --
needs direct searches for the new states in order to be probed.
The discussion follows that of two Higgs doublet models of section \ref{ss.2hdm}. 
We recall here the main features for convenience, following the recent study of ref. \cite{Craig:2015jba}: exact alignment $\delta = 0$ implies BR$_{H \to hh}$ = BR$_{H \to VV}$ = BR$_{ A \to Zh}$ = BR$_{H^\pm \to W^\pm h}$ = 0, but contrary to the $h$-$\phi$ singlet case some couplings to SM fermions survive, allowing to probe the existence of $H$ in resonant searches in $t\bar{t}$, $b\bar{b}$, $\ell^+\ell^-$ and $\gamma\gamma$.
For larger values of $m_H$, the $t\bar{t}$ channel opens and dominates the branching ratio. The 100 TeV studies of heavy Higgs production in association with SM fermions~\cite{Hajer:2015gka, Craig:2016ygr} 
indicates that 5-10 TeV masses can be probed, which likely applies to this NMSSM scenario as well, see Fig.~\ref{fig:exclusion_normal}. 

\begin{figure}[t]
\centering
\resizebox{0.47\textwidth}{!}{%
\includegraphics{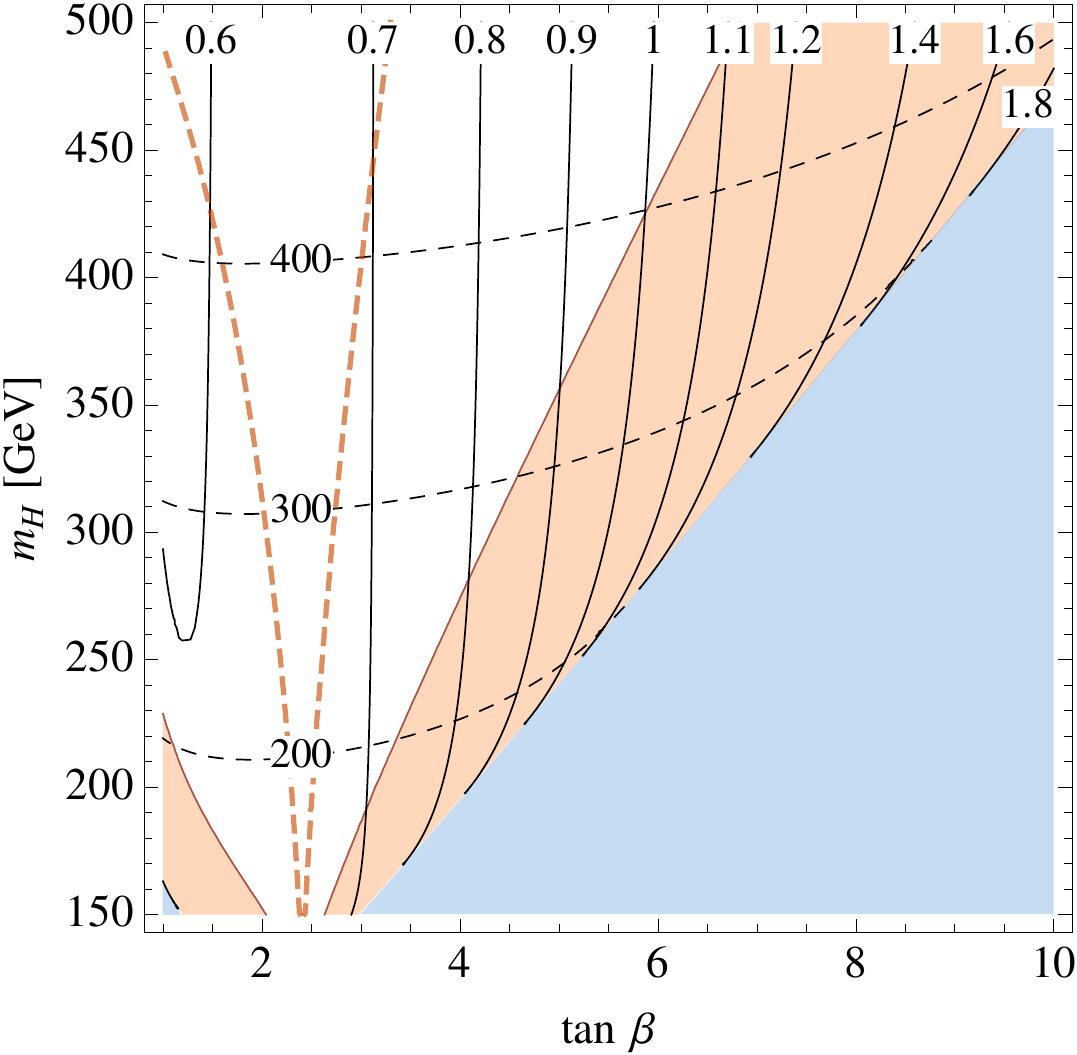}
}\hspace{.41 cm}
\resizebox{0.45\textwidth}{!}{%
\includegraphics{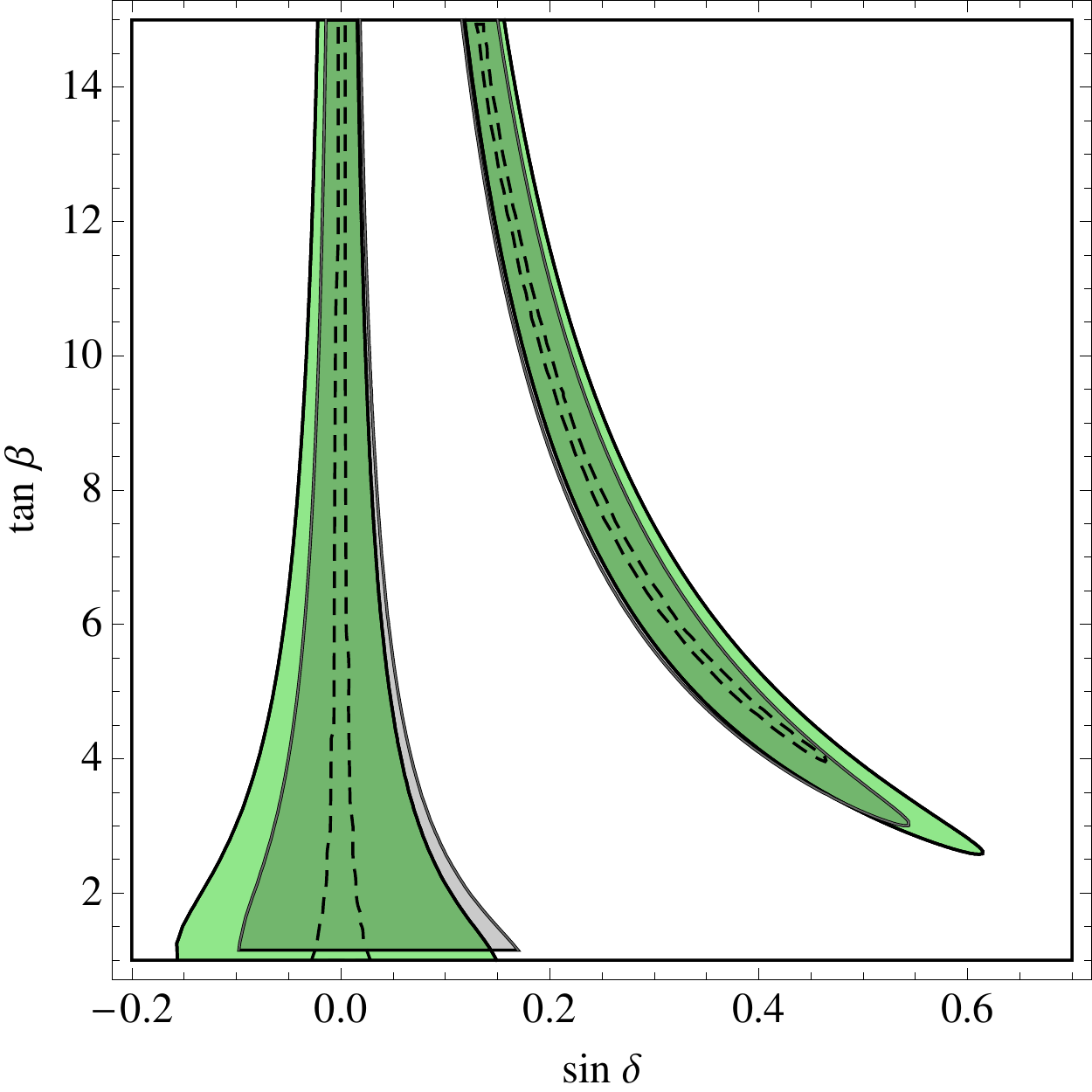}}
\caption{Left: parameter space of the NMSSM scenario with a decoupled scalar $S$, with radiative corrections to the Higgs mass fixed at $\Delta = 75 \ \gev$.
Isolines of $\lambda$ (solid black), and of $m_{H^\pm}$ (dashed black). 95\% CL exclusions from Higgs coupling measurements at LHC8 (shaded pink), expected reach at LHC14 with 300 fb$^{-1}$ (pink dashed). Blue: unphysical regions. The entire shown parameter space is likely to be excluded by direct searches for heavy Higgs doublets at 100 TeV, see Fig.~\ref{fig:exclusion_normal}.  
Right: Higgs couplings fit (95\% CL allowed regions) for $\sin^2\gamma = 0$ (green) and $\sin^2\gamma = 0.15$ (grey) from the LHC run 1, and projection for LHC14 (dashed, $\gamma = 0$). Taken from ref.~\cite{Buttazzo:2014nha}.}
\label{fig:NMSSM_hH}       
\end{figure}

Going back to a fully mixed situation, where all the states $h$, $H$ and $\phi$ are kept in the spectrum, demands to work with more parameters. Numerical scans are usually employed for this purpose, and they allow to interconnect the phenomenology of the Higgs scalar sector with that of other SUSY particles. We do not explore this case further, and refer the reader to the recent studies \cite{King:2014xwa,Carena:2015moc} for a discussion.
Also, in case some signal is observed, it will be important to explain it in a fully mixed situation. For this purpose, analytical relations such as the ones presented in refs \cite{Barbieri:2013hxa, Barbieri:2013nka} would provide a useful guidance.

Finally, we mention that the case of an extra Higgs lighter than 125 GeV is motivated and still partially unexplored. This is true especially for a singlet like state $\phi$, since flavour bounds on $m_{H^\pm}$ pose serious challenges to having $m_H < 125$ GeV.

\paragraph{Twin Higgs}
\label{sss.twin}
Another motivated scenario where we expect a scalar singlet at accessible energies is the Twin Higgs model (TH). As discussed in \sssref{neutralnaturalness}, the TH is a well-motivated solution to the Hierarchy problem with uncolored top partners and many possible discovery channels. Here we discuss how the extended Higgs sector may be directly probed.


Amongst the twin states the mirror Higgs, which is a SM singlet, can be singly produced via its mixing with the SM Higgs, and accessible at present and future colliders.  In order to describe the main phenomenology, we focus on the linearised TH model \cite{Craig:2015pha}, where the scalar potential consists of only two degrees of freedom, the 125 GeV Higgs and the mirror Higgs $\sigma$. Notice that the presence of this extra singlet is a feature of any TH construction, and therefore it constitutes a natural signature. Using the label A for our SM sector and B for the twin sector, we have
\begin{equation}
V(H_A,H_B) =
\kappa \big(|H_A|^4 + |H_B|^4\big) + m^2 \big(|H_A|^2 - |H_B|^2\big) +
\lambda_* \Big( |H_A|^2 + |H_B|^2 - \frac{f_0^2}{2} \Big)^2.
\label{eq:V_TH_tot}
\end{equation}
The first term in the potential breaks the SU(4) global symmetry but leaves $Z_2$ intact, the second term softly breaks $Z_2$ as needed to achieve a separation between the two VEVs. The last one parametrises a spontaneous symmetry breaking SU(4)/SU(3). It is a combination of the spontaneous symmetry breaking and the soft breaking of the $Z_2$ symmetry that realises the TH mechanism. 

From a phenomenological point of view, the relevant parameters are $\lambda_*$ and $f_0$, where the size of $\lambda_*$ is required to be small, or the mirror Higgs will get a mass ($m_\sigma \sim \sqrt{\lambda_*} f_0$) of the order of the cut-off of the model invalidating our phenomenological picture. The particle $\sigma$ is often called the radial-mode of the corresponding symmetry breaking pattern.

The two Higgs doublets $H_A$ and $H_B$ are charged under the SM and twin weak interactions, respectively. Therefore, in the unitary gauge, six Goldstone bosons are eaten by the gauge bosons of the two sectors, leaving only two scalar degrees of freedom the SM Higgs, $h$, and $\sigma$. 
In the interaction basis they have a mass mixing. Trading two of the four parameters for the electro-weak VEV and Higgs mass, one can compute the mixing parameter between the two states
\begin{equation}\label{singammaTWIN}
\sin^2\gamma= \frac{v^2}{f^2}- \frac{m_h^2}{m_\sigma^2 -m_h^2}\Big(1- 2 \frac{v^2}{f^2}\Big),
\end{equation}
where $m_h$ and $m_\sigma$ are the physical masses, while in terms of the parameters in eq. \eqref{eq:V_TH_tot}, $v^2 = \frac{\kappa  \lambda_*\,f_0^2 - (\kappa+2  \lambda_*)\,m^2}{\kappa(\kappa + 2 \lambda_*)}$, $f^2 = f_0^2 \frac{2 \lambda_*}{2\lambda_*+\kappa}$.

This is very similar to the spectrum of a simple singlet-extension of the SM, given that the mixing angle enters the $\sigma$ signal strengths, and it also controls the leading and model independent contributions to the electro-weak $S$ and $T$ parameters. Differently from the NMSSM or other weakly coupled extensions, however, in this case the mixing angle does not vanish in the large-$m_\sigma$ regime, but it approaches a constant proportional to $v^2/f^2$. This is reminiscent of the pNGB nature of the Higgs in the Twin Higgs model.
This simple scenario can be meaningfully constrained by means of indirect and direct measurement. While precision Higgs measurements are only sensitive to $\sin\gamma$ in \eqref{singammaTWIN}, the direct searches of $\sigma$ depend also on the branching ratios in its possible decay channels. 

We now discuss direct searches for the radial mode $\sigma$.
Through the mixing~\eqref{singammaTWIN}, $\sigma$ inherits all the decay mode of a SM Higgs with mass $m_\sigma$. However, $\sigma$ has model independent couplings to the twin electro-weak gauge bosons.
We expect $\sigma$ to have a mass that scales with the parameter $f$, therefore in the $O(500-1000\ \mathrm{GeV})$ range, where all the decay channels are practically open. It mostly decays to vector bosons of the SM and twin sector, the latter contribute to the invisible decay width. In the large mass limit, the branching ratios of $\sigma$ are fixed by the equivalence theorem to be
\begin{equation}
\mathrm{BR}_{\mathrm{twin}-VV}\simeq \frac{3}{7},\quad \mathrm{BR}_{hh}\simeq\mathrm{BR}_{ZZ}\simeq \frac{1}{2}\mathrm{BR}_{WW} \simeq \frac{1}{7}.
\end{equation}
The above equation has two immediate consequences: $i)$ the dominant decay channel is diboson (including double Higgs), as expected for an electro-weak singlet in the TeV region;
$ii)$ the additional invisible decay channel to twin dibosons dilutes the branching fractions for the visible channels, and contributes to the widening of the width (especially in the large mass limit). $i)$ and $ii)$, together with the scaling of $m_\sigma$, then suggest that the largest impact of direct searches is expected for weakly coupled scenarios, with $m_\sigma \sim f$, in a region of moderate $f$.

These considerations are reflected in Figure \ref{fig:twin}, where, in the plane of the only two free parameters, $(f,m_\sigma)$, we show the impact of Higgs coupling determination and recast of direct searches in the diboson channel. For reference, we fix the invisible branching ratio to its asymptotic value of 3/7.  As expected, direct searches will provide a strong probe of the scenario considered for $m_\sigma \sim f$, while for much higher values indirect constraints are expected to dominate\footnote{In this Twin Higgs model, the trilinear Higgs coupling $g_{hhh}$ is fixed in terms of $m_\sigma$ and $f$, and does not substantially deviate from its SM value.}. This is an important example of complementarity between proposed future lepton and hadron colliders, which together will be able to probe the natural Twin Higgs parameter space with TeV-scale mirror Higgs vevs and masses.

\begin{figure}[t]
\begin{center}
\includegraphics[width=.495\textwidth]{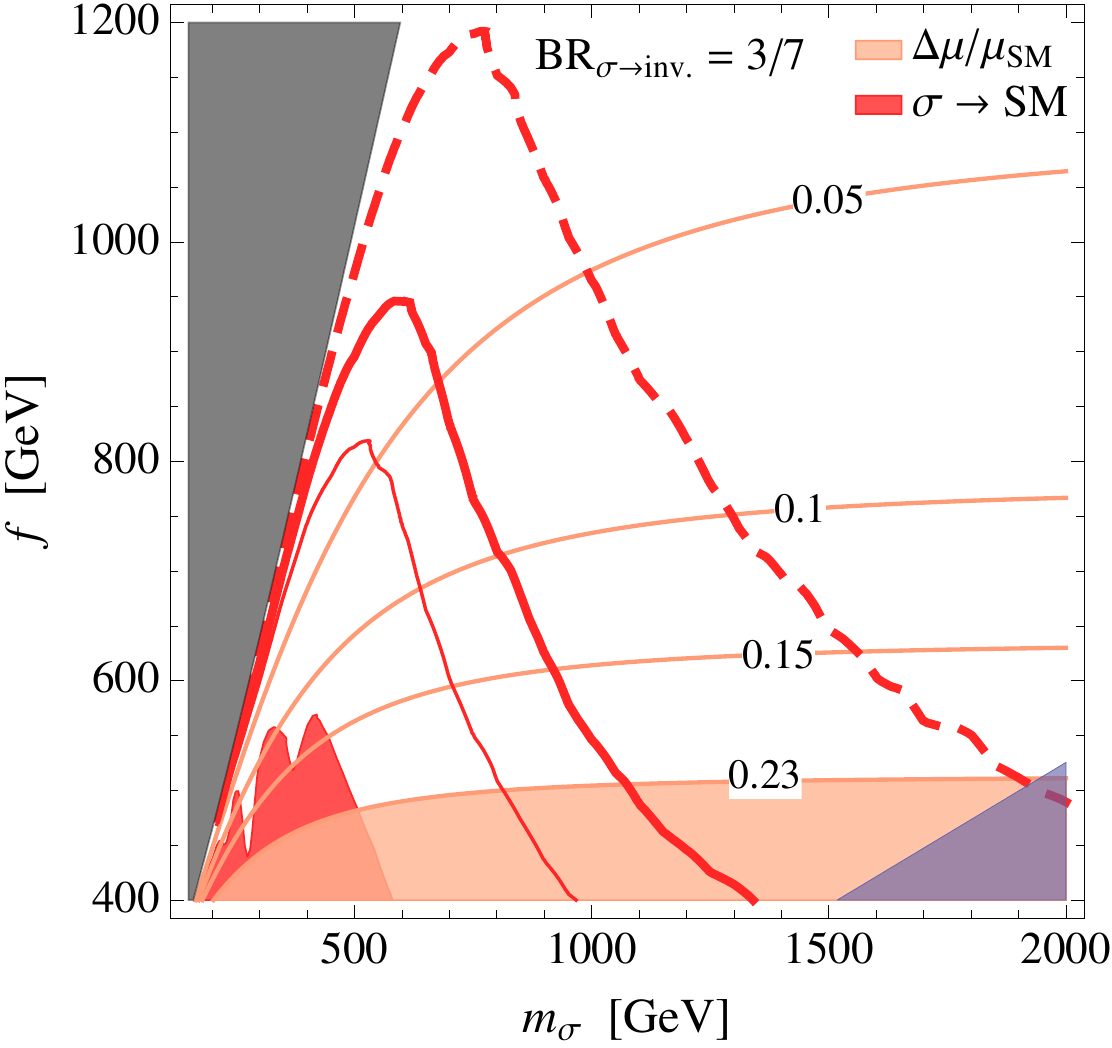}\hfill%
\includegraphics[width=.495\textwidth]{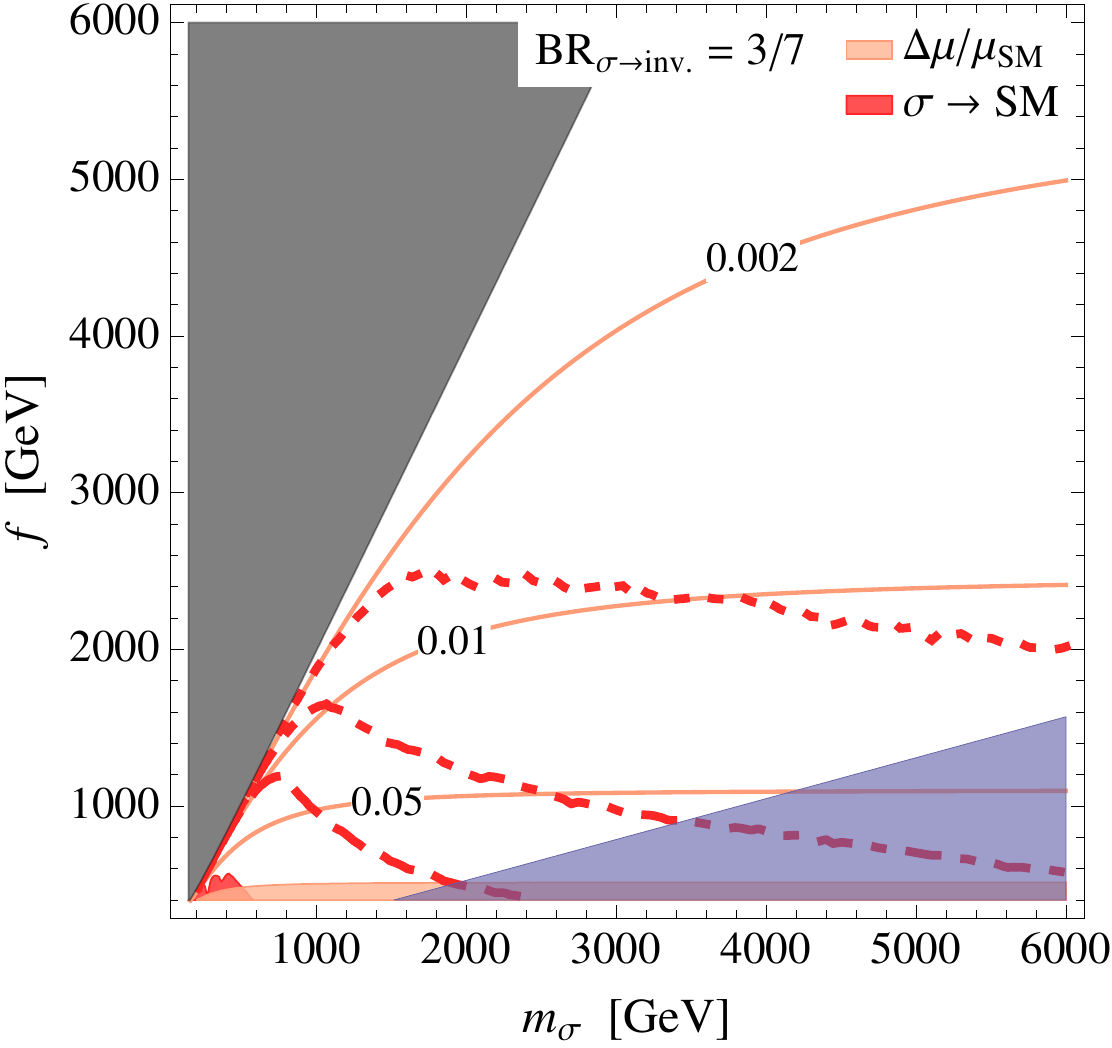}
\caption{\label{fig:twin}\small Parameter space of the scalar singlet in Twin Higgs scenarios in the plane of the mass of the radial mode $m_\sigma$ and the mirror Higgs vev $f$.
Pink lines: contours of $s_\gamma^2$. A lepton collider like ILC250 (TLEP) will likely be able to exclude $s_\gamma^2$ at the 0.05 (0.008) level \cite{Dawson:2013bba, Gomez-Ceballos:2013zzn}. Red lines: expected reach of the LHC13 (continuous thin), LHC14 (continuous), HL-LHC (dashed), HE-LHC (dot-dashed), and FCC-hh with 3~ab$^{-1}$ (dotted). Shaded regions: excluded at 95\% C.L. by direct searches (red), excluded by Higgs couplings (pink), $\Gamma_\sigma > m_\sigma$ (blue), unphysical parameters (grey). BR$_{\phi \to hh}$ fixed to its asymptotic value 3/7 for reference. Figure taken from ref. \cite{Buttazzo:2015bka}.}
\end{center}
\end{figure}



\clearpage
\subsubsection*{Acknowledgements}
The research of D.Buttazzo was supported in part by the Swiss
National Science Foundation (SNF) under contract 200021-159720.
The work of C.-Y.C is supported by NSERC, Canada. Research at the
Perimeter Institute is supported in part by the Government of Canada
through NSERC and by the Province of Ontario through MEDT.
D.C. is supported by National Science Foundation grant
No. PHY-1315155 and the Maryland Center for Fundamental Physics.
The work of Q-H.C. is supported in part by the National Science
Foundation of China under Grand No. 11175069, No. 11275009 and
No. 11422545.
NG was supported by the Swiss National Science Foundation under
contract PZ00P2\_154829. 
TG is supported by the DOE grant DE-SC0011842 at the University of
Minnesota.
AK is supported by the British Science and Technology Facilities 
Council and by the Buckee Scholarship at Merton College.
The work of M.L.M. is supported by the European Research
Council advanced grant 291377 {\it ``LHCtheory'': Theoretical predictions and
analyses of LHC physics: advancing the precision frontier}. 
The work of R.N.M. is supported in part by the US National
Science Foundation Grant No. PHY-1315155.
F.P. is supported by the U.S. DOE grants DE-FG02-91ER40684 and
DE-AC02-06CH11357.
MP is supported by the U.S. National Science Foundation grant
PHY-1316222.
M.J.R-M. was supported in part under U.S. Department of Energy
contract DE-SC0011095. 
F.S.~is supported by the European Research Council ({\sc ERC}) under the EU
Seventh Framework Programme (FP7/2007-2013)/{\sc ERC} Starting Grant (agreement
n.\ 278234 --- `{\sc NewDark}' project).
The work of M.S. is supported by the National Fund for Scientific Research
(F.R.S.-FNRS Belgium), by the IISN ``MadGraph'' convention 4.4511.10,
by the IISN ``Fundamental interactions'' convention 4.4517.08, and in
part by the Belgian Federal Science Policy Office through the
Interuniversity Attraction Pole P7/37.
The work of P.T. has received funding from the European
Union Seventh Framework programme for research and innovation under
the Marie Curie grant agreement N. 609402-2020 researchers: Train
to Move (T2M).
The work of TMPT is supported in part by NSF Grant No.~PHY-1316792
The work of P.S.B.D. was supported by the DFG with grant RO
 2516/5-1. Y.C.Z. would like to thank the IISN and Belgian
 Science Policy (IAP VII/37) for support. 
 The work of G.B., J.C., J.G., T.J.,
 T.S., R.S. is supported  by Department of Science \&
 Technology, Government of INDIA under the Grant Agreement number
 IFA12-PH-34 (INSPIRE Faculty Award),  Polish National Center of
 Science (NCN) under the Grant Agreement number
 DEC-2013/11/B/ST2/04023 and Science and Engineering Research Canada
 (NSERC).
The work of S-F.S. and F.K. was supported by US Department of Energy
under Grant DE-FG02-04ER-41298. F.K. also acknowledges support from
the Fermilab Graduate Student Research Program in Theoretical Physics
operated by Fermi Research Alliance, LLC under Contract
No. DE-AC02-07CH11359 with the United States Department of Energy.
The work of J.K.B. is supported by a Rhodes Scholarship.
D.B., J.A.F. and C.I. are supported by the STFC.
J.Rojo and N. H. are supported by an European Research Council Starting
Grant ``PDF4BSM''. 
J.Rojo is supported by an STFC Rutherford Fellowship and Grant
ST/K005227/1 and ST/M003787/1. 
The research of WA, WB, and GZ is supported by the ERC consolidator
grant 614577 ``HICCUP: High Impact Cross Section Calculations for
Ultimate Precision''.
For part of this research the computing resources from the
Rechenzentrum Garching were used.

\bibliographystyle{report}
\bibliography{report}

\end{document}